# A Class of Models with the Potential to Represent Fundamental Physics

Stephen Wolfram

*A class of models intended to be as minimal and structureless as possible is introduced. Even in cases with simple rules, rich and complex behavior is found to emerge, and striking correspondences to some important core known features of fundamental physics are seen, suggesting the possibility that the models may provide a new approach to finding a fundamental theory of physics*

# 1 | Introduction

Quantum mechanics and general relativity—both introduced more than a century ago—have delivered many impressive successes in physics. But so far they have not allowed the formulation of a complete, fundamental theory of our universe, and at this point it seems worthwhile to try exploring other foundations from which space, time, general relativity, quantum mechanics and all the other known features of physics could emerge.

The purpose here is to introduce a class of models that could be relevant. The models are set up to be as minimal and structureless as possible, but despite the simplicity of their construction, they can nevertheless exhibit great complexity and structure in their behavior. Even independent of their possible relevance to fundamental physics, the models appear to be of significant interest in their own right, not least as sources of examples amenable to rich analysis by modern methods in mathematics and mathematical physics.

But what is potentially significant for physics is that with exceptionally little input, the models already seem able to reproduce some important and sophisticated features of known fundamental physics—and give suggestive indications of being able to reproduce much more.

Our approach here is to carry out a fairly extensive empirical investigation of the models, then to use the results of this to make connections with known mathematical and other features of physics. We do not know *a priori* whether any model that we would recognize as simple can completely describe the operation of our universe—although the very existence of physical laws does seem to indicate some simplicity. But it is basically inevitable that if a simple model exists, then almost nothing about the universe as we normally perceive it—including notions like space and time—will fit recognizably into the model.

And given this, the approach we take is to consider models that are as minimal and structureless as possible, so that in effect there is the greatest opportunity for the phenomenon of emergence to operate. The models introduced here have their origins in network-based models studied in the 1990s for [1], but the present models are more minimal and structureless. They can be thought of as abstracted versions of a surprisingly wide range of types of mathematical and computational systems, including combinatorial, functional, categorical, algebraic and axiomatic ones.

In what follows, sections 2 through 7 describe features of our models, without specific reference to physics. Section 8 discusses how the results of the preceding sections can potentially be used to understand known fundamental features of physics.

An informal introduction to the ideas described here is given in [2].

# 2 | Basic Form of Models

## 2.1 Basic Structure

At the lowest level, the structures on which our models operate consist of collections of relations between identical (but labeled) discrete elements. One convenient way to represent such structures is as graphs (or, in general, hypergraphs). The elements are the nodes of the graph or hypergraph. The relations are the (directed) edges or hyperedges that connect these elements.

For example, the graph

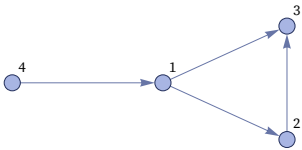

corresponds to the collection of relations

{{1, 2}, {1, 3}, {2, 3}, {4, 1}}

The order in which these relations are stated is irrelevant, but the order in which elements appear within each relation is considered significant (and is reflected by the directions of the edges in the graph). The specific labels used for the elements (here 1, 2, 3, 4) are arbitrary; all that matters is that a particular label always refer to the same element.

## 2.2 First Example of a Rule

The core of our models are rules for rewriting collections of relations. A very simple example of a rule is:

{{$x$, $y$}} → {{$x$, $y$}, {$y$, $z$}}

Here $x$, $y$ and $z$ stand for any elements. (The elements they stand for need not be distinct; for example, $x$ and $y$ could both stand for the element 1.) The rule states that wherever a relation that matches {$x$,$y$} appears, it should be replaced by {{$x$,$y$},{$y$,$z$}}, where $z$ is a new element. So given {{1, 2}} the rule will produce {{1,2},{2,□}} where □ is a new element. The label for the new element could be anything—so long as it is distinct from 1 and 2. Here we will use 3, so that the result of applying the rule to {{1,2}} becomes:

{{1, 2}, {2, 3}}

If one applies the rule again, it will now operate again on {1,2}, and also on {2,3}. On {1,2} it again gives {{1,2},{2,□}}, but now the new node □ cannot be labeled 3, because that label is already taken—so instead we will label it 4. When the rule operates on {2,3} it gives {{2,3},{3,□}}, where again □ is a new node, which can now be labeled 5. Combining these gives the final result:

{{1, 2}, {2, 4}, {2, 3}, {3, 5}}

(We have written this so that the results from {{1,2}} are followed by those from {{2,3}}—but there is no significance to the order in which the relations appear.)

In graphical terms, the rule we have used is:

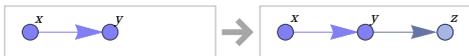

and the sequence of steps is:

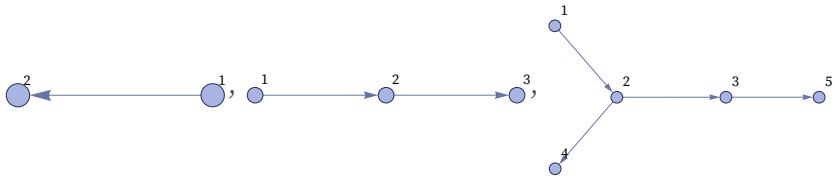

It is important to note that all that matters in these graphs is their connectivity. Where nodes are placed on the page in drawing the graph has no fundamental significance; it is usually just done to make the graphs as easy to read as possible.

Continuing to apply the same rule for three more steps gives:

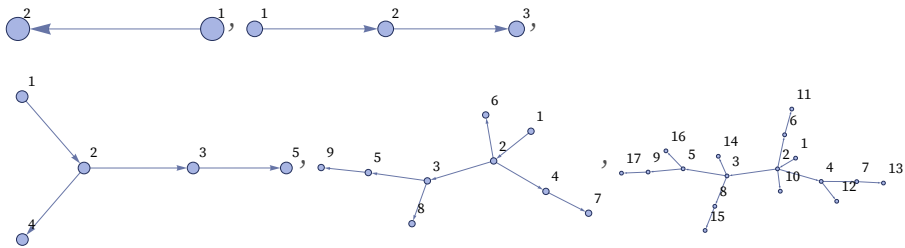

Laying out nodes differently makes it easier to see some features of the graphs:

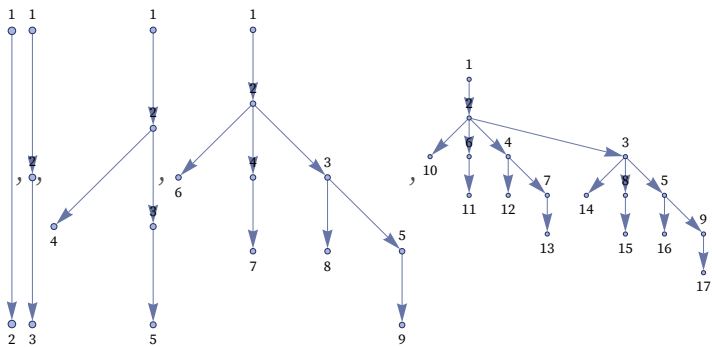

Continuing for a few more steps with the original layout gives the result:

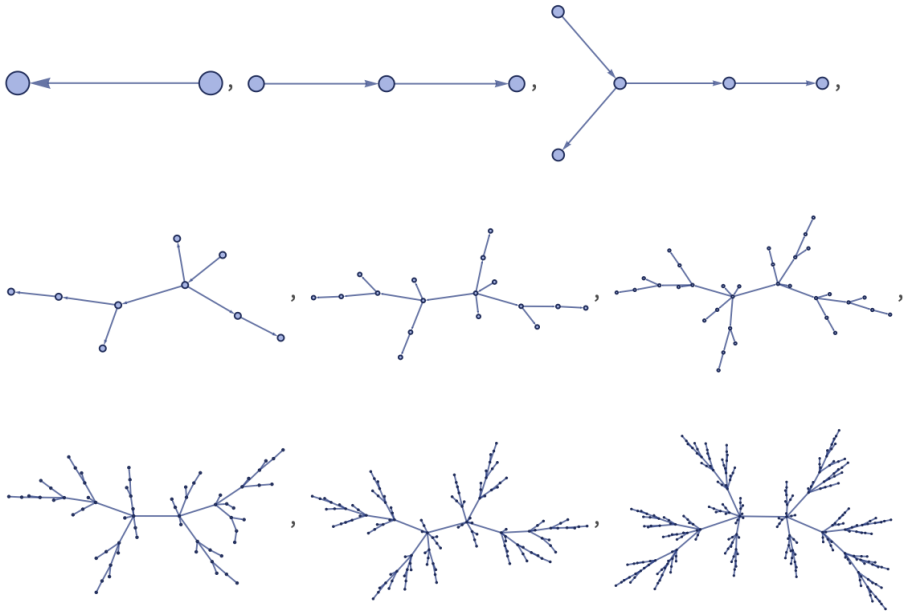

Showing the last 3 steps with the other layout makes it a little clearer what is going on:

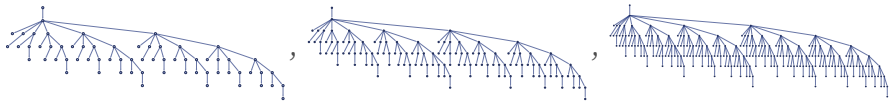

The rule is generating a binomial tree, with $2^n$ edges (relations) and $2^{n+1}$ nodes (distinct elements) at step $n$ (and with Binomial[$n$, $s$ –1] nodes at level $s$).

## 2.3  A Slightly Different Rule

Since order within each relation matters, the following is a different rule:

{{x, y}} → {{z, y}, {y, x}}

This rule can be represented graphically as:

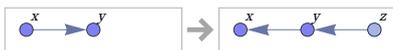

Like the previous rule, running this rule also gives a tree, but now with a somewhat different structure:

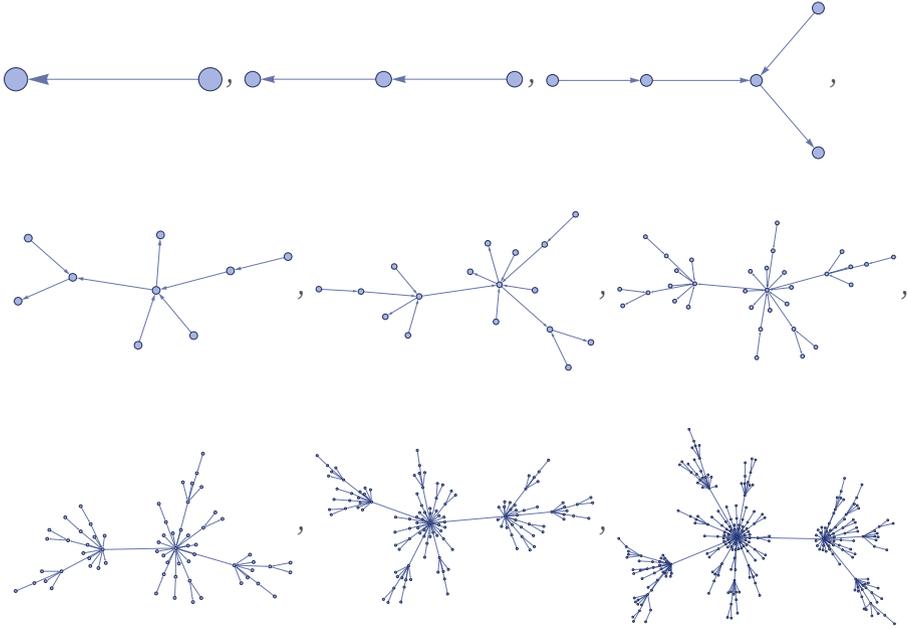

With the other rendering from above, the last 3 steps here are:

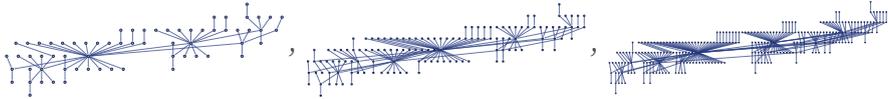

## 2.4 Self-Loops

A relation can contain two identical elements, as in {0,0}, corresponding to a self-loop in a graph. Starting our first rule from a single self-loop, the self-loop effectively just stays marking the original node:

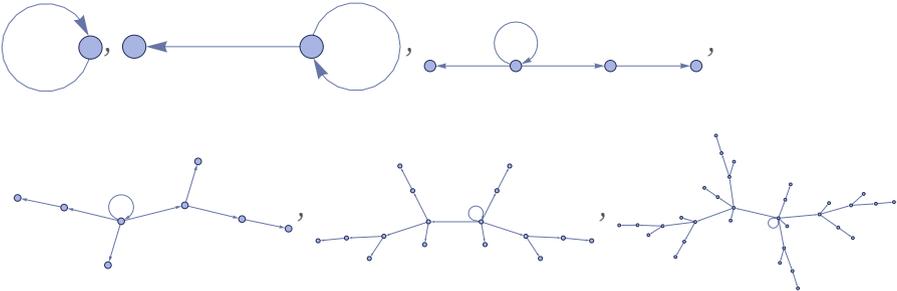

However, with for example the rule:

{{x, y}} → {{y, z}, {z, x}}

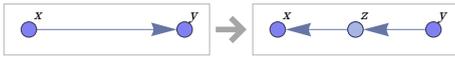

the self-loop effectively "takes over" the system, "inflating" to a $2^n$ – gon:

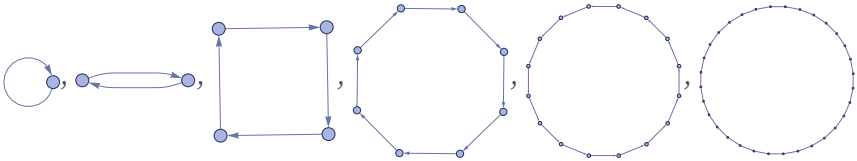

The rule can also contain self-loops. An example is

{{x, x}} → {{y, y}, {y, y}, {x, y}}

represented graphically as:

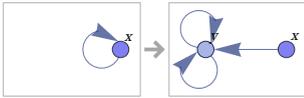

Starting from a single self-loop, this rule produces a simple binary tree:

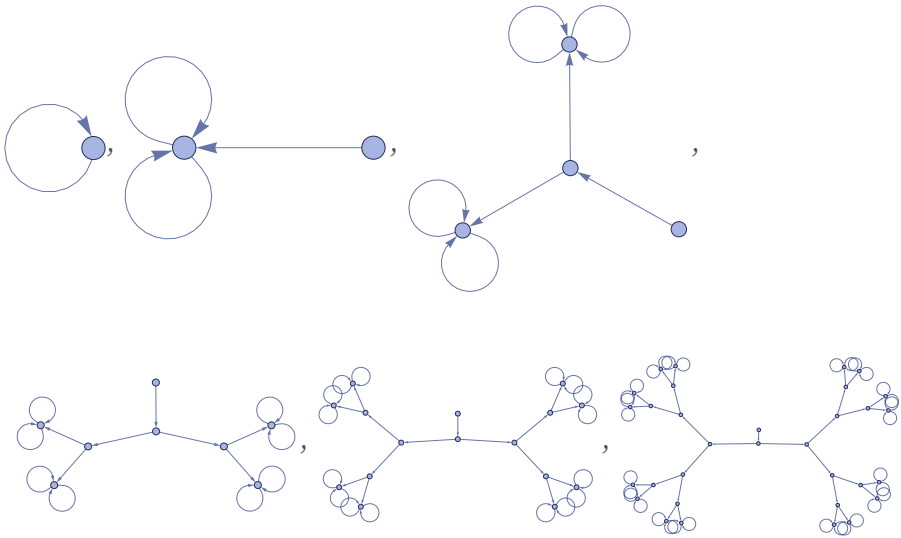

## 2.5 Multiedges

Rules can involve several copies of the same relation, corresponding to multiedges in a graph. A simple example is the rule:

{{x, y}} → {{x, z}, {x, z}, {y, z}}

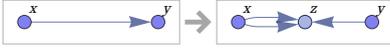

Running this rule produces a structure with $3^n$ edges and $\frac{1}{6}(3^n + 3)$ nodes at step $n$:

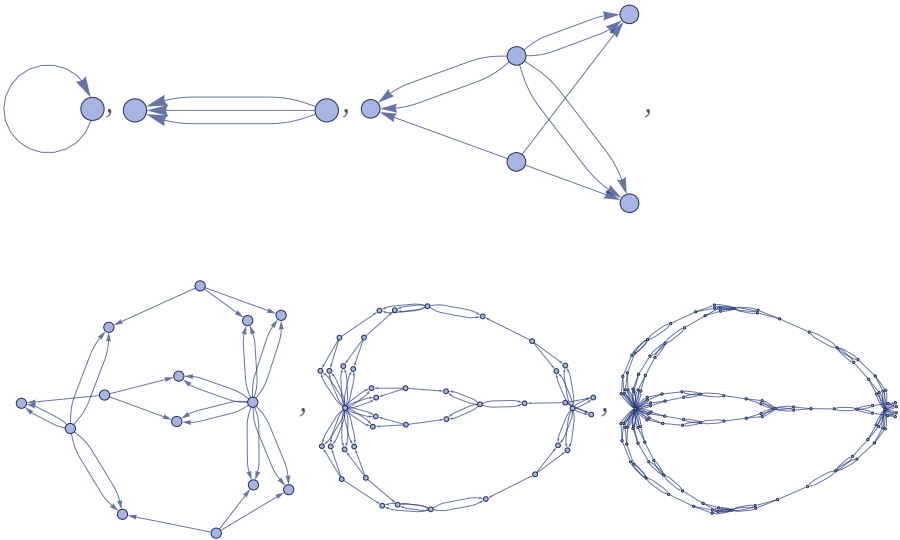

Rules can both create and destroy multiedges. The rule

{{x, y}} → {{x, z}, {z, w}, {y, z}}

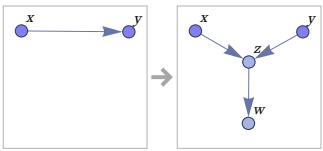

generates a multiedge after one step, but then destroys it:

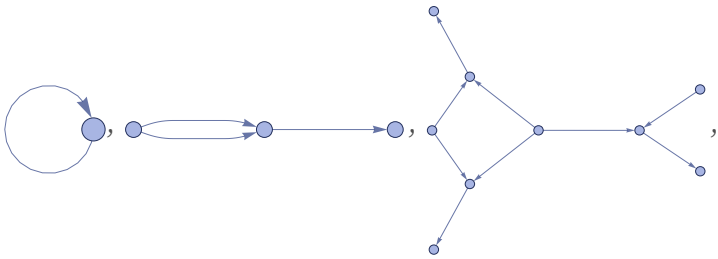

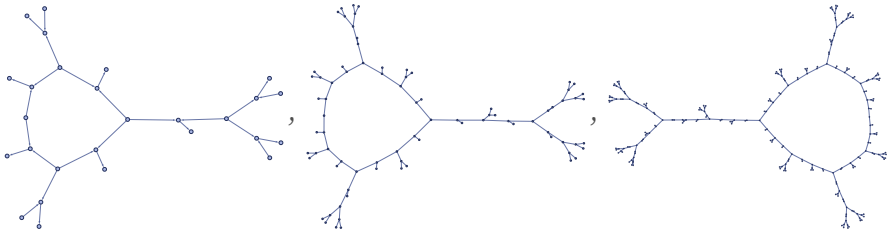

## 2.6 The Representation of Rules

The examples we have discussed so far all contain only relations involving two elements, which can readily be represented as ordinary directed graphs. But in the class of models we consider, it is also possible to have relations involving other numbers of elements, say three.

As an example, consider:

{{1, 2, 3}, {3, 4, 5}}

which consists of two ternary relations. Such an object can be represented as a hypergraph consisting of two ternary hyperedges:

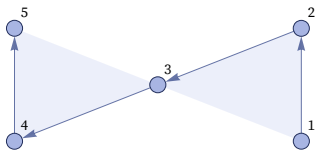

Because our relations are ordered, the hypergraph is directed, as indicated by the arrows around each hyperedge.

Note that hypergraphs can contain full or partial self-loops, as in the example of

{{1, 1, 1}, {1, 2, 3}, {3, 4, 4}}

which can be drawn as:

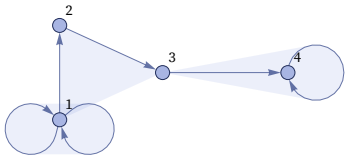

Rules can involve *k*-ary relations. Here is an example with ternary relations:

{{*x*, *y*, *z*}} → {{*x*, *y*, *w*}, {*y*, *w*, *z*}}

This rule can be represented as:

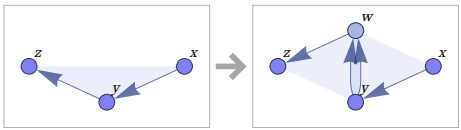

Starting from a single ternary self-loop, here are the first few steps obtained with this rule:

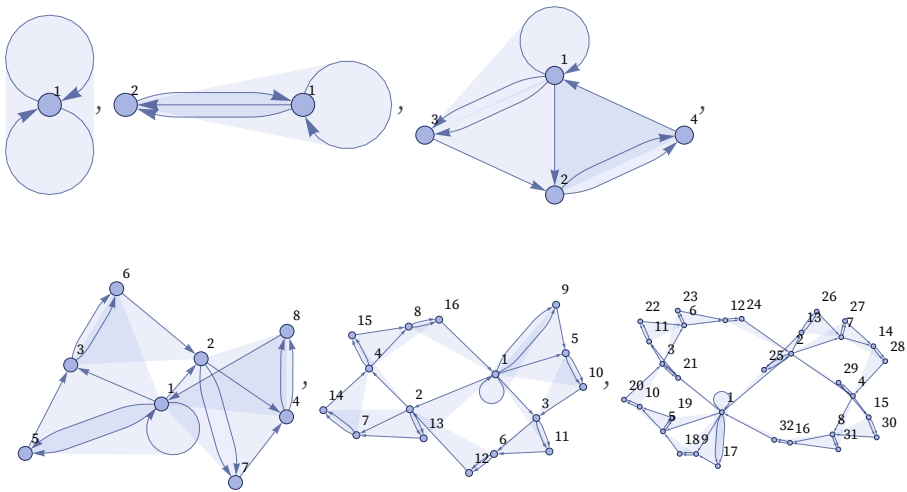

Continuing with this rule gives the following result:

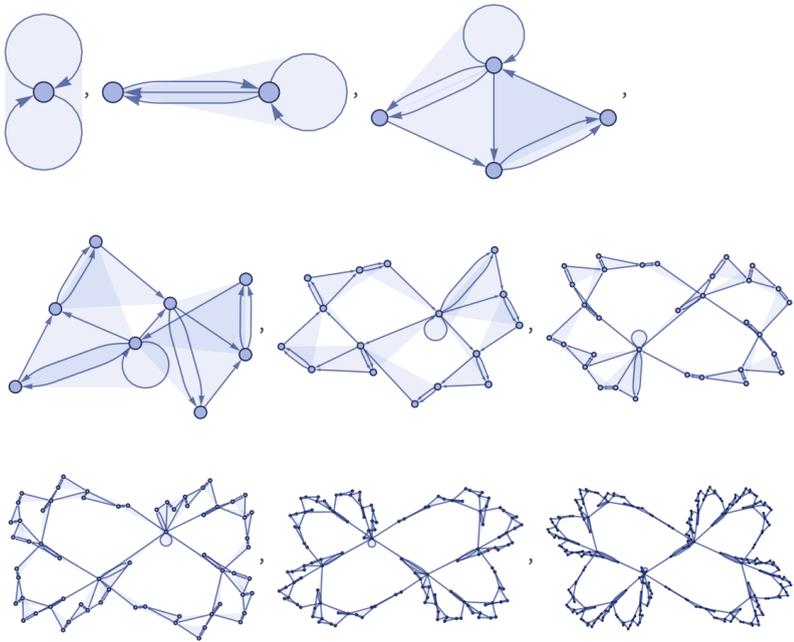

It is worth noting that in addition to having relations involving 3 or more elements, it is also possible to have relations with just one element. Here is an example of a rule involving unary relations:

{{x}} → {{x, y}, {y}, {y}}

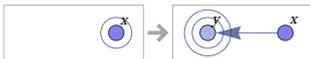

Starting from a unary self-loop, this rule leads to a binary tree with double-unary self-loops as leaves:

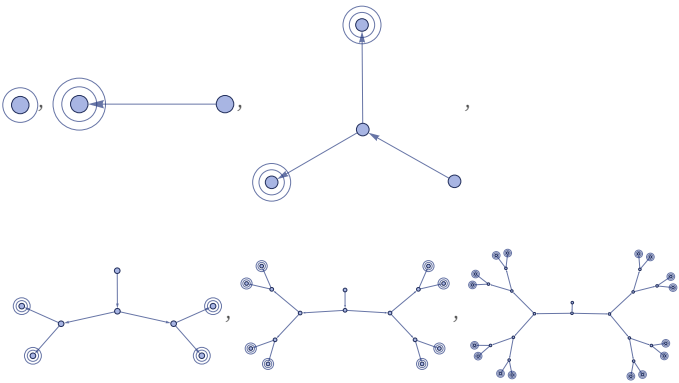

## 2.7 Rules Depending on More Than One Relation

A crucial simplifying feature of the rules we have considered so far is that they depend only on one relation, so that in a collection of relations, the rule can be applied separately to each relation (cf. [1:p82]). Put another way, this means that all the rules we have considered always transform single edges or hyperedges independently.

But consider a rule like:

$\{\{x, y\}, \{x, z\}\} \to \{\{x, y\}, \{x, w\}, \{y, w\}, \{z, w\}\}$

This can be represented graphically as:

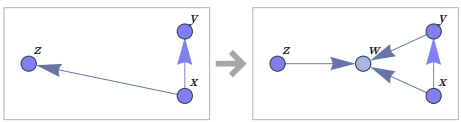

Here is the result of running the rule for several steps:

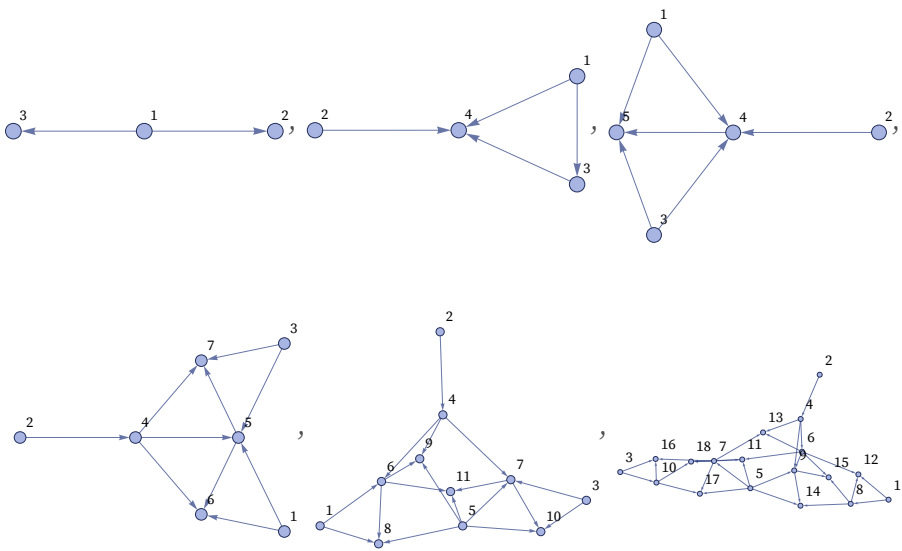

Here is the result for 10 steps:

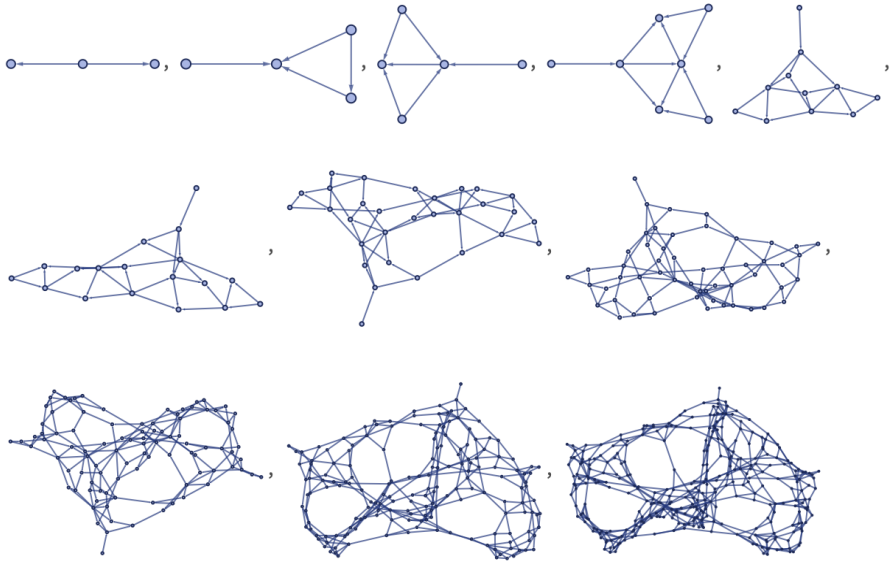

Despite the simplicity of the underlying rule, the structure that is built (here after 15 steps, and involving 6974 elements and 13,944 relations) is complex:

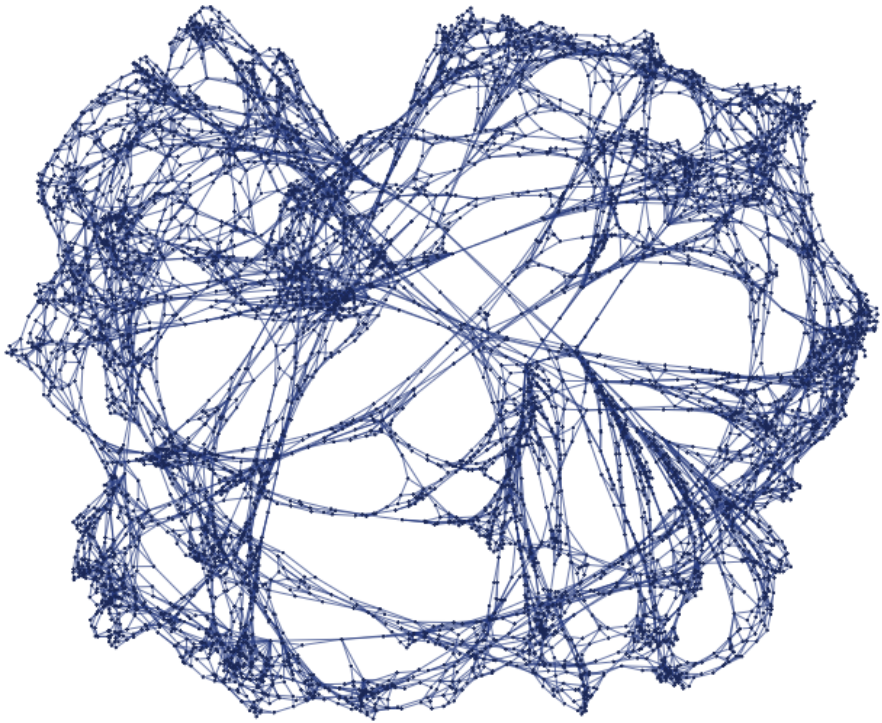

In getting this result, we are, however, glossing over an important issue that will occupy us extensively in later sections, and that potentially seems intimately connected with foundational features of physics.

With a rule that just depends on a single relation, there is in a sense never any ambiguity in where the rule should be applied: it can always separately be used on any relation. But with a rule that depends on multiple relations, ambiguity is possible.

Consider the configuration:

{{1, 2}, {1, 3}, {1, 4}, {1, 4}}

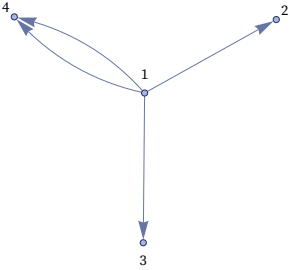

The rule

{{$x$, $y$}, {$x$, $z$}} → {{$x$, $y$}, {$x$, $w$}, {$y$, $w$}, {$z$, $w$}}

can be applied here in two distinct, but overlapping, ways. First, one can take:

{$x$ → 1, $y$ → 2, $z$ → 3}

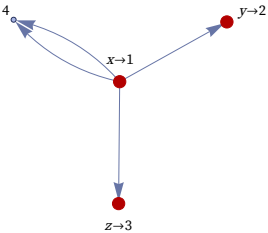

giving the result:

{{1, 3}, {1, 5}, {2, 5}, {3, 5}, {1, 4}, {1, 4}}

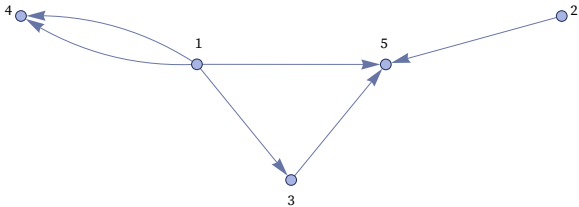

But one can equally well take:

$\{x \to 1, y \to 3, z \to 4\}$

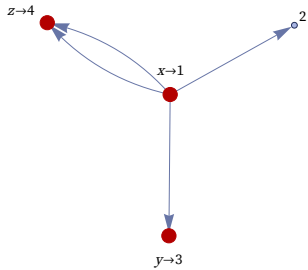

giving the inequivalent result:

$\{\{1, 2\}, \{1, 4\}, \{1, 5\}, \{3, 5\}, \{4, 5\}, \{1, 4\}\}$

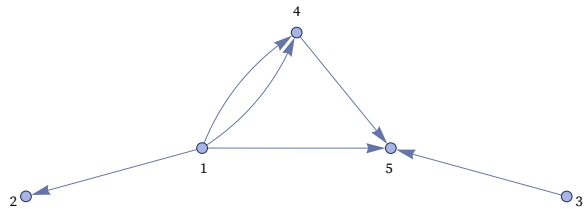

With a rule that just depends on a single relation, there is an obvious way to define a single complete step in the evolution of the system: just make it correspond to the result of applying the rule once to each relation. But when the rule involves multiple relations, we have seen that there can be ambiguity in how it is applied (cf. [1:p501]), and one consequence of this is that there is no longer an obvious unique way to define a single complete step of evolution. For our purposes at this point, however, we will take each step to be what is obtained by scanning the configuration of the system, and finding the largest number of non-overlapping updates that can be made (cf. [1:p487]). In other words, in a single step, we update as many edges (or hyperedges) as possible, while never updating any edge more than once.

For now, this will give us a good indication of what kind of typical behavior different rules can produce. Later, we will study the results of all possible updating orders. And while this will not affect our basic conclusions about typical behavior, it will have many important consequences for our understanding of the models presented here, and their potential relevance to fundamental physics.

## 2.8 Termination

We have seen that there can be several ways to apply a particular rule to a configuration of one of our systems. It is also possible that there may be no way to apply a rule. This can happen trivially if the evolution of the system reduces the number of relations it contains, and at some point there are simply no relations left. It can also happen if the rule involves, say, only *k*-ary relations, but there are no *k*-ary relations in the configuration of the system.

In general, however, a rule can continue for any number of steps, but then get to a configuration where it can no longer apply. The rule below, for example, takes 9 steps to go from {{0,0,0},{0,0}} to a configuration that contains only a single 3-edge, and no 2-edges that match the pattern for the rule:

{{*x*, *y*, *z*}, {*u*, *x*}} → {{*x*, *u*, *v*}, {*z*, *y*}, {*z*, *u*}}

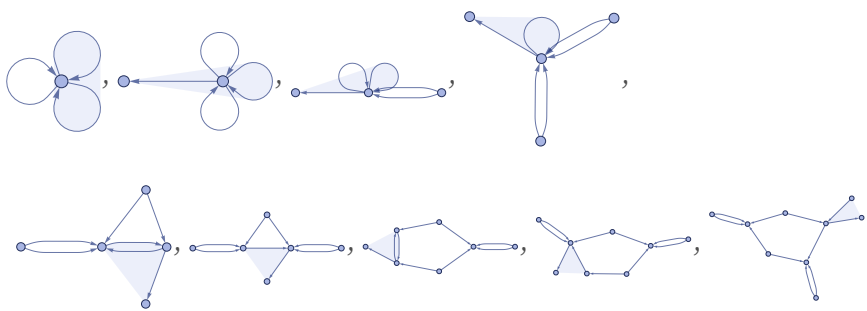

It can be arbitrarily difficult to predict if or when a particular rule will "halt", and we will see later that this is to be expected on the basis of computational irreducibility [1:12.6].

## 2.9 Connectedness

All the rules we have seen so far maintain connectedness. It is, however, straightforward to set up rules that do not. An obvious example is:

{{*x*, *y*}} → {{*y*, *y*}, {*x*, *z*}}

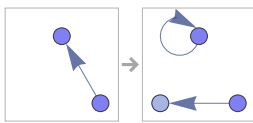

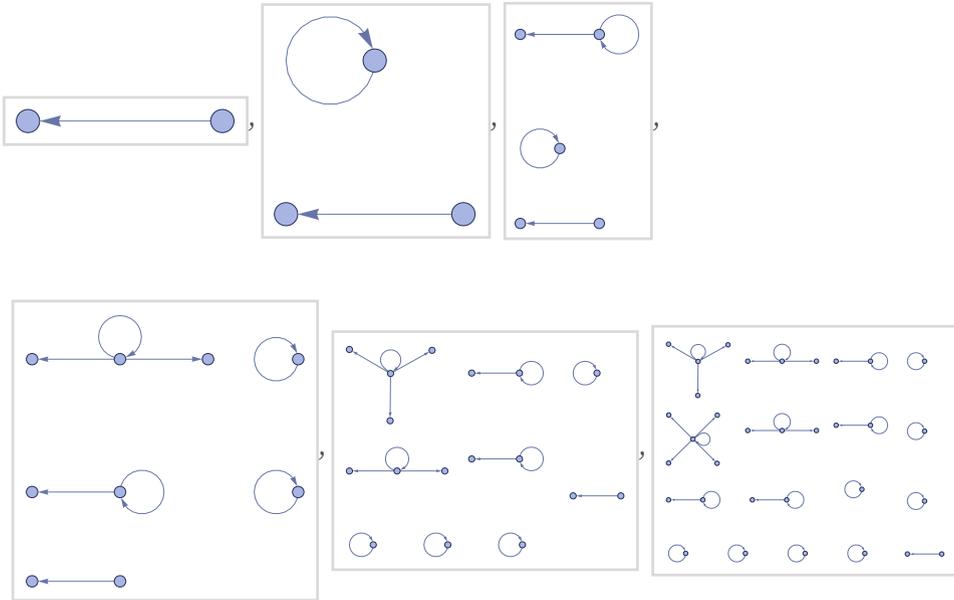

At step $n$, there are $2^{n+1}$ components altogether, with the largest component having $n + 1$ relations.

Rules that are themselves connected can produce disconnected results:

$\{\{x, y\}\} \to \{\{x, x\}, \{z, x\}\}$

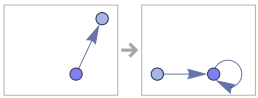

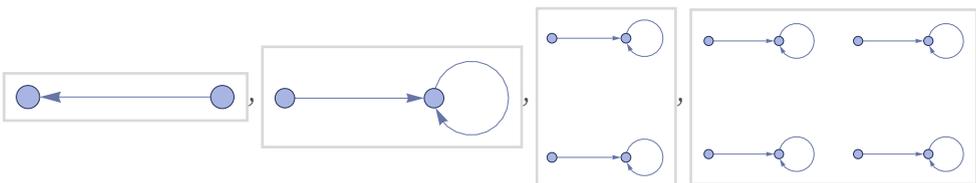

Rules whose left-hand sides are connected in a sense operate locally on hypergraphs. But rules with disconnected left-hand sides (such as $\{\{x\},\{y\}\}\to\{\{x,y\}\}$) can operate non-locally and in effect knit together elements from anywhere—though such a process is almost inevitably rife with ambiguity.

# 3 | Typical Behaviors

## 3.1 The Representation of Rules

Having introduced our class of models, we now begin to study the general distribution of behavior in them. Like with cellular automata [1:c2] and other kinds of systems defined by what can be thought of as simple computational rules [1:c3, c4, c5], we will find great diversity in behavior as well as unifying trends.

Any one of our models is defined by a rule that specifies transformations between collections of relations. It is convenient to introduce the concept of the "signature" of a rule, defined as the number of relations of each arity that appear on the left and right of each transformation.

Thus, for example, the rule

{{$x$, $y$}, {$x$, $z$}} → {{$x$, $y$}, {$x$, $w$}, {$y$, $w$}, {$z$, $w$}}

has signature $2_2 \to 4_2$ (and involves a total of 4 distinct elements). Similarly, the rule

{{$a$, $a$, $b$}, {$c$, $d$}} → {{$b$, $b$, $d$}, {$a$, $e$, $d$}, {$b$, $b$}, {$c$, $a$}}

has signature $1_3 1_2 \to 2_3 2_2$ (and involves 5 distinct elements).

So far, we have always used letters to indicate elements in a rule, to highlight the fact that these are merely placeholders for the particular elements that appear in the configuration to which the rule is applied. But in systematic studies it is often convenient just to use integers to represent elements in rules, even though these are still to be considered placeholders (or pattern variables), not specific elements. So as a result, the rule just mentioned can be written:

{{1, 1, 2}, {3, 4}} → {{2, 2, 4}, {1, 5, 4}, {2, 2}, {3, 1}}

It is important to note that there is a certain arbitrariness in the way rules are written. The names assigned to elements, and the order in which relations appear, can both be rearranged without changing the meaning of the rule. In general, determining whether two presentations of a rule are equivalent is essentially a problem of hypergraph isomorphism. Here we will give rules in a particular canonical form obtained by permuting names of elements and orders of relations in all possible ways, numbering elements starting at 1, and using the lexicographically first form obtained. (This form has the property that **DeleteDuplicates**[**Flatten**[{*lhs*,*rhs*}]] is always a sequence of successive integers starting at 1.)

Thus for example, both

{{1, 1}, {2, 4, 5}, {7, 5}} → {{3, 8}, {2, 7}, {5, 4, 1}, {4, 6}, {5, 1, 7}}

and

{{7, 3}, {4, 4}, {8, 5, 3}} → {{3, 4, 7}, {5, 6}, {8, 7}, {3, 5, 4}, {1, 2}}

would be given in the canonical form

{{1, 2, 3}, {4, 4}, {5, 3}} → {{3, 2, 4}, {3, 4, 5}, {1, 5}, {2, 6}, {7, 8}}

From the canonical form, it is possible to derive a single integer to represent the rule. The basic idea is to get the sequence **Flatten**[{*lhs,rhs*}] (in this case {1, 2, 3, 4, 4, 5, 3, 3, 2, 4, 3, 4, 5, 1, 5, 2, 6, 7, 8} ) and then find out (through a generalized pairing or "tupling" function [3]) where in a list of all possible tuples of this length this sequence occurs [4]. In this example, the result is 310528242279018009.

But unlike for systems like cellular automata [5][1:p53][6] or Turing machines [1: p888][7] where it is straightforward to set up a dense sequence of rule numbers, only a small fraction of integers constructed like this represent inequivalent rules (most correspond to non-canonical rule specifications).

In addition—for example for applications in physics—one is usually not even interested in all possible rules, but instead in a small number of somehow "notable" rules. And it is often convenient to refer to such notable rules by "short codes". These can be obtained by hashing the canonical form of the rule, but since hashes can collide, it is necessary to maintain a central repository to ensure that short codes remain unique. In our Registry of Notable Universes [8], the rule just presented has short code wm8678.

## 3.2 The Number of Possible Rules

Given a particular signature, one may ask how many distinct possible canonical rules there are with that signature. As a first step, one can ask how many distinct elements can occur in the rule. If the rule signature has terms $n_{i k_i}$ on both left and right, the maximum conceivable number of distinct elements is $\sum n_i k_i$. (For example, a possible canonical $2_2 \to 2_2$ rule is {{1,2},{3,4}}→{{5,6},{7,8}}.)

But for many purposes we will want to impose connectivity constraints on the rule. For example, we may want the hypergraph corresponding to the relations on the left-hand side of the rule to be connected [9], and for elements in these relations to appear in some way on the right. Requiring this kind of "left connectivity" reduces the maximum conceivable number of distinct elements to $\sum_{i \in \text{LHS}} n_i(k_i-1) + \sum_{i \in \text{RHS}} n_i k_i$ (or 6 for $2_2 \to 2_2$). (If the right-hand side is also required to be a connected hypergraph, the maximum number of distinct elements is $1 + \sum n_i(k_i-1)$, or 5 for $2_2 \to 2_2$.)

Given a maximum number of possible elements m, an immediate upper bound on the number of rules is $m^{\sum n_i k_i}$. But this is usually a dramatic overestimate, because most rules are not canonical. For example, it would imply 1,679,616 left-connected $2_2 \to 2_2$ rules, but actually there are only 562 canonical such rules.

The following gives the number of left-connected canonical rules for various rule signatures (for $n_1 \to \text{anything}$ there is always only one inequivalent left-connected rule):

| $1_2 \to 1_2$ | 11 |
|---|---|
| $1_2 \to 2_2$ | 73 |
| $1_2 \to 3_2$ | 506 |
| $1_2 \to 4_2$ | 3740 |
| $1_2 \to 5_2$ | 28 959 |
| $2_2 \to 1_2$ | 64 |
| $2_2 \to 2_2$ | 562 |
| $2_2 \to 3_2$ | 4702 |
| $2_2 \to 4_2$ | 40 405 |
| $2_2 \to 5_2$ | 353 462 |
| $3_2 \to 1_2$ | 416 |
| $3_2 \to 2_2$ | 4688 |
| $3_2 \to 3_2$ | 48 554 |
| $4_2 \to 1_2$ | 3011 |
| $4_2 \to 2_2$ | 42 955 |
| $5_2 \to 1_2$ | 23 211 |

| $1_3 \to 1_3$ | 178 |
|---|---|
| $1_3 \to 2_3$ | 9373 |
| $1_3 \to 3_3$ | 637 568 |
| $1_3 \to 4_3$ | 53 644 781 |
| $1_3 \to 5_3$ | $5.4 \times 10^9$ |
| $2_3 \to 1_3$ | 8413 |
| $2_3 \to 2_3$ | 772 696 |
| $2_3 \to 3_3$ | 79 359 764 |
| $2_3 \to 4_3$ | $9.2 \times 10^9$ |
| $2_3 \to 5_3$ | $1.2 \times 10^{12}$ |
| $3_3 \to 1_3$ | 568 462 |
| $3_3 \to 2_3$ | $8.4 \times 10^7$ |
| $3_3 \to 3_3$ | $1.4 \times 10^{10}$ |
| $4_3 \to 1_3$ | $4.9 \times 10^7$ |
| $4_3 \to 2_3$ | $1.1 \times 10^{10}$ |
| $5_3 \to 1_3$ | $5.3 \times 10^9$ |

| $1_4 \to 1_4$ | 3915 |
|---|---|
| $1_4 \to 2_4$ | 2 022 956 |
| $1_4 \to 3_4$ | $1.7 \times 10^9$ |
| $1_4 \to 4_4$ | $2.1 \times 10^{12}$ |
| $1_4 \to 5_4$ | $\approx 4 \times 10^{15}$ |
| $2_4 \to 1_4$ | 1 891 285 |
| $2_4 \to 2_4$ | $2.3 \times 10^9$ |
| $2_4 \to 3_4$ | $3.5 \times 10^{12}$ |
| $2_4 \to 4_4$ | $\approx 9 \times 10^{15}$ |
| $2_4 \to 5_4$ | $\approx 3 \times 10^{19}$ |
| $3_4 \to 1_4$ | $1.6 \times 10^9$ |
| $3_4 \to 2_4$ | $3.8 \times 10^{12}$ |
| $3_4 \to 3_4$ | $\approx 1 \times 10^{16}$ |
| $4_4 \to 1_4$ | $2.1 \times 10^{12}$ |
| $4_4 \to 2_4$ | $\approx 9 \times 10^{15}$ |
| $5_4 \to 1_4$ | $\approx 4 \times 10^{15}$ |

Although the exact computation of these numbers seems to be comparatively complex, it is possible to obtain fairly accurate lower-bound estimates in terms of Bell numbers [10]. If one ignores connectivity constraints, the number of canonical rules is bounded below by BellB[∑$n_i\, k_i$]/∏$n_i$!. Here are some examples comparing the estimate with exact results both for the unconstrained and left-connected cases:

| | estimate | unconstrained | left–connected |
|---|---|---|---|
| $1_1 \to 2_1$ | 2.5 | 4 | 1 |
| $1_2 \to 2_2$ | 102 | 117 | 73 |
| $1_2 \to 3_2$ | 690 | 877 | 506 |
| $1_3 \to 2_3$ | 10 574 | 10 848 | 9373 |
| $2_2 \to 2_2$ | 1035 | 1252 | 562 |
| $2_2 \to 3_2$ | 9665 | 12 157 | 4702 |
| $2_2 \to 4_2$ | 87 783 | 117 121 | 40 405 |

Based on the estimates, we can say that the number of canonical rules typically increases faster than exponentially as either $n_i$ or $k_i$ increase. (For 5≤$n$≤10 874, one finds $2^n$ < BellB[$n$] < $2^{n \log n}$, and for larger $n$, $2^n$<BellB[$n$]<$n^n$.)

Note that given an estimate for unconstrained rules, an estimate for the number of left-connected rules can be found from the fraction of randomly sampled unconstrained rules that are left connected. For signature $1_p \to 1_q$, the number of unconstrained canonical rules is BellB[$p+q$], but given the constraint of left-connectedness there is only ever one canonical rule in this case. When there are no connectivity constraints, the number of canonical rules for signature $a \to b$ is the same as for signature $b \to a$. With the constraint of left connectivity, the number of $1_2 \to 5_2$ rules is slightly larger than $5_2 \to 1_2$ rules, because there are fewer constraints in the former case.

For any given signature, we can ask how many distinct elements occur in different canonical rules. Here are histograms for a few cases:

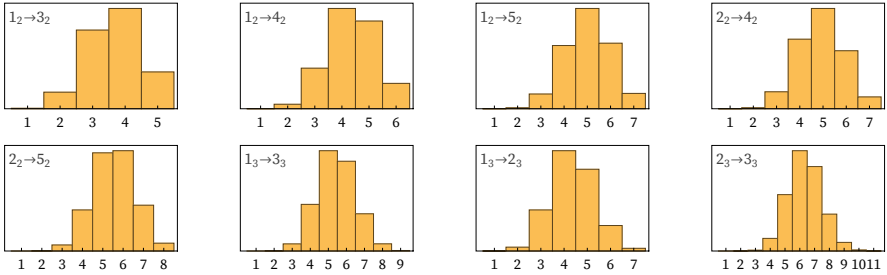

There are a number of features of rules that are important when using them in our models. First, if there are relations of arity $k$ on the left-hand side of the rule, there must be relations of the same arity on the right-hand side if those relations are not just going to be inert under that rule. Thus, for example, a rule with signature $a_2 \to b_3$ will never (on its own) apply more than once, regardless of the values of $a$ and $b$.

In addition, if a rule is going to have a chance of leading to growth, the number of relations of some arity on the right-hand side must be greater than the number of that arity on the left.

These two constraints, however, do not always apply if a complete rule involves several individual rules. Thus, for example, a complete rule containing individual rules with signatures $2_2 \to 3_2\ 2_1,\ 2_2\ 1_1 \to 1_2$ can show growth, and can involve all relations. Note that since canonicalization is independent between different individual rules, the total number of possible inequivalent complete rules is just the product of the number of possible individual inequivalent rules.

When investigating cases where a large number of inequivalent rules are possible, it will often be convenient to do random sampling. If one picks a random rule first, and then canonicalizes it, some canonical rules will be significantly more common than others. But it is possible to pick with equal probability among canonical rules by choosing an integer between 1 and the total number of rules, then decoding this integer as discussed above to give the canonical rule.

## 3.3 Initial Conditions

In addition to enumerating rules, we can also consider enumerating possible initial conditions. Like each side of a rule, these can be characterized by sequences $n_1{}^{k_1}\, n_2{}^{k_2}\ldots$ which give the number of relations $n_i$ of arity $k_i$.

The only possible inequivalent $1_2$ initial conditions are {{1,1}}, corresponding a graph consisting of a single self-loop, and {{1,2}}, consisting of a single edge. The possible inequivalent connected $2_2$ initial conditions are:

{{{1, 1}, {1, 1}}, {{1, 1}, {1, 2}}, {{1, 1}, {2, 1}},
 {{1, 2}, {1, 2}}, {{1, 2}, {2, 1}}, {{1, 2}, {1, 3}}, {{1, 2}, {2, 3}}, {{1, 2}, {3, 2}}}

These correspond to the graphs:

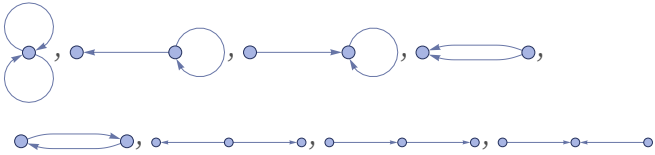

The possible inequivalent $1_3$ initial conditions are:

{{{1, 1, 1}}, {{1, 1, 2}}, {{1, 2, 1}}, {{1, 2, 2}}, {{1, 2, 3}}}

These correspond to the hypergraphs:

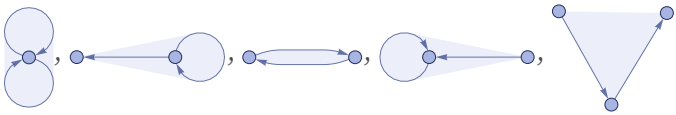

There are 102 inequivalent connected $2_3$ initial conditions. Ignoring ordering of relations, these correspond to hypergraphs with the following structures:

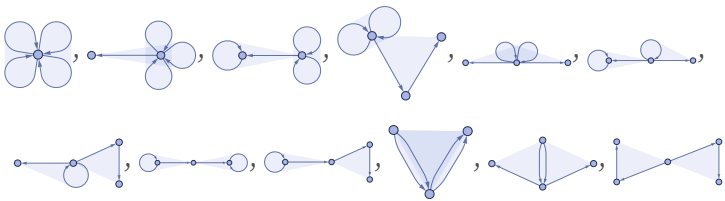

Ignoring connectivity, the number of possible inequivalent $n_1$ initial conditions is **PartitionsP**[$n$], the number of $1_n$ ones is **BellB**[$n$], while the number of $n_2$ ones can be derived using cycle index polynomials (see also [11:A137975]). The number of inequivalent connected initial conditions for various small signatures is as follows (with essentially the same Bell number estimates applying as for rules) [12]:

| | | | | | | | | | | |
|---|---|---|---|---|---|---|---|---|---|---|
| $1_2$ | 2 | $7_2$ | 40211 | $3_3$ | 3268 | $2_4$ | 2032 | $3_5$ | $2.3\times10^8$ |
| $2_2$ | 8 | $8_2$ | 293370 | $4_3$ | 164391 | $3_4$ | 678358 | $4_5$ | $2.1\times10^{12}$ |
| $3_2$ | 32 | $9_2$ | 2255406 | $5_3$ | $1.1\times10^7$ | $4_4$ | $4.2\times10^8$ | $1_6$ | 203 |
| $4_2$ | 167 | $10_2$ | 18201706 | $6_3$ | $9.0\times10^8$ | $5_4$ | $4.1\times10^{11}$ | $2_6$ | 2089513 |
| $5_2$ | 928 | $1_3$ | 5 | $7_3$ | $7.1\times10^{10}$ | $1_5$ | 52 | $3_6$ | $1.1\times10^{11}$ |
| $6_2$ | 5924 | $2_3$ | 102 | $1_4$ | 15 | $2_5$ | 57109 | $1_7$ | 877 |

A rule can only apply to a given initial condition if the initial condition contains at least enough relations to match all elements of the left-hand side of the rule. In other words, for a rule with signature $n_k \to \ldots$ there must be at least $n$ $k$-ary relations in the initial condition.

One way to guarantee that a rule will be able to apply to an initial condition is to make the initial condition in effect be a copy of the left-hand side of the rule, for example giving an initial condition {{1,2},{1,3}} for a rule with left-hand side {{*x*,*y*},{*x*,*z*}}. But the initial condition that in effect has the most chance to match is what is in many ways the simplest possible initial condition: the "self-loop" one where all elements are identical, or in this case {{0,0},{0,0}}. In what follows we will usually use such self-loop initial conditions Table[0,*n*,*k*].

## 3.4  Rules Depending on a Single Unary Relation

The very simplest possible rules are ones that transform a single unary relation, for example the $1_1 \to 2_1$ rule:

{{*x*}} → {{*x*}, {*y*}}

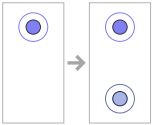

This rule generates a disconnected hypergraph, containing $2^n$ disconnected unary hyperedges at step *n*:

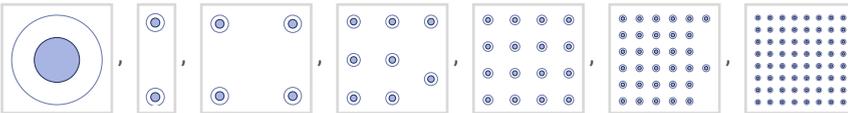

To get less trivial behavior, one must introduce at least one binary relation. With the $1_1 \to 1_2 1_1$ rule

{{*x*}} → {{*x*, *y*}, {*x*}}

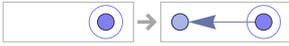

one just gets a figure with progressively more binary-edge "arms" being added to central unary hyperedge:

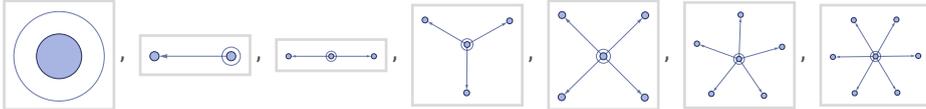

The rule

{{*x*}} → {{*x*, *y*}, {*y*}}

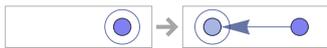

produces a growing linear structure, progressively "extruding" binary edges from the unary hyperedge:

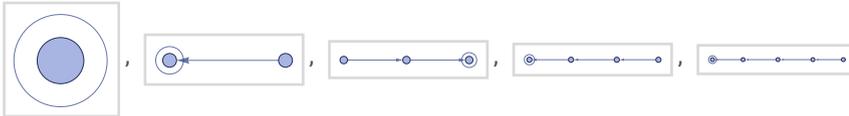

With two unary relations and one binary relation (signature $1_1 \to 1_2 2_1$) there are 16 possible rules; after 4 steps starting from a single unary relation, these give:

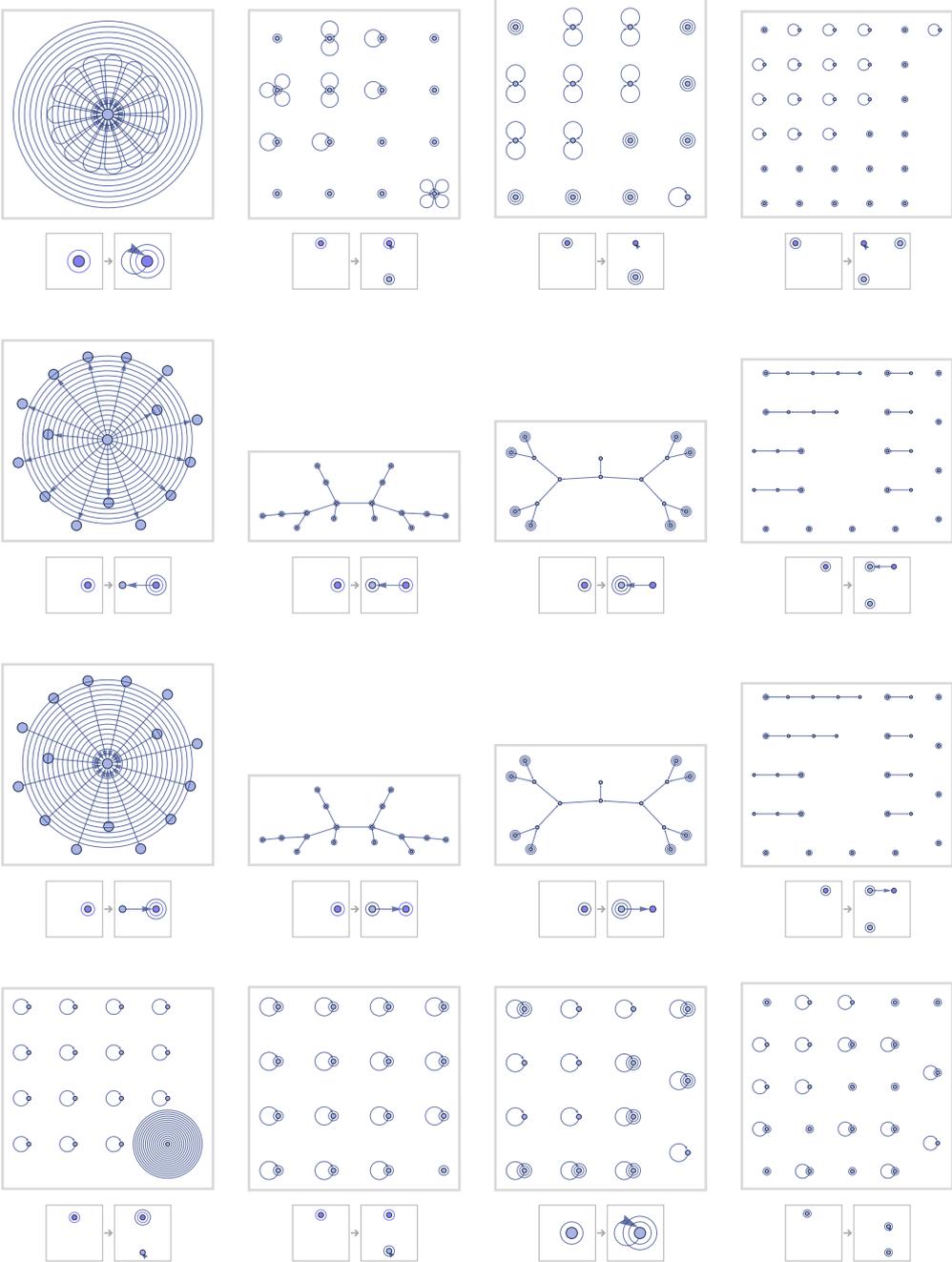

Many lead to disconnected hypergraphs; four lead to binary trees with structures we have already seen. ({{x}}→{{x,y},{x},{y}} is a $1_2 \to 1_2 2_1$ rule that gives the same result as the very first $1_2 \to 2_2$ rule we saw.

Rules for a single unary relation can never give structures more complex than trees, though the morphology of the trees can become slightly more elaborate:

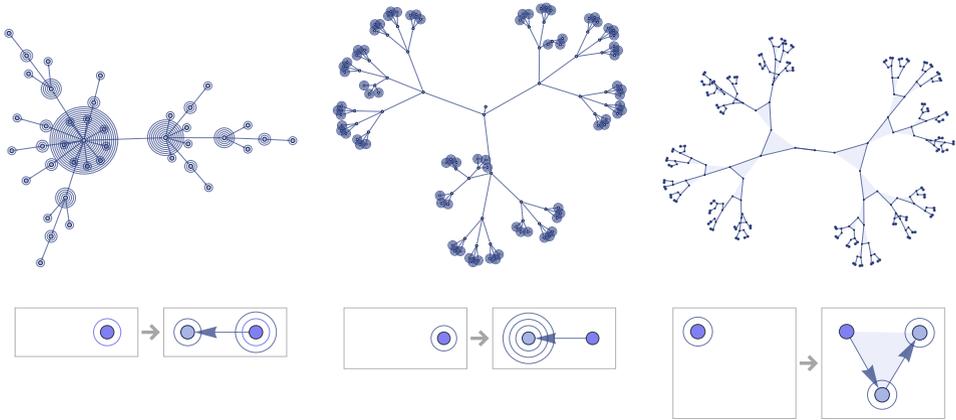

## 3.5 Rules Depending on a Single Binary Relation

There are 73 inequivalent left-connected $1_2 \to 2_2$ rules, but none lead to structures more complex than trees. Starting each from a single self-loop, the results after 5 steps are (note that even a connected rule like $\{\{x,y\}\} \to \{\{x,z\},\{z,x\}\}$ can give a disconnected result):

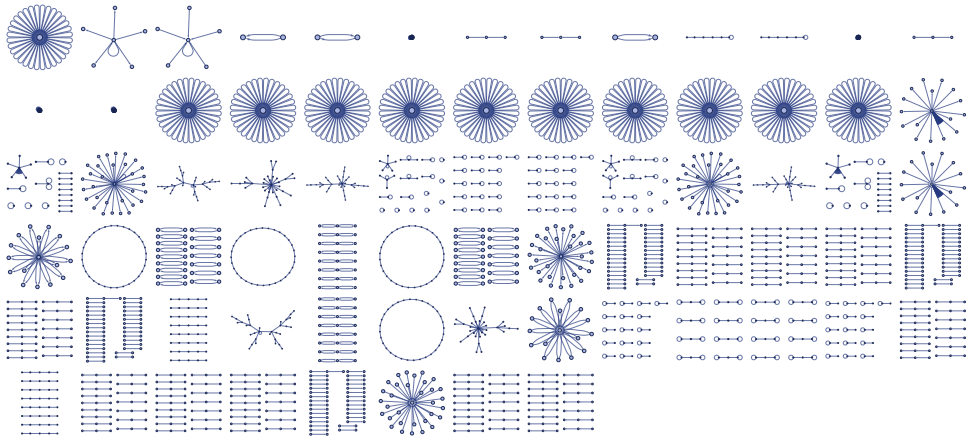

With an initial condition consisting of a square graph, the following very similar results are obtained:

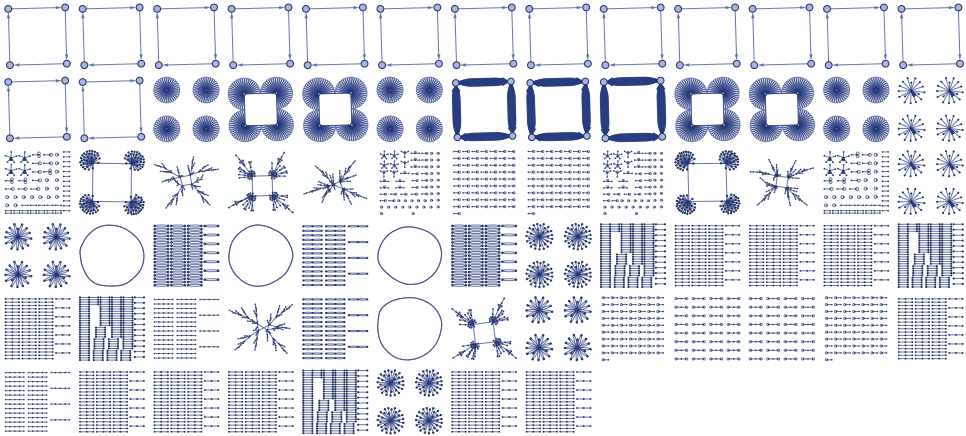

There are 506 inequivalent left-connected $1_2 \to 3_2$ rules. Running all these rules for 5 steps starting from a single self-loop, and keeping only distinct connected results, one gets (note that similar-looking results can differ in small-scale details):

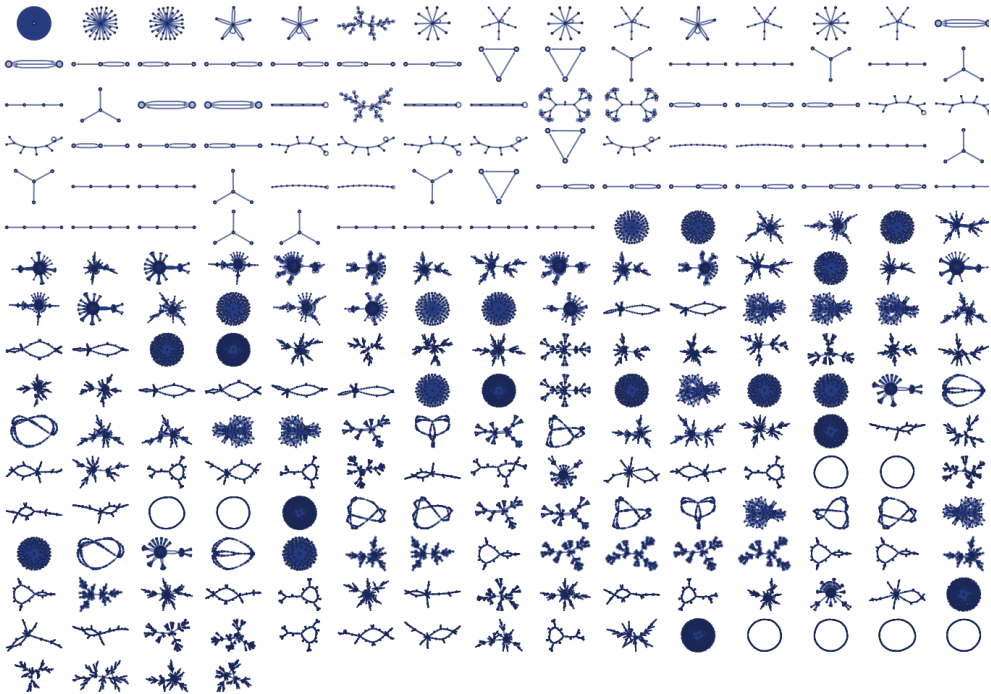

Several distinct classes of behavior are visible. Beyond simple lines, loops, trees and radial "bursts", there are nested ("cactus-like") graphs such as

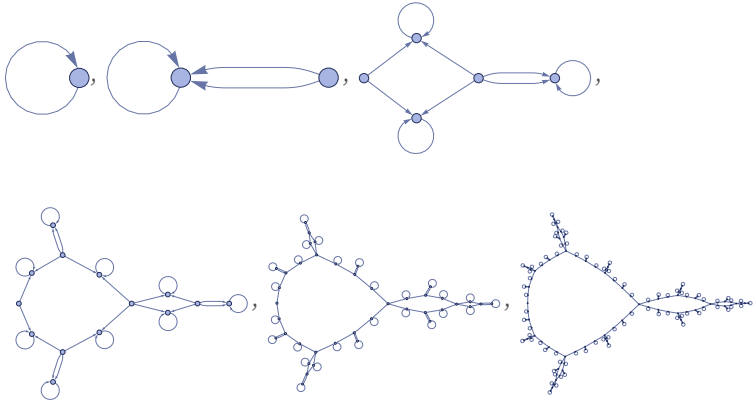

obtained from the rule

{{x, y}} → {{z, z}, {x, z}, {y, z}}

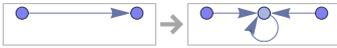

The only slightly different rule

{{x, y}} → {{x, x}, {x, y}, {z, x}}

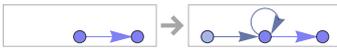

gives a rather different structure:

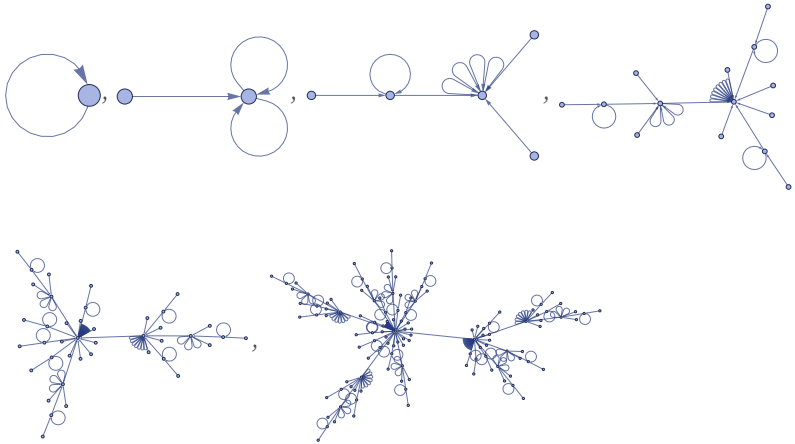

A layered rendering makes the behavior slightly clearer:

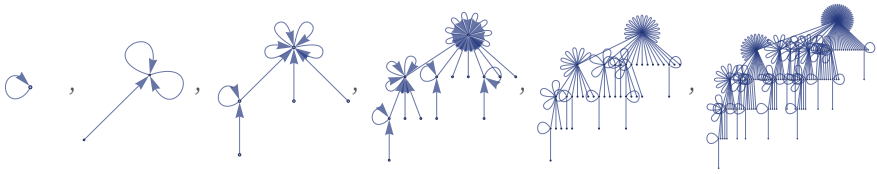

Another notable rule similar to one we saw in the previous section is:

{{x, y}} → {{x, z}, {x, z}, {z, y}}

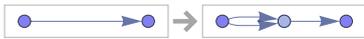

From a single edge this gives:

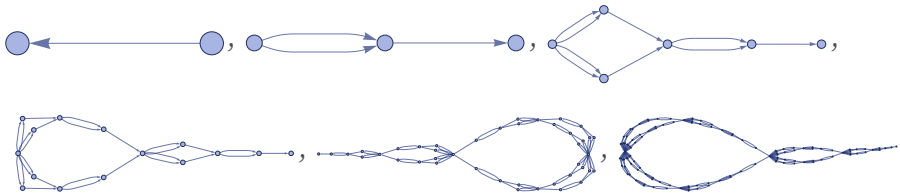

Starting from a single self-loop gives a more complex topological structure (and copies of this structure appear when the initial condition is more complex):

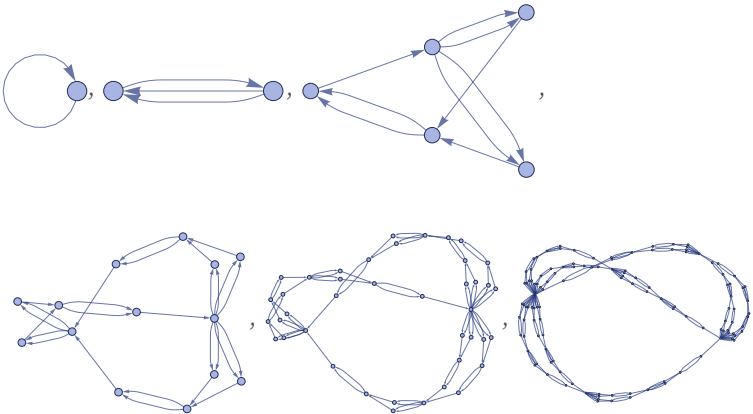

Another notable $1_2 \to 2_2$ rule is:

{{x, y}} → {{x, y}, {y, z}, {z, x}}

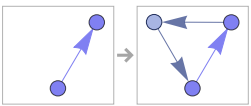

which produces an elaborately filled-in structure:

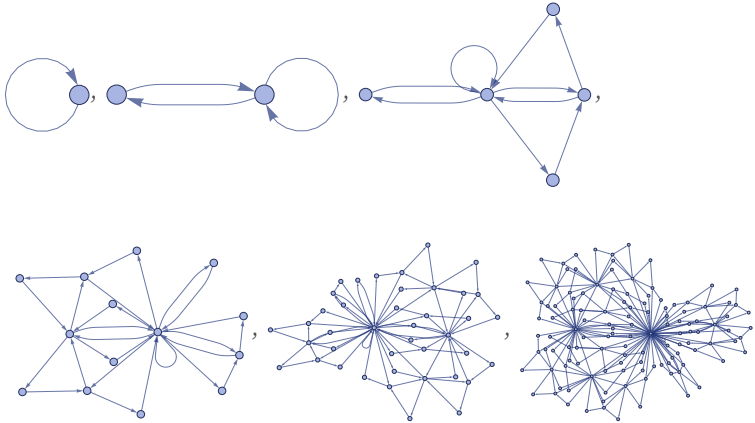

After 8 steps, the structure has the form:

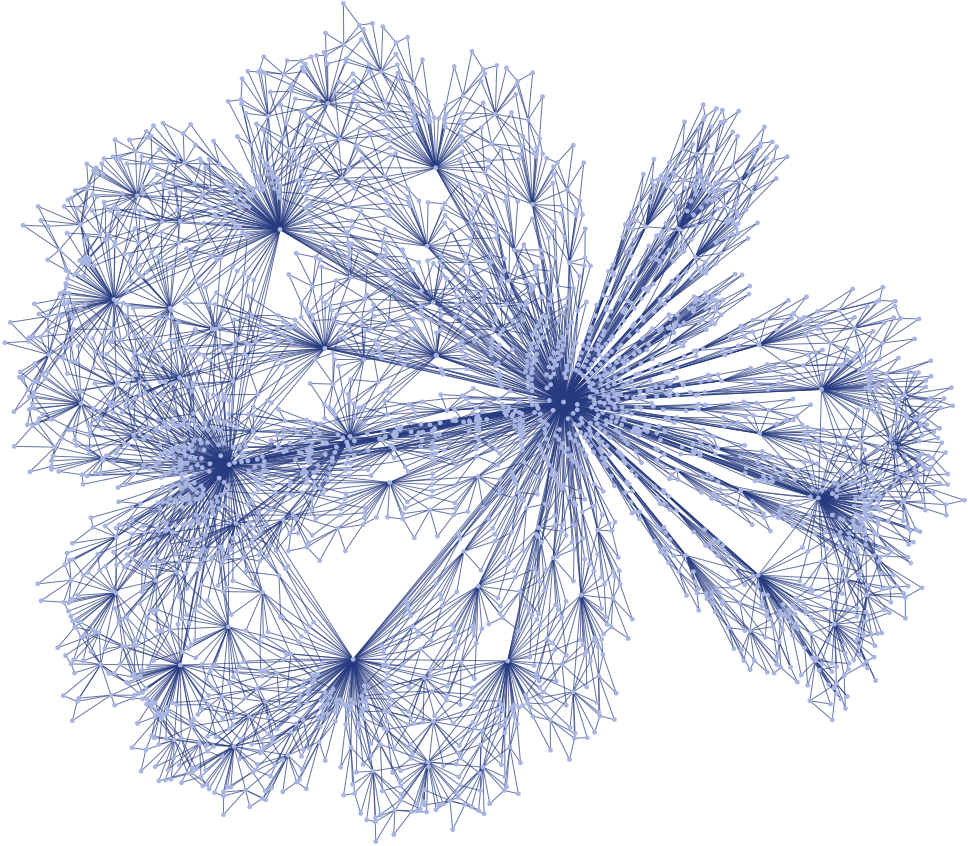

After $t$ steps, there are $3^{t-1}$ nodes, and $\frac{1}{6}(3^t + 1)$ edges. The graph diameter is $2t - 1$ if directions of edges are taken into account, and $t - 1$ if they are not. The maximum degree of any vertex is $2^t$—and all vertices have degrees of the form $2^s$, with the number of vertices of degree $2^s$ being proportional to $3^{t-s}$.

Starting from a single edge makes it slightly easier to understand what is going on:

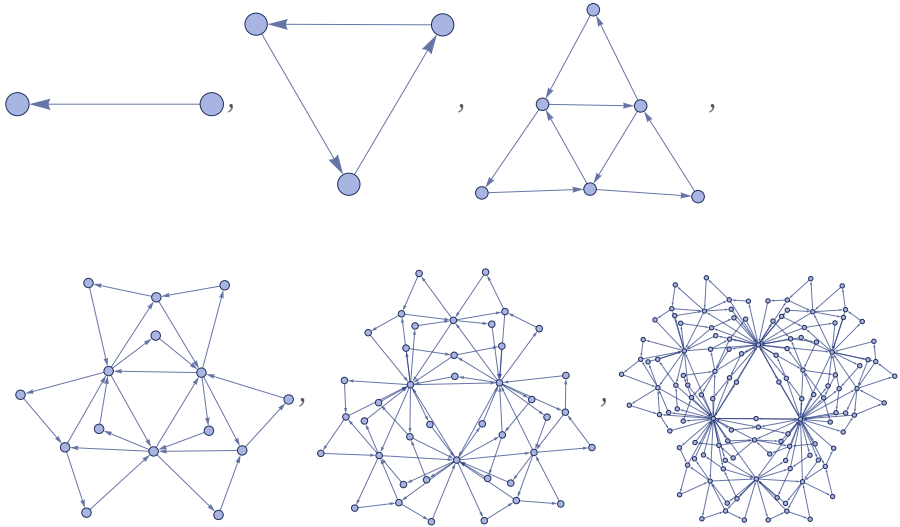

As the rule indicates, every edge of every triangle "sprouts" a new triangle at every step, in effect producing a sequence of "frills upon frills". But even though this may seem complicated, the whole structure basically corresponds just to a ternary tree in which each node is replaced by a triangle:

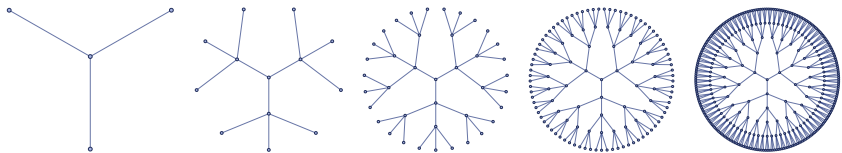

Starting from a single self loop, all $1_2 \to 3_2$ rules give after n steps a number of relations that is either constant, or goes like $2t - 1$, $2^t - 1$ or $3^{t-1}$.

For $1_2 \to 4_2$, there are 3740 distinct left-connected rules. As suggested by the random cases below, their behavior is typically similar to $1_2 \to 3_2$ rules, though the forms obtained can be somewhat more elaborate:

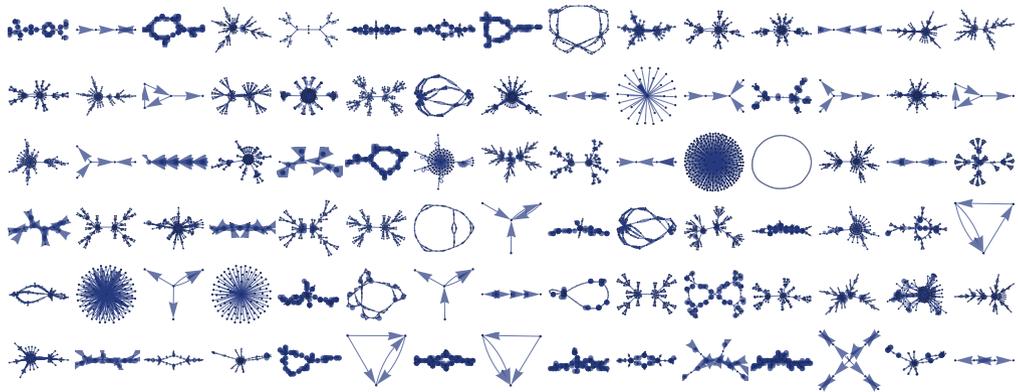

For example, the rule

{{x, y}} → {{y, z}, {y, z}, {z, y}, {z, x}}

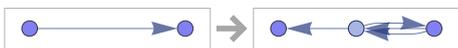

gives the following:

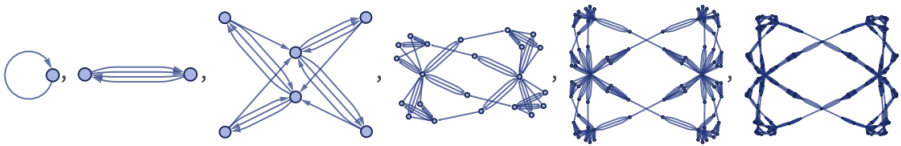

The rule

{{x, y}} → {{z, w}, {w, z}, {z, x}, {z, y}}

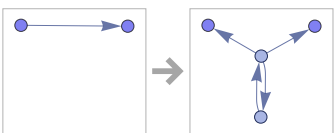

gives a nested form:

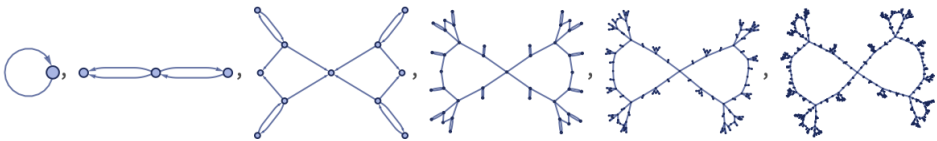

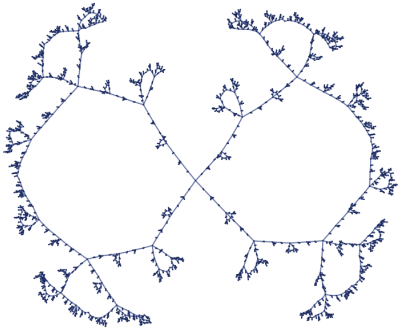

The rule

{{x, y}} → {{y, z}, {y, w}, {z, w}, {z, x}}

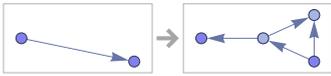

gives

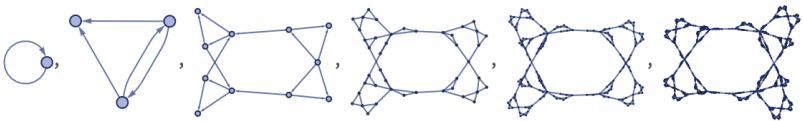

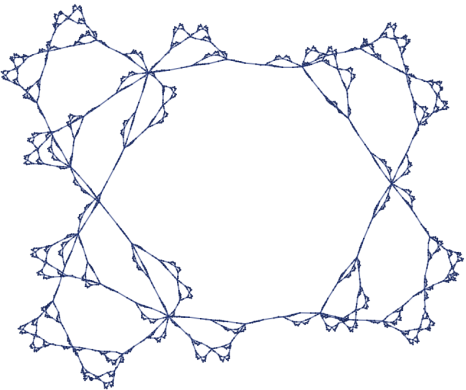

while the similar rule

{{x, y}} → {{x, z}, {x, w}, {z, w}, {z, y}}

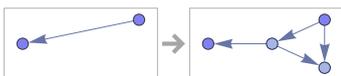

gives

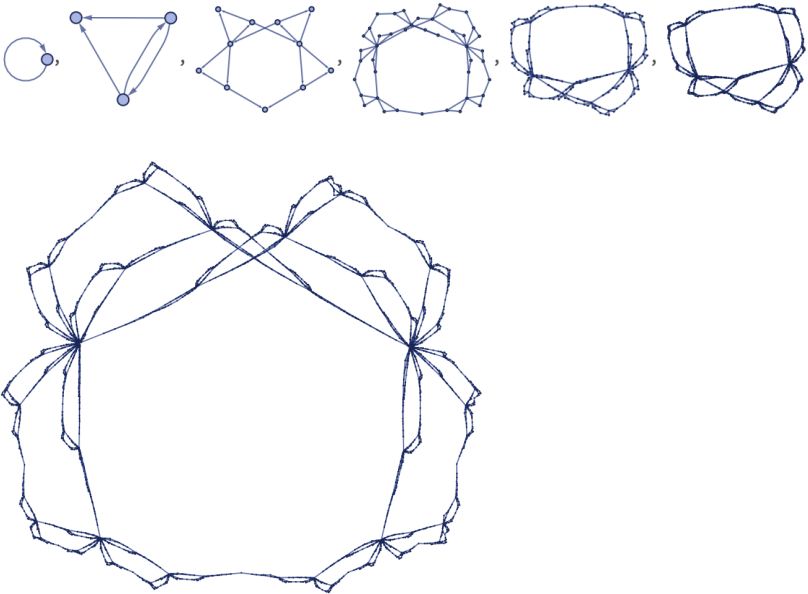

Successive steps in effect just fill in this shape, which seems somewhat irregular when rendered in 2D, but appears more regular if rendered in 3D.

Another rule with a simple structure when rendered in 3D is

$\{\{x, y\}\} \to \{\{y, z\}, \{y, z\}, \{z, x\}, \{z, x\}\}$

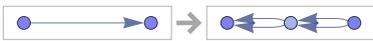

which yields:

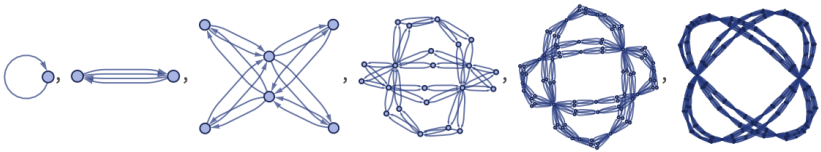

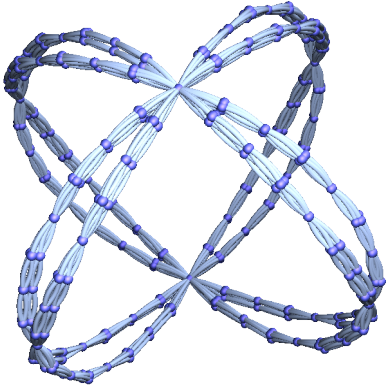

The outputs from $1_2 \to 4_2$ rules all grow either linearly (for example, like $3\,t - 2$), or exponentially, asymptotically like $2^t$, $3^t$ or $4^t$. The number of relations after $t$ steps is always given by a linear recurrence relation; for the rule $\{\{x,x\}\}\to\{\{x,x\},\{x,x\},\{x,y\},\{x,y\}\}$ the recurrence is f[t]=3f[t−1]−2f[t−2] (with f[1]=1, f[2]=4), giving size $\frac{1}{2}$ $(3{\times}2^t - 4)$.

## 3.6  Rules Depending on One Ternary Relation

There are 9373 inequivalent left-connected $1_3 \to 2_3$ rules. Here are typical examples of their behavior after 5 steps, starting from a single ternary self-loop:

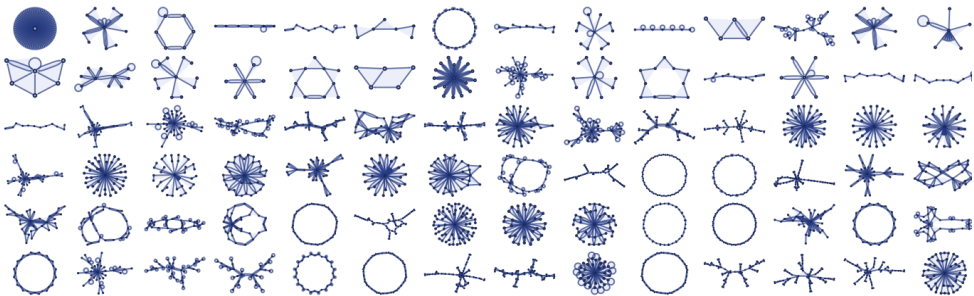

Here are results from a few of these rules after 10 steps:

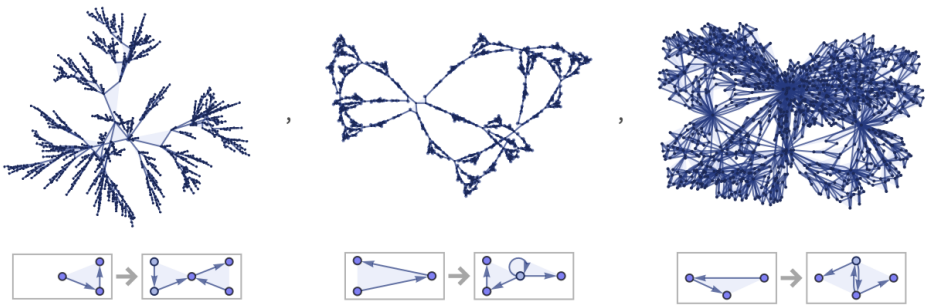

The number of relations in the evolution of $1_3 \to 2_3$ rules can grow in a slightly more complicated way than for $1_2 \to n_2$ rules. In addition to linear and $2^t$ growth, there is also, for example, quadratic growth: in the rule

{{$x, x, y$}} → {{$y, y, z$}, {$x, y, x$}}

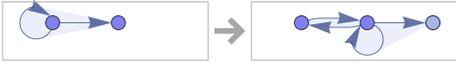

each existing "arm" effectively grows by one element each step, and there is one new arm generated, yielding a total size of $\sum_{k=2}^{t} k = \frac{1}{2}(t^2 + t - 2)$:

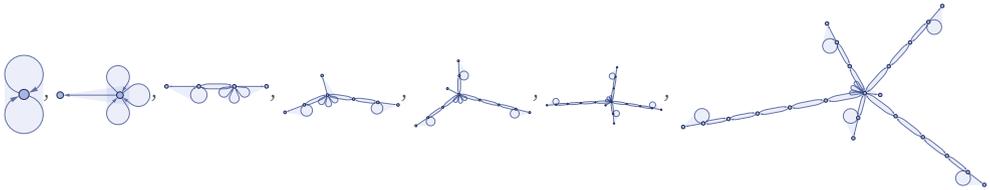

The rule

{{$x, x, y$}} → {{$y, y, y$}, {$x, y, z$}}

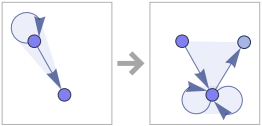

yields a Fibonacci tree, with size Fibonacci[$t$+2]–1 ~ $\phi^t$:

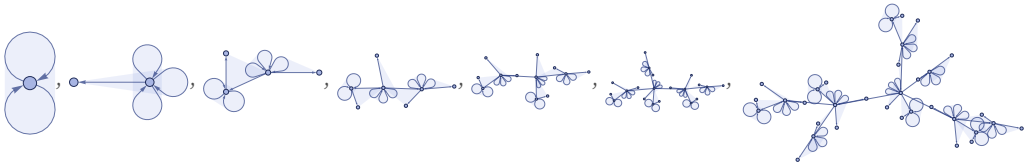

$1_3 \to 2_3$ rules can produce results that look fairly complex. But it is a consequence of their dependence only on a single relation that once such rules have established a large-scale structure, later updates (which are necessarily purely local) can in a sense only embellish it, not fundamentally change it:

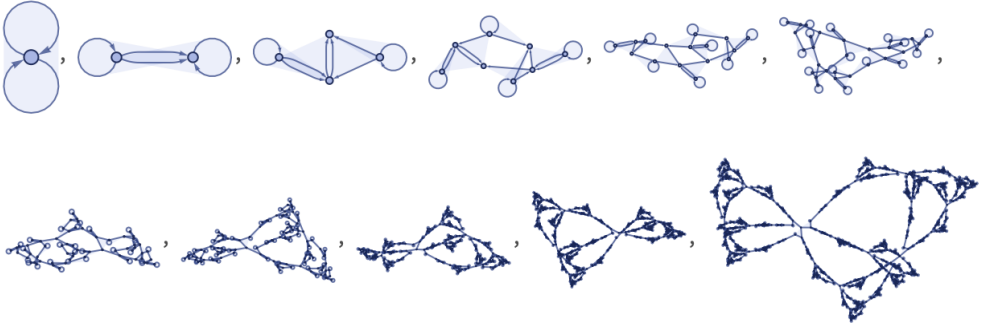

There are 637,568 inequivalent left-connected $1_3 \to 3_3$ rules; here are samples of their behavior:

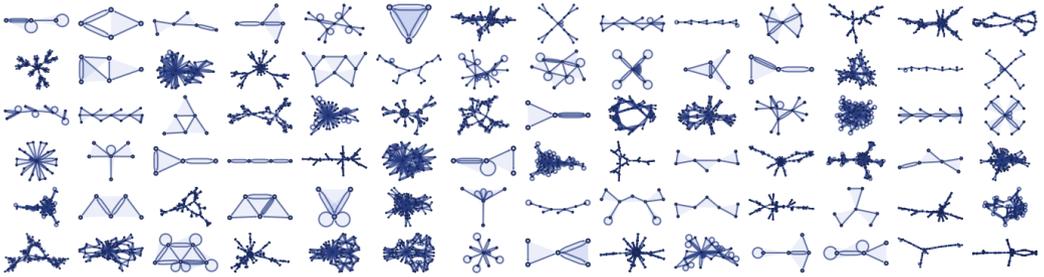

The results can be more elaborate than for $1_3 \to 2_3$ rules—as the following examples illustrate—but remain qualitatively similar:

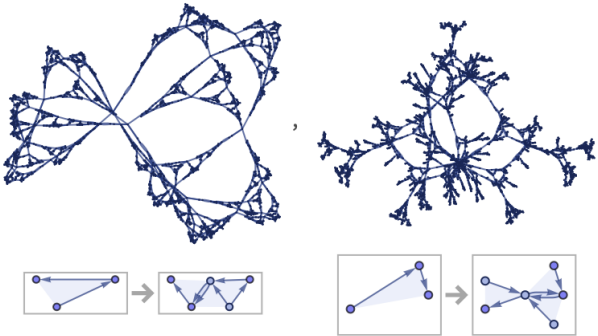

One notable $1_3 \rightarrow 3_3$ rule (that we will discuss below) in a sense directly implements the recursive formation of a nested Sierpiński pattern:

$\{\{x, y, z\}\} \rightarrow \{\{x, u, v\}, \{z, v, w\}, \{y, w, u\}\}$

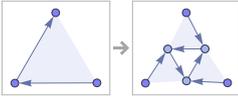

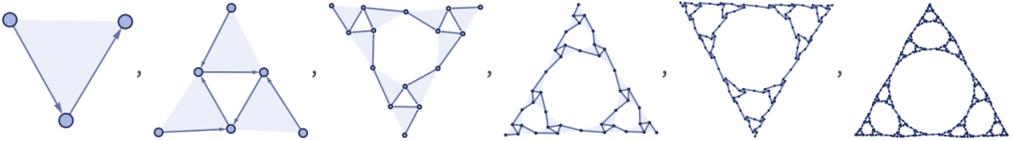

## 3.7 Rules Depending on More Than One Relation: The $2_2 \rightarrow 3_2$ Case

The smallest nontrivial signature that can lead to growth (and therefore unbounded evolution) is $2_2 \rightarrow 3_2$. There are 4702 distinct left-connected rules with this signature. Here is a random sample of the behavior they generate, starting from a double self-loop $\{\{0,0\},\{0,0\}\}$ and run for 8 steps:

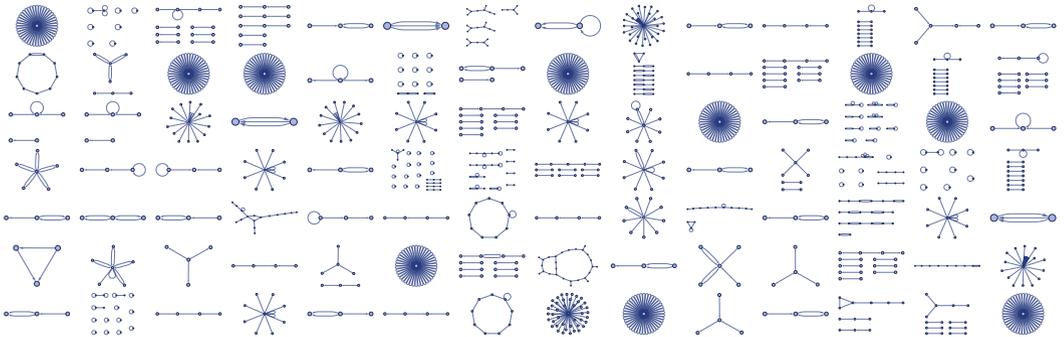

Restricting to connected cases, there are 291 distinct outputs involving more than 10 relations after 8 steps:

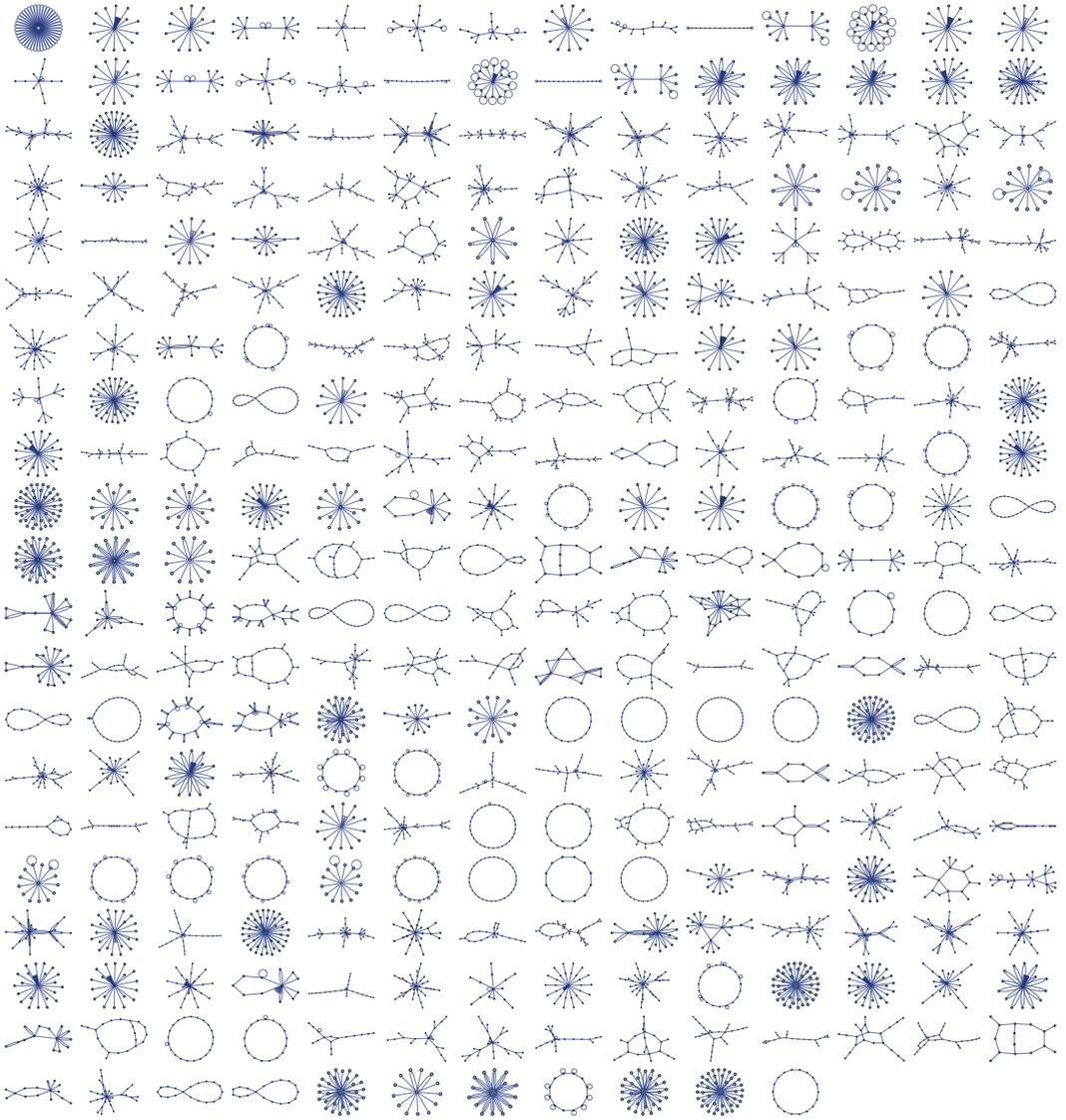

The overall behavior we see here is very similar to what we saw with rules depending only on a single relation. But there is a new issue now to be addressed. With rules depending only on a single relation there is never any ambiguity about where the rule should be applied. But with rules that depend on more than one relation, there can be ambiguity, and the results one gets can potentially depend on the order in which updating is done.

Consider the rule

{{x, y}, {x, z}} → {{x, w}, {y, w}, {z, w}}

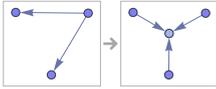

With our standard updating order, the result of running this rule for 30 steps is:

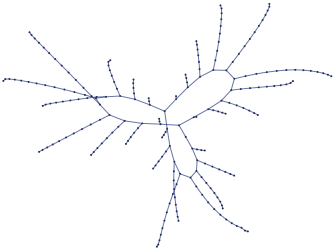

But with 6 different choices of random updating orders one gets instead:

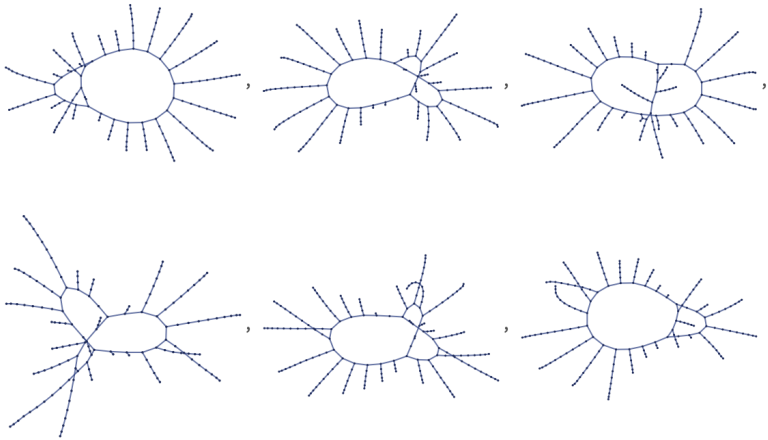

None of these graphs are isomorphic, but all of them are qualitatively similar. Later on, we will discuss in detail the consequences of different updating orders, and their potentially important implications for physics. But for now, suffice it to say that at a qualitative level different updating orders typically lead to similar behavior.

As an example of something of an exception, consider the $2_2 \to 3_2$ rule shown in the array above:

{{x, y}, {y, z}} → {{x, w}, {w, z}, {z, x}}

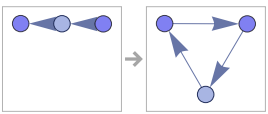

With our standard updating order, this rule behaves as follows, yielding complicated-looking results with about $1.5^n$ relations after $n$ steps:

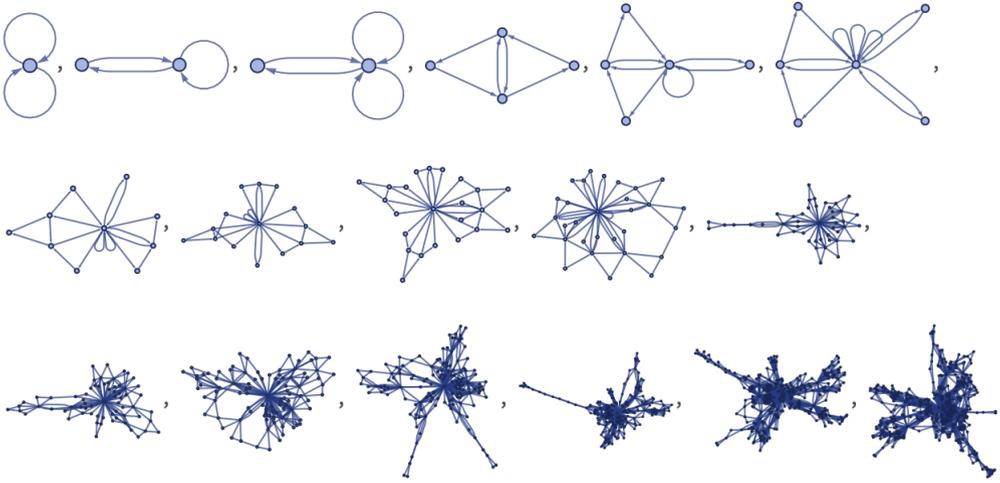

But with random updating order, the behavior is typically quite different. Here are six examples of results obtained after 10 steps—and all of them are disconnected:

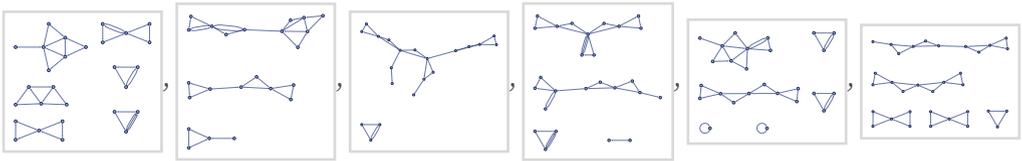

## 3.8 Rules with Signature $2_2 \to 4_2$

For $2_2 \to 4_2$, there are 40,405 inequivalent left-connected rules. Of these, about 36% stay connected when they evolve. Starting from two self-loops {{0,0},{0,0}}, and running for 8 steps, here is a sample of the 4000 or so distinct behaviors that are produced:

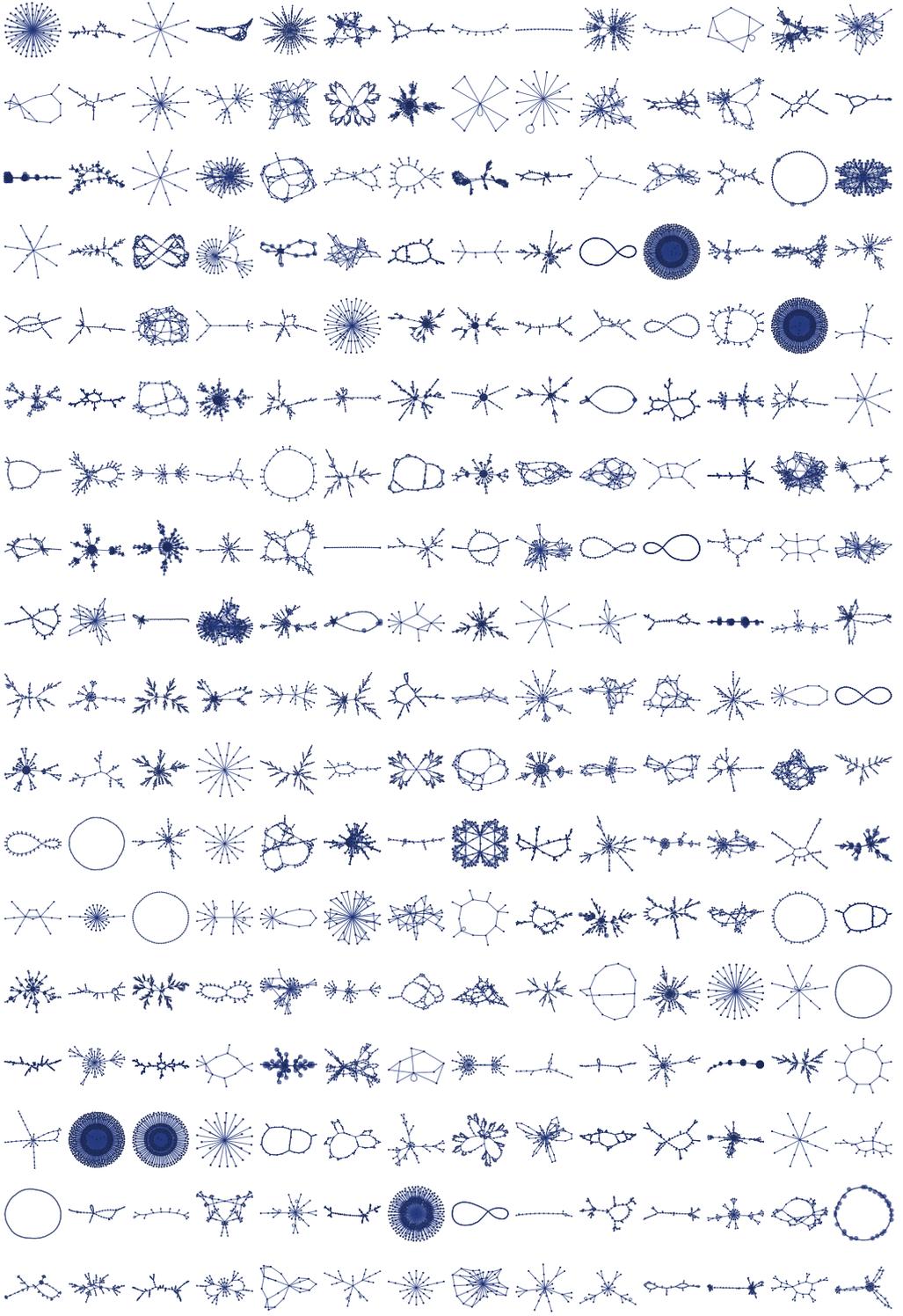

Most of these rules show the same kinds of behaviors we have seen before. But there is one major new kind of behavior that is observed: in a little less than 1% of all cases, the rules produce globular structures that in effect continually add various forms of cross-connections. Here are a few examples (notably, even though $2_2 \to 4_2$ rules can involve up to 7 distinct elements, these rules all involve just 4):

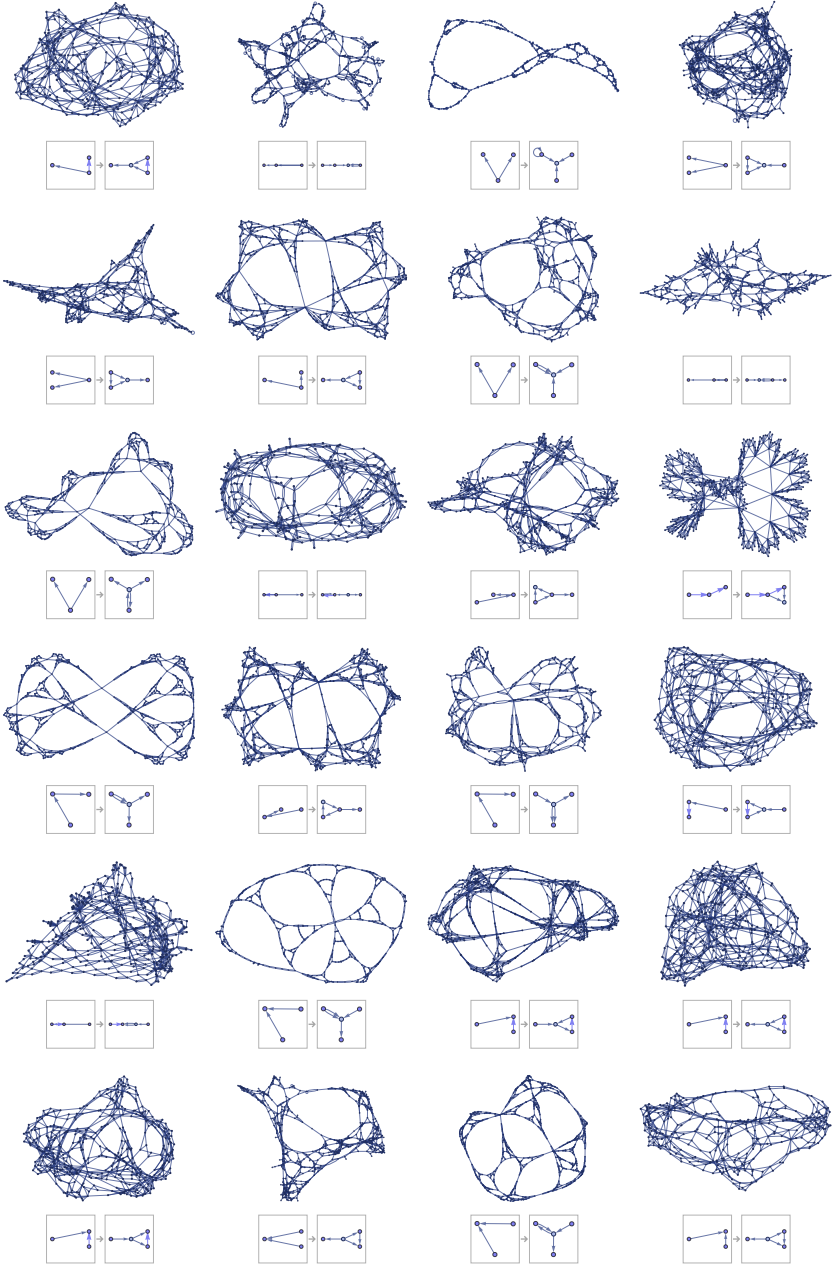

We will study these kinds of structures in more detail later. Note that the specific forms shown here depend on the underlying updating order used—though for example random orders typically seem to give similar results. It is also the case that the detailed visual layout of graphs can affect the impression of these structures; we will address this in the next section when we discuss various forms of quantitative analysis.

It is remarkable how complex the structures are that can be created even from very simple rules. Here are three examples (with short codes wm5583, wm4519, wm2469) shown in more detail:

{{x, y}, {x, z}} → {{y, z}, {y, w}, {z, w}, {w, x}}

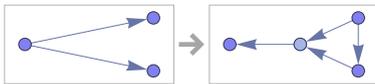

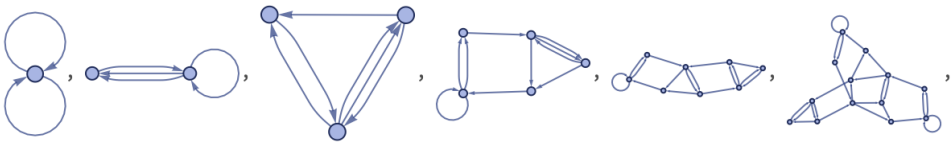

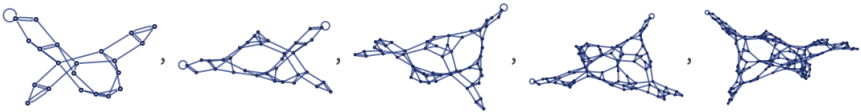

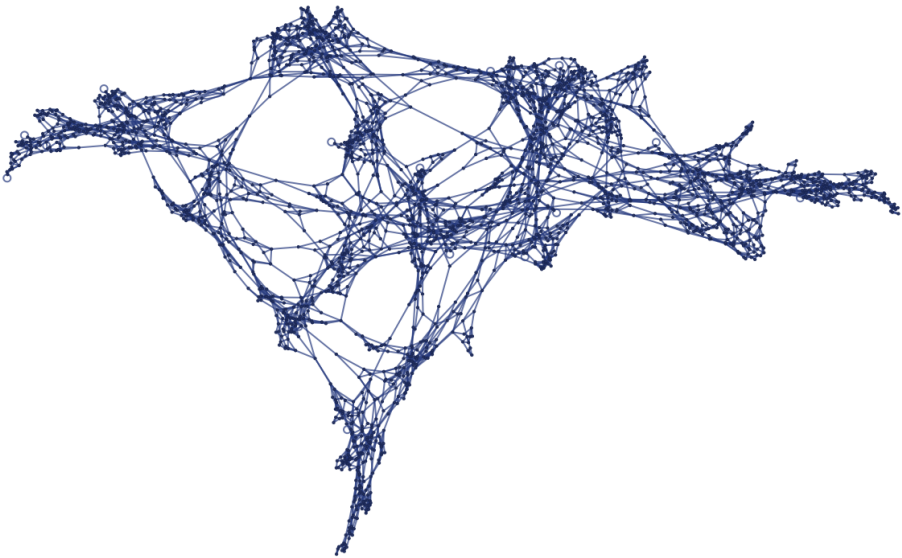

$\{\{x, y\}, \{y, z\}\} \rightarrow \{\{x, y\}, \{y, x\}, \{w, x\}, \{w, z\}\}$

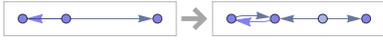

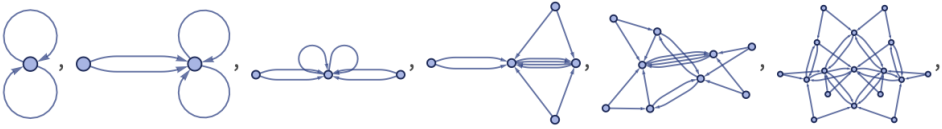

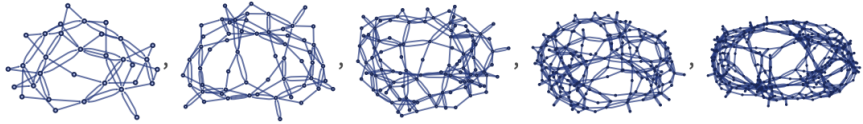

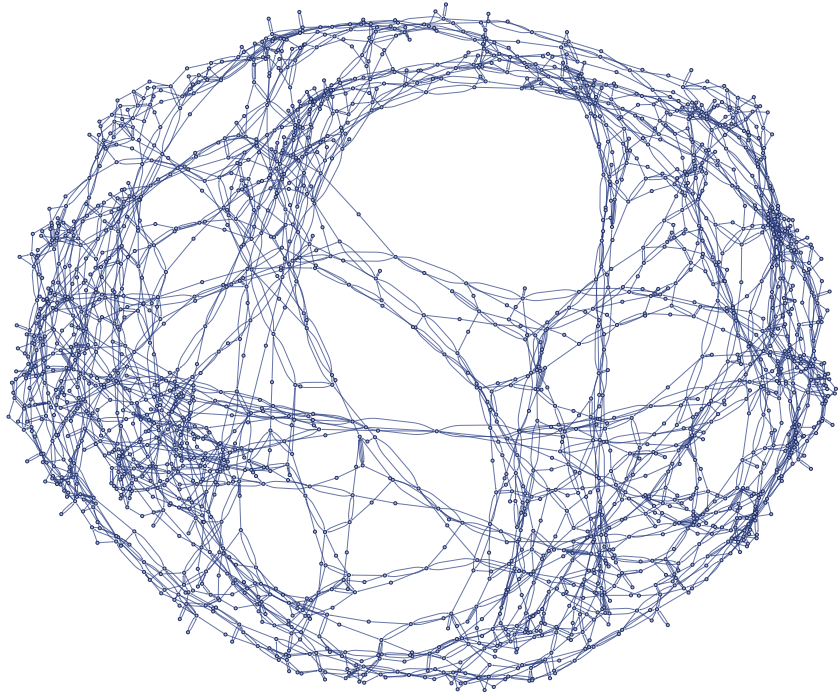

$\{\{x, y\}, \{y, z\}\} \rightarrow \{\{w, y\}, \{y, z\}, \{z, w\}, \{x, w\}\}$

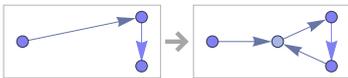

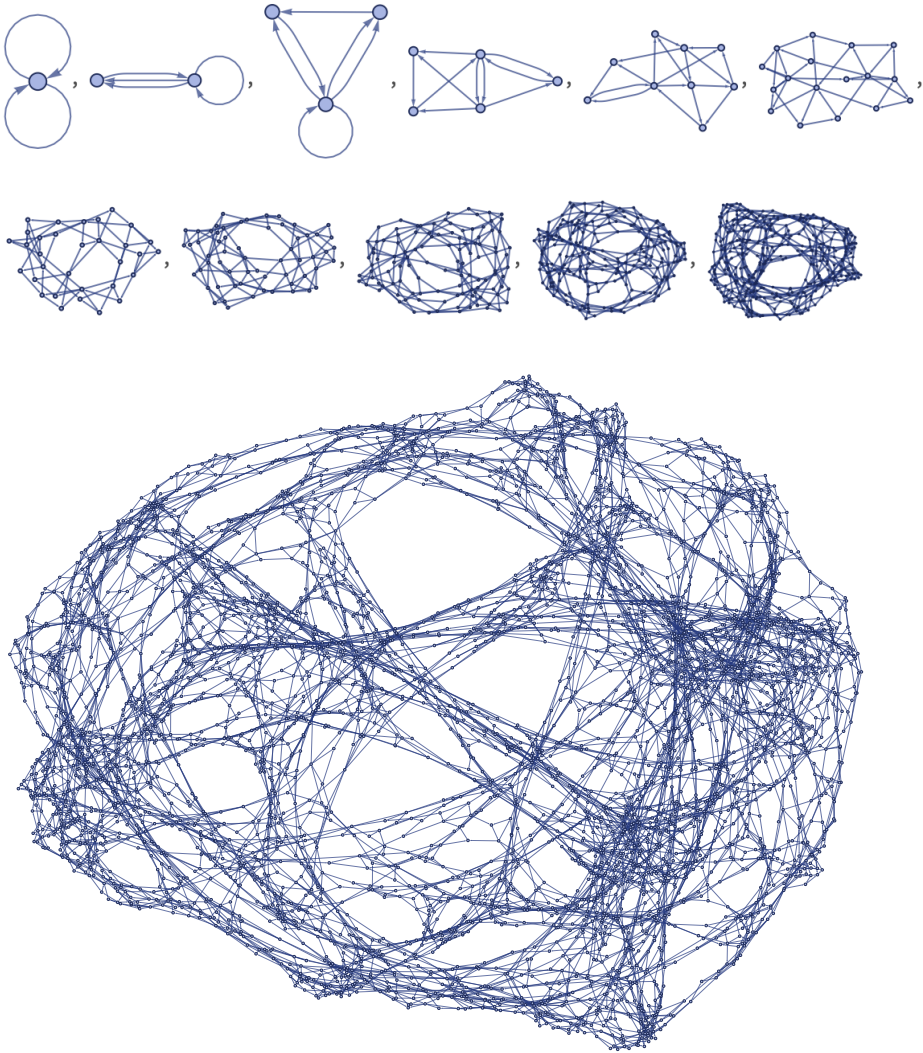

Much as we have seen in other systems such as cellular automata [1], there seems to be no simple way to deduce from the rules from our systems here what their behavior will be. And indeed even seemingly very similar rules can give dramatically different behavior, sometimes simple, and sometimes complex.

## 3.9  Binary Rules with Signatures Beyond $2_2 \to 4_2$

Going from signature $2_2 \to 3_2$ to signature $2_2 \to 4_2$ brought us the phenomenon of globular structures. Going to signature $2_2 \to 5_2$ and beyond does not seem to bring us any similarly widespread significant new form of behavior. The fraction of rules that yield connected results decreases, but among connected results, similar fractions of globular structures are seen, with examples from $2_2 \to 5_2$ including:

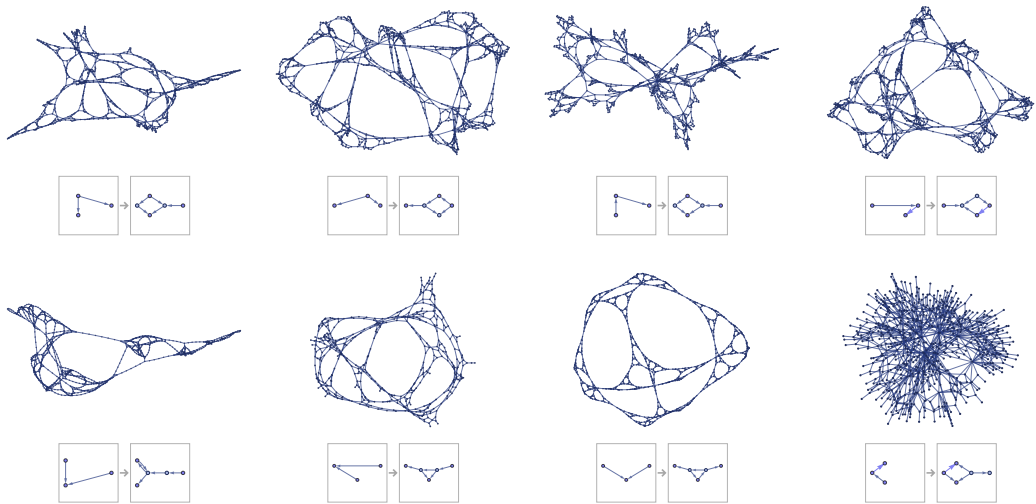

The last rule shown here has a feature that is seen in a few $2_2 \to 4_2$ rules, but is more prominent in $2_2 \to 5_2$ rules: the presence of many "dangling ends" that at least visually obscure the structure. To see the structure better, one can take the evolution of this rule

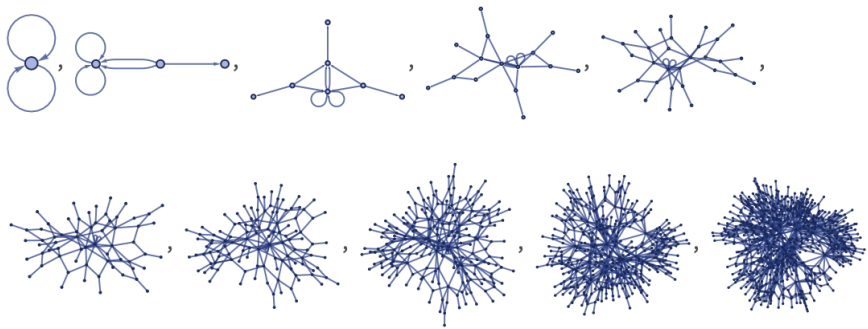

and effectively just "edit" the graphs obtained at each step, removing all dangling ends:

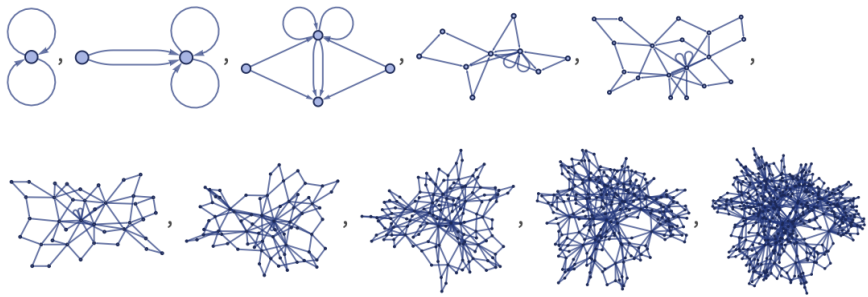

In addition to increasing the number of relations on the right-hand side of the rule, one can also increase the number on the left. For example, one can consider $3_2 \to 4_2$ rules. These much more often lead to termination than $2_2 \to \ldots$ rules, and appear to produce results generally similar to $2_2 \to 3_2$ rules.

$3_2 \to 5_2$ rules also produce globular structures, though more rarely than $2_2 \to 4_2$ rules, and with slower growth. A few examples are:

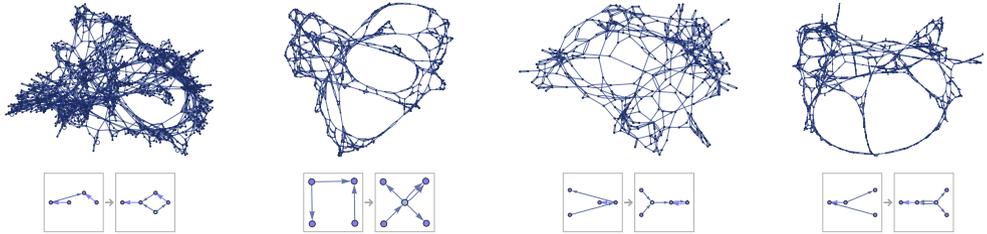

## 3.10  Rules Depending on Two Ternary Relations: The $2_3 \to 3_3$ Case

There are 79,359,764 inequivalent left-connected $2_3 \to 3_3$ rules. The fraction of these rules showing continued growth is considerably smaller than for $2_2 \to \ldots$ rules. But here is a typical sample of growth rules (note that different rules are run for different numbers of steps to achieve a roughly balanced level of detail):

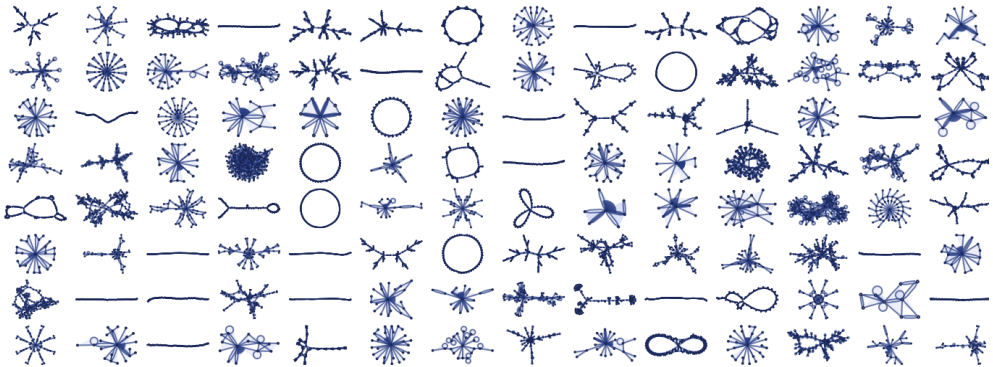

And even though there are only 3 relations on the right-hand side (rather than the 4 in $2_2 \to 4_2$) these rules can produce globular structures. Some examples are:

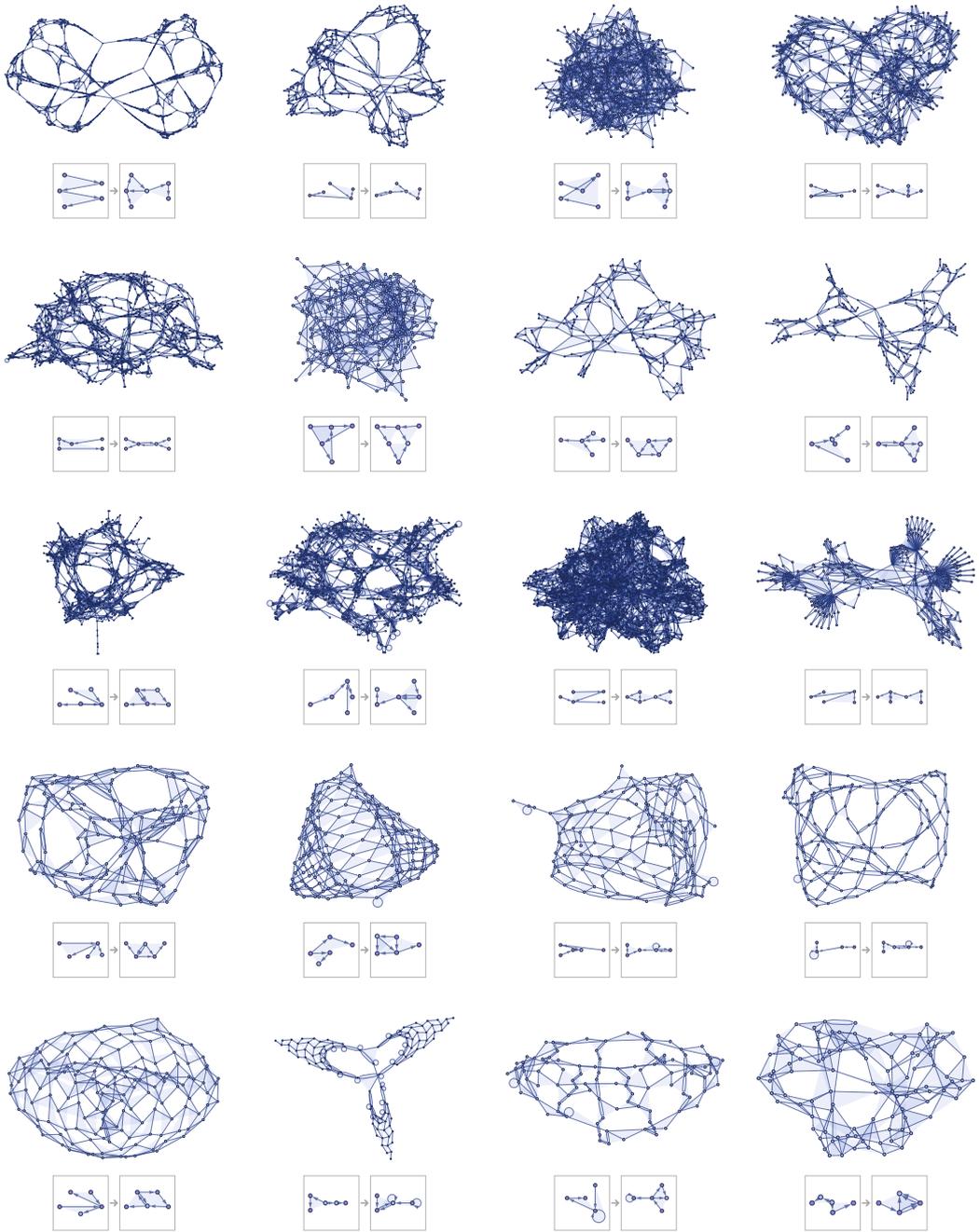

A new phenomenon exhibited by $2_3 \to 3_3$ rules is the formation of globular structures by what amounts to slow grow. This is exemplified by a rule like:

$\{\{x, y, z\}, \{x, u, v\}\} \to \{\{x, w, u\}, \{v, w, y\}, \{w, y, z\}\}$

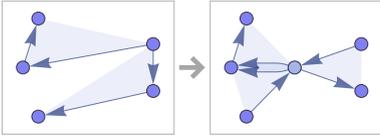

This rule progressively builds up a structure by growing only in one place at a time (the position of the surviving self-loop):

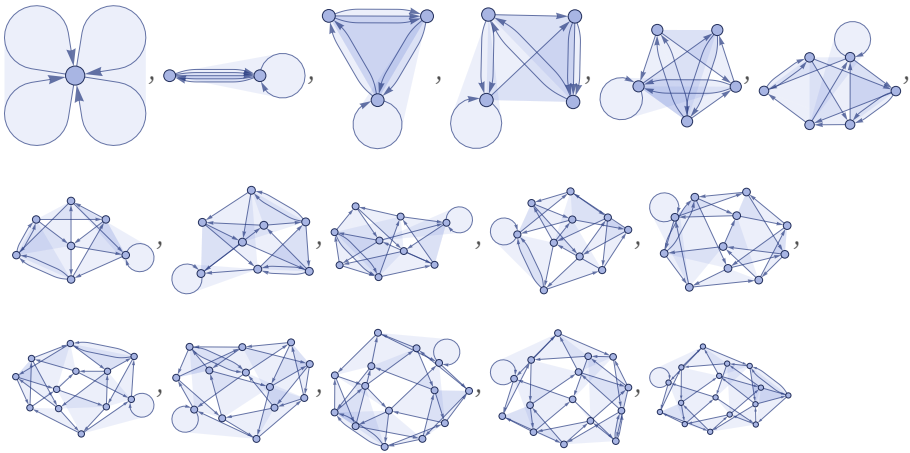

After 1000 steps the rule has produced this structure containing 1000 ternary relations (plus the 2 already present in the initial condition):

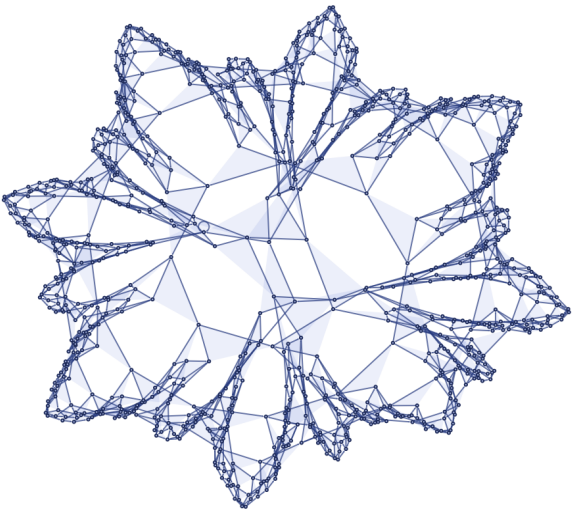

Another example of slow growth occurs in the rule

{{x, x, y}, {z, u, x}} → {{u, u, z}, {v, u, v}, {v, y, x}}

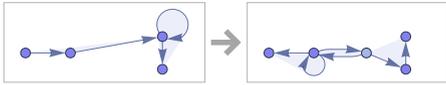

which after 1000 steps generates:

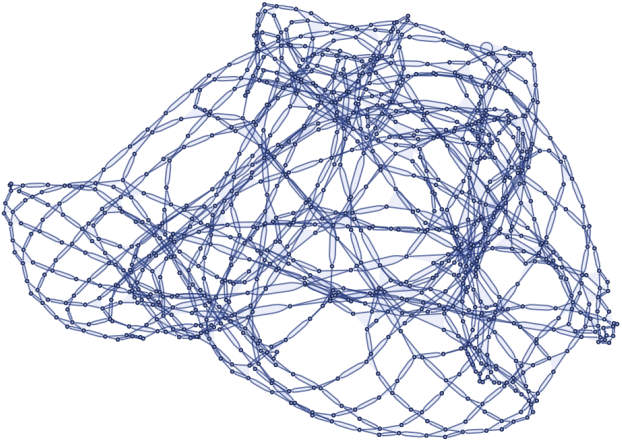

Note the presence here of regions of square grids. These occur even more prominently in the rule

{{x, y, z}, {u, y, v}} → {{w, z, x}, {z, w, u}, {x, y, w}}

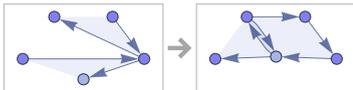

which after 500 steps produces:

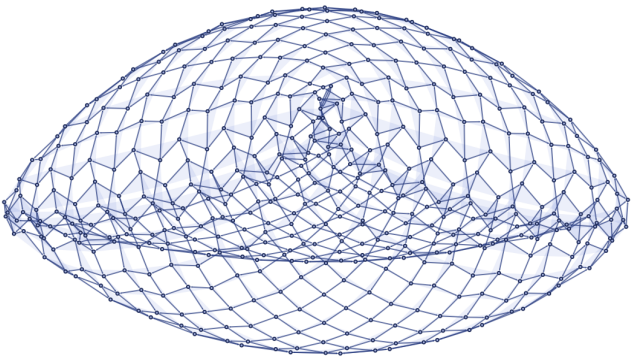

As we will discuss in the next section, the grid here becomes quite explicit when the hypergraph is rendered in 3D. Notice that the grid is not evident even after 20 steps in the evolution of the rule; it takes longer to emerge:

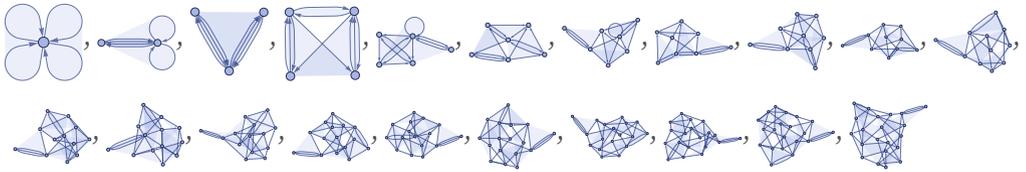

Once again, though, the rule adds just a single relation at each generation; in effect the grid is being "knitted" one node at a time.

The emergence of a grid is still easier to see in the rule:

{{*x*, *y*, *z*}, {*x*, *u*, *v*}} → {{*z*, *z*, *w*}, {*w*, *w*, *v*}, {*u*, *v*, *w*}}

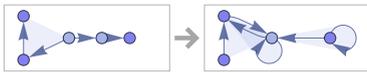

which after 200 steps yields:

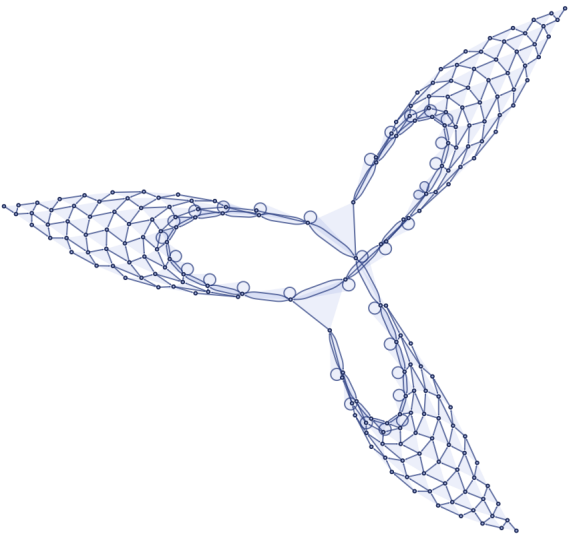

Once again, the "knitting" of this form is far from obvious in the first 20 steps of evolution:

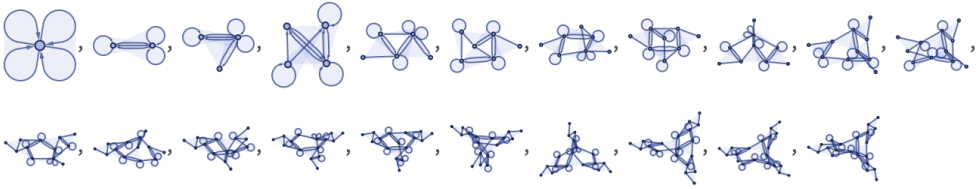

Just sometimes, however, the behavior is quite easy to trace, as in this particularly direct example of "knitting":

$\{\{x, y, y\}, \{z, x, u\}\} \rightarrow \{\{y, v, y\}, \{y, z, v\}, \{u, v, v\}\}$

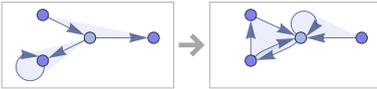

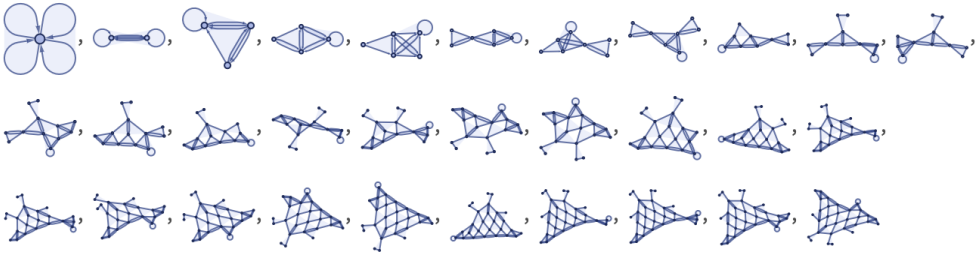

which after 200 steps yields:

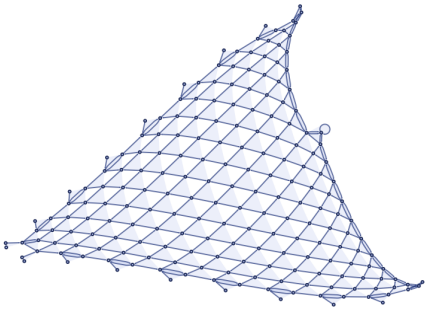

As a different example of slow growth, consider the rule

$\{\{x, y, y\}, \{y, z, u\}\} \rightarrow \{\{u, z, z\}, \{u, x, v\}, \{y, u, v\}\}$

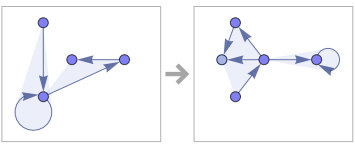

After 200 steps this rule gives

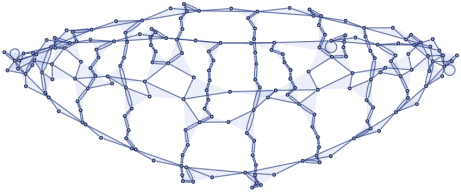

while after 500 steps it gives:

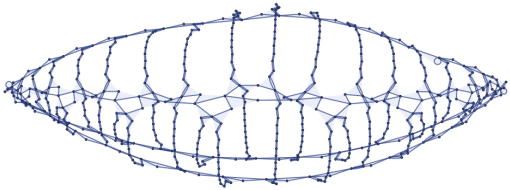

Looking at all 79 million or so $2_3 \to 3_3$ rules in canonical order, one finds that rules with slow growth are quite rare and are strongly localized to about 10 broad regions in the space of possible rules. Of rules with slow growth, only a few percent form nontrivial globular structures. And of these, perhaps 10% exhibit obvious lattice-like patterns.

The pictures below show additional examples. Note that—as we will discuss later—many of the patterns here are best visualized in 3D.

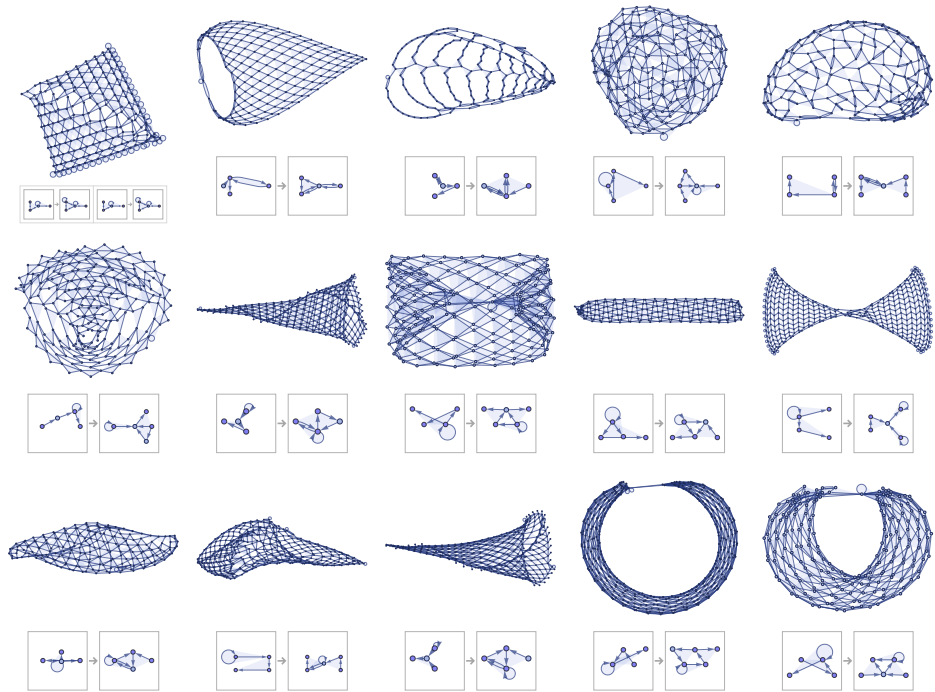

## 3.11  Rules Involving More Ternary Relations

There are about 9 billion inequivalent left-connected $2_3 \to 4_3$ rules. About 20% lead to connected results, and of these about half show continued growth. Here is a random sampling of the behavior of such rules:

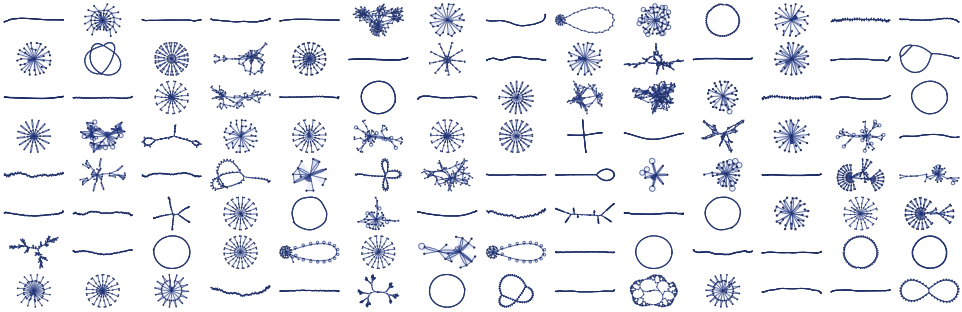

The fraction of complex behavior appears to be no higher than for $2_3 \to 3_3$ rules, and no obvious major new phenomena are seen. Much like in systems such as cellular automata (and as suggested by the Principle of Computational Equivalence [1:c12]), above some low threshold, adding complexity to the rules does not appear to add complexity to the typical behavior produced.

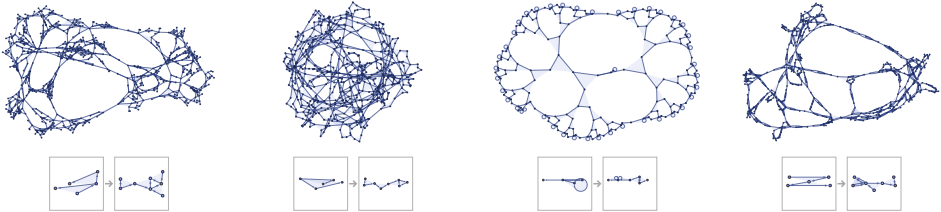

The trend continues with $3_3 \to 4_3$ rules, with one notable feature here being an increased propensity for rules to yield results that become disconnected, though only after many steps. The general difficulty of predicting long-term behavior is illustrated for example by the evolution of this $3_3 \to 5_3$ rule, sampled every 10 steps:

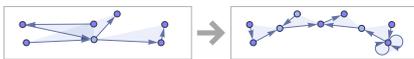

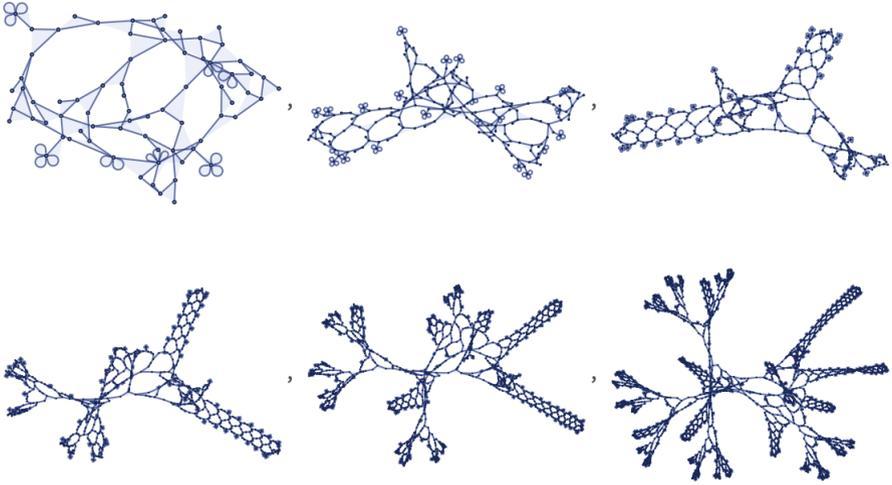

## 3.12 Rules with Mixed Arity

So far essentially all the rules we have considered have "pure signatures" of the form $m_k \to n_k$ for some arity $k$. Continued growth is never possible unless the right-hand side of a rule contains some relations with the same arity as appear on the left. But, for example, it is perfectly possible to have growth in rules with signatures like $1_2 \to 2_2 2_1$. Such rules produce unary relations, which can serve as "markers" for the application of the rule, but cannot themselves affect how or where the rule is used:

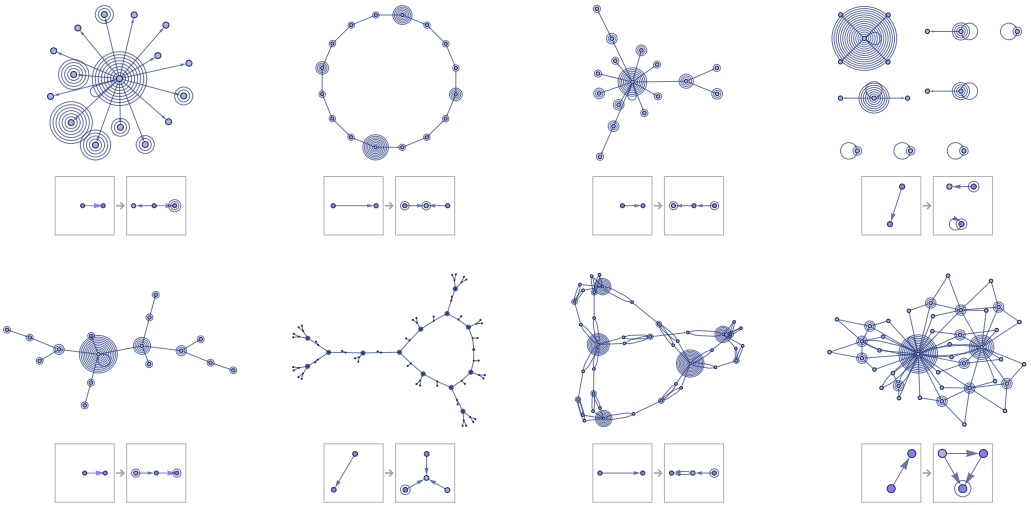

The 634 rules with signature $1_2 \to 1_3 1_2$ all show very simple behavior (as do the 2212 rules with signature $1_2 \to 1_3 1_2 1_1$), with not even trees being possible. But among the 7652 $1_2 \to 1_3 2_2$ rules there are not only many trees, but also closed structures such as:

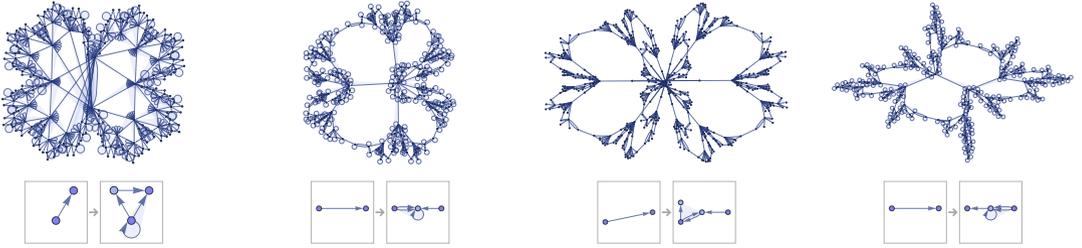

Previously we had only seen structures like the first one above in rules that depend on more than one relation. But as this illustrates, such structures can be produced even with just a single relation on the left:

{{x, y}} → {{x, x, y}, {x, z}, {z, y}}

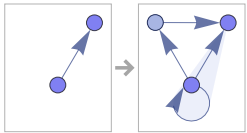

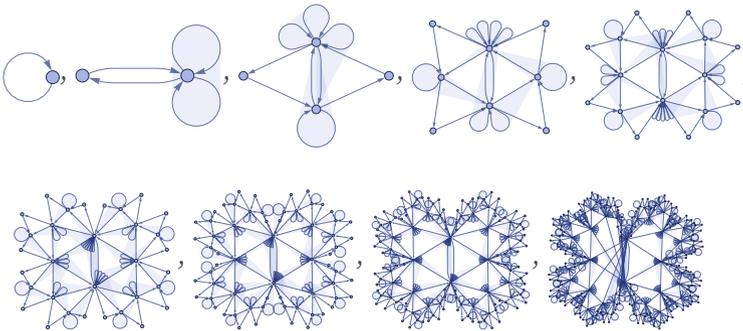

The 44,686 rules with signature $1_2 \to 2_3 1_2$ cannot even produce trees. Rules with signature $1_3 \to 2_3 1_2$ can produce trees, as well as closed structures similar to those seen in $1_2 \to 1_3 2_2$ rules.

A minimal way to add mixed arity to the left-hand sides of rule is to introduce unary relations—but the presence of these seems to inhibit the production of any more complex forms of behavior.

Looking at mixed binary and ternary left-hand sides, none of the 1,141,692 rules with signature $1_3 1_2 \to 1_3 2_2$ seem to produce even trees. But rules with signature $1_3 1_2 \to 2_3 2_2$ readily produce structures such as:

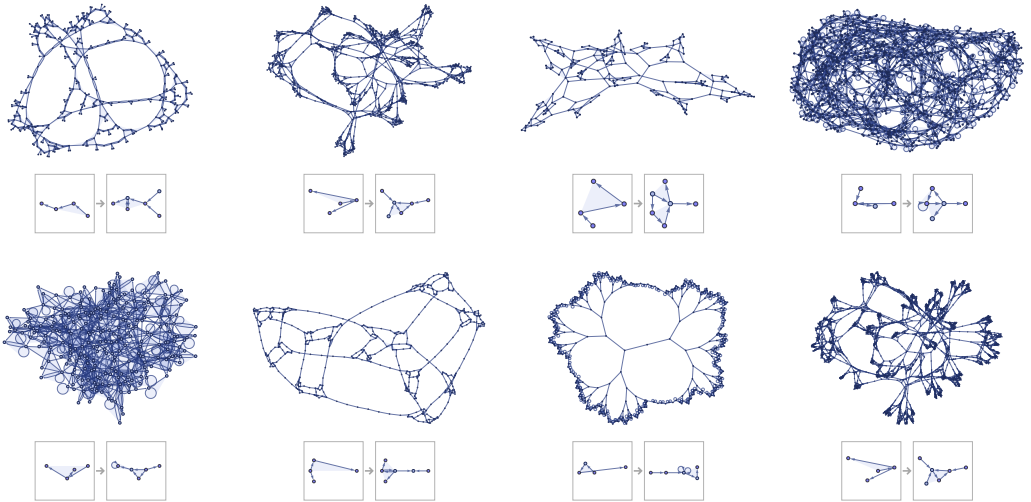

One can go on and look at rules with higher signatures, and probably the most notable finding is that—in keeping with the Principle of Computational Equivalence [1:c12]—the overall behavior seen does not appear to change at all. Here are nevertheless a few examples of slightly unusual behavior found in $2_31_2 \to 3_32_2$ and $2_31_2 \to 4_34_2$ rules:

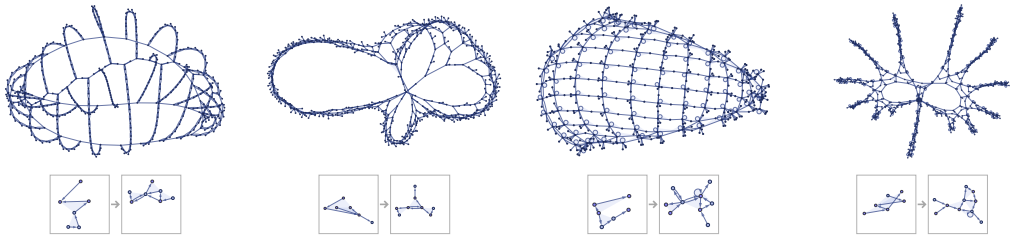

## 3.13 Multiple Transformation Rules

So far we have always considered having just a single possible transformation rule which can be used wherever it applies. It is also possible to have multiple transformation rules which are used wherever they apply. A single transformation rule can either increase or decrease the number of relations, but must do the same every time it is used. With multiple transformations, some can increase the number of relations while others decrease it.

As a minimal example, consider the rule

{{{x, x}} → {{y, x}, {x, z}}, {{x, y}, {y, z}} → {{x, x}}}

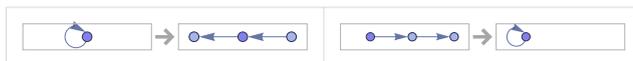

On successive steps, this rule simply alternates between two cases:

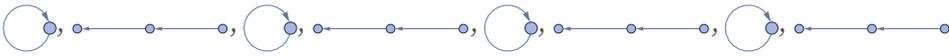

As another example, consider the rule

{{{x, x}} → {{y, x}, {y, x}, {z, x}}, {{x, y}, {z, y}} → {{y, y}}}

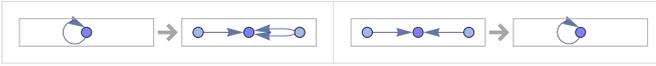

This rule produces results that alternately grow and shrink on successive steps:

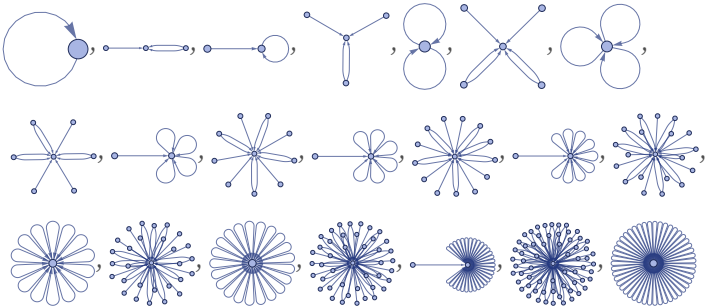

It is fairly common with multiple transformation rules to find that one transformation is occasionally applied. But at least with our standard updating order, it is difficult to find rules in which, for example, the total size of the results varies in anything but a fairly regular way from step to step.

## 3.14  Rules Involving Disconnected Pieces

We have mostly restricted ourselves so far to cases where the results generated by a rule remain connected. But in fact if one looks at all possible rules the majority generate disconnected pieces, or at least can do so for certain initial conditions. Among the 73 rules with signature $1_2 \to 2_2$, only 33 generate connected results starting from initial condition {{0,0}} (and a further 10 terminate from this initial condition).

(Note that we are ignoring order in determining connectivity, so that, for example, the relation {1,2} is considered connected not only to {2,3} but also to {1,3}. Translating binary relations like these into directed edges in a graph, this means we are considering weak connectivity, or, equivalently, we are looking only at the undirected version of the graph.)

Most $1_2 \rightarrow 2_2$ rules that yield disconnected results essentially just produce exponentially more copies of the same structure:

$\{\{x, y\}\} \rightarrow \{\{y, z\}, \{y, z\}\}$

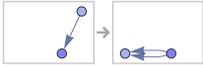

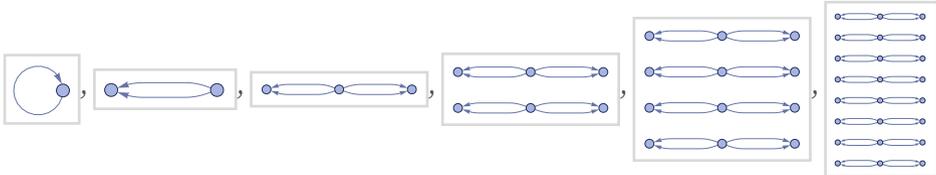

(Note that this rule is an example of one that yields disconnected results even though the rule itself is not disconnected.)

A few rules show slightly more complicated behavior. Examples are (wm575, wm879):

$\{\{x, y\}\} \rightarrow \{\{y, y\}, \{x, z\}\}$

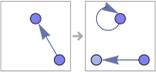

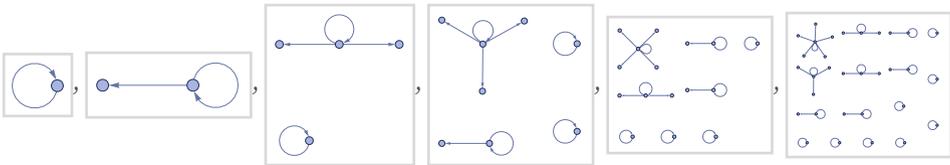

$\{\{x, y\}\} \rightarrow \{\{x, x\}, \{y, z\}\}$

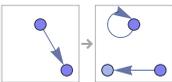

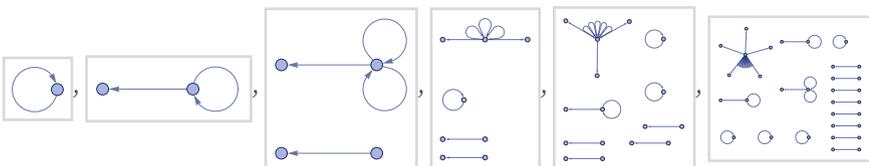

Both these rules still show exponentially increasing numbers of connected components. In the first case, at step $t$ there are components of all sizes 1 through $t$, in exponentially decreasing numbers. In the second case, the size of the largest component is

1, 2, 2, 3, 4, 6, 9, 14, 22, 35, 56, 90, 145, 234, 378, ...

or asymptotically $\sim \phi^n$ (it follows the recurrence f[$n$]=2f[$n$–1]–f[$n$–3]).

Note that if one tracks only the largest component, one gets a sequence of results that could only be generated by a rule involving several separate transformations (in this case {{x,x}}→{{x,y},{x,x}} and {{x,y}}→{{x,x}}). In general, with a single transformation, the total number of relations must either always increase or always decrease. But if there are disconnected pieces, and one tracks, say, only the largest component, one can get a sequence of results that can both increase and decrease in size.

As an example, consider the rule:

{{x, y}, {x, z}} → {{y, z}, {z, y}, {x, w}}

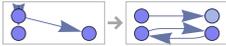

Evolving this rule with our standard updating order gives:

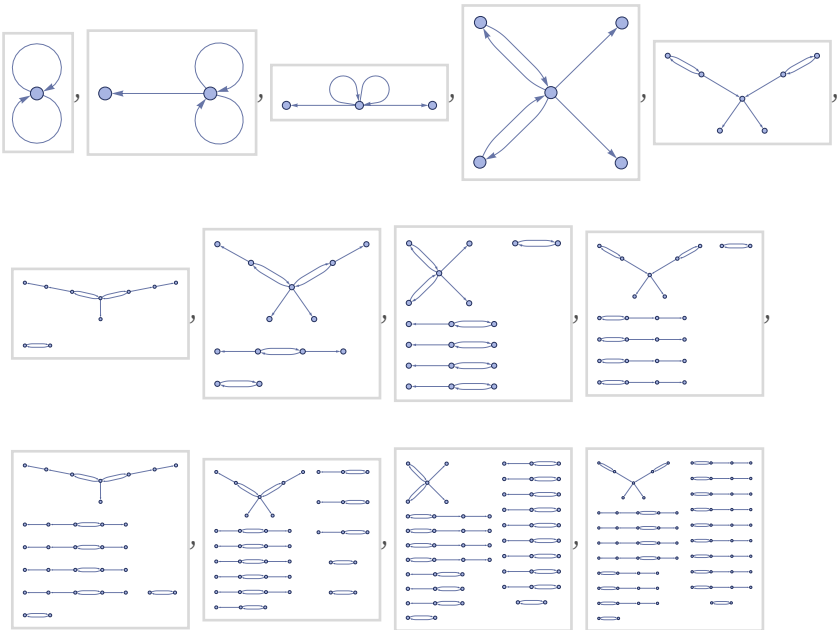

The total number of relations increases roughly exponentially. But tracing only the largest component, we see that it oscillates in size, eventually settling into the cycle 5,8,9,8:

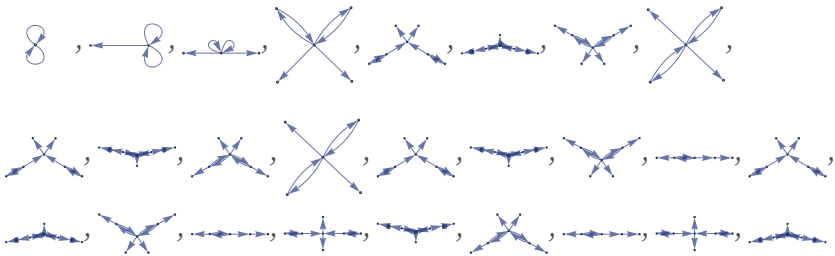

Note that this result is quite specific to the use of our standard updating order. A random updating order, for example, will typically give larger results for the largest component, and no cycle will normally be seen.

It is quite common to see rules that sometimes yield connected results, and sometimes do not. (In fact, proving that a given rule in a given case can never generate disconnected components can be arbitrarily difficult.) Sometimes there can be a large component with a complex structure, with small disconnected pieces occasionally getting "thrown off". Consider for example the rule:

{{x, y}, {y, z}} → {{x, w}, {w, x}, {z, x}}

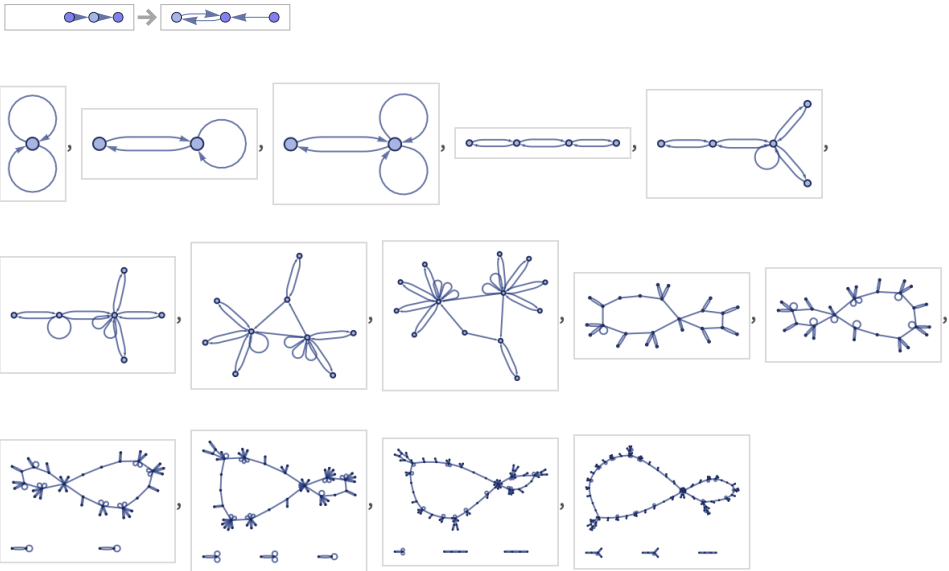

With the standard updating order, it remains connected for 10 steps, then suddenly starts throwing off small disconnected pieces.

As a more elaborate example, consider the rule:

{{x, y, z}, {u, v, z}} → {{y, w, u}, {w, x, y}, {u, y, x}}

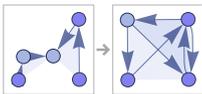

This remains connected for 16 steps, then starts throwing off disconnected pieces:

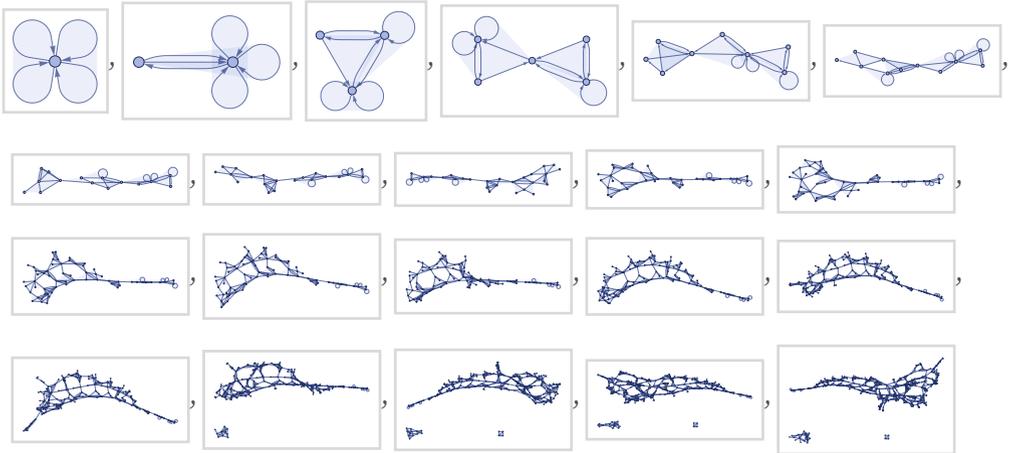

With a rule like this, once components become disconnected, they can in a sense never interact again; their evolutions become completely separate. The only way for disconnected components to interact is to have a rule which itself has a disconnected left-hand side.

For example, a rule like

{{*x*}, {*y*}} → {{*x*, *y*}}

will collect even completely disconnected unary relations, and connect pairs of them into binary relations:

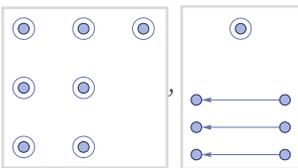

Connected unary relations (i.e. such as {1}, {1}, ...) can end up in the same component, but the result depends critically on the order in which updates are done:

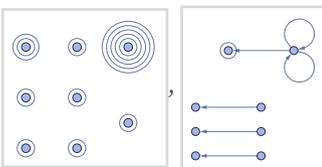

For now, we will not consider any further rules with disconnected left-hand sides—and the extreme nonlocality they represent.

## 3.15 Termination

Not all rules continue to evolve forever from a given initial state. Instead they can reach a fixed point where the rule no longer applies. If the rule depends only on a single relation, this can only happen at the very first step. But if the rule depends on multiple relations, it can happen after multiple steps. Among the 4702 rules with signature $2_2 \to 3_2$, 1788 rules eventually reach a fixed point starting from a self-loop initial condition, at least using our standard updating order. Their "halting time" decreases roughly exponentially, with the maximum being 7 steps, achieved by the rule:

$\{\{x, y\}, \{z, y\}\} \to \{\{y, u\}, \{u, x\}, \{v, z\}\}$

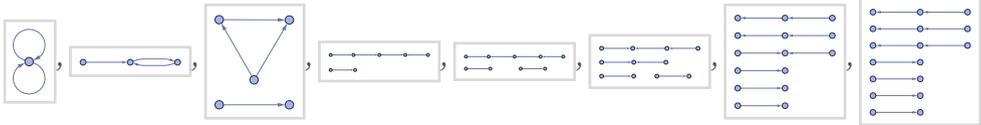

The longest halting time for which connectedness is maintained is 3 steps, achieved for example by:

$\{\{x, y\}, \{y, z\}\} \to \{\{y, z\}, \{y, u\}, \{v, z\}\}$

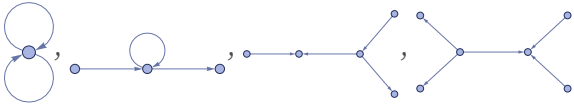

Among the 40,405 $2_2 \to 4_2$ rules, 10,480 evolve to fixed points starting from self-loops. The maximum halting time is 13 steps; the maximum maintaining connectedness is 6 steps, achieved by:

$\{\{x, x\}, \{y, x\}\} \to \{\{y, y\}, \{y, z\}, \{z, x\}, \{w, z\}\}$

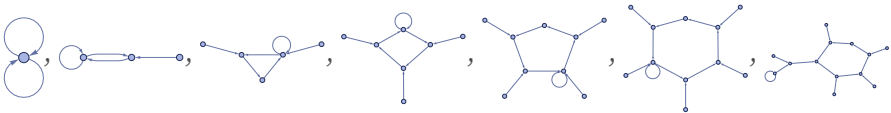

Among the 353,462 $2_2 \to 5_2$ rules, 67,817 (or about 19%) evolve to fixed points. The maximum halting time is 24 steps; the maximum maintaining connectedness is 10 steps, achieved for example by

$\{\{x, x\}, \{y, x\}\} \to \{\{y, y\}, \{y, z\}, \{y, z\}, \{z, x\}, \{w, z\}\}$

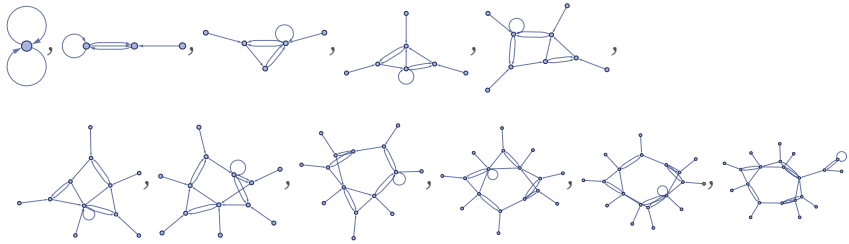

Among $2_3 \to 3_3$ rules

$\{\{x, y, z\}, \{x, u, v\}\} \to \{\{y, x, w\}, \{w, u, s\}, \{v, z, u\}\}$

has halting time 20:

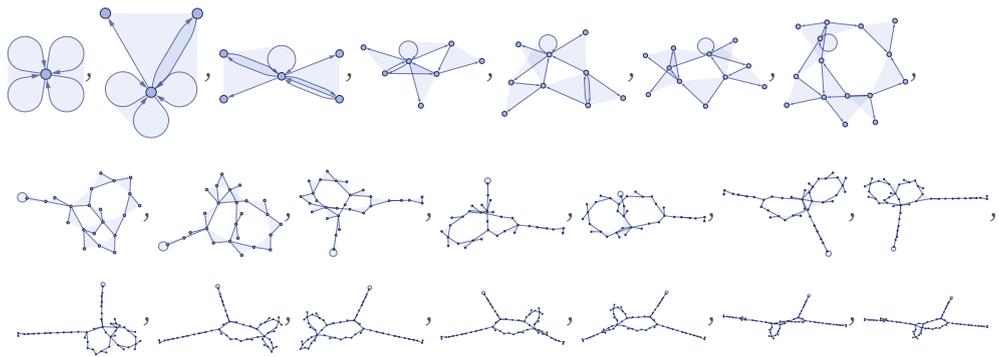

## 3.16 The Effect of Initial Conditions

For rules that depend on only a single relation, adding relations to initial conditions always just leads to replication of identical structures, as in these examples for the rule

$\{\{x, y\}\} \to \{\{y, z\}, \{z, x\}\}$

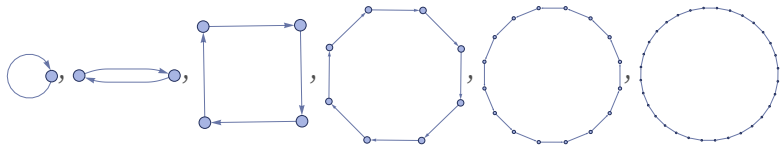

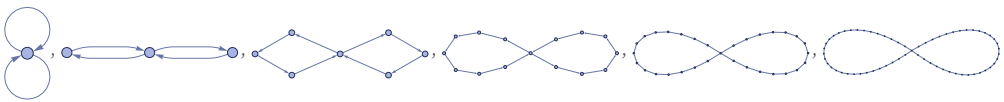

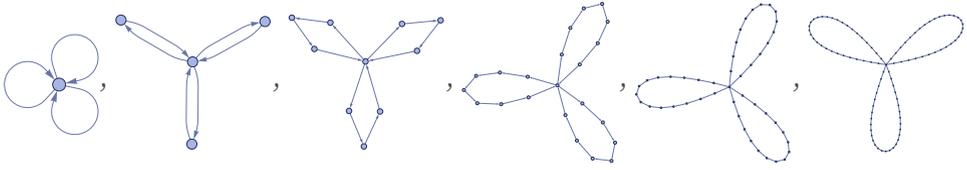

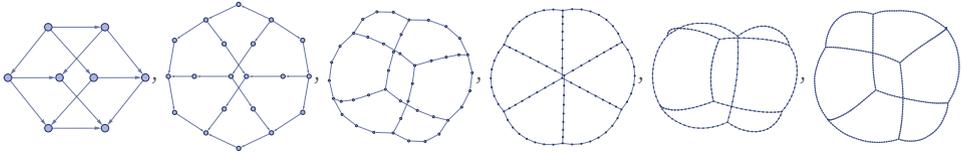

Sometimes, however, the layout of hypergraphs for visualization can make the replication of structures a little less obvious, as in this example for the rule

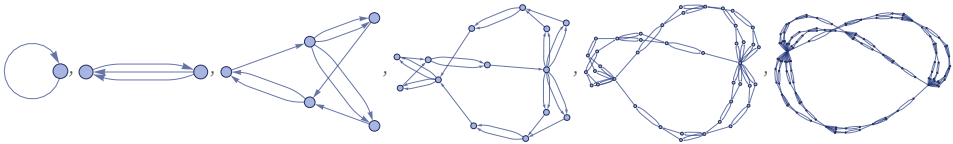

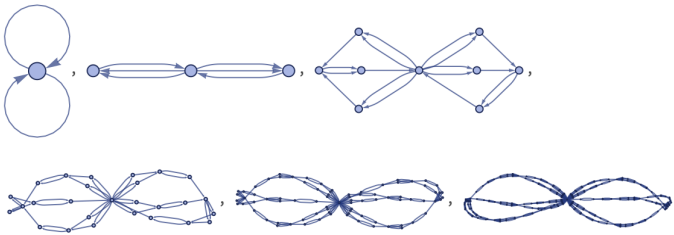

For rules depending on more than one relation, initial conditions can have more important effects. Starting the rule

{{*x*, *y*}, {*y*, *z*}} → {{*w*, *z*}, {*w*, *z*}, {*x*, *w*}, {*y*, *z*}}

from all 8 inequivalent 2-relation and all 32 inequivalent 3-relation initial conditions, one sees quite a range of behavior:

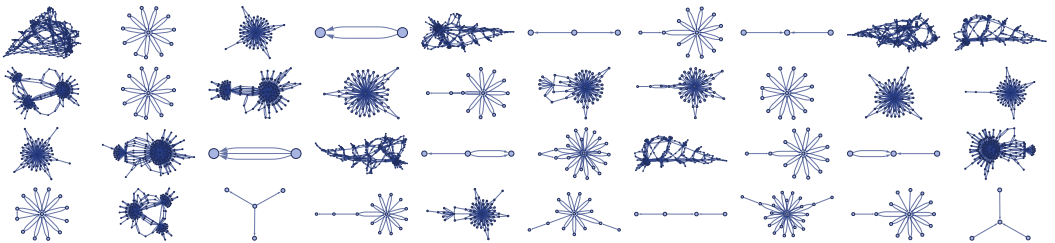

But in other rules—particularly many of those such as

{{*x*, *y*}, {*x*, *z*}} → {{*x*, *y*}, {*x*, *w*}, {*y*, *w*}, {*z*, *w*}}

that yield globular structures—different initial conditions (so long as they lead to growth at all) produce behavior that is different in detail but similar in overall features, a bit like what happens in class 3 cellular automata such as rule 30 [1:p251]):

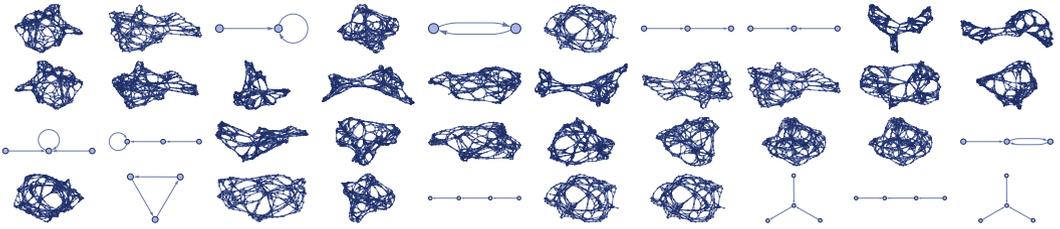

For the evolution of a rule to not just immediately terminate, the left-hand side of the rule must be able to match the initial conditions given (and so must be a sub-hypergraph of the initial conditions). This is guaranteed if the initial conditions are in effect just a copy of the left-hand side. But the most "fertile" initial conditions, with the most possibility for different matches, are always self-loops: in particular, $n$ $k$-ary self-loops for a rule with signature $n_k \to \ldots$. And in what follows, this is the form of initial conditions that we will most often use.

One practical issue with self-loop initial conditions, however, is that they can make it visually more difficult to tell what is going on. Sometimes, for example, initial conditions that lead to slightly less activity, or enforce some particular symmetry, can help. Note, however, that in the evolution of rules that depend on more than one relation, there may be no way to preserve symmetry, at least with any specific updating order (see section 6). Thus, for example, the rule

$\{\{x, y\}, \{x, z\}\} \to \{\{x, y\}, \{x, w\}, \{y, w\}, \{z, w\}\}$

with our standard updating order gives:

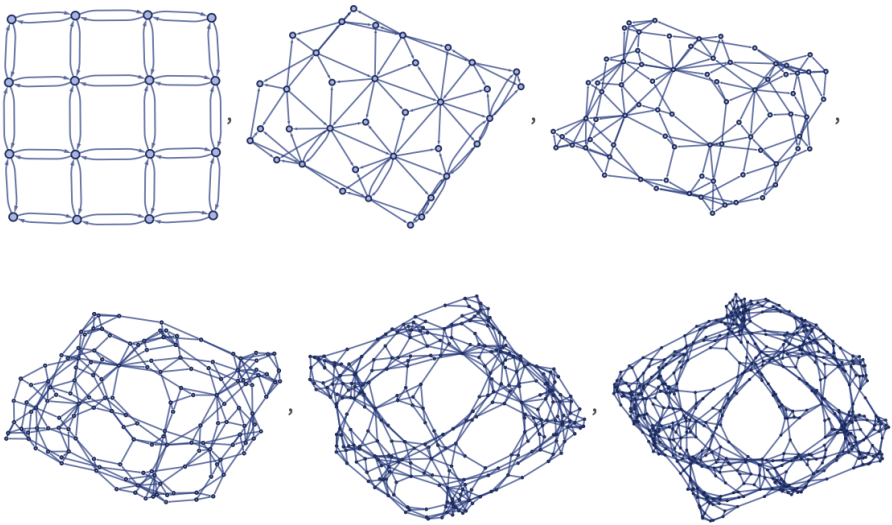

Another feature of initial conditions is that they can affect the connectivity of the results from a rule. Thus, for example, even in the case of the rule above that generates a grid, initial conditions consisting of different numbers of 3-ary self-loops lead to differently connected results:

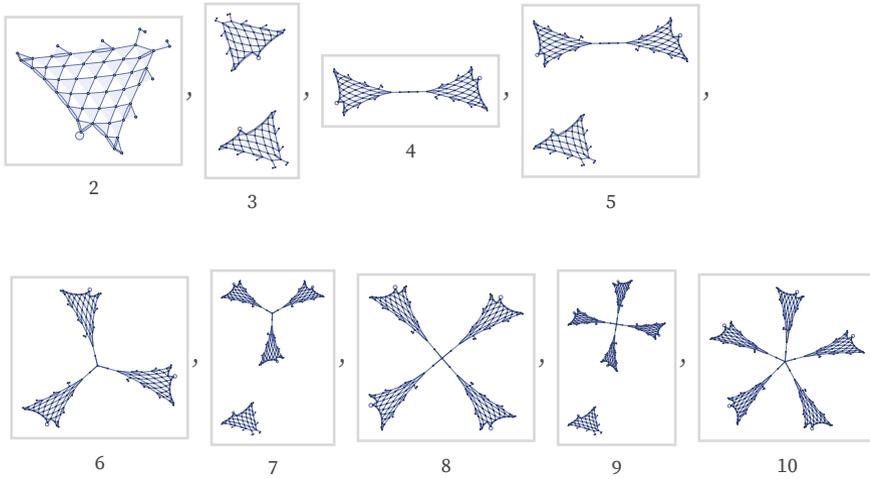

## 3.17 Behavior without Growth

Essentially all the rules we have considered so far have been set up to add relations. But with extended initial conditions, it makes sense also to consider rules that maintain a fixed number of relations.

Rules with signatures like $1_2 \to 1_2$ that depend only on one relation cannot give rise to nontrivial behavior. But rules with signature $2_2 \to 2_2$ (of which a total of 562 are inequivalent) already can.

Consider the rule

$\{\{x, y\}, \{y, z\}\} \to \{\{y, x\}, \{z, x\}\}$

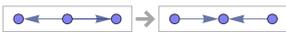

Starting from a chain of 5 binary relations, this is the behavior of the rule:

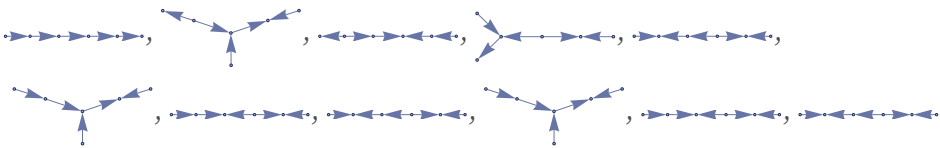

Given that only a finite number of elements and relations are involved, the total number of possible states of the system is finite, so it is inevitable that the evolution of the system must eventually repeat. In the particular case shown here, the repetition period is only 3 steps. (Note that the detailed behavior—and the repetition period—can depend on the updating order used.)

In general, the total number of possible states of the system is given by the number of distinct hypergraphs of a certain size. One can then construct a state transition graph for these states under a rule. Here is the result for the rule above with the 32 distinct connected $3_2$ hypergraphs (note that with this rule, the hypergraphs always remain connected):

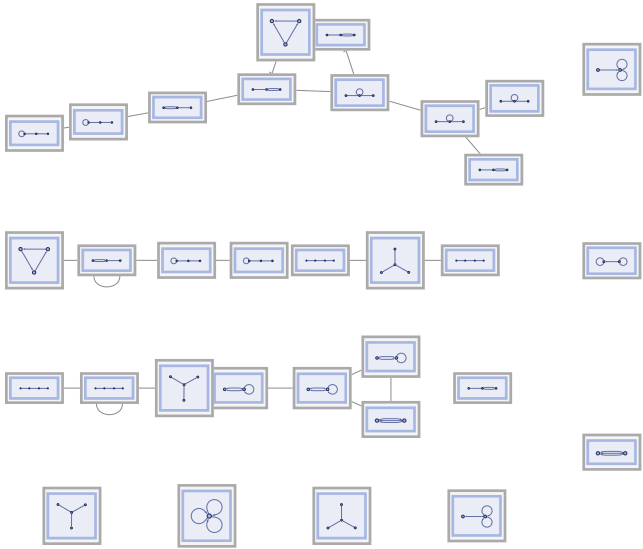

The result for all 928 $5_2$ hypergraphs is:

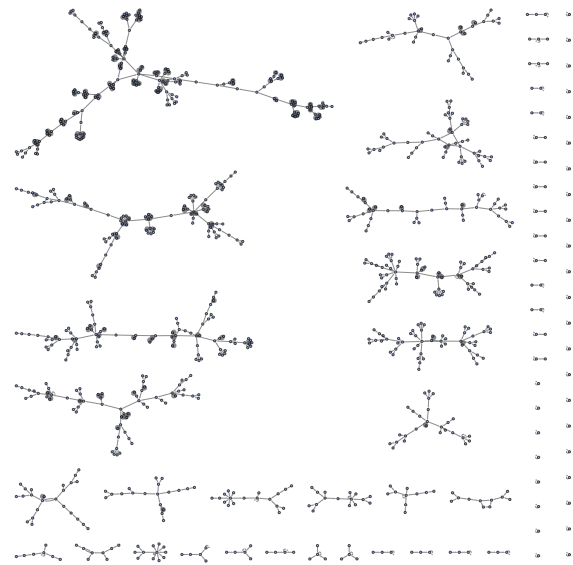

This graph contains trees corresponding to transients, leading into cycles. The maximum cycle length in this case is 5. But when the size of the system increases, the lengths of cycles can increase rapidly (cf. [1:6.4]). The length is bounded by the number of distinct $n_k$ hypergraphs, which grows faster than exponentially with $n$. The plot below shows the lengths of cycles and transients in the rule above for initial conditions consisting of progressively longer chains of relations:

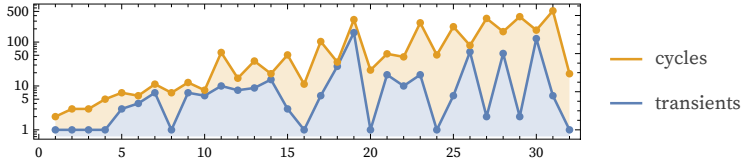

## 3.18 Random Rules and Overall Classification of Behavior

Here are samples of random rules with various signatures (only connected results are included):

| $2_2 \to 3_2$ | |
| $2_2 \to 4_2$ | |
| $2_2 \to 5_2$ | |
| $2_2 \to 6_2$ | |
| $2_2 \to 7_2$ | |
| $2_2 \to 8_2$ | |
| $2_2 \to 9_2$ | |
| $2_2 \to 10_2$ | |

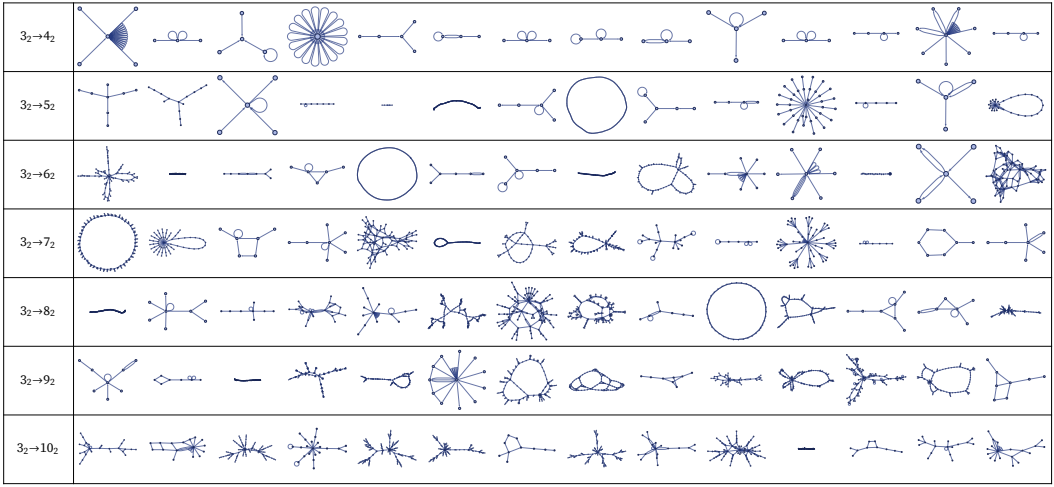

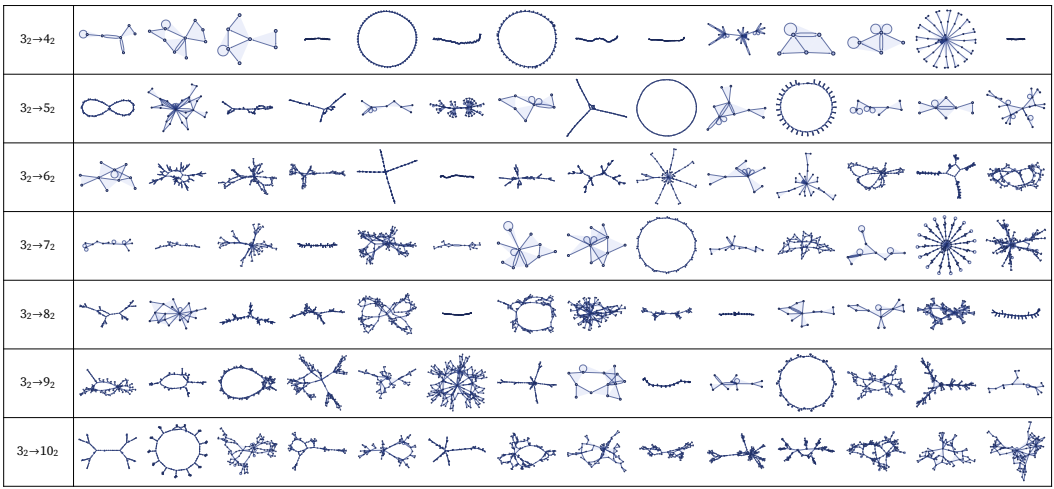

As expected from the Principle of Computational Equivalence [1:c12], above a low threshold more complex rules do not generally lead to more complex behavior, although the frequencies of different kinds of behavior do change somewhat.

At a basic visual level, one can identify several general types of behavior:

• Line-like: elements are connected primarily in sequences (lines, circles, etc.)
• Radial: most elements are connected to just a few core elements
• Tree-like: elements repeatedly form independent branches
• Globular: more complex, closed structures

Inevitably, these types of behavior are neither mutually exclusive, nor precisely defined. There are certainly specific graph-theoretic and other methods that could be used to discriminate different types, but there will always be ambiguous cases (and sometimes it will even be formally undecidable what category something is in). But just like for cellular automata—or for many systems in the empirical sciences—classifications can still be useful in practice even if their definitions are not unique or precise.

As an alternative to categorical classification, one can also consider systematically arranging behaviors in a continuous feature space (e.g. [13]). The results inevitably depend on how features are extracted. Here is what happens if one takes images like the ones above, and directly applies a feature extractor trained on images of picturable nouns in human language [14]:

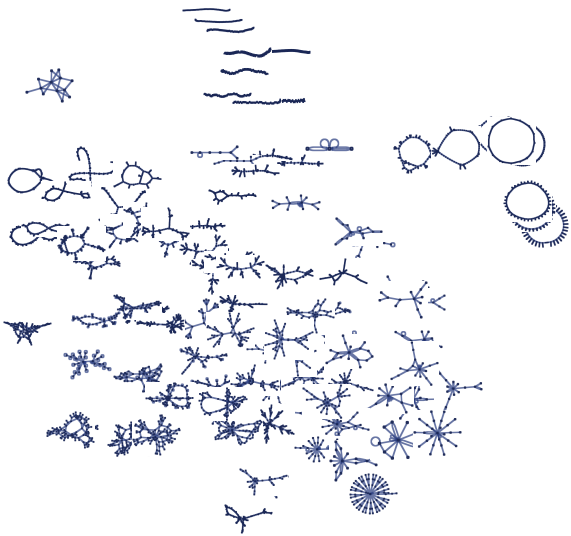

Here is the fairly similar result based on feature extraction of underlying adjacency matrices:

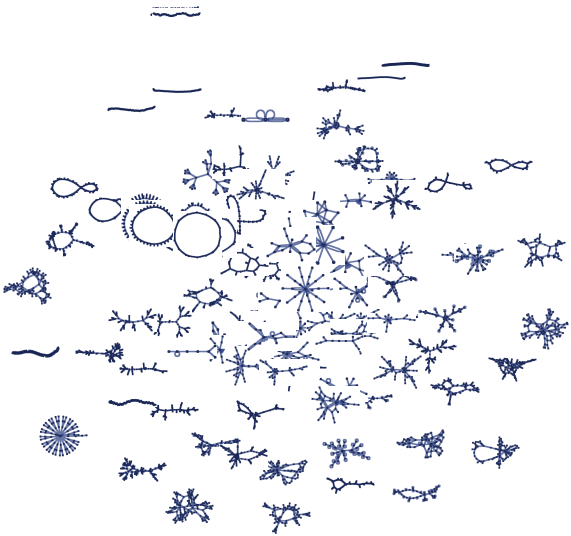

In addition to characterizing the behavior of individual rules, one can also ask to what extent behavior is clustered in rule space. Here are samples of what happens if one starts from particular $2_2 \to 7_2$ rules, then looks at a collection of "nearby" rules that differ by one element in one relation:

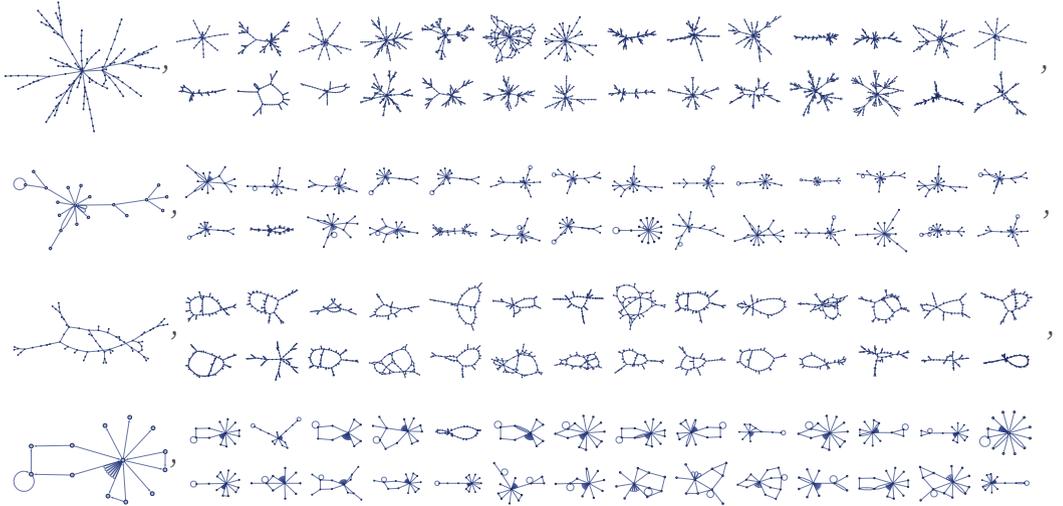

And what we see is that even though there are only 68 million or so $2_2 \to 7_2$ rules, changing one element (out of 14) still usually gives a rule whose overall behavior is similar.

# 4 | Limiting Behavior and Emergent Geometry

## 4.1 Recognizable Geometry

Particularly for potential applications to fundamental physics, it will be of great importance to understand what happens if we run our models for many steps—and to find ways to characterize overall behavior that emerges. Sometimes the characterization is easy. One gets a loop with progressively more links:

{{x, y}} → {{y, z}, {z, x}}

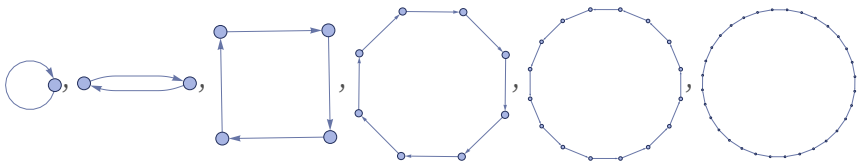

Or one gets a tree with progressively more levels of branching:

{{x}} → {{x, y}, {y}, {y}}

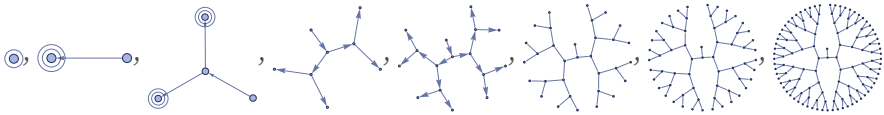

But what about a case like the following? Is there any way to characterize the limiting behavior here?

{{x, y}, {x, z}} → {{x, y}, {x, w}, {y, w}, {z, w}}

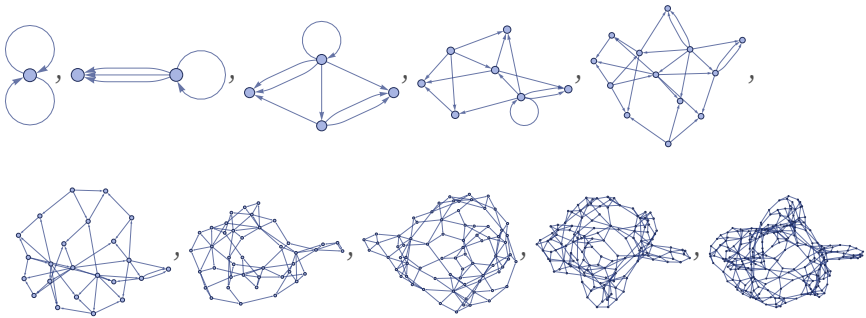

It turns out that a rare phenomenon that we saw in the previous section gives a critical clue. Consider the rule:

{{*x*, *y*, *y*}, {*z*, *x*, *u*}} → {{*y*, *v*, *y*}, {*y*, *z*, *v*}, {*u*, *v*, *v*}}

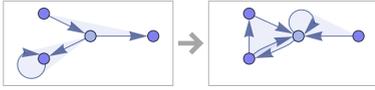

Looking at the first 10 steps it is not clear what it will do:

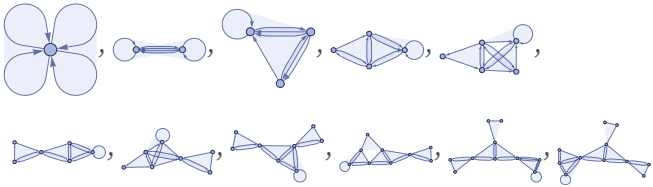

But showing the results every 10 steps thereafter it starts to become clearer:

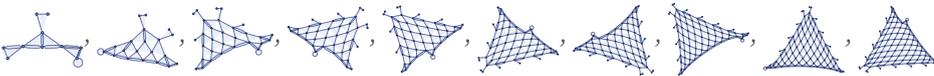

And after 1000 steps it is very clear: the rule has basically produced a simple grid:

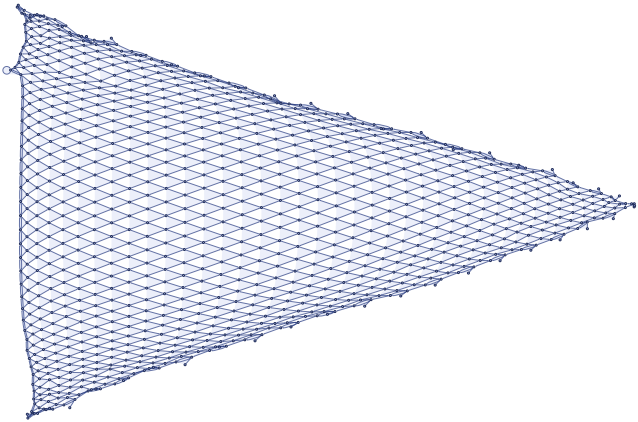

Geometrical though this looks, it is important to understand that at a fundamental level there is no geometry in our model: it just involves abstract collections of relations. Our visualization methods make it evident, however, that the pattern of these relations corresponds to the pattern of connections in a grid.

In other words, from the purely combinatorial structure of the model, what we can interpret as geometrical structure has emerged. And if we continue running the model, the grid in our picture will get finer and finer, until eventually it approximates a triangular piece of continuous two-dimensional space.

Consider now the rule

{{*x*, *x*, *y*}, {*x*, *z*, *u*}} → {{*u*, *u*, *z*}, {*y*, *v*, *z*}, {*y*, *v*, *z*}}

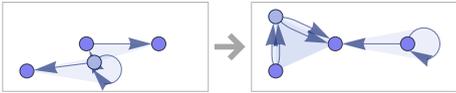

Looking at the first 10 steps of evolution it is again not clear what will happen:

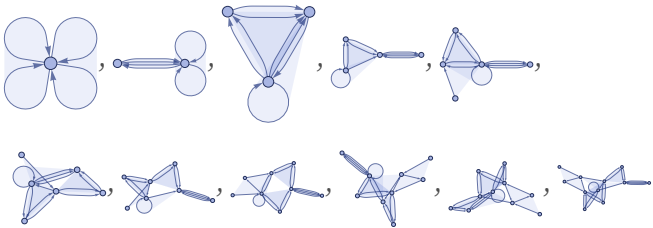

But after 1000 steps a definite geometric structure has emerged:

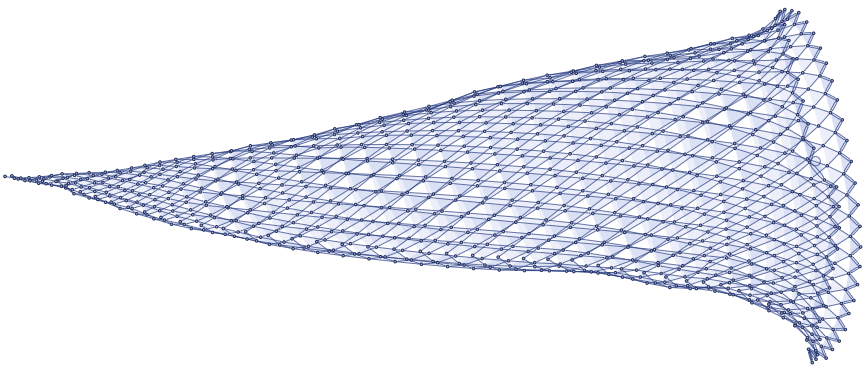

There is evidence of a grid, but now it is no longer flat. Visualizing in 3D makes it clearer what is going on: the grid is effectively defining a 2D surface in 3D:

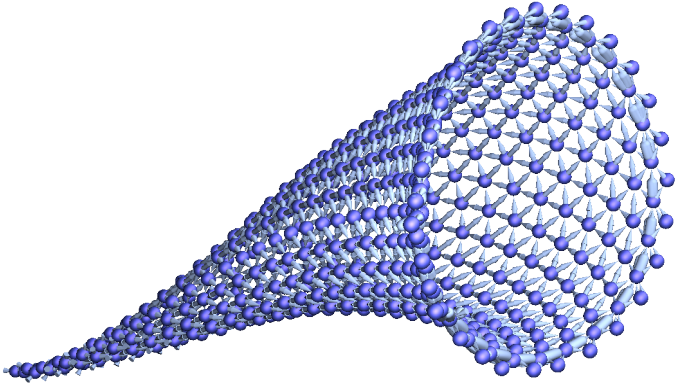

To make its form clearer, we can go for 2000 steps, and include an approximate surface reconstruction [15]:

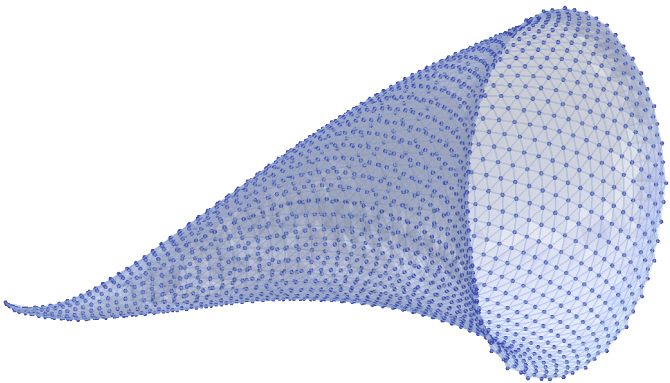

The result is that we can identify that in the limit this rule can be characterized as creating what is essentially a cone.

Other rules produce other shapes. For example, the rule

{{*x*, *y*, *z*}, {*u*, *y*, *v*}} → {{*w*, *z*, *x*}, {*z*, *w*, *u*}, {*x*, *y*, *w*}}

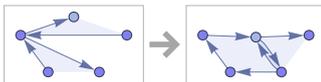

gives after 1000 steps:

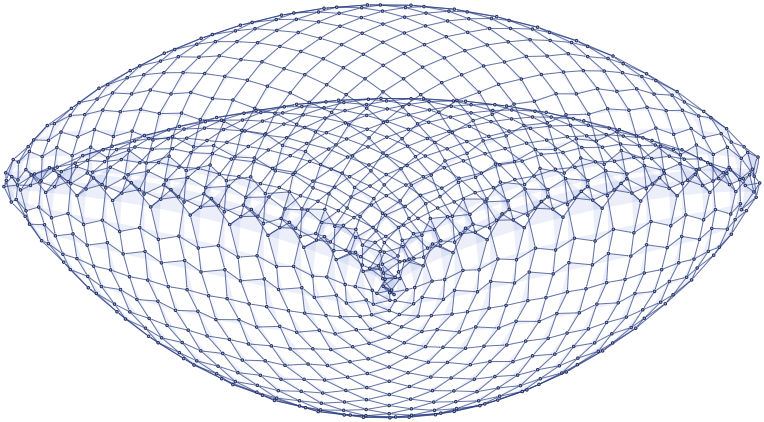

The structure is clearer when visualized in 3D:

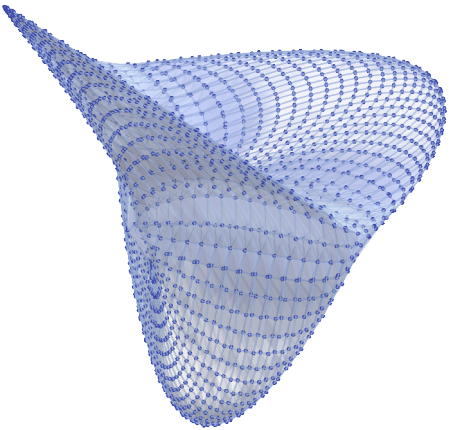

Despite its smooth form, there does not seem to be a simple mathematical characterization of this surface. (Its three-lobed structure means it cannot be an ordinary algebraic surface [16]; it is similar but not the same as the surface $r = \sin(\phi)$ in spherical coordinates.)

Changing the initial condition from 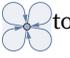 to 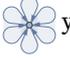 yields the rather different surface:

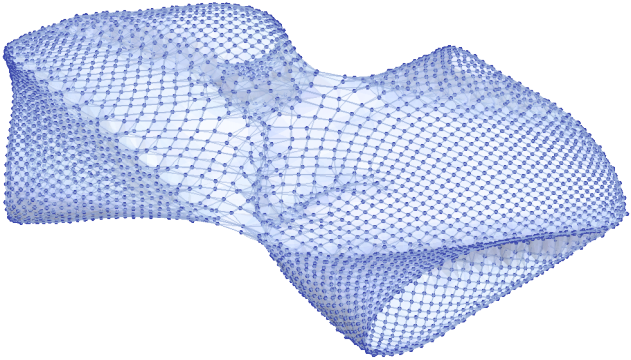

This rule gives a closer approximation to a sphere, though there is a definite threefold structure to be seen:

{{*x*, *y*, *y*}, {*x*, *z*, *u*}} → {{*u*, *v*, *v*}, {*v*, *z*, *y*}, {*x*, *y*, *v*}}

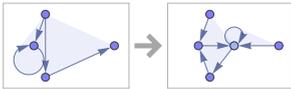

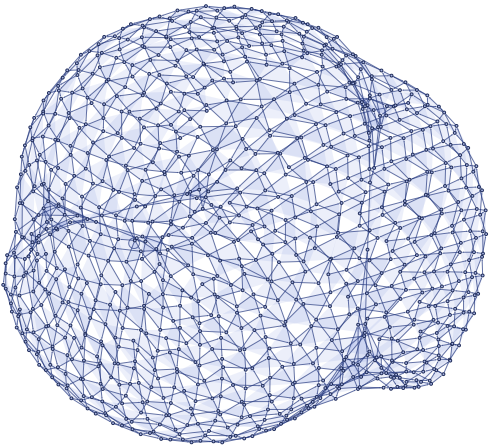

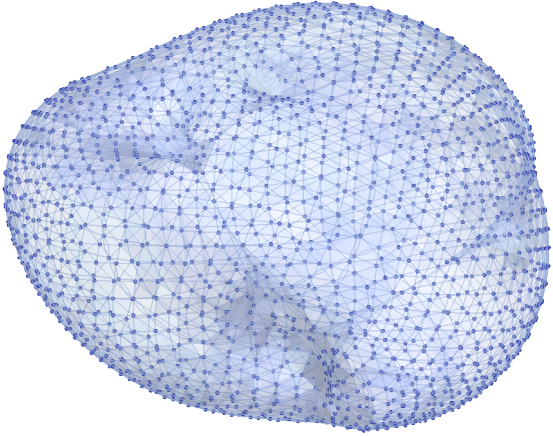

Simpler forms, such as cylindrical tubes, also appear:

{{x, x, y}, {z, u, y}} → {{u, u, y}, {x, v, y}, {x, v, z}}

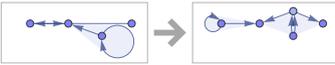

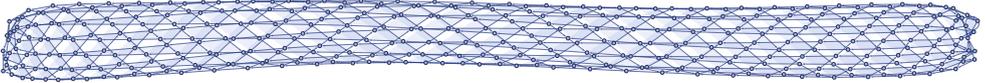

It is worth pausing for a moment to consider in what sense the limiting object here "is" a tube. What we ultimately have is a collection of relations which define a hypergraph. But there is an obvious measure of distance on the hypergraph: how many relations you have to follow to go from one element to another. So now one can ask whether there is a way to make this hypergraph distance between elements correspond to an ordinary geometrical distance. Can one assign positions in space to the elements so that the spatial distances between them agree with the hypergraph distances?

The answer in this case is that one can—by placing the elements at a lattice of positions on the surface of a cylinder in three-dimensional space. (And, conveniently, it so happens that our visualization method for hypergraphs basically automatically does this.) But it is important to realize that such a direct correspondence with an easy-to-describe surface is a rare and special feature of the particular rule used here.

Consider the rule:

{{x, x, y}, {x, z, u}} → {{u, u, v}, {v, u, y}, {z, y, v}}

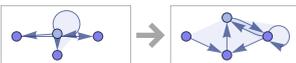

After 1000 steps, this rule produces:

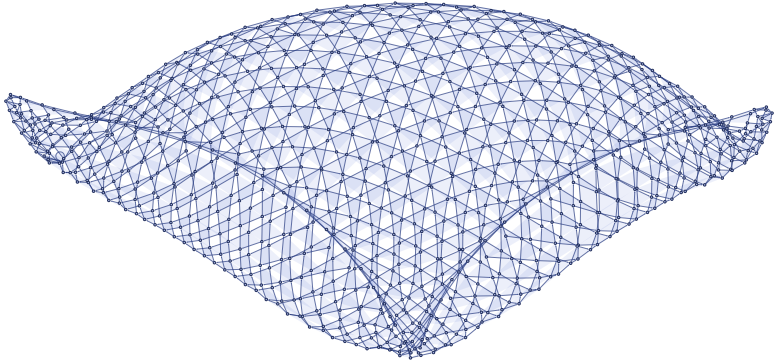

In 3D, this can be visualized as:

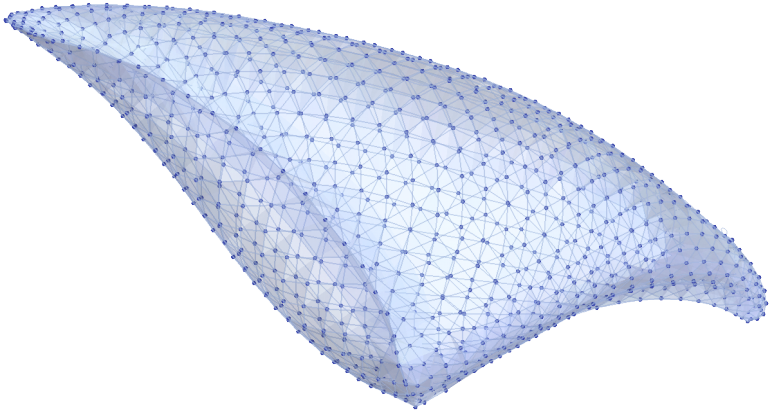

There are many subtle issues here. First, at every step the rule adds more elements, and in principle this could change the emergent geometry. But it appears that after enough steps, there is a definite limiting shape. Unlike in the case of a cylinder, however, it is much less clear how to assign spatial coordinates to different elements. It does not help that the limiting shape does not appear to have a completely smooth surface; instead there are places at which it appears to form cusps (reminiscent of an orbifold [17]).

There are rules that give more obvious "singularities"; an example is:

{{x, x, y}, {y, z, u}} → {{v, v, u}, {v, u, x}, {z, y, v}}

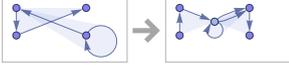

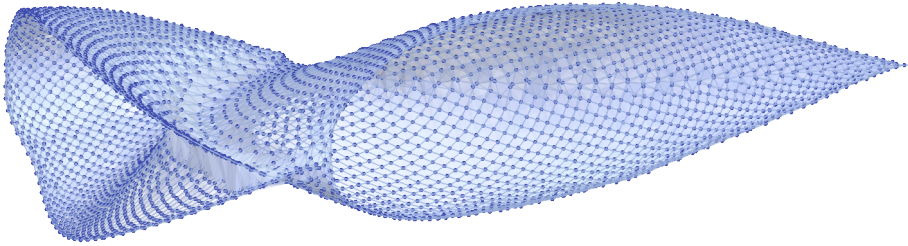

Some rules produce surfaces with complex folds:

{{x, x, y}, {z, u, x}} → {{z, z, v}, {y, v, x}, {y, w, v}}

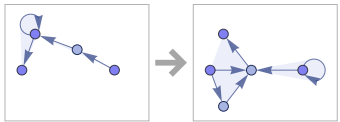

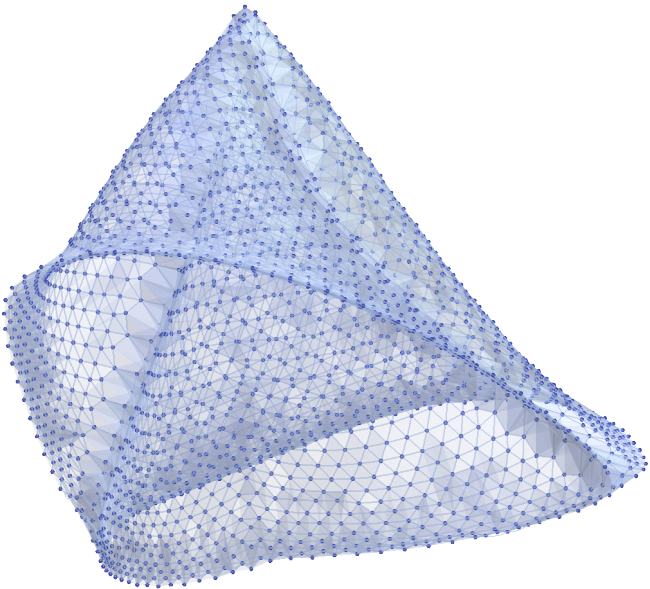

It is also perfectly possible for the emergent geometry to have nontrivial topology. This rule produces a (strangely twisted) torus:

{{*x*, *x*, *y*}, {*z*, *u*, *x*}} → {{*x*, *x*, *z*}, {*u*, *v*, *x*}, {*y*, *v*, *z*}}

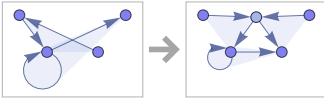

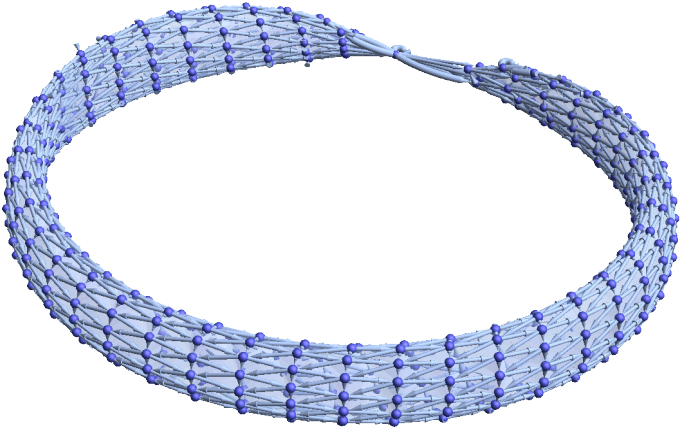

All the emergent geometries we have seen so far in effect involve a regular mesh. But this rule, instead uses a mixture of triangles, quadrilaterals and pentagons to cover a region:

{{*x*, *y*, *x*}, {*x*, *z*, *u*}} → {{*u*, *v*, *u*}, {*v*, *u*, *z*}, {*x*, *y*, *v*}}

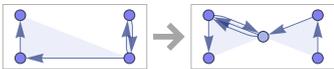

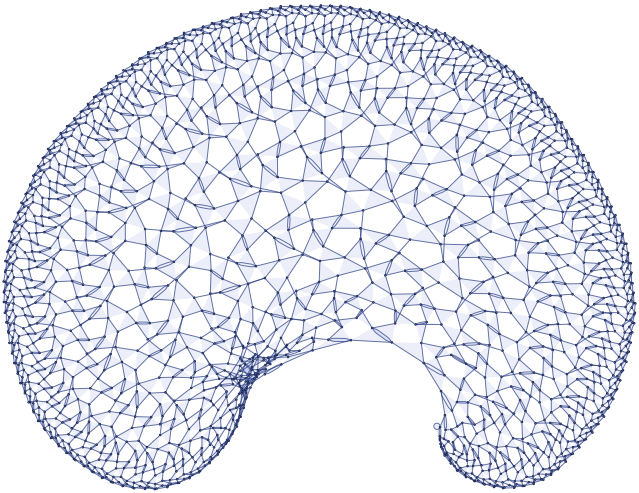

## 4.2 Hyperbolic Space

Among all possible rules, the formation of geometrical shapes of the kind we have just been discussing is very rare. Slightly more common is the type of behavior that we see in a rule like:

{{x, y}, {y, z}} → {{w, x}, {w, y}, {x, y}, {y, z}}

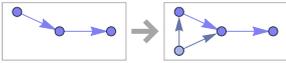

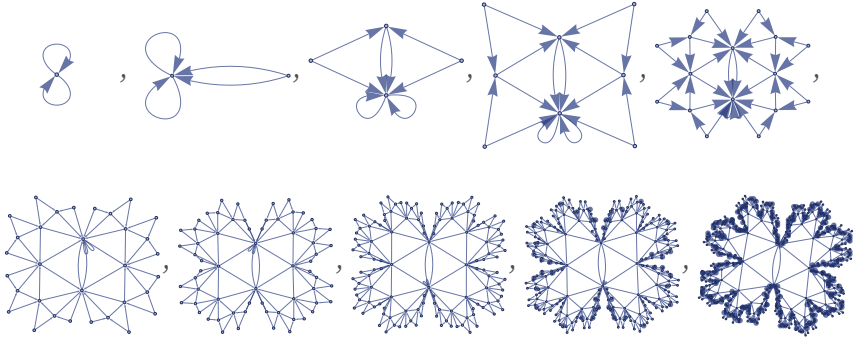

Essentially the same behavior also occurs in a mixed-arity rule with a single relation on the left-hand side:

{{x, y}} → {{z, y, x}, {y, z}, {z, x}}

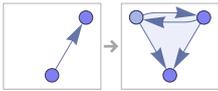

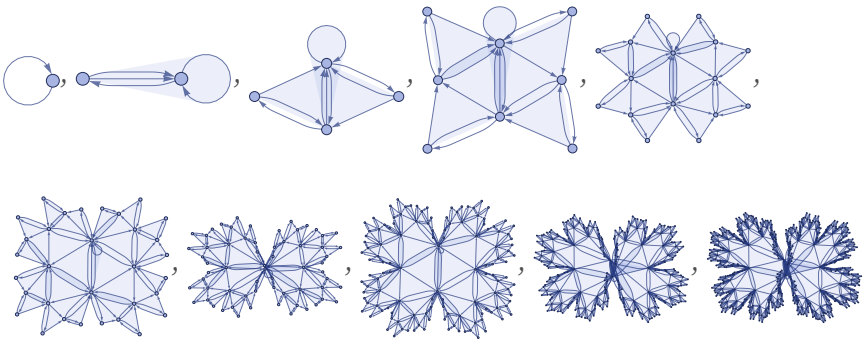

We can think of the structure that is produced as being like a binary tree of triangles:

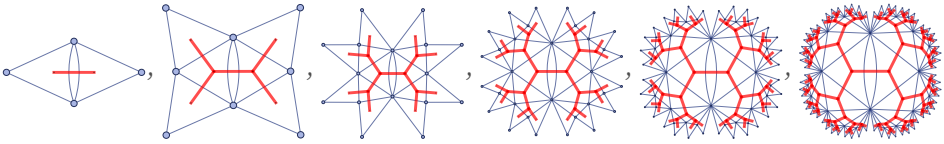

The same structure can be produced from an Apollonian circle packing (e.g. [18][1: p985]):

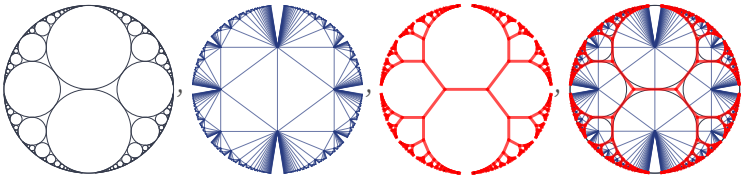

If each triangle is required to have the same area, the structure can be rendered in 2D as:

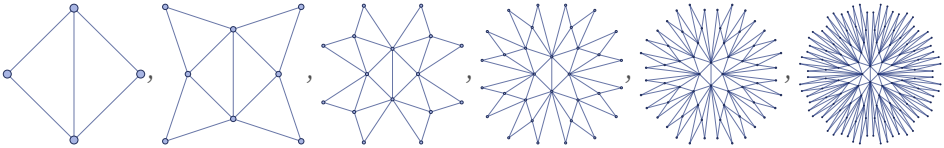

If we tried to render this with every triangle roughly the same size, then even in 3D the best we could do would be to have something that crinkles up more and more at the edge, like an idealized lettuce leaf:

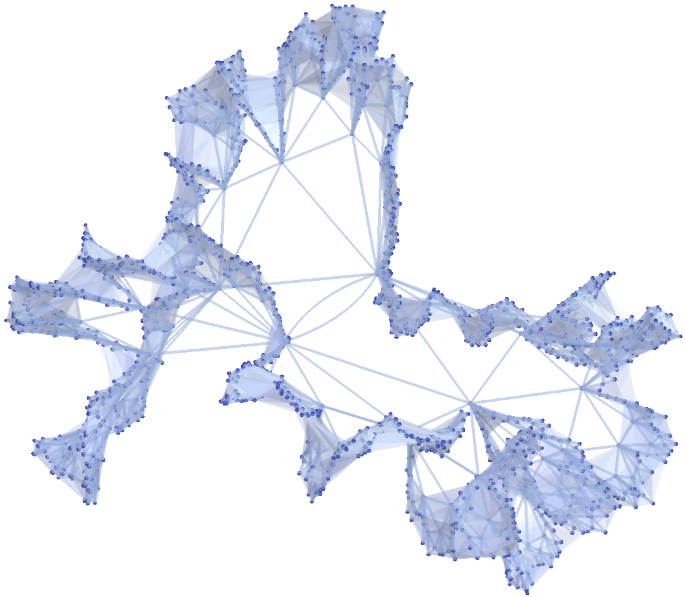

But just as we can think of the grids we discussed before as being regularly laid out in ordinary 2D or 3D space, so now we can think of the object we have here as being regularly laid out in a hyperbolic space [19][20] of constant negative curvature.

In particular, the object corresponds to an infinite-order triangular tiling of the hyperbolic plane (with Schläfli symbol {3,∞}). There are a variety of ways to visualize the hyperbolic plane. One example is the Poincaré disk model in which hyperbolic-space straight lines are rendered as arcs of circles orthogonal to the boundary:

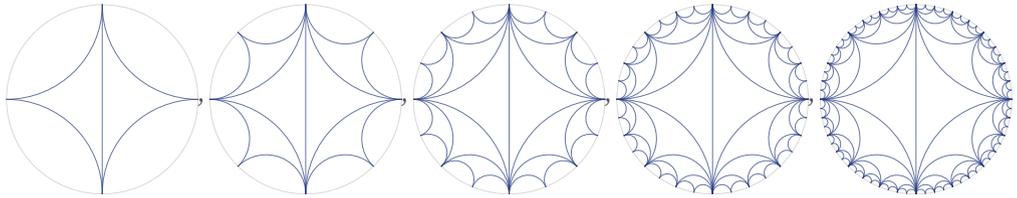

(The particular graph here happens to be the Farey graph [21].)

## 4.3 Geometry from Subdivision

The grids and surfaces that we saw above were all produced by rules that end up executing a laborious "knitting" process in which they add just a single relation at each step. But it is also possible to generate recognizable geometric forms more quickly—in effect by a process of repeated subdivision.

Consider the $2_3 1_2 \to 4_3 4_2$ rule:

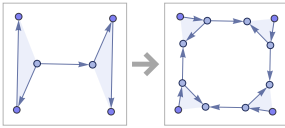

At each step, this rule doubles the number of relations—and quickly produces a structure with a definite emergent geometrical form:

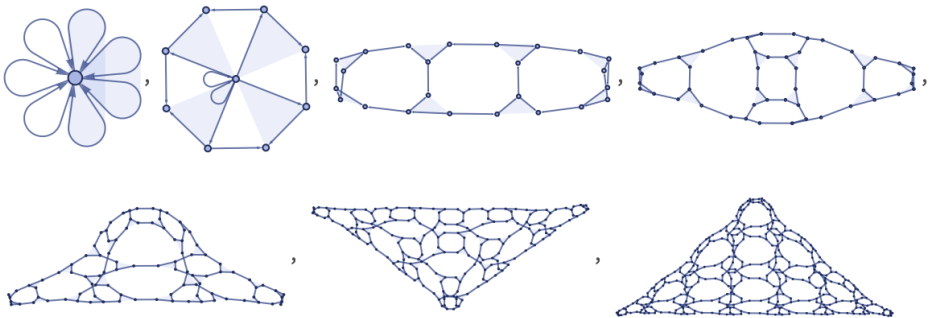

After 10 steps the rule has generated 2560 relations, in the following structure:

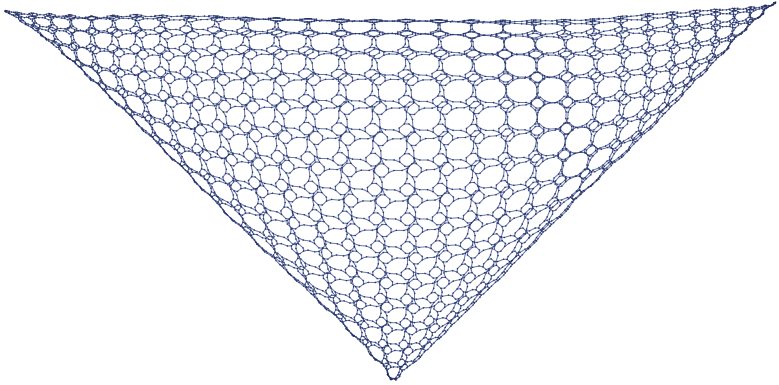

Visualized in 3D, this becomes:

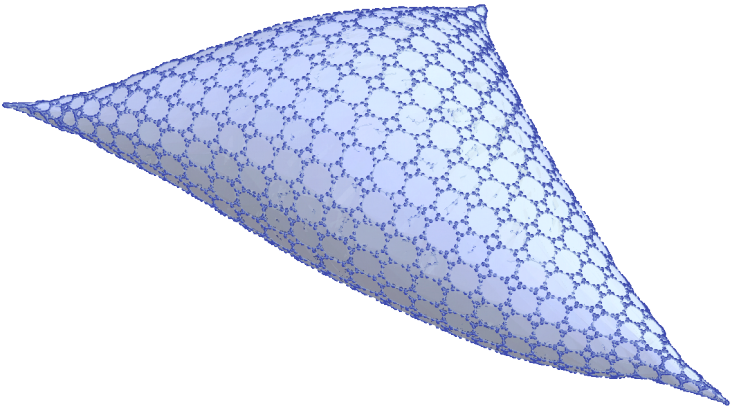

Once again, this corresponds to a smooth surface, but with 3 cusps. The surface is defined not by a simple triangular grid, but instead by an octagon-square ("truncated square") tiling—that in this case becomes twice as fine at every step.

Changing the initial conditions can give a somewhat different structure:

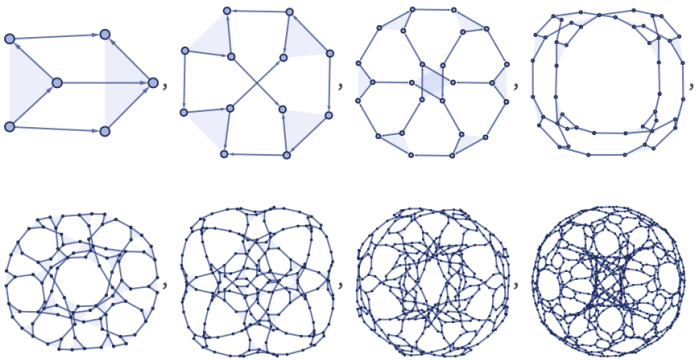

Visualized in 3D after 10 steps (and reconstructing less of the surface), this becomes:

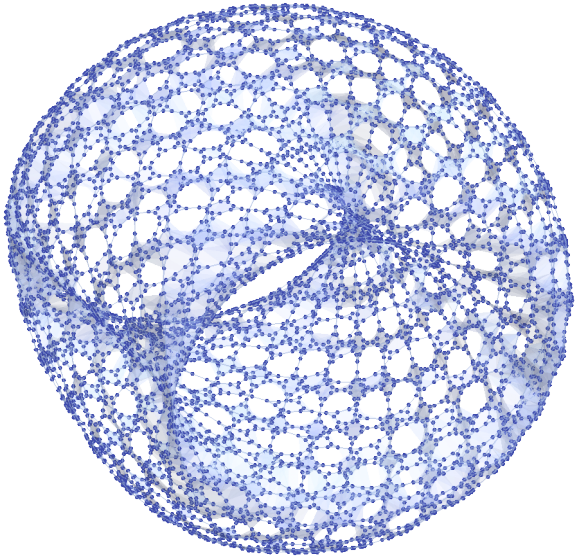

## 4.4 Nested Patterns

Consider the $1_3 \to 3_3$ rule:

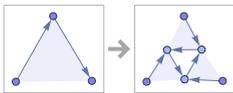

Starting from a single ternary relation with three distinct elements {{1,2,3}}, this gives a classic Sierpiński triangle structure:

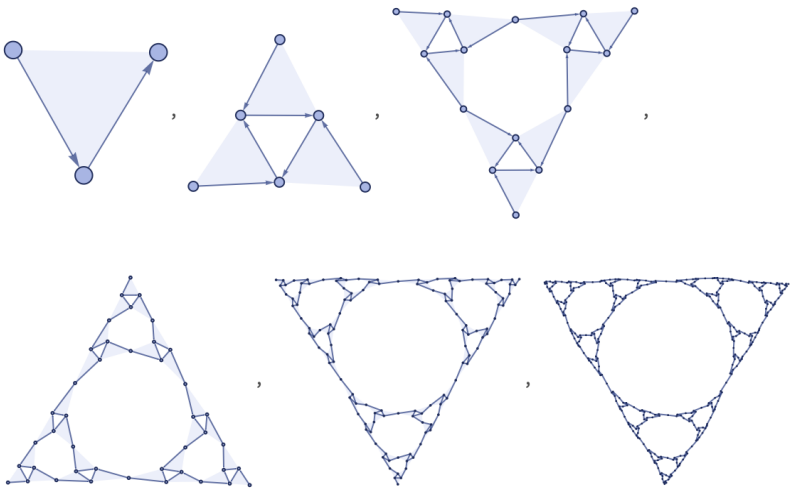

Starting instead from a ternary self loop {{0,0,0}} one gets what amounts to a tetrahedron of Sierpiński triangles:

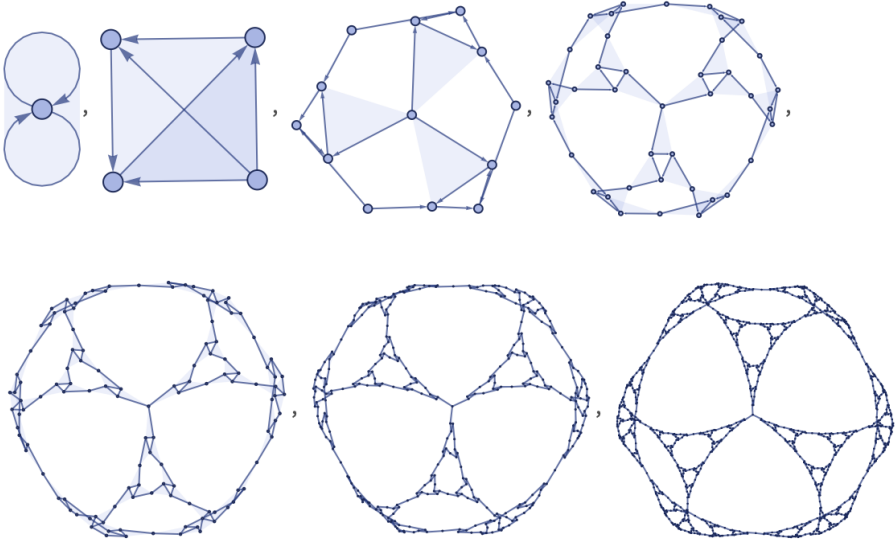

This is exactly the same as one would get by starting with a tetrahedron graph, and repeatedly replacing every trivalent vertex with a triangle of vertices [1:p509]:

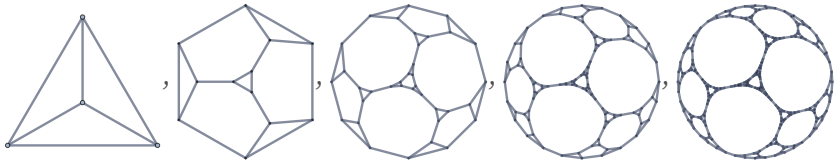

In an ordinary Sierpiński triangle, the points on the edges have different neighborhoods from those in the interior. But in the structure shown here, all points have the same neighborhoods (so there is an isometry).

Many of the rules we have used have completely different behavior if the order of elements in their relations are changed. But in this case the limiting shape is always the same, regardless of ordering, as in these examples:

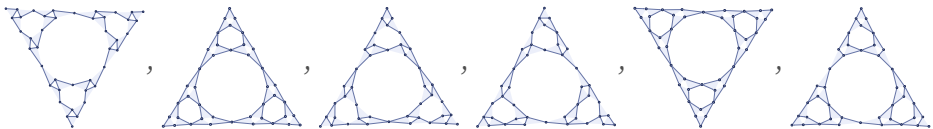

The rule we have discussed so far in this section in a sense directly implements the recursive construction of nested patterns [1:5.4]. But the formation of nested patterns is also a common feature of the limiting behavior of many rules that do not exhibit any such obvious construction.

As an example, consider the $1_3 \to 2_3$ rule

$\{\{x, y, z\}\} \to \{\{z, w, w\}, \{y, w, x\}\}$

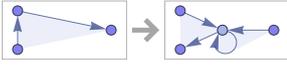

This rule effectively constructs a nested sequence of self-similar "segments":

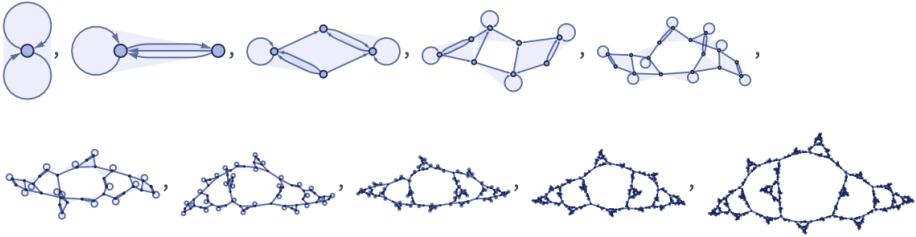

Similar behavior is seen in rules with binary relations, such as the $1_2 \to 4_2$ rule:

$\{\{x, y\}\} \to \{\{z, w\}, \{z, x\}, \{w, x\}, \{y, w\}\}$

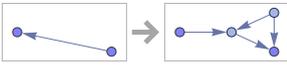

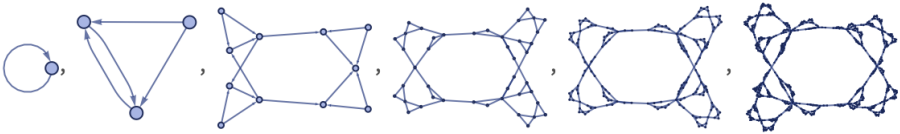

A clear "naturally occurring" Sierpiński pattern appears in the limiting behavior of the $2_2 \to 4_2$ rule

$\{\{x, y\}, \{z, y\}\} \to \{\{y, w\}, \{y, w\}, \{w, x\}, \{z, w\}\}$

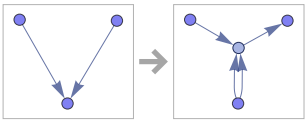

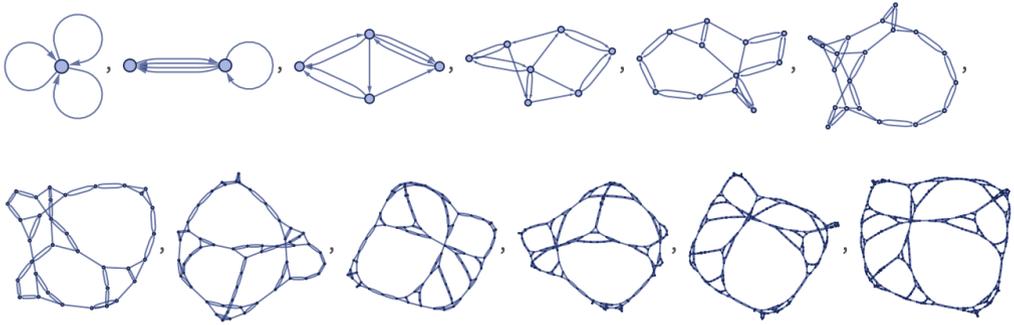

After 15 steps, the rule yields:

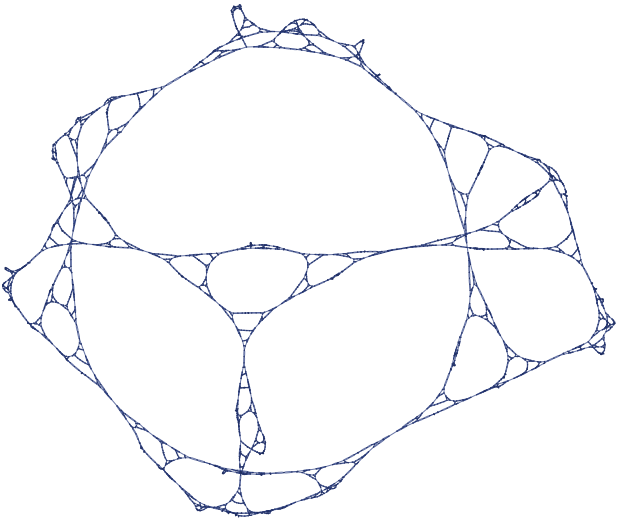

## 4.5 The Notion of Dimension

In traditional geometry, a basic feature of any continuous space is its dimension. And we have seen that at least in certain cases we can characterize the limiting behavior of our models in terms of the emergence of recognizable geometry—with definite dimension. So this suggests that perhaps we might be able to use a notion of dimension to characterize the limiting behavior of our models even when we do not readily recognize traditional geometrical structure in them.

For standard continuous spaces it is straightforward to define dimension, normally in terms of the number of coordinates needed to specify a position. If we make a discrete approximation to a continuous space, say with a progressively finer grid, we can still identify dimension in terms of the number of coordinates on the grid. But now imagine we only have a connectivity graph for a grid. Can we deduce what dimension it corresponds to?

We might choose to draw the grids so they lay out according to coordinates, here in 1-, 2- and 3-dimensional Euclidean space:

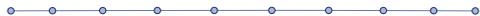

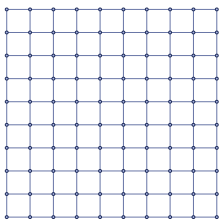

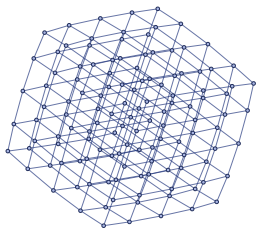

But these are all the same graph, with the same connectivity information:

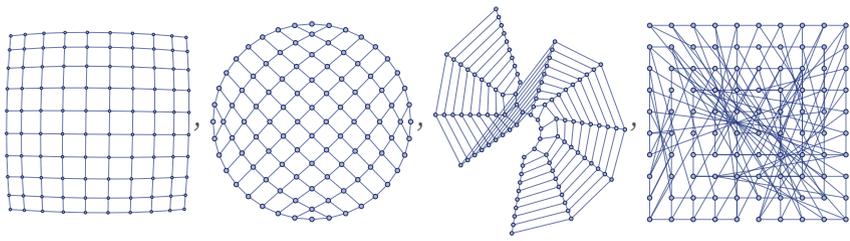

So just from intrinsic information about a graph—or, more accurately, from information about a sequence of larger and larger graphs—can we deduce what dimension of space it might correspond to?

The procedure we will follow is straightforward (cf. [1:p479][22]). For any point $X$ in the graph define $V_r(X)$ to be the number of points in the graph that can be reached by going at most graph distance $r$. This can be thought of as the volume of a ball of radius $r$ in the graph centered at $X$.

For a square grid, the region that defines $V_r(X)$ for successive $r$ starting at a point in the center is:

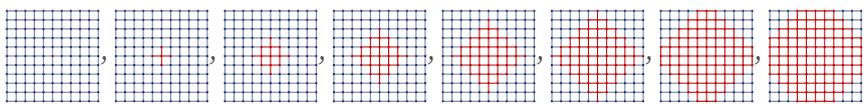

For an infinite grid we then have:

$V_r = 2r^2 + 2r + 1$

For a 1D grid the corresponding result is:

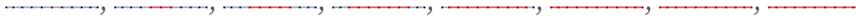

$V_r = 2r + 1$

And for a 3D grid it is:

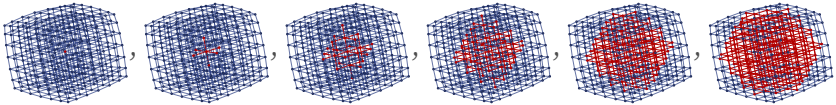

$V_r = \dfrac{4r^3}{3} + 2r^2 + \dfrac{8r}{3} + 1$

In general, for a $d$-dimensional cubic grid (cf. [1:p1031]) the result is a terminating hypergeometric series (and the coefficient of $z^d$ in the expansion of $(z+1)^r/(z-1)^{r+1}$):

$${}_2F_1(-d, r+1; -d+r+1; -1) \binom{r}{d} = \dfrac{2^d}{d!} r^d + \dfrac{2^{d-1}}{(d-1)!} r^{d-1} + \dfrac{2^{d-2}(d+1)}{3(d-2)!} r^{d-2} + \ldots$$

But the important feature for us is that the leading term—which is computable purely from connectivity information about the graph—is proportional to $r^d$.

What will happen for a graph that is less regular than a grid? Here is a graph made by random triangulation of a 2D region:

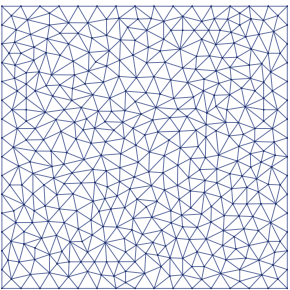

And once again, the number of points reached at graph distance $r$ grows like $r^2$:

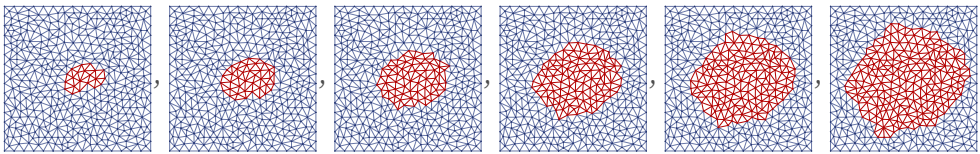

In ordinary $d$-dimensional continuous Euclidean space, the volume of a ball is exactly

$$\frac{\pi^{d/2}}{(d/2)!} r^d$$

And we should expect that if in some sense our graphs limit to $d$-dimensional space, then in correspondence with this, $V_r$ should always show $r^d$ growth.

There are, however, many subtle issues. The first—immediately evident in practice—is that if our graph is finite (like the grids above) then there are edge effects that prevent $r^d$ growth in $V_r$ when the radius of the ball becomes comparable to the radius of the graph. The pictures below show what happens for a grid with side length 11, compared to an infinite grid, and the $r^d$ term on its own:

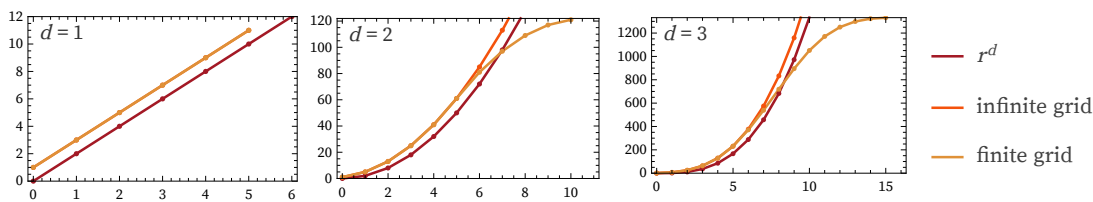

One might imagine that edge effects would be avoided if one had a toroidal grid graph such as:

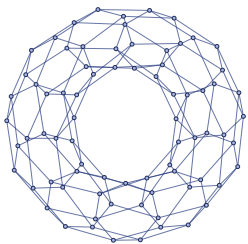

But actually the results for $V_r(X)$ for any point on a toroidal graph are exactly the same as those for the center point in an ordinary grid; it is just that now finite-size effects come from paths in the graph that wrap around the torus.

Still, so long as $r$ is small compared to the radius of the graph—but large enough that we can see overall $r^d$ growth—we can potentially deduce an effective dimension from measurements of $V_r$.

In practice, a convenient way to assess the form of $V_r$, and to make estimates of dimension, is to compute log differences as a function of $r$:

$$\Delta(r) = \frac{\log(V_{r+1}) - \log(V_r)}{\log(r+1) - \log(r)}$$

Here are results for the center points of grid graphs (or for any point in the analogous toroidal graphs):

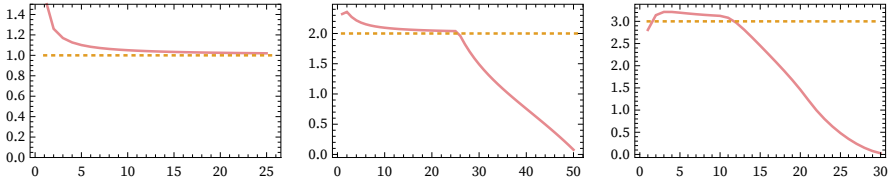

The results are far from perfect. For small *r* one is sensitive to the detailed structure of the grid, and for large *r* to the finite overall size of the graph. But, for example, for a 2D grid graph, as the size of the graph is progressively increased, we see that there is an expanding region of values of *r* at which our estimate of dimension is accurate:

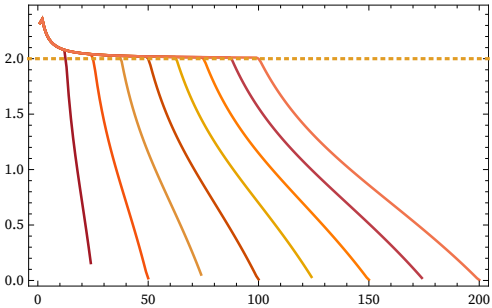

A notable feature of measuring dimension from the growth rate of $V_r(X)$ is that the measurement is in some sense local: it starts from a particular position $X$. Of course, in looking at successively larger balls, $V_r(X)$ will be sensitive to parts of the graph progressively further away from $X$. But still, the results can depend on the choice of $X$. And unless the graph is homogeneous (like our toroidal grids above), one will often want to average over at least a range of possible positions $X$. Here is an example of doing such averaging for a collection of starting points in the center of the random 2D graph above. The error bars indicate $1\sigma$ ranges in the distribution of values obtained from different points $X$.

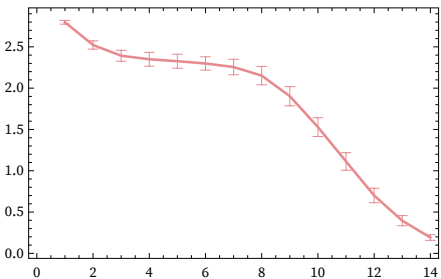

So far we have looked at graphs that approximate standard integer-dimensional spaces. But what about fractal spaces [23]? Let us consider a Sierpiński graph, and look at the growth of a ball in the graph:

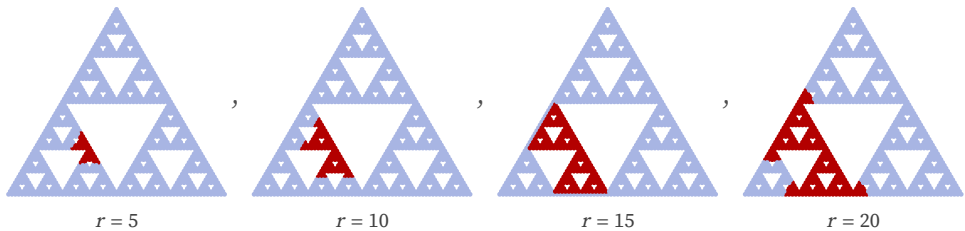

Estimating dimension from $V_r(X)$ averaged over all points we get (for graphs made from 6 and 7 recursive subdivisions):

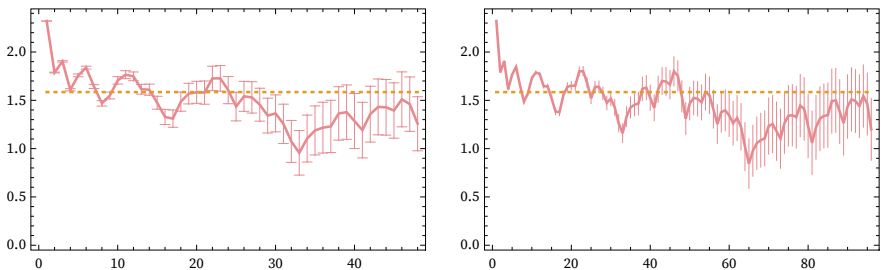

The dotted line indicates the standard Hausdorff dimension $\log_2(3) \approx 1.58$ for a Sierpiński triangle [23]. And what the pictures suggest is that the growth rate of $V_r$ approximates this value. But to get the exact value we see that in addition to everything else, we will need average estimates of dimension over different values of $r$.

In the end, therefore, we have quite a collection of limits to take. First, we need the overall size of our graph to be large. Second, we need the range of values of $r$ for measuring $V_r$ to be small compared to the size of the graph. Third, we need these values to be large relative to individual nodes in the graph, and to be large enough that we can readily measure the leading order growth of $V_r$—and that this will be of the form $r^d$. In addition, if the graph is not homogeneous we need to be averaging over a region $X$ that is large compared to the size of inhomogeneities in the graph, but small compared to the values of $r$ we will use in estimating the growth of $V_r$. And finally, as we have just seen, we may need to average over different ranges of $r$ in estimating overall dimension.

If we have something like a grid graph, all of this will work out fine. But there are certainly cases where we can immediately tell that it will not work. Consider, for example, first the case of a complete graph, and second of a tree:

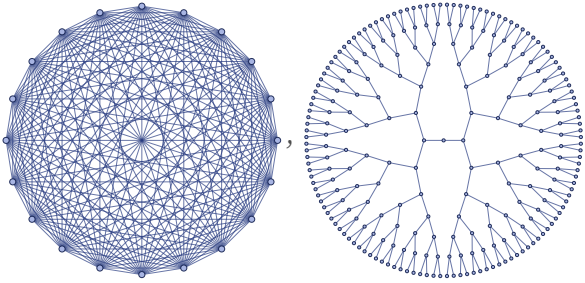

For a complete graph there is no way to have a range of *r* values "smaller than the radius of graph" from which to estimate a growth rate for $V_r$. For a tree, $V_r$ grows exponentially rather than as a power of *r*, so our estimate of dimension $\Delta(r)$ will just continually increase with *r*:

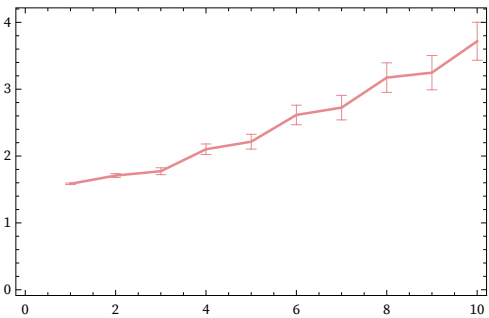

But notwithstanding these issues, we can try applying our approach to the objects generated by our models. As constructed, these objects correspond to directed graphs or hypergraphs. But for our current purposes, we will ignore directedness in determining distance, effectively taking all elements in a particular *k*-ary relation—regardless of their ordering—to be at unit distance from each other.

As a first example, consider the $2_3 \to 3_3$ rule we discussed above that "knits" a simple grid:

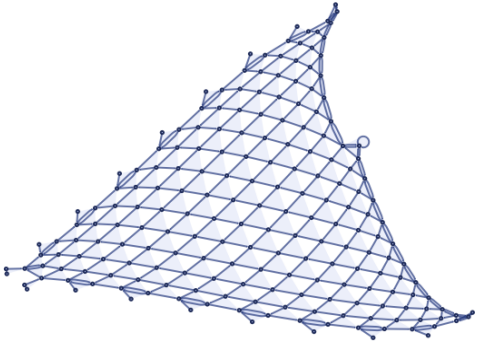

As we run the rule, the structure it produces gets larger, so it becomes easier to estimate the growth rate of $V_r$. The picture below shows $\Delta(r)$ (starting at the center point) computed after successively more steps. And we see that, as expected, the dimension estimate appears to converge to value 2:

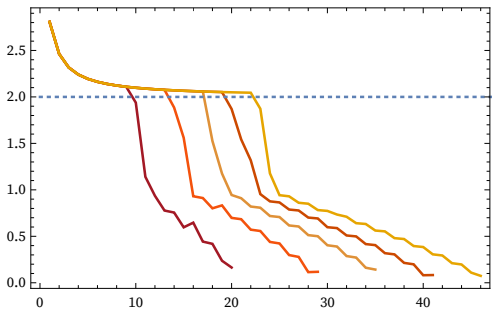

It is worth mentioning that if we did not compute $V_r(X)$ by starting at the center point, but instead averaged over all points, we would get a less useful result, dominated by edge effects:

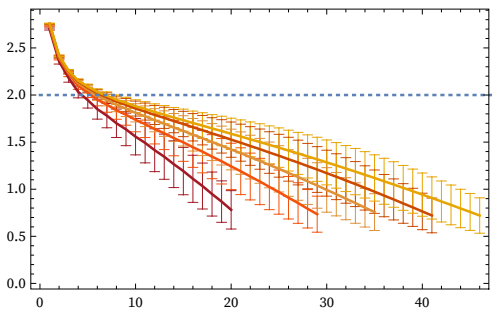

As a second example, consider the $2_3 \to 3_3$ rule that slowly generates a somewhat complex kind of surface:

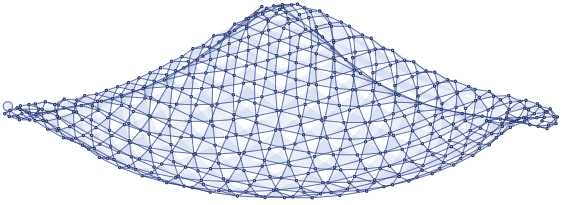

As we run this longer, we see what appears to be increasingly close approximation to dimension 2, reflecting the fact that even though we can best draw this object embedded in 3D space, its intrinsic surface is two-dimensional (though, as we will discuss later, it also shows the effects of curvature):

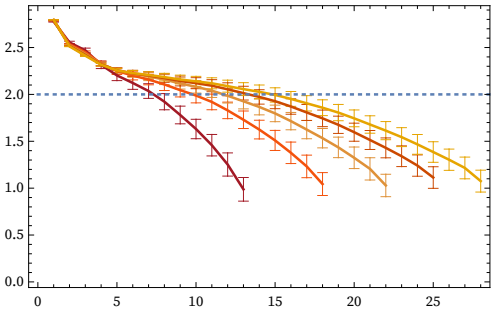

The successive dimension estimates shown above are spaced by 500 steps in the evolution of the rule. As another example, consider the $2_3 1_2 \to 4_3 4_2$ rule, in which geometry emerges rapidly through a process of subdivision:

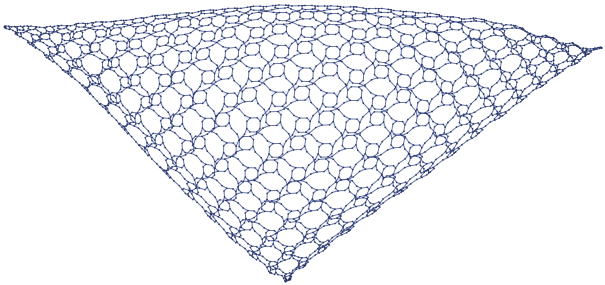

These are dimension estimates for all of the first 10 steps in the evolution of this rule:

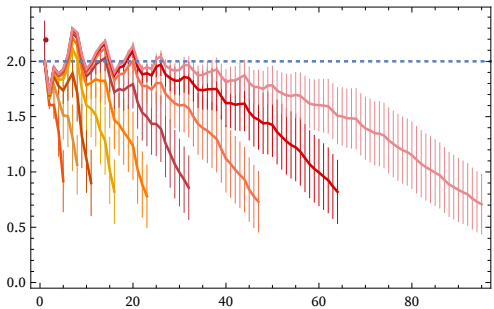

We can also validate our approach by looking at rules that generate obviously nested structures. An example is the $2_2 \to 4_2$ rule that produces:

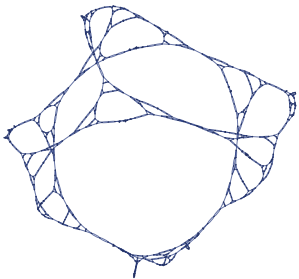

The results for each of the first 15 steps show good correspondence to dimension $\log_2(3) \approx 1.58$:

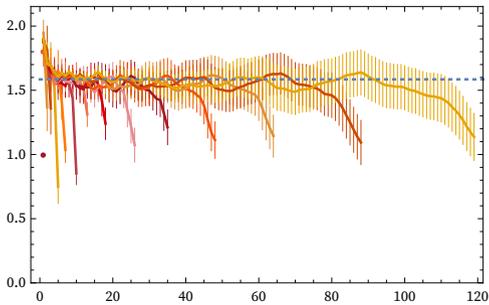

## 4.6 Dimension-Related Characterizations

Having seen how our notion of dimension works in cases where we can readily recognize emergent geometry, we now turn to using it to study the more general limiting behavior of our models.

As a first example, consider the $2_2 \to 4_2$ rule

$\{\{x, y\}, \{x, z\}\} \to \{\{x, y\}, \{x, w\}, \{y, w\}, \{z, w\}\}$

which generates results such as (with about $1.84^t$ relations at step $t$):

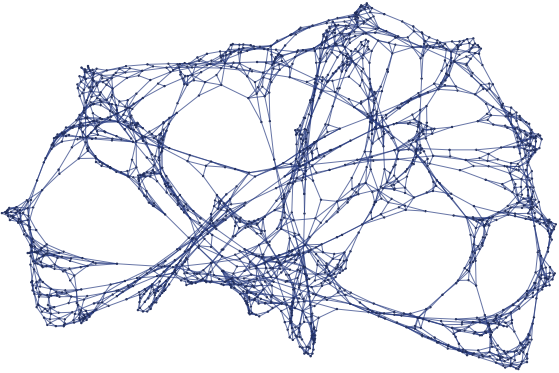

If we attempt to reconstruct a surface from successive steps in the evolution of this rule, no clearly recognizable geometry emerges:

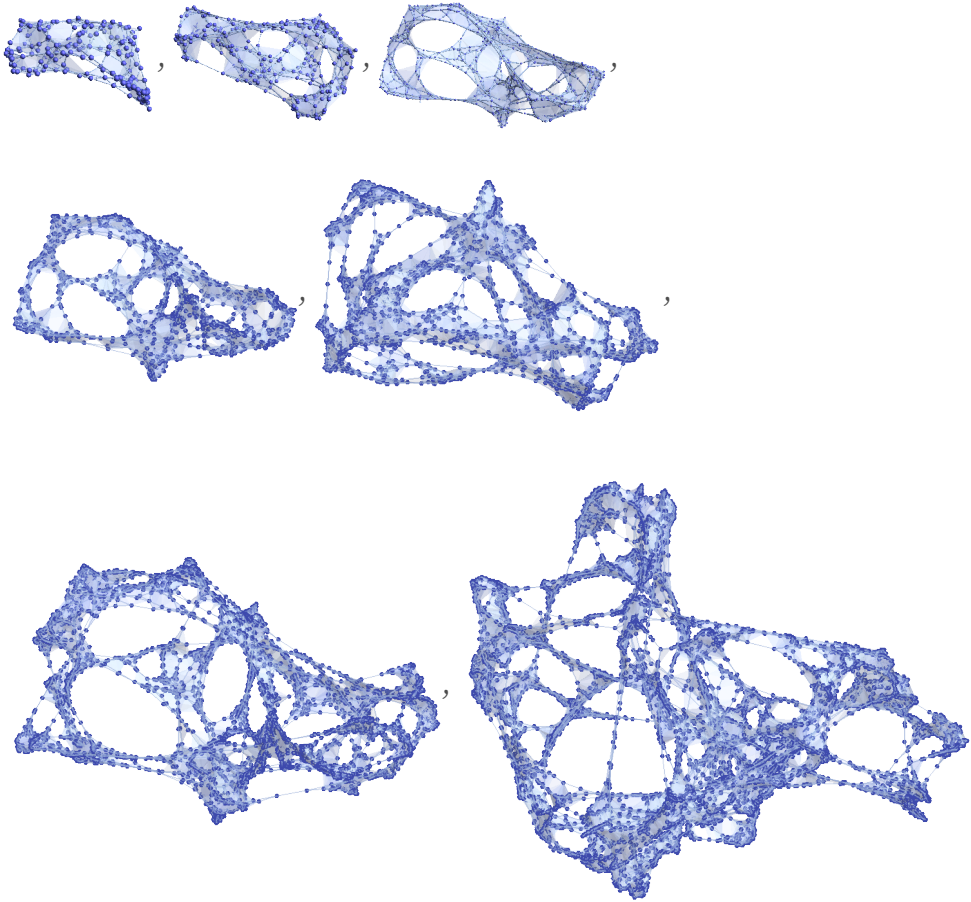

But instead we can try to characterize the results using $V_r(X)$ and our notion of dimension. We compute $V_r(X)$ as we do elsewhere: by starting at a point in the structure and constructing successively larger balls:

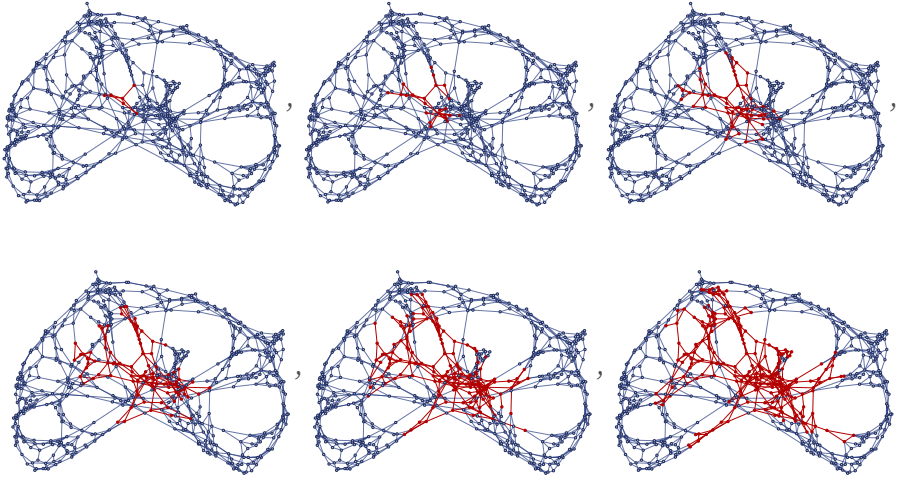

Computing the $\Delta(r)$ for all points over the first 16 steps of evolution gives:

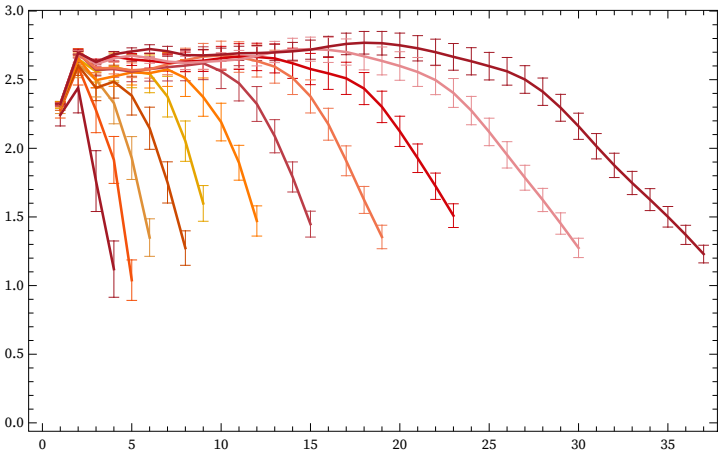

The most important feature of this plot is that it suggests $\Delta(r)$ might approach a definite limit as the number of steps increases. And from the increasing region of flatness there is some evidence that perhaps $V_r$ might approach a stable $r^d$ form, with $d \approx 2.7$, suggesting that in the limit this rule might produce some kind of emergent geometry with dimension around 2.7.

What about other rules? Here are some examples for rules we have discussed above:

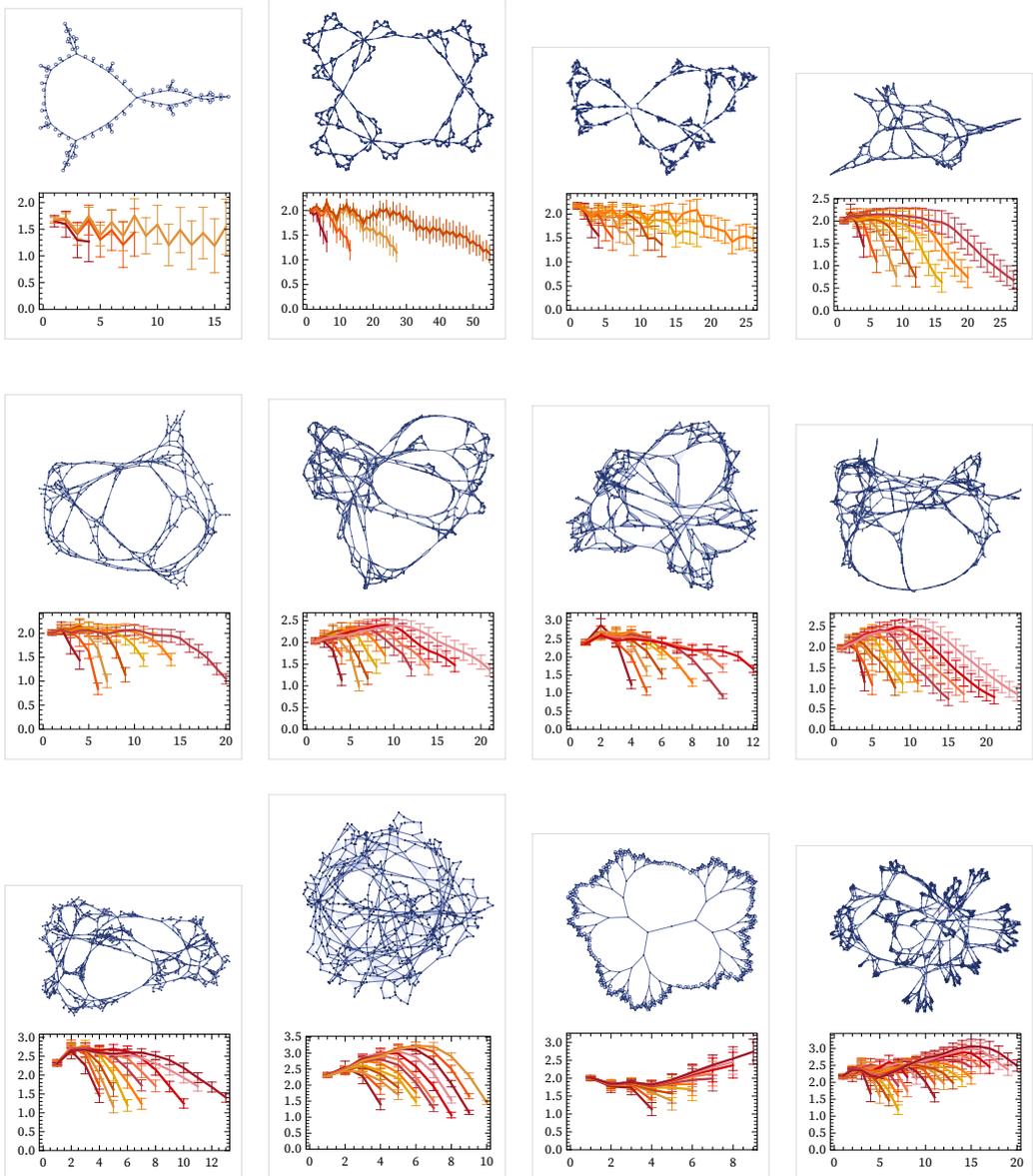

Some rules do not show convergence, at least over the number of steps sampled here. Other rules show quite stable limiting forms, often with a flat region which suggests a structure with definite dimension. Sometimes this dimension is an integer, like 1 or 2; often it is not. Still other rules seem to show linear increase in log differences of $V_r$, implying an exponential form for $V_r$ itself, characteristic of tree-like behavior.

## 4.7 Curvature

In ordinary plane geometry, the area of a circle is $\pi r^2$. But if the circle is drawn on the surface of a sphere of radius $a$, the area of the spherical region enclosed by the circle is instead:

$$2\pi a^2 \left(1 - \cos\left(\frac{r}{a}\right)\right) = \pi r^2 \left(1 - \frac{r^2}{12\,a^2} + \frac{r^4}{360\,a^4} - \ldots\right)$$

In other words, curvature in the underlying space introduces a correction to the growth rate for the area of the circle as a function of radius. And in general there is a similar correction for the volume of a $d$-dimensional ball in a curved space (e.g. [24][1:p1050]):

$$\frac{\pi^{d/2}}{(d/2)!}\, r^d \left(1 - \frac{r^2}{6\,(d+2)}\, R + O(r^4)\right)$$

where here $R$ is the Ricci scalar curvature of the space [25][26][27]. (For example, for the $d$-dimensional surface of a $(d+1)$-dimensional sphere of radius $a$, $R = \frac{(d-1)\,d}{a^2}$.)

Now consider the sequence of "sphere" graphs:

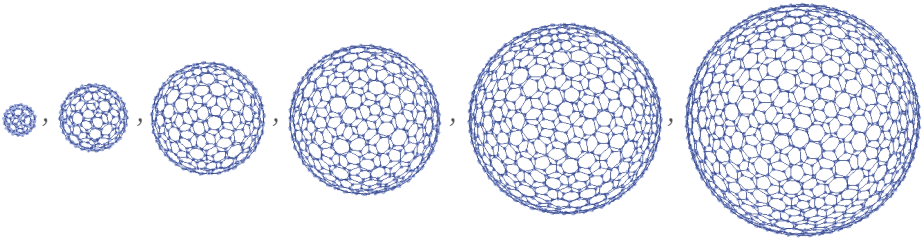

We can compute $V_r$ for each of these graphs. Here are the log differences $\Delta(r)$ (the error bars come from the different neighborhoods associated with hexagonal and pentagonal "faces" in the graph):

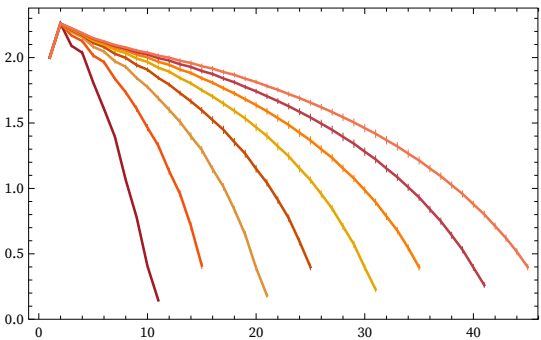

We immediately see the effect of curvature: even though in the limit the graphs effectively define 2D surfaces, the presence of curvature introduces a negative correction to pure $r^2$ growth in $V_r$. (Somewhat confusingly, there is only one scale defined for the kind of "pure sphere" graphs shown here, so they all have the same curvature, independent of size.)

A torus, unlike a sphere, has no intrinsic surface curvature. So torus graphs of the form

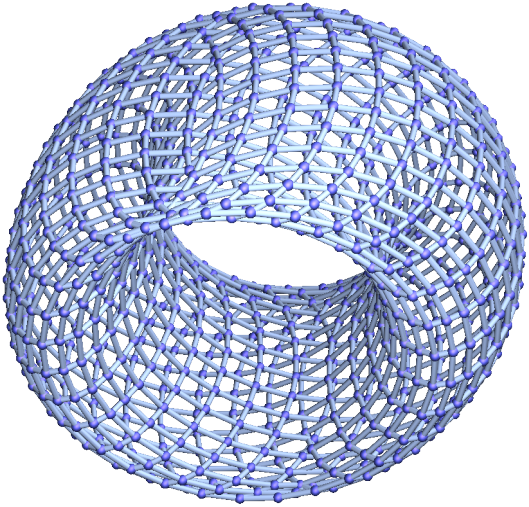

give flat log differences for $V_r$:

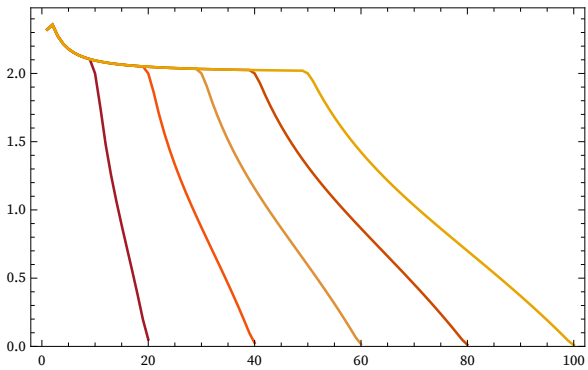

A graph based on a tiling in hyperbolic space

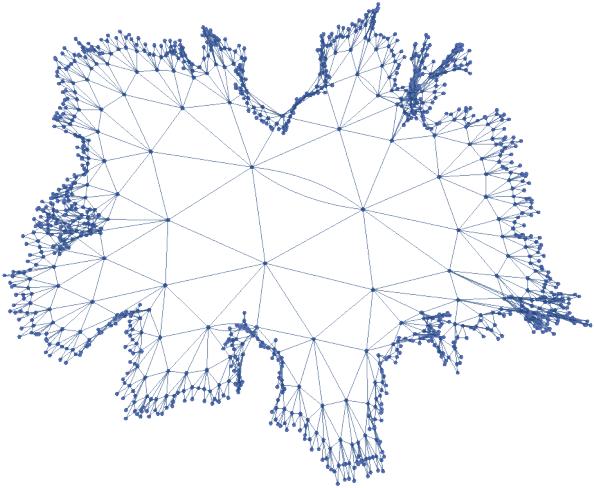

has negative curvature, so leads to a positive correction to $V_r$:

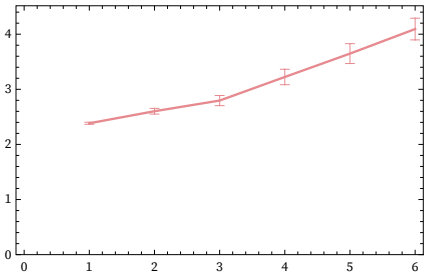

(One can imagine getting other examples by taking 3D objects and putting meshes on their surfaces. And indeed if the meshes are sufficiently faithful to the intrinsic geometry of the surfaces—say based on their geodesics—then the $V_r(X)$ for the connectivity graphs of these meshes [28] will reflect the intrinsic curvatures of the surfaces. In practical computational geometry, though, meshes tend to be based on things like coordinate parametrizations, and so do not reflect intrinsic geometry.)

Many structures produced by our models exhibit curvature. There are cases of negative curvature:

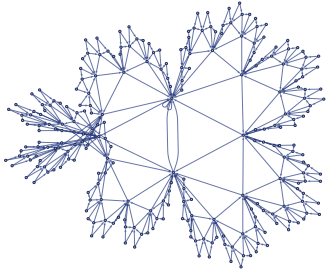 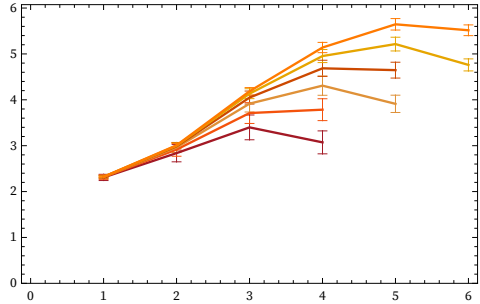

As well as positive curvature:

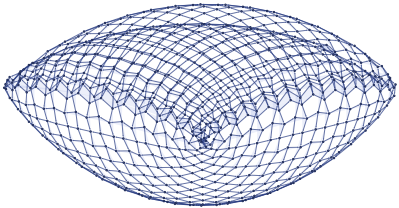 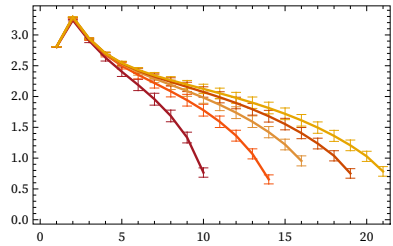

The most obvious examples of nested structures have fractional dimension, but no curvature:

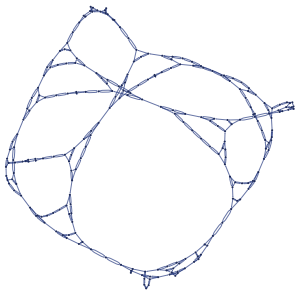 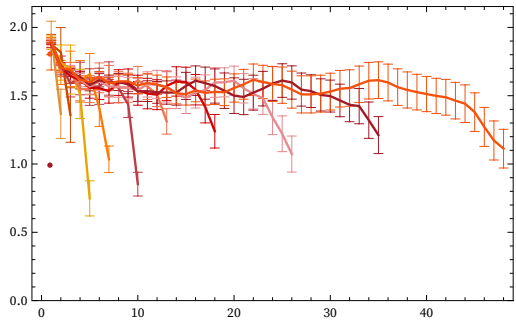

But even though it is not well characterized using ideas from traditional calculus, there is every reason to expect that the limits of our models can exhibit a combination of fractional dimension and curvature.

In general, though, there is no obvious constraint on the possible limiting form of $V_r$. Curvature can be thought of as associated with the $O(r^2)$ term in a Taylor expansion of $V_r$ about $r = 0$, after factoring out $r^d$. But there is nothing to say that the leading behavior of $V_r$ should match a form like $r^d$. In addition to exponentials like $\lambda^r$ it could show an infinite collection of intermediate asymptotic scales, like $2^{r \log(r)}$ or $r^{\sqrt{r}}$ [29][30].

## 4.8 Homogeneity and Local Graph Neighborhoods

In studying $V_r$ we are looking at the total size of the neighborhood up to distance $r$ around a point in a graph. But what about the actual local structure of the neighborhood?

In general, it can be different for every point on the graph. Thus, for example, in obtained from 10 steps of the rule $\{\{x,y\},\{x,z\}\}\to\{\{x,z\},\{x,w\},\{y,w\},\{z,w\}\}$ the collection of distinct range-1 neighborhoods (with their counts) is:

$\{\rhd\!\!-\to 139, \bowtie\to 29, \perp\to 25, \triangleleft\!\rhd\to 19, \curlyvee\to 19, \bowtie\!\!-\to 14, \diamond\!\!-\to 10, \bowtie\to 10, \bowtie\!\!\!\star\to 7,$
$\star\to 7, \diamond\!\!-\to 7, \bowtie\!\!\star\to 7, \perp\!\!\!\star\to 5, \star\!\!\star\to 4, \bowtie\!\!\!\star\to 3, \circledast\to 2, \star\!\!\!\star\to 2, \perp\!\!\!\!\star\to 2, \triangledown\to 2,$
$\circledast\to 1, \circledast\to 1, \star\!\!\!\star\to 1, \star\!\!\!\star\to 1, \star\!\!\!\star\to 1, \star\!\!\!\star\to 1, \star\!\!\!\star\to 1, \star\!\!\!\star\to 1, \star\!\!\!\star\to 1, -\!\!-\to 1\}$

The corresponding result after 12 steps is:

$\{\rhd\!\!-\to 478, \bowtie\to 94, \perp\to 79, \triangleleft\!\rhd\to 67, \curlyvee\to 62, \bowtie\!\!-\to 55, \bowtie\to 36,$
$\diamond\!\!-\to 26, \diamond\!\!-\to 24, \bowtie\!\!\star\to 23, \star\!\!\star\to 20, \star\!\!\star\to 18, \star\!\!\star\to 13, \perp\!\!\!\star\to 13, \star\!\!\!\star\to 10,$
$\bowtie\!\!\!\star\to 9, \star\!\!\!\star\to 7, \star\!\!\!\star\to 7, \circledast\to 6, \star\!\!\!\star\to 4, \star\!\!\!\star\to 3, \star\!\!\!\star\to 3, \circledast\to 2, \star\!\!\!\star\to 2,$
$\circledast\to 2, \circledast\to 2, \star\!\!\!\star\to 2, \triangledown\to 2, \circledast\to 1, \circledast\to 1, \circledast\to 1, \circledast\to 1, \circledast\to 1,$
$\circledast\to 1, \circledast\to 1, \star\!\!\!\star\to 1, \star\!\!\!\star\to 1, \star\!\!\!\star\to 1, \star\!\!\!\star\to 1, \rhd\!\!-\to 1, -\!\!-\to 1\}$

And it seems that for this rule the distribution of different forms for a given range of neighborhood generally stabilizes as the number of steps increases. (It may be possible to characterize it as limiting to an invariant measure in the space of possible hypergraphs, perhaps with some related entropy (cf. [1:p958][31]).)

One sees the same kind of stabilization for most rules, though, for example, in a case like

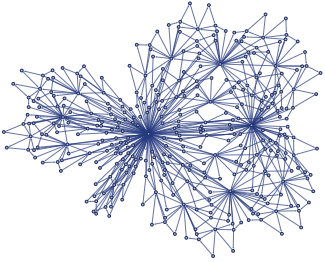

from the rule {{x,y}}→{{x,y},{y,z},{z,x}} one always gets some neighborhoods with new forms at each step:

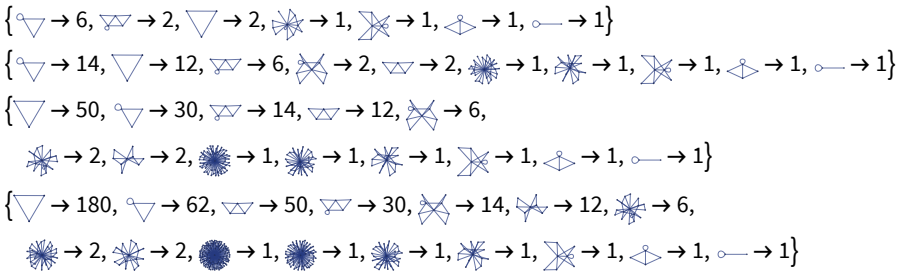

In general, the presence of many identical neighborhoods reflects a certain kind of approximate symmetry or isometry of the emergent geometry of the system.

In a torus graph, for example, the symmetry is exact, and all local neighborhoods of a given range are the same:

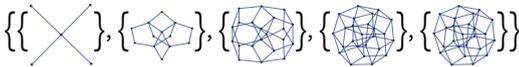

The same is true for a 3D torus graph:

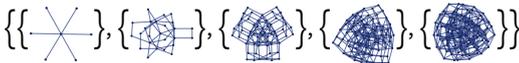

For a sphere graph not every point has the exact same local neighborhood, but there are a limited number of neighborhoods of a given range:

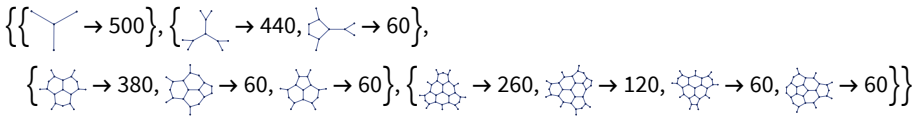

And from the dual graph it becomes clear that these are associated with hexagonal and pentagonal "faces":

$$\{\{\text{⬡} \to 240, \text{⬠} \to 12\}, \{\text{⬡} \to 180, \text{⬡} \to 60, \text{⬡} \to 12\},$$
$$\{\text{⬡} \to 60, \text{⬡} \to 60, \text{⬡} \to 60, \text{⬡} \to 60, \text{⬠} \to 12\}\}$$

For a (spherical) Sierpiński graph, there are also a limited number of neighborhoods of a given range:

$$\{\{\text{▷—} \to 108\}, \{\text{▷—◁} \to 108\}, \{\text{▷—◁} \to 72, \text{▷—◁} \to 36\}, \{\text{△} \to 72, \text{▷—◁} \to 36\}\}$$

Whenever every local neighborhood is essentially identical, $V_r(X)$ will have the same form for every point $X$ in a graph or hypergraph. But in general $V_r(X)$ (and the log differences $\Delta_r(X)$) will depend on $X$. The picture below shows the relative values of $\Delta_r(X)$ at each point in the structure we showed above:

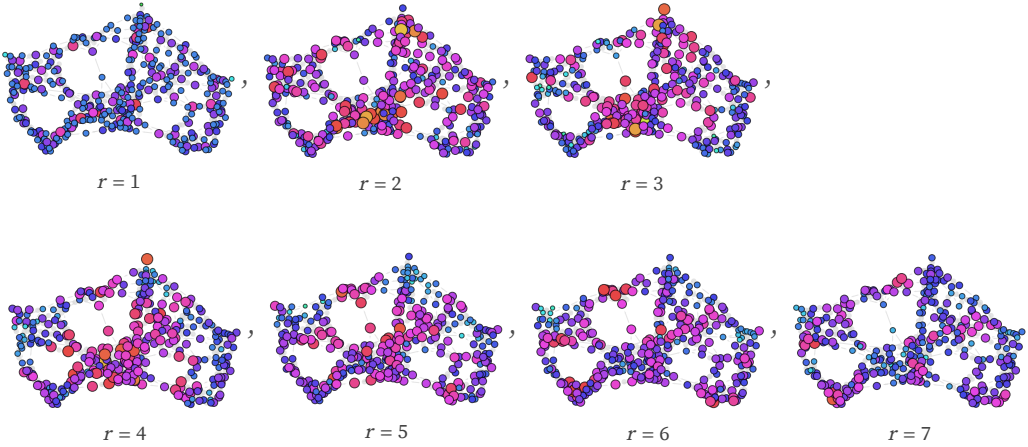

We can also compute the distribution of values for $\Delta_r(X)$ across the structure, as a function of $r$:

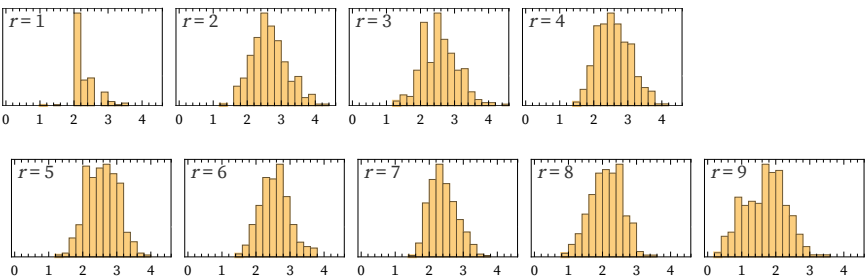

Both these pictures indicate a certain statistical uniformity in $V_r(X)$. This is also seen if we look at the evolution of the distribution of $\Delta_r(X)$, here shown for the specific value $r = 6$, for steps 8 through 16:

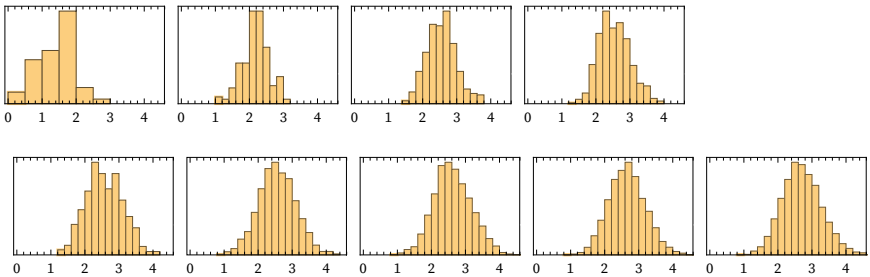

## 4.9 Adjacency Matrices and Age Distributions

We have made explicit visualizations of the connectivity structures of the graphs (and hypergraphs) generated by our models. But an alternative approach is to look at adjacency matrices (or tensors). In our models, there is a natural way to index the nodes in the graph: the order in which they were created. Here are the adjacency matrices for the first 14 steps in the evolution of the rule $\{\{x,y\},\{x,z\}\} \to \{\{x,z\},\{x,w\},\{y,w\},\{z,w\}\}$ discussed above:

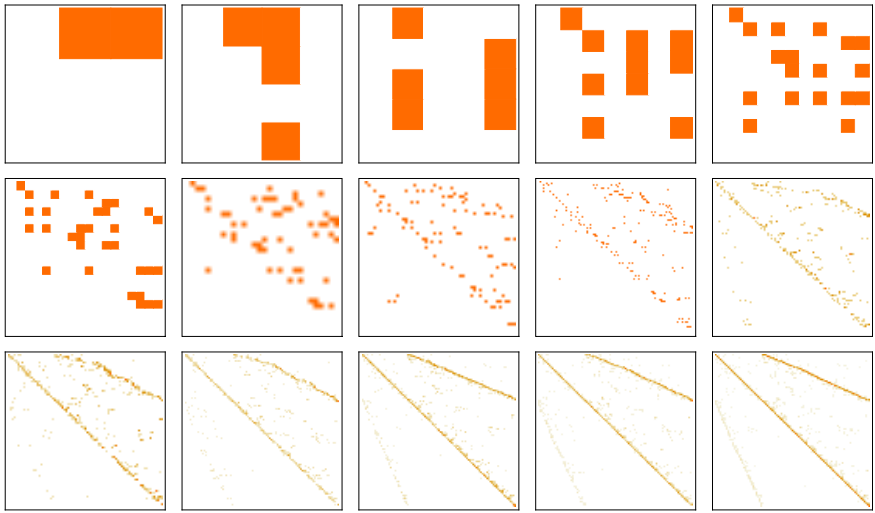

It is notable that even though these adjacency matrices grow by roughly a factor of 1.84 at each step, they maintain many consistent features—and something similar is seen in many other rules.

Our models evolve by continually adding new relations, and for example in the rule we are currently considering, there are roughly exponentially more relations at each step. The result, as shown below for step 14, is that at a given step the relations that exist will almost all be from the most recent step (shown in red):

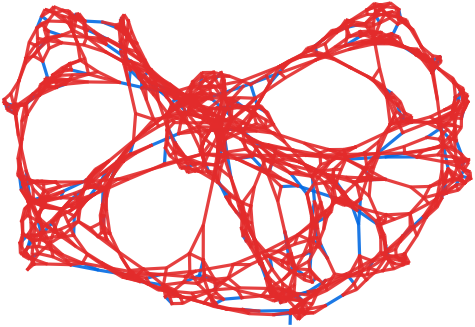

Other rules can show quite different age distributions. Here are age distributions for a few rules that "knit" their structures one relation at a time:

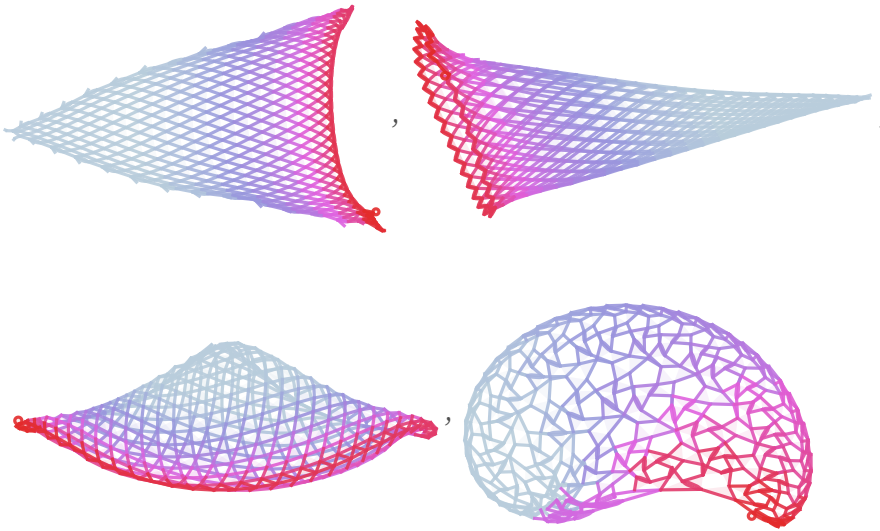

## 4.10 Other Graph Properties

There are many graph and hypergraph properties that can be studied for the output of our models. Here we primarily give examples for the rule {{x,y},{x,z}}→{{x,z},{x,w},{y,w},{z,w}} discussed above.

A basic question is how the numbers of vertices and edges (elements and relations) grow with successive steps. Plotting on a logarithmic scale suggests eventually roughly exponential growth in this case:

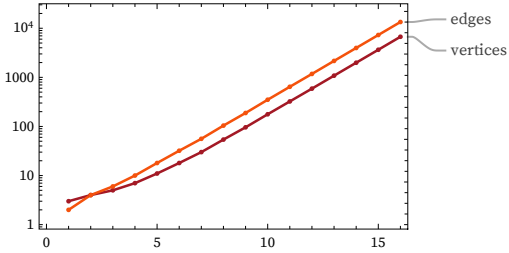

We can also compute the growth of the graph diameter (greatest distance between vertices) and graph radius:

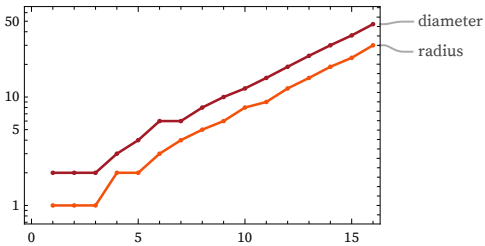

If one assumes that the total vertex count $V$ is related to diameter $D$ by $V = D^d$, then plotting $d$ gives (to be compared to dimension approaching ≈2.68 computed from the growth of $V_r$):

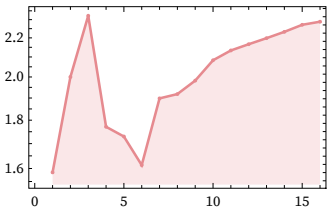

There are many measures of graph structure which basically support the expectation that after many steps, the outputs from the model somehow converge to a kind of statistically invariant "equilibrium" state:

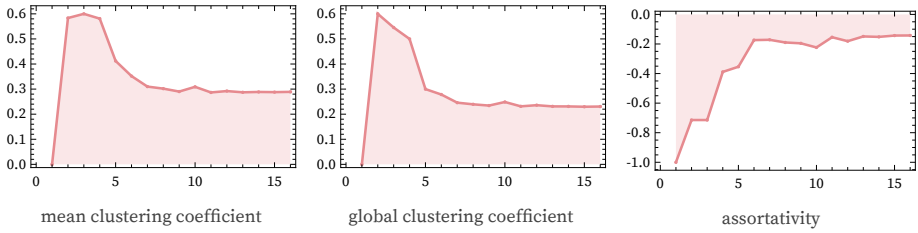

mean clustering coefficient, global clustering coefficient, assortativity

Some centrality measures [32][33] start (here at step 10) somewhat concentrated, but rapidly diffuse to be much more broadly distributed:

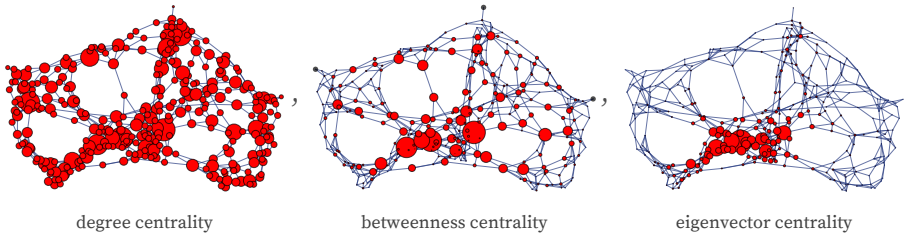

degree centrality, betweenness centrality, eigenvector centrality

There are local features of the graph that are closely related to $V_1(X)$ and $V_2(X)$:

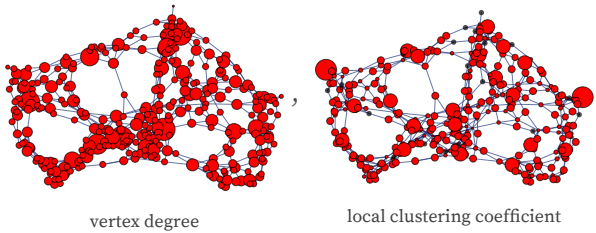

vertex degree, local clustering coefficient

Another feature of our graphs to study is their cycle structure. At the outset, our graphs give us only connectivity information. But one way to imagine identifying "faces" that could be used to infer emergent topology is to look at the fundamental cycles in the graph:

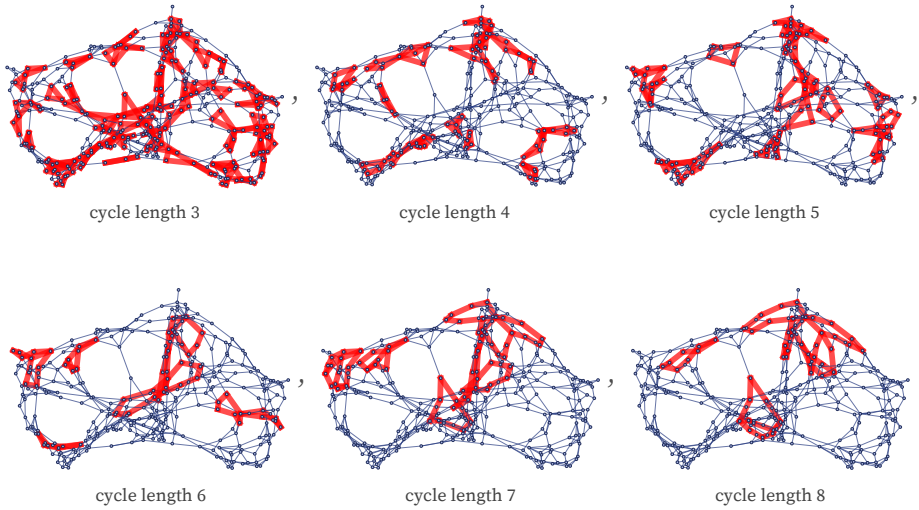

| cycle length 3 | cycle length 4 | cycle length 5 |
| cycle length 6 | cycle length 7 | cycle length 8 |

In this particular graph, there are altogether 320 fundamental cycles, with the longest one being of length 24. The distribution of cycle lengths on successive steps once again seems to approach an "equilibrium" form:

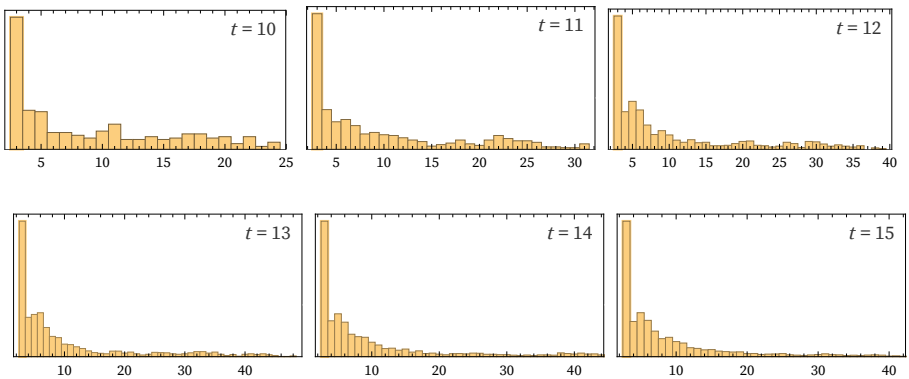

One way to probe overall properties of a graph is to consider the evolution of some dynamical process on the graph. For example, one could run a totalistic cellular automaton with values at nodes of the graph. Another possibility is to solve a discretized PDE. For example, having computed a graph Laplacian [34] (or its higher order analogs) one can determine the distribution of eigenvalues, or the eigenmodes, for a particular graph [35]. The density of eigenvalues is then closely related to $V_r$ and our estimates of dimension and curvature.

## 4.11 Graph Properties Conserved by Rules

Many rules (at least when they exhibit complex behavior) seem to lead to statistically similar behavior, independent of their initial conditions. But there could still be disjoint families of states that can be reached from different initial conditions, perhaps characterized by different graph or hypergraph invariants.

As one example, we can ask whether there are rules that preserve the planarity of graphs. All rules with signature $1_2 \to 2_2$ inevitably do this. A rule like

$\{\{x, y\}\} \to \{\{x, y\}, \{y, z\}, \{z, x\}\}$

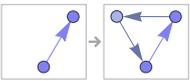

might not at first appear to:

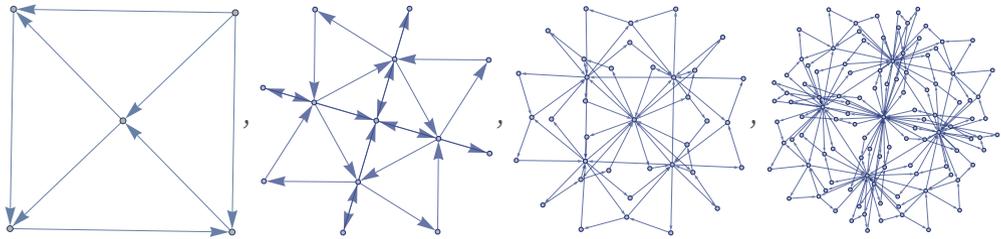

But a different graph layout shows that actually all these graphs are planar [36]:

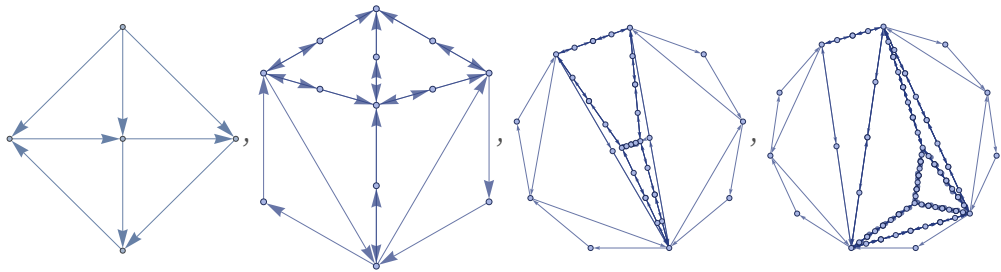

Among larger rules, many still preserve planarity. But for example,
{{x,y},{x,z}}→{{x,z},{x,w},{y,w},{z,w}} does not, since it transforms the planar graph

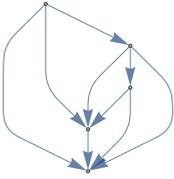

to the nonplanar one:

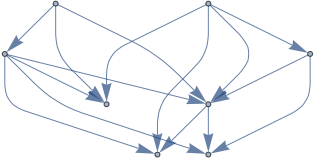

In general, a graph is planar so long as it does not contain as a subgraph either of [37]

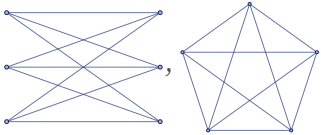

so a rule preserves planarity if (and only if) it never generates either of these subgraphs.

Planarity is one of a class of properties of graphs that are preserved under deletion of vertices and edges, and contraction of edges. Another such property is whether a graph can be drawn without crossings on a 2D surface of any specific genus $g$ [38]. It turns out [38] that for any such property it is known that there are in principle only a finite number of subgraphs that can "obstruct" the property—so if a rule never generates any of these, it must preserve the property.

## 4.12 Apparent Randomness and Growth Rates

The phenomenon of intrinsic randomness generation is an important and ubiquitous feature of computational systems [1:7.5][39]—the rule 30 cellular automaton [1:2.1] being a quintessential example. In our models the phenomenon definitely often occurs, but two issues make it slightly more difficult to identify.

First, there is considerable arbitrariness in the way we choose to present or visualize graphs or hypergraphs—so it is more difficult to tell whether apparent randomness we see is a genuine feature of our system, or just a reflection of some aspect of our presentation or visualization method.

And second, there may be many possible choices of updating orders, and the specific results we get may depend on the order we choose. Later we will discuss the phenomenon of causal invariance, and we will see that there are causal graphs that can be independent of updating order. But for now, we can consider our updating process to just be another deterministic procedure added to the rules of our system, and we can ask about apparent randomness for this combined system.

And to avoid the arbitrariness of different graph or hypergraph presentations, we can look at graph or hypergraph invariants, which are the same for all isomorphic graphs or hypergraphs, independent of their presentation or visualization.

The most obvious invariants to start with are the total numbers of elements and relations (nodes and edges) in the system. For rules that involve only a single relation on the left-hand side, it is inevitable that these numbers must be determined by a linear recurrence (cf. [1:p890]). (For $1_k \to n_k$ rules, up to a $k$-term linear recurrence may be involved.)

For rules that involve more than one relation more complicated behavior is common. Consider for example the rule:

$\{\{x, y\}, \{x, z\}\} \to \{\{y, w\}, \{w, x\}, \{z, w\}\}$

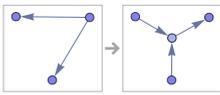

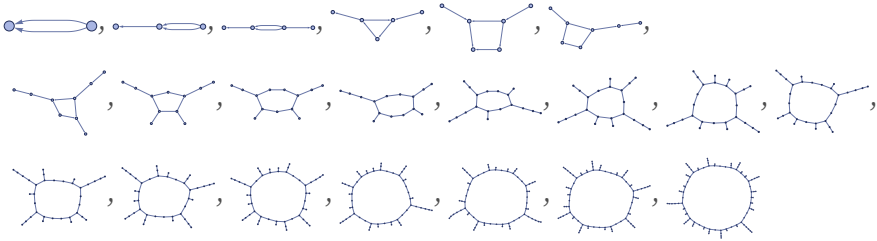

The number of relations generated over the first 50 steps (using our standard updating order) is:

{2, 3, 4, 5, 6, 7, 9, 11, 13, 15, 18, 22, 26, 30, 35, 42, 50, 58, 67, 79, 94, 110, 127, 148, 175, 206, 239, 277, 325, 383, 447, 518, 604, 710, 832, 967, 1124, 1316, 1544, 1801, 2093, 2442, 2862, 3347, 3896, 4537, 5306, 6211, 7245, 8435, 9845}

Taking third differences yields:

{0, 0, 0, 1, −1, 0, 0, 1, 0, −1, 0, 1, 1, −1, −1, 1, 2, 0, −2, 0, 3, 2, −2, −2, 3, 5, 0, −4, 1, 8, 5, −4, −3, 9, 13, 1, −7, 6, 22, 14, −6, −1, 28, 36, 8, −7, 27, 64}

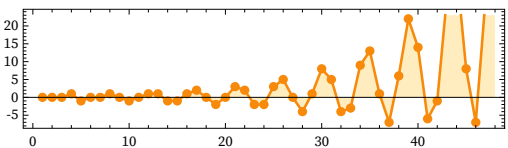

One can consider many other invariants, including counts of in and out degrees of elements, and counts of cycles. In general, one can construct invariant "fingerprints" of hypergraphs—and the typical observation is that for rules whose behavior seems complex, almost all of these will exhibit extensive apparent randomness.

## 4.13 Statistical Mechanics

Features like dimension and curvature can be used to probe the consistent large-scale structure of the limiting behavior of our models. But particularly insofar as our models generate apparent randomness, it also makes sense to study their statistical features. We discussed above the overall distribution of values of $V_r(X)$ and $\Delta_r(X)$. But we can also consider fluctuations and correlations.

For example, we can look at a 2-point correlation function
$S_r(s) = (\langle V_r(X)\, V_r(Y) \rangle - \langle V_r(X) \rangle^2)/\langle V_r(X) \rangle^2$ for points $X$ and $Y$ separated by graph distance $s$. For a uniform graph such as the torus graph, $S_r(s)$ always vanishes. For the buckyball approximation to the sphere that we used above, $S_r(s)$ shows peaks at the distances between "pentagons" in the graph.

For the rule $\{\{x,y\},\{x,z\}\} \rightarrow \{\{x,z\},\{x,w\},\{y,w\},\{z,w\}\}$, $S_r(s)$ steadily expands the region of $s$ over which it shows positive correlations, and perhaps (at least for larger $r$, indicated by redder curves) approaches a limiting form:

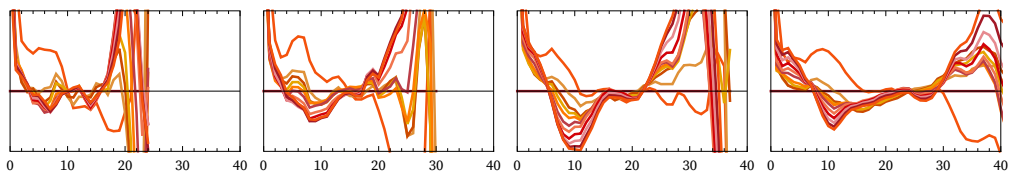

It is conceivable that for this or other rules there might be systematic rescalings of distance and number of steps that would lead to fixed limiting forms.

In statistical mechanics, it is common to think about the ensemble of all possible states of a system—and for example to discuss evolution from all possible initial conditions. But typical systems in statistical mechanics can basically be discussed in terms of a fixed number of degrees of freedom (either coordinates or values).

For our models, there is no obvious way to apply the rules but, for example, to limit the total number of relations—making it difficult to do analysis in terms of ensembles of states.

One can certainly imagine the set of all possible hypergraphs (and even have Ramsey-theory-style results about it), but this set does not appear to have the kind of geometry or structure that has typically been necessary for results in statistical mechanics or dynamical systems theory. (One could however potentially think in terms of a distribution of adjacency matrices, limiting to graphon-like functions [40] for infinite graphs.)

## 4.14 The Effect of Perturbations

Imagine that at some step in the evolution of a rule one reverses a single relation. What effect will it have? Here is an example for the rule {{x,y},{x,z}}→{{x,z},{x,w},{y,w},{z,w}}. The first row is the original evolution; the second is the evolution after reversing the relation:

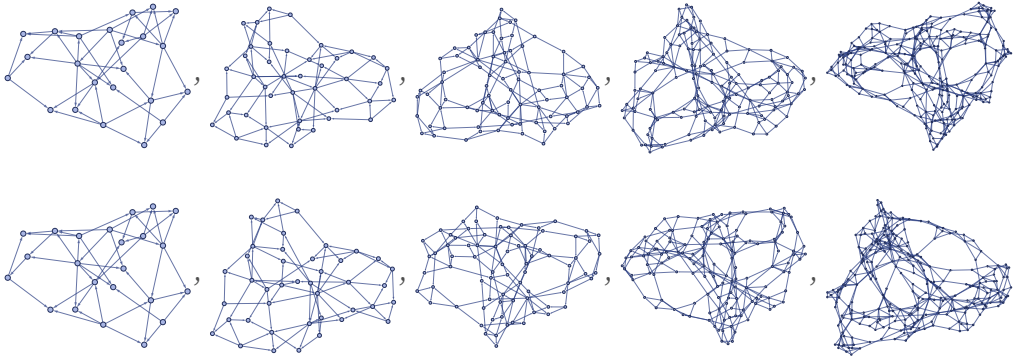

We can illustrate the effect by coloring edges in the first row of graphs that are different in the second one (taking account of graph isomorphism) [41]:

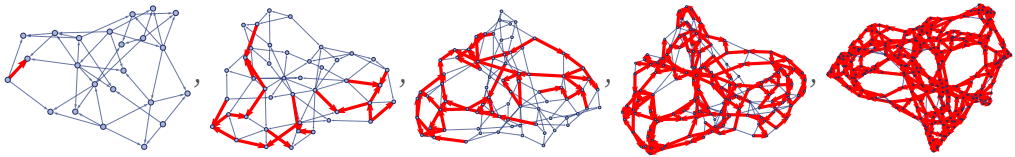

Visualizing the second and third graphs in 3D makes it more obvious that the changed edges are mostly connected:

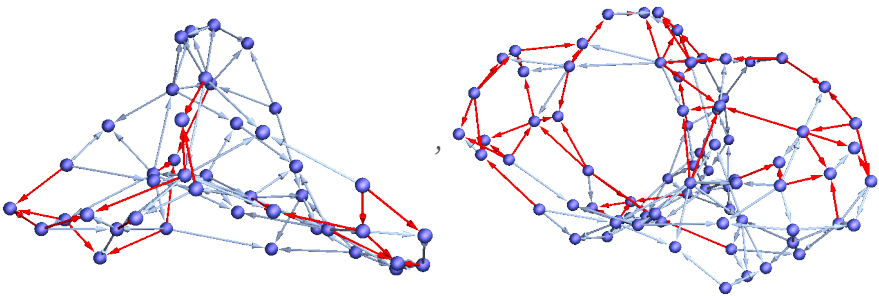

It takes only a few steps before the effect of the change has spread to essentially all parts of the system. (In this particular case, with the updating order used, about 20% of edges are still unaffected after 5 steps, with the fraction slowly decreasing, even as the number of new edges increases.)

In rules with fairly simple behavior, it is common for changes to remain localized:

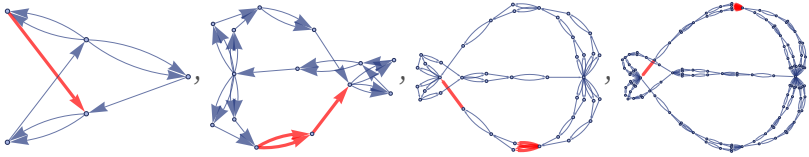

However, when complex behavior occurs, changes tend to spread. This is analogous to what is seen, for example, in the much simpler case of class 2 versus class 3 cellular automata [31][1:6.3]:

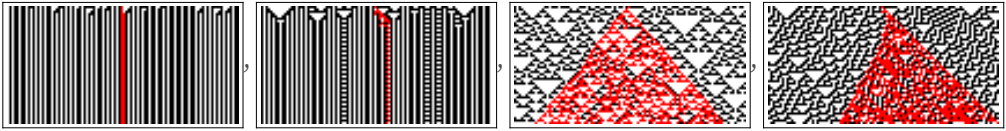

Cellular automata are also known [31] to exhibit the important phenomenon of class 4 behavior—in which there is a discrete set of localized "particle-like" structures through which changes typically propagate:

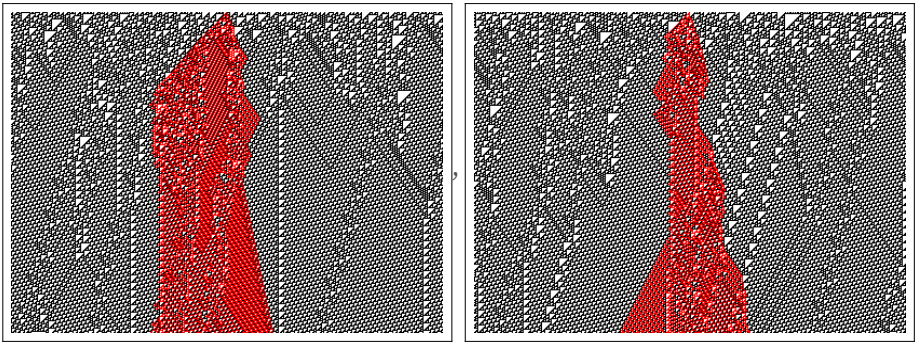

In cellular automata, there is a fixed lattice on which local rules operate, making it straightforward [1:6.3] to identify the region that can in principle be affected by a change in initial conditions. In the models here, however, everything is dynamic, and so even the question of what parts can in principle be affected by a change in initial conditions is nontrivial.

As we will discuss at length later, however, it is always possible to trace which updating events in a particular evolution depend on which others, and which relations are associated with these. The result will always be a superset of the actual effect of a change in the initial condition:

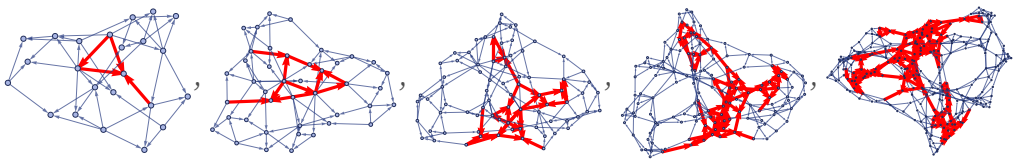

We discussed above the quantity $V_r(X)$ obtained by "statically" looking at the number of nodes in a hypergraph reached by going graph distance $r$—in effect computing the volume of a ball of radius $r$ in the hypergraph. By looking at the dependence of updating events in $t$ successive steps of evolution, we can define another quantity $C_t(X)$ which in effect measures the volume of a cone of dependencies in the evolution of the system.

$V_r(X)$ is in a sense a quantity that is "applied" to the system from outside; $C_t(X)$ is in a sense intrinsic. But as we will discuss later, $V_r(X)$ is in some sense an approximation to $C_t(X)$—and particularly when we can reasonably consider the evolution of a model to have reached some kind of "equilibrium", $V_r(X)$ will provide a useful characterization of the "state" of a model.

## 4.15 Geodesics

Given any two points in a graph or hypergraph one can find a (not necessarily unique) shortest path (or "geodesic") between them, as measured by the number of edges or hyperedges traversed to go from one point to the other. Here are a few examples of such geodesics:

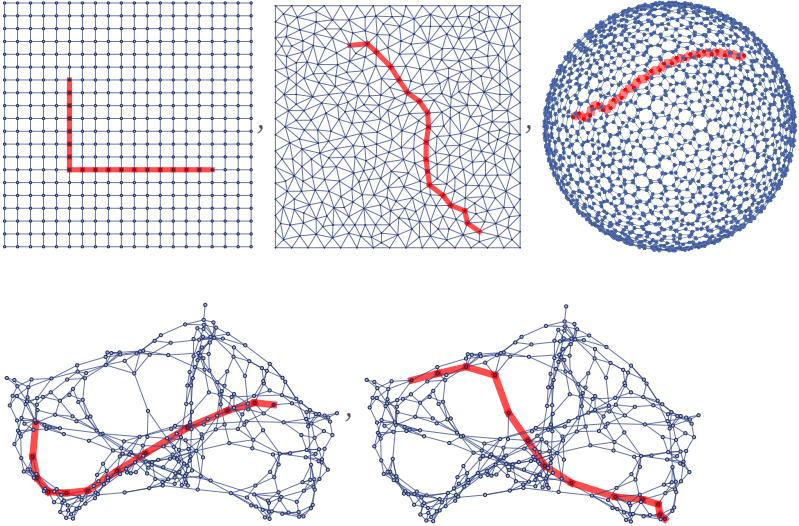

A geodesic in effect defines the analog of a straight line in a graph or hypergraph, and by analogy with the way geodesics work in continuous spaces, we can use them to probe emergent geometry.

For example, in the case of positive curvature, we can expect that nearby geodesics diverge, while in the case of negative curvature they converge:

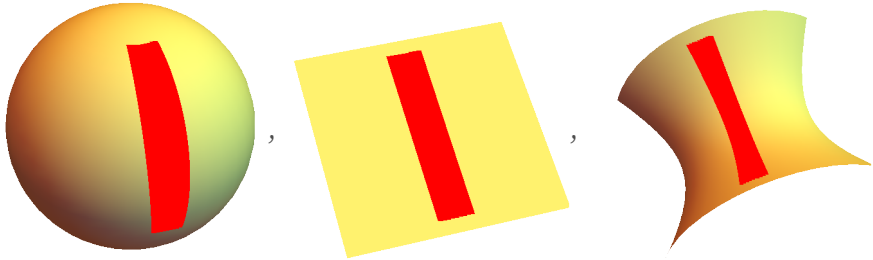

One can see the same effect in sufficiently large graphs (although it can be obscured by regularities in graphs which lead to large numbers of "degenerate" geodesics, all of the same length):

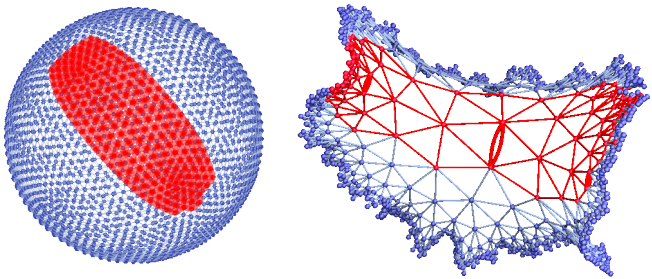

We saw before that the growth rate of the volume $V_r(X)$ of a ball centered at some point $X$ in a graph could be identified as giving a measure of the Ricci scalar curvature $R$ at $X$. But we can now consider tubes formed from balls centered at each point on a geodesic. And from the growth rates of volumes of these tubes we will be able to measure what can be identified as different components of curvature associated with the Ricci tensor (cf. [1:p1048][27][42]).

In a continuous space (or, more precisely, on a Riemannian manifold) the infinitesimal volume element at a point $X$ is given in terms of the metric tensor $g$ by $\sqrt{\det g(X)}$. If we look at a nearby point $X + \delta x$ we can expand in a power series in $\delta x$ (e.g. [43]):

$$\sqrt{\det g(X + \delta x)} = \sqrt{\det g(X)} \left(1 - \frac{1}{6} \sum R_{ij}(X)\, \delta x^i \delta x^j + O(\delta x^3) + ....\right)$$

where $R_{ij}$ is the Ricci tensor and the $\delta^i$ (contravariant vectors) are orthogonal components of $\delta x$ (say along axes defined by some coordinate system).

If we integrate over a ball of radius $r$ in $d$ dimensions, we recover our previous formula for the volume of a ball

$$V_r(X) = \int \sqrt{\det g(X+\delta x)}\, d^d\delta x = \frac{\pi^{d/2}}{(d/2)!} r^d \left(1 - \frac{r^2}{6(d+2)} \sum R_i^{\ i} + O(r^4)\right)$$

where $R = \sum R_i^{\ i}$ is the Ricci scalar curvature.

But now let us consider integrating over a tube of radius $r$ that goes a distance $\delta$ along a geodesic starting at $X$. Then we get a formula for the volume of the tube (cf. [44])

$$\bar{V}_{r,\delta x}(X) = \frac{\pi^{\frac{d-1}{2}}}{\left(\frac{d-1}{2}\right)!} r^{d-1}\, \delta x \left(1 - \left(\frac{d-1}{d+1}\right)(R - \sum R_{ij}\, \hat{\delta x}^i \hat{\delta x}^j)\, r^2 + O(r^3 + r^2\delta x) + \ldots \right)$$

where the $\hat{\delta x}^i$ are components of unit vectors along the geodesic.

There is now a direct analog in our hypergraphs: just as we measured the growth rates of geodesic balls to find Ricci scalar curvature, we can now measure growth rates of geodesic tubes to probe full Ricci curvature.

To construct an example, consider a graph formed from a mesh on the surface of an ellipsoid. (It is important that this mesh is intrinsic to the surface, with each mesh element corresponding to about the same surface area—and that the mesh does not just come from, say, a standard $\theta, \phi$ coordinate grid.)

As a first step, consider balls of progressively larger radii at different points on the ellipsoid mesh graph:

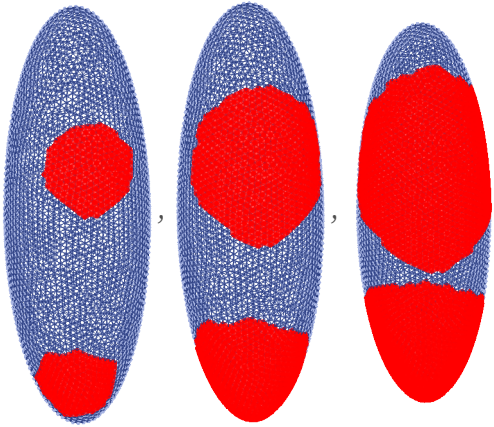

In the region of higher curvature near the tip, the area of the ball for a given radius is smaller, reflecting higher values of the Ricci scalar curvature $R$ there:

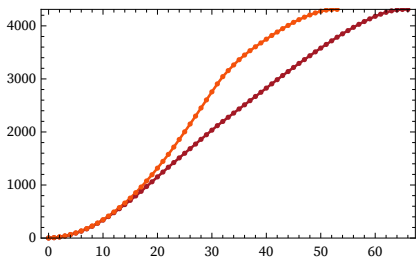

But now consider tubes around geodesics on the ellipsoid mesh graph. Instead of measuring the scalar curvature $R$, these instead in effect measure components of the Ricci tensor along these geodesics.

To measure all the components of the Ricci tensor, we could consider not just a tube but a bundle of geodesics, and we could look at the sectional curvature associated with deformations of the shape of this bundle. Or, as an alternative, we could consider tubes along not just one, but two geodesics through a given point. But in both cases, the analogy with the continuous case is easiest if we can identify something that we can consider an orthogonal direction.

One way to do this on a graph is to start from a particular geodesic through a given point, then to look at all other geodesics through that point, and work out which ones are the largest graph distance away. These show sequences of progressively more distant geodesics (as measured by the graph distance to the original geodesic of their endpoints):

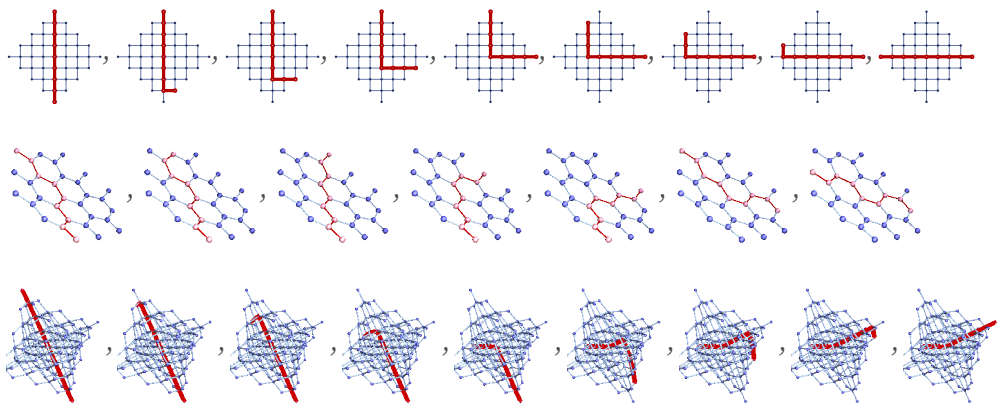

In general there may be many choices of these geodesics—and in a sense these correspond to different local choices of coordinates. But given particular choices of geodesics we can imagine using them to form a grid.

Looking at growth rates of volumes on this grid then gives us results not just about the Ricci tensor, but also about the Riemann tensor, about parallel transport and about covariant derivatives (cf. [27]).

The examples we have shown so far all involve graphs that have a straightforward correspondence with familiar geometry. But exactly the same methods can be used on the kinds of graphs and hypergraphs that arise from our models. This shows tubes of successively larger radii along two different geodesics:

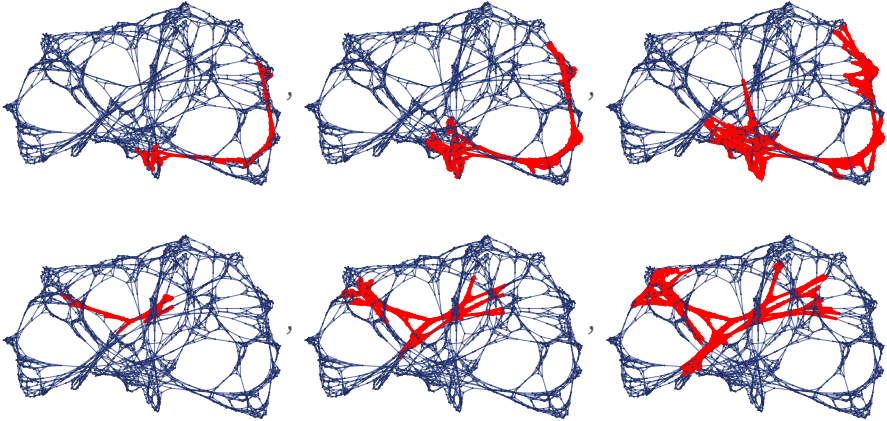

In the limit of a large number of steps, we can measure the volumes of tubes like these to compute approximations to projections of the Ricci tensor—and for example determine the level of isotropy of the emergent geometry of our models.

## 4.16  Functions on Graphs

Traditional Riemannian manifolds are full of structure that our hypergraphs do not have. Nevertheless, we are beginning to see that there are analogs of many ideas from geometry and calculus on manifolds that can be applied to our hypergraphs—at least in some appropriate limit as they become sufficiently large.

To continue the analogy, consider trying to define a function on a hypergraph. For a scalar function, we might just assign a value to each node of the hypergraph. And if we want the function to be somehow smooth, we should make sure that nearby nodes are assigned similar values.

But what about a vector function? An obvious approach is just to assign values to each directed edge of the hypergraph. And given this, we can find the component in a direction corresponding to a particular geodesic just by averaging over all edges of the hypergraph along that geodesic. (To recover results for continuous spaces, we must take all sorts of potentially intricate limits.)

(At a slightly more formal mathematical level, to define vectors in our system, we need some analog of a tangent space. On manifolds, the tangent space at a point can be defined in terms of the equivalence class of geodesics passing through that point. In our systems, the obvious analog is to look at the edges around a point, which are exactly what any geodesic through that point must traverse.)

For a rank-$p$ tensor function, we can assign values to $p$ edges associated either with a single node, or with a neighborhood of nearby nodes. And, once again, we can compute "projections" of the tensor in particular "directions" by averaging values along $p$ geodesics.

The gradient of a scalar function $\nabla f$ at a particular point $X$ can be defined by starting at $X$ and seeing along what geodesic the (suitably averaged) values decrease fastest, and at what rate. The results of this can then be assigned to the edges along the geodesic so as to specify a vector function.

The divergence of a vector function $\nabla \cdot \vec{f}$ can be defined by looking at a ball in the hypergraph, and asking for the total of the values of the function on all hyperedges in the ball. The analog of Gauss's theorem then becomes a fairly straightforward "continuity equation" statement about sums of values on edges inside and at the surface of part of a hypergraph.

## 4.17  Manifolds and Model Spaces

We saw above a rule which generates a sequence of hypergraphs like this:

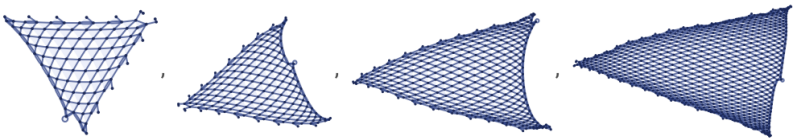

We can think of this after an infinite number of steps as giving an infinitely fine mesh—or in effect a structure which limits to a two-dimensional manifold. In standard mathematics, the defining feature of a manifold is that it is locally like Euclidean space (in some number of dimensions $d$) [45]. By using Euclidean space as a model many things can be defined and computed about manifolds (e.g. [26]).

Some of our models here yield emergent geometry whose limit is an ordinary manifold. But the question arises of what mathematical structures might be appropriate for describing the limiting behavior of other cases. Is there perhaps some other kind of model space whose properties can be transferred?

It is tempting to try to start from Euclidean space (or $\mathbb{R}^n$), and define some subset such as a Cantor set. But it seems more likely to be fruitful to start from convenient discrete structures, and see how their limits might correspond to what we have. One important feature of Euclidean space is its uniformity: every point is in a sense like every other, even if different points can be labeled by different coordinates.

So this suggests that by analogy we could consider graphs (and hypergraphs) whose vertices all have the same graph neighborhood (vertex transitive graphs). Several obvious infinite examples are the limits of:

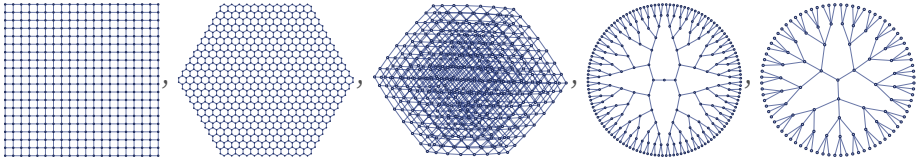

Any uniform tessellation or regular tree provides an example. One might think that another example would be graphs formed by uniform application of a vertex substitution rule—such as "spherical Sierpiński graphs" starting from a tetrahedron, dodecahedron or buckyball graph:

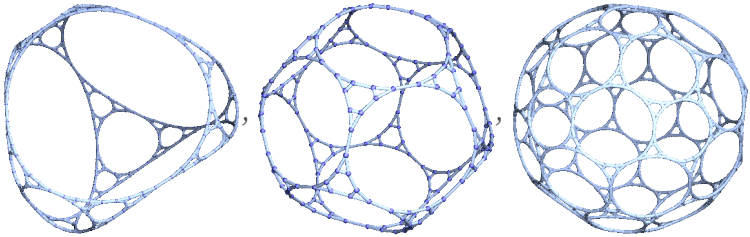

But (as mentioned in 4.8) in these graphs not every vertex has exactly the same neighborhood, at least if one goes beyond geodesic distance 2. The number of distinct neighborhoods does, however, grow fairly slowly, suggesting that it may be possible to consider such graphs "quasi vertex transitive" (in rough analogy to quasiconformal).

But one important class of graphs that are precisely vertex transitive are Cayley graphs of groups—and indeed the infinite tessellation and tree graphs above are all examples of these. (Note that not all vertex-transitive graphs are Cayley graphs; the Petersen graph is an example [46]. It is also known that there are infinite vertex-transitive graphs that are not Cayley graphs [47], and are not even "close" to any such graphs [48].)

In a Cayley graph for a group, each node represents an element of the group, and each edge is labeled with a generator of the group. (Different presentations of the group—with different choices of generators and relations—can have slightly different Cayley graphs, but their infinite limits can be considered the same.) Each point in the Cayley graph can then be labeled (typically not uniquely) by a word in the group, specified as a product of generators of the group.

One can imagine progressively building up a Cayley graph by looking at longer and longer words. In the case of a free group with two generators A and B, this yields:

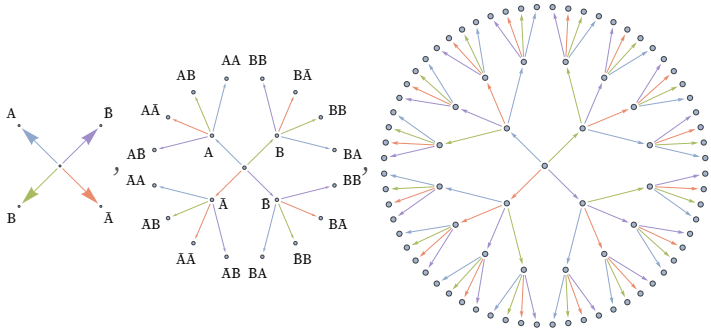

If one adds the relation AB = BA, defining an Abelian group, the Cayley graph is instead a grid, with "coordinates" given by the numbers of As and Bs (or their inverses ):

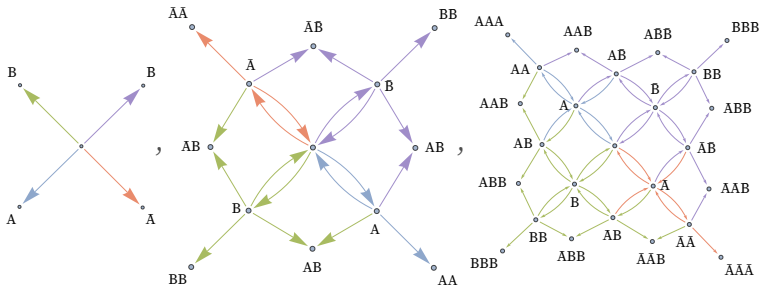

Here are the Cayley graphs for the first few symmetric and alternating (finite) groups:

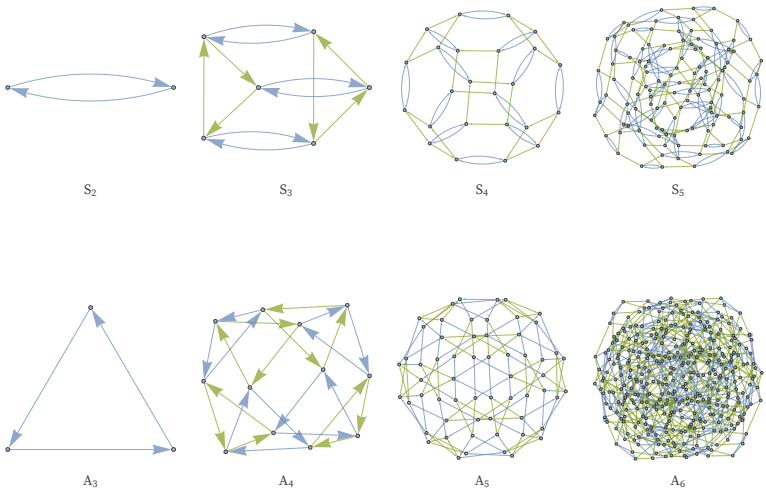

For a given infinite Cayley graph, one can compute a limiting $V_r$ just as we have for hypergraphs. If one picks a finite number of generators and relations at random, one will usually get a Cayley graph that has a basically tree-like structure, with a $V_r$ that grows exponentially. For nilpotent groups, however, $V_r$ always has polynomial growth—an example being the Heisenberg group $H_3$ whose Cayley graph is the limit of [49]:

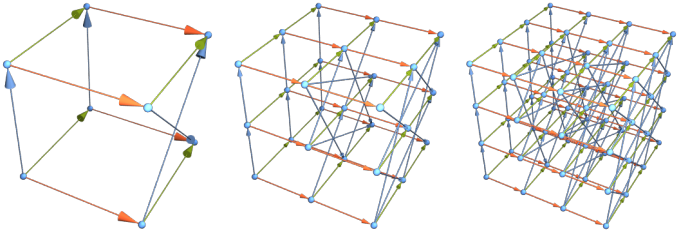

There are also groups known that yield growth intermediate between polynomial and exponential [50]. There do not, however, appear to be groups that yield fractional-power growth, corresponding to finite but fractional dimension.

It is possible that one could view the evolution of one of our models as being directly analogous to the growth of the Cayley graph for a group—or at least somehow approximated by it. As we discussed above, the hypergraphs generated by most of our systems are not, however, uniform, in the sense that the structures of the neighborhoods around different points in the hypergraph can be different. But this does not mean that a Cayley graph could not provide a good (at least approximate) local model for a part of the hypergraph. And if this connection could be made, there might be useful results from modern geometric group theory that could be applied, for example in classifying different kinds of limiting behaviors of our systems.

On a sufficiently small scale, any manifold is defined to be like Euclidean space. But if one goes to a slightly larger scale, one needs to represent deviations from Euclidean space. And a convenient way to do this is again to consider model spaces. The most obvious is a sphere (or in general a hyperellipsoid)—and this is what gives the notion of curvature. Quite what the appropriate analog even of this is in fractional dimensional space is not clear, but it would potentially be useful in studying our systems. And when there is a possibility for change in dimension as well as change in curvature, the situation is even less clear.

# 5 | The Updating Process for String Substitution Systems

## 5.1 String Substitution Systems

The basic concept of our models is to define rules for updating collections of relations. But for a particular collection of relations, there are often multiple ways in which a given rule can be applied, and there is considerable subtlety in the question of what effects different choices can have.

To begin exploring this, we will first consider in this section the somewhat simpler case of string substitution systems (e.g. [1:3.5][51]). String substitution systems have arisen in many different settings under many different names [1:p893], but in all cases they involve strings whose elements are repeatedly replaced according to fixed substitution rules.

As a simple example, consider the string substitution system with rules {A→AB,B→BA}. Starting with A and repeatedly applying these rules wherever possible gives a sequence of results beginning with:

{A, AB, ABBA, ABBABAAB, ABBABAABBAABABBA, ABBABAABBAABABBABAABABBAABBABAAB}

The application of these particular rules is simple and unambiguous. At each step, every occurrence of A or B is independently replaced, in a way that does not depend on its neighbors. One can visualize the process as a tree:

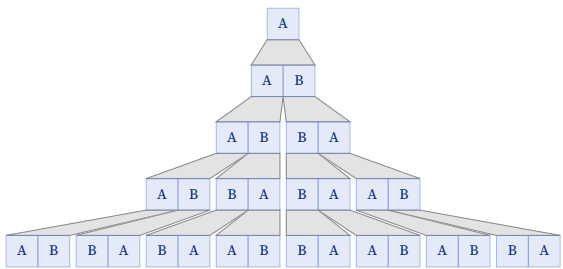

At step $n$, there are $2^n$ elements, and the $k^{\text{th}}$ element is determined by whether the number of 1s in the base-2 decomposition of $k$ is even or odd. (This particular case corresponds to the Thue–Morse sequence.)

The evolution of the string substitution system can still be represented by a tree even if the replacements are not the same length. The rules {A→B, B→AB} yield the "Fibonacci tree":

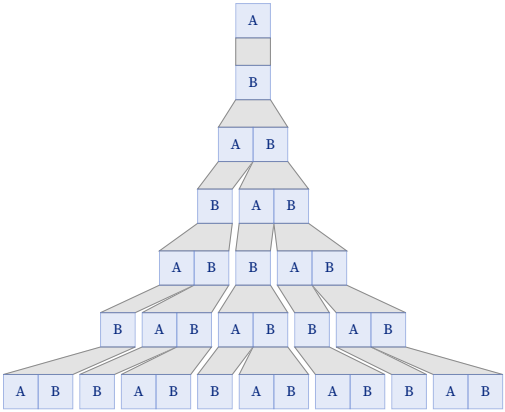

In this case, the number of elements on the $t^{th}$ step is the $t^{th}$ Fibonacci number. For any neighbor-independent substitution system, the number of elements on the $n^{th}$ step is determined by a linear recurrence, and usually, but not always, grows exponentially. (Equivalently, the number of elements of each type on the $n^{th}$ step can be determined from the $n^{th}$ power of the transition matrix.)

But consider now a substitution system with rules {A→BBB, BB→A}. If one starts with A, the first step is unambiguous: just replace A by BBB. But now there are two possible replacements for BBB: either replace the first BB to get AB, or replace the second one to get BA.

We can represent all the different possibilities as a multiway system [1:5.6] in which there can be multiple outcomes at each step, and there are multiple possible paths of evolution (here shown over the course of 5 steps):

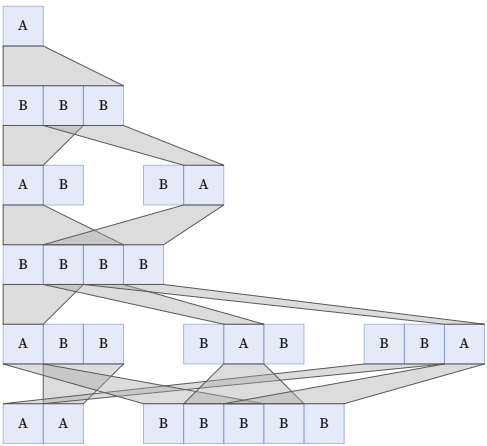

We can represent the possible paths of 5 steps of evolution between states of the system by a graph:

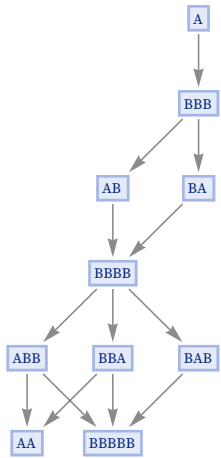

In effect each path through this graph represents a different possible history for the system, based on applying a different sequence of possible updates.

By adding something extra to our model, we can of course force a particular history. For example, we could consider sequential substitution systems (analogous to search-and-replace in a text editor) in which we always do only the first possible replacement in a left-to-right scan of the state reached at each step. With this setup we get the following history for the system shown above:

{A, BBB, AB, BBBB, ABB, BBBBB, ABBB, BBBBBB, ABBBB, BBBBBBB, ABBBBB}

An alternative strategy (analogous, for example, to the operation of `StringReplace` in the Wolfram Language) is to scan from left to right, but rather than just doing the first possible replacement at each step, instead keep scanning after the first replacement, and also carry out every subsequent replacement that can independently be done. With this "maximum scan" strategy, the sequence of states reached in the example above becomes:

{A, BBB, AB, BBBB, AA, BBBBBB, AAA, BBBBBBBBB, AAAAB, BBBBBBBBBBBBB, AAAAAAB}

(The first deviation occurs at BBBB. After replacing the first BB, the maximum scan strategy can continue and replace the second BB as well, thereby in effect "skipping a step" in the multiway evolution graph shown above.)

Note that in both the strategies just described, the evolution obtained can depend on the order in which different replacements are stated in the rule. With the rule {BB→A, A→BBB} instead of {A→BBB, BB→A}, the sequential substitution system updating scheme yields:

{A, BBB, AB, BBBB, ABB, AA, BBBA, ABA, BBBBA, ABBA, AAA}

instead of:

{A, BBB, AB, BBBB, ABB, BBBBB, ABBB, BBBBBB, ABBBB, BBBBBBB, ABBBBB}

Given a particular multiway system, one can ask whether it can ever generate a given string. In other words, does there exist any sequence of replacements that leads to a given string?

In the case of {A→BBB, BB→A}, starting from A, the strings B and BB cannot be generated, though with this particular rule, all other strings eventually can be generated (it takes 5 ($k - 1$) steps to get all $2^k$ strings of length $k$).

One of the applications of multiway systems is as an idealization of derivations in equational logic [1:p777], in which the rules of the multiway system correspond to axioms that define transformations between equivalent expressions in the logical system. Starting from a state corresponding to a particular expression, the states generated by the multiway system are expressions that are ultimately equivalent to the original expression. The paths in the multiway system are then chains of transformations that represent proofs of equivalences between expressions—and the problem of whether a particular equivalence between expression holds is reduced to the (still potentially very difficult) problem of determining whether there is a path in the multiway system that connects the states corresponding to these expressions.

Thus, for example, with the transformations {A→BBB,BB→A}, it is possible to see that A can be transformed to AAA, but the path required is 10 steps long:

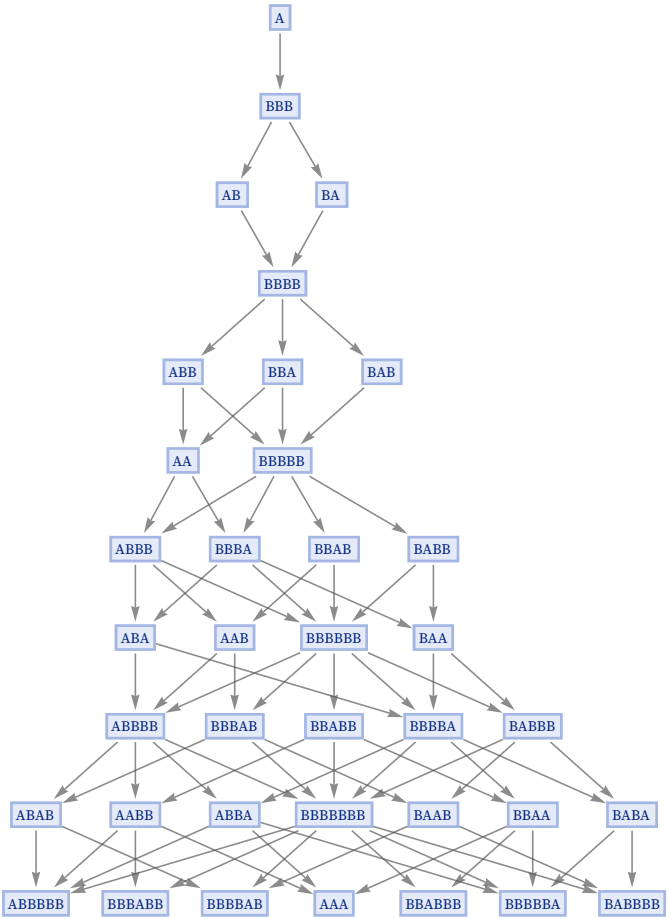

In general, there is no upper bound on how long a path may be required to reach a particular string in a multiway system, and the question of whether a given string can ever be reached is in general undecidable [52][53][1:p778].

## 5.2 The Phenomenon of Causal Invariance

Consider the rule {BA→AB}, starting from BABABA:

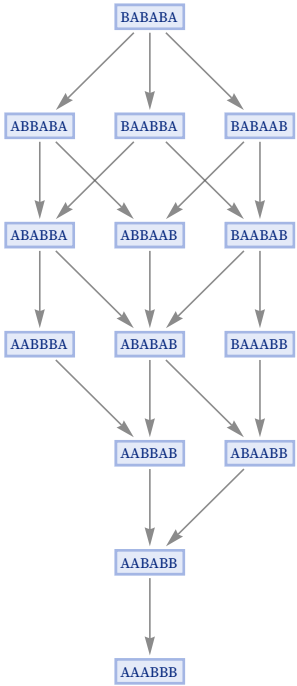

As before, there are different possible paths through this graph, corresponding to different possible histories for the system. But now all these paths converge to a single final state. And in this particular case, there is a simple interpretation: the rule is effectively sorting As in front of Bs by repeatedly doing the transposition BA→AB. And while there are multiple different possible sequences of transpositions that can be used, all of them eventually lead to the same answer: the sorted state AAABBB.

There are many practical examples of systems that behave in this kind of way, allowing operations to be carried out in different orders, generating different intermediate states, while always leading to the same final answer. Evaluation or simplification of (parenthesized) arithmetic (e.g. [54]), algebraic or Boolean expressions are examples, as is lambda function evaluation [55].

Many substitution systems, however, do not have this property. For example, consider the rule {AB→AA,AB→BA}, again starting from BABABA. Like the sorting rule, after a limited number of steps this rule gets into a final state that no longer changes. But unlike the sorting rule, it does not have a unique final state. Depending on what path is taken, it goes in this case to one of three possible final states:

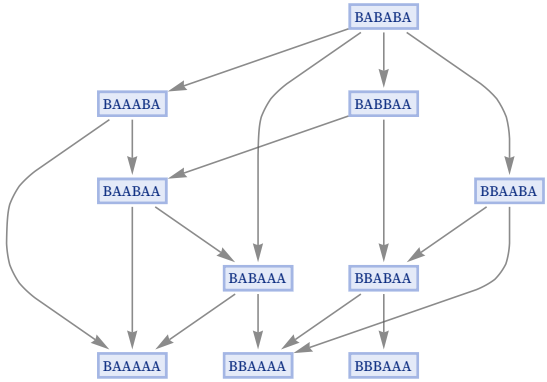

But what about systems that do not "terminate" at particular final states? Is there some way to define a notion of "path independence" [55]—or what we will call "causal invariance" [1:9.10]—for these?

Consider again the rule {A→BBB,BB→A} that we discussed above:

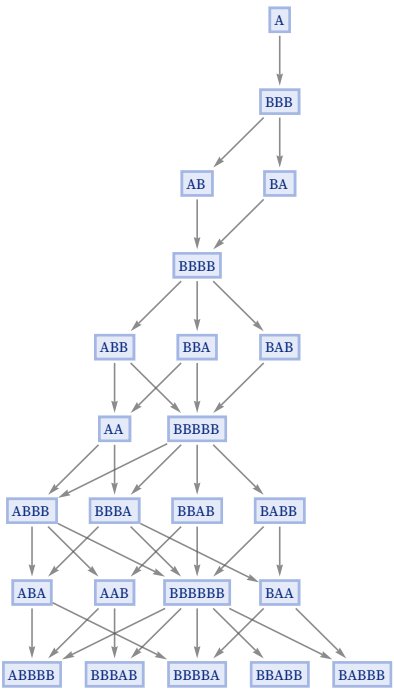

The state BBB at step 2 has two possible successors: AB and BA. But after another step, AB and BA converge again to the state BBBB. And in fact the same kind of thing happens throughout the graph: every time two paths diverge, they always reconverge after just one more step. This means that the graph in effect consists of a collection of "diamonds" of edges [56]:

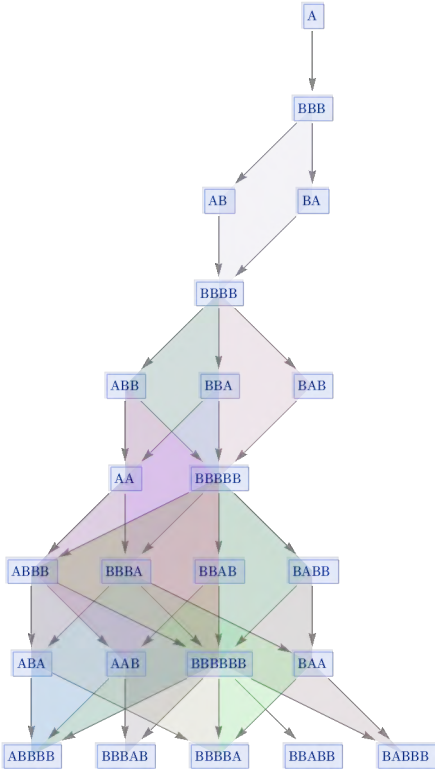

There is no need, however, for reconvergence to happen in just one step. Consider for example the rule {A→AA, AA→AB}:

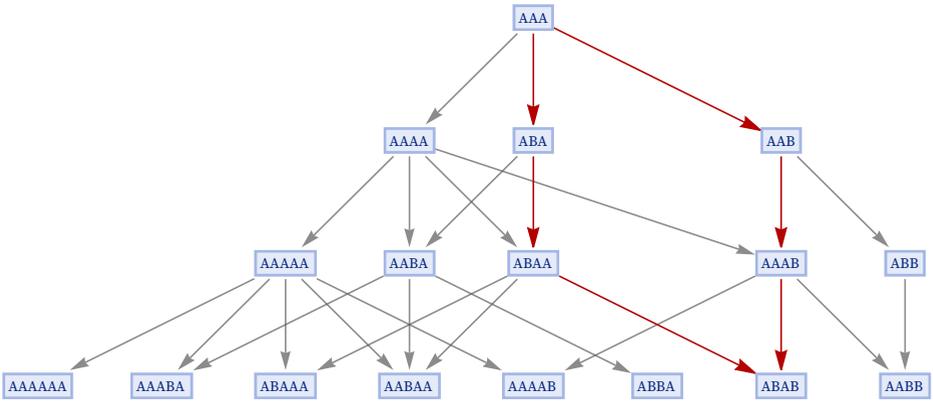

As the picture indicates, there are paths from AAA leading to both ABA and AAB—but these only reconverge again (to ABAB) after two more steps. (In general, it can take an arbitrary number of steps for reconvergence to occur.)

Whether a system is causal invariant may depend on its initial conditions. Consider, for example, the rule AA→AAB. With initial condition AABAA the rule is causal invariant, but with initial condition AAA it is not

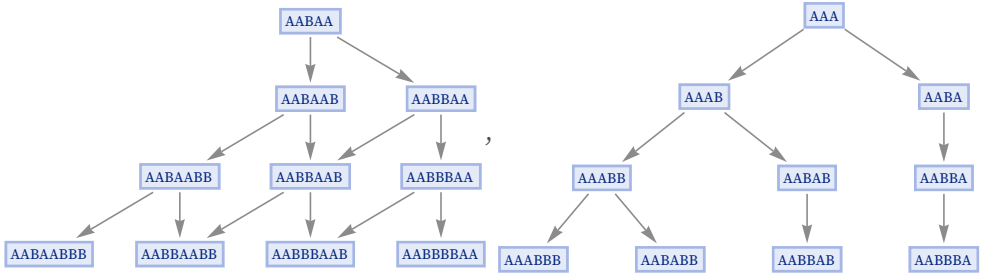

When a system is causal invariant for all possible initial conditions, we will say that it is totally causally invariant. (This is essentially the confluence property discussed in the theory of term-rewriting systems.) Later, we will discuss how to systematically test for causal invariance—and we will see that it is often easier to test for total causal invariance than for causal invariance for specific initial conditions.

Causal invariance may at first seem like a rather obscure property. But in the context of our models, we will see in what follows that it may in fact be the key to a remarkable range of fundamental features of physics, including relativistic invariance, general covariance, and local gauge invariance, as well as the possibility of objective reality in quantum mechanics.

## 5.3 States Graphs

If there are multiple successors to a particular state in a substitution system one thing to do would be just to assume that each of these successors is a new, unique state. The result of this will always be to produce a tree of states, here shown for the rule {A→BBB, BB→A}:

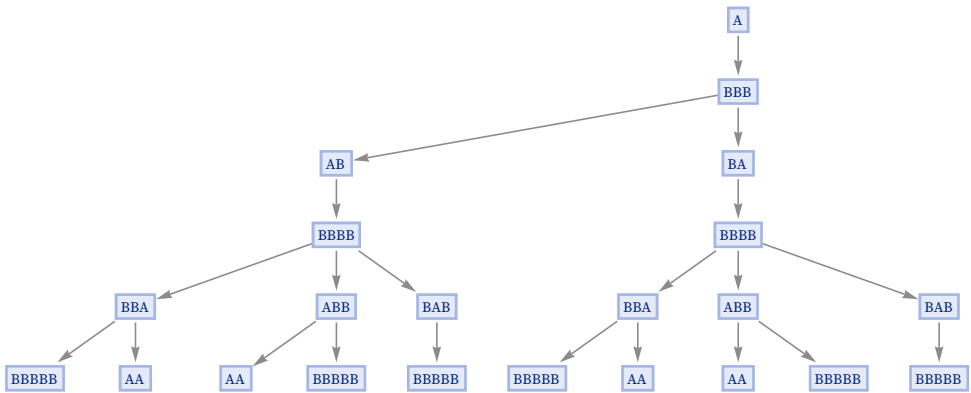

But in our construction of multiway systems, we assume that actually the states produced are not all unique, and instead that states at a given step consisting of the same string can be merged, thereby reducing the tree above to the directed graph:

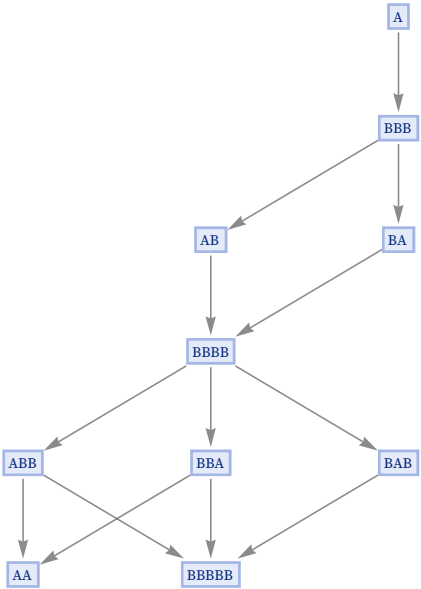

But if we are going to merge identical states, why do so only at a given step? Why not merge identical states whenever they occur in the evolution of the system? After all, given the setup, a particular state—wherever it occurs—will always evolve in the same way, so in some sense it is redundant to show it multiple times.

The particular rule {A→BBB,BB→A} that we have just used as an example has the special feature that it always "makes progress" and never repeats itself—with the result that a given string only ever appears once in its evolution. Most rules, however, do not have this property.

Consider for example the rule {AB→BAB, BA→A}. Starting from ABA, here is our normal "evolution graph":

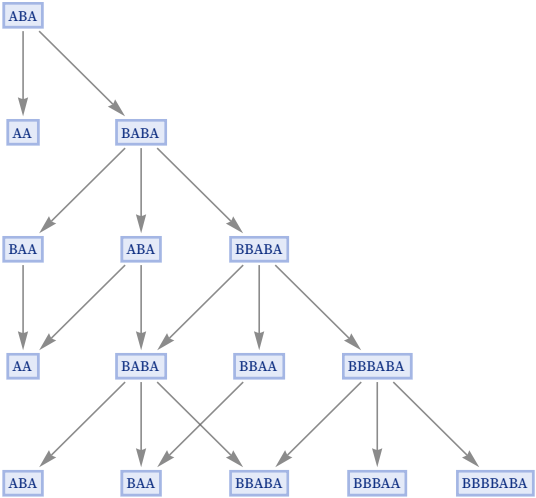

Notice that in this evolution even the original state ABA appears again, both at step 3 and at step 5—and each time it appears, it necessarily makes a complete repeat of the same evolution graph. To remove this redundancy, we can make a graph in which we effectively merge all instances of a given state, so that we show each state only once, connecting it to whatever states it evolves to under the rule:

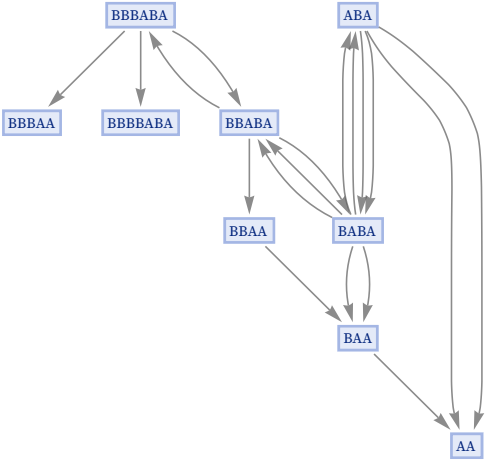

But in this graph there are, for example, two connections from ABA to BABA, because this transformation happens twice in the 4 steps of evolution that we are considering. But in a sense this multiplicity again always gives redundant information, so in what we will call our "states graph" [1:p209], we only ever keep one connection between any given pair of states, so that in this case we get:

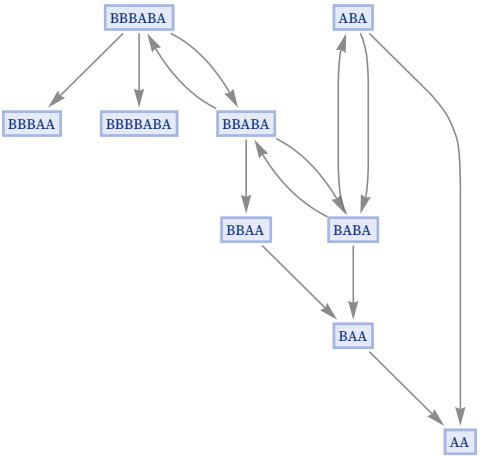

There is one further subtlety in the construction of a states graph. The graph only records which state can be transformed into which other: it does record how many different replacements could be applied to achieve this. In the rule we just showed, it is never possible to have different replacements on a single string yield the same result.

Consider the rule {A→AA, A→B} starting with AA. There are, for example, two different ways that this rule can be applied to AA to get AAA:

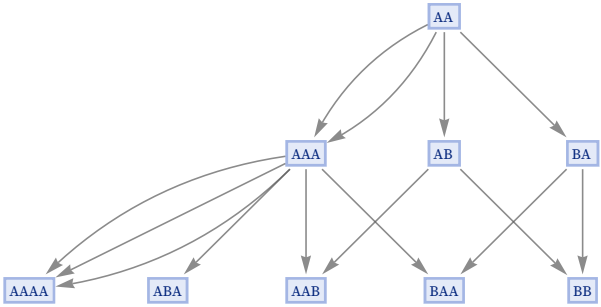

In our standard states graph, however, we show only that AA is transformed to AAA, and we do not record how many different possible replacements can achieve this:

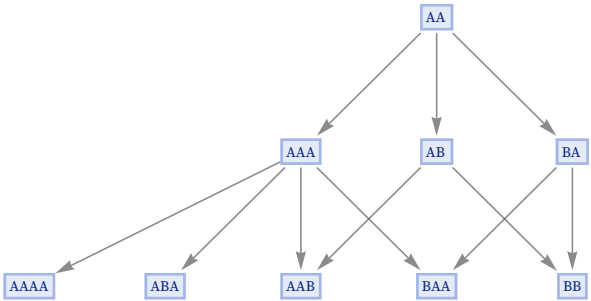

The degree of compression achieved in going from evolution graphs to states graphs can be quite dramatic. For example, for the rule {BA→AB, AB→BA} the evolution graph is:

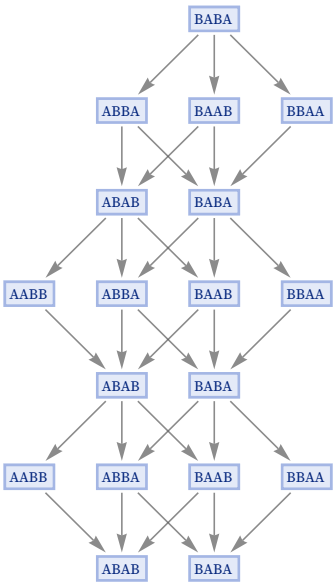

and the states graph is:

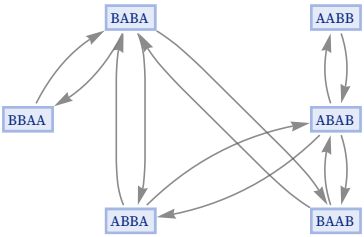

or in our standard rendering:

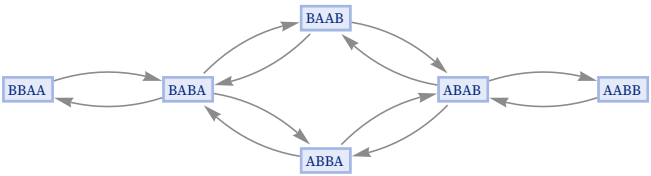

Note that causal invariance works the same in states graphs as it does in evolution graphs: if a rule is causal invariant, any two paths that diverge must eventually reconverge.

## 5.4 Typical Multiway Graph Structures

Before considering further features of the updating process, it is helpful to discuss the typical kinds of multiway graphs that are generated from string substitution systems. Much as for our primary models based on general relations (hypergraphs), we can assign signatures to string substitution system rules based on the lengths of the strings in each transformation. For example, we will say that the rule {A→BBB,BB→A} has signature 2: 1→3, 2→1, where the initial 2 indicates the number of distinct possible elements (here A and B) that occur in the rule.

With a signature of the form $k$: $n_1 \to n_2$, $n_3 \to n_4$, … there are nominally $k^{\sum n_i}$ possible rules. However, many of these rules are equivalent under renaming of elements or reversal of strings. Taking this into account, the number of inequivalent possible rules for various cases is:

| | k=2 | k=3 | k=4 |
|---|---|---|---|
| 1 → 1 | 2 | 2 | 2 |
| 1 → 2 | 3 | 4 | 4 |
| 1 → 3 | 6 | 10 | 11 |
| 2 → 2 | 6 | 10 | 11 |
| 1 → 1, 1 → 1 | 8 | 14 | 15 |
| 1 → 4 | 10 | 25 | 31 |
| 2 → 3 | 10 | 25 | 31 |
| 1 → 1, 1 → 2 | 12 | 28 | 35 |
| 1 → 5 | 20 | 70 | 107 |
| 2 → 4 | 20 | 70 | 107 |
| 3 → 3 | 20 | 70 | 107 |
| 1 → 1, 1 → 3 | 24 | 82 | 123 |
| 1 → 1, 2 → 2 | 20 | 70 | 107 |
| 1 → 2, 1 → 2 | 20 | 70 | 107 |
| 1 → 2, 2 → 1 | 20 | 70 | 107 |
| 1 → 1, 1 → 1, 1 → 1 | 32 | 122 | 187 |

| | k=2 | k=3 | k=4 |
|---|---|---|---|
| 1 → 6 | 36 | 196 | 379 |
| 2 → 5 | 36 | 196 | 379 |
| 3 → 4 | 36 | 196 | 379 |
| 1 → 1, 1 → 4 | 40 | 205 | 395 |
| 1 → 1, 2 → 3 | 40 | 205 | 395 |
| 1 → 2, 1 → 3 | 40 | 205 | 395 |
| 1 → 2, 3 → 1 | 40 | 205 | 395 |
| 1 → 3, 2 → 1 | 40 | 205 | 395 |
| 1 → 2, 2 → 2 | 36 | 196 | 379 |
| 1 → 1, 1 → 1, 1 → 2 | 48 | 244 | 459 |
| 1 → 7 | 72 | 574 | 1451 |
| 2 → 6 | 72 | 574 | 1451 |
| 3 → 5 | 72 | 574 | 1451 |
| 1 → 1, 1 → 5 | 80 | 610 | 1515 |
| 4 → 4 | 72 | 574 | 1451 |
| 1 → 1, 2 → 4 | 72 | 574 | 1451 |

| | k=2 | k=3 | k=4 |
|---|---|---|---|
| 1 → 2, 1 → 4 | 72 | 574 | 1451 |
| 1 → 2, 4 → 1 | 72 | 574 | 1451 |
| 1 → 4, 2 → 1 | 72 | 574 | 1451 |
| 1 → 1, 3 → 3 | 80 | 610 | 1515 |
| 1 → 3, 1 → 3 | 80 | 610 | 1515 |
| 1 → 3, 3 → 1 | 80 | 610 | 1515 |
| 1 → 2, 2 → 3 | 72 | 574 | 1451 |
| 1 → 2, 3 → 2 | 72 | 574 | 1451 |
| 1 → 3, 2 → 2 | 72 | 574 | 1451 |
| 2 → 1, 2 → 3 | 72 | 574 | 1451 |
| 1 → 1, 1 → 1, 1 → 3 | 96 | 730 | 1771 |
| 2 → 2, 2 → 2 | 72 | 574 | 1451 |
| 1 → 1, 1 → 1, 2 → 2 | 80 | 610 | 1515 |
| 1 → 1, 1 → 2, 1 → 2 | 80 | 610 | 1515 |
| 1 → 1, 1 → 2, 2 → 1 | 80 | 610 | 1515 |
| 1 → 1, 1 → 1, 1 → 1, 1 → 1 | 128 | 1094 | 2795 |

Rules with signature $k: 1 \to n$ must always be totally causal invariant. If they are started from strings of length 1, their states graphs can never branch. However, with strings of length more than 1, branching can (but may not) occur, as in this example of A→AB started from three strings of length 2:

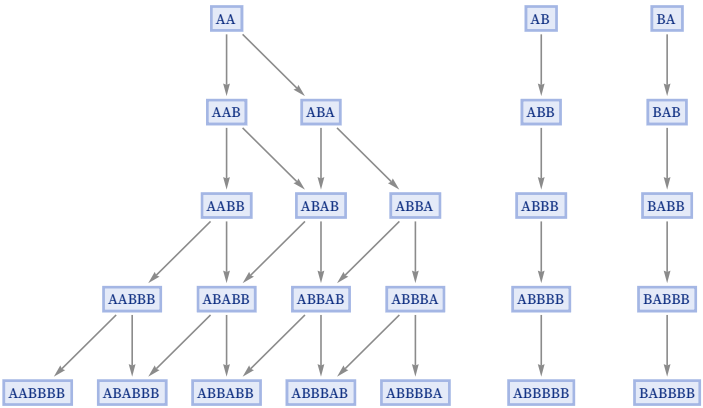

With initial conditions AAA and AAAA, this rule produces the following states graphs:

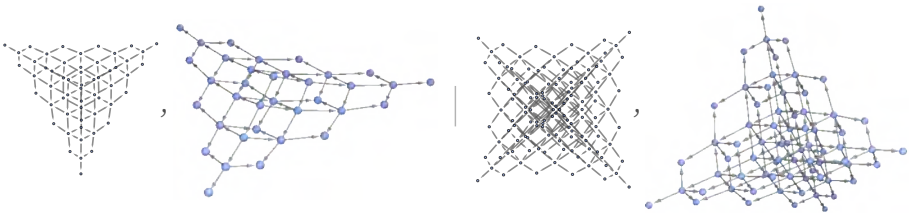

Even the rule A→AA can produce states graphs like these if it has initial conditions that contain Bs as "separators":

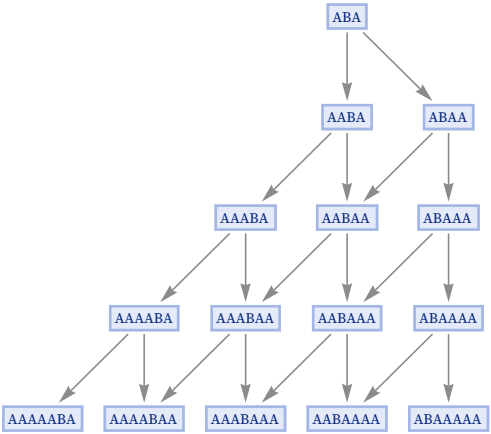

And as a seemingly even more trivial example, the rule A→B with an initial condition containing $n$ As gives a states graph corresponding to an $n$-dimensional cube (with a total of $2^n$ nodes):

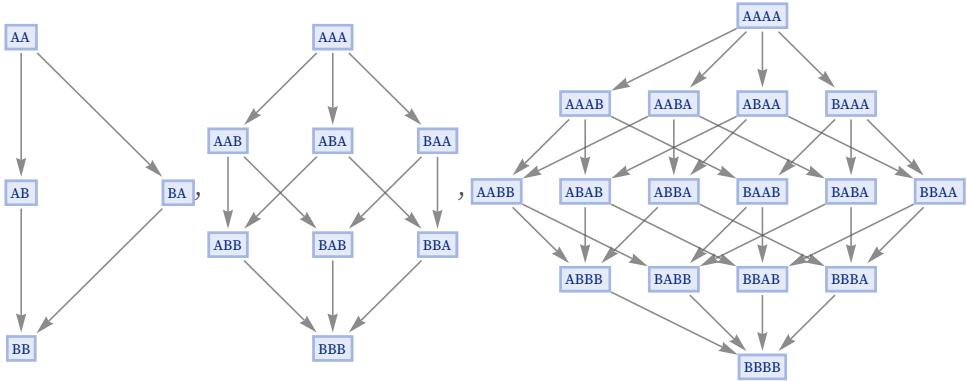

With multiple rules, one can get tree-like structures, with exponentially increasing numbers of states. The simplest case is the 2: 0→1, 0→1 rule {""→A, ""→B} starting with the null string "":

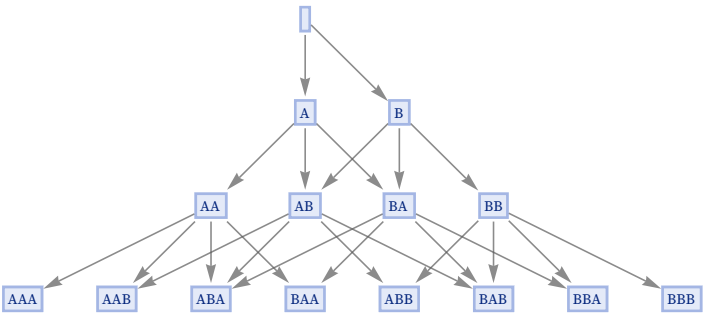

With multiple rules, even with single-symbol left-hand sides, causal invariance is no longer guaranteed. Of the 8 inequivalent 2: 1→1, 1→1 rules, all are totally causal invariant, although the rule {A→B, B→A} achieves this through cyclic behavior:

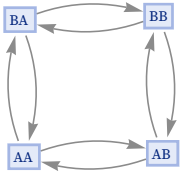

Among the 14 inequivalent 3: 1→1, 1→1 rules, all but two are totally causal invariant. The exceptions are the cyclic rule, and also the rule {A→B, A→C}, which in effect terminates before its B and C branches can reconverge:

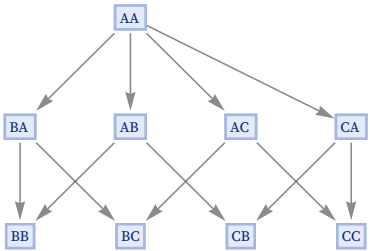

The 12 inequivalent 2: 1→2, 1→1 rules yield the following states graphs when run for 4 steps starting from all possible length-3 strings of As and Bs:

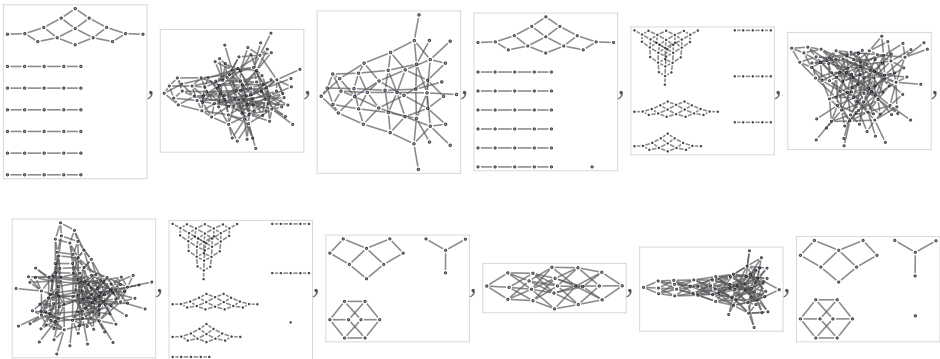

All but three of these rules are totally causal invariant. Among causal invariant ones are:

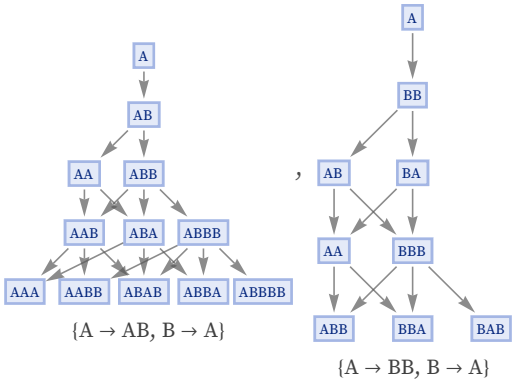

After more steps these yield:

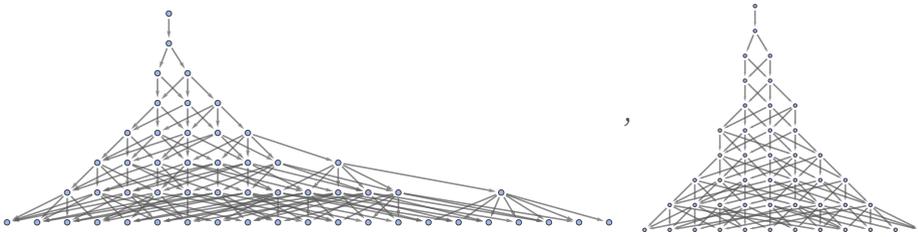

Examples of non-causal invariant 2: 1→2, 1→1 rules are:

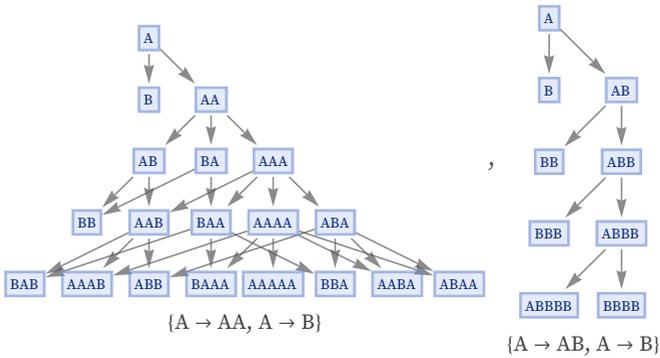

After more steps these yield:

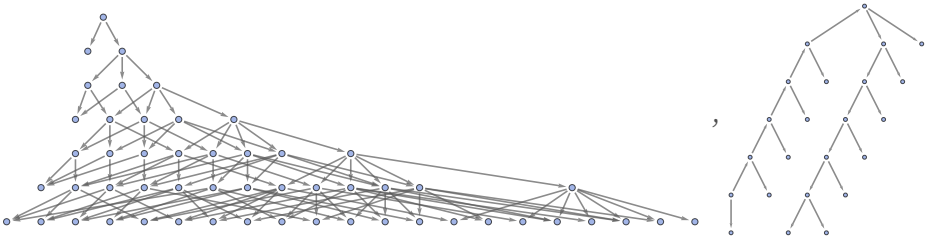

When we look at rules with larger signatures, the vast majority at least superficially show the same kinds of behavior that we have already seen.

Like for the hypergraphs from our models that we considered in previous sections, we can study the limiting structure of states graphs generated in multiway systems, and see what emergent geometry they may have. And in analogy to $V_r(X)$ for hypergraphs, we can define a quantity $M_t(S)$ which specifies the total number of distinct states reached in the multiway system after $t$ steps of evolution starting from a state $S$.

For the rule {A→AB} mentioned above, the geometry of the multiway graph obtained by starting from $n$ As is effectively a regular $n$-dimensional grid:

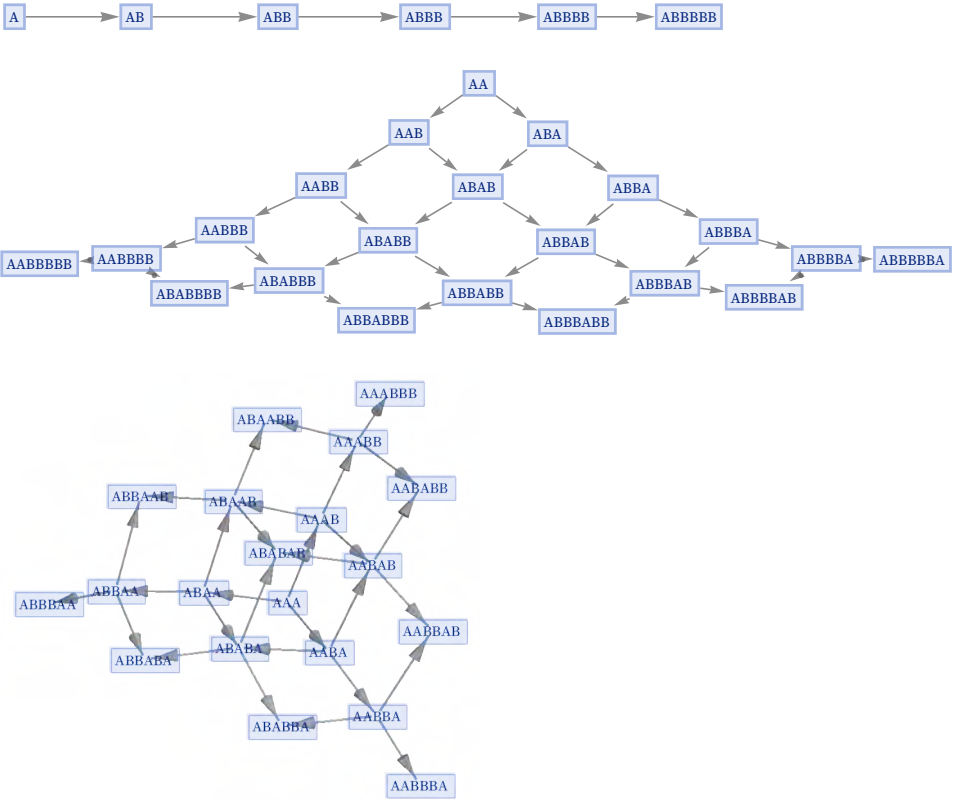

$M_t(A^n)$ in this case is then an $n$-fold nested sum of $t$ 1s:

$$M_t(A^n) = \prod_{k=0}^{n-1} \frac{(t-k)}{n!} \sim \frac{t^n}{n!}\left(1 - \binom{n}{2}\frac{n}{t} + \ldots\right)$$

Some rules create states graphs that are trees, with $M_t \sim m^t$. Other rules do not explicitly give trees, but still give exponentially increasing $M_t$. An example is {A→AB,B→A}, for which $M_t$ is a sum of Fibonacci numbers, and successive states graphs can be rendered as:

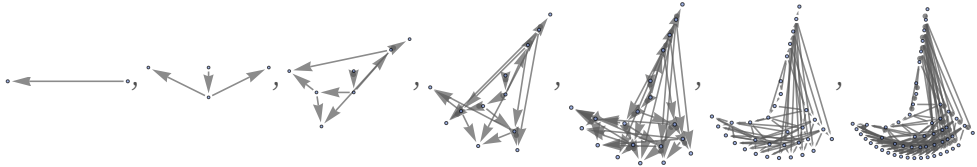

Note that rules with both polynomial and exponential growth in $M_t$ can exhibit causal invariance.

It is not uncommon to find rules with fairly long transients. A simple example is the totally causal invariant rule {A→BBB,BBBB→A}. When started from $n$ As this stabilizes after $7n$ steps, with $\sim 2.8^n$ states, with a states graph like:

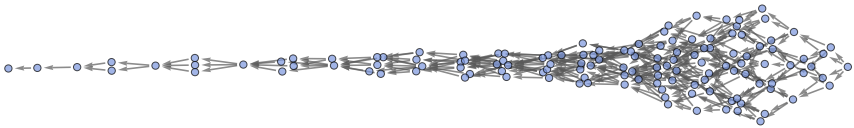

Rules typically seem to have either asymptotically exponential or asymptotically polynomial $M_t$. (This may have some analogy with what is seen in the growth of groups [57][58][22].) Among rules with polynomial growth, it is typical to see fairly regular grid-like structures. An example is the rule

{AB → A, ABA → BBAABB}

which from initial condition ABAA gives states graph:

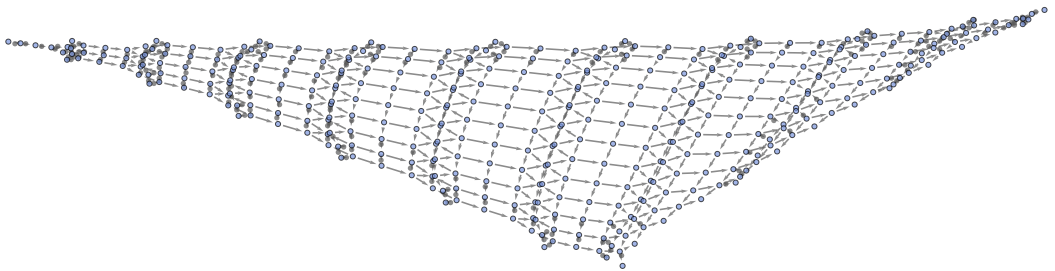

With the slightly different initial condition ABAAA, the states graph has a more elaborate structure:

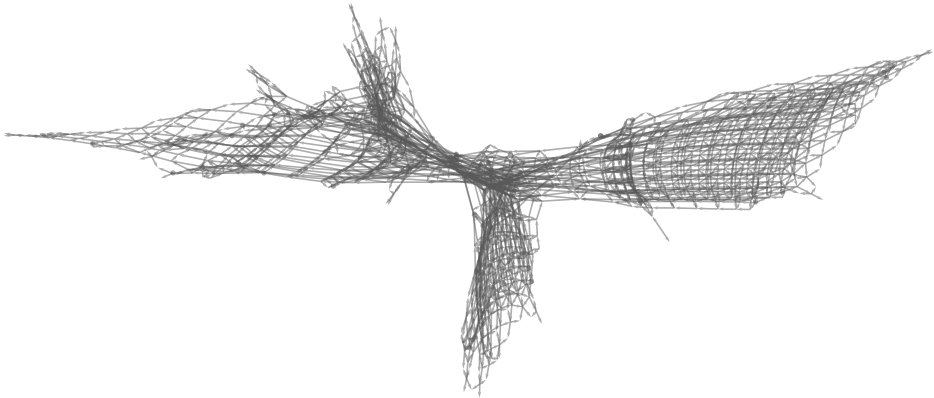

and the form of $M_t$ is more complicated (though $\sim t^2$ and ultimately quasiperiodic); the second differences of $M_t$ in this case are:

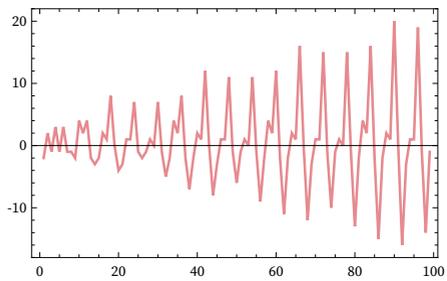

Another example of a somewhat complex structure occurs in the rule

{ABA → BBAA, BAA → AAB}

that was discussed in [1:p 205]. Starting from BABBAAB it gives the states graph:

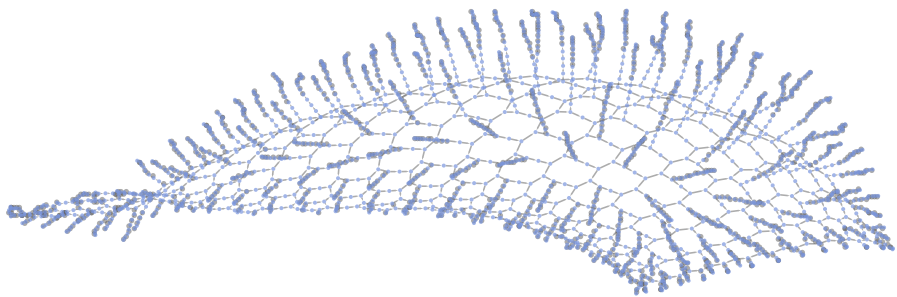

## 5.5 Testing for Causal Invariance

Causal invariance implies that regardless of the order in which updates are made, it is always possible to get to the same outcome. One way this can happen is if different updates can never interfere with each other. Consider the rule {A→AA, B→BB}. Every time one sees an A, it can be "doubled", and similarly with a B. But these doublings can happen independently, in any order. Looking at an evolution graph for this rule we see that at each state there can be an "A replacement" applied, or a "B replacement". There are two branches of evolution depending on which replacement is chosen, but one can always eventually reach the same outcome, regardless of which choice was made—and as a result the rule is causal invariant:

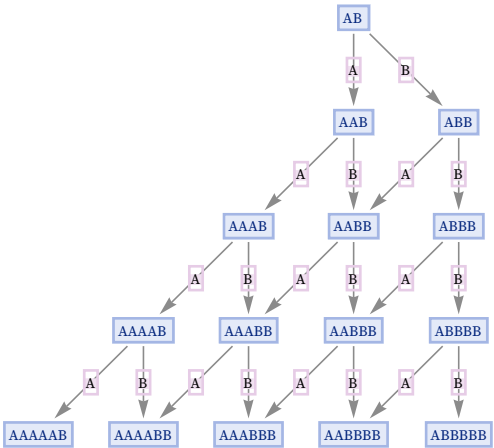

The replacements A→AA and B→BB trivially cannot interfere because their left-hand sides involve different elements. But a more general way to guarantee that interference cannot occur is for the left-hand sides of replacements not to be able to overlap with each other—or with themselves.

In general, the set of strings of As and Bs up to length 5 that do not overlap themselves are (up to A,B interchange and reversal): [1:p1033][59]

{A, AB, AAB, AAAB, AABB, AAAAB, AAABB, AABAB}

These can be formed into pairs in the following ways:

{{A, B}, {AABAB, AABB}, {AABB, ABABB}, {AAABB, AABAB}, {AAABB, ABABB}}

The first triples with no overlaps are of length 6. An example is {AAABB, ABAABB, ABABB}. And whenever there is a set of strings with no overlaps being used as the left-hand side of replacements, one is guaranteed to have a system that is totally causal invariant.

It is also perfectly possible for the right-hand sides of rules to be such that the system is totally causal invariant even though their left-hand sides overlap. An example is the simple rule {A→AA, AA→A}:

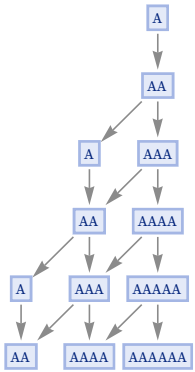

At every branch point in a multiway system, one can identify all pairs of strings that can be generated. We will call such pairs of strings branch pairs (they are often called "critical pairs" [60]). And the question of whether a system is causal invariant is then equivalent to the question of whether all branch pairs that can be generated will eventually resolve, in the sense that there is a common successor for both members of the pair [61][62][63][64].

Consider for example the rule {A→AB, B→A}:

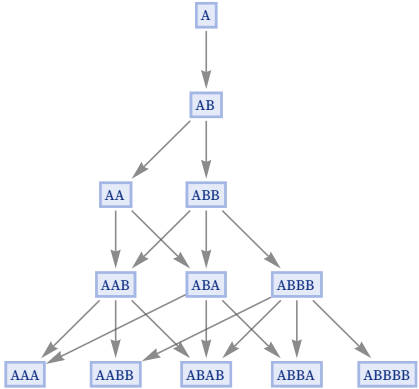

After two steps, a branch pair appears: {AA, ABB}. But after just one more step it resolves. However, two more branch pairs are generated: {AAB, ABA} and {ABA, ABBB}. But after another step, these also resolve. And in fact in this system all branch pairs that are ever generated resolve (actually always in just one step), and so the system is causal invariant.

In general, it can, however, take more than one step for a branch pair to resolve. The simplest case involving resolution after two steps involves the rule:

{A → B, AB → AA}

The state AB generates the branch pair {BB, AA}:

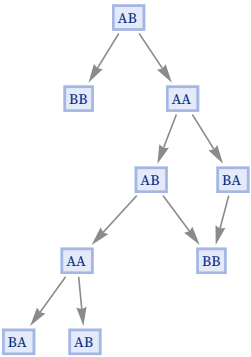

In the evolution graph, we do not see a resolution of this branch pair. But looking at the states graph, we see that the branch pair does indeed resolve in two steps, though with BB being a terminating state:

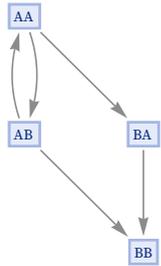

The same kind of thing can also happen without a terminating state. Consider for example

{A → AA, AB → BA}

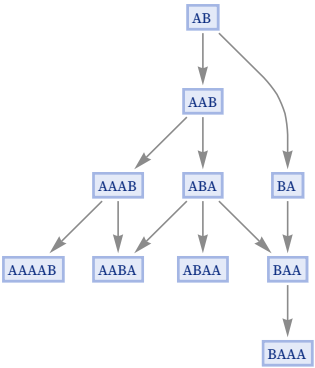

where the branch pair {AAB, BA} takes 2 steps to resolve.

In the rule

{A → AA, AAB → BA}

it takes 3 steps for the branch pair {AAAB, BA} to resolve:

```
                    AAB
                     ↓
                    AAAB
                   ↙   ↓
              AAAAB   ABA
               ↓  ↓    ↓
         BA  AAAAAB  AABA
        ↙↓   ↓       ↓
  AAAAAAB  BAA      AAABA
    ↓      ↓         ↓              ABAA
  AAAAAAAB AAAABA                    ↓
                        AABAA     ABAAA
                         ↓  ↓      ↓
              BAAA  AAABAA  AABAAA ABAAAA
               ↓
             BAAAA
               ↓
            BAAAAA
```

Things can get quite complicated even with simple rules. For example, in the rule

{A → AA, AA → BAB}

the branch pair {AAA, BAB} resolves after 4 steps. The following is essentially a proof of this:

```
   AAA          BAB
    ↓            ↓
   AAAA         BAAB
    ↓            ↓
   AAAAA        BAAAB
    ↓            ↓
  AAABAB       BAAAAB
       ↘        ↓
           BABABAB
```

But finding this in the 67-node 4-step states graph is quite complicated:

In the rule

{A → AAB, ABBA → A}

it takes 7 steps for the branch pair

{A, AABBBA}

to resolve to the common successor AAAABBBAAB. The 7-step states graph involved has 5869 nodes, and the proof that the branch pair resolves is:

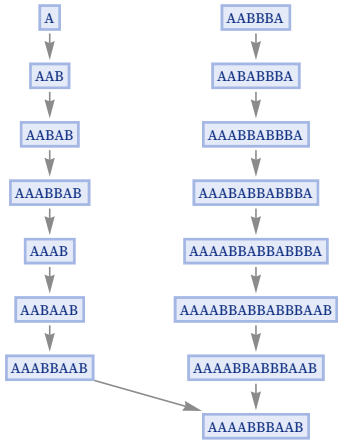

In general, there is no upper bound on how long it may take a branch pair to resolve, or for example how long its common successor—or intermediate strings involved in reaching it—may be. Here are the simplest rules with two distinct elements that take successively longer to resolve (the last column gives the size of the states graph when resolution is found):

| 1 | {A → A} | A → {A, A} → A | 2 |
|---|---|---|---|
| 2 | {A → B, AB → AA} | AB → {BB, AA} → BB | 5 |
| 3 | {A → AA, A → BAB} | A → {AA, BAB} → BABABAB | 22 |
| 4 | {A → AA, A → BAAB} | A → {AA, BAAB} → BAABAABAAB | 121 |
| 5 | {A → AA, A → BAABB} | A → {AA, BAABB} → BAABBAABBAABBABB | 515 |
| 6 | {A → AAB, ABAA → A} | ABAA → {AABBAA, A} → AABBAABB | 1664 |
| 7 | {A → AAB, ABBA → A} | ABBA → {AABBBA, A} → AAAABBBAAB | 2401 |
| 8 | {A → AA, AA → BABBB} | AA → {AAA, BABBB} → BABBBABBBABBBBBBBB | 759 |
| 9 | {A → B, BB → A, AAAA → B} | AAAA → {BAAA, B} → B | 30 |
| 10 | {A → AA, AA → BABBBB} | AA → {AAA, BABBBB} → BABBBBABBBBABBBBBBBBBBBBBBBB | 4020 |
| 11 | {A → AA, AAA → BABBB} | AAA → {AAAA, BABBB} → BABBBABBBABBBBBBBB | 2294 |
| 12 | {A → AA, AA → B, BBBB → A} | AA → {AAA, B} → B | 405 |
| 13 | {A → B, BB → A, AAAAAB → A} | AAAAAB → {BAAAAB, A} → A | 94 |
| 14 | {A → AB, BAA → A, BBB → A} | BBBAA → {AAA, BBA} → AA | 2698 |
| 15 | {A → AA, BBB → A, AAAA → B} | AAAA → {AAAAA, B} → B | 430 |
| 16 | {A → AA, AAA → B, BBBB → A} | AAA → {AAAA, B} → B | 906 |

Note that—like the first case of a 2-step resolution that we showed above—quite a few of these longest-to-resolve rules actually terminate, with a member of the branch pair being their final output. Thus, for example, the last case listed is really just a reflection of the fact that with this rule, AAAA takes 16 steps to reach the termination state B:

{AAAA, AAAAA, AAAAAA, AAAAAAA, AAAAAAAA, AAAAAAAAA, AAAAAAAAAA, AAAAAAAAAAA, AAAAAAAAAAAA, AAAAAAAAAAAAA, AAAAAAAAAAAAAA, AAAAAAAAAAAAB, AAAAAAAABB, AAAAABBB, AABBBB, AAA, B}

(With 3 distinct elements similar results are seen; the first time a shorter example is seen is {A→BAC,A→AABA} at resolution-length 6.)

Despite the existence of long-to-resolve cases, most branch pairs in most rules in practice resolve quickly: the fraction that take $\tau$ steps seems to decrease roughly like $2^{-\tau}$. But there is still in a sense an arbitrarily long tail—and the general problem of determining whether a branch pair will resolve is known to be formally undecidable (e.g. [65]).

One interesting feature of causal invariance testing is that (while still in principle undecidable) it is in some ways easier to test for total causal invariance than to test for partial causal invariance for specific initial conditions. The reason is that if a rule is going to be totally causal invariant then there is a certain core set of branch pairs that must resolve, and if all these resolve then the rule is guaranteed to be totally causal invariant. This core set of branch pairs is derived from the possible overlaps between left-hand sides of replacements, and are the successors of the minimal "unifications" of these left-hand sides, formed by minimally overlapping the strings.

Consider the rule:

{AA → A, AB → BAA}

The only possible overlap between the left-hand sides is A. This leads to the minimal unification AAB, which yields the branch pair {ABAA,AB}:

```
        AAB
       /   \
    ABAA    AB
```

From this branch pair we construct a states graph:

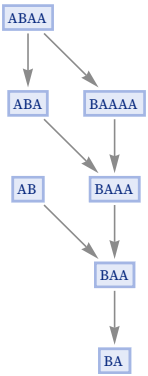

And from this we see that the branch pair resolves in 3 steps. And because this is the only branch pair that can arise through overlaps between the left-hand sides, this resolution now establishes the total causal invariance of this rule.

In the rule

{AB → BA, BAA → A}

there are two possible overlaps between the left-hand sides: A and B. These lead to two different minimal unifications: BAAB and ABAA. And these two unifications yield two branch pairs, {BABA, AB} and {BAAA, AA}. But now we can establish that both of these resolve, thus showing the total causal invariance of the original rule.

In general, one might in principle have to continue for arbitrarily many steps to determine if a given branch pair resolves. But the crucial point here is that because the number of possible overlaps (and therefore unifications) is fine, there are only a finite number of branch pairs one needs to consider in order to determine if a rule is totally causal invariant. There is no need to look at branch pairs that arise from larger strings than the unifications; any additional elements are basically just "padding" that cannot affect the basic interference between replacements that leads to a breakdown of causal invariance.

Looking at branch pairs from all possible unifications is a way to determine total causal invariance—and thus to determine whether a rule will be causal invariant from all possible initial conditions. But even if a rule is not totally causal invariant, it may still be causal invariant for particular initial conditions—effectively because whatever branch pairs might break causal invariance simply never occur in evolution from those initial conditions.

In practice, it is fairly common to have rules that are causal invariant for some initial conditions, but not others. In effect, there is some "conservation law" that keeps the rule away from branch pairs that would break causal invariance—and that keeps the rule operating with some subset of its possible states that happen to yield causal invariance.

## 5.6 The Frequency of Causal Invariance

The plots below show the fractions of rules found to be totally causal invariant [66], as a function of the total number of elements they involve, for the cases of $k = 2$ (A, B) and $k = 3$ (A, B, C):

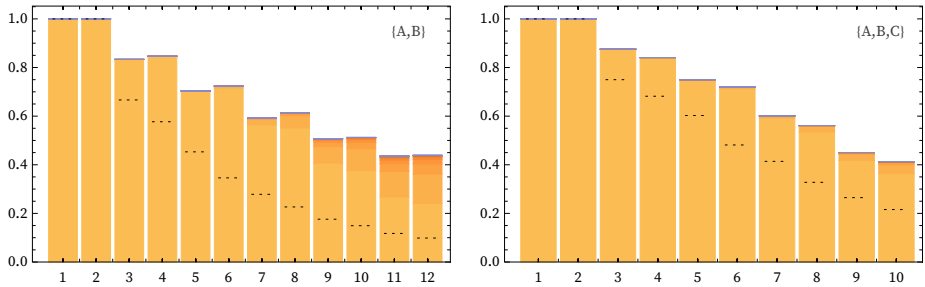

The darker colors indicate larger numbers of steps to resolve branch pairs. (There is some uncertainty in these plots—conceivably as much as 9% for 10 total elements with $k = 3$—since in some cases the states graph became too big to compute before it could be determined whether all branch pairs resolved.)

The dotted lines indicate rules are in a sense inevitably causal invariant because their left-hand sides involve strings that do not overlap themselves or each other, thereby guaranteeing total causal invariance. Rules such as AA→AAA are causal invariant despite having overlapping left-hand sides because their right-hand sides in a sense give the same result whatever overlap occurs.

Ignoring the structure of rules, one can just ask what fraction of strings are non-overlapping [67]. Out of the total of $k^n$ possible strings with length $n$ containing $k$ distinct elements the number that do not overlap themselves is given by [1:p1033]:

a[0] = 1; a[n_] := k a[n – 1] – If[EvenQ[n], a[n/2], 0]

This yields the following fractions (for the limit see e.g. [68][@Finch: 5.17]):

| $n$ | $k = 2$ | $k = 3$ | $k = 4$ | $k = 5$ |
|---|---|---|---|---|
| 2 | 0.500 | 0.667 | 0.750 | 0.800 |
| 3 | 0.500 | 0.667 | 0.750 | 0.800 |
| 4 | 0.375 | 0.593 | 0.703 | 0.768 |
| 5 | 0.375 | 0.593 | 0.703 | 0.768 |
| 6 | 0.313 | 0.568 | 0.691 | 0.762 |
| 7 | 0.313 | 0.568 | 0.691 | 0.762 |
| 8 | 0.289 | 0.561 | 0.689 | 0.760 |
| 9 | 0.289 | 0.561 | 0.689 | 0.760 |
| 10 | 0.277 | 0.558 | 0.688 | 0.760 |
| ∞ | 0.268 | 0.557 | 0.688 | 0.760 |

One can also look at how many possible sets of *s* strings of length up to *n* allow no overlaps with themselves or each other [1:p1033]. The numbers and fractions for $k = 2$ are as follows:

| $n$ | $s = 1$ | $s = 2$ | $s = 3$ | $s = 4$ | $s = 5$ |
|---|---|---|---|---|---|
| 2 | 2 (0.5) | 2 (0.2) | 2 (0.1) | 2 (0.057) | 2 (0.036) |
| 3 | 4 (0.5) | 4 (0.11) | 4 (0.033) | 4 (0.012) | 4 (0.0051) |
| 4 | 6 (0.38) | 6 (0.044) | 6 (0.0074) | 6 (0.0015) | 6 (0.00039) |
| 5 | 12 (0.38) | 20 (0.038) | 28 (0.0047) | 36 (0.00069) | 44 (0.00012) |
| 6 | 20 (0.31) | 54 (0.026) | 104 (0.0023) | 170 (0.00022) | 252 (0.000024) |
| 7 | 40 (0.31) | 220 (0.027) | 728 (0.002) | 1788 (0.00015) | 3672 (0.000012) |
| 8 | 74 (0.29) | 798 (0.024) | 4806 (0.0017) | 19 708 (0.00011) | 62 668 ($6.6 \times 10^{-6}$) |

To get a sense of the distribution of non-overlapping strings, one can make an array that shows which pairs of strings (ordered lexicographically) do not allow overlaps. Here are the results for $k = 2$ and $k = 3$ for strings respectively up to length 6 and length 4, showing clear structure in the space of possible strings [51]:

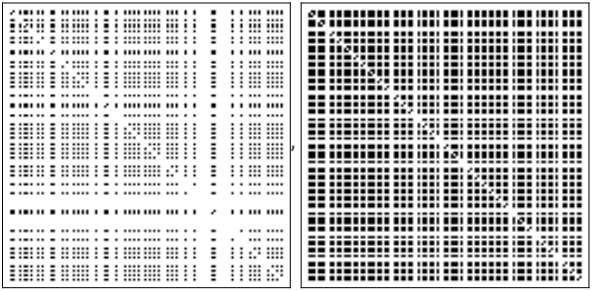

## 5.7 Events and Their Causal Relationships

So far the nodes in our graphs have always been states generated by substitution systems. But we can also introduce nodes to represent the "updating events" associated with replacements performed on strings. Here is the result for evolution according to the rule {AB→BAB, BA→A} starting from ABA—with each event node indicating the string replacement to which it corresponds:

We can also show this as a states graph, where we have merged instances of the same state that occur at different steps:

States are connected through events. But how are events connected? Given two events the key question to ask is whether they are causally related. Does one event depend on the other—in the sense that all or part of its input comes from the output of the other event?

Looking at the graph above, for example, the event $\substack{AB\\BAB}A$ depends on $\substack{BA\\A}BA$ because $\substack{AB\\BAB}A$ uses as input the A that arises as output from $\substack{BA\\A}BA$. On the other hand, $A\substack{BA\\A}$ does not depend on $\substack{BA\\A}BA$ because the BA it consumes was not generated by $\substack{BA\\A}BA$.

We can add this dependency information to the evolution graph by putting dotted lines between events that are causally related:

We can also do this in the states graph, in which we have merged instances of states from different steps:

We can redraw this graph without the layering that puts the initial state at the top. We have added an "initialization event" ABA " to indicate the creation of the initial condition:

[Figure: Evolution graph with state nodes (blue) and event nodes (yellow), starting from ABA and branching through various states including BABA, BBABA, BBAA, BAA, AA, etc.]

If we want to focus on causal relationships, we can now drop the state nodes altogether, and get a multiway causal graph that represents possible causal relationships between events:

[Figure: Multiway causal graph showing only event nodes (yellow) connected by red directed edges.]

This causal graph is dual to our original evolution graph in the sense that edges in the original evolution graph correspond to events—which now become nodes in the causal graph. Similarly, each edge in the causal graph is associated with some state which appears as a node in the evolution graph.

We can get a sense of "possible causal histories" of our system by arranging the multiway causal graph in layers:

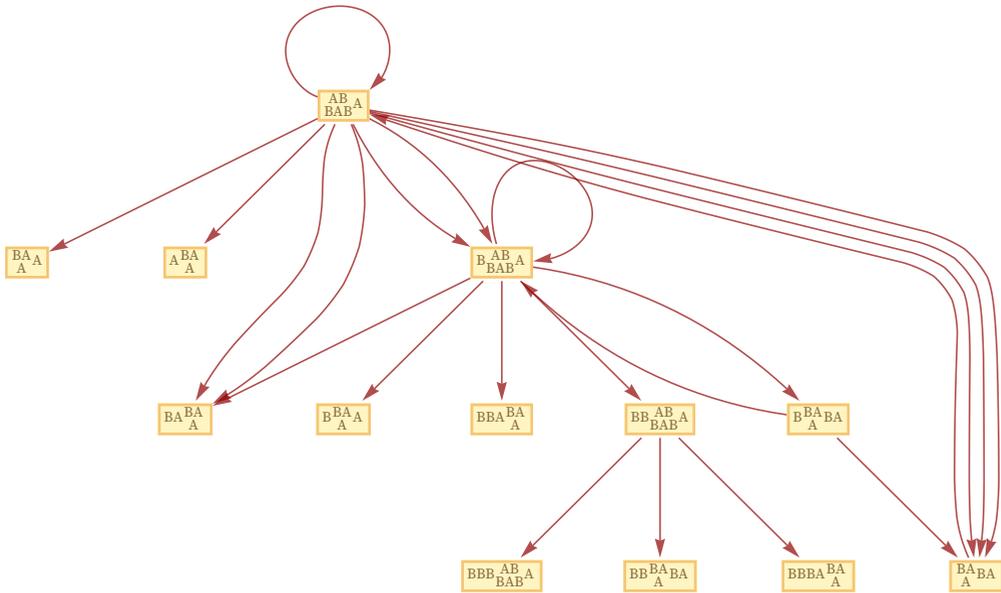

Just like every possible path through the multiway evolution graph gives a possible sequence of states that can occur in the evolution of a system, so also every possible path through the multiway causal graph gives a possible sequence of events that can occur in the evolution of the system.

After 10 steps, the graph in our example has become quite complicated:

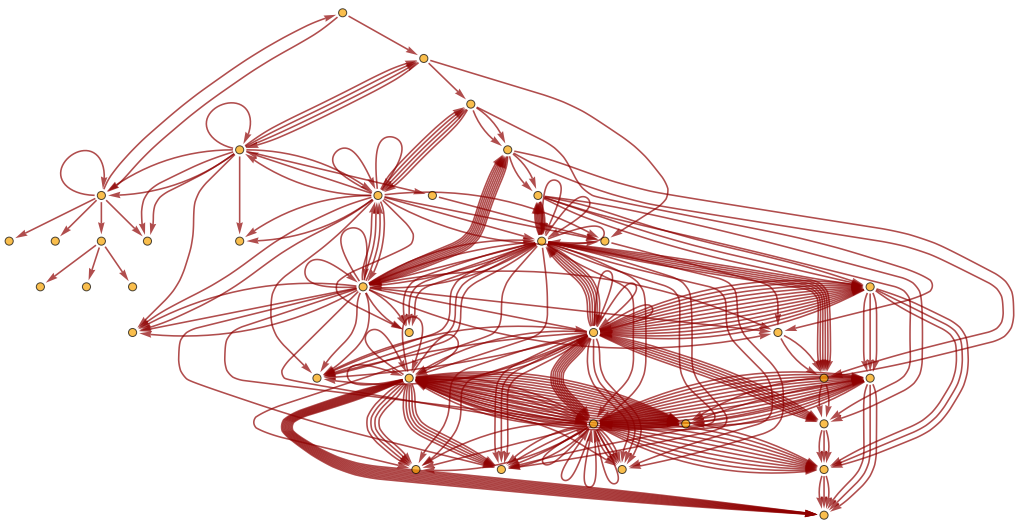

## 5.8 Causal Graphs for Particular Updating Sequences

The multiway causal graph that we have just constructed shows the causal relationships for all possible paths of evolution in a multiway system. But what if we just pick a single path of evolution? Instead of looking at the results for every possible updating order, let us pick a particular updating order.

For example, for the rule above we could pick "sequential updating", in which at each step, first for AB→BAB and then for BA→A, we scan the string from left to right, doing the first replacement we can [1:3.6]. This leads to a specific single path of evolution for the system (now drawn across the page):

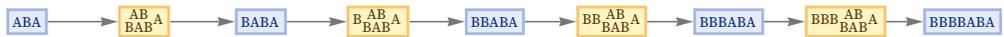

We can show the causal relationships between the events in this evolution:

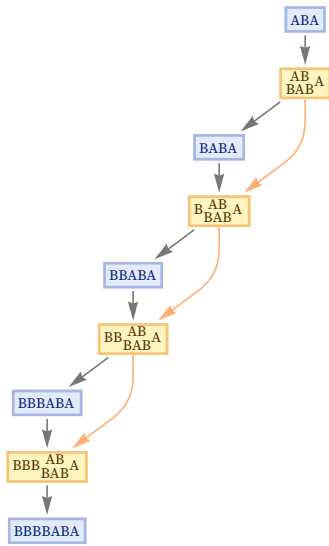

And we can generate a causal graph—which is very simple:

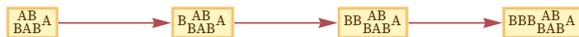

But now let us consider a different updating scheme, in which now as we scan the string, we try at each position both AB→BAB and BA→A, then do the first replacement we can. This procedure again leads to a specific single path of evolution, but it is a different one from before:

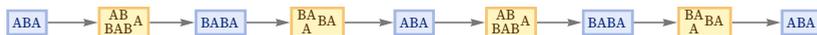

This particular path just involves an alternation between the states ABA and BABA, so the states graph is a cycle:

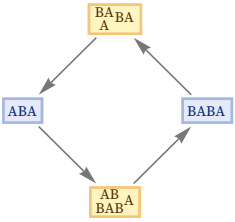

Including causal relationships here we get:

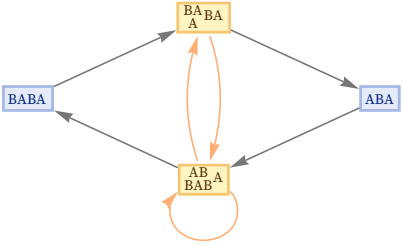

The resulting causal graph is then:

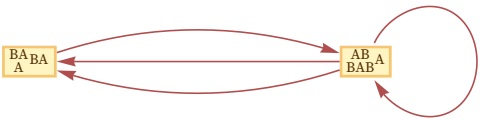

This causal graph is quite different from the one we got for the previous updating scheme. But both individual causal graphs necessarily occur in the whole multiway causal graph:

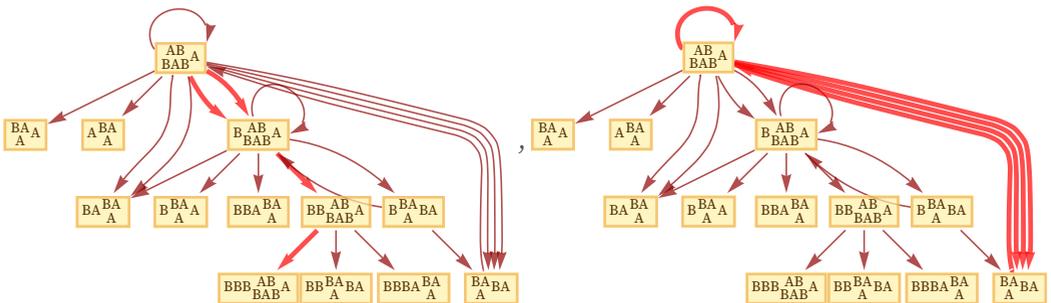

Given the multiway causal graph, we can explicitly find all possible individual causal graphs (not showing individual loop configurations separately):

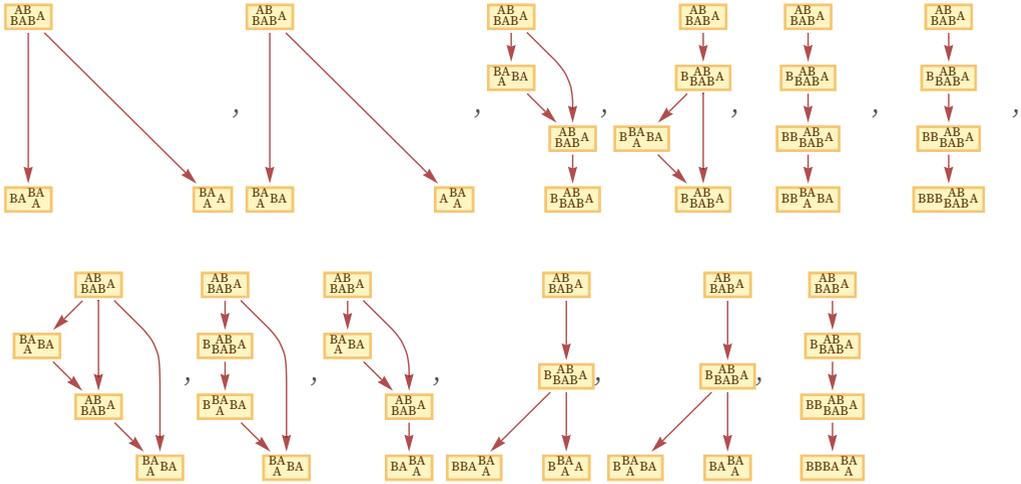

And what this shows is that at least with the particular rule we are looking at, there are many different classes of evolution paths, that can lead to distinctly different individual causal graphs—and therefore causal histories.

## 5.9 The Significance of Causal Invariance

One might think that all rules would work like the one we just studied, and would give different causal graphs depending on what specific path of evolution one chose, or what updating scheme one used. But a crucial fact that will be central to the potential application of our models to physics is this is not the case.

Instead, whenever a rule is causal invariant, it does not produce many different causal graphs. Instead, whatever specific path of evolution one choses, the rule always yields an exactly equivalent causal graph.

The rule we just studied is not causal invariant. But consider instead the simple causal invariant rule {BA→AB} evolving from the state BBBAA:

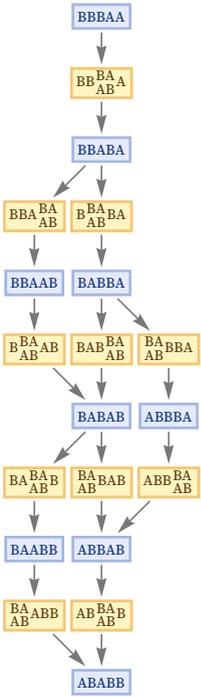

Adding in causal relationships this becomes:

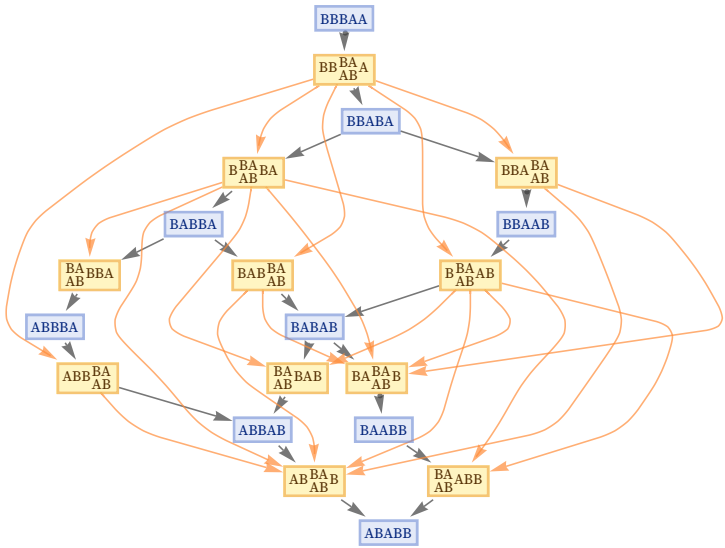

The corresponding multiway causal graph is:

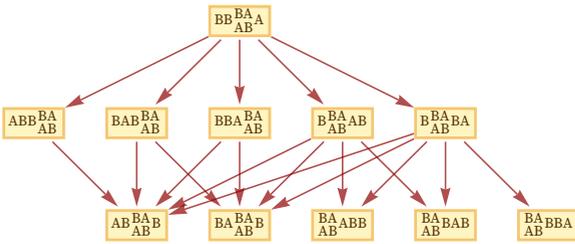

But now consider the individual causal graphs, corresponding to different possible paths of evolution. From the multiway causal graph we can extract all of these, and the result is:

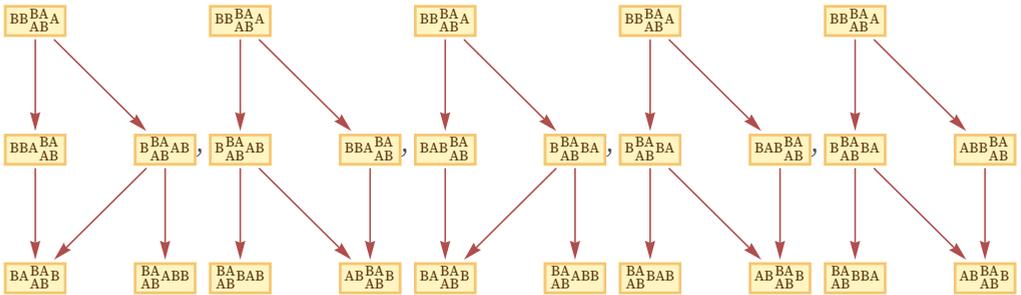

There are five different cases. But the remarkable fact is that they all correspond to isomorphic graphs. In other words. even though the specific sequence of states that the system visits is different in each case, the network of causal relationships is always exactly the same, regardless of what path is followed.

And this is a general feature of any causal invariant system. The underlying evolution rules for the system may allow many different paths of evolution—and many different sequences of states. But when the rules for the system are causal invariant, it means that the network of relationships between updating events is always the same. Depending on the particular order in which updates are done, one can see different paths of evolution; but if one looks only at event updates and their relationships, there is just one thing the system does.

Here is a slightly larger example for the rule {BA→AB} starting from BBBBAAAA. In each case a different random updating order is used, leading to a different sequence of states. But the final causal graphs representing the causal relationships between events are exactly the same:

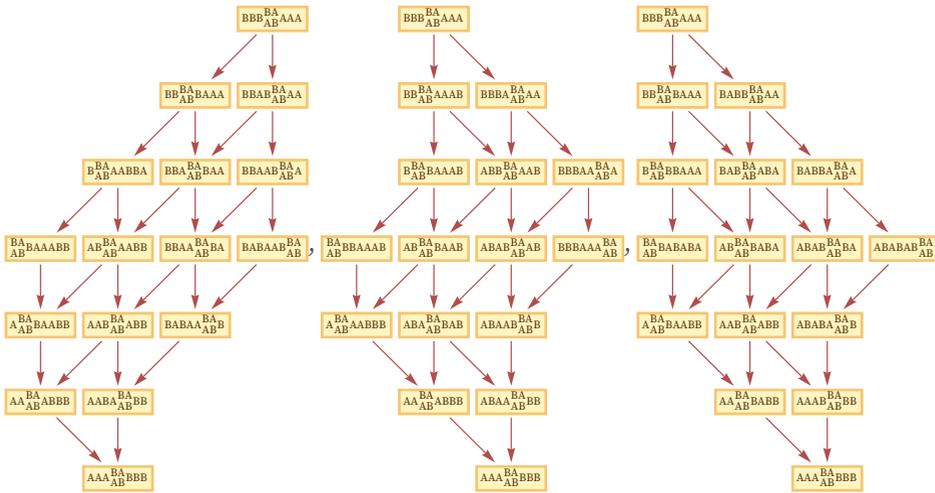

## 5.10 Causal Foliations and Causal Cones

We have discussed causal invariance in terms of path independence in multiway systems. But we can also explore it in terms of specific evolution histories for the underlying substitution system. Consider the rule {BA→AB}. Here is a representation of one way it can act on a particular initial string:

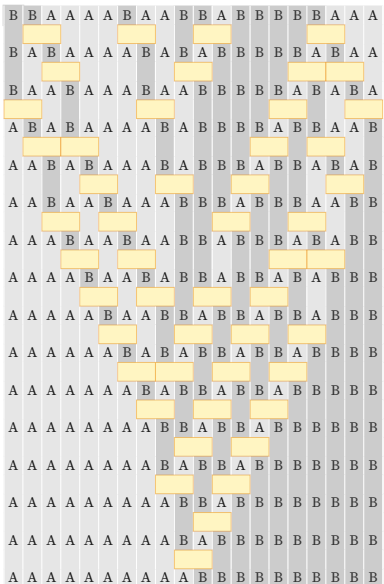

In a multiway system, each path represents a particular sequence of updates that occur one after another. But here in showing the action of {BA→AB} we are choosing to draw several updates on the same row. If we want, we can think of these updates as being done in sequence, but since they are all independent, it is consistent to show them as we do.

If we annotate the picture by showing causal connections between updates, we see the causal graph for the evolution—and we see that the updates we have drawn on the same row are indeed independent: they are not connected in the causal graph:

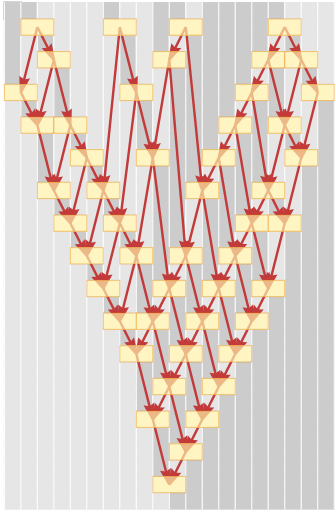

The picture above in effect uses a particular updating order. The pictures below show three possible random choices of updating orders. In each case, the final result of the evolution is the same. The intermediate steps, however, are different. But because our rule is causal invariant, the causal graph of causal relationships always has exactly the same form:

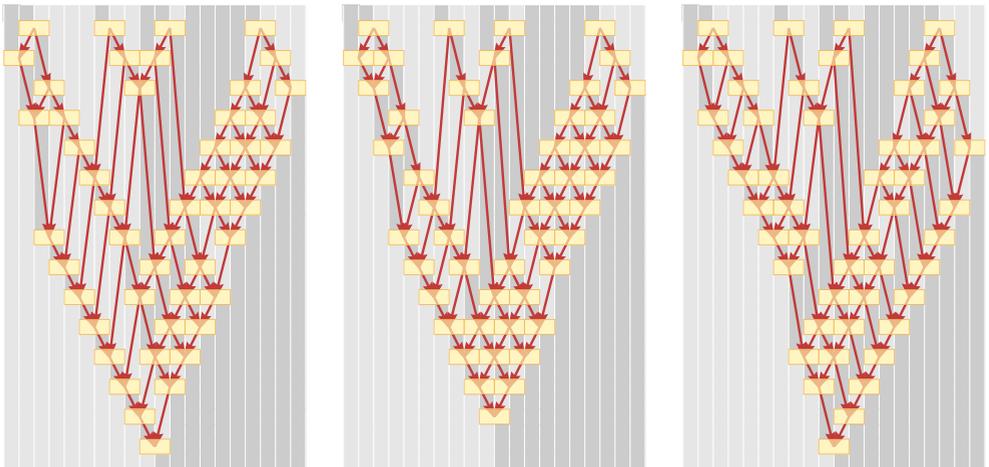

In picking updating orders there is always one constraint: no update can happen until the input for it is ready. In other words, if the input for update *V* comes from the output of update *U*, then *U* must already have happened before *V* can be done. But so long as this constraint is satisfied, we can pick whatever order of updates we want (cf. [1:9.10]).

It is sometimes convenient, however, to think in terms of "complete steps of evolution" in which all updates that could yet be done have been done. And for the particular rule we are currently discussing, we can readily do this, separating each "complete step" in the pictures below with a red line:

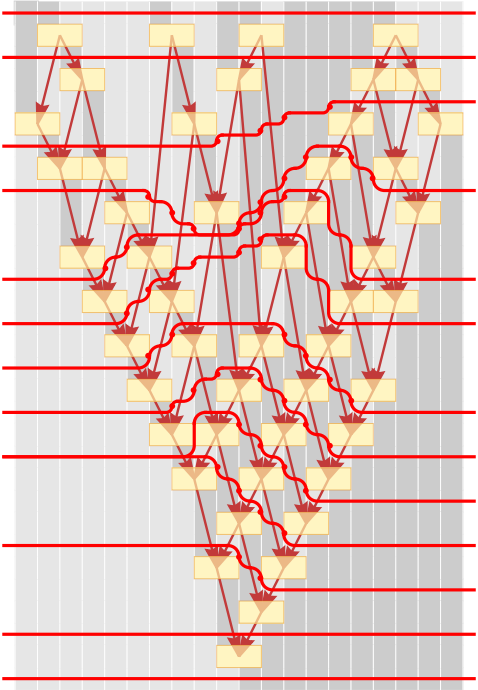

Where the dotted lines go depends on the update order we choose. But because of causal invariance it is always possible to draw them to delineate steps in which a collection of independent updates occur.

Each choice of how to assign updates to steps in effect defines a foliation of the evolution. We will call foliations in which the updates at each step are causally independent "causal foliations". Such causal foliations are in effect orthogonal to the connections defined by the causal graph. (In physics, the analogy is that the causal foliations are like foliations of spacetime defined by a sequence of spacelike hypersurfaces, with connections in the causal graph being timelike.)

The fact that our underlying substitutions (in this case just BA→AB) involve neighboring elements implies a certain locality to the process of evolution. The consequence of this is that we can meaningfully discuss the "spatial" spreading of causal effects. For example, consider tracing the causal connections from one event in our system:

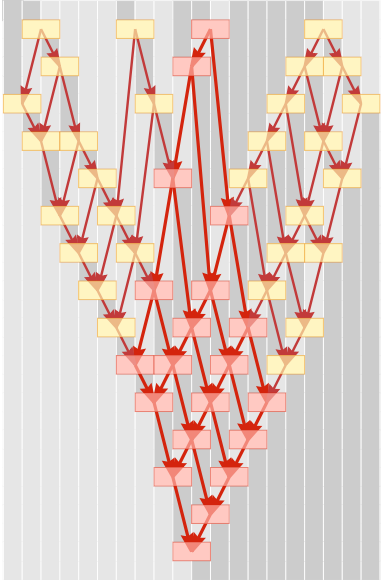

In effect there is a cone (here just two-dimensional) of elements in the system that can be affected. We will call this the causal cone for the evolution. (In physics, the analogy is a light cone.) If we pick a different updating order, the causal cone is distorted. But viewed in terms of the causal foliations, it is exactly the same:

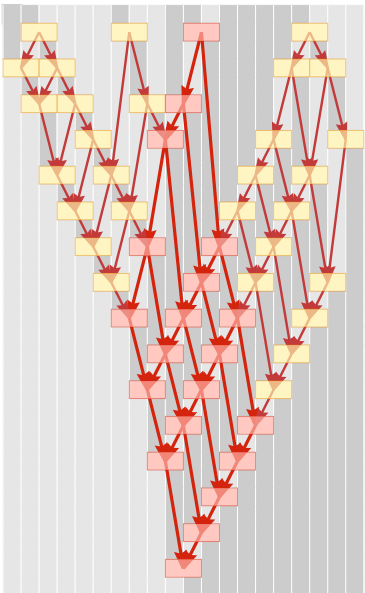

This is the result purely in terms of the causal graph:

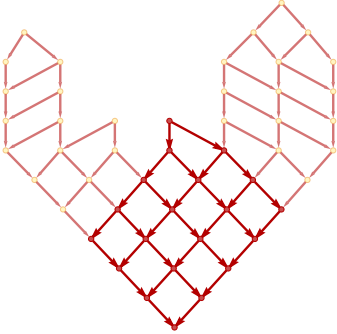

Now let us turn things around. Imagine we have a causal graph. Then we can ask how it relates to an actual sequence of states generated by a particular path of evolution. The pictures below show how we can arrange a causal graph so that its nodes—corresponding to events—appear at positions down the page that correspond to a particular causal foliation and thus a particular path of evolution:

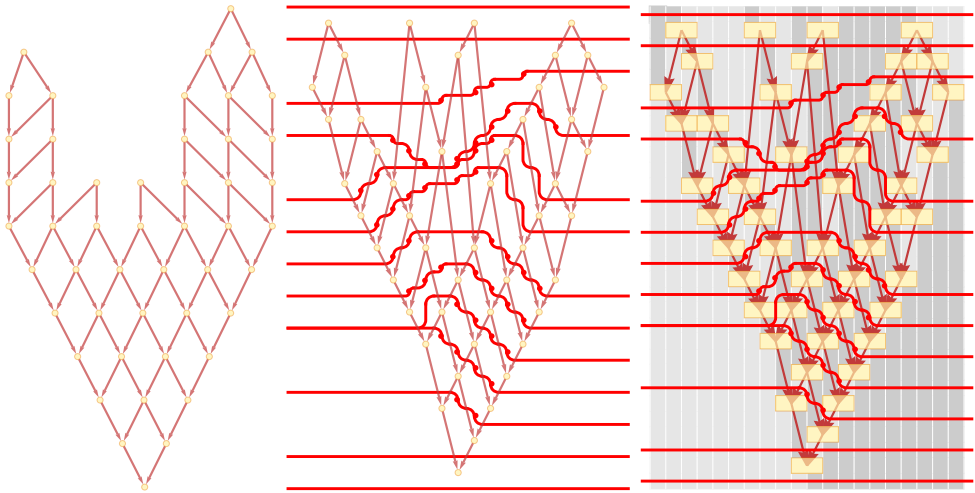

It is worth noticing that at least for the rule we are using here the intrinsic structure of the causal graph defines a convenient foliation in which successive events are simply arranged in layers:

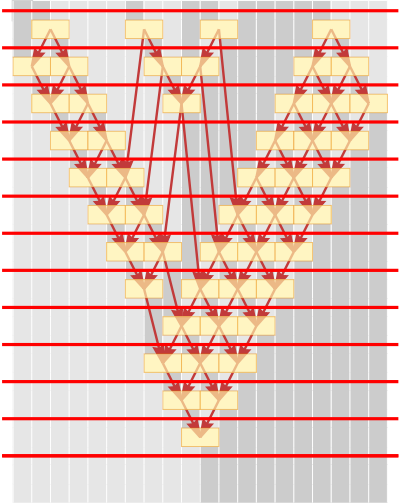

The rule BA→AB that we are using has many simplifying features. But the concepts of causal foliations and causal cones are general, and will be important in much of what follows.

As it happens, we have already implicitly seen both ideas. The "standard updating order" for our main models defines a foliation (similar, in fact, to the last one we showed here, in which in some sense "as much gets done as possible" at each step)—though the foliation is only a causal one if the rule used is causal invariant.

In addition, in 4.14 we discussed how the effect of a small change in the state in one of our models can spread on subsequent steps, and this is just like the causal cone we are discussing here.

## 5.11 Causal Graphs for Infinite Evolutions

One of the simplifying features of the rule BA→AB discussed in the previous subsection is that for any finite initial condition, it always evolves to a definite final state after a finite number of steps—so it is possible to construct a complete multiway causal graph for it, and for example to verify that all the causal graphs for specific paths of evolution are identical.

But consider the rule {A→BB,B→A}:

[Graph showing evolution tree starting from A, branching through BB, AB, BA, AA, BBB, ABB, BBA, BAB, ABA, AAB, BBBB, BAA]

This rule is causal invariant, but never evolves to a definite state, and instead keeps growing forever. Including events in the evolution we get:

[Graph showing the evolution with events included, from A through multiple branching states to ABA, BBBB, AAB, BAA]

The corresponding multiway causal graph is:

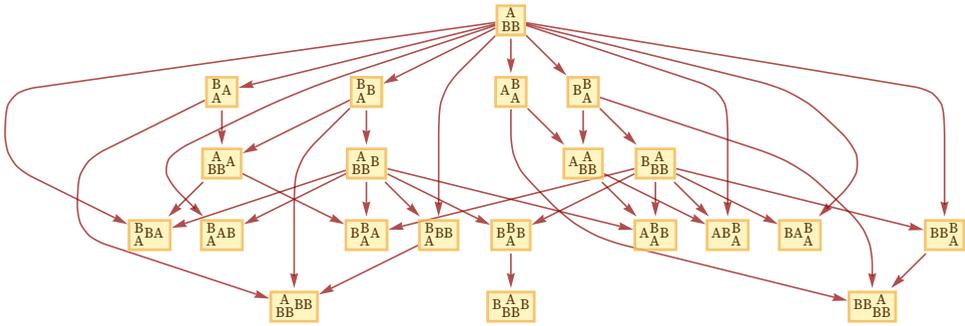

And now if we extract possible individual causal graphs from this, we get:

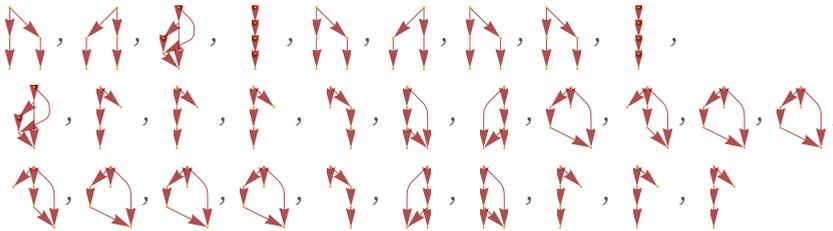

These look somewhat similar, but they are not directly equivalent. And the reason for this has to do with how we are "counting steps" in the evolution of our system. If we evolve for longer, the effect becomes progressively less important. Here are causal graphs generated by a few different randomly chosen specific sequences of updates (each corresponding to a specific path through the multiway system):

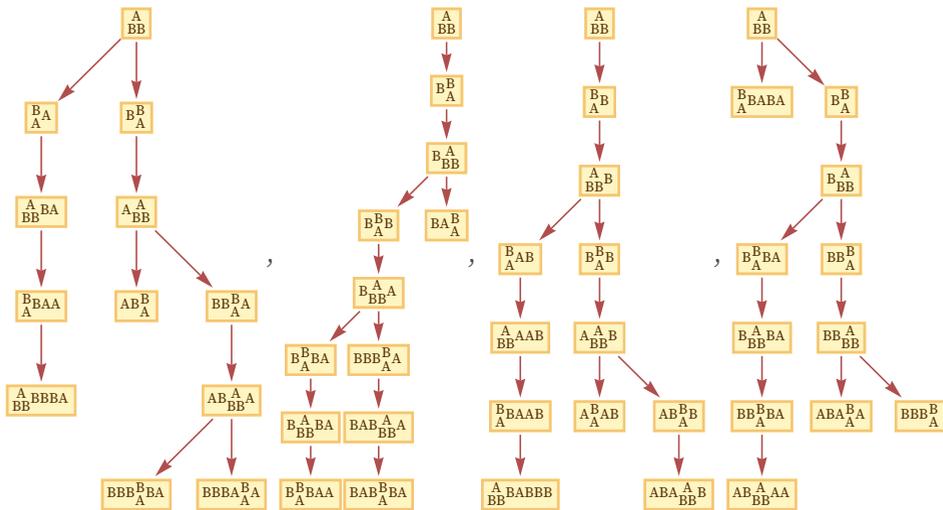

Here are the corresponding results after a few more updates:

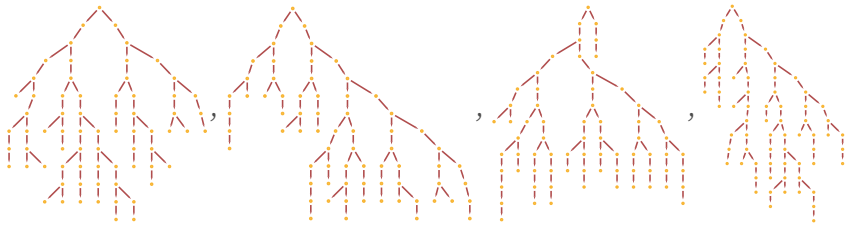

This is a different rendering:

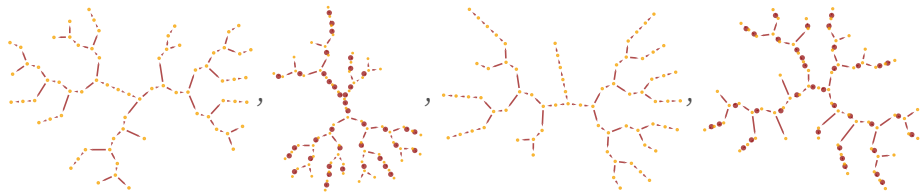

And this is what happens after many more updates (with a somewhat more systematic ordering):

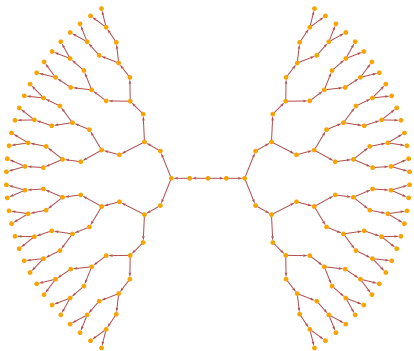

If we continued for an infinite number of updates, all these would give the same result—and the same infinite causal graph, just as we expect from causal invariance. But in the particular cases we are showing, they are being cut off in different ways. And this is directly related to the causal foliations we discussed in the previous subsection.

Here are examples of evolution with specific choices of updating orders:

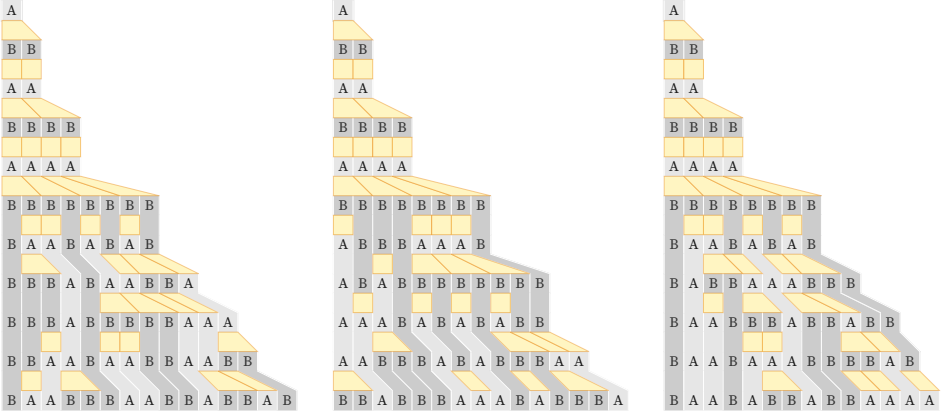

Adding causal graphs we get:

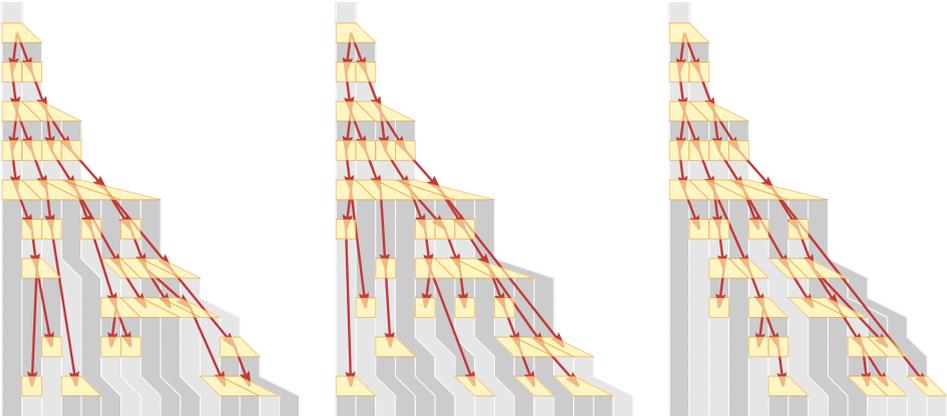

Here is what happens if we continue these for longer:

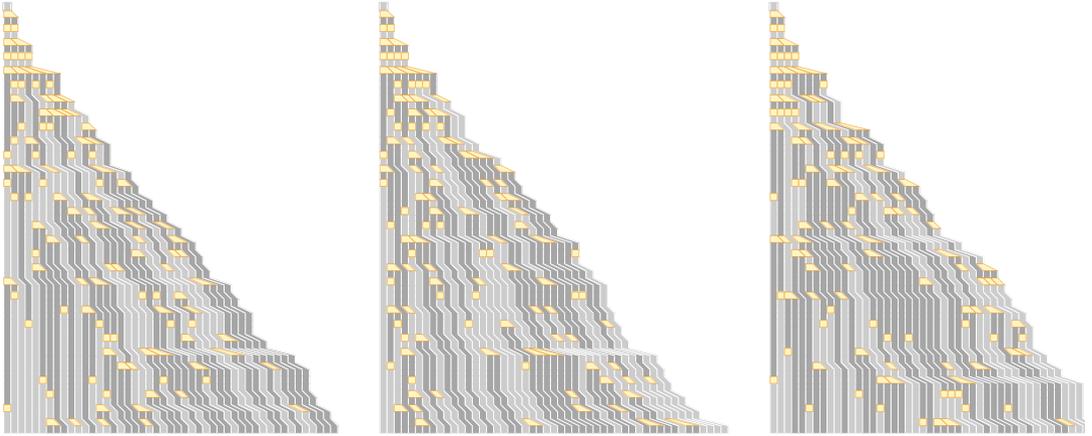

And here are the causal graphs that correspond to these evolutions:

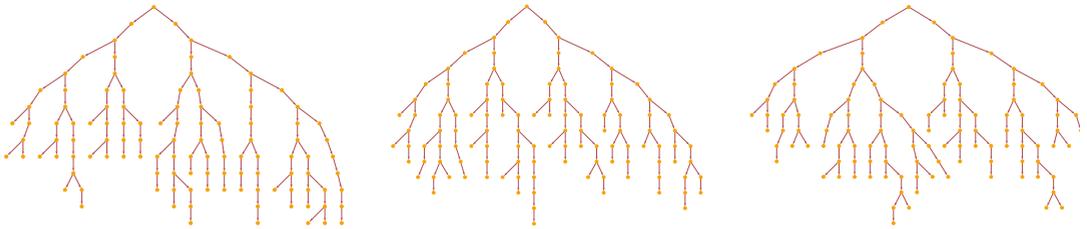

## 5.12 Typical Causal Graphs

Any causal invariant system always ultimately has a unique causal graph. The graph can be found by analyzing any possible evolution for the system, with any updating scheme—though for visualization purposes, it is usually useful to use an updating scheme where as much happens as possible at each step.

The trivial causal invariant rule A→A starting from A has causal graph:

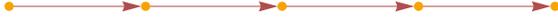

Starting from a string of 10 As it has causal graph:

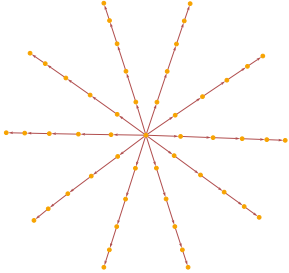

A→AA has a causal graph starting from A that is a binary tree

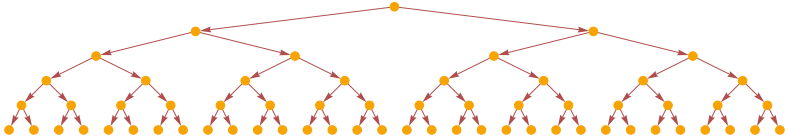

which can also be rendered:

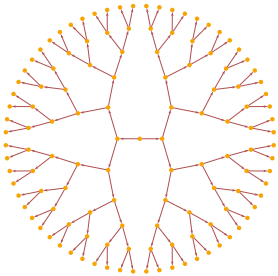

The rule

{A → A, A → AA}

starting from A gives a "two-step binary tree" with $2^{\frac{t}{2}}$ nodes at level $t$:

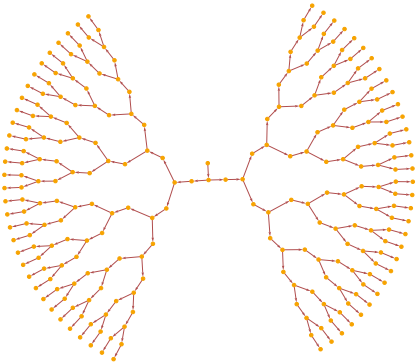

One does not have to go beyond rules involving just a single element (all of which are causal invariant) to find a range of causal graph structures. For example, here are all the forms obtained by rules allowing up to 6 instance of a single element A, with initial condition AA:

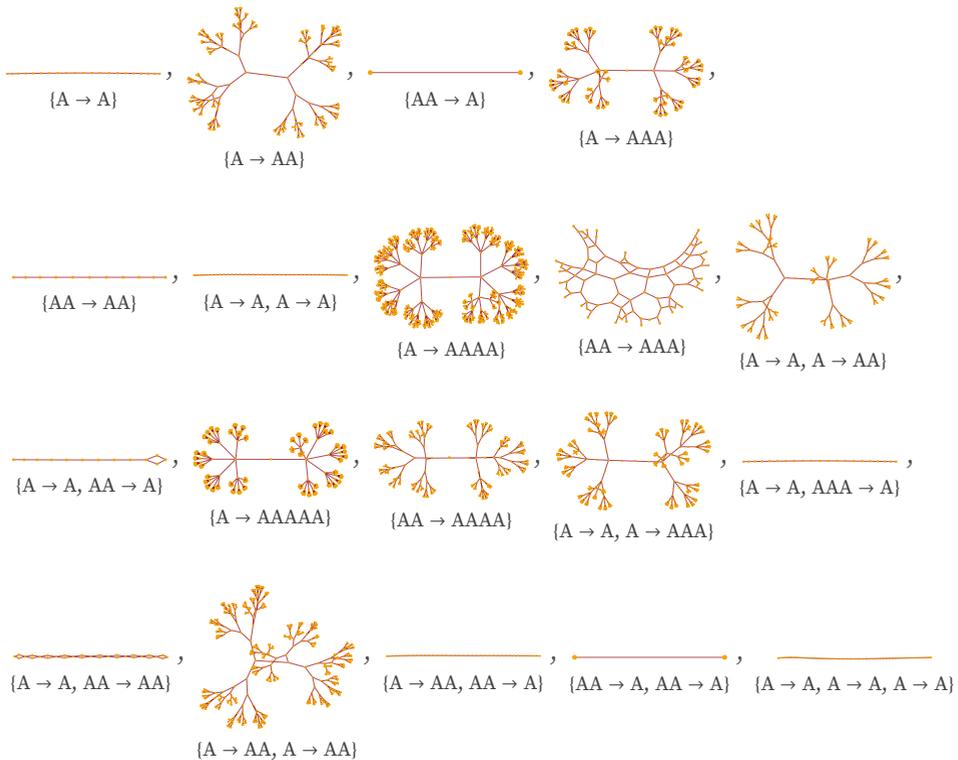

A notable case is the rule:

{AA → AAA}

Shown in layered form, the first few steps give:

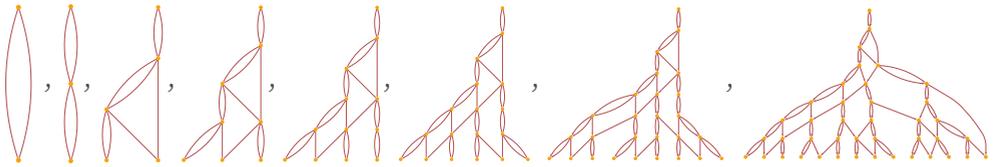

After a few more steps, this can be rendered as:

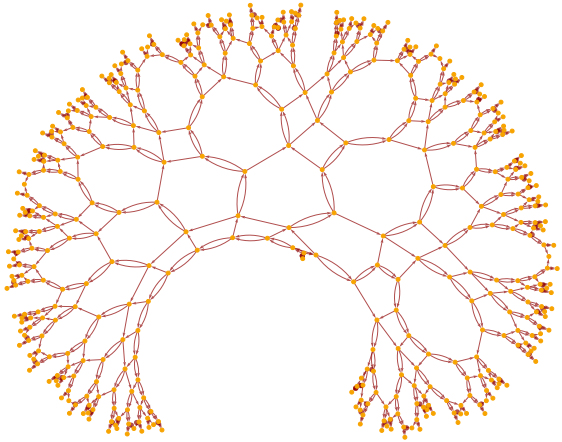

Running the underlying substitution system AA→AAA updating as much as possible at each step (the **StringReplace** scheme), one gets strings with successive lengths

{2, 3, 4, 6, 9, 13, 19, 28, 42, 63, 94, 141, 211, 316, 474, 711,
   1066, 1599, 2398, 3597, 5395, 8092, 12 138, 18 207, 27 310, 40 965}

which follow the recurrence:

a[1] = 2;  a[n_] := If[EvenQ[n], 3 n / 2, (3 n − 1) / 2]

Other rules of the form $A^p \to A^q$ for non-commensurate $p$ and $q$ give similar results, analogous to tessellations in hyperbolic space:

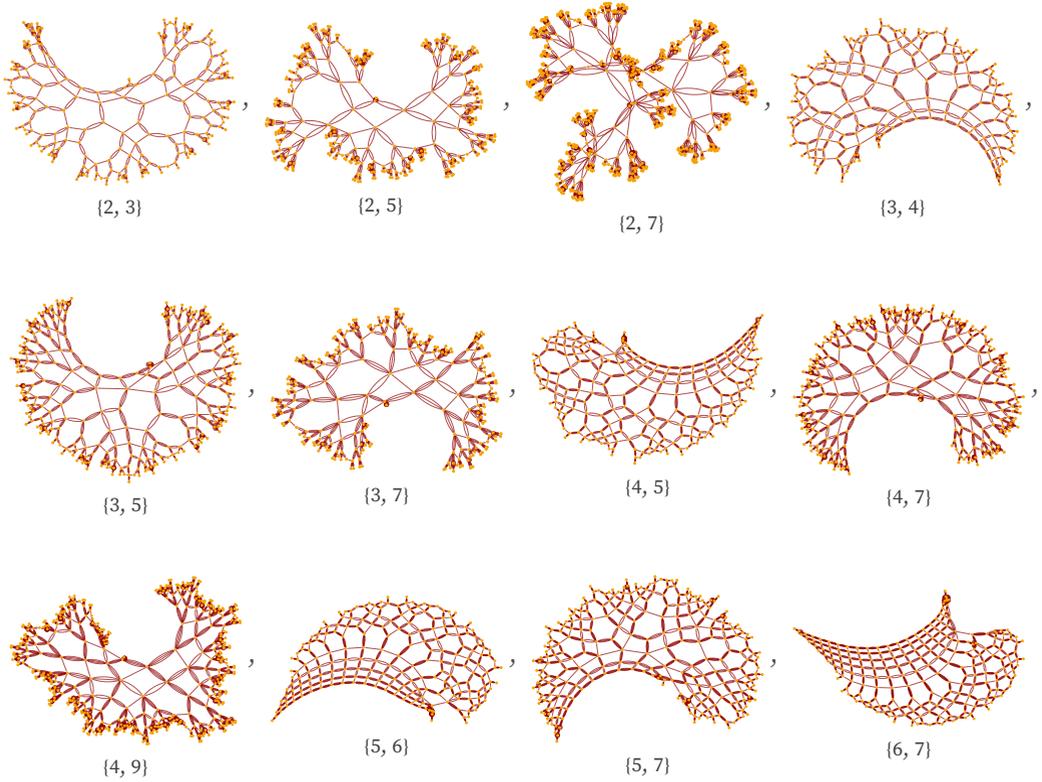

Rules that involve multiple replacements can give similar behavior even starting from a single A:

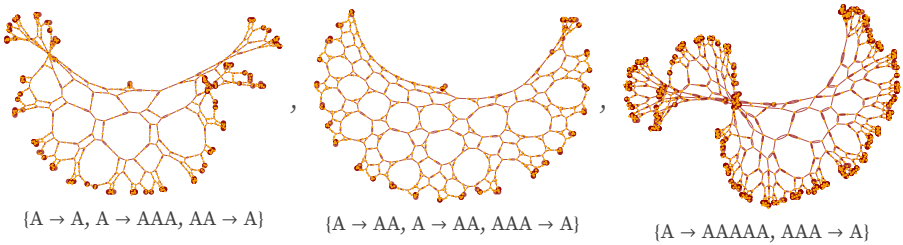

Rules just containing only As cannot progressively grow to produce ordinary tilings. One can get these with the "sorting rule"

{BA → AB}

which when started with 20 BAs yields:

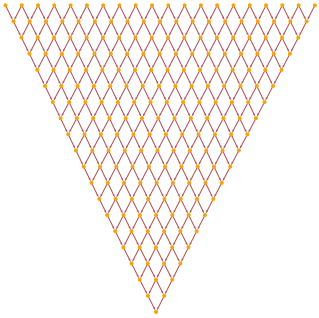

There are also rules which "grow" grid-like tilings. For example, the rule

{A → AB, BB →  BB}

starting from a single A produces

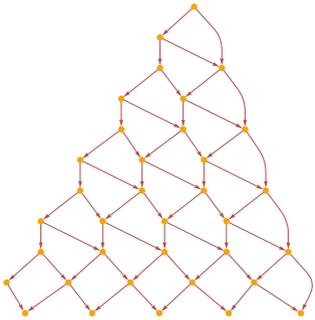

which is equivalent to a square grid:

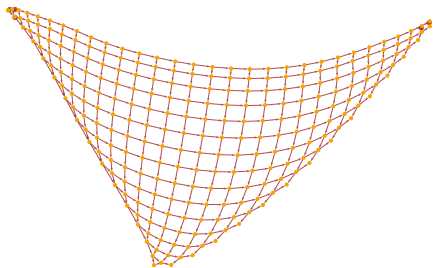

There is also a simple rule that generates essentially a hexagonal grid:

{A → B, B → AB, BA → A}

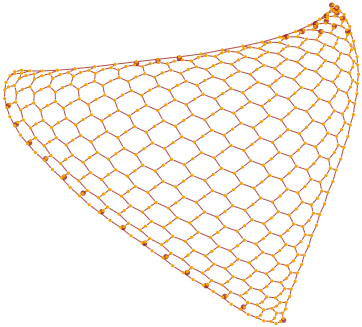

Other forms of causal graphs produced by simple causal invariant substitution systems include (starting from A, AB or ABA):

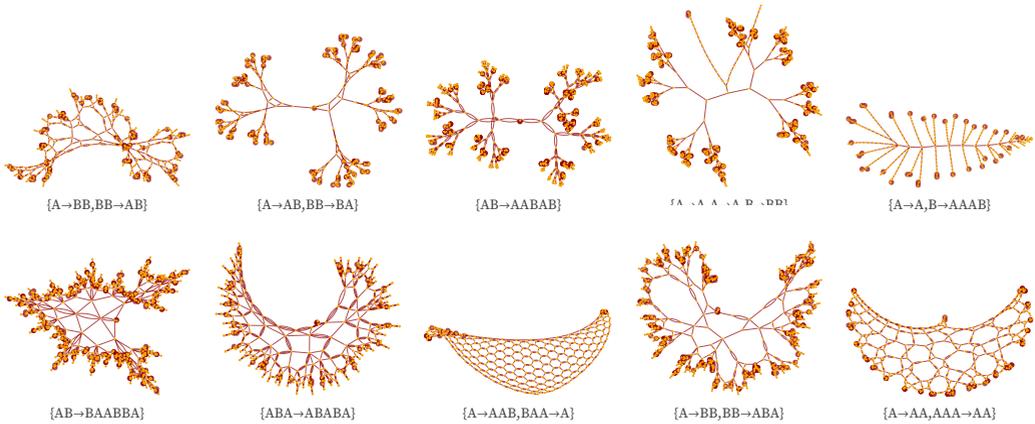

When rules terminate they yield finite causal graphs. But these can often be quite complicated. For example, the rule

{A → BBB, BBBB → A}

started from strings consisting of from 1 to 6 As yields the following finite causal graphs:

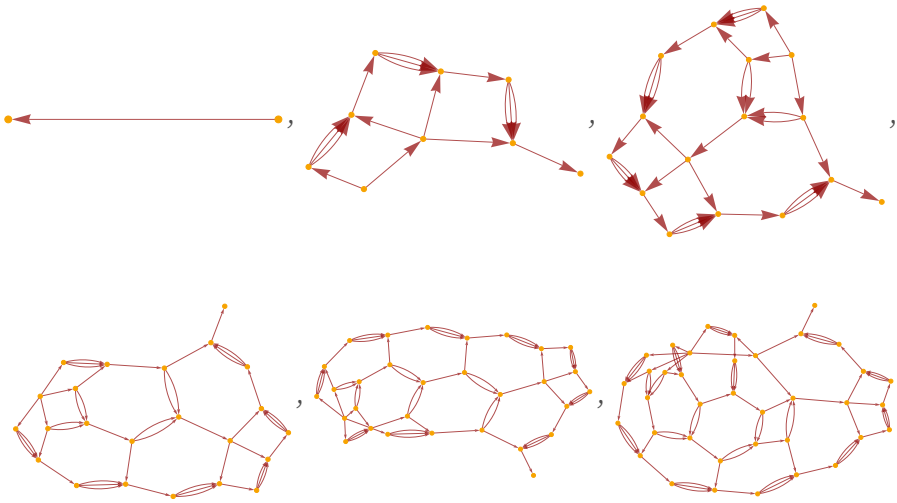

With a string of 50 As, the rule gives the finite causal graph:

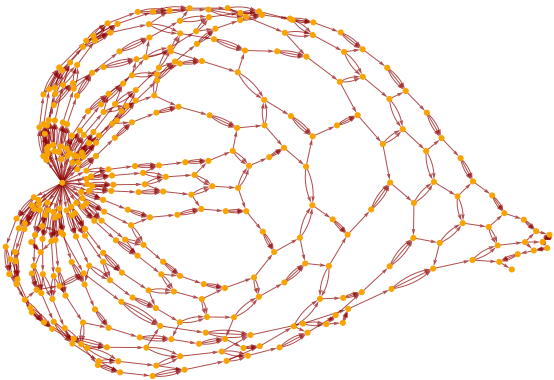

Compared to the hypergraphs we studied in previous sections, or even the multiway graphs from earlier in this section, the causal graphs here may seem to have rather simple structures. But there is a good reason for this. While there can be many updating events in the evolution of a string substitution system, all of them are in a sense arranged on the same one-dimensional structure that is the underlying string. And since the updating rules we consider involve strings of limited length, there is inevitably a linear ordering to the events along the string. This greatly simplifies the possible forms of causal graphs that can occur, for example requiring them always to remain planar. In the next section, we will see that for our hypergraph-based models—which have no simplifying underlying structure—causal graphs can be considerably more complex.

## 5.13 Limits of Causal Graphs

Much as we did for hypergraphs in section 4, we can consider the limiting structures of causal graphs after a large number of steps. And much as for hypergraphs, we can potentially describe these limiting structures in terms of emergent geometry. But one difference from what we did for hypergraphs is that for causal graphs, it is essential to take account of the directedness of their edges. It is still perfectly possible to have a limit that is like a manifold, but now to measure its properties we must generate the analog of cones, rather than balls.

Consider for example the simple directed grid graph:

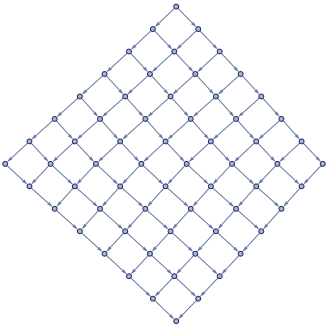

Now consider starting from a particular node, and constructing progressively larger "cones":

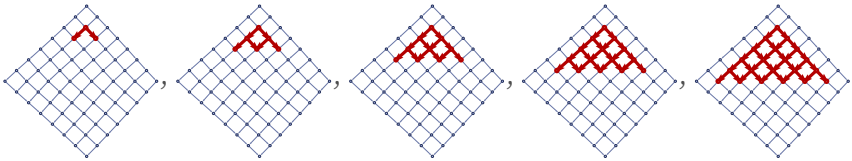

We can call the number of nodes in this cone after $t$ steps $C_t$. In this case the result (for $t$ below the diameter of the graph) is:

$$C_t = \frac{1}{2} t(1 + t)$$

And in the limit of large graphs, we will have:

$$C_t \sim t^2$$

We can also set up a 3D directed grid graph

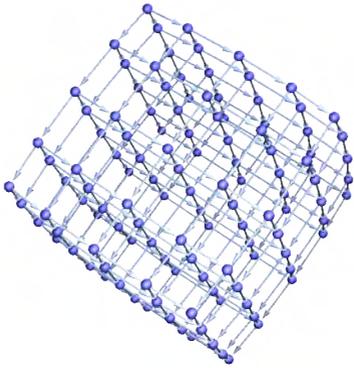

and generate a similar cone

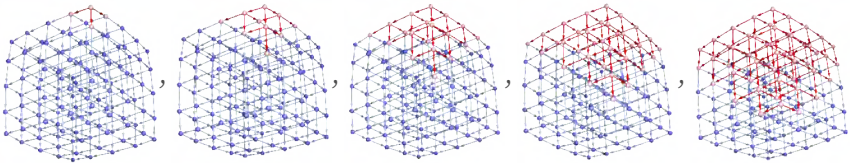

for which now $C_t \sim t^3$.

In general, we can think about the limits of these grid graphs as generating a $d$-dimensional "directed space". There is also nothing to prevent having cyclic versions, such as and in general a family of graphs that are going to behave like $d$-dimensional directed space in the limit will have $C_t \sim t^d$.

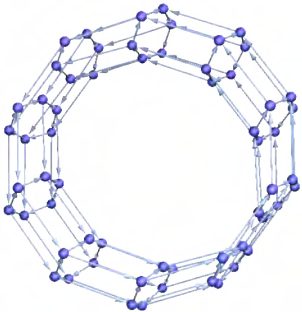

In direct analogy to what we did with hypergraphs in section 4, we can compute $C_t$ for causal graphs, and then estimate effective dimension by looking at its growth rate. (There are some additional subtleties, though, because whereas at any given step in the evolution of the system, $V_r$ can be computed for any $r$ for any point in a hypergraph, $C_t$ can be computed only until $t$ "reaches the edge of the causal graph" from that starting point—and later we will see that the cutoff can also depend on the foliation one uses.)

Consider the substitution system:

{A → AB, BB → BB}

This generates the causal graph:

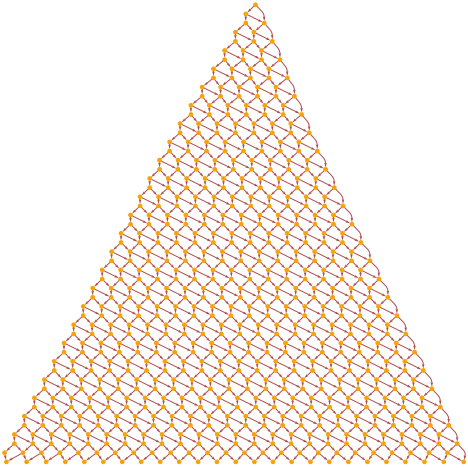

The log differences of $C_t$ averaged over all points for causal graphs obtained from 10 through 100 steps of evolution have the form (the larger error bars for larger $t$ in each case are the result of fewer starting points being able to contribute):

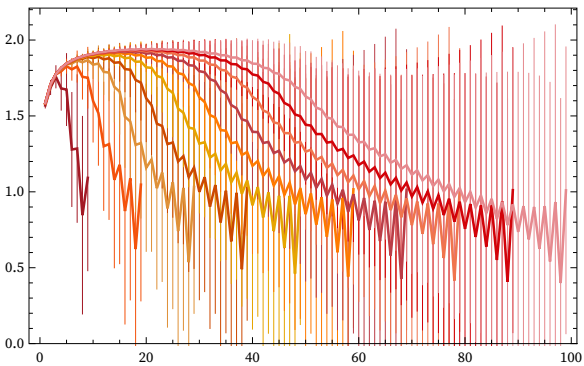

As expected, for $t$ small compared to the number of steps, the limiting estimated dimension is 2.

Most of the other causal graphs shown in the previous subsection do not have finite dimension, however. For example, for the rule AA→AAA the causal graph has the form

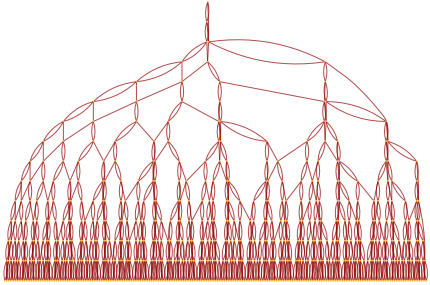

which increases exponentially, with $C_t \sim (\frac{3}{2})^t$.

The limits of the grid graphs we showed above essentially correspond to flat $d$-dimensional directed space. But we can also consider $d$-dimensional directed space with curvature. Although we cannot construct a complete sphere graph that is consistently directed, we can construct a partial sphere graph:

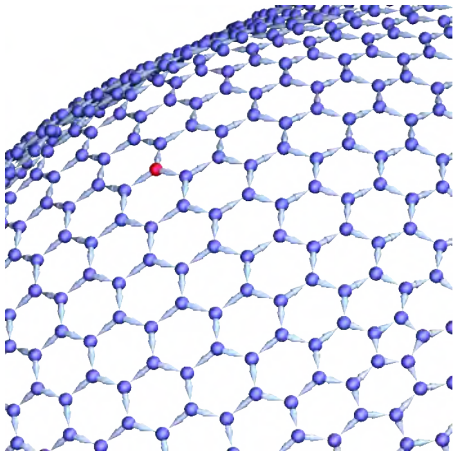

In a layered rendering, this is:

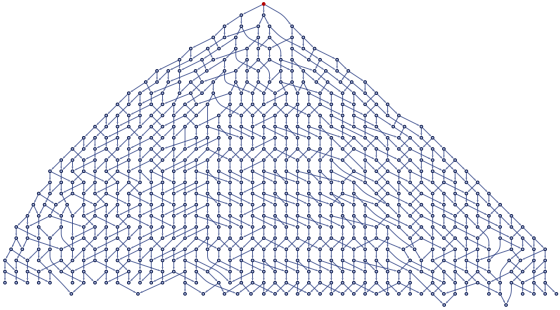

Once again, we can compute the log differences of $C_t$ (the similarity of sphere graphs means that using larger versions does not change the result):

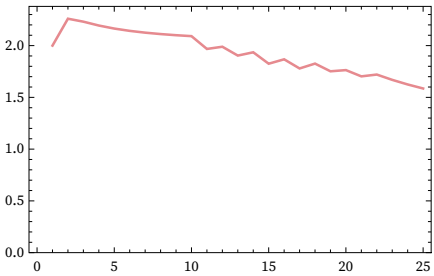

The systematic deviation from the $d = 2$ result is—like in the hypergraph case—a reflection of curvature.

## 5.14 Foliations and Coordinates on Causal Graphs

One way to describe a causal graph is to say that it defines the partial ordering of events in a system—or, in other words, it is a representation of a poset. (The actual graph is essentially the Hasse diagram of the poset (e.g. [69]).) Any particular sequence of updating events can then be thought of as a particular total ordering of the events.

As a simple example, consider a grid causal graph (as generated, for example, by the rule BA→AB):

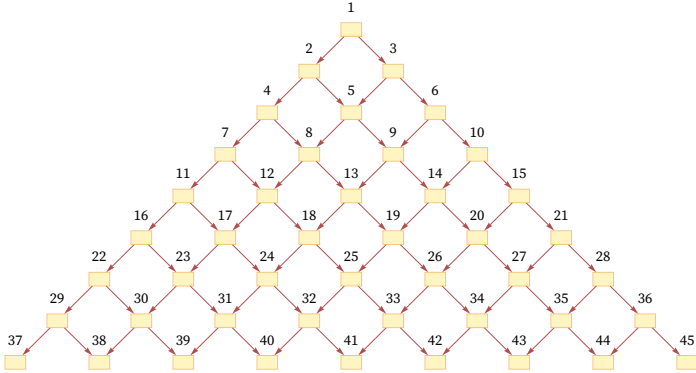

One total ordering of events consistent with all the causal relations in the graph is a "breadth-first scan" [70]:

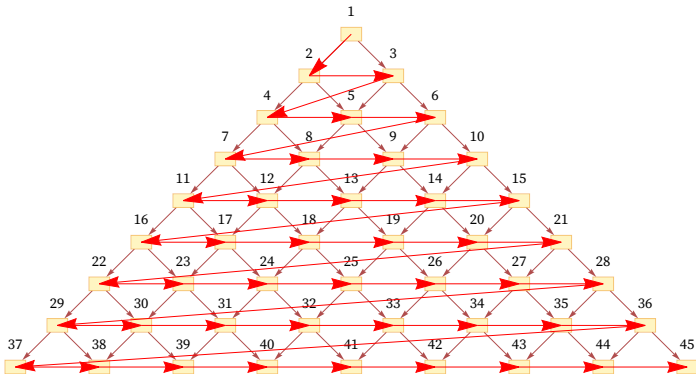

But another possible ordering is a "depth-first scan" [71]:

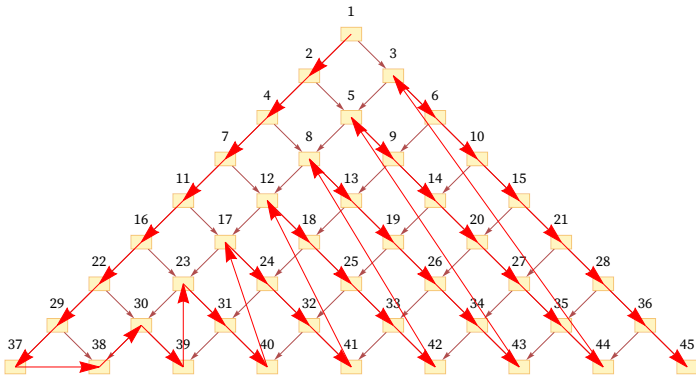

Another conceivable ordering would be:

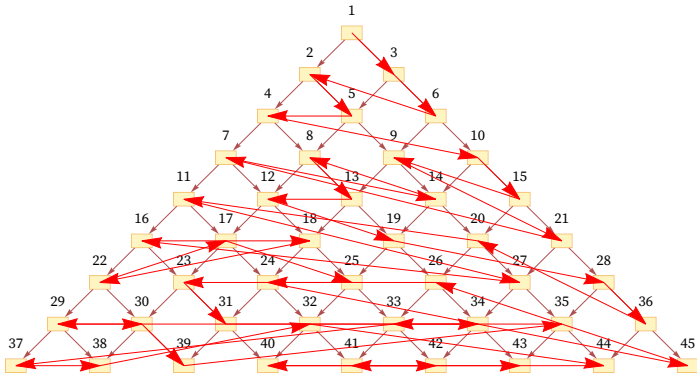

And in general there are many orderings consistent with the relations defined by the causal graph. For the particular graph shown here, out of the $n!$ conceivable orderings of nodes 1 through $n$, the orderings consistent with the causal relations correspond to possible Young tableaux, and the number of them is equal to the number of involutions (self-inverse permutations) of $n$ elements (e.g. [11:A000085]) which is asymptotically a little larger than $\left(\frac{n}{e}\right)^{-\frac{n}{2}}$ times $n!$.

Here are the possible causal orderings that visit nodes 1 through 6 above (out of all 6! = 720 orderings):

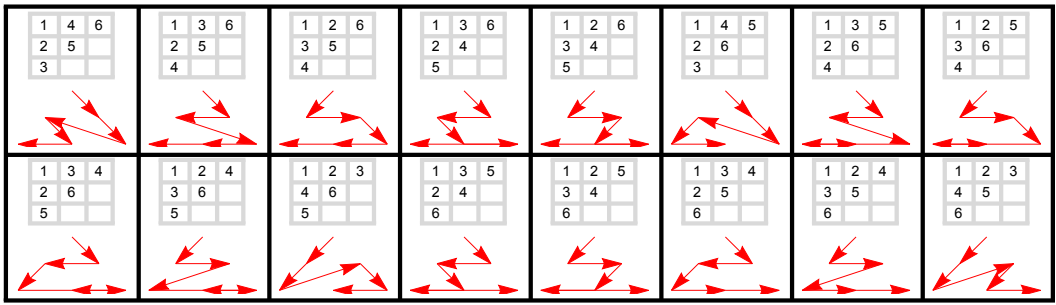

And here are all 76 possible causal orderings that visit any six nodes starting with node 1:

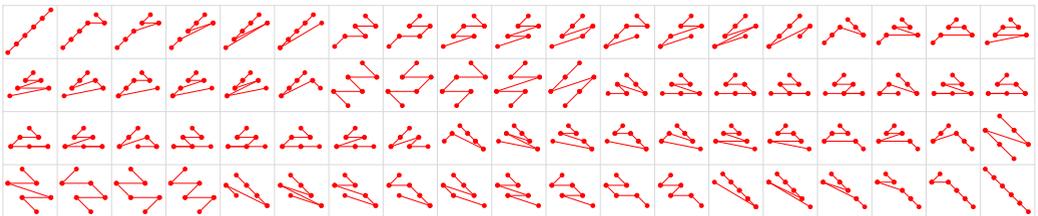

But while a great many total orderings are in principle possible, if one wants, for example, to think about large-scale limits, one usually wants to restrict oneself to orderings that can specified by ("reasonable") foliations.

The idea of a foliation is to define a sequence of slices with the property that events on successive slices must occur in the order of the slices, but that events within a slice can occur in any order. So, for example, an immediate possible foliation of the causal graph above is just:

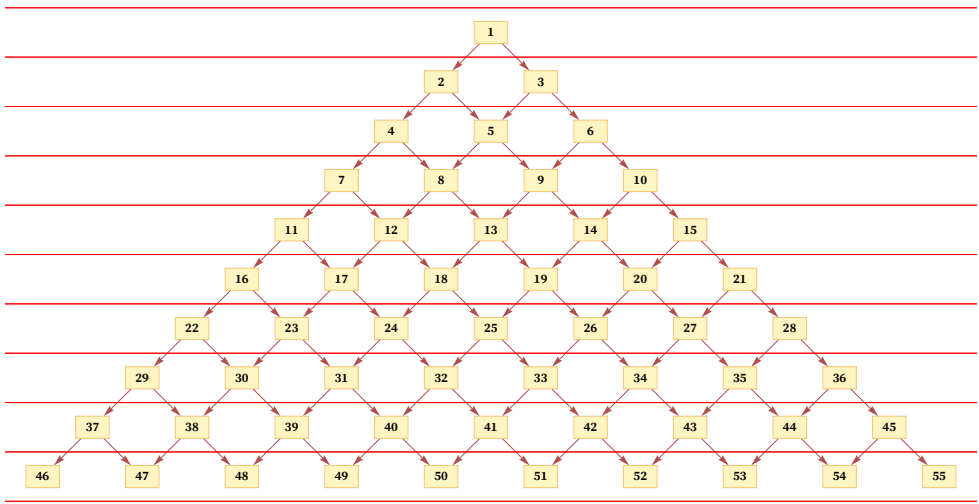

This foliation specifies that event 1 must happen first, but then events 2 and 3 can happen in any order, followed by events 4, 5 and 6 in order, and so on.

But another consistent foliation takes diagonal slices (with actual locations of events in the diagram being thought of as the centers of the boxes):

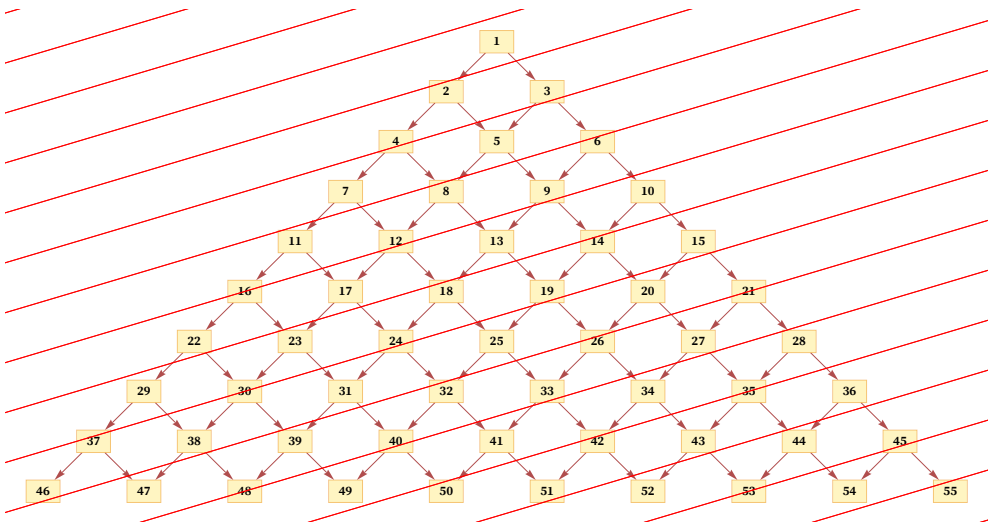

And so long as the diagonals are not steeper than the connections in the causal graph, this foliation will again lead to orderings that are consistent with the partial order defined by the causal graph. (For example, here event 1 must occur first, followed by event 2, followed by events 3 and 4 in any order, and so on.)

Particularly if the diagonals are steeper, multiple events will often happen in a single slice (as we see with events 4 and 7 here):

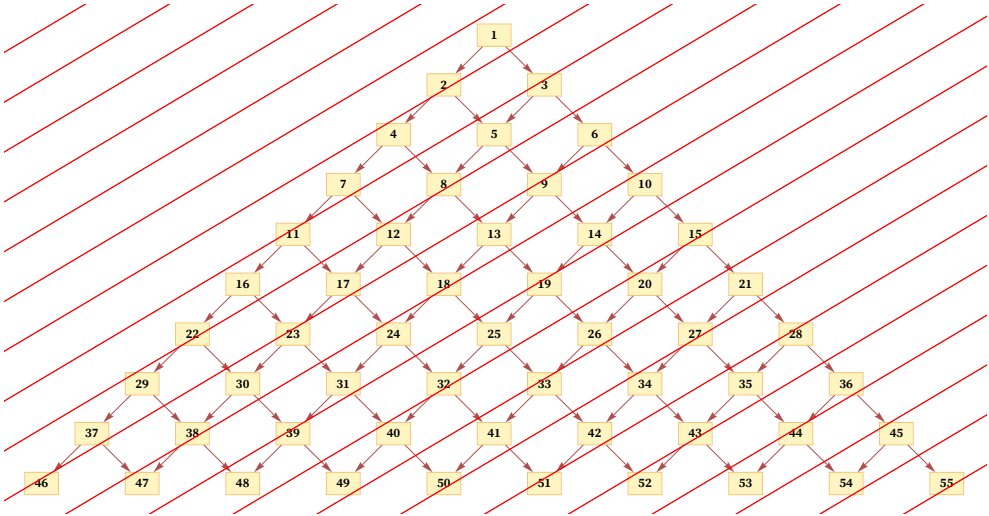

Now let us consider how this relates to taking the limit of a large number of events. If it were not for the directedness of the graph, we could do as we did in 4.17, and just imagine a process of refinement that leads to a manifold. But the Euclidean space that is the model for the manifold does not immediately have a way to capture the directedness of the graph, and so we need to do a little more.

But this is a place where foliations help. Because within a slice of a foliation we have events that can happen in any order. And at least for our string substitution system, the events can be thought of in the limit as being on a one-dimensional manifold, with a coordinate related to position on the string. And then there is just a second coordinate that is the index of the slices in the foliation.

But if the limit of our causal graph is a continuous space, we should be able to have a consistent notion of "distance between events". For events that are "out of order", the distance should be undefined (or perhaps infinite). But for other events, we should be able to compute the distance in terms of the coordinate (say *t*) that indexes the slices in the foliation and the coordinate (say *x*) within the slices.

Given the particular setup of diagonal slices on a causal graph that is a grid, there is a unique distance function that is independent of the angle of the slices, which can be expressed in terms of the coordinate differences $\Delta t$ and $\Delta x$:

$$\sqrt{\Delta t^2 - \Delta x^2}$$

This function is exactly the standard Minkowski metric for a Lorentzian manifold [72][73], and we will encounter it again in section 8 when we discuss potential connections to physics. But here the metric is simply an abstract way to express distance in the limit of our causal graphs for string substitution systems.

What happens if we use a different foliation? For example, a foliation like the following also leads to orderings of events that are consistent with the partial ordering required by the causal graph:

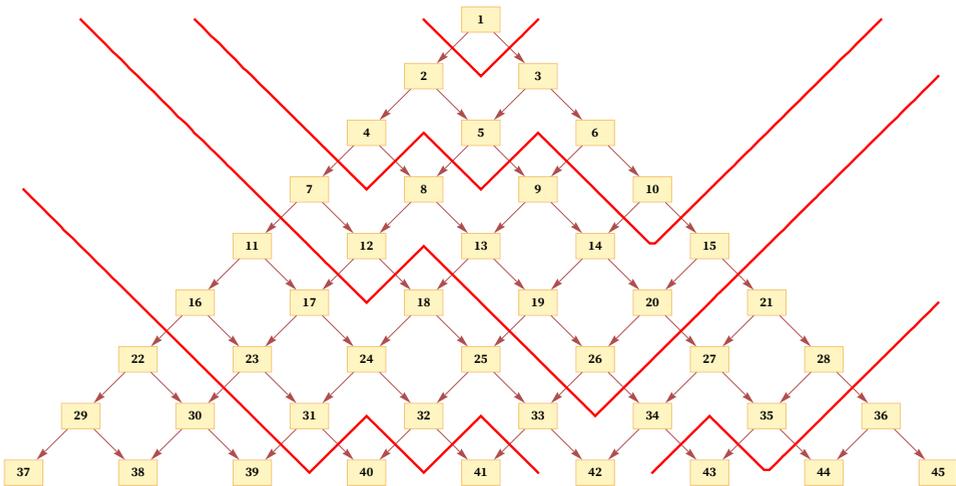

The process of limiting to a manifold is more complicated here. We can start by defining a "lapse function" $\alpha(t,x)$ (in analogy with the ADM formalism of general relativity [74][75]) which effectively says "how thick" each slice of the foliation is at each position. (If we also wanted to skew our foliations, we could include a "shift vector" as well.) And in the limit we can potentially define a distance by integrating $\sqrt{\alpha(t,x)^2 \, \delta t^2 - \delta x^2}$ along the shortest path from one point to another.

In a sense, however, even by imagining that there is a reasonable function $\alpha(t,x)$ that depends on real variables $t$ and $x$ we are implicitly assuming that our foliation has a certain simplicity and structure—and is not trying to reproduce some of the more circuitous total orderings at the beginning of this subsection.

But in a sense the question of what type of foliation we need to consider depends on what we want to use it for. And in making potential connections with physics, foliations will in effect be how we parametrize observers. And as soon as we assume that observers are limited in their computational capabilities, this puts constraints on the types of foliations we need to consider.

## 5.15 The Concept of Branchial Graphs

Causal graphs provide one kind of summary of the evolution of a system, based on capturing the causal relationships between events. What we call branchial graphs provide another kind of summary, based on capturing relationships between states on different branches of a multiway system. And whereas causal graphs capture relationships between events at different steps in the evolution of a system, branchial graphs capture relationships between states on different branches at a given step. And in a sense they define a map for exploring branchial space in a multiway system.

One might perhaps have imagined that states on different branches of a multiway system would be completely independent. But when causal invariance is present they are definitely not—because for example whenever they split (to form a branch pair), they will always merge again.

Consider the multiway evolution graph (for the rule {A→AB,B→A}):

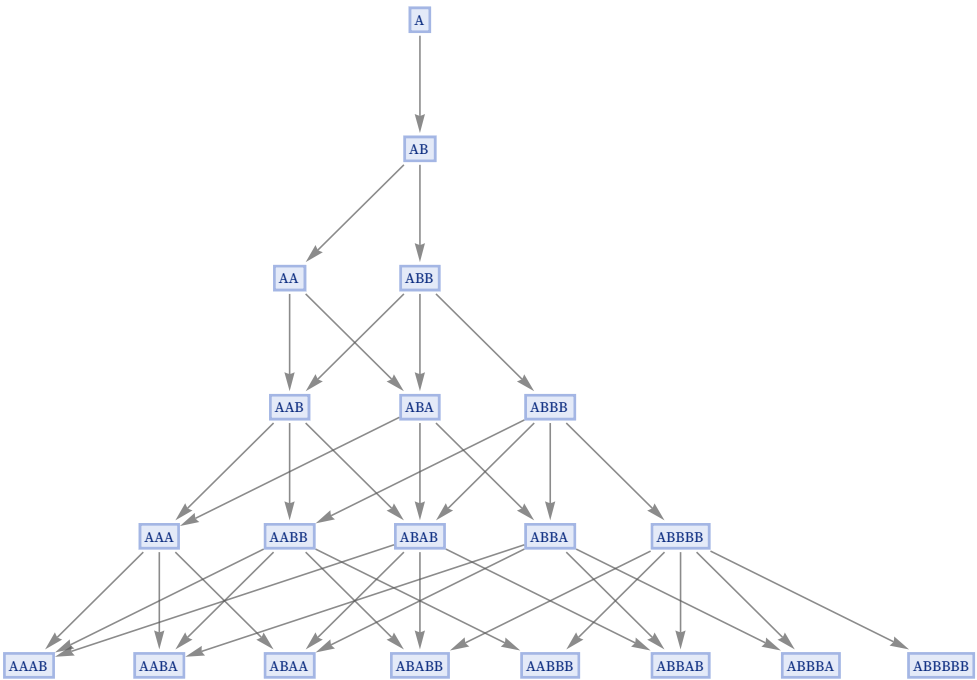

Look at the second-to-last step shown. This contains the following unresolved branch pairs:

{{AAA, AABB}, {AAA, ABAB}, {AAA, ABBA}, {AABB, ABAB}, {AABB, ABBA}, {AABB, ABBBB}, {ABAB, ABBA}, {ABAB, ABBBB}, {ABBA, ABBBB}}

The two states in each of these branch pairs are related, in the sense that they diverged from a common ancestor—and will converge to a common successor. We form the branchial graph by connecting the states that appear in newly unresolved branch pairs at a given step. For the steps shown in the evolution graph above, the successive branchial graphs are:

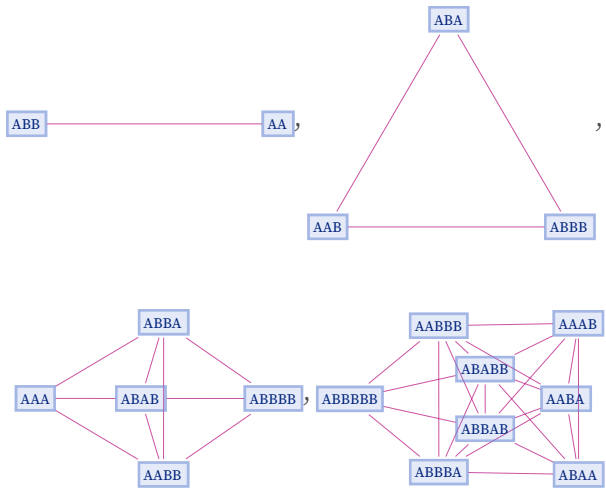

For the next few steps, the states branchial graphs are:

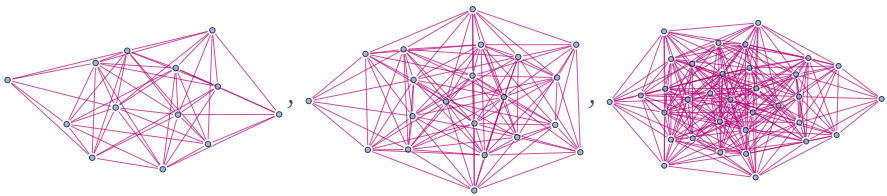

For the rule shown here, the number of nodes at the $t^{\text{th}}$ step is the $t^{\text{th}}$ Fibonacci number, or $\sim \phi^t$ for large $t$. The graphs are highly connected, but far from completely so. The number of edges $\sim 2^t$, while the diameter is $\lfloor \frac{t}{2} \rfloor$. The degree of transitivity (i.e. whether $X$ being connected to $Y$ and $Y$ being connected $Z$ implies $X$ being connected to $Z$) [76] gradually decreases with $t$. As a measure of uniformity, one can look at the local clustering coefficient [77] (which measures to what extent there are local complete graphs):

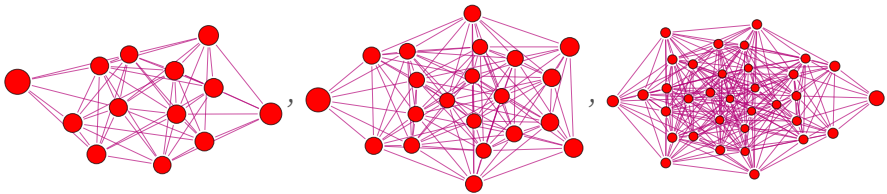

The graphs have somewhat complex vertex degree distributions (here shown after 10 and 15 steps):

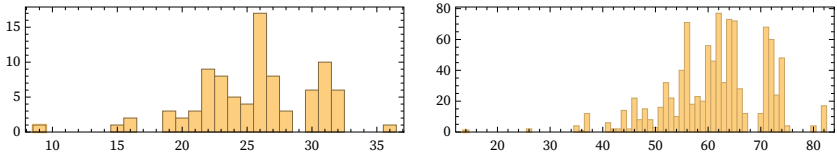

For the rule we are showing here, the branchial graph turns out to have a particularly simple interpretation as a map of the level of common ancestry of states. By definition, if two states are directly connected on the branchial graph, it means they had an immediate common ancestor one step before. But what does it mean if two states are graph distance 2 apart?

The particular rule shown here has the property that it is causal invariant, but also that all branch pairs resolve in just one step. And from this it follows that states that are distance 2 apart on the branchial graph must have a common ancestor 2 steps back. And in general the distance on the branchial graph is equal to the number of steps one must go back before one gets to a common ancestor.

The following histograms show the distribution of graph distances between all pairs of states at steps 10 and 15—or alternatively, the distribution of how many steps it has been since the states had a common ancestor (for this rule the mean increases roughly like $\frac{t}{6}$):

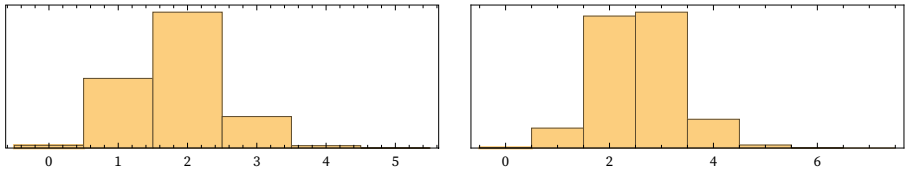

We are defining the branchial graph to be based on looking at branch pairs that are unresolved at a particular step, and are new at that step. But we could generalize to consider a "thickened" branchial graph that includes all branch pairs that have been new within the past $m$ steps. By doing this we can capture common ancestry of states even when they are associated with branch pairs that take up to $m$ steps to resolve—but when we do this it is at the cost of having many additions to the graph associated with branch pairs that have "come and gone" within our thickening depth.

It should be noted that any possible interpretation of branchial graphs in terms of common ancestors depends on having causal invariance. Absent causal invariance, there is no guarantee that states with common ancestors will even be connected in the branchial graph. As an extreme example, consider the rule:

{A → AB, A → AC}

The multiway graph for this rule is a tree

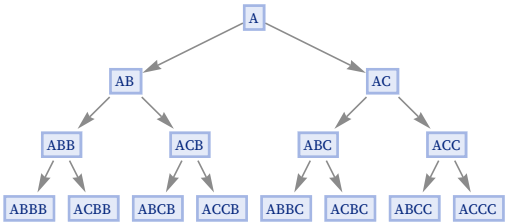

and its branchial graph on successive steps just consists of a collection of disconnected pieces:

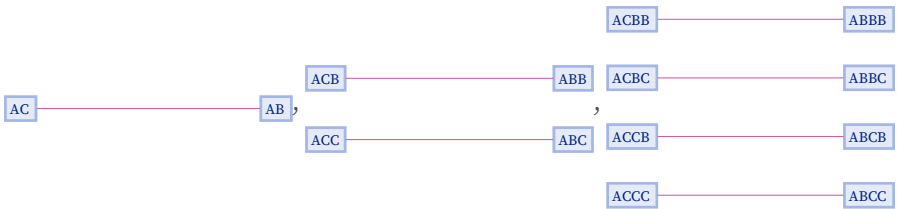

By thickening the branchial graph by *m* steps, one could capture *m* steps of common ancestry. And in general one could imagine infinitely thickening the graph, so that one looks all the way back to the initial conditions. But the branchial graph one would get in this way would essentially just be a copy of the whole multiway graph.

When a rule has branch pairs that take several steps to resolve, it is possible for the branchial graph to be disconnected, even when the rule is causal invariant. Consider for example the rule

{A → AA, A → BAB}

in which the branch pair {BAAB,BBABB} takes 3 steps to resolve. The first few branchial graphs for this rule are:

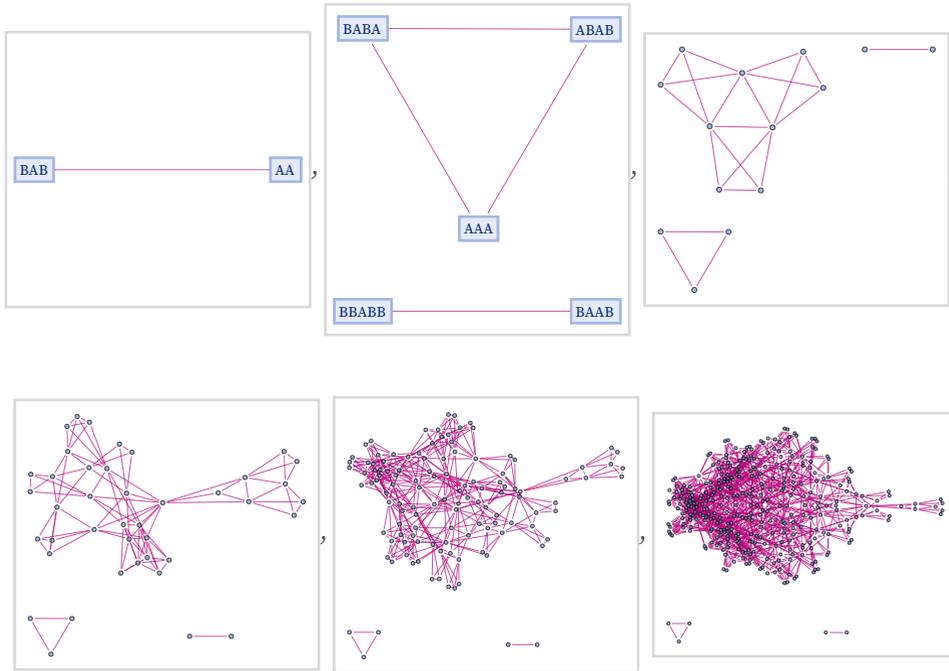

It is fairly common to get branchial graphs in which a few disconnected pieces are "thrown off" in the first few steps, and never recombine. Sometimes, however, disconnected pieces can recombine. The rule

{A → BB, BB → AA}

starting from initial condition A yields the sequence of branchial graphs:

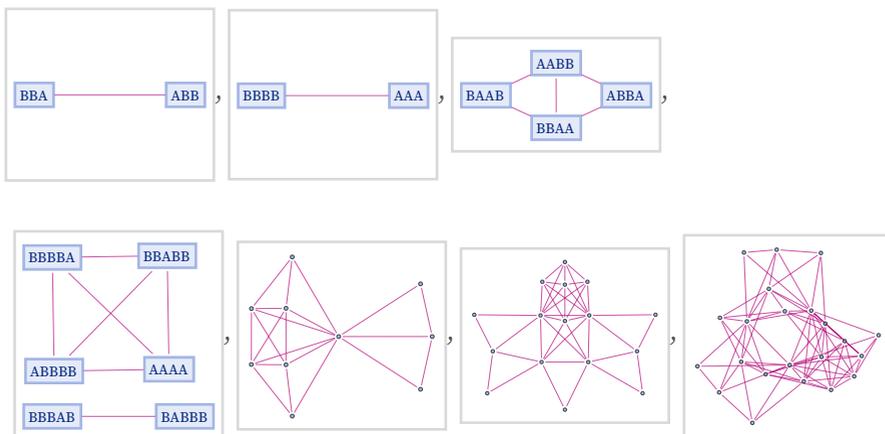

Sometimes the branchial graph can break into many components. This happens for the rule

{AA → AABB}

starting from initial condition AA:

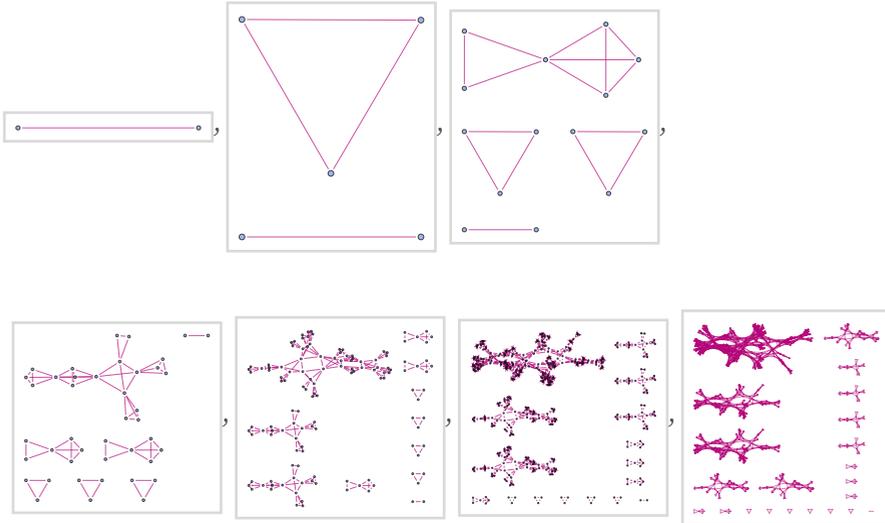

The causal graph for this rule reveals that there is also causal disconnection in this case:

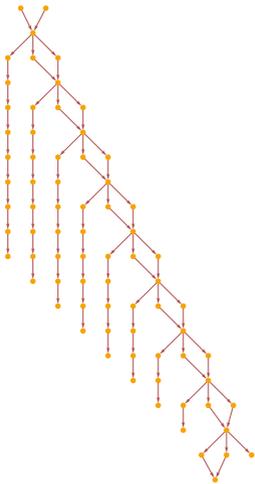

## 5.16 Typical Forms of Branchial Graphs

As a first example, consider the (causal invariant) rule which effectively just creates either A or B from nothing:

{ → A, → B}

At step $t$, this rule produces all $2^t$ possible strings. Its multiway way graph is:

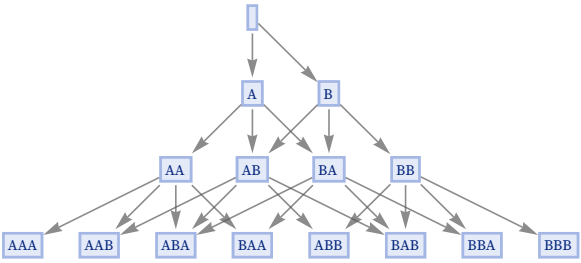

The succession of branchial graphs is then:

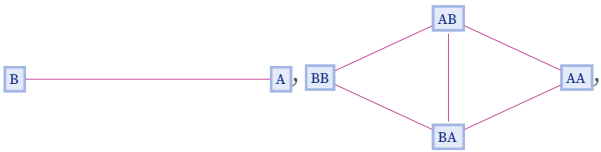

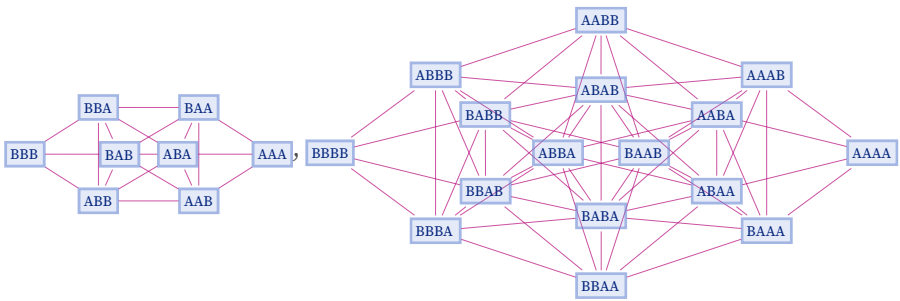

The graph on step $t$ has $2^t$ nodes and $2^{t-2}(t^2 - t + 4) - 1$ edges. The distance on the graph between two states is precisely the difference in the total number of As (or Bs) between them, plus 1—so combining states which differ only through ordering of A and B the last graph becomes:

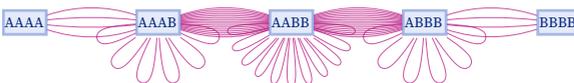

With the rule

{ → A, → B, → C}

the sequence of branchial graphs is

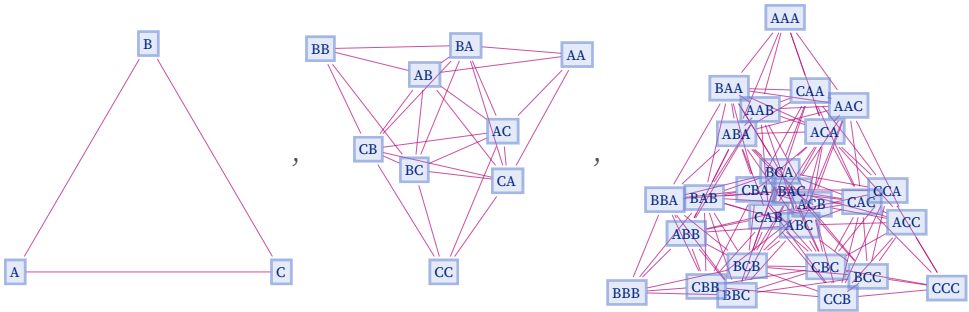

and in the last case combining states which differ only in the order of elements, one gets:

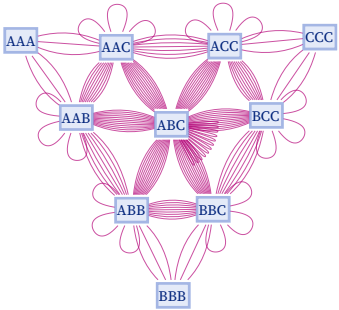

Note that no rule involving only As can have a nontrivial branchial graph, since all branch pairs immediately resolve.

Consider now the rule:

{A → AB}

As mentioned in 5.4, with initial condition AA this rule gives a multiway graph that corresponds to a 2D grid:

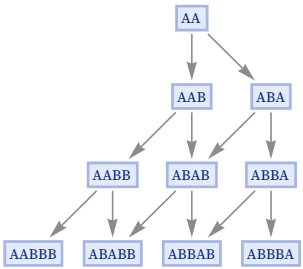

The corresponding branchial graphs are 1D:

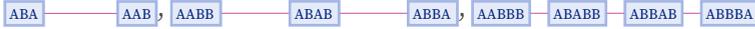

With initial condition AAA, the multiway graph is effectively a 3D grid, and the branchial graph is a 2D grid:

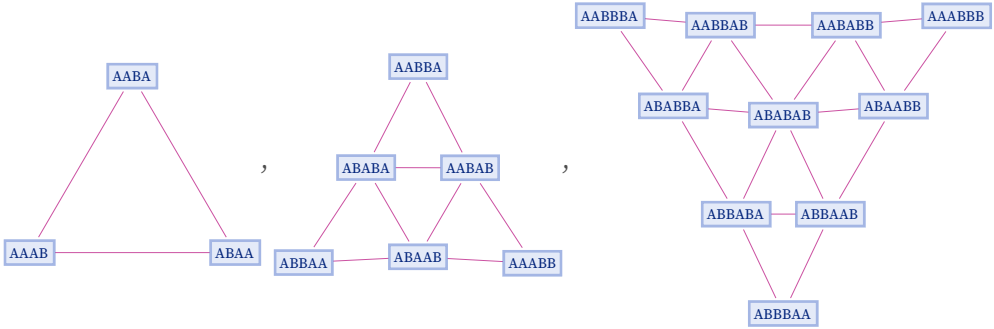

Some rules produce only finite sequences of branchial graphs. For example, the rule

{A → B}

with initial condition AAAA yields what are effectively sections through a cube oriented on its corner:

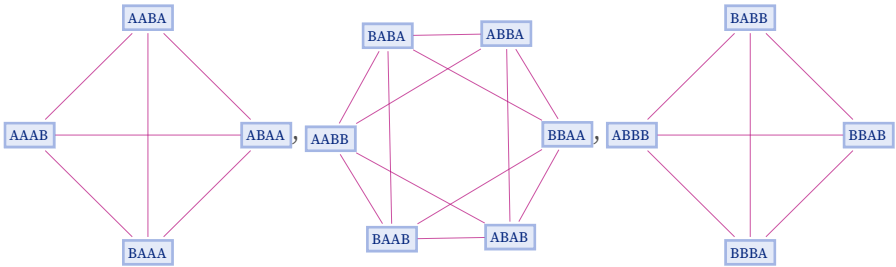

As another example producing a finite sequence of branchial graphs, consider the rule:

{BA → AB}

Starting from BBBAAA it gives:

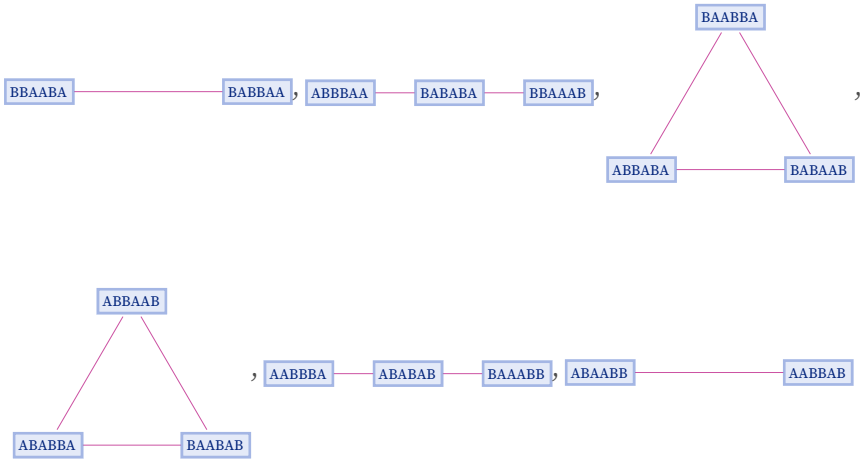

Starting from BBBBBAAAAA it gives:

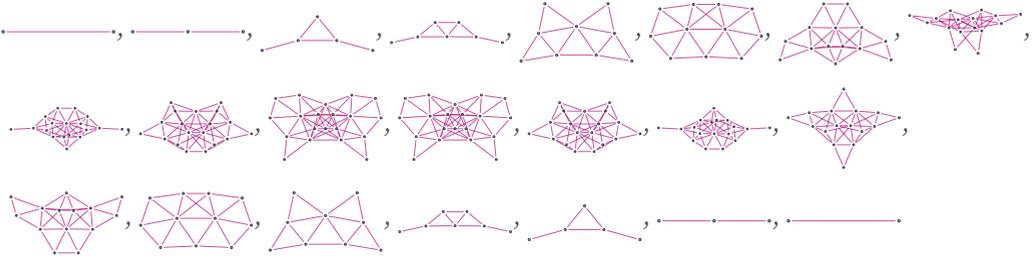

One can think of this as showing the "shapes" of successive slices through the multiway system for the evolution of this rule.

As another example of a rule yielding an infinite sequence of branchial graphs, consider:

{A → AAB}

This yields the following branchial graphs:

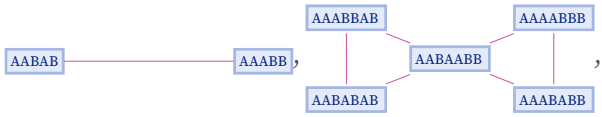

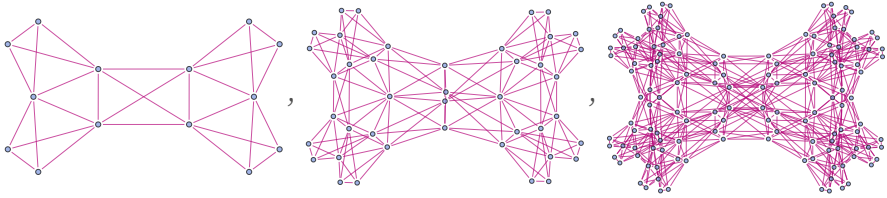

In a 3D rendering, the graph on the next step is:

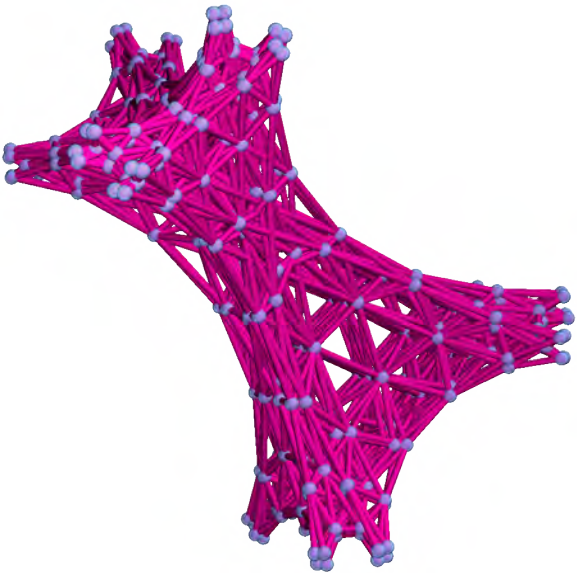

The following are the distinct forms of branchial graphs obtained from rules involving a total of up to 6 As and Bs (starting from a single A):

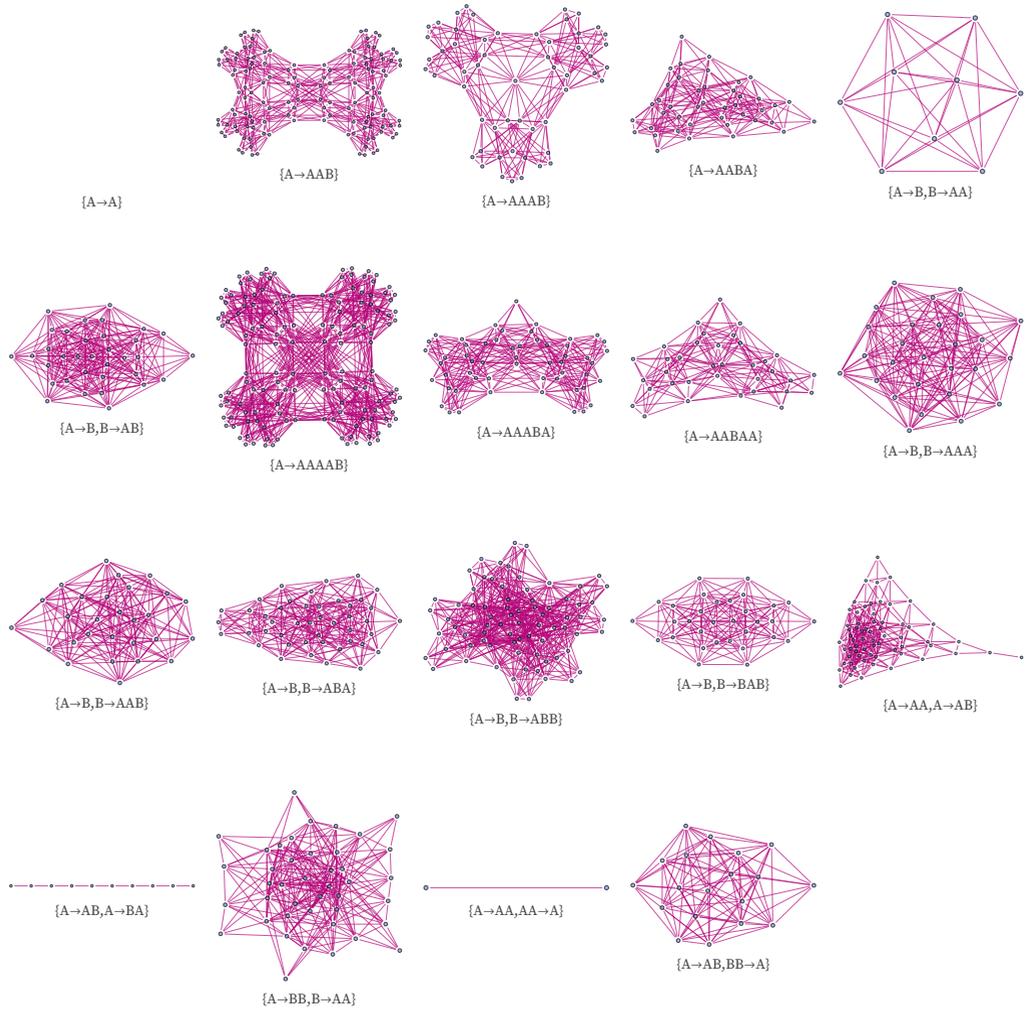

We have seen that branchial graphs can form regular grids. But many branchial graphs have much higher levels of connectivity. No branchial graph can continue to be a complete graph (with all neighbors having distance 1) for more than a limited number of steps. However, the diameters of branchial graphs do tend to grow slowly, and on step $t$ they can be no larger than $t$. Some branchial graphs show linear or polynomial growth with the number of steps in vertex and edge count, but many show exponential growth.

In analogy to what we did for hypergraphs and causal graphs, we can define a quantity $B_b$ which measures the number of nodes in the branchial graph reached by going out to graph distance $b$ from a given node.

Consider for example the rule:

{ → A, → B}

The sequence of forms for $B_b$ as a function of $b$ on successive steps is:

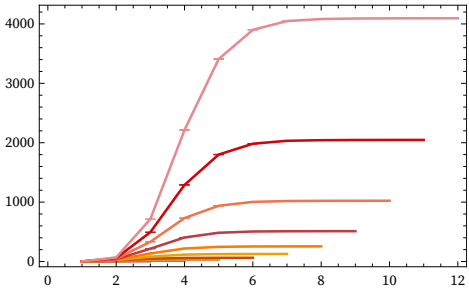

At step $t$, the diameter of the graph is just $t$, and $B_{b=t} = 2^t$. For smaller $b$, the ratios of the $B_b$ for given $b$ at successive steps $t$ steadily decrease, perhaps suggesting ultimately less-than-exponential growth:

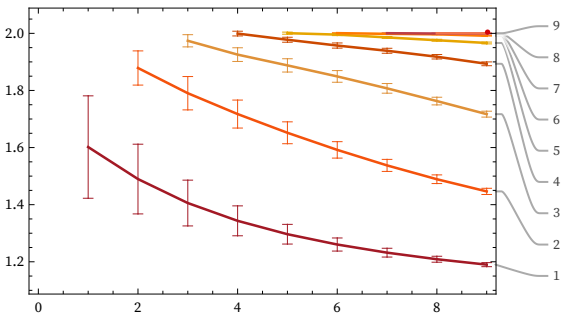

One can ask what the limit of a branchial graph after a large number of steps may be. As an initial possible model, consider graphs representing $n$-cubes in progressively higher dimensions:

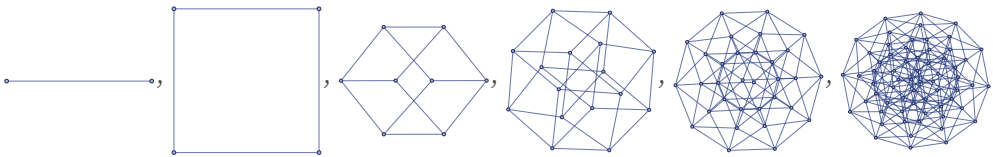

The graph distances between nodes in these graphs are exactly the same as the Euclidean distances between the $2^n$ possible tuples of 0s and 1s (here shown in distance matrices arranged in lexicographic order):

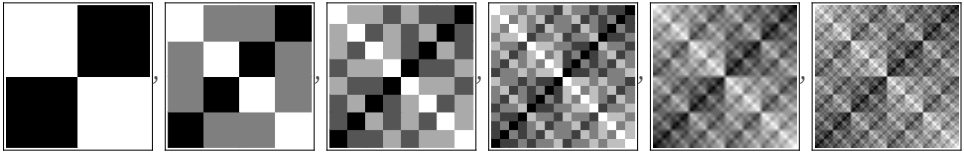

The values of $B_b$ in this case can be found from [11:A008949]:

Accumulate[Table[Binomial[n, k], {k, 0, n}]]

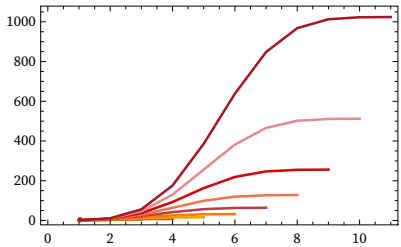

This shows the ratios of $B_b$ for given $b$ for successive $n$:

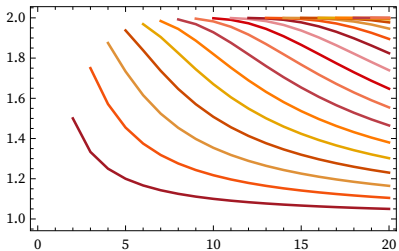

Much as one can consider progressively larger grid graphs as limiting to a manifold, so perhaps one may consider higher and higher "dimensional" cube graphs as limiting to a Hilbert space.

It is also conceivable that limits of branchial graphs may be related to projective spaces [78]. As one potential connection, one can look at discrete models of incidence geometry [79]. For example, with integers representing points and triples representing lines, the Fano plane

{{1, 2, 3}, {1, 4, 5}, {1, 6, 7}, {2, 4, 6}, {2, 5, 7}, {3, 4, 7}, {3, 5, 6}}

is a discrete model of the projective plane. One can consider the sequence of such objects as hypergraphs

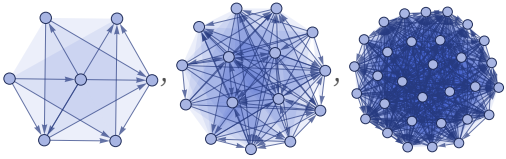

and representing both points and lines here as nodes, these correspond to the graphs:

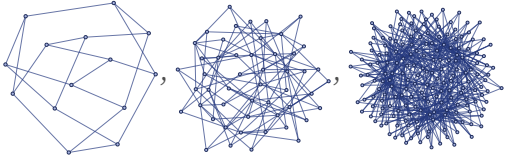

But for such graphs one finds that $B_b$ has a very different form from typical branchial graphs:

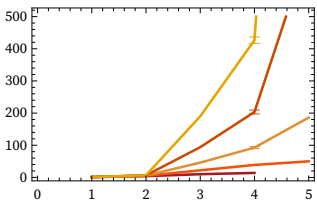

An alternative approach to connecting with discrete models of projective space is to think in terms of lattice theory [69][80][81]. A multiway graph can be interpreted as a lattice (in the algebraic sense), with its evolution defining the partial order in the lattice. The states in the multiway system are elements in the lattice, and the meet and join operations in the lattice correspond to finding the common ancestors and common successors of states.

The analogy with projective geometry is based on thinking of states in the multiway system (which correspond to elements in the lattice) as points, connected by lines that correspond to their evolution in the multiway system. Points are considered collinear if they have the same common successor. But (assuming the multiway system starts from a single state), causal invariance is exactly the condition that any set of points will eventually have a common successor—or in other words, that all lines will eventually intersect, suggesting that the multiway graph is indeed in some sense a discrete model of projective space—so that branchial graphs may also be models of projective Hilbert spaces.

## 5.17 Foliations of the Multiway Graph and the Structure of Branchial Space

Just as states that occur at successive steps in the evolution of our underlying systems can be thought of as associated with successive slices in a foliation of the causal graph, so also branchial graphs can be thought of as being associated with successive slices in a foliation of the multiway graph.

As we discussed above, different foliations of the causal graph define different relative orderings of updating events within our underlying system. But we can now think about this at a higher level and consider foliations of the multiway graph, that in effect define different relative orders of updating events on different branches of the multiway system. A foliation of the causal graph in effect defines how we should line up our notion of "time" for events in different parts of our underlying system; a foliation of the multiway graph now also defines how we should line up our notion of "time" for events on different branches of the multiway system.

For example, with the rule

{ A → AB}

starting from AA, we can define the following foliation of the multiway graph:

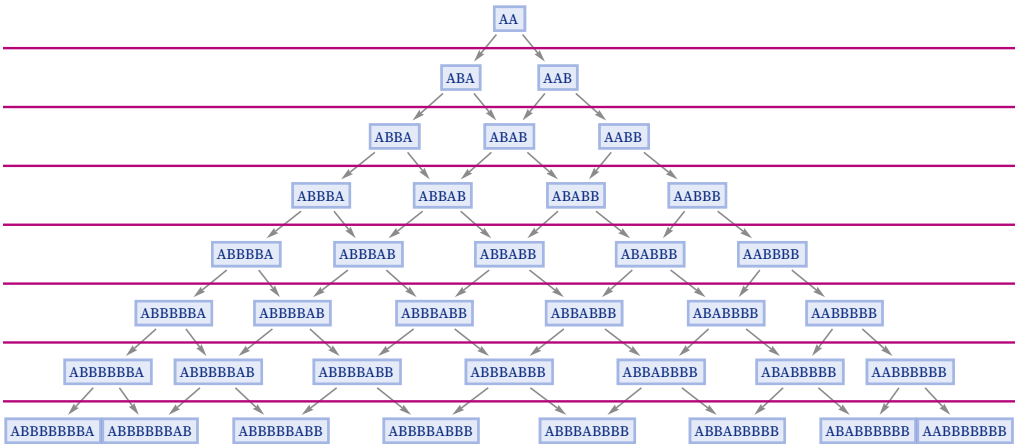

This yields the branchial graphs:

| ABA | — | AAB |
|---|---|---|

| AABB | — ABAB — | ABBA |

| AABBB — ABABB — ABBAB — ABBBA |

| AABBBB — ABABBB — ABBABB — ABBBAB — ABBBBA |

| AABBBBB — ABABBBB — ABBABBB — ABBBABB — ABBBBAB — ABBBBBA |

| AABBBBBB — ABABBBBB — ABBABBBB — ABBBABBB — ABBBBABB — ABBBBBAB — ABBBBBBA |

| AABBBBBBB — ABABBBBBB — ABBABBBBB — ABBBABBBB — ABBBBABBB — ABBBBBABB — ABBBBBBAB — ABBBBBBBA |

Just as the causal graph defines a partial order for events, so now the multiway graph defines a partial order for states. And so long as it is consistent with this partial order, we can pick any total order for the states. And we can parametrize some of these total orders as foliations.

For the particularly simple case shown here, an alternative foliation consistent with the partial order defined by the multiway system is:

And if we use this foliation, we get a different sequence of branchial graphs, now no longer connected:

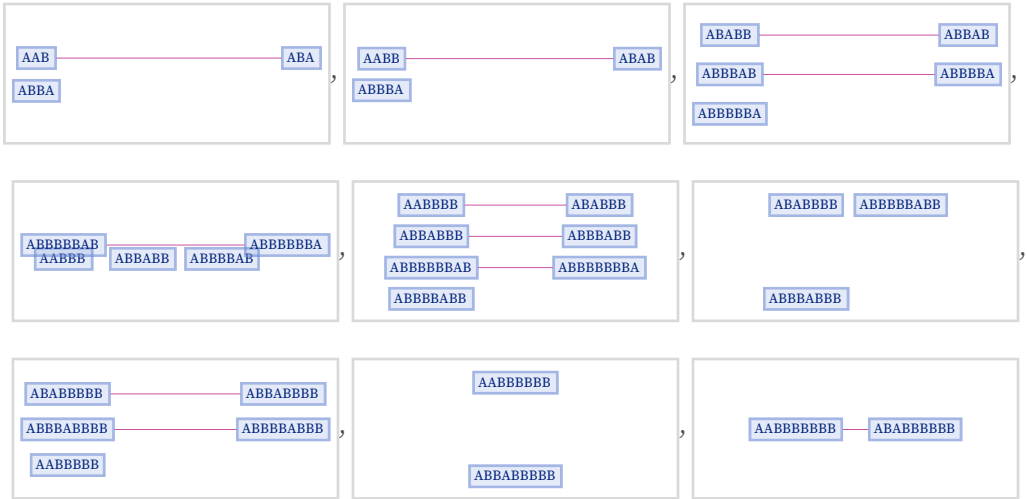

This example is particularly obvious in its analogy with the causal graphs we discussed above. But what makes this work is a special feature of the branchial graphs in this case: the fact that the states that appear in them can in effect be assigned simple 1D "coordinates". In the original foliation with unslanted ("one event per slice") slices, the "coordinate" is effectively just the position of the second A in the string. And with respect to the "space" defined by this "coordinate", the branchial graphs can readily be laid out in one dimension, and we can readily set up "slanted" foliations, just like we did for causal graphs.

With the underlying systems we are discussing in this section being based on strings of elements, it is inevitable that there will be 1D coordinates that can be assigned to the events that occur in the causal graph. But nothing like this need be true of the branchial graph, and indeed most branchial graphs have a significantly more complex structure.

Consider the same rule as above, but now started from AAA. The multiway graph in this case—with a foliation indicated—is then:

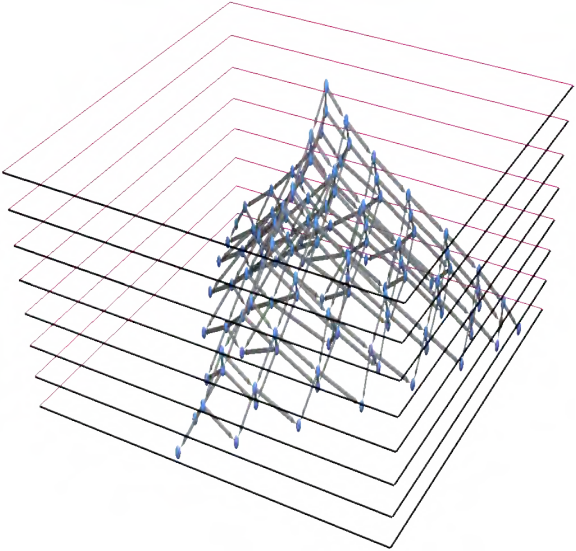

The branchial graphs now have a two-dimensional structure—with the positions of the second and third As in each string providing potential "coordinates":

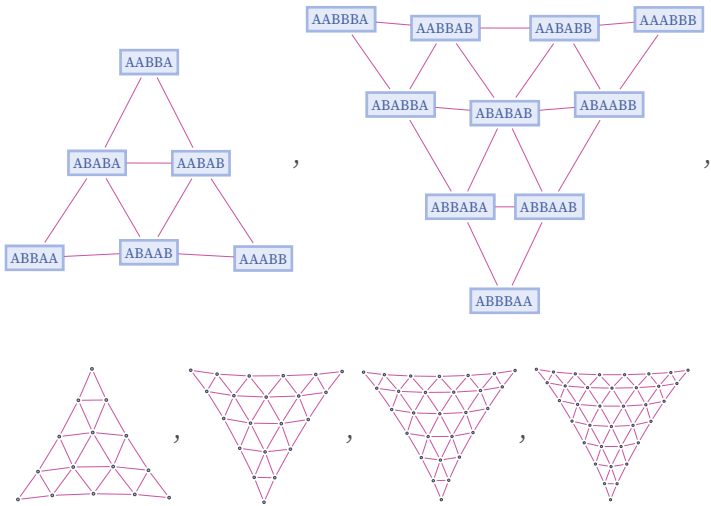

But consider now the rule

{BA → AB}

started from BABABABA. The multiway graph in this case is:

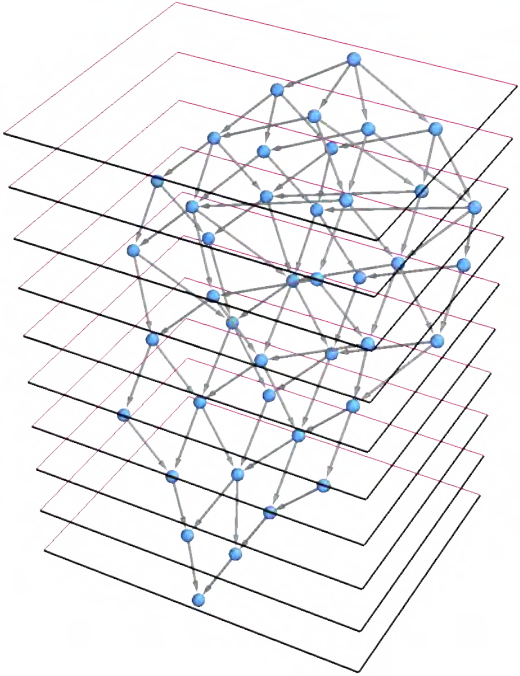

And the sequence of branchial graphs based on the foliation above is now:

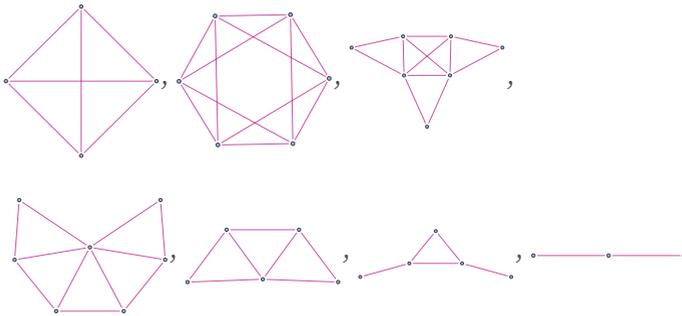

But here it is already much less clear how to assign "coordinates" in "branchial space", or how to create a meaningful family of foliations of the multiway graph.

In thinking about multiway graphs and their foliations there is another complication that can arise for some rules. Consider two versions of the multiway graph for the rule

{AB → BAB, BA → A}

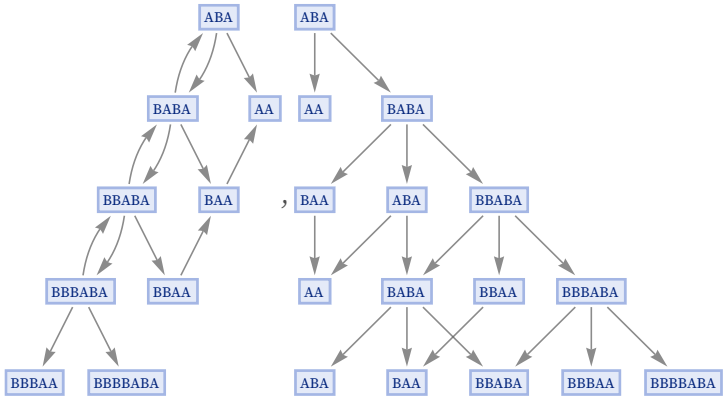

In the first version, every distinct state is shown only once. But in the second case, the evolution is "partially unrolled" to show separately different instances of the same state, produced after different numbers of updating events. With a foliation whose slices correspond to the layers in the renderings above, the first version of the multiway system yields the branchial graphs:

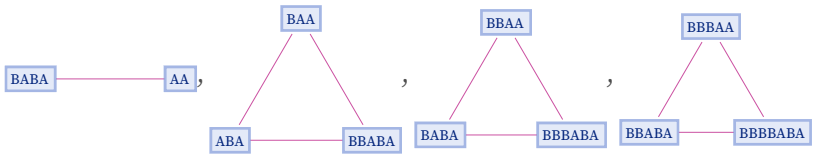

The second version, however, yields different branchial graphs:

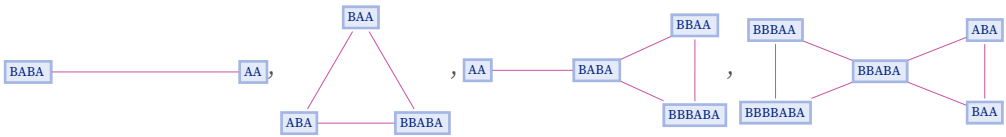

To some extent this difference is just like taking a different foliation. But there is more to it, because the second version of the multiway graph actually defines different ordering constraints than the first one. In the second version, there is a true partial ordering, defined by the directed edges in the multiway graph. But in the first version, there can be loops, and so no strict partial order is defined. (We will discuss this phenomenon more in 6.9.)

## 5.18 The Relationship between Graphs, and the Multiway Causal Graph

In the course of this section, we have seen various ways of describing and relating the possible behaviors of systems. In many ways the general is the combined multiway evolution and multiway causal graph.

For the rule

{A → AB}

starting from AA this graph has the form:

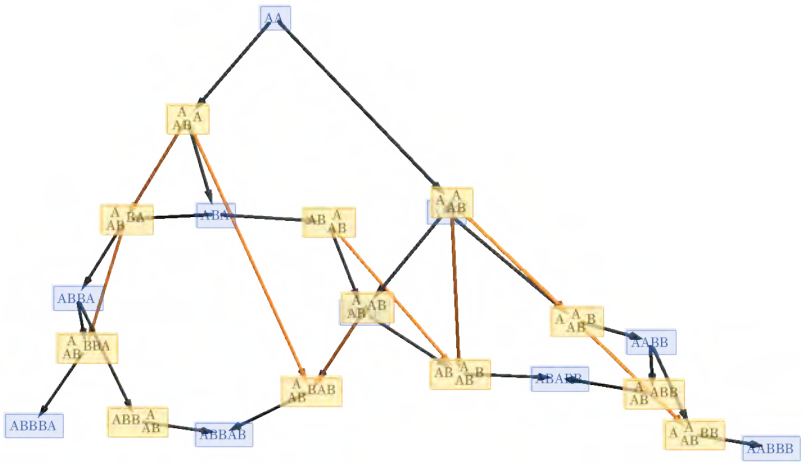

Continuing for another step, we have:

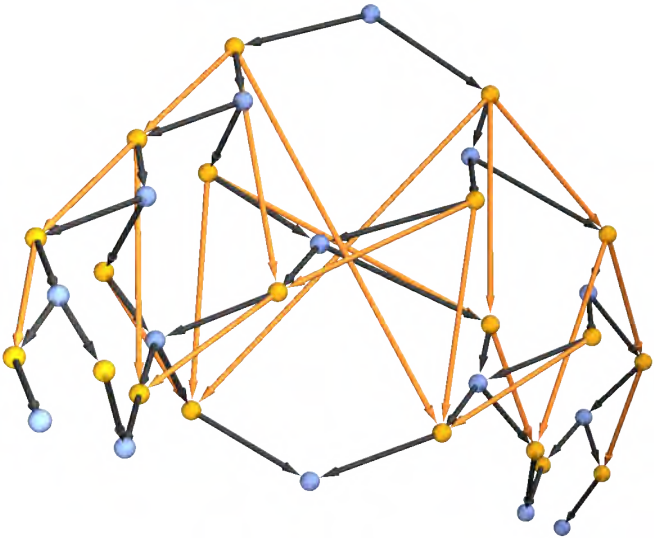

There are several different kinds of descriptions that we can derive from this graph. The standard multiway graph gives the evolution relationship between states:

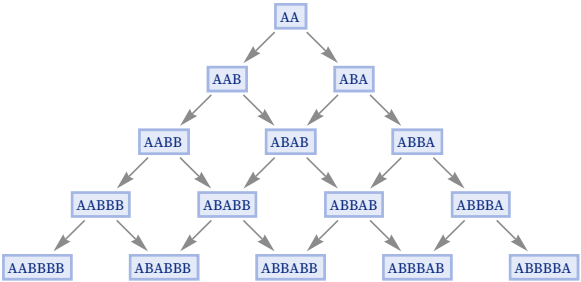

Each possible path through this graph corresponds to a possible evolution history for the system.

The multiway causal graph gives the causal relationships between all events that can happen on any branch. The full multiway causal graph for the rule shown here is infinite. But truncating to show only the part contained in the graph above, one gets:

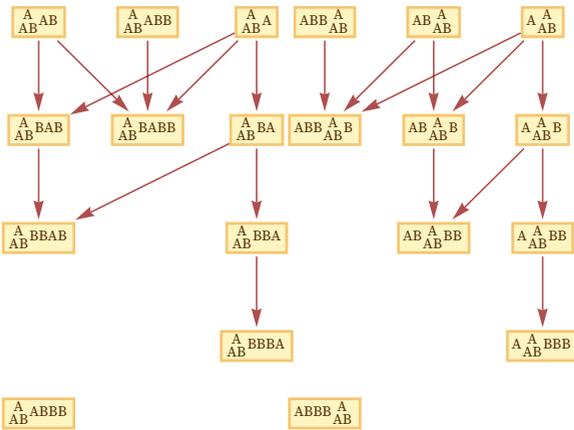

Continuing for more steps one gets:

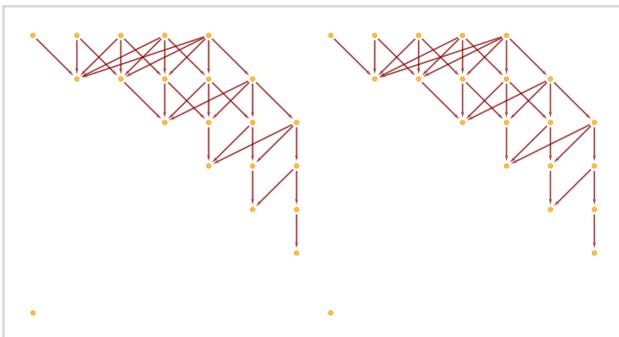

From the multiway causal graph, one can project out the specific causal graph for each possible evolution history, corresponding to each possible branch in the multiway system.

But for rules like the one shown here that have the property of causal invariance, every one of these specific causal graphs (at least if extended far enough) must have exactly the same structure. For the particular rule shown here, this structure is extremely simple:

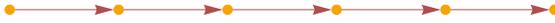

(In effect, the nodes here are "generic events" in the system, and could be labeled just by copies of the underlying local rule.)

The multiway graph and multiway causal graph effectively give two different "vertical views" of the original graph—using respectively states as nodes and and events as nodes. But an alternative is to view the graph in terms of "horizontal slices". To get such slices we have to do foliations.

But now if we look at horizontal slices associated with states, we get the branchial graphs, which for this rule with this initial condition are rather trivial:

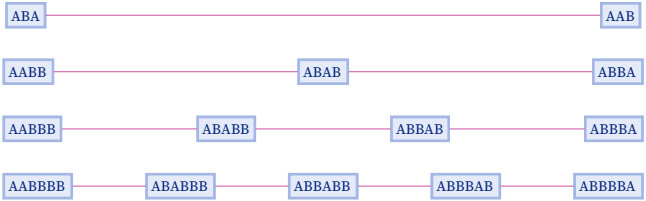

In principle we could also ask about horizontal slices associated with events. But by construction the events analog of the branchial graph must just consist of a collection of complete graphs.

However, a particular sequence of slices through any particular causal graph defines an actual sequence of states for the underlying system, and thus a possible evolution history, such as:

{AA, ABAB, ABBABB, ABBBABBB, ABBBBABBBB}

As a slightly cleaner example with similar behavior, consider the rule:

{A → AB, A → BA}

The combined multiway evolution and multiway causal graph in this case is

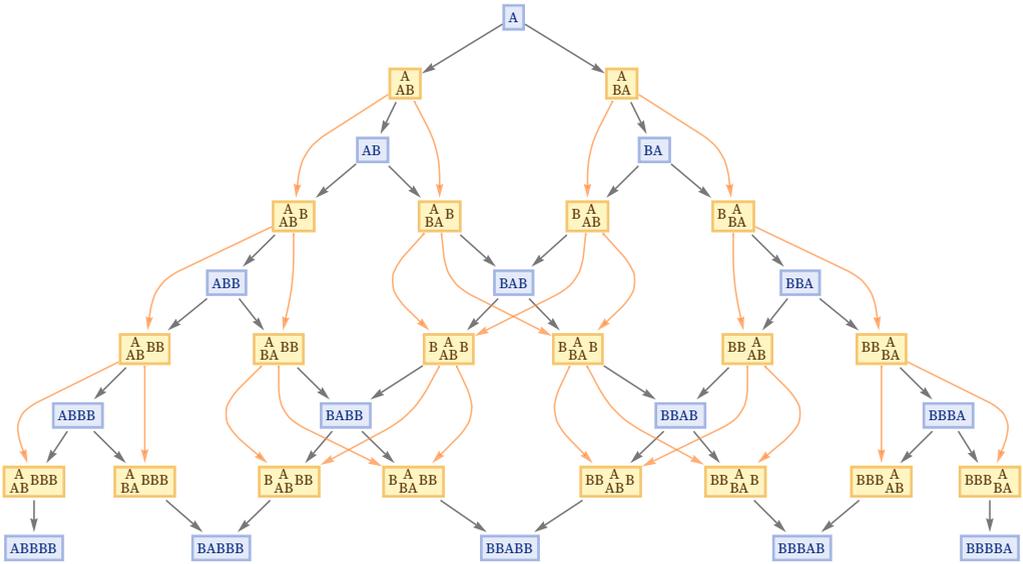

and the individual multiway evolution and causal graphs are both regular 2D grids:

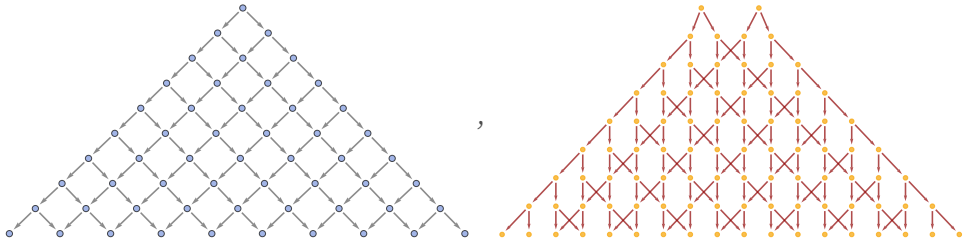

The rules we have used as an example so far have behavior that is in a sense fairly trivial. But consider now the related rule with slightly less trivial behavior:

{A → AB, B → A}

For this rule, the combined multiway evolution and causal graph has the form:

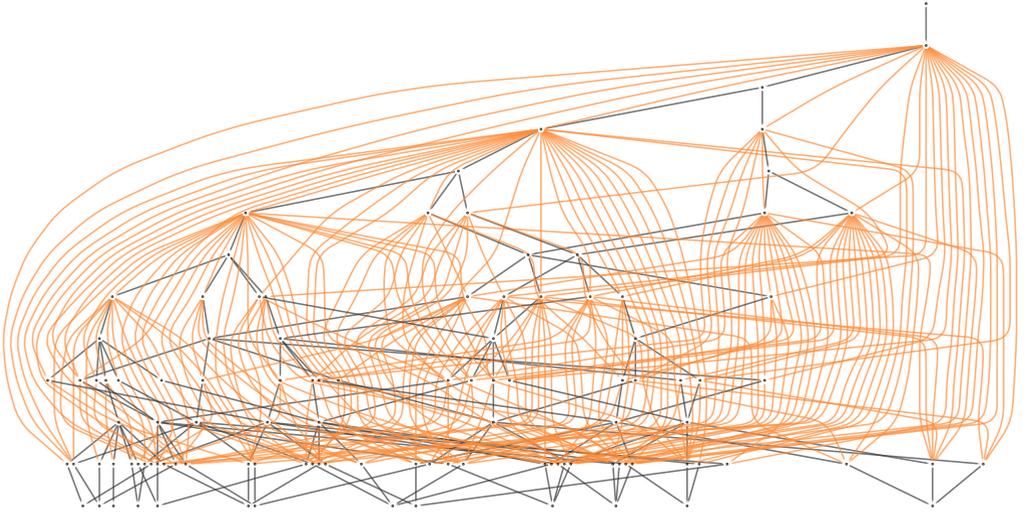

On their own, the state evolution graph and the event causal graph have the forms:

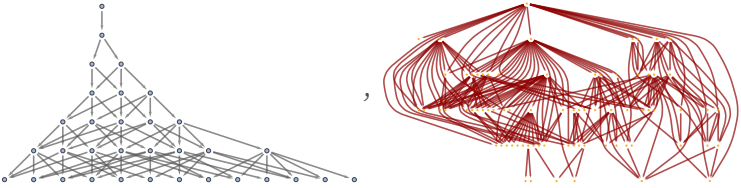

The sequence of branchial graphs is:

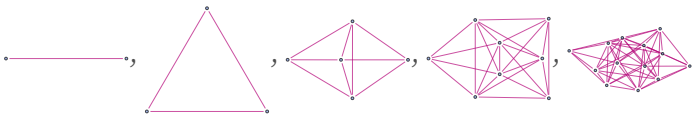

This rule is causal invariant, and so the multiway causal graph decomposes into many identical copies of causal graphs for all individual possible paths of evolution. In this case, these graphs all have the form:

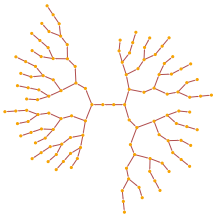

But even though the multiway causal graph can be decomposed into identical pieces, it still contains more information than any of them. Because in effect it describes not only "spatial" causal relationships between events happening in different places in the underlying string, but also "branchial" causal relationships between events happening on different branches of the multiway system.

And just like for other graphs, we can study the large-scale structure of multiway causal graphs. We can define a quantity $C_t^M$ which is the multiway analog of the cone volume $C_t$ for individual causal graphs. For the rule shown here, the various graph growth rates (as computed with our standard foliation) have the forms:

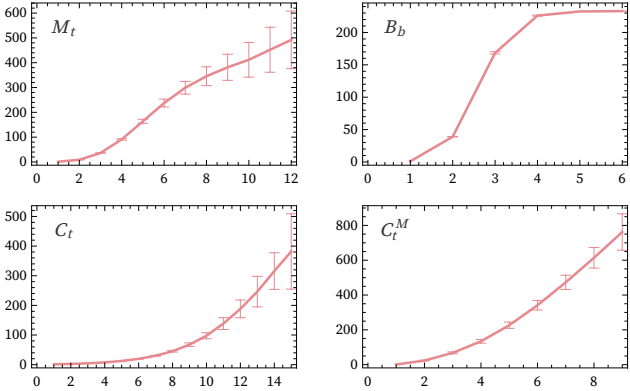

As another example, consider the "sorting" rule

{BA → AB}

starting from BABABABA. The combined multiway evolution and causal graph has the (terminating) form:

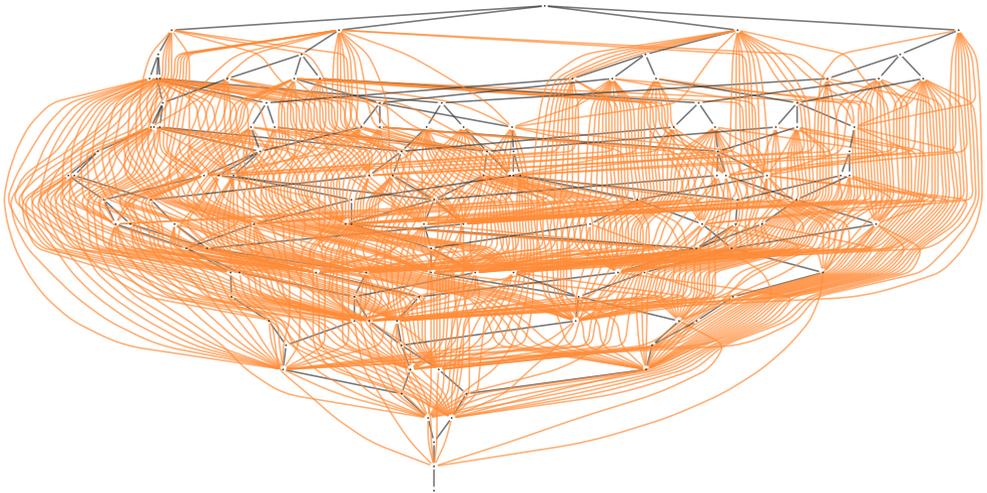

The multiway evolution and causal graphs on their own are:

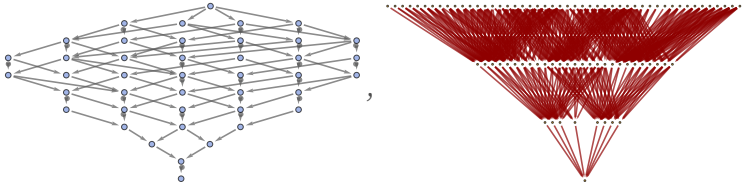

The branchial graphs are

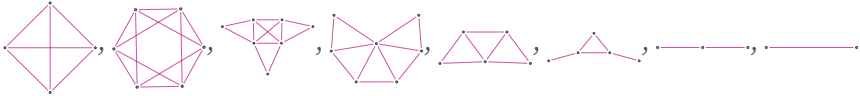

and the causal graph for a single (finite) path of evolution is:

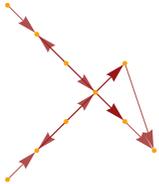

Earlier in this section we looked at the multiway evolution graphs generated by all 12 inequivalent 2: 1→2, 1→1 rules. The pictures below now compare these with the multiway causal graphs for the same rules (starting from all possible length-3 strings of As and Bs, and run for 4 steps of our standard multiway foliation):

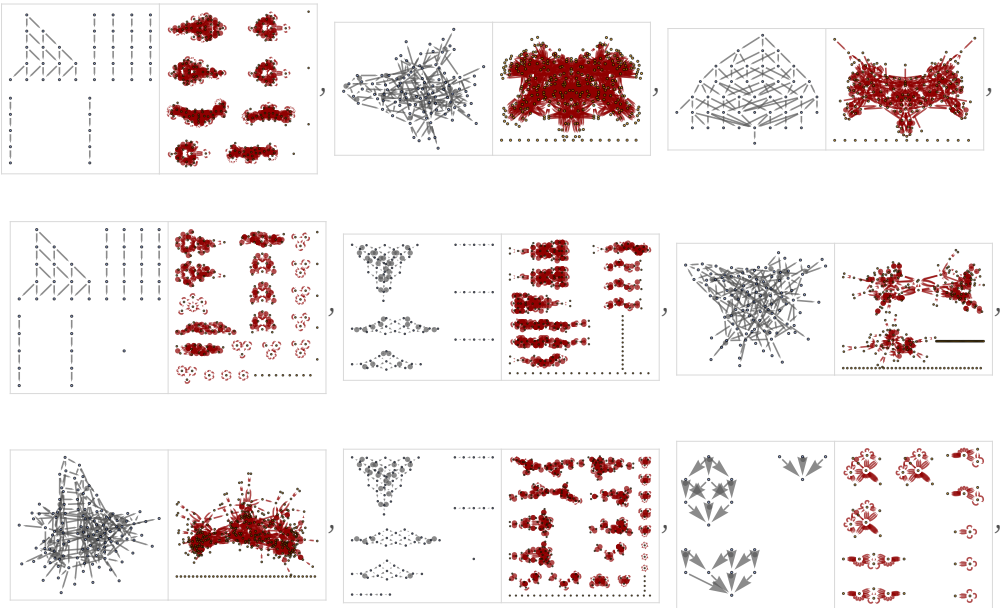

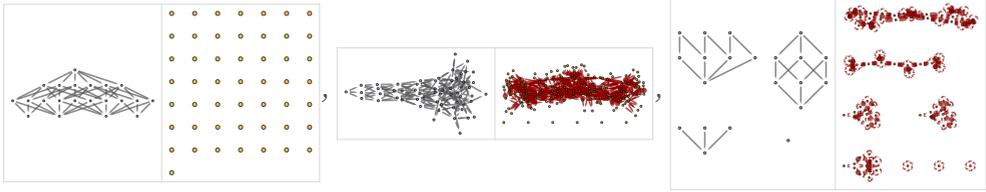

The multiway causal graph is in some ways a kind of dual to the multiway evolution graph—resulting in many similarities among the graphs in the pictures above. Like $M_t$ for multiway evolution graphs, $C_t^M$ for multiway causal graphs typically seems to grow either polynomially or exponentially.

But even in a case like the rule

{AA → AAA}

where the causal graph for a single evolution has the fairly regular form

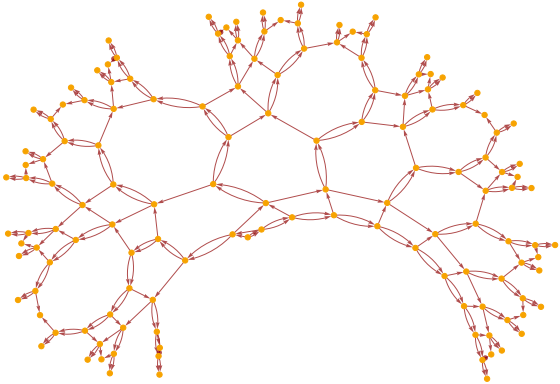

the full multiway causal graph is quite complex. This shows how it builds up over the first few steps (in our standard multiway foliation):

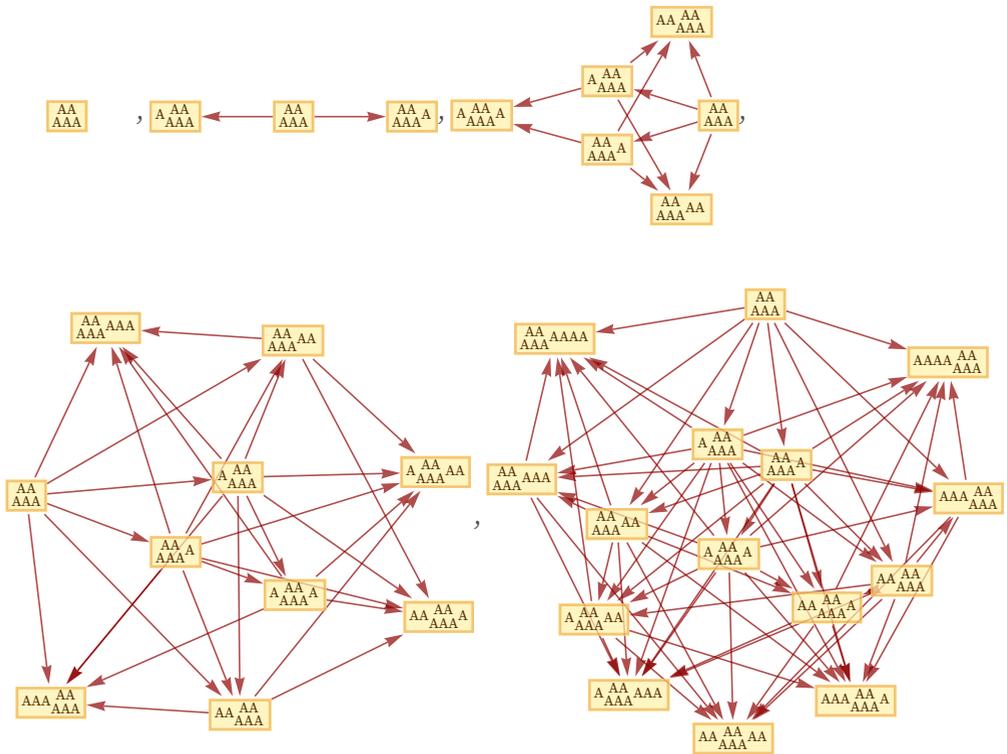

And here are 3D renderings after 8 and 9 steps:

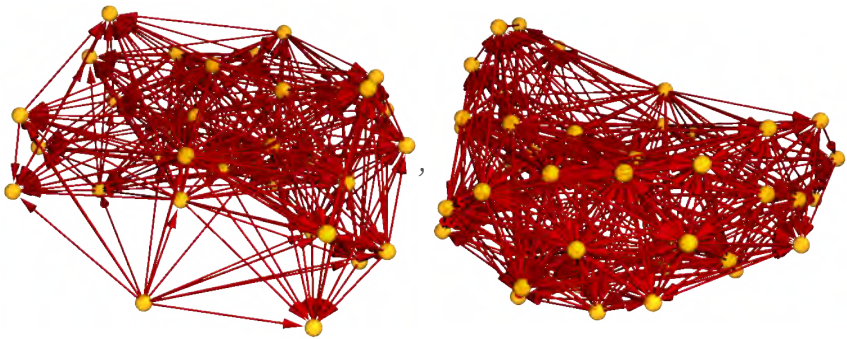

## 5.19 Weighted Multiway Graphs

In a multiway system, there are in general multiple paths that can produce the same state. But in our usual construction of the multiway graph, we record only what states are produced, not how many paths can do it.

Consider the rule

{A → AA, A → A}

**227**

The full form of its multiway graph—including an edge for every possible event—is:

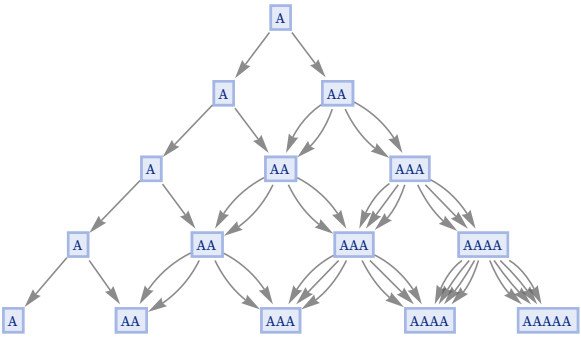

Here is the same graph, with a count included at each node for the number of distinct paths from the root that reach it:

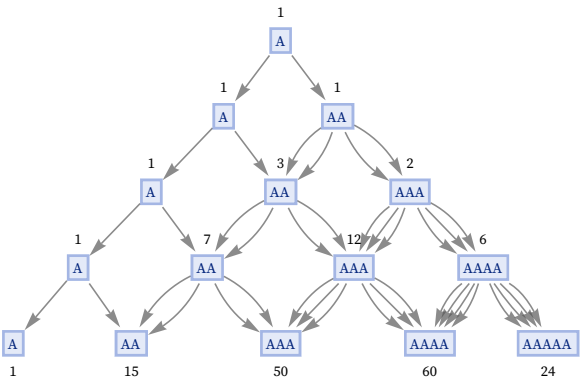

An alternative weighting scheme might be to start with weight 1 for the initial state, then at each state we reach, to distribute the weight to its successors, dividing it equally among possible events:

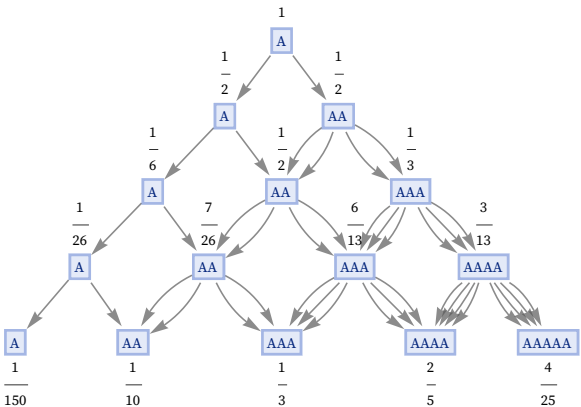

This approach has the feature that it gives normalized weights (summing to 1) at each successive layer in a graph like this. But in general the approach is not robust, and if we even took a different foliation through the graph above, the weights on each slice would no longer be normalized. In addition, if we were to combine identical states from different steps, we would not know what weights to assign. Pure counting of paths, however, still works even in this case, although any normalization has to be done only after all the counts are known:

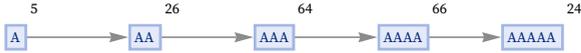

Note that even the counting of paths becomes difficult to define if there is a loop in the multiway graph—though one can adopt the convention that one counts paths only to the first encounter with any given state:

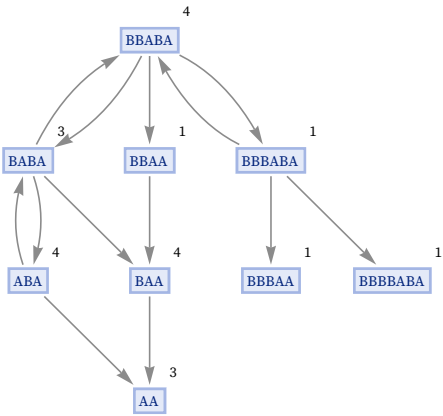

Weights on the multiway graph can also be inherited by branchial graphs. Consider for example the rule

{A → AB, B → BA}

The multiway graph for this rule, weighted with path counts, is:

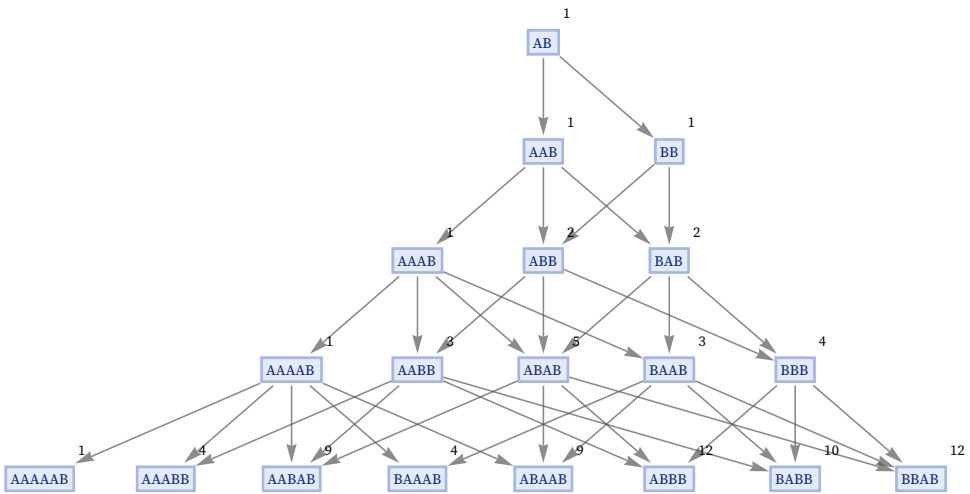

The corresponding weighted branchial graphs are:

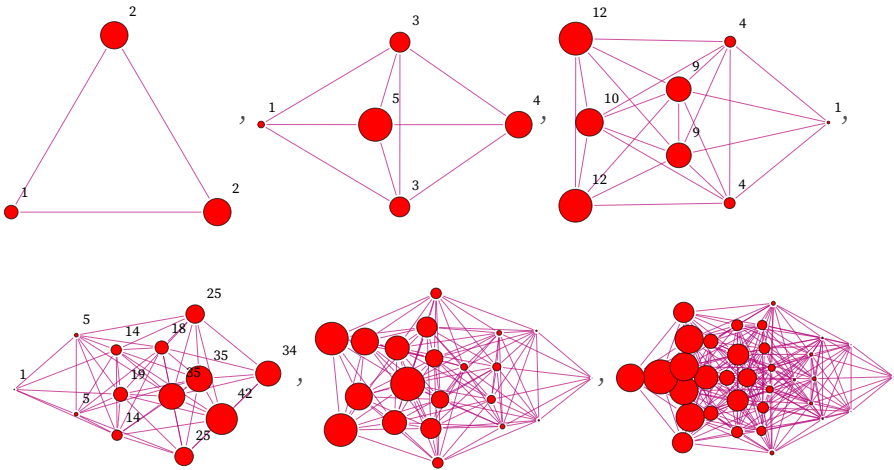

The weights in effect define a measure on the branchial graph. A case with a particular straightforward limiting measure is the rule:

{A → AB}

With initial condition A this gives weights that reproduce Pascal's triangle, and yield a limiting Gaussian:

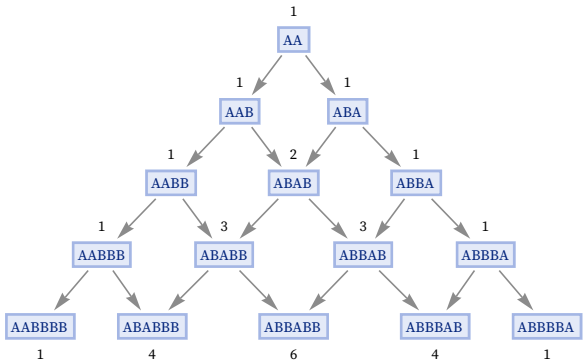

With initial condition AAA, the weights in the branchial graph limit to a 2D Gaussian:

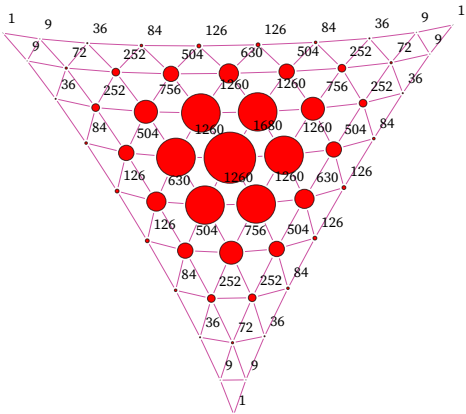

In general, after sufficiently many steps one can expect that the weights will define an invariant measure, although a complexity is that the branchial graph will typically continue to grow. As one indication of the limiting measure, one can compute the distribution of values of the weights.

The results for the rule {A→B,B→AB} above illustrate slow convergence to a limiting form:

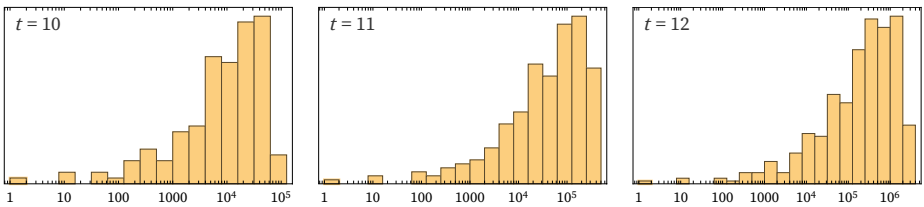

We discussed in a previous subsection probing the structure of the branchial graph by computing the number of nodes $B_b$ at most graph distance $b$ from a given point. We can now generalize this to computing a path-weighted quantity $B_b^\mu$ (cf. [1:p959]). At least for simple multiway graphs, this may be related in the limit to the results of solving a PDE on the multiway graph.

## 5.20 Effective Causal Invariance

In any causal invariant rule, all branch pairs that are generated must eventually resolve. But by looking at how many branch pairs are resolved—and unresolved—at each step, we can get a sense of "how far" a rule is from causal invariance.

Consider the causal invariant rule:

{A → AB, B → A}

With this rule, all branch pairs resolve in one step

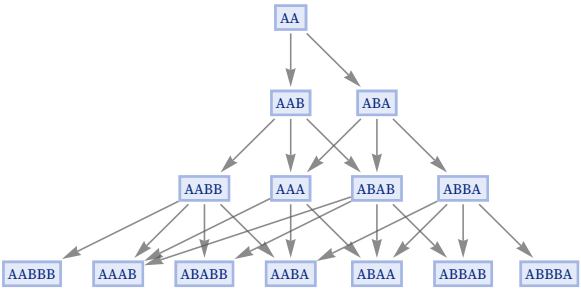

and the total number of branch pairs that have already resolved on successive steps is (roughly $2^t$):

{0, 6, 22, 66, 174, 420, 951, 2053, 4273, 8643}

But now consider the slightly different—and non-causal-invariant—rule:

{A → AB, BA → BB}

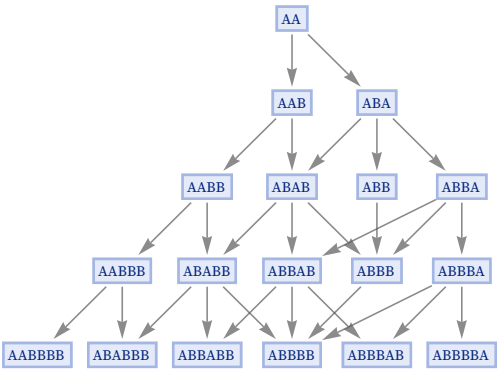

The number of resolved branch pairs goes up on successive steps—in this case quadratically:

{1, 5, 11, 19, 29, 41, 55, 71, 89, 109}

But now there is a "residue" of new unresolved branch pairs at each step that reflect the lack of causal invariance

{4, 6, 8, 10, 12, 14, 16, 18, 20, 22}

There are other rules in which the deviation from causal invariance is in a sense larger. Consider the rule:

{AA → AAB, AA → B}

The multiway graph for this rule shows various "dead ends"

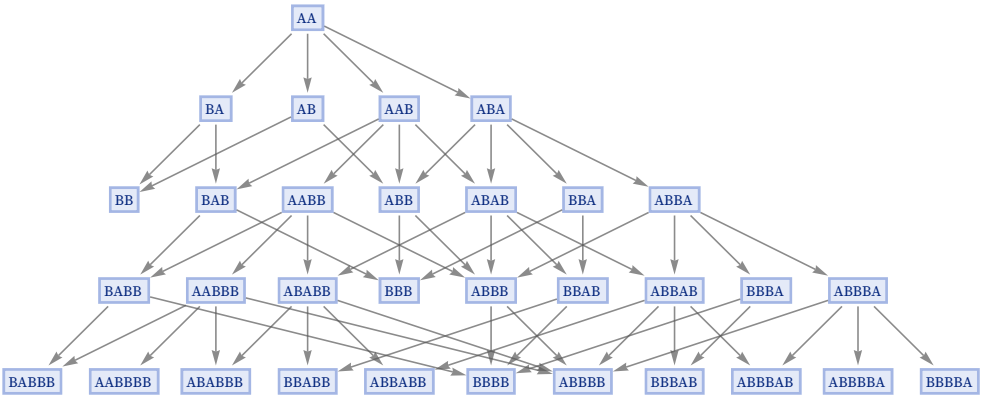

and now the number of resolved branch pairs is

{5, 15, 30, 50, 75, 105, 140, 180, 225, 275}

while the number of unresolved ones grows at a similar rate:

{14, 23, 33, 44, 56, 69, 83, 98, 114, 131}

When a system is not causal invariant, what it means is that in a sense the system can reach states from which it cannot ever "get back" to other states. But this suggests that by extending the rules for the system, one might be able to make it causal invariant.

Consider the rule

{A → AA, A → B}

with multiway graph:

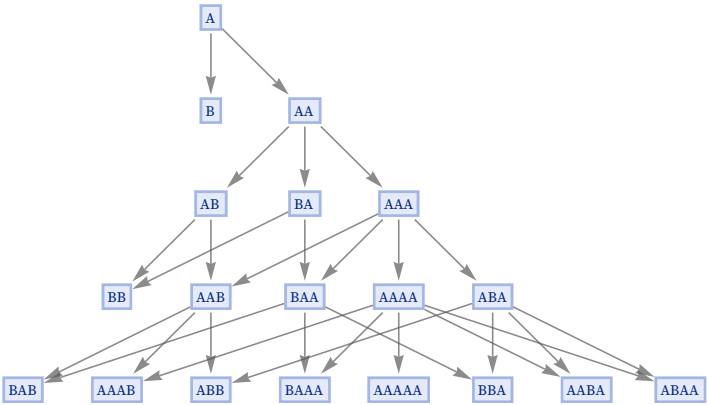

With this rule, the branch pair {B, AA} never resolves. But now let us just extend the rule by adding B→AA:

{A → AA, A → B, B → AA}

The multiway graph becomes

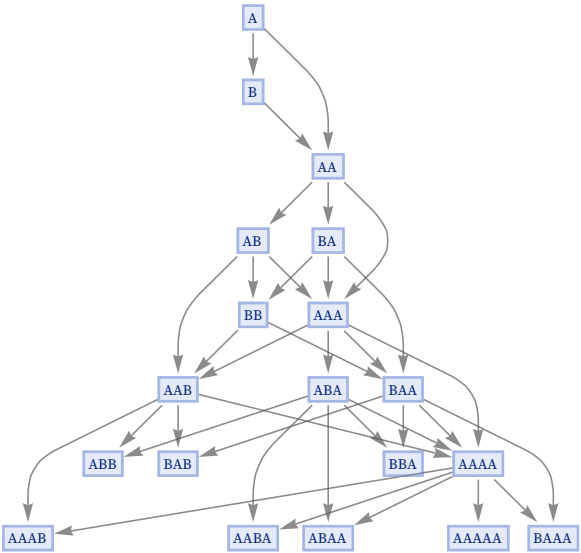

and the rule is now causal invariant.

In general, given a rule that is not causal invariant, we can consider extending it by including transformations that relate strings in branch pairs that do not resolve. For the rule

{AA → AAB, AA → B}

discussed above it turns out that the minimal additions to achieve causal invariance are:

{AB → AAAB, BA → AB}

Having added these transformations, the multiway graph now begins:

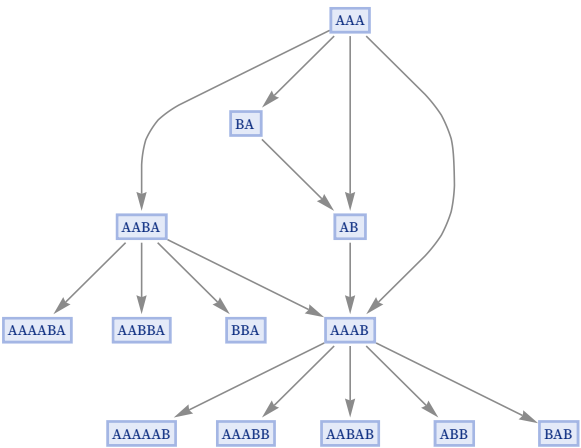

After more steps, the multiway graph is:

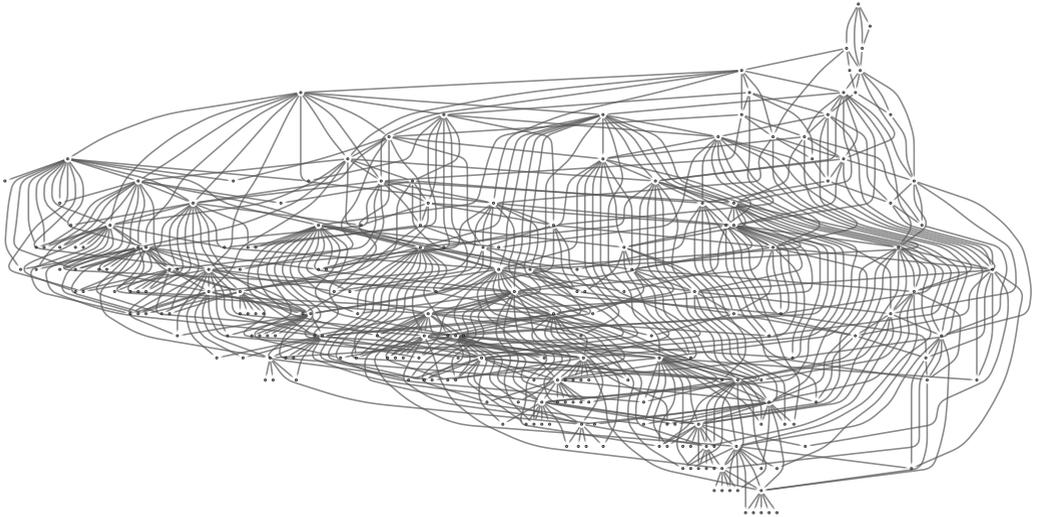

This kind of "completion" procedure of adding "relations" in order to achieve what amounts to causal invariance is familiar in automated theorem proving [82] and related areas [60]. For our purposes we can think of it as a kind of "coarse graining" of our systems, in which the additional rules in effect define equivalences between states that would otherwise be distinct.

If a particular multiway graph terminates after a finite number of steps, then it is always possible to add enough completion rules to the system to ensure causal invariance [63][64]. But if the multiway graph grows without bound, this may not be possible. Sometimes one may succeed in adding enough completions to achieve causal invariance for a certain number of steps, only to have it fail after more steps. And in general, like determining whether branch pairs will resolve, there is ultimately no upper bound on how long one may have to wait, making the problem of completion ultimately formally undecidable [83].

But if (as is often the case) one only has to add a small number of completion rules to make a system causal invariant, then one can take this to mean that the system is not far from causal invariance, so that it is likely that many of the large-scale properties of the completion of the system will be shared by the system itself.

## 5.21 Generational Evolution

Systems like cellular automata always update every element at every step in their evolution. But in string substitution systems (as well as in our hypergraph-based models), the presence of overlaps between possible updating events typically means that there is no single, consistent way to do this kind of parallel updating. Nevertheless, in studying our models in earlier sections, we often used "steps" of evolution in which we updated as many elements as we consistently could. And we can also apply this kind of "generation-based" updating to string substitution systems.

Consider the rule:

{A → AB, B → A}

We can construct the multiway graph for this rule by considering how one state is produced from another by a single updating event, corresponding to a single application of one of the transformations in the rule:

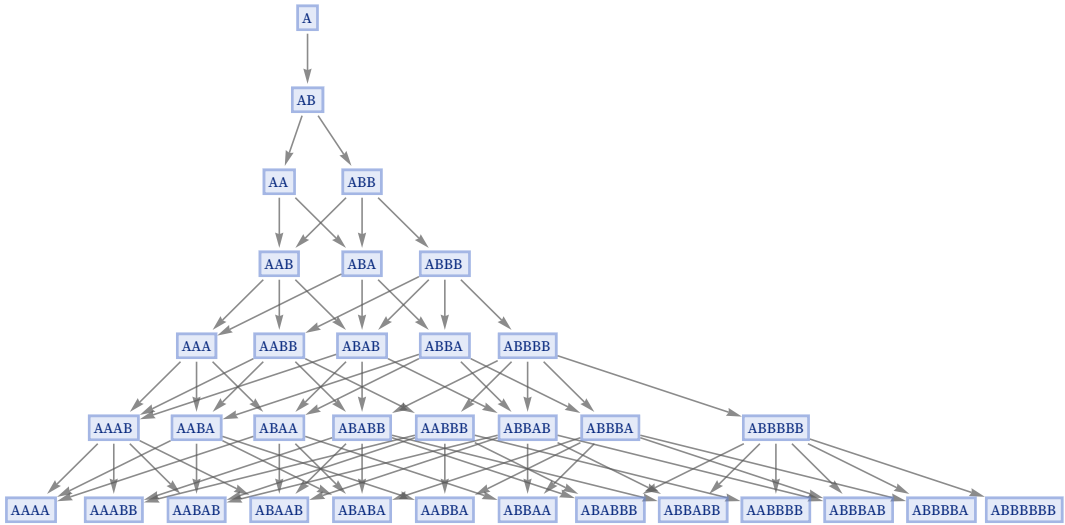

But we can also consider producing a "generational multiway graph" in which we do as many non-overlapping updates as possible on any given string. For this particular rule, doing this is straightforward, since every A and every B in the string can be transformed separately.

But the result is now a radically simplified multiway graph, in which there is just a single path of evolution:

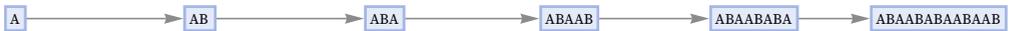

The "generational steps" here involve an increasing number of update events, as we can see from this rendering of the evolution:

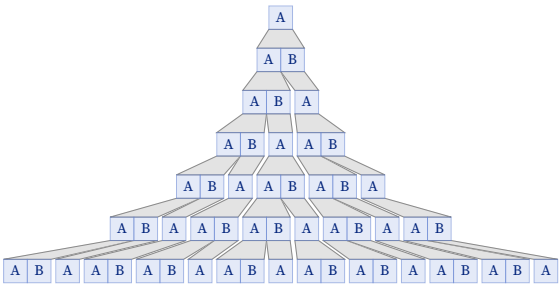

All the states obtained at generational steps do appear somewhere in the full multiway graph, but the full graph also contains many additional states—that, among other things, can be thought of as representing all possible intermediate stages of generational steps:

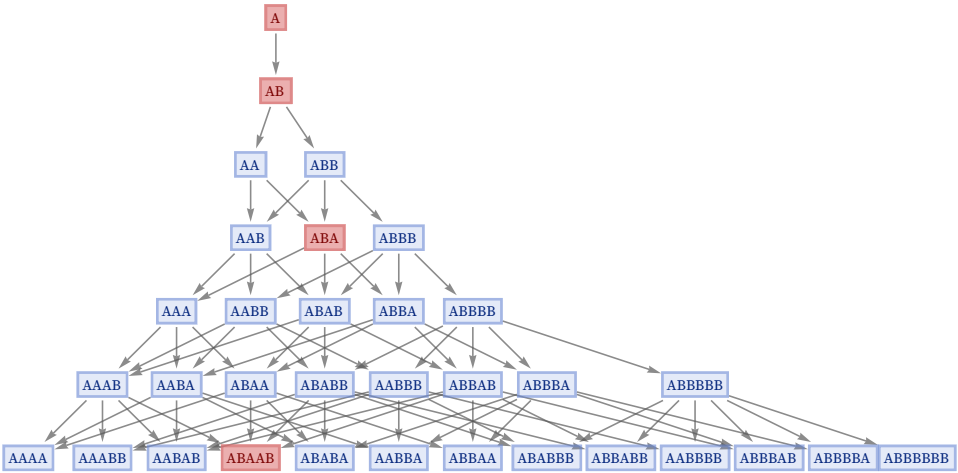

For a rule like

{A → AB}

the generational steps

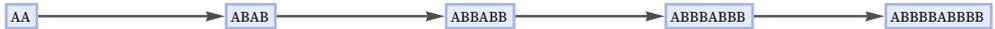

correspond to a particularly simple trajectory through the full multiway graph:

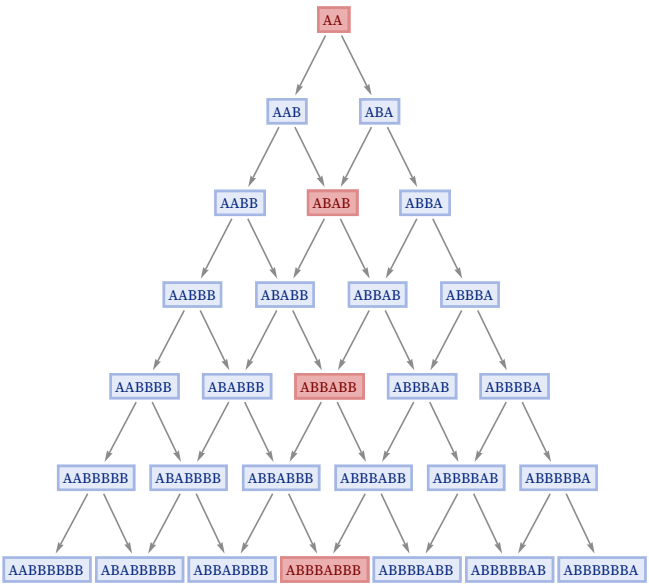

It is not always the case that the generational multiway graph involves just a single path. Consider the rule:

{A → AB, A → B}

The ordinary multiway graph for this rule starting from AA is:

And the generational multiway graph is now:

The generational multiway graph is always in a sense a compression of the full multiway graph. And one way to think of it is as being derived from the full multiway graph by combining sequences of edges when they correspond to updating events that do not overlap on the string.

But there is also another view of generational evolution. Consider a branchial graph from the full multiway graph above (this branchial graph is derived from the layered foliation shown):

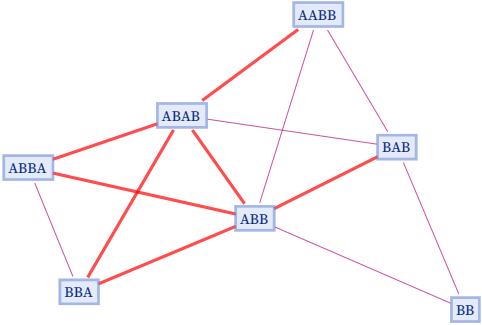

The branch pairs in this branchial graph (shown as adjacent nodes) can be thought of as being of two kinds. The first are produced by applying different rules to a single part of a string (e.g. ABA→{ABBA,BBA}). And the second (highlighted in the graph above) by applying rules to different parts of a string (e.g. ABA→{ABBA,ABAB}).

In the full multiway graph, no distinction is made between these two kinds of branch pairs, and the graph includes both of them. But in a generational multiway system, strings in the second kind of branch pairs can be combined.

And indeed this provides another way to construct a generational multiway system: look at branchial graphs and take pairs of strings corresponding to "spatially" disjoint updating events, and then knit these together to form generational steps. And if there is only one way to do this for each branchial graph, one will get a single path of generational evolution. But if there are multiple ways, then the generational multiway graph will be more complicated.

For the rule

{A → AB, B → A}

that we discussed above, the sequence of branchial graphs is

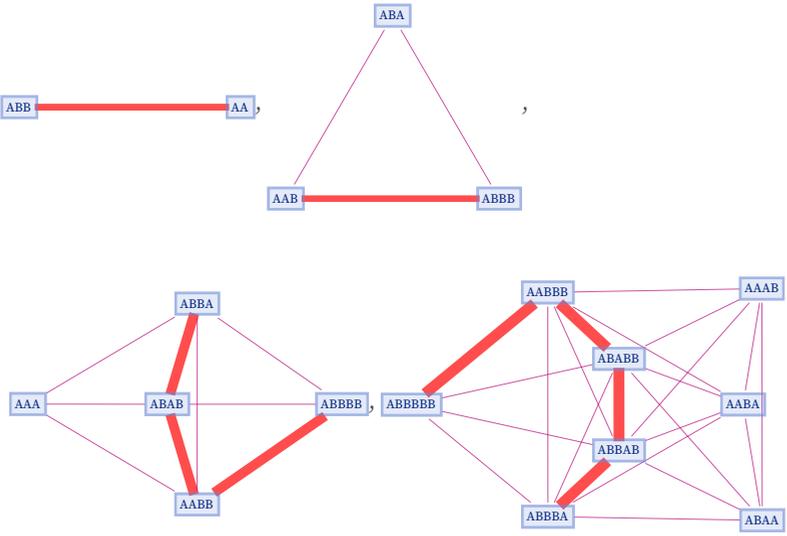

and it is readily possible to assemble "string fragments" (such as those highlighted) to produce the states at successive generational steps:

{{AA}, {ABAB}, {ABBABB}, {ABBBABBB}}

The branchial graphs are determined by the foliation of the multiway graph that one uses. But given a foliation, one can then assemble strings corresponding to generational steps using the procedure above.

There is always an alternative, however: instead of combining strings from a branchial graph to produce the state for a generational step, one can always in a sense just wait, and eventually the full multiway system will have done the necessary sequence of updates to produce the complete state for the generational step.

Whenever there are no possible overlaps in the application of rules, the generational multiway graph must always yield a single path of history. But there is also another feature of such rules: they are guaranteed to be causal invariant. There are, however, plenty of rules that are causal invariant even though they allow overlaps—in a sense because their right-hand sides also appropriately agree. And for such rules, the generational multiway graph may have multiple paths of history.

A simple example is the rule:

{A → AB, A → BA}

This rule is causal invariant, and starting from A yields the full multiway graph:

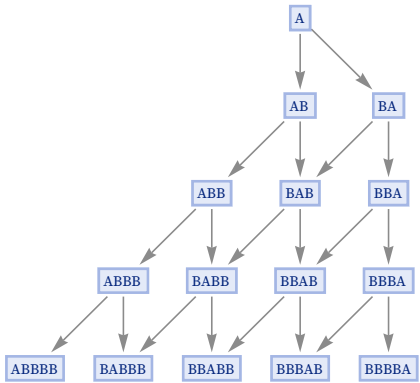

Its generational multiway graph is actually identical in this case—because all the rule ever does is to apply one transformation or the other to the single A that appears:

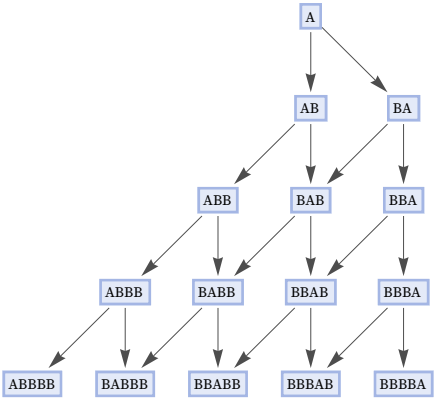

Starting from AA, however, the rule yields the full multiway graph

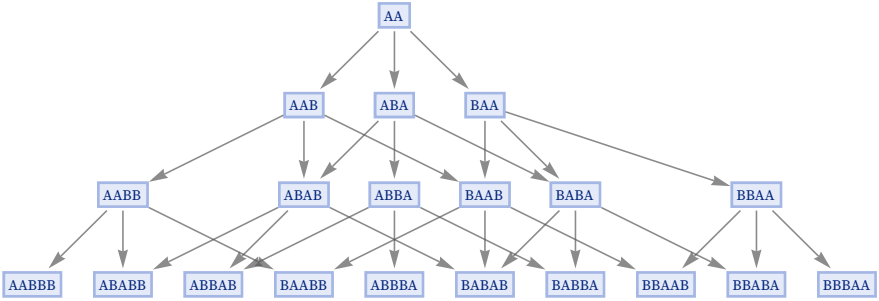

and the generational multiway graph

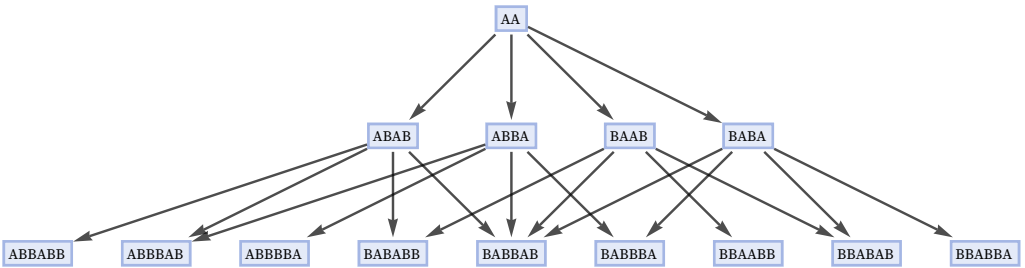

In an alternative rendering, these graphs after a few more steps become, respectively:

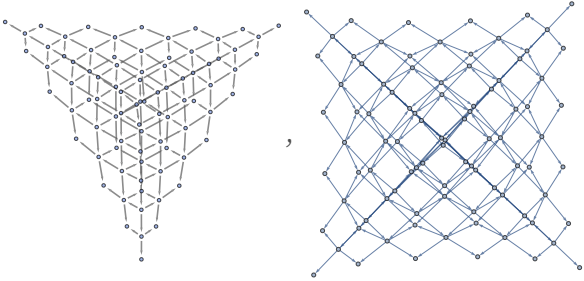

Note that in both cases, the number of states reached after $t$ steps grows like $t^2$ (in the first case it is $\frac{1}{2} t (t + 1)$; in the second case exactly $t^2$).

In the case of a rule like

{A → AB, A → BB}

the presence of many potential overlaps in where updates can be applied makes many of the possible states in the full multiway graph also appear in the generational multiway graph (in the limit about 64% of all possible states are generational results):

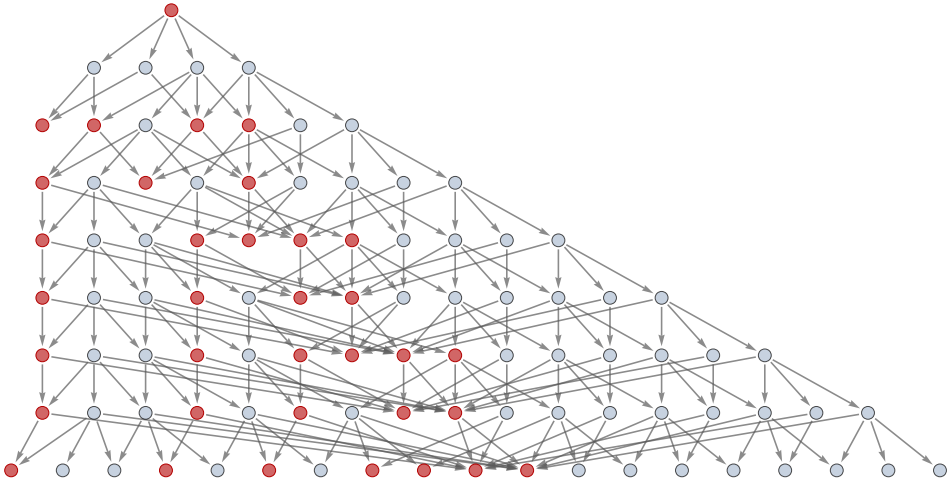

Generational multiway graphs share many features with full multiway graphs. For example, generational multiway graphs can also show causal invariance—and indeed unless strings grow too fast, any deviation from causal invariance must also appear in the generational multiway graph.

The basic construction of ordinary multiway graphs ensures that the number of states $M_t$ after $t$ steps can grow at most exponentially with $t$. In a generational multiway graph, there can be faster growth.

Consider for example the rule (whose full multiway graph grows in a Fibonacci sequence $\approx \phi^t$):

{A → AA, A → B}

Its full multiway graph grows in a Fibonacci sequence $\approx \phi^t$:

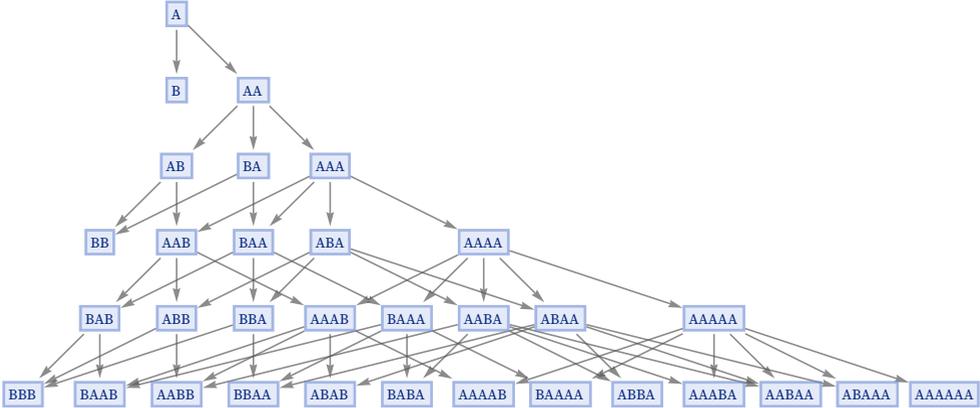

But its generational multiway graph grows much faster. After 3 generational steps it has the form

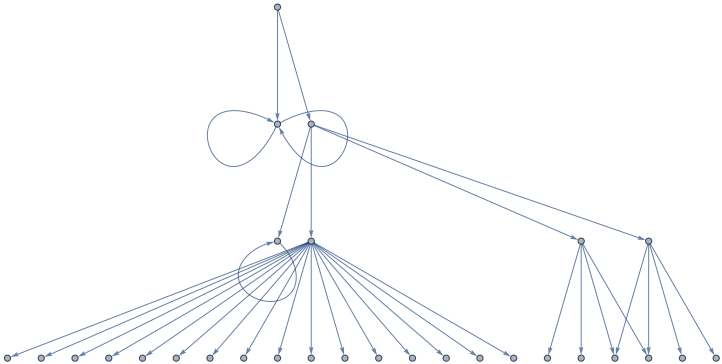

while after 4 steps (in a different rendering) it is:

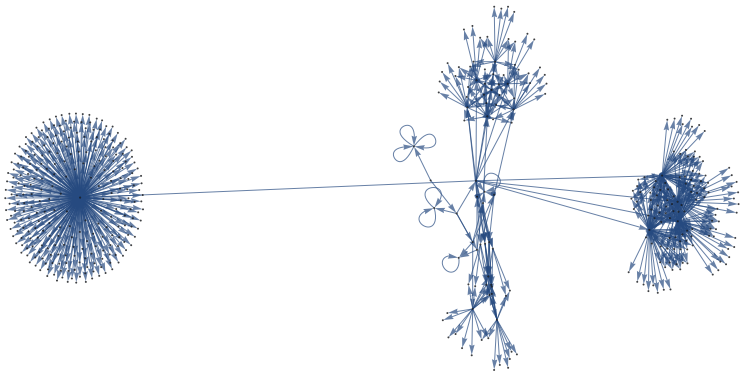

The number of states reached in successive steps is:

{1, 2, 5, 24, 455, 128 702}

Although there is a distribution of lengths for the strings, say, at steps 4 and 5

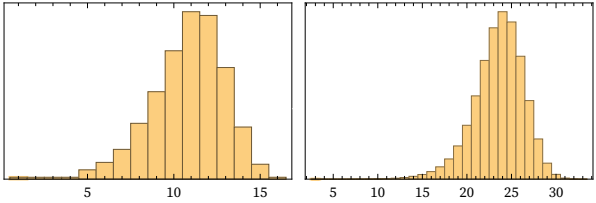

the fact that the maximum string length at generational step $t$ is $2^{t-1}$—combined with the lack of causal invariance for this rule—-allows for double exponential growth in the number of possible states with generational steps. In fact, with this particular rule, by step $t$ almost all sequences of up to $2^{t-1}$ Bs and AAs have appeared (the missing fractions on steps 3, 4, 5 are 0.13, 0.078, 0.017) so at step $t$ the total number of states approaches $2^{2^{t-1}}$

# 6 | The Updating Process in Our Models

## 6.1 Updating Events and Causal Dependence

Consider the rule:

{{*x*, *y*}, {*x*, *z*}} → {{*x*, *y*}, {*x*, *w*}, {*y*, *w*}, {*z*, *w*}}

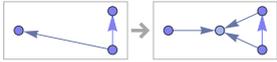

When we discussed this rule previously, we showed the first few steps in its evolution as:

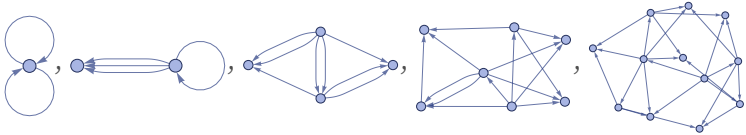

But to understand the updating process in our models in more detail, it is helpful to "look inside" these steps, and see the individual updating events of which they are comprised:

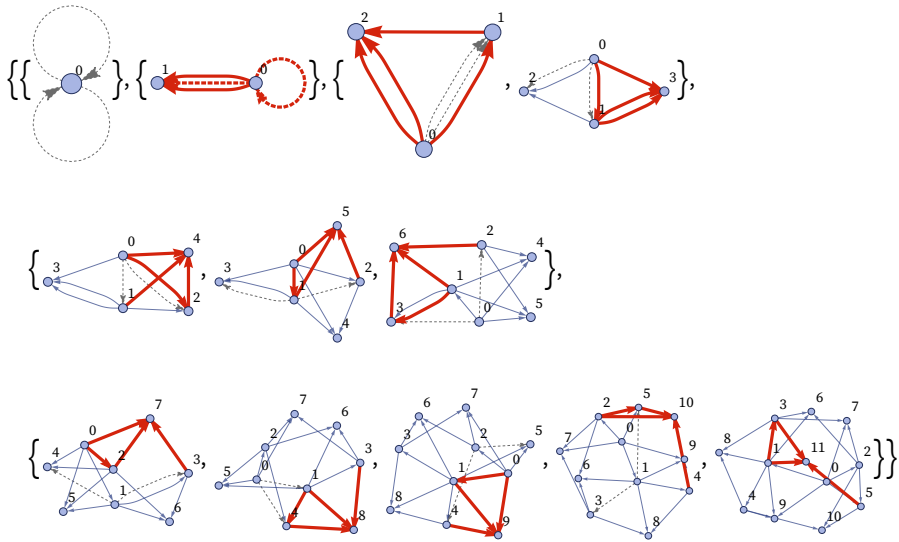

The $2_2 \to 4_2$ signature of the rule means that in each updating event, two relations are destroyed, and four new ones are created. In the pictures above, new relations from each event are shown in red; the ones that will disappear in the next event are shown dotted. The elements are numbers in the sequence they are created.

There are in general many possible sequences of updating events that are consistent with the rule. But in making the pictures above (and in much of our discussion in previous sections), we have used our "standard updating order", in which each step in the overall evolution in effect includes as many non-overlapping updates as will "fit". (In more detail, what is done is that in each overall step, relations are scanned from oldest to newest, in each case using them in an update event so long as this can be done without using any relation that has already been updated in this overall step.)

Our models—and the hypergraphs on which they operate—are in many ways more difficult to handle than the string-based systems we discussed in the previous section. But one way in which they are simpler is that they more directly expose causal relationships between events. To see if an event B depends on an event A, all we need do is to see whether ele-ments that were involved in A are also involved in B.

Looking at the sequence of updates above, therefore, we can immediately construct a causal graph:

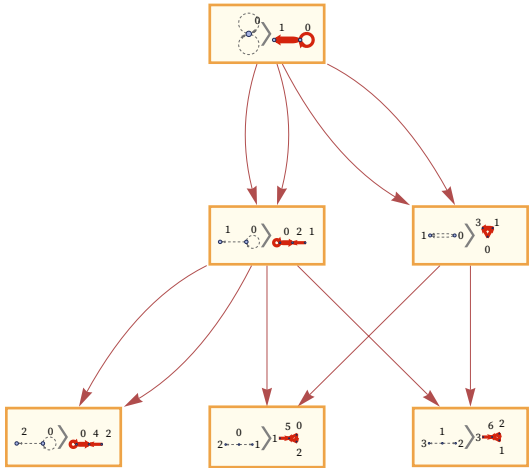

Another feature of our models is that every element can be thought of as having a unique "lineage", in that it was created by a particular updating event, which in turn was the result of some other updating event, and so on. When we introduced our models in section 2, we just said that any element created by applying a rule should be new and distinct from all others. If we were implementing the model, this might then make us imagine that the element would have a name based on some global counter, or a UUID.

But there is another, more deterministic (as well as more local and distributed) alternative: think of each new element as being a kind of encapsulation of its lineage (analogous to a chain of pointers, or to a hash like in blockchains [84] or Git). In the evolution above, for example, we could describe element 10 by just saying it was created as part of the relation {2,10} from the relations {{2,4},{2,5}} (as the second part of the output of an update that uses them)—but then we could say that these relations were in turn created from earlier rela-tions, and so on recursively, all the way back to the initial state of the system:

| |
|---|
| {{2, 10}} |
| {{{2, 4}, {2, 5}} ▷ 2} |
| {{{{0, 1}, {0, 2}} ▷ 4, {{0, 2}, {0, 1}} ▷ 3} ▷ 2} |
| {{{{{0, 0}, {0, 1}} ▷ 1, {{0, 0}, {0, 1}} ▷ 2} ▷ 4, {{{0, 0}, {0, 1}} ▷ 3, {{0, 1}, {0, 1}} ▷ 1} ▷ 3} ▷ 2} |
| {{{{{{0, 0}, {0, 0}} ▷ 1, {{0, 0}, {0, 0}} ▷ 2} ▷ 1, {{{0, 0}, {0, 0}} ▷ 1, {{0, 0}, {0, 0}} ▷ 2} ▷ 2} ▷ 4, {{{{0, 0}, {0, 0}} ▷ 1, {{0, 0}, {0, 0}} ▷ 2} ▷ 3, {{{0, 0}, {0, 0}} ▷ 3, {{0, 0}, {0, 0}} ▷ 4} ▷ 1} ▷ 3} ▷ 2} |

The final expression here can also be written as:

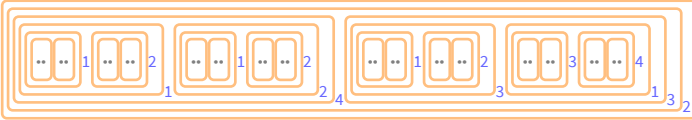

Roughly what this is doing is specifying an element (in this case the one we originally labeled simply as 10) by giving a symbolic representation of the path in the causal graph that led to its creation. And we can then use this to create a unique symbolic name for the element. But while this may be structurally interesting, when it comes to actually using an element as a node in a hypergraph, the name we choose to use for the element is irrelevant; all that matters is what elements are the same, and what are different.

## 6.2  Multiway Systems for Our Models

Just like for the string substitution systems of section 5, we can construct multiway systems [1:5.6]  for our models, in which we include a separate path for every possible updating event that can occur:

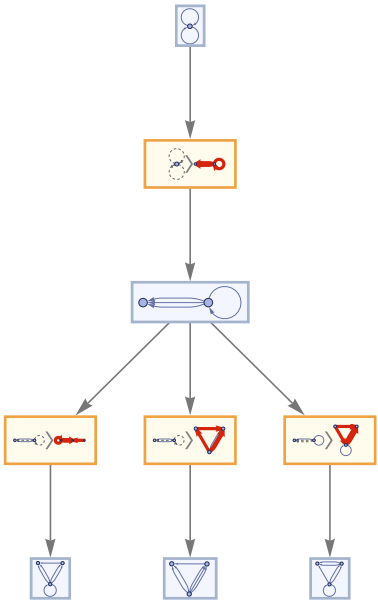

For string systems, it is straightforward to determine when states in the system should be merged: one just has see whether the strings corresponding to them are identical. For our systems, it is more complicated: we have to determine whether the hypergraphs associated with states are isomorphic [85], in the sense that they are structurally the same, independent of how their nodes might be labeled.

Continuing one more step with our rule, we see some cases of merging:

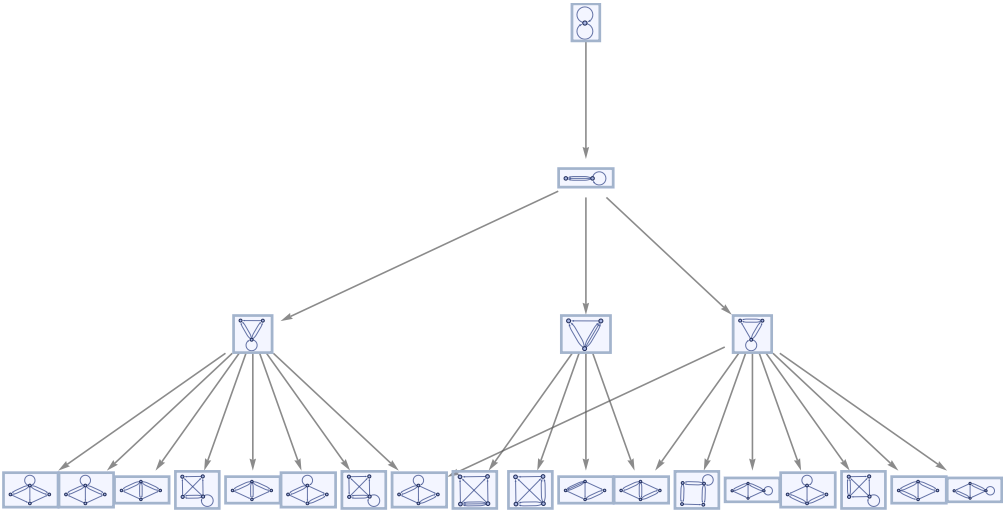

Here is an alternative rendering, now also showing the particular path obtained by following our "standard updating order":

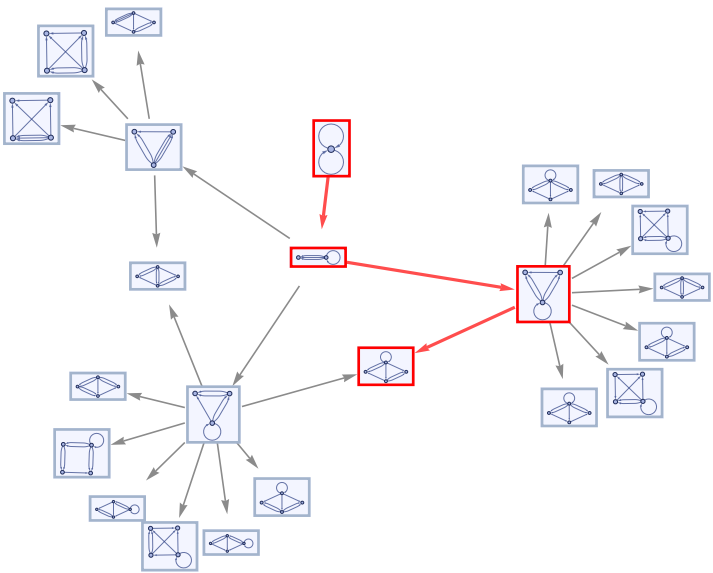

In general, each path in the multiway system corresponds to a possible sequence of updating events—here shown along with the causal relationships that exist between them:

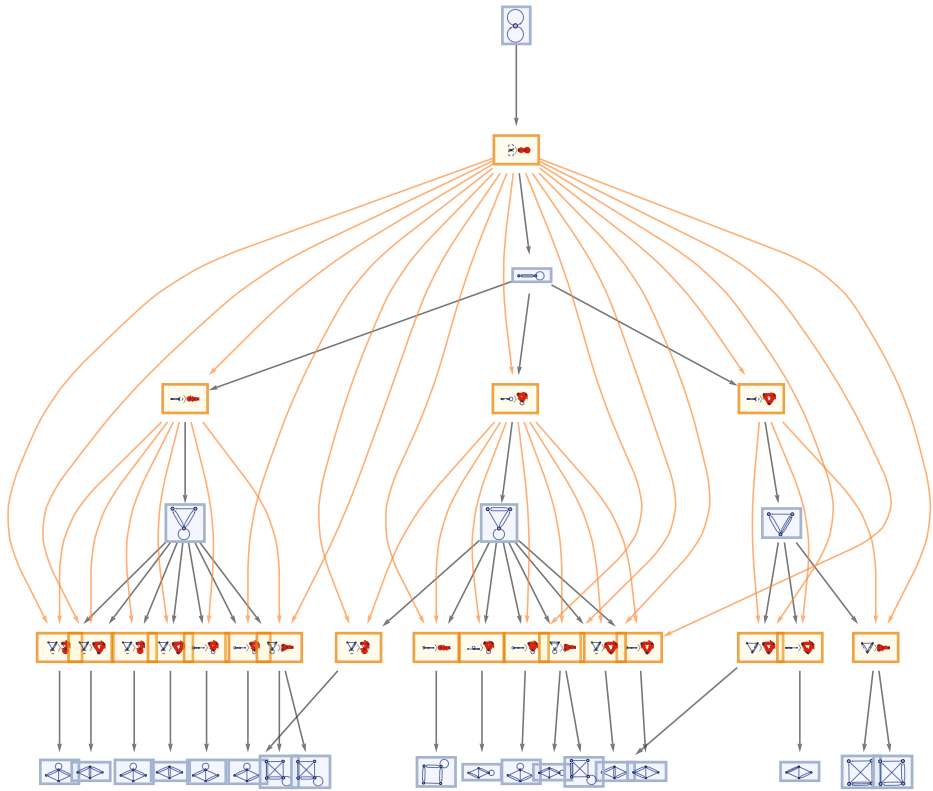

## 6.3 Causal Invariance

Like string substitution systems, our models can have the important feature of causal invariance [1:9.9]. In analogy with neighbor-independent string substitution systems, causal invariance is guaranteed if there is just a single relation on the left-hand side of a rule.

Consider for example the rule:

$\{\{x, y\}\} \to \{\{x, y\}, \{y, z\}\}$

Starting from a single self-loop, this gives the multiway system:

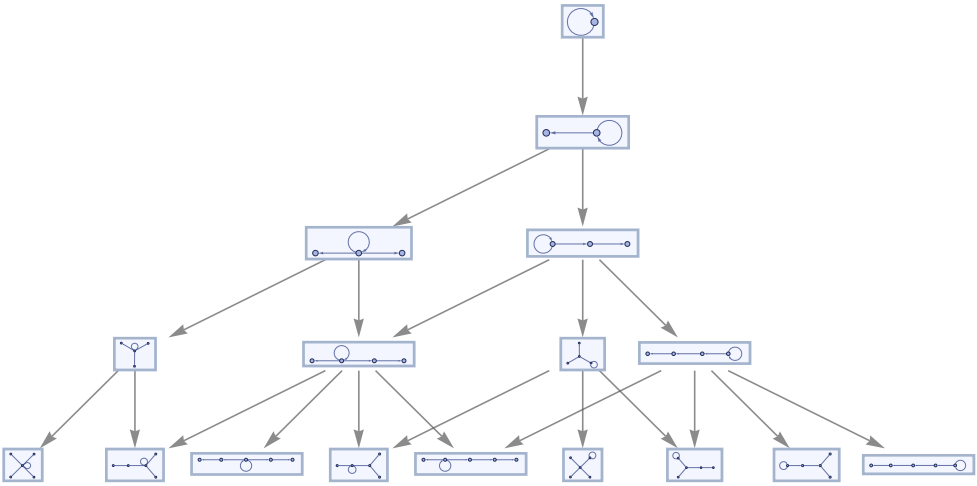

As implied by causal invariance, every pair of paths that diverge must reconverge. And looking at a few more steps, we can see that in fact with this particular rule, branches always recombine after just one step:

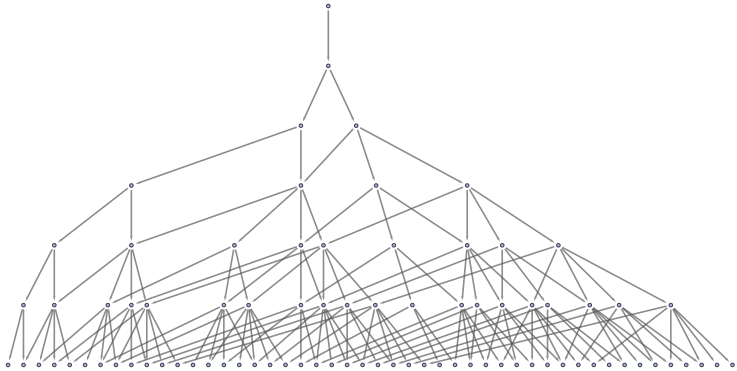

The different paths here lead to hypergraphs that look fairly different. But causal invariance implies that every time there is divergence, there must always eventually be reconvergence.

And for some rules, different paths give hypergraphs that do look very similar. An example is the rule

$\{\{x, y\}\} \to \{\{y, z\}, \{z, x\}\}$

where the hypergraphs produced on different paths differ only by the directions of their hyperedges:

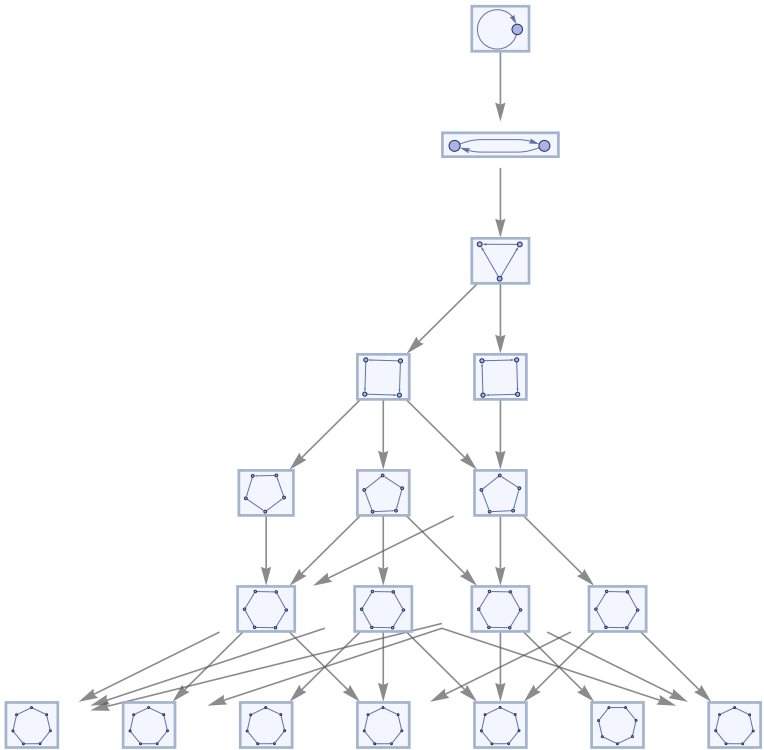

For rules that depend on more than one relation, causal invariance is not guaranteed, and in fact is fairly rare. Of the 4702 inequivalent $2_2 \to 3_2$ rules, perhaps 5% are causal invariant.

In some cases, the causal invariance is rather trivial. For example, the rule

$\{\{x, y\}, \{x, y\}\} \to \{\{z, z\}, \{z, z\}, \{y, z\}\}$

leads to a multiway graph that only allows one path of evolution:

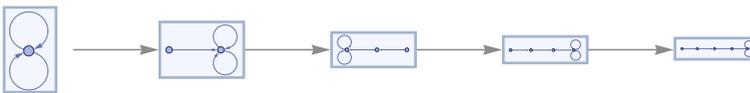

A less trivial example is the rule

$\{\{x, y\}, \{z, y\}\} \to \{\{x, w\}, \{y, w\}, \{z, w\}\}$

which yields the multiway system:

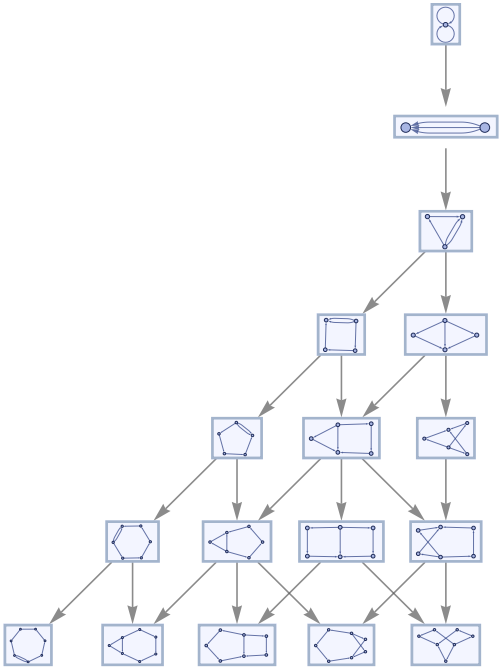

With our standard updating order, this rule eventually produces forms like

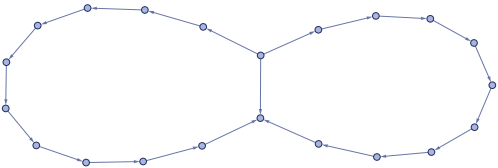

but the multiway system shows that other structures are also possible.

As another example, consider the rule

{{x, y}, {z, y}} → {{x, z}, {y, z}, {w, z}}

which with our standard updating order gives:

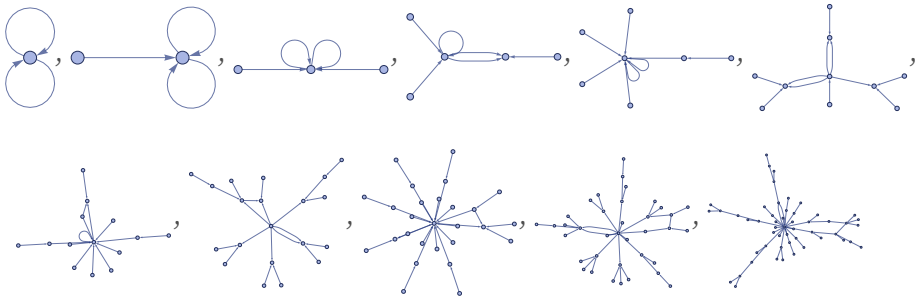

The multiway system for this rule branches rapidly

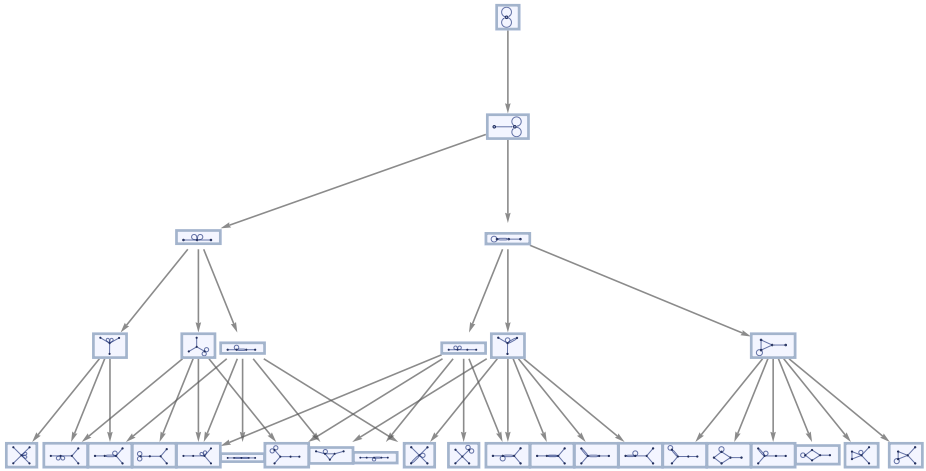

but every pair of branches still reconverges in one step.

The rule

{{x, y}, {x, z}} → {{y, w}, {y, z}, {w, x}}

provides an example of causal invariance in which branches can take 3 steps to converge:

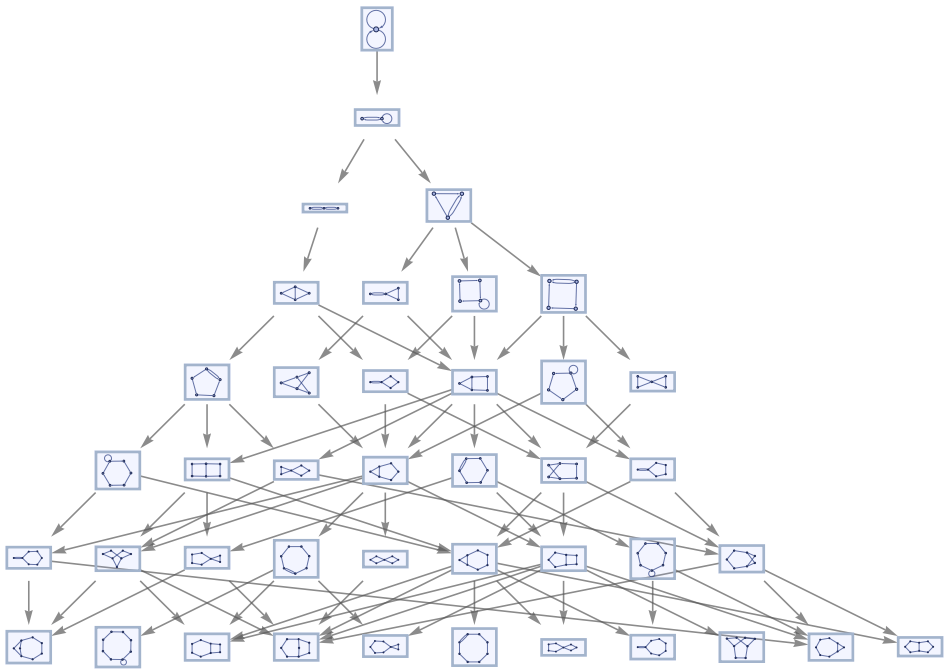

Among all possible rules, causal invariance is much more common for rules that generate disconnected hypergraphs. It is also perhaps slightly less common for rules with ternary relations instead of binary ones.

## 6.4 Testing for Causal Invariance

Testing for causal invariance in our models is similar in principle to the case of strings. Failure of causal invariance is again the result of branch pairs that do not resolve. And just like for strings, it is possible to test for total causal invariance by determining whether a certain finite set of core branch pairs resolve. (Again in analogy with strings, the finiteness of this set is a consequence of the finiteness of the hypergraphs involved in our rules.)

The core branch pairs that we need to test represent the minimal cases of overlap between left-hand sides of rules—or, in a sense, the minimal unifications of the hypergraphs that appear. For the two hypergraphs

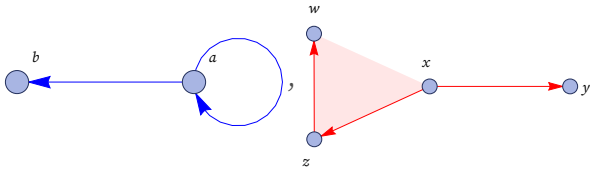

there are two possible unifications (where the purple edge shows the overlap):

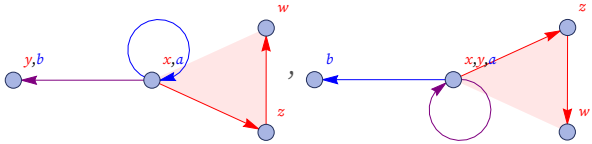

For a single rule with left-hand side

{{x, y}, {x, z}}

the core branch pairs arise from unifications associated with the possible self-overlaps of this small hypergraph. Representing two copies of the hypergraph as

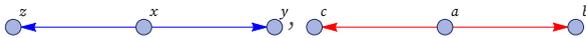

the possible unifications are:

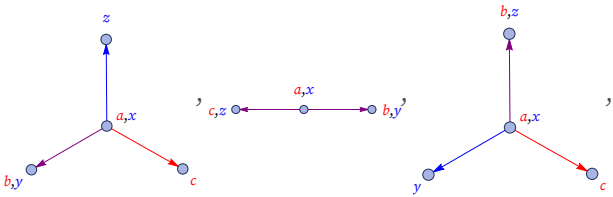

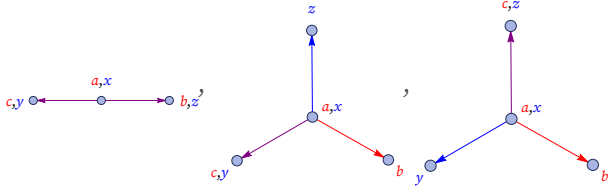

In the case of strings, all that matters is what symbols appear within the unification. In the case of hypergraphs, one also has to know how the unification in effect "attaches", and so one has to distinguish different labelings of the nodes.

Starting from the unifications, one applies the rule to find what branch pairs can be produced. These branch pairs form the core set of branch pairs for the rule—and determining whether the rule is causal invariant then becomes a matter of finding out whether these branch pairs resolve.

In the case of

$\{\{x, y\}, \{x, z\}\} \to \{\{x, y\}, \{x, w\}, \{y, w\}, \{z, w\}\}$

the application of the rule to the unifications above yields the following 58 core branch pairs:

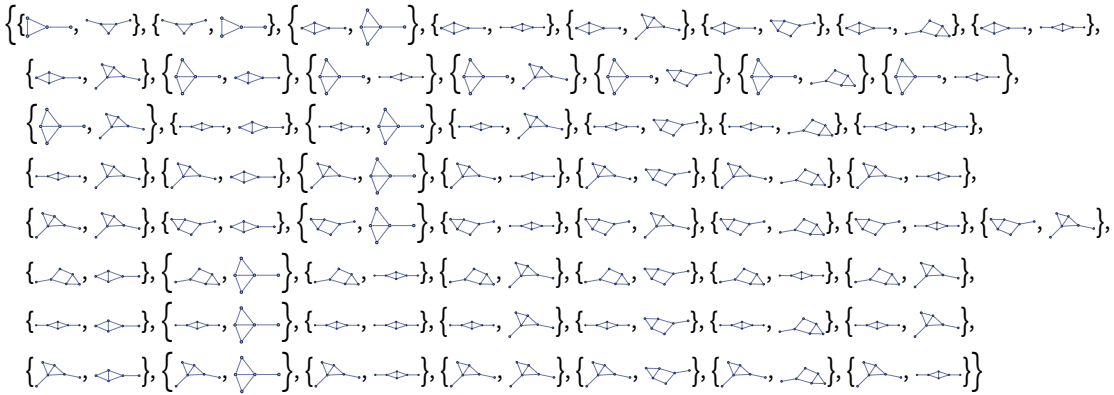

Running the rule for one step yields resolutions for 6 of these branch pairs:

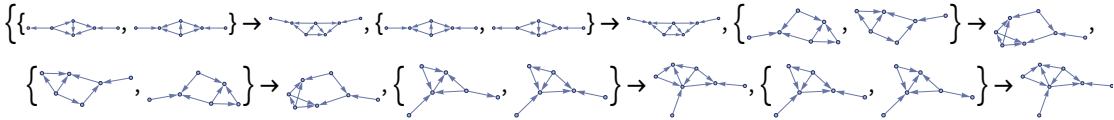

Running for another step resolves no additional branch pairs.

Will the rule turn out to be causal invariant in the end? As a comparison, consider the rule

$\{\{x, y\}, \{x, z\}\} \to \{\{y, w\}, \{y, z\}, \{w, x\}\}$

discussed in the previous subsection. This rule starts with 14 core branch pairs:

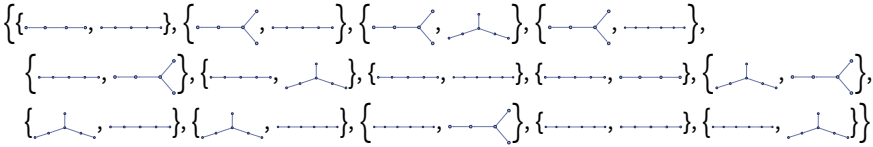

After one step, 6 of them resolve:

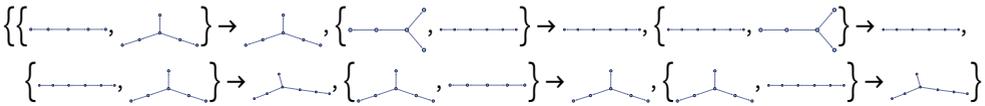

Then after another step the 8 remaining ones resolve, establishing that the rule is indeed causal invariant:

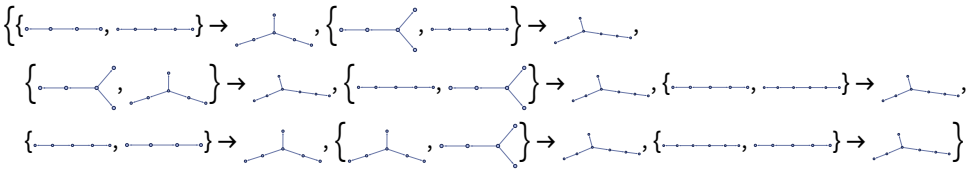

But in general there is no upper bound on the number of steps it can take for core branch pairs to resolve. Perhaps the fact that so many additional branch pairs are generated at each step in the rule $\{\{x,y\},\{x,z\}\} \to \{\{x,z\},\{x,w\},\{y,w\},\{z,w\}\}$ makes it seem unlikely that they will all resolve, but ultimately this is not clear.

And even if the rule does not show total causal invariance, it is still perfectly possible that it will be causal invariant for the particular set of states generated from a certain initial condition. However, determining this kind of partial causal invariance seems even more difficult than determining total causal invariance.

Note that if one looks at all 4702 $2_2 \to 3_2$ rules, the largest number of core branch pairs for any rule is 554; the largest number that resolve in 1 step is 132, and the largest number that remain unresolved is 430.

## 6.5 Causal Graphs for Causal Invariant Rules

An important consequence of causal invariance is that it establishes that a rule produces the same causal graph independent of the particular order in which update events occurred. And so this means, for example, that we can generate causal graphs just by looking at evolution with our standard updating order.

For rules that depend on only one relation, the causal graph is always just a tree

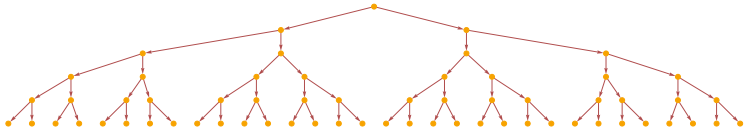

regardless of whether the structure generated is also a tree

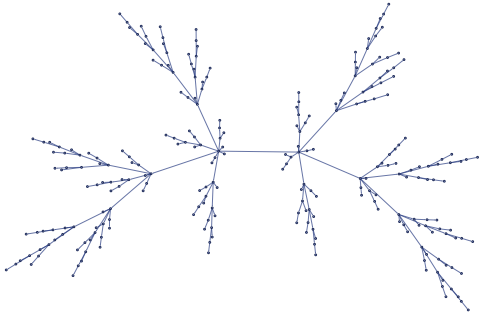

or has a more compact form:

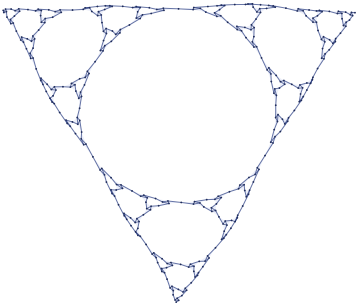

But as soon as a rule depends on more than one relation, the causal graph can immediately be more complicated. For example, consider even the rule:

{{*x*}, {*x*}} → {{*x*}, {*x*}, {*x*}}

The multiway system for this rule shows that only one path is possible (immediately demonstrating causal invariance):

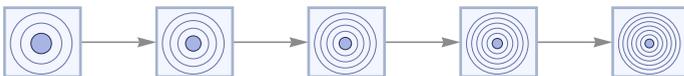

But the causal relationships between steps are not so straightforward

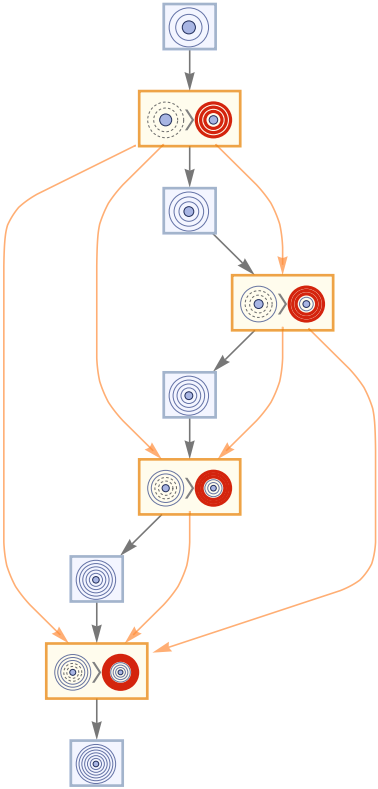

and after 15 steps the causal graph has the form

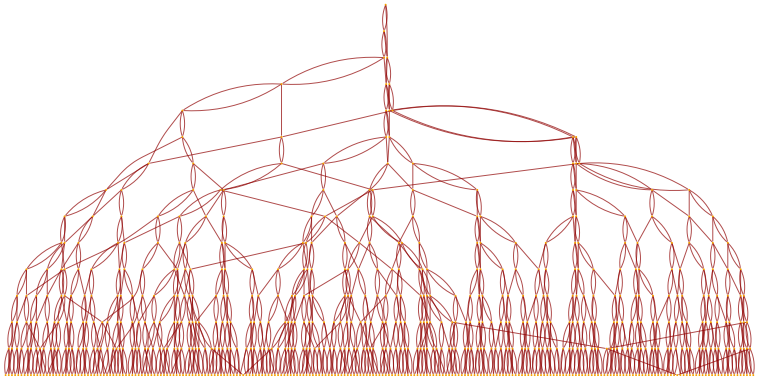

or in an alternative rendering:

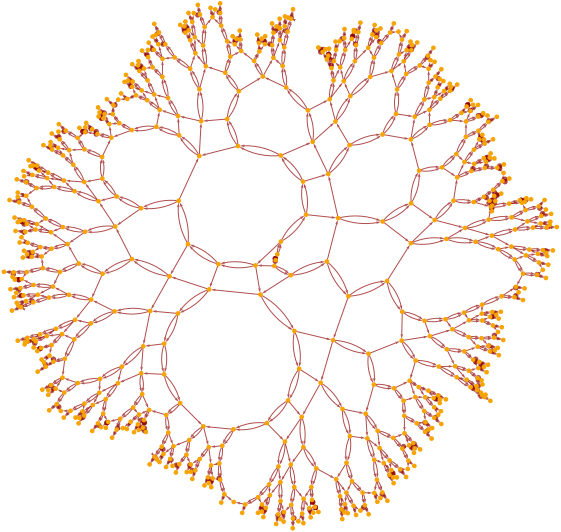

The fact that the multiway system is nontrivial does not mean that the causal graph for a particular rule evolution will be nontrivial. Consider for example a causal invariant rule that we discussed above:

$\{\{x, y\}, \{z, y\}\} \to \{\{x, w\}, \{y, w\}, \{z, w\}\}$

The multiway system for this rule, with causal connections shown, is:

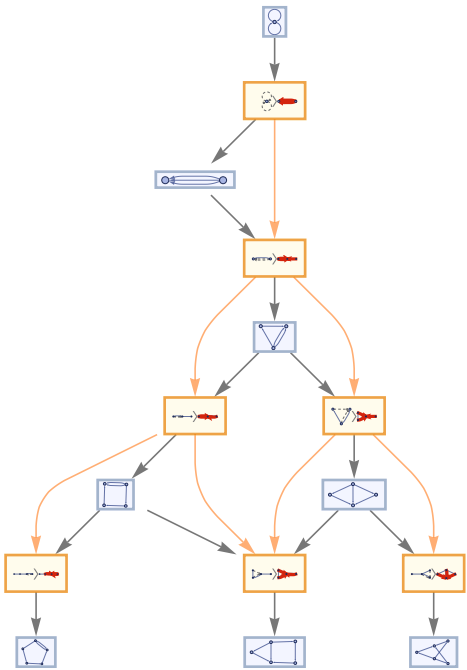

This yields the multiway causal graph:

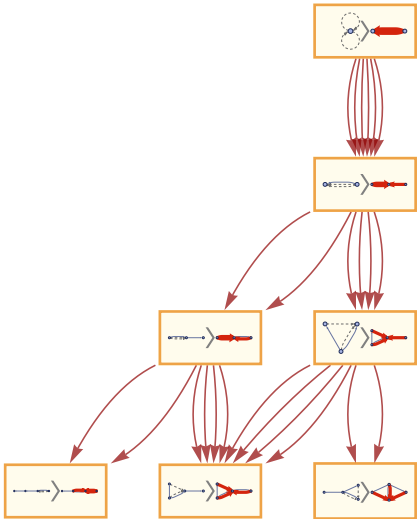

But the causal graph for any individual evolution is just:

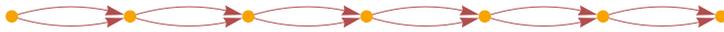

For the causal invariant rule (also discussed above)

$\{\{x, y\}, \{z, y\}\} \to \{\{x, z\}, \{y, z\}, \{w, z\}\}$

the multiway system after 5 steps has the form:

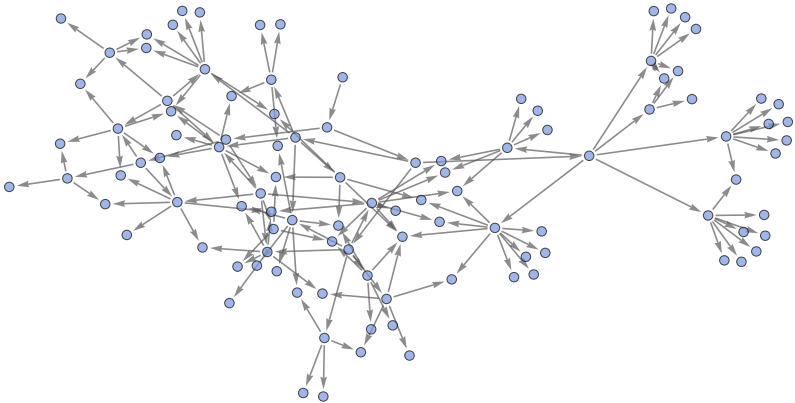

After 20 steps of evolution with our standard updating order gives:

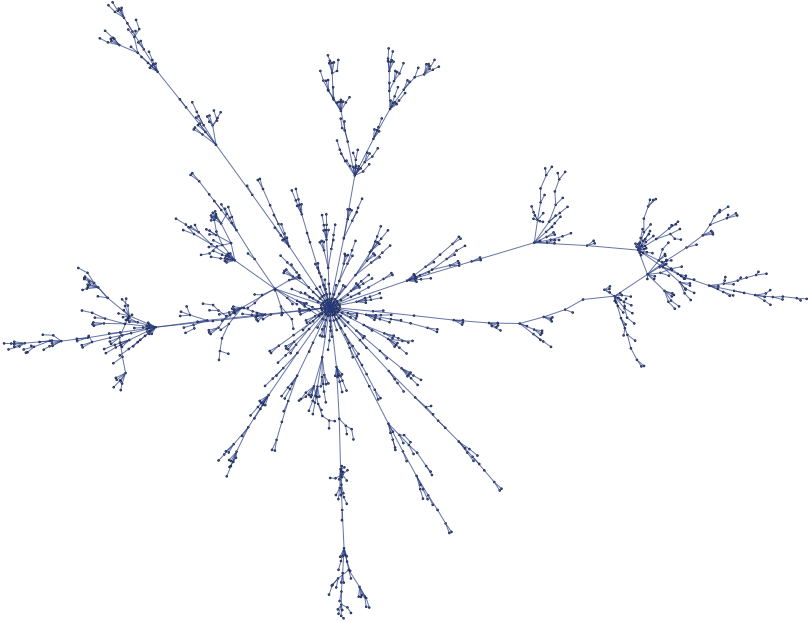

The causal graph for this rule after 10 steps is

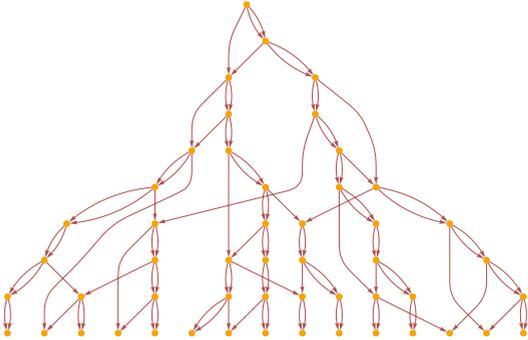

and after 20 steps, in a different rendering, it becomes:

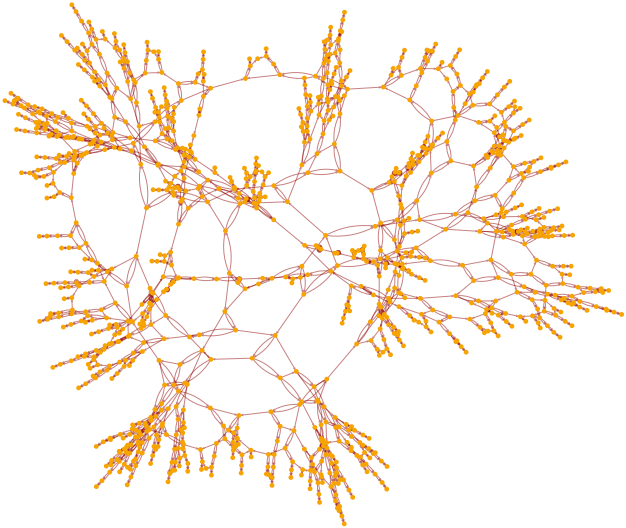

As another example, consider the rule (also discussed above):

{{*x*, *y*}, {*x*, *z*}} → {{*y*, *w*}, {*y*, *z*}, {*w*, *x*}}

The multiway system for this rule (with events included) has the form:

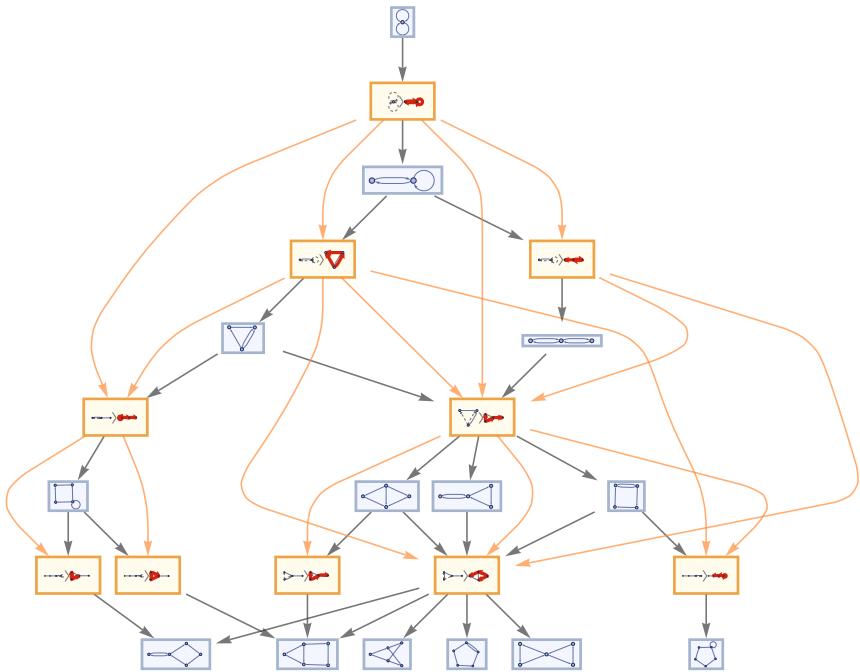

After 20 steps, the causal graph is:

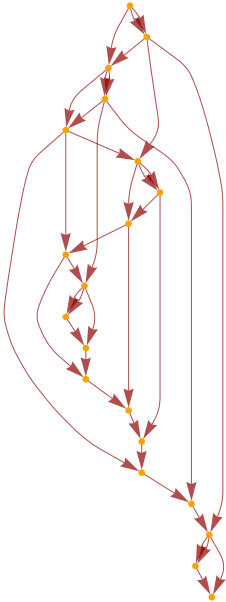

After 100 steps it is:

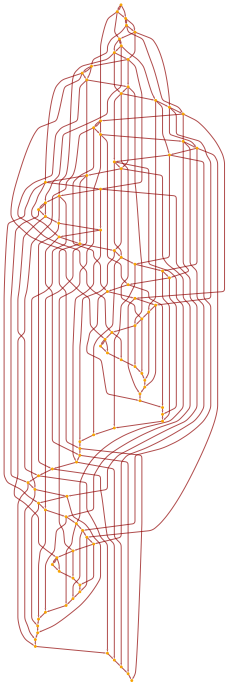

After 500 steps, in an alternative rendering, a grid-like structure emerges (the directed edges point outward from the center):

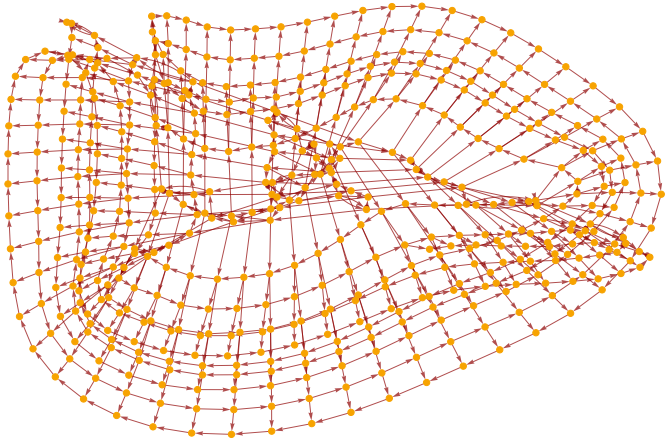

After 5000 steps, rendering the graph in 3D with surface reconstruction reveals an elaborate effective geometry:

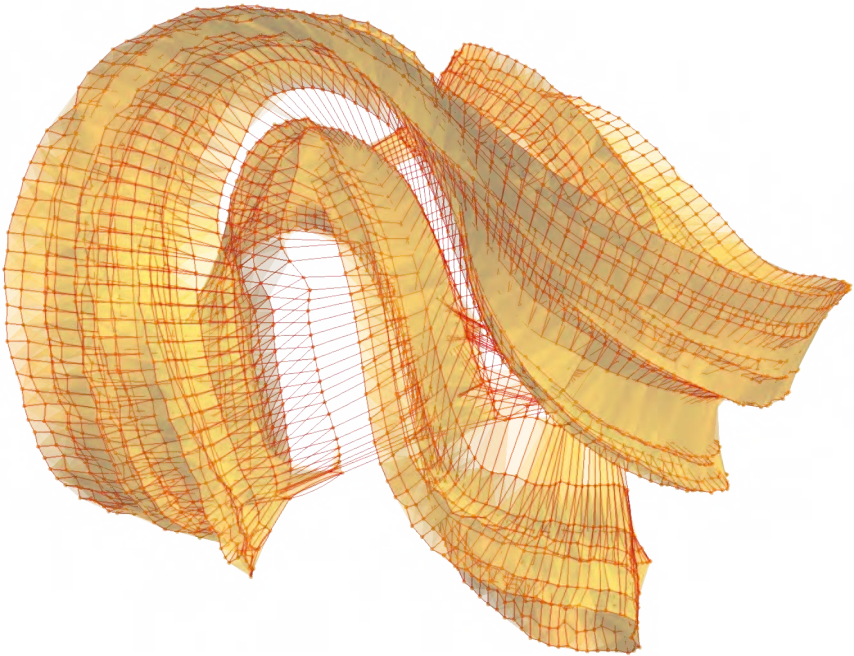

## 6.6 The Role of Causal Graphs

Even if we do not know that a rule is causal invariant, we can still construct a causal graph for it based on a particular updating order—and often different updating orders will give at least similar causal graphs.

Thus, for example, for the rule

{{*x*, *y*}, {*x*, *z*}} → {{*x*, *y*}, {*x*, *w*}, {*y*, *w*}, {*z*, *w*}}

applying our standard updating order for 5 steps gives the causal graph

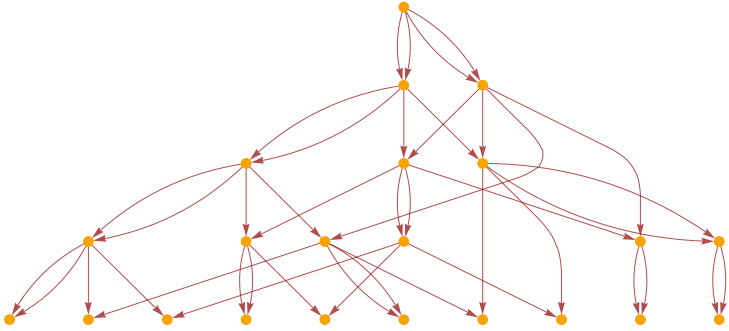

Continuing for 10 steps, we get:

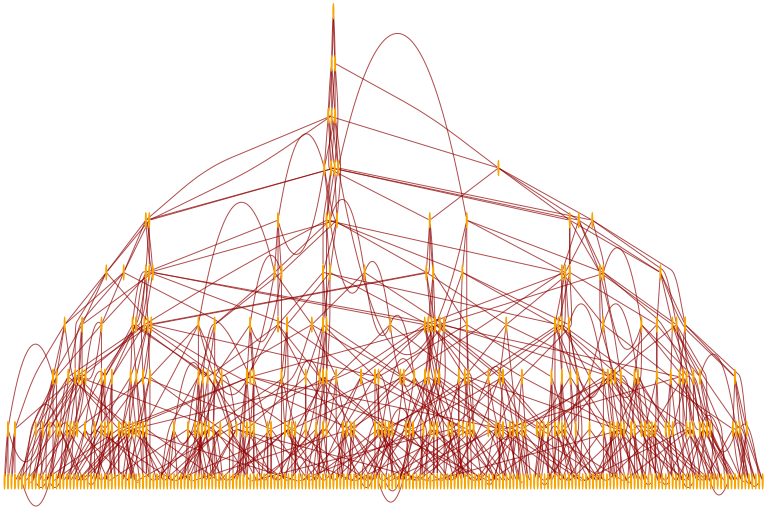

This can also be rendered as:

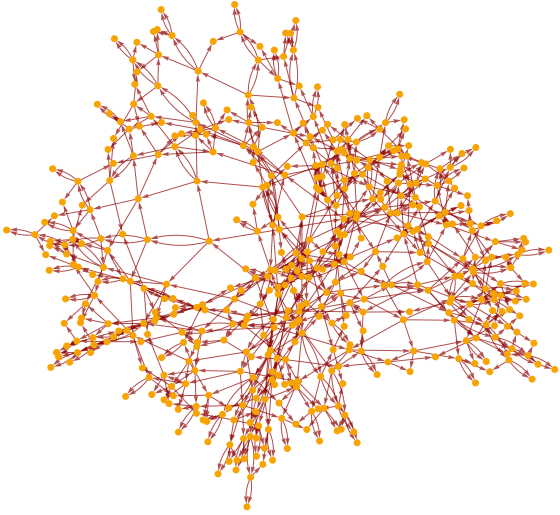

After 15 steps, there are 10,346 nodes:

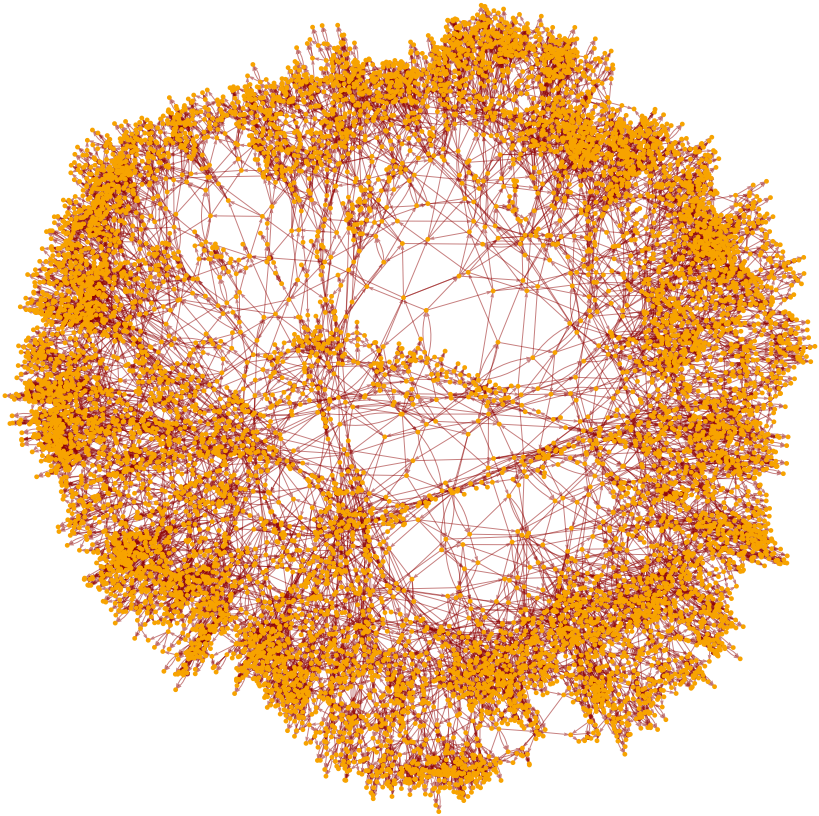

In effect the successive steps in the evolution of the system correspond to successive slices through this causal graph. In the case of causal invariant rules, any possible updating order must correspond to a possible causal foliation of the graph. But here we can at least say that the foliation obtained by looking at successive layers starting from the root corresponds to successive steps of evolution with our standard updating order.

For any system whose evolution continues infinitely, the causal graph will ultimately be infinite. But by slicing the graph as we have above, we are effectively showing the events that contribute to forming the state of the system after 15 steps of evolution (in this case, with our standard updating order):

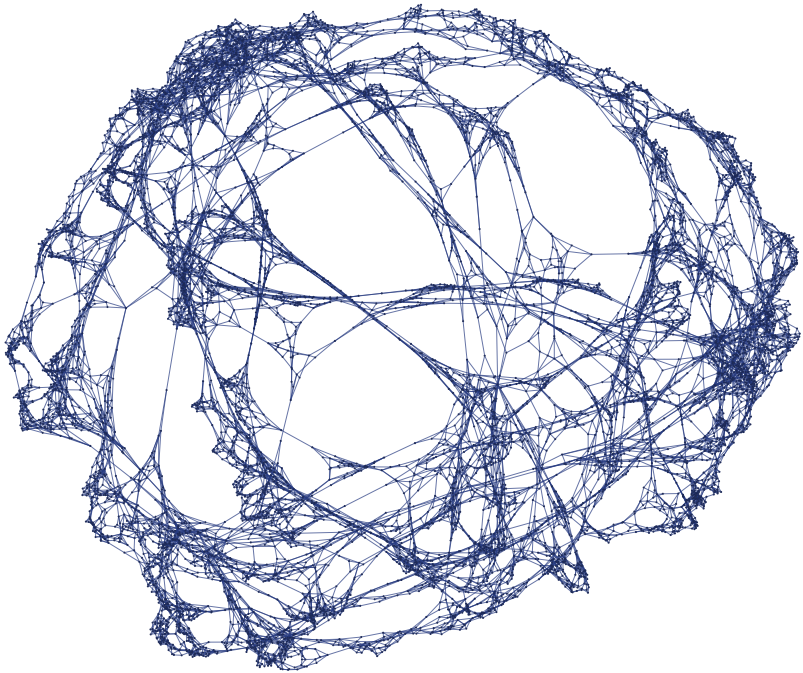

(Note that with this particular rule, the vast majority of relations that appear at step 15 were added specifically at that step, so in a sense most of the state at step 15 is associated just with the slice of the causal graph at layer 15.)

## 6.7 Typical Causal Graphs

As we discussed in 5.12, causal graphs for string substitution systems tend to have fairly simple structures. Causal graphs for our models tend to be considerably more complicated, and even among $2_2 \rightarrow 3_2$ rules considerable diversity is observed. A typical random sample of different forms is:

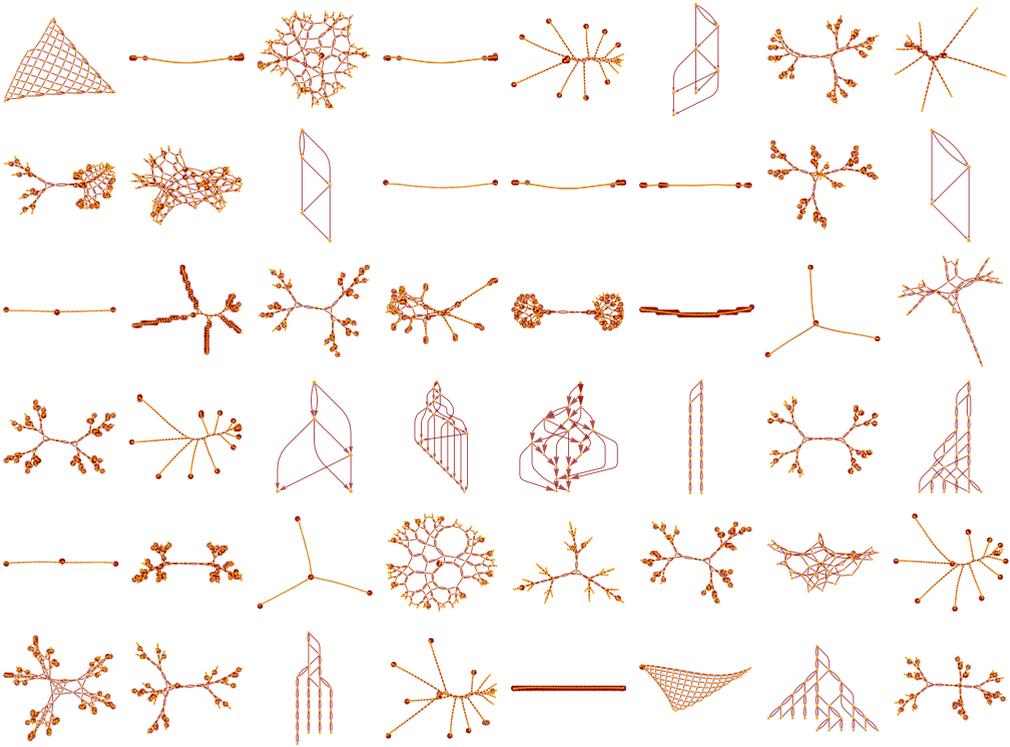

Even rules whose states seem quite simple can produce quite complex causal graphs:

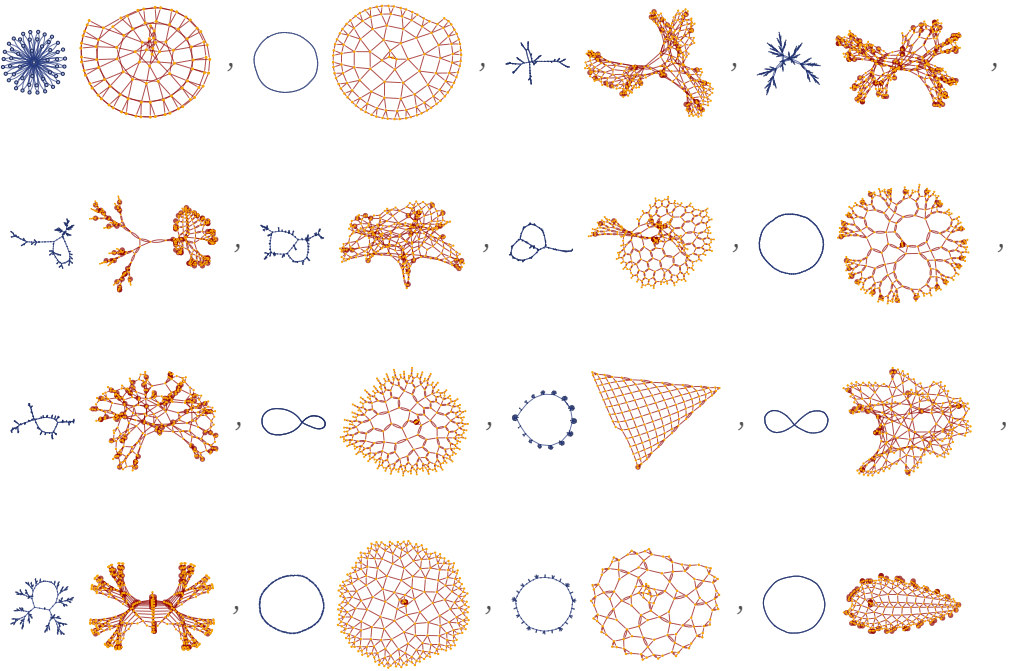

As a first example, consider the rule:

{{x, x}, {x, y}} → {{y, y}, {z, y}, {x, z}}

At each step, the self-loop just adds a relation, and effectively moves around the growing loop:

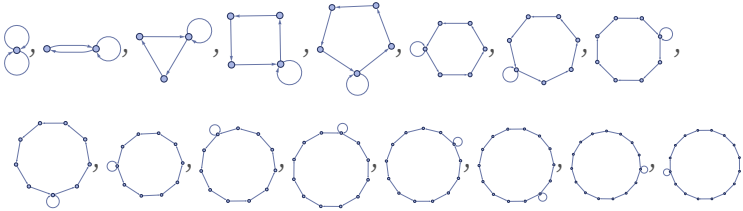

The causal graph captures the causal connections created by the self-loop encountering the same relations again after it goes all the way around:

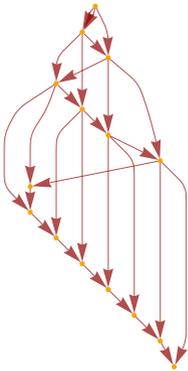

The structure gets progressively more complicated:

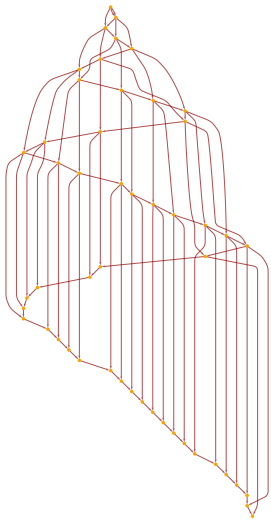

Re-rendering this gives

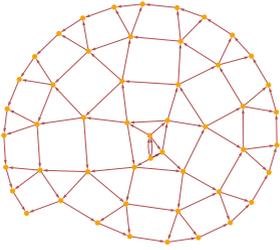

or after 500 steps:

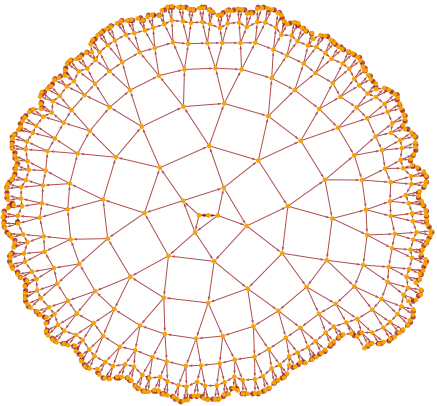

As another example, consider the rule:

$\{\{x, y\}, \{x, z\}\} \rightarrow \{\{y, w\}, \{y, x\}, \{w, x\}\}$

Here are the first 25 steps in its evolution (using our standard updating order):

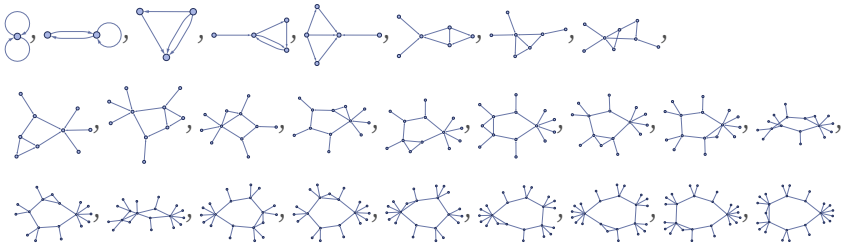

After a few steps all that happens is that there is a small structure that successively moves around the loop creating new "hairs". The causal graph (here shown after 25 steps) captures this process:

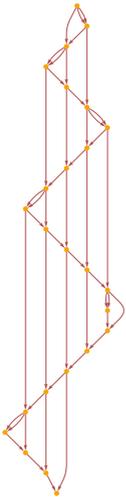

An alternative rendering shows that a grid structure emerges:

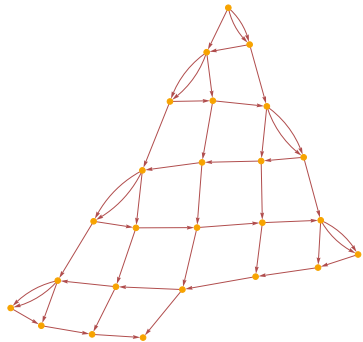

Here are the corresponding results after 100 steps:

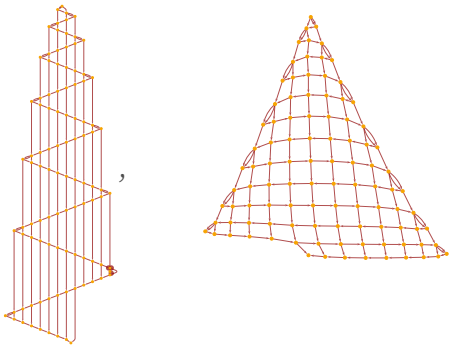

As a somewhat different example, consider the rule:

{{*x*, *y*}, {*y*, *z*}} → {{*x*, *w*}, {*x*, *y*}, {*w*, *z*}}

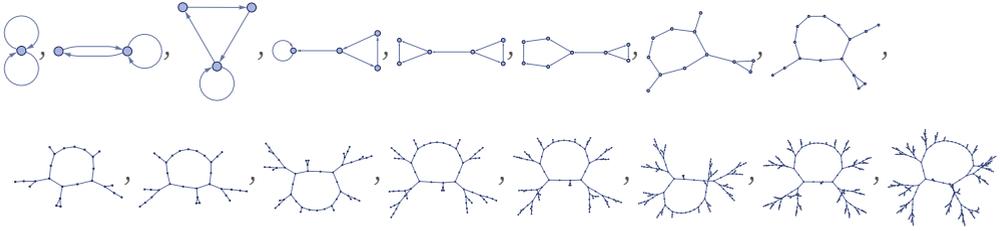

After the same number of steps, one can effectively see the separate trees in the causal graph:

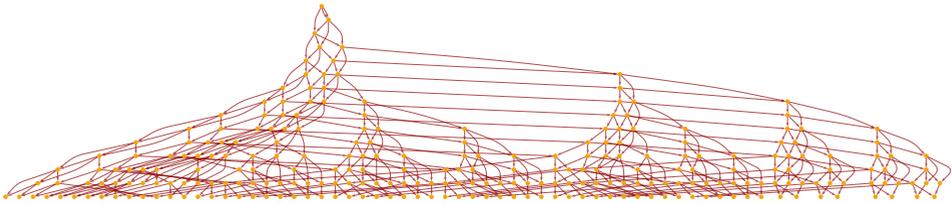

Re-rendering the causal graph, it has a structure that is quite similar to the actual state of the system:

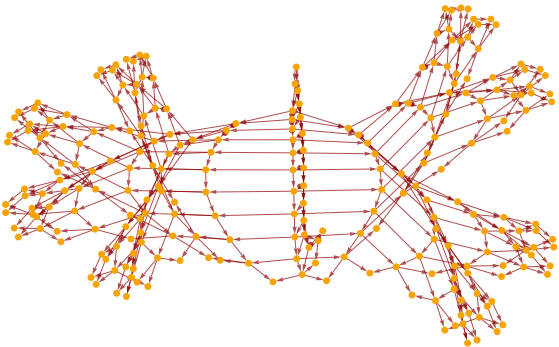

Continuing for a few more steps, a definite tree structure emerges:

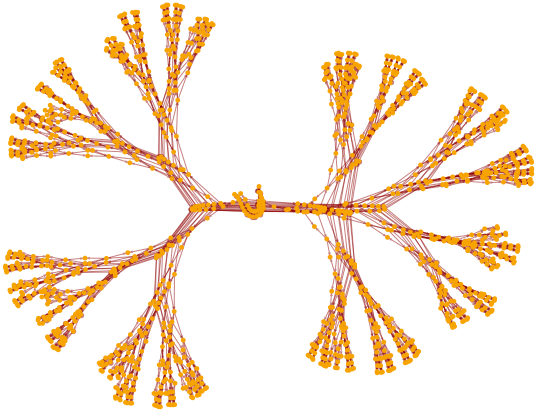

It is not uncommon for a causal graph to "look like" the actual hypergraph generated by one of our models. For example, rules that produce globular structures tend to produce similar "globular" causal graphs (here shown for three $2_2 \to 4_2$ rules from section 3):

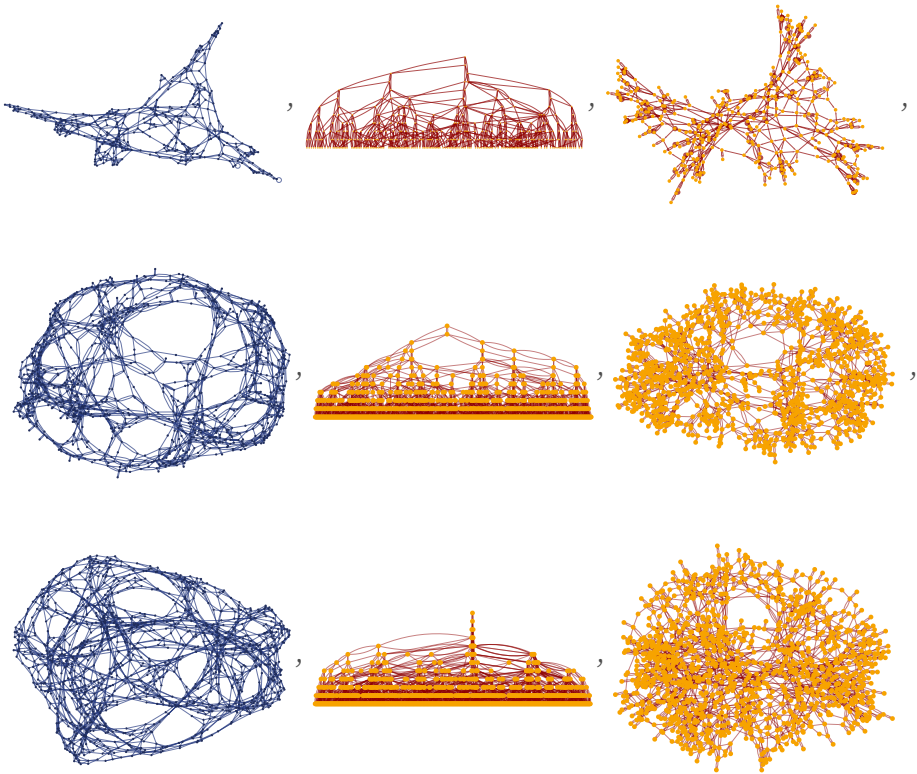

Rules that exhibit slow growth often yield either grid-like or "hyperbolic" causal graphs (here shown for some $2_3 \to 3_3$ rules from section 3):

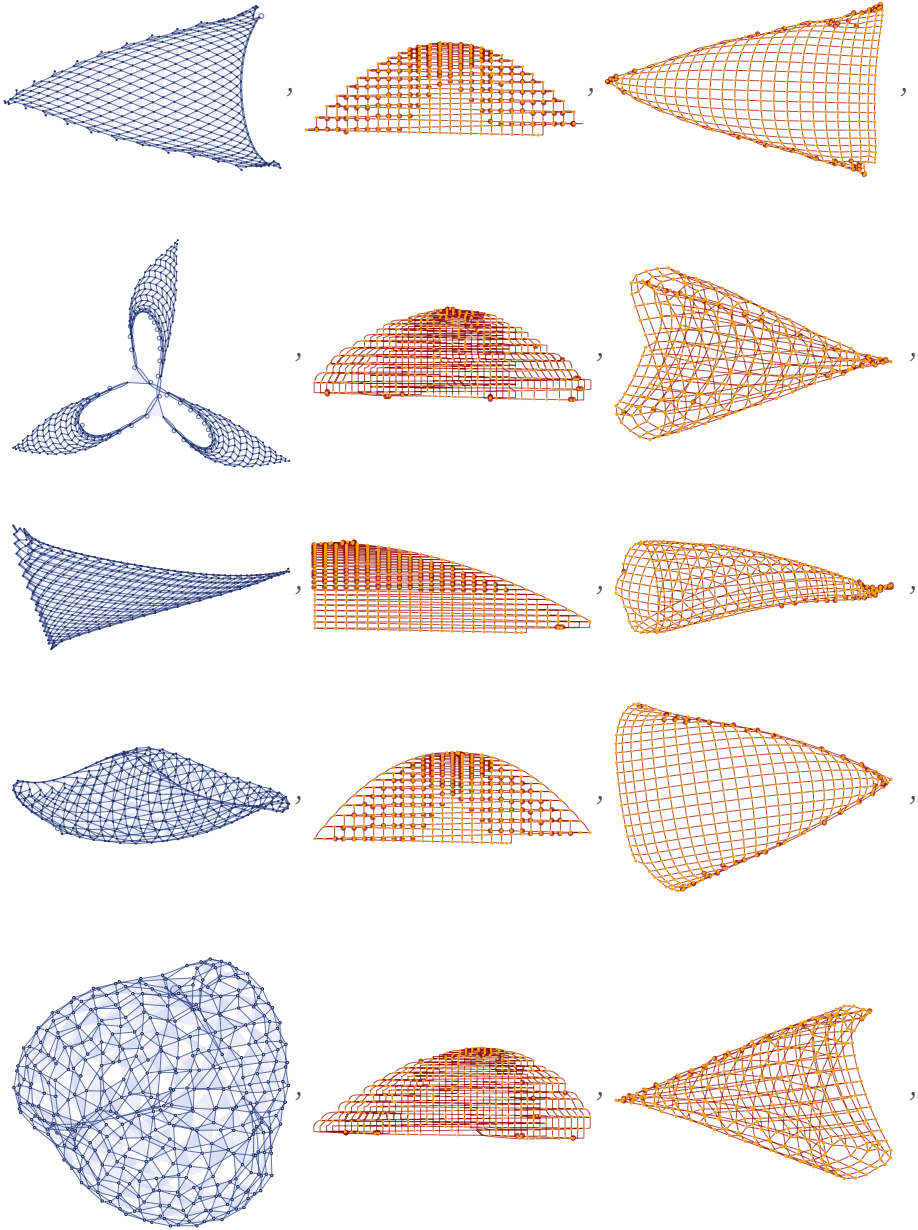

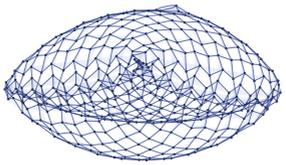 , 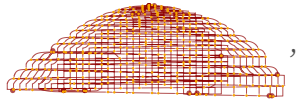 , 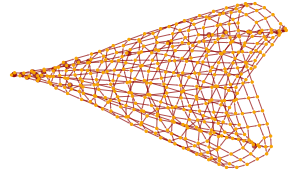 ,

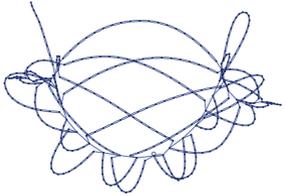 , 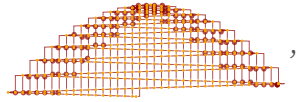 , 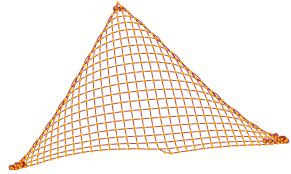 ,

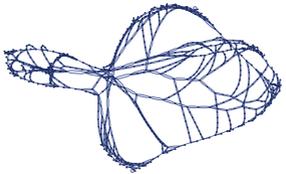 , 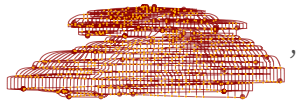 , 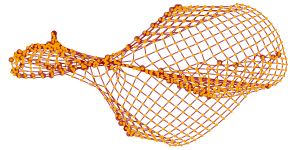 ,

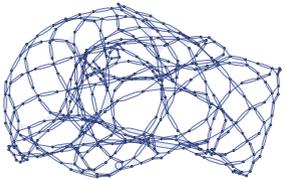 , 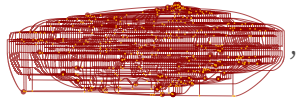 , 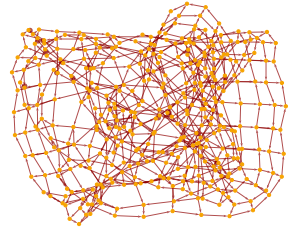 ,

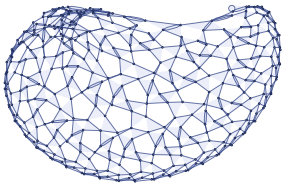 , 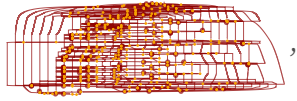 , 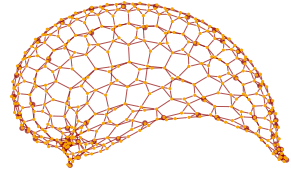 ,

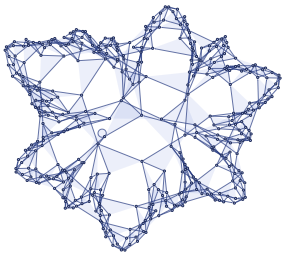 , 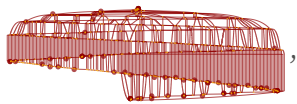 , 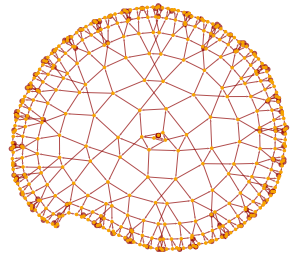 ,

A typical source of grid-like causal graphs [1:p489] is rules where in a sense only one thing ever happens at a time, or, in effect, the rules operate like a mobile automaton [1:3.3] or a Turing machine, with a single active element. As an example, consider the rule (see 3.10):

$\{\{x, y, y\}, \{z, x, u\}\} \to \{\{y, v, y\}, \{y, z, v\}, \{u, v, v\}\}$

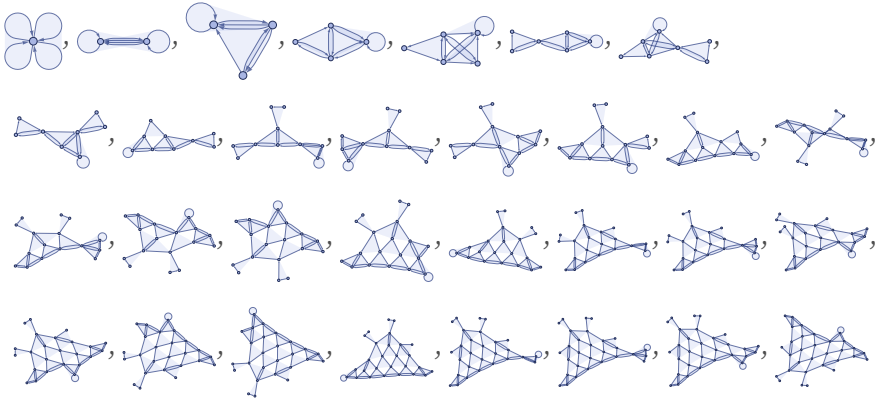

Updates can only occur at the position of the self-loop, which progressively "moves around", "knitting" a grid pattern. The causal graph captures the fact that "only one thing happens at a time":

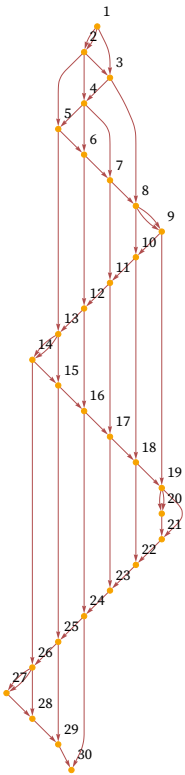

But what is notable is that if we ask about the overall causal relationships between events, we realize that even events that happened many steps apart in the evolution as shown here are actually directly causally connected, because in a sense "nothing else happened in between". Re-rendering the causal graph illustrates this, and shows how a grid is built up:

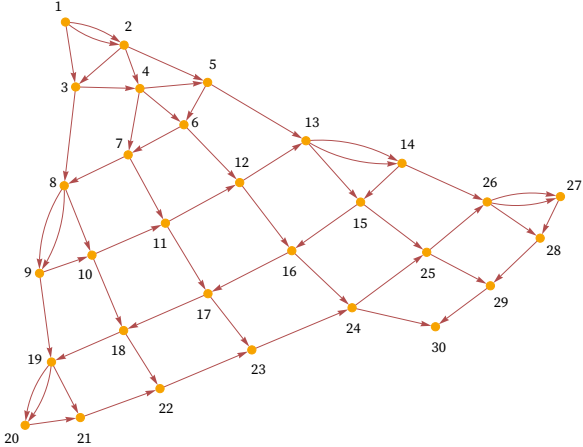

Sometimes the actual growth process can be more complicated, as in the case of the rule $\{\{x, y, y\}, \{z, y, u\}\} \to \{\{v, z, v\}, \{v, u, u\}, \{u, v, x\}\}$

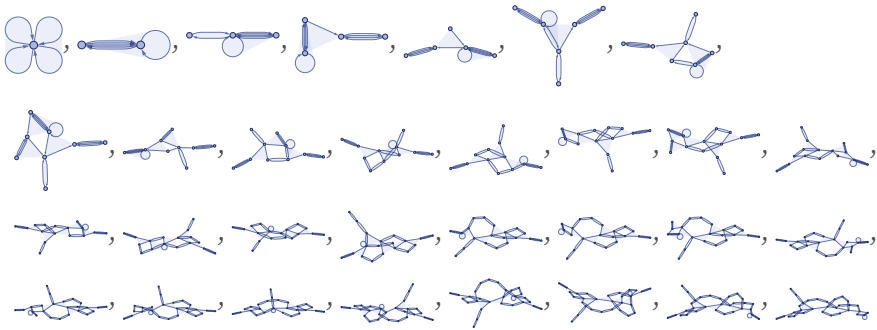

After 200 steps this yields:

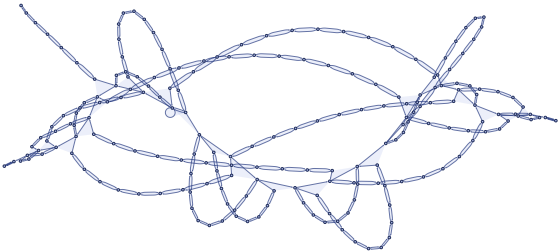

And after 1000 steps it gives:

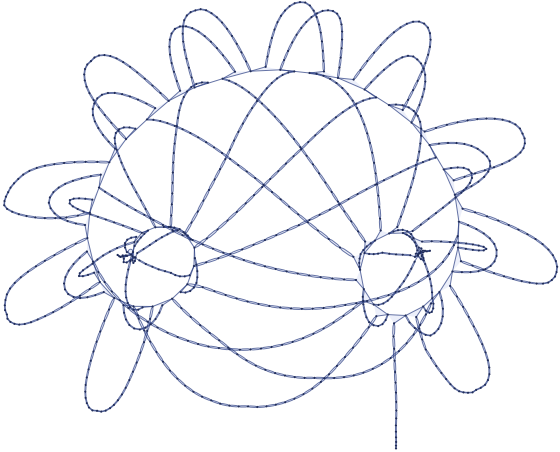

But despite this elaborate structure, the causal graph is very simple:

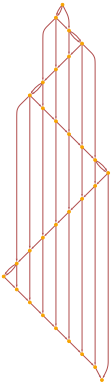

After 200 steps, the grid structure is clear:

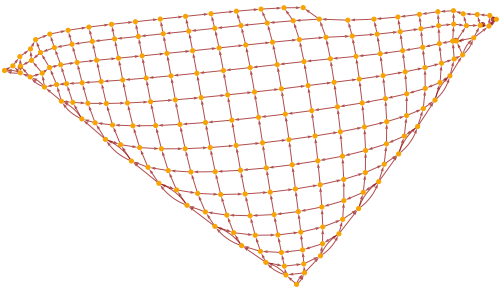

Sometimes the causal graph can locally be like a grid, while having a more complicated overall topological structure. Consider for example the rule:

{{x, y, y}, {z, x, u}} → {{y, z, y}, {z, u, u}, {y, u, v}}

After 200 steps this gives:

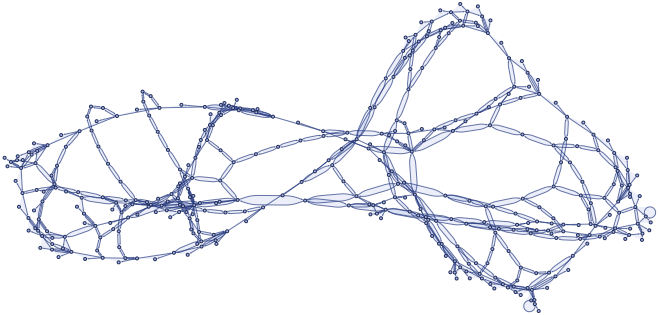

The corresponding causal graph is:

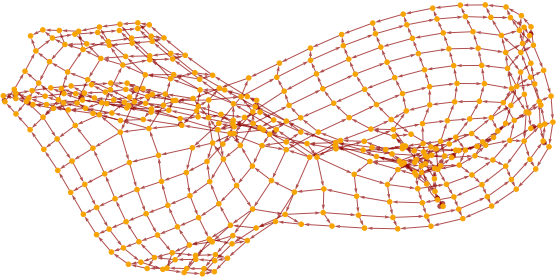

After 1000 steps with surface reconstruction this gives:

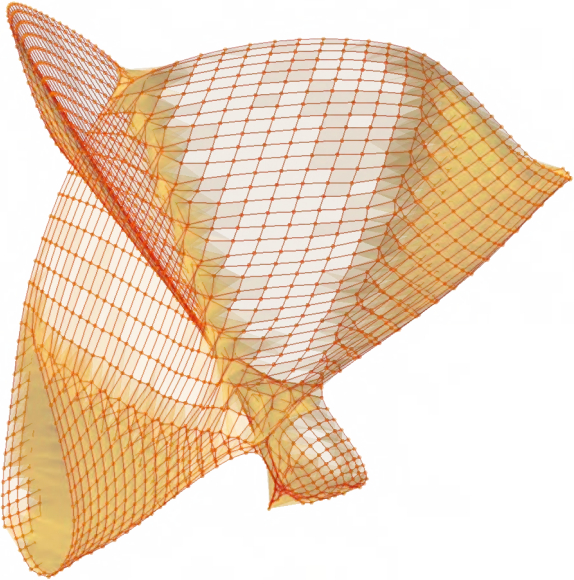

Rules (such as those with signature $2_2 \to 2_2$) that cannot exhibit growth inevitably terminate or repeat, thus leading to causal graphs that are either finite or repetitive—but may still have fairly complex structure. Consider for example the rule (compare 3.15):

$\{\{x, y\}, \{y, z\}\} \to \{\{z, x\}, \{z, y\}\}$

Evolution from a chain of 9 relations leads to a 31-step transient, then a 9-step cycle:

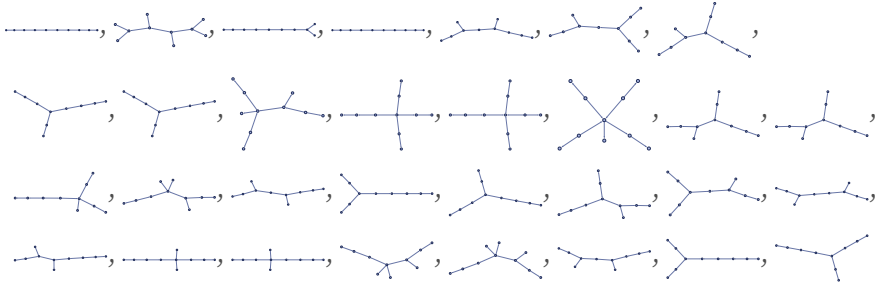

The first 30 layers in the causal graph are:

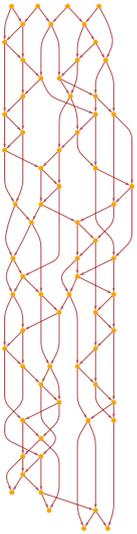

In an alternative rendering, the graph is:

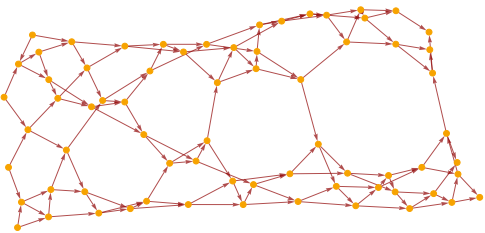

After 50 more steps, the repetitive structure becomes clear:

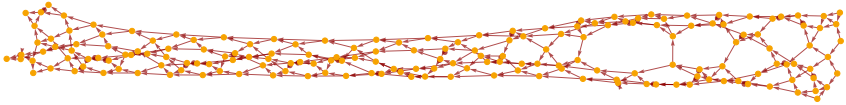

Sometimes the structure of the causal graph may be very much a reflection of the updating order used. Consider for example the rather trivial "identity" rule:

{{x, y}, {y, z}} → {{x, y}, {y, z}}

Starting with a chain of 3 relations, this shows update events according to our standard updating order (note that the same relation can be both created and destroyed at a particular step):

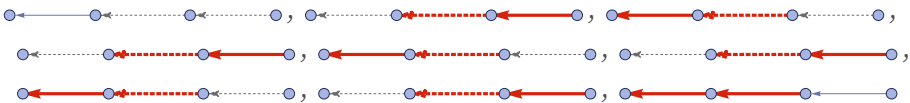

The corresponding causal graph is:

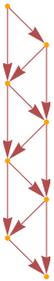

For a chain of length of 21 the causal graph consists largely of independent regions—except for the connection created by updates fitting differently at different steps:

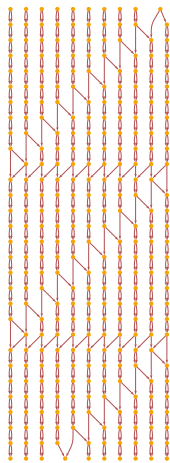

Re-rendering this gives a seemingly elaborate structure:

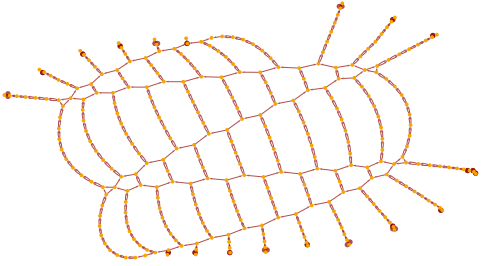

After 100 steps, though, its repetitive character becomes clear:

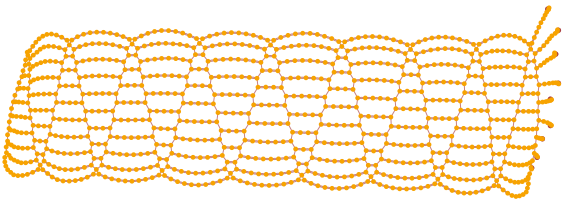

Note that if the initial condition is a ring rather than a chain, one gets

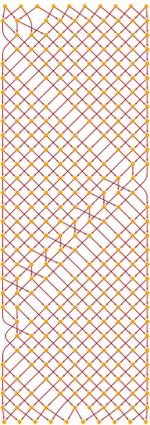

together with the tube-like structure:

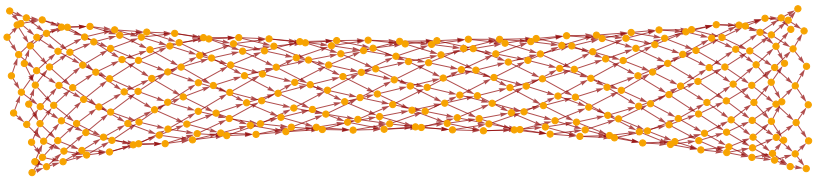

## 6.8 Large-Scale Structure of Causal Graphs

In section 5 we used the cone volume $C_t$ to probe the large-scale structure of causal graphs generated by string substitution systems. Now we use $C_t$ to probe the large-scale structure of causal graphs generated by our models.

Consider for example the rule

$\{\{x, y\}, \{x, z\}\} \rightarrow \{\{x, y\}, \{x, w\}, \{y, w\}, \{z, w\}\}$

We found in section 4 that after a few steps, the volumes $V_r$ of balls in the hypergraphs generated by this rule grow roughly like $r^{2.6}$, suggesting that in the limit the hypergraphs behave like a finite-dimensional space, with dimension $\approx 2.6$.

The pictures below show the log differences in $V_r$ and $C_t$ for this rule after 15 steps of evolution:

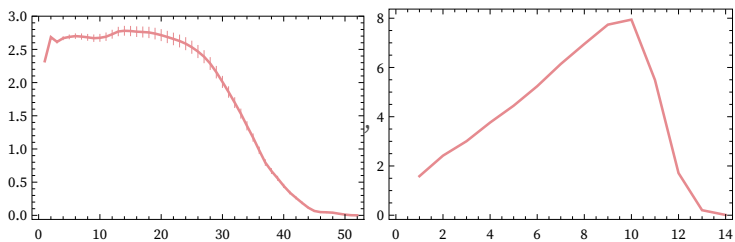

The linear increase in this plot implies exponential growth in $C_t$ and indeed we find that for this rule:

$C_t \sim 2.2^t$

This exponential growth—compared with the polynomial growth of $V_r$—implies that expansion according to this rule is in a sense sufficiently rapid that there is increasing causal disconnection between different parts of the system.

The other three $2_2 \rightarrow 4_2$ globular-hypergraph-generating rules shown in the previous subsection show similar exponential growth in $C_t$, at least over the number of steps of evolution tested.

A rule such as

$\{\{x, y, y\}, \{x, z, u\}\} \rightarrow \{\{u, v, v\}, \{v, z, y\}, \{x, y, v\}\}$

whose hypergraph and causal graph (after 500 steps) are respectively

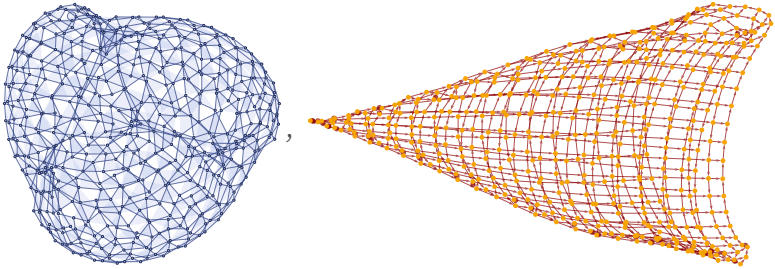

gives the following for the log differences of $V_r$ and $C_t$ after 10,000 steps:

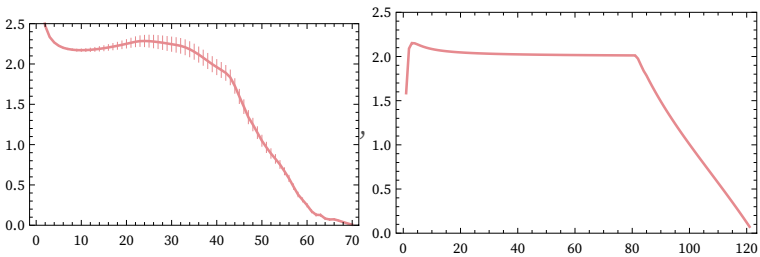

This implies that for this rule the hypergraphs it generates and its causal graph both effectively limit to finite-dimensional spaces, with the hypergraphs having dimension perhaps slightly over 2, and the causal graph having dimension 2.

Consider now the rule:

$\{\{x, y, x\}, \{x, z, u\}\} \to \{\{u, v, u\}, \{v, u, z\}, \{x, y, v\}\}$

The hypergraph and causal graph (after 1500 steps) for this rule are respectively:

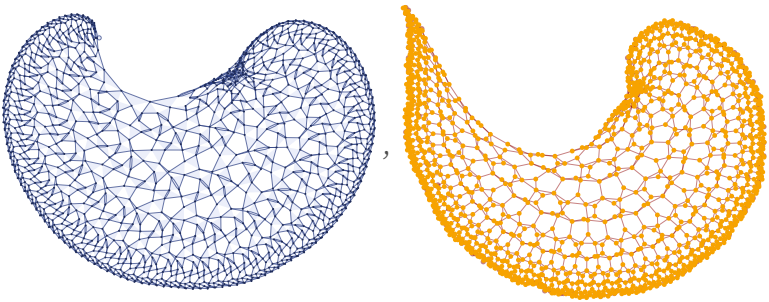

The log differences of $V_r$ and $C_t$ after 10,000 steps are then:

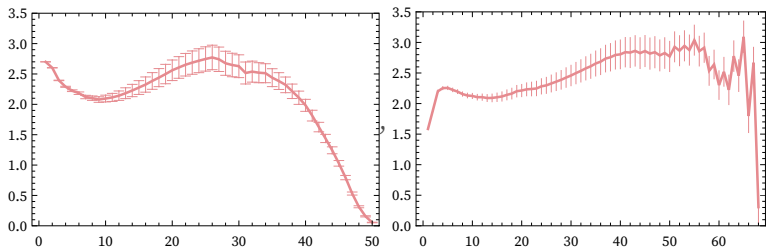

Both suggest limiting spaces with dimension 2, but with a certain amount of (negative) curvature.

## 6.9 Foliations of Causal Graphs

At least in a causal invariant system, the structure of the causal graph is always the same, and it defines the causal relationships that exist between updating events. But in relating the causal graph to actual underlying evolution histories for the system, we need to specify how we want to foliate the causal graph—or, in effect, how we want to define "steps" in the evolution of the system.

As an example, consider the rule:

$\{\{x, y\}, \{z, y\}\} \to \{\{x, z\}, \{y, z\}, \{w, z\}\}$

(This rule is probably not causal invariant, but this fact will not affect our discussion here.) The most obvious foliation for the causal graph basically follows our standard updating order:

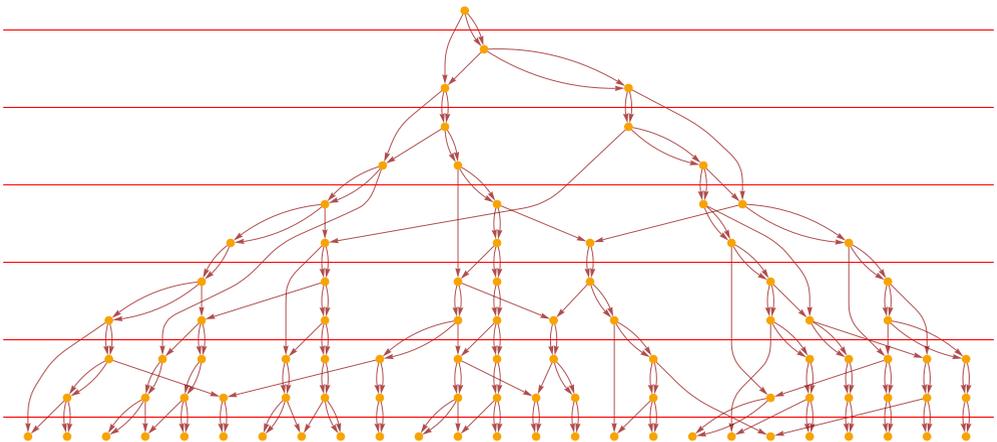

But this is not the only foliation we can use. In fact, we can divide the graph into slices in any way, so long as the slices respect the causal relationships defined by the graph, in the sense that within a slice the causal relationships allow the events to occur in any order, and between successive slices events must occur in the order of the slices. And with these criteria, for example, another possible foliation is:

With the first foliation shown above, the hypergraphs from what we consider to be the first "few steps" in the evolution of the underlying rule are:

But the second foliation in effect has a different (and coarser) definition of "steps", and with this foliation the first few steps would be:

When we discussed foliations in the context of string substitution systems, there were a number of simplifying features in our discussion. First, the underlying system fundamentally involved a linear string of elements. And second, the main causal graph we actually considered was a simple grid.

With a rule like

{{x, y, y}, {y, z}} → {{x, y}, {y, z, z}}

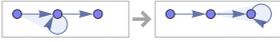

we can also get a simple grid causal graph (and this rule happens to be causal invariant). With the obvious foliation

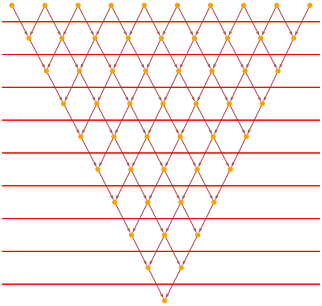

the steps in the evolution of the underlying system from a particular initial condition are:

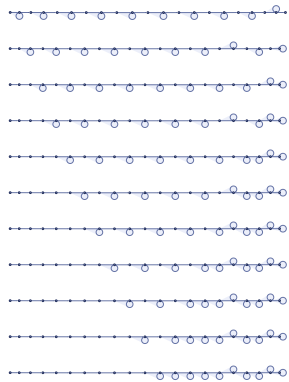

But given the grid structure of the causal graph, we can use the same diagonal slice method for generating foliations that we did in 5.14. And for example with the foliation

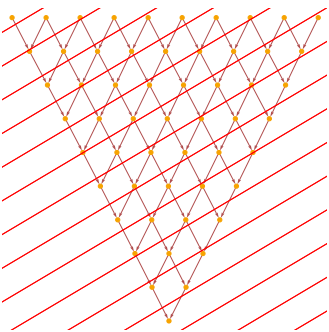

there are more steps involved in the evolution of the system:

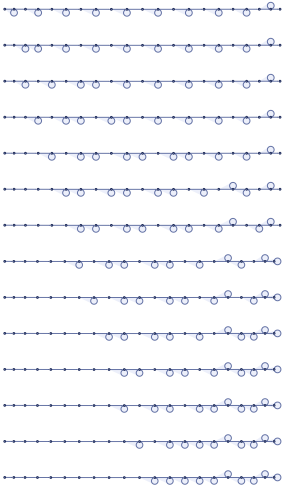

But when the causal graph does not have such a simple structure, the definition of foliations can be much more complicated. When the causal graph at least in some statistical sense limits to a sufficiently uniform structure, it should be possible to set up foliations that are analogous to the diagonal slices. And even in other cases, it will often be possible to set up foliations that can be described, for example, by the kind of lapse functions we discussed in 5.14.

But there is one issue that can make it impossible to set up any reasonable "progressive" foliation of a causal graph at all, and that is the issue of loops. This issue is actually already present even in the case of string substitution systems (and even causal invariant ones). Consider for example the rule:

{AA → A, A → AA}

Starting from AA the multiway causal graph for this rule is:

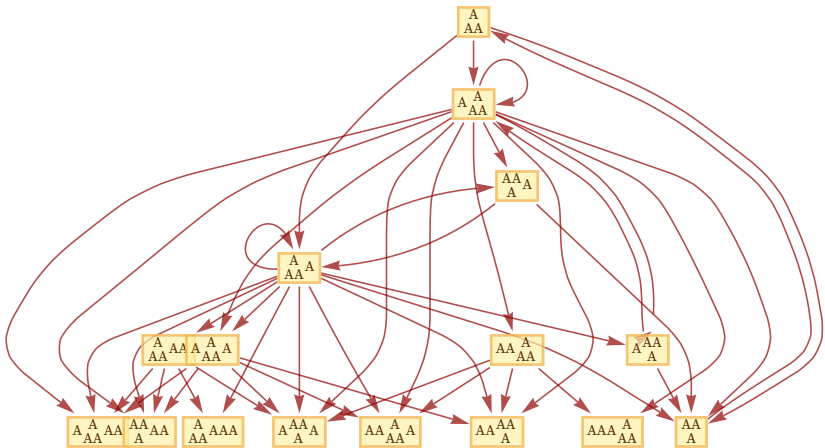

But note here the presence of several loops. And looking at the states graph in this case one can see where these loops come from: they are reflections of the fact that in the evolution of the system, there are states that can repeat—and where in a sense a state can return to its past.

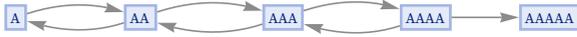

Whenever this happens, there is no way to make a progressive foliation in which events in future slices systematically depend only on events in earlier slices. (In the continuum limit, the analog is failure of strong hyperbolicity [86]; loops are the analog of closed timelike curves (e.g. [75])) (Self-loops also cause trouble for progressive foliations by forcing events to happen repeatedly within a slice, rather than only affecting later slices.)

The phenomenon of loops is quite common in string substitution systems, and already happens with the trivial rule A→A. It also happens for example with a rule like:

{AB → BAB, BA → A}

Starting with ABA, this gives the causal graph

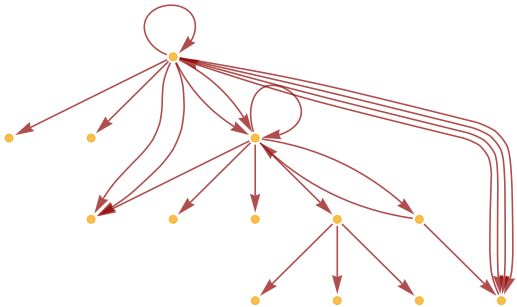

and has a states graph:

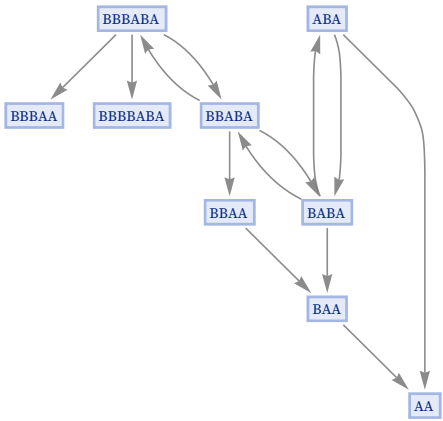

Loops can also happen in our models. Consider for example the very simple rule:

{{{x}, {x}} → {{x}}, {{x}} → {{x}, {x}}}

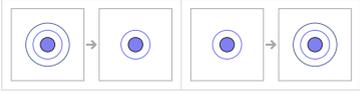

The multiway graph for this rule is:

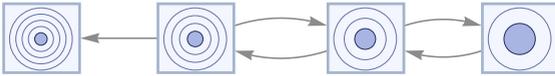

This contains loops, as does the corresponding causal graph:

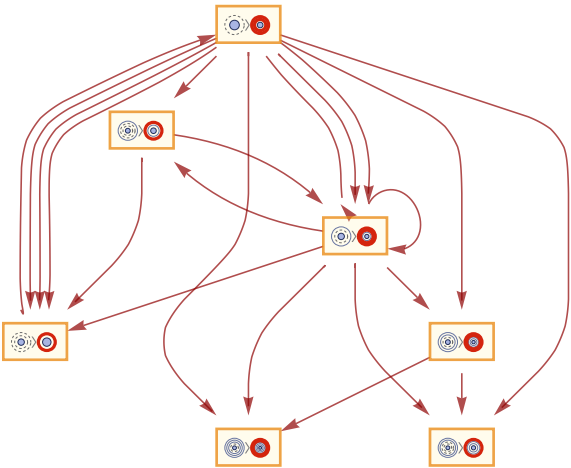

(Note that the issue discussed in 6.1 of when we consider states "identical" as opposed to "equivalent" can again arise here. There are similar issues when we consider finite-size systems where the whole state inevitably repeats—and where in principle we can define a cyclic analog of our foliations.

## 6.10 Causal Disconnection

In 2.9 we discussed the fact that some rules—even though the rules themselves are connected—can lead to hypergraphs that are disconnected. And—unless one is dealing with rules with disconnected left-hand sides—any hypergraphs that are disconnected must also be causally disconnected.

As a simple example of what can happen, consider the rule:

{{x, y}} → {{y, z}, {y, z}}

The evolution of this rule quickly leads to disconnected hypergraphs:

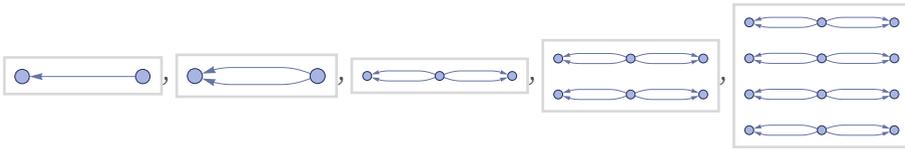

The corresponding causal graph is a tree:

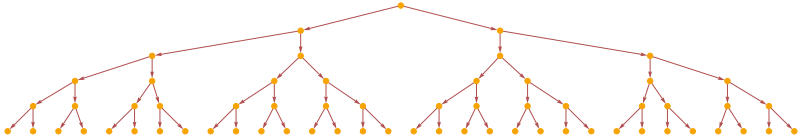

In this particular case, the different branches happen to correspond to isomorphic hypergraphs, so that in our usual way of creating a multiway graph, this rule leads to a connected multiway graph, which even shows causal invariance:

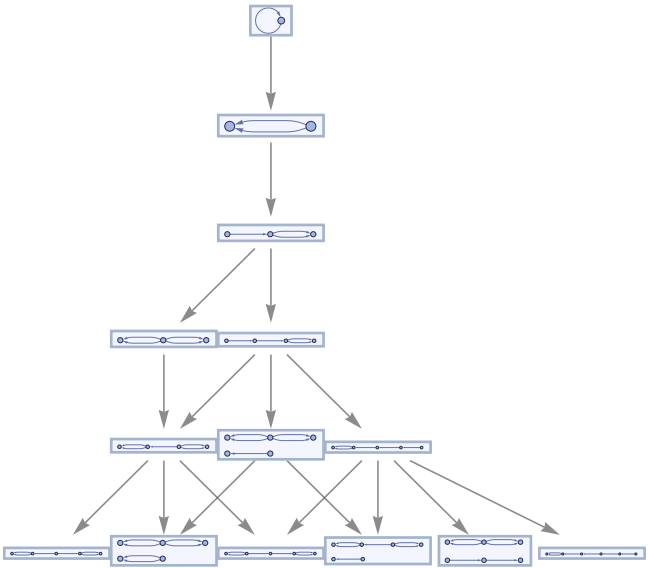

(Note that causal invariance is in a sense easier to achieve with disconnected hypergraphs, because there is no possibility of overlap, or of ambiguity in updates.)

In the case of an extremely simple rule like

$\{\{x\}\} \to \{\{y\}, \{z\}\}$

the evolution is immediately disconnected

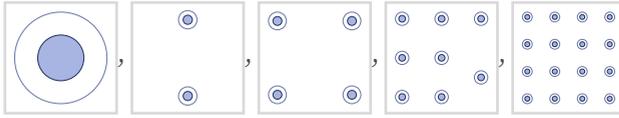

the causal graph is a tree

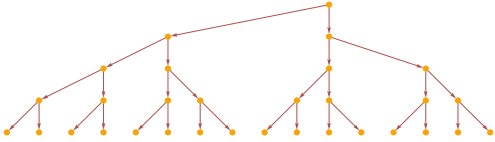

but the multiway graph consists of a simple "counting" sequence of states:

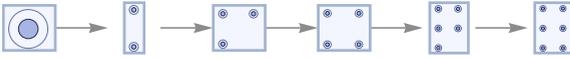

In other rules, the disconnected pieces are not isomorphic, and the multiway graph can split. An example where this occurs is the rule:

$\{\{x, y\}, \{x, z\}\} \to \{\{x, x\}, \{y, u\}, \{u, v\}\}$

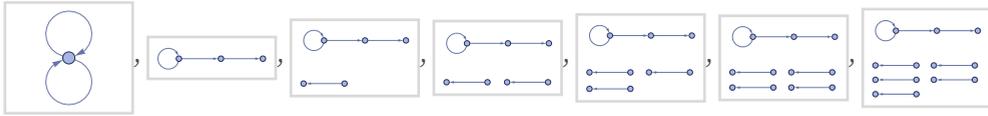

The multiway graph in this case is a tree:

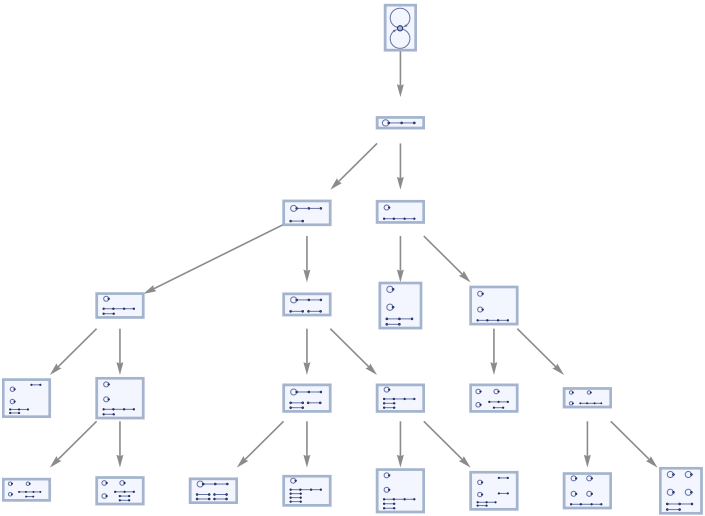

The multiway causal graph, however, does not have an exponential tree structure, but instead effectively just has one branch for each disconnected component in the hypergraph:

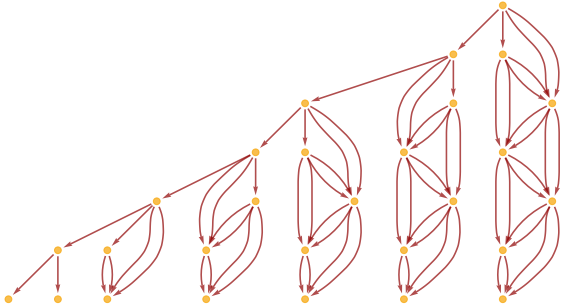

As a result, for this rule the ordinary causal graph has a simple, sequential form:

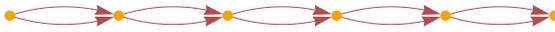

As a related example, consider the rule:

$\{\{x, y\}, \{y, z\}\} \to \{\{u, v\}, \{v, x\}, \{x, y\}\}$

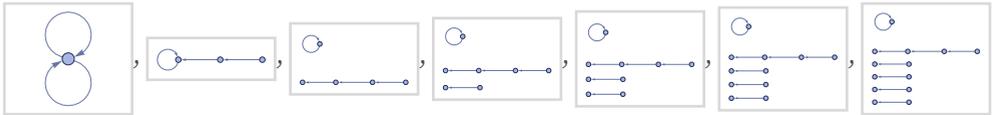

In this case, the multiway graph has the two-branch form

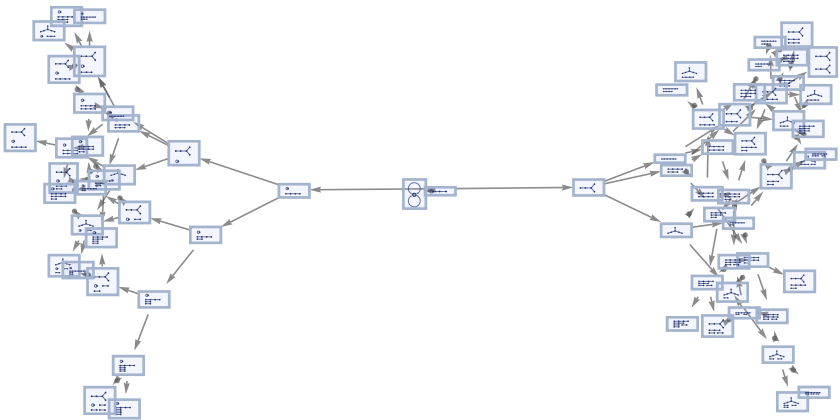

and the multiway causal graph has the similarly two-branch form

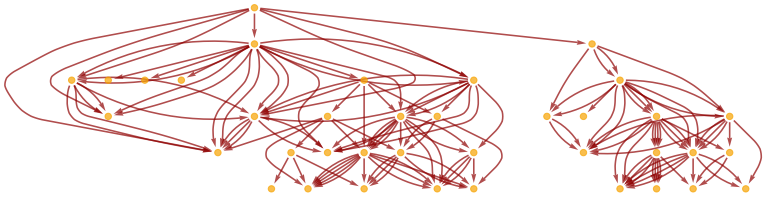

though the ordinary causal graph is still just:

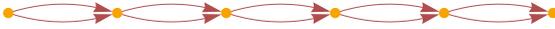

Sometimes there can be multiple branches in both the multiway graph and the ordinary causal graph. An example occurs in the rule

$\{\{\{x, y\}, \{x, z\}\} \to \{\{y, y\}, \{z, u\}, \{z, v\}\}\}$

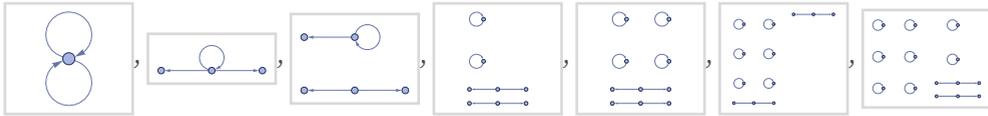

where the multiway graph is

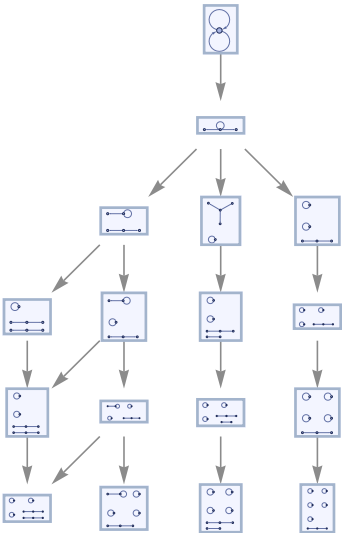

the multiway causal graph is

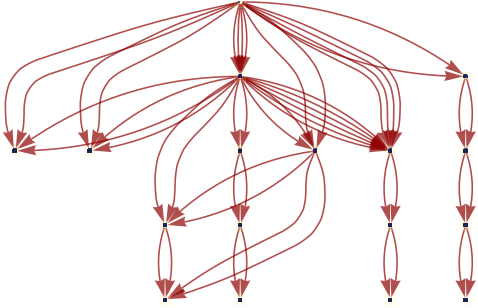

and the ordinary causal graph is:

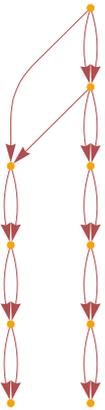

Is it possible to have both an infinitely branching multiway graph, and an infinitely branching ordinary causal graph? One of the issues is that in general it can be undecidable whether this is ultimately infinite branching. Consider for example the rule:

{{{x, x}, {x, y}} → {{y, y}, {y, y}, {x, z}}}

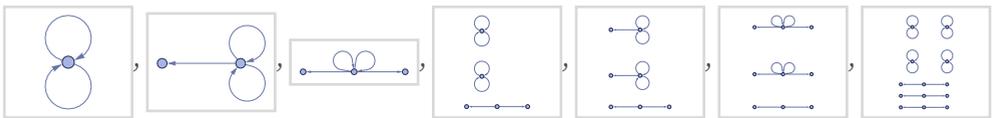

The ordinary causal graph for this rule has the form

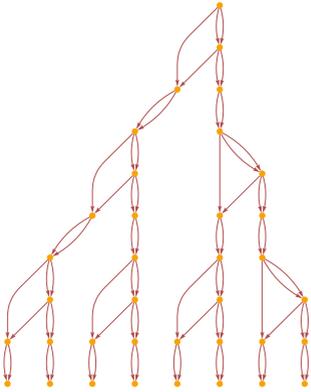

or in a different rendering after more steps:

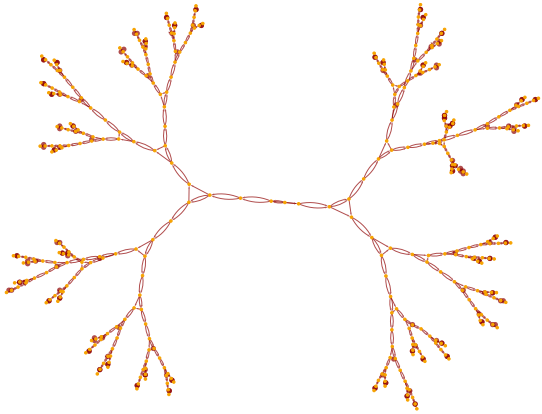

The multiway causal graph in this case is:

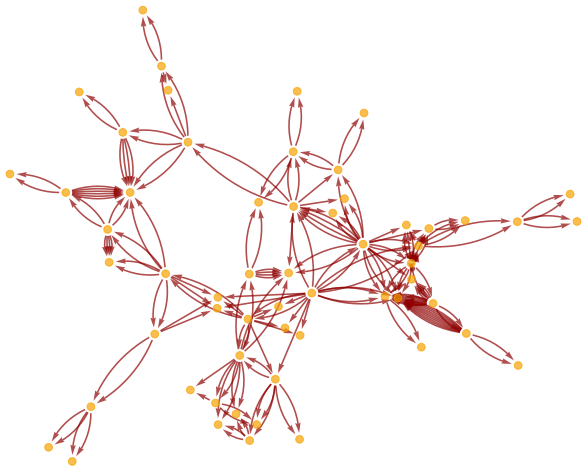

But now the multiway graph is:

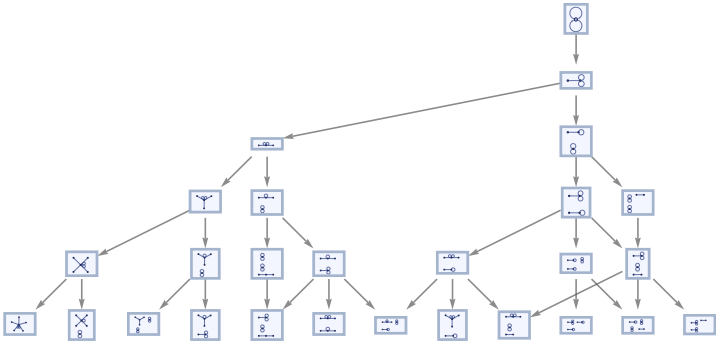

Continuing for more steps yields:

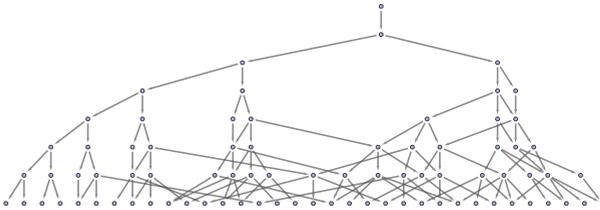

But while it is fairly clear that this multiway graph does not show causal invariance, it is not clear whether it will branch forever or not.

As a similar example, consider the rule:

$\{\{x, y\}, \{y, z\}\} \to \{\{x, x\}, \{x, y\}, \{w, y\}\}$

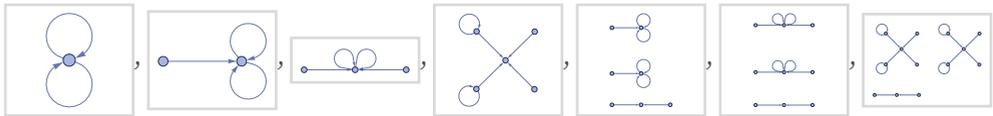

This yields the same ordinary causal graph as the previous rule

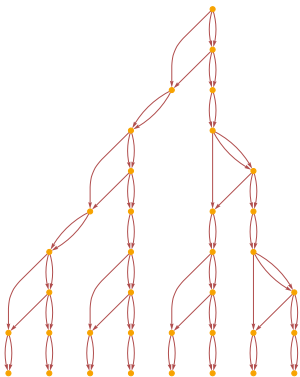

but now its multiway graph has a form that appears slightly more likely to branch forever:

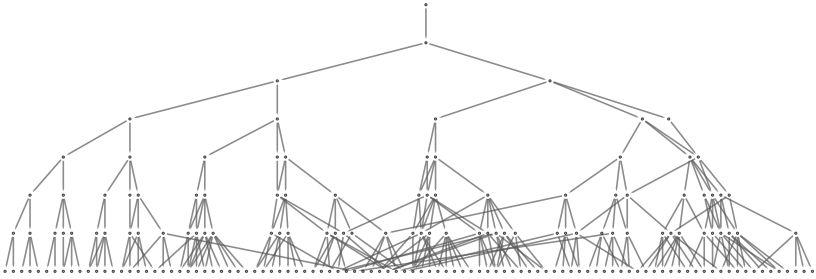

All the examples we have seen so far involve explicit disconnection of hypergraphs. However, it is also possible to have causal disconnection even without explicit disconnection of hypergraphs. As a very simple example, consider the rule:

{{x, x}} → {{x, x}, {x, x}}

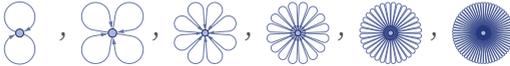

The causal graph in this case is

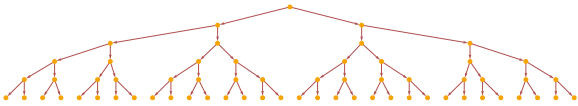

although the multiway graph is just:

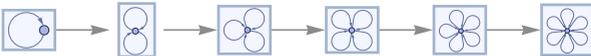

For the rule

{{x, y}} → {{x, x}, {x, z}}

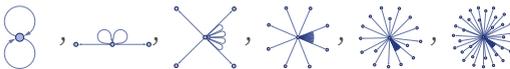

the causal graph is again a tree

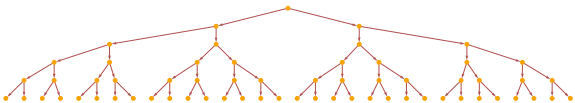

but now the multiway graph is:

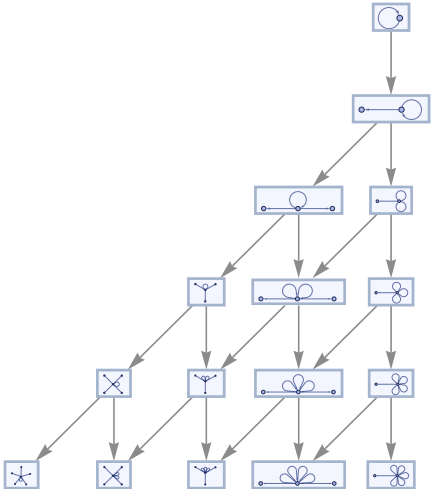

The rule

{{*x*, *y*}} → {{*x*, *y*}, {*y*, *z*}}

gives exactly the same causal graph, but now its hypergraph is a tree:

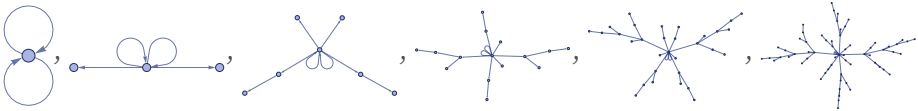

Like some of the rules shown above, its multiway graph is somewhat complex:

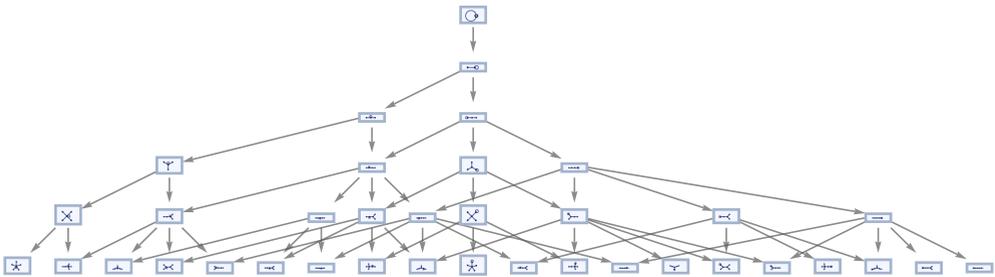

It is actually fairly common to have causal graphs that look like the corresponding hypergraphs in the case of rules where effectively only one update happens at a time. An example occurs in the case of the rule (see 3.10):

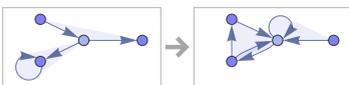

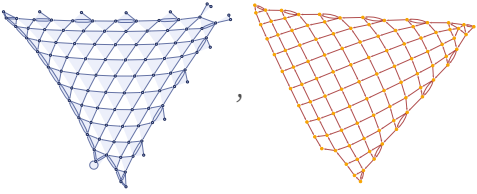

So far, we have only considered fairly minimal conditions. But as soon as it is possible to have multiple independent events occur in the initial conditions, it is also possible to get completely disconnected causal graphs. (Note that if an "initial creation event" to create the initial conditions was added, then the causal graphs would again be connected.) As an example of disconnected causal graphs, consider the rule

{{*x*}} → {{*x*}, {*x*}}

with an initial condition consisting of connected unary relations:

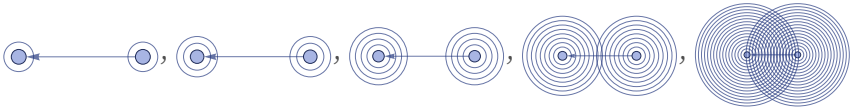

This rule yields a disconnected causal graph:

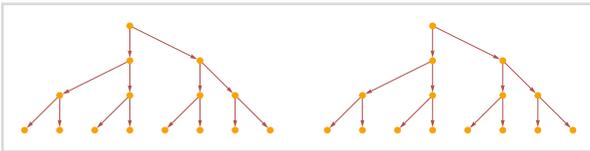

The multiway graph in this case is connected, and shows causal invariance:

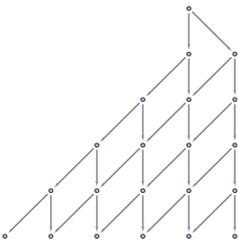

Sometimes the relationship between disconnection in the hypergraph and the existence of disconnected causal graphs can be somewhat complex. This shows results for the rule above with initial conditions consisting of increasing numbers of self-loops:

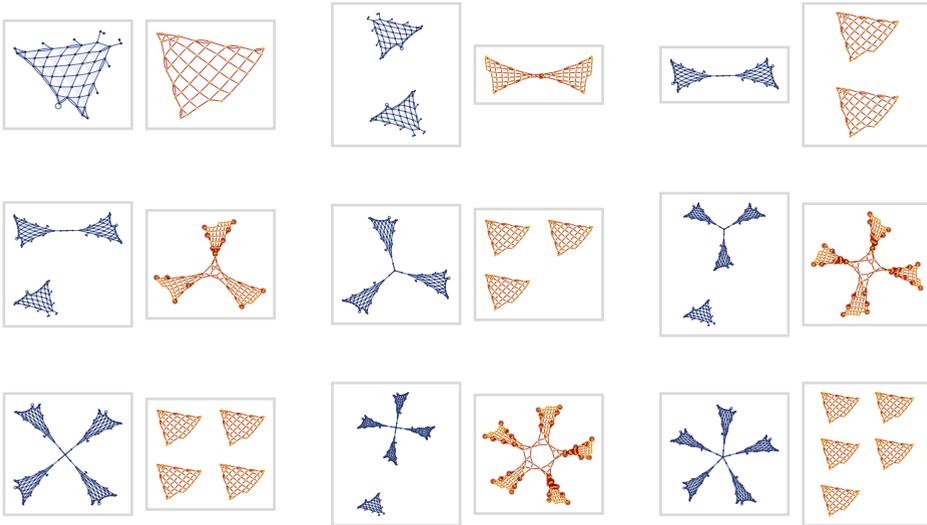

Even with rather simple rules, the forms of branching in causal graphs can be quite complex—even when the actual hypergraphs remain simple. Here are a few examples:

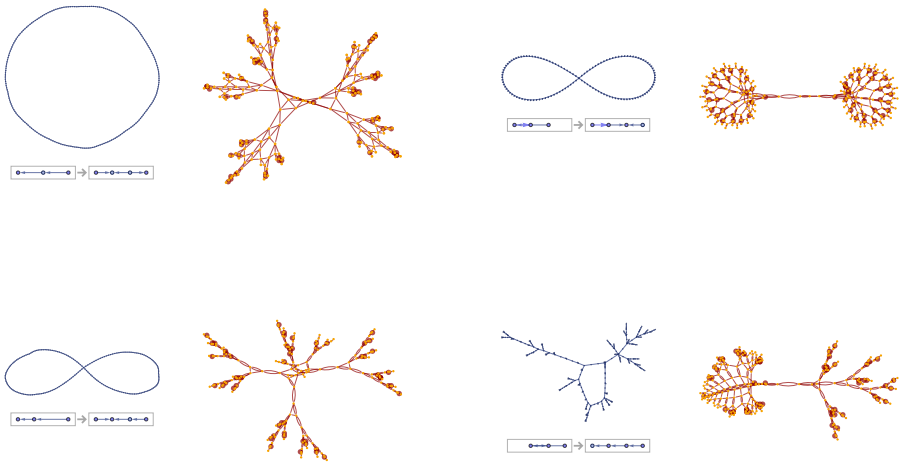

## 6.11 Global Symmetries and Conservation Laws

Given the rule (stated here using numbers rather than our usual letters)

{{1, 2, 3}, {3, 4, 5}} → {{6, 7, 1}, {6, 3, 8}, {5, 7, 8}}

if we reverse the elements in each relation we get:

{{3, 2, 1}, {5, 4, 3}} → {{1, 7, 6}, {8, 3, 6}, {8, 7, 5}}

But the canonical version of this rule is:

{{1, 2, 3}, {3, 4, 5}} → {{6, 7, 1}, {6, 3, 8}, {5, 7, 8}}

In graphical form, the rule and its transform are:

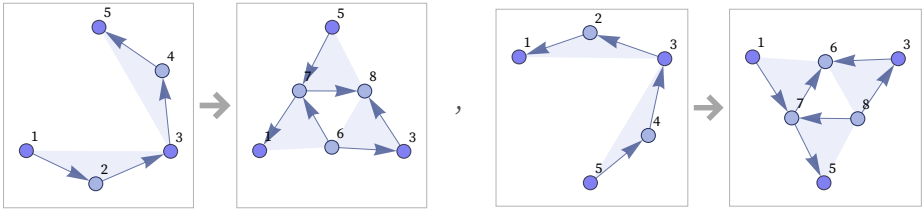

Most rules would not be left invariant under a reversal of each relation. For example, the rule

{{1, 2, 3}, {2, 4, 5}} → {{5, 6, 4}, {6, 5, 3}, {7, 8, 5}}

yields after reversal of each relation

{{3, 2, 1}, {5, 4, 2}} → {{4, 6, 5}, {3, 5, 6}, {5, 8, 7}}

but the canonical form of this is

{{1, 2, 3}, {4, 3, 5}} → {{4, 1, 6}, {2, 6, 1}, {1, 7, 8}}

which is not the same as the original rule.

If a rule is invariant under a symmetry operation such as reversing each relation, it implies that the rule commutes with the symmetry operation. So given a rule $R$ and a symmetry operation $\Theta$, this means that for any state $S$, $R\,(\Theta\,S)$ must be the same as $\Theta\,(R\,S)$.

With the symmetric rule above, evolving from a particular initial state gives:

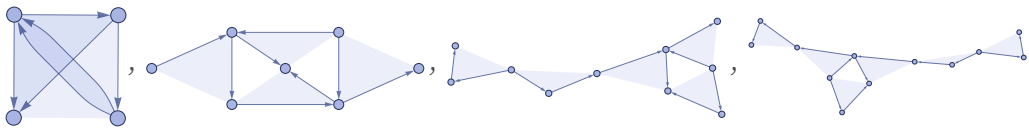

But now reversing the relations in the initial state gives essentially the same evolution, but with states whose relations have been reversed:

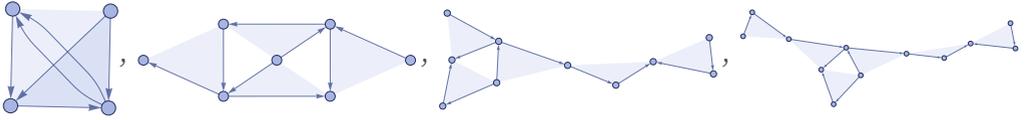

For the nonsymmetric rule above, evolution from a particular initial state gives:

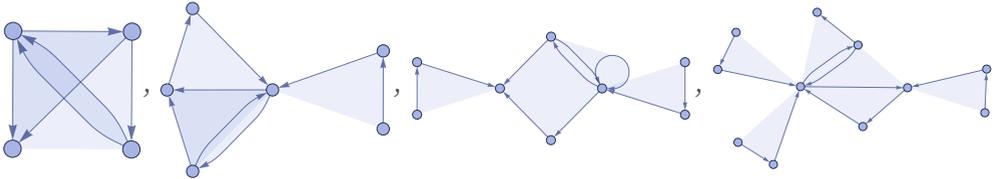

But if one now reverses the relations in the initial state, the evolution is completely different:

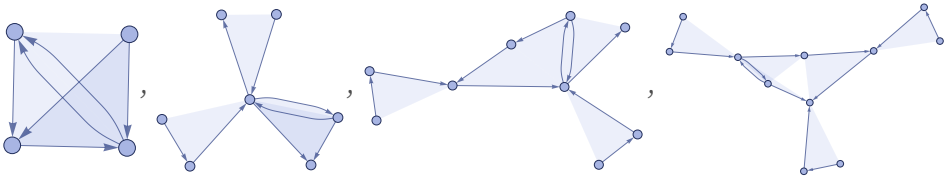

For rules with binary relations, the only symmetry operation that can operate on relations is reversal, corresponding to the permutation {2,1}. Of the 73 distinct $1_2 \to 2_2$ rules, 11 have this symmetry. Of the 4702 $2_2 \to 3_2$ rules, 92 have the symmetry. Of the 40,405 $2_2 \to 4_2$ rules, 363 have the symmetry. Those with the most complex behavior are:

{{1, 2}, {2, 3}} → {{1, 2}, {1, 4}, {2, 3}, {4, 3}}

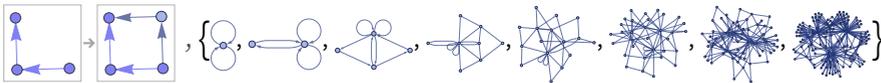

{{1, 2}, {2, 3}} → {{1, 4}, {1, 3}, {4, 5}, {5, 3}}

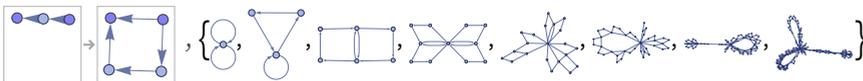

For rules with ternary relations, there are six distinct symmetry classes corresponding to the six subgroups of the symmetric group $S_3$: no invariance, invariance under transposition of two elements (3 cases of $S_2$) ({1,3,2}, {3,2,1} or {2,1,3} only), invariance under cyclic rotation ($A_3$) ({2,3,1} and {3,1,2}), or invariance under any permutation (full $S_3$). Here are the numbers of rules of various signatures with these different symmetries:

|  | $1_3 \to 1_3$ | $1_3 \to 2_3$ | $1_3 \to 3_3$ | $2_3 \to 1_3$ | $2_3 \to 2_3$ | $2_3 \to 3_3$ | $3_3 \to 1_3$ |
|---|---|---|---|---|---|---|---|
| none | 114 | 8520 | 627 072 | 7662 | 759 444 | 79 170 508 | 559 602 |
| $S_2$ (each of 3) | 20 | 282 | 3475 | 248 | 4413 | 63 028 | 2933 |
| $A_3$ | 2 | 4 | 46 | 4 | 8 | 131 | 40 |
| $S_3$ | 2 | 3 | 25 | 3 | 5 | 41 | 21 |

Examples of rules with full $S_3$ symmetry include (compare 7.2):

{{1, 2, 3}, {2, 3, 1}} → {{2, 2, 4}, {4, 3, 3}, {1, 4, 1}}

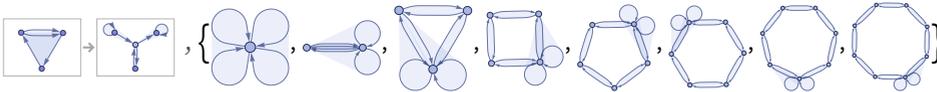

{{1, 2, 3}, {2, 3, 1}} → {{4, 4, 2}, {4, 1, 4}, {3, 4, 4}}

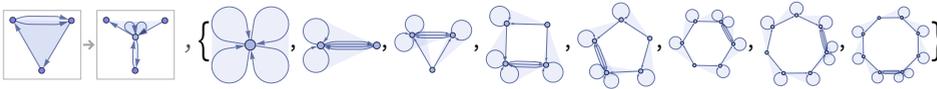

An example of a rule with only cyclic ($A_3$) symmetry is:

{{1, 2, 3}, {2, 3, 1}} → {{1, 1, 4}, {4, 2, 2}, {3, 4, 3}}

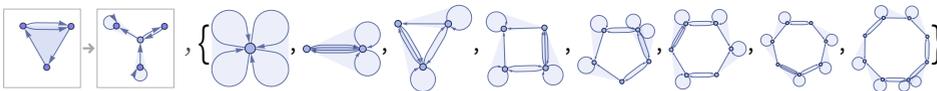

The existence of symmetry in a rule has implications for its multiway graph, effectively breaking its state transition graph into pieces corresponding to different cosets (compare [1:p963]). For example, starting from all 102 distinct 2-element ternary hypergraphs, the first completely symmetric rule above gives multiway system:

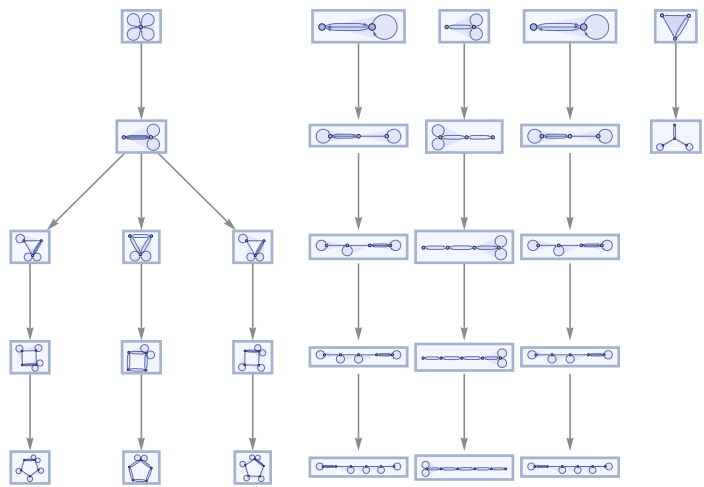

A somewhat simpler example of a completely symmetric rule is:

{{1, 2, 3}, {2, 3, 1}} → {{1, 2, 3}, {2, 3, 1}, {3, 1, 2}}

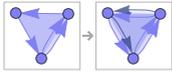

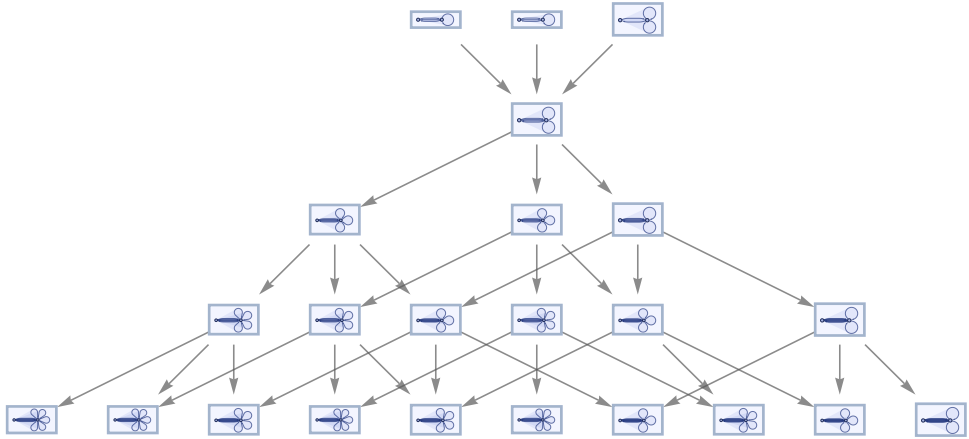

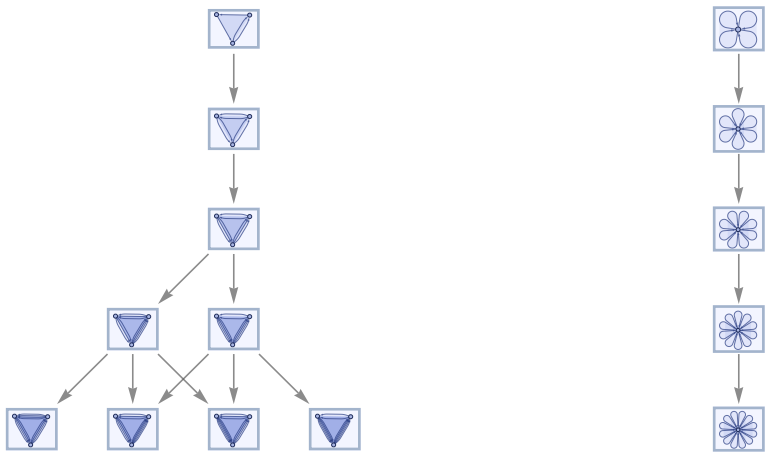

This rule has a simple conservation law: it generates new relations but not new elements. And as a result its multiway graph breaks into multiple separate components.

In general one can imagine many different kinds of conservation laws, some associated with identifiable symmetries, and some not. To get a sense of what can happen, let us consider the simpler case of string substitution systems.

The rule (which has reversal symmetry)

{BA → AB, AB → BA}

gives a multiway graph which consists of separate components distinguished by their total numbers of As and Bs:

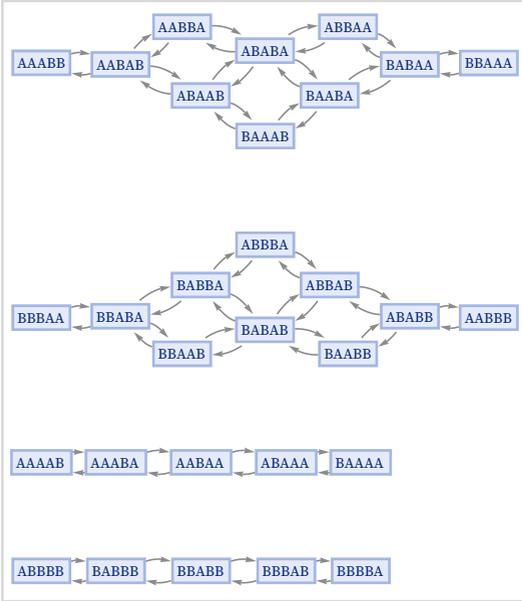

The rule

{AA → BB, BB → AA}

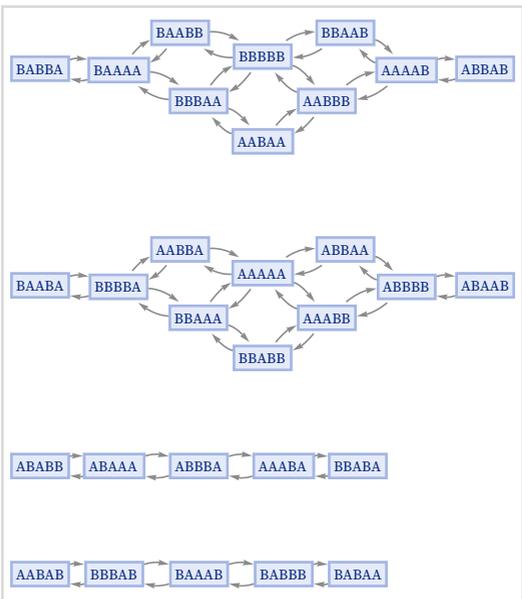

gives the same basic structure, but now what distinguishes the components is the difference in the number of ABs vs. BAs that occur in each string. In both these examples, the number of distinct components increases linearly with the length of the strings.

The rule

{AA → BB, AB → BA}

already gives exactly two components, one with an even number of Bs, and one with an odd number:

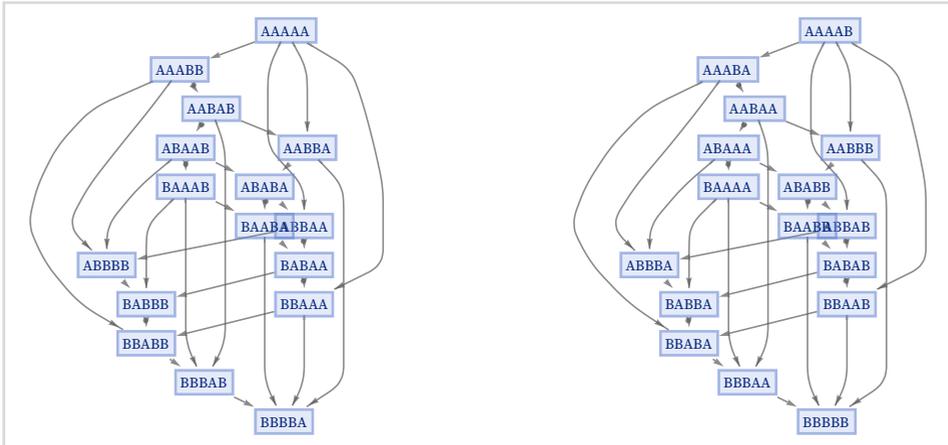

The rule

{AB → AA, BB → BA}

also gives two components, but now these just correspond to strings that start with A or start with B.

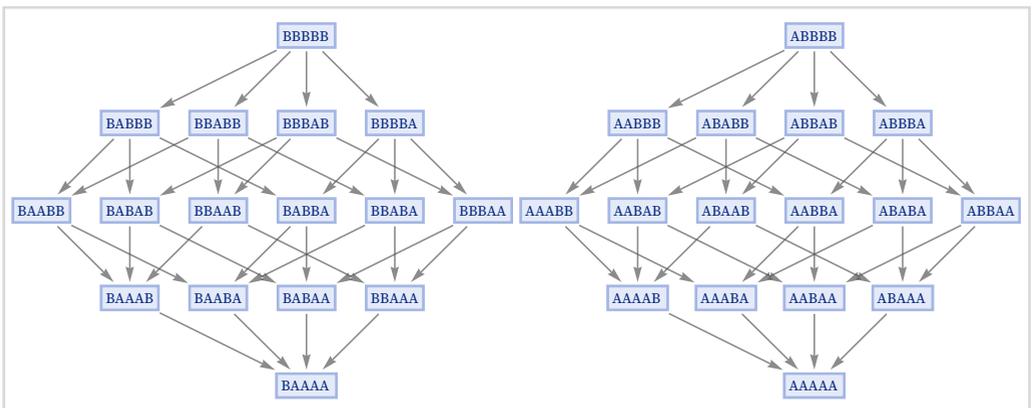

## 6.12 Local Symmetries

In the previous subsection, we considered symmetries associated with global transformations made on all relations in a system. Here we will consider symmetries associated with local transformations on relations involved in particular rule applications.

Every time one does an update with a given rule, say

$\{\{x, y\}, \{z, y\}\} \to \{\{x, x\}, \{y, y\}, \{z, w\}\}$

one needs to match the "variables" that appear on the left-hand side with actual elements in the hypergraph. But in general there may be multiple ways to do this. For example, with the hypergraph

$\{\{1, 2\}, \{3, 2\}\}$

one could either match

$\{x \to 1, y \to 2, z \to 3\}$

or:

$\{z \to 1, y \to 2, x \to 3\}$

The possible permutations of matches correspond to the automorphism group of the hypergraph that represents the left-hand side of the rule.

For $2_2$ hypergraphs of which $\{\{x,y\},\{y,z\}\}$ is an example, there are only two possible automorphism groups: the trivial group (i.e. no invariances), and the group $S_2$ (i.e. permutations $\{2,3\}$, $\{1,3\}$ or $\{1,2\}$).

Here are automorphism groups for binary and ternary hypergraphs with various signatures. In each case the group order is included, as are a couple of sample hypergraphs:

| | | |
|---|---|---|
| E | 1 | 5 |
| $\mathbb{Z}_2$ | 2 | 3 |

$2_2$

| | | |
|---|---|---|
| E | 1 | 25 |
| $\mathbb{Z}_2$ | 2 | 4 |
| $\mathbb{Z}_3$ | 3 | 1 |
| $S_3$ | 6 | 2 |

$3_2$

| | | |
|---|---|---|
| E | 1 | 120 |
| $\mathbb{Z}_2$ | 2 | 38 |
| $\mathbb{Z}_2\times\mathbb{Z}_2$ | 4 | 2 |
| $\mathbb{Z}_4$ | 4 | 1 |
| $S_3$ | 6 | 4 |
| $S_4$ | 24 | 2 |

$4_2$

| | | | |
|---|---|---|---|
| E | 1 | 81 | |
| $\mathbb{Z}_2$ | 2 | 21 | |

$2_3$

,

| | | | |
|---|---|---|---|
| E | 1 | 3042 | |
| $\mathbb{Z}_2$ | 2 | 204 | |
| $\mathbb{Z}_3$ | 3 | 10 | |
| $S_3$ | 6 | 12 | |

$3_3$

,

| | | | |
|---|---|---|---|
| E | 1 | 155 542 | |
| $\mathbb{Z}_2$ | 2 | 8427 | |
| $\mathbb{Z}_3$ | 3 | 30 | |
| $\mathbb{Z}_2 \times \mathbb{Z}_2$ | 4 | 179 | |
| $\mathbb{Z}_4$ | 4 | 15 | |
| $S_3$ | 6 | 174 | |
| $D_4$ | 8 | 12 | |
| $S_4$ | 24 | 12 | |

$4_3$

If the right-hand side of a rule has at least as high a symmetry as the left-hand side, then any possible permutation of matches of elements will lead to the same result—which means that the same update will occur, and the only consequence will be a potential change in path weightings in the multiway graph.

But if the right-hand side of the rule has lower symmetry than the left-hand side (i.e. its automorphism group is a proper subgroup), then different permutations of matches can lead to different outcomes, on different branches of the multiway system. It may still nevertheless be the case that some permutations will lead to identical outcomes—and this will happen whenever the canonical form of the rule is the same after a permutation of the elements on the left-hand side (cf. [87]).

Thus for example the rule

{{x, y}, {z, y}} → {{x, x}, {y, y}, {z, w}}

is invariant under any of the permutations

{{1, 2, 3}, {2, 1, 3}, {3, 1, 2}, {3, 2, 1}}

of the elements corresponding to {x, y, z}. Note that the permutations that appear here do not form a group. To compose multiple such transformations one must take account of relabeling on the right-hand side as well as the left-hand side.

For the 3138 $2_2 \to 3_2$ rules that involve 3 elements on the left, the following lists the 10 of 64 subsets of the 6 possible permutations that occur:

| | | |
|---|---|---|
| 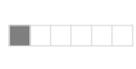 | 1164 | 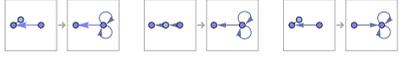 |
| 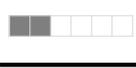 | 808 | 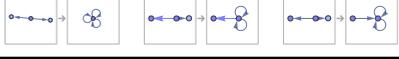 |
| 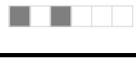 | 113 | 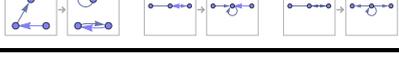 |
| 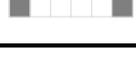 | 808 | 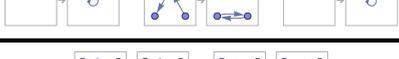 |
| 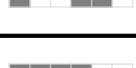 | 2 | 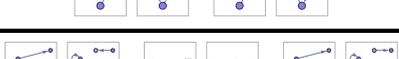 |
| 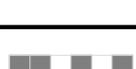 | 16 | 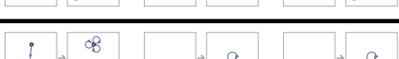 |
| 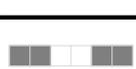 | 94 | 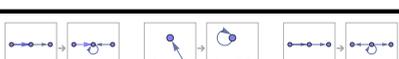 |
| 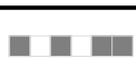 | 97 | 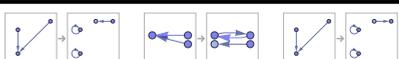 |
| 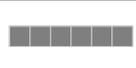 | 19 | 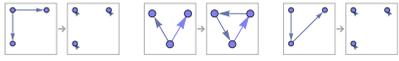 |
| 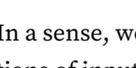 | 17 | 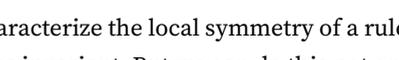 |

In a sense, we can characterize the local symmetry of a rule by determining what permutations of inputs it leaves invariant. But we can do this not only for a single update, but for a sequence of multiple updates. In effect, all we have to do is to form a power of the rule, and then apply the same procedure as above.

There are several ways to put a notion of powers (or in general, products) of rules. As one example, we can consider situations in which a rule is applied repeatedly to an overlapping set of elements—so that in effect the successive rule applications are causally connected.

In this case—much as we did for testing total causal invariance—we need to work out the unifications of the possible initial conditions. Then we effectively just need to trace the multiway evolution from each of these unified initial conditions.

Consider for example the rule:

$\{\{x, y\}\} \to \{\{x, y\}, \{y, z\}\}$

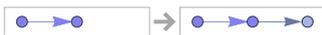

The "square" of this rule is:

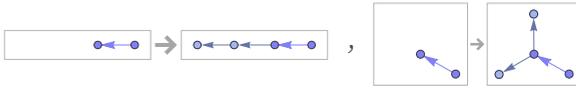

And its cube is:

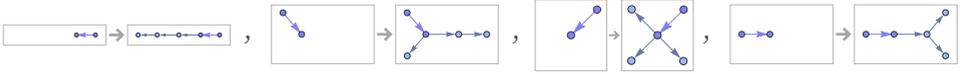

The multiway graph for the original rule after 4 updates is:

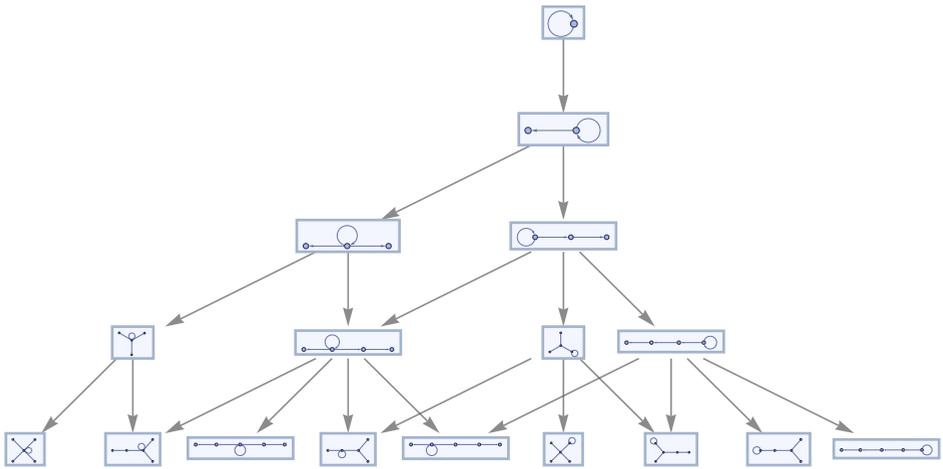

The "square" of the rule generates the same states in only 2 updates:

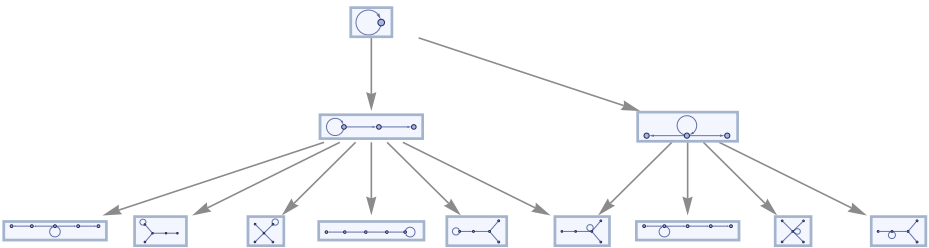

We can use the same approach to find the "square" of a rule like:

$\{\{x, y\}, \{x, z\}\} \to \{\{x, y\}, \{x, w\}, \{y, w\}, \{z, w\}\}$

The result is:

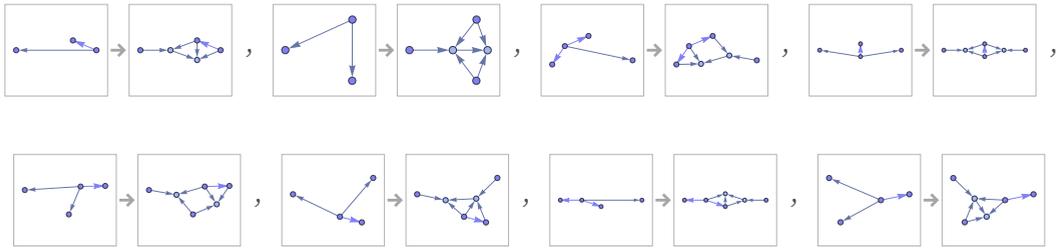

Using this for actual evolution gives the result:

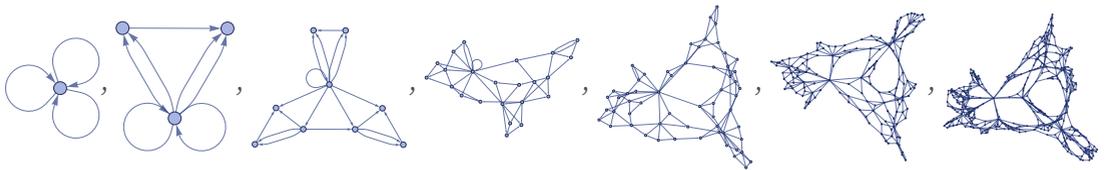

And now that we can compute the power of a rule, we have a way to compute the effective symmetry for multiple updates according to a rule. In general, after *t* updates we will end up with a collection of permutations of variables that leave the effective "power rule" invariant.

But if we now consider increasingly large values of *t*, we can ask whether the collections of permutations we get somehow converge to a definite limit. In direct analogy to the way that our hypergraphs can limit to manifolds, we may wonder whether these collections of permutations could limit to a Lie group (cf. [88]).

As a simple example, say that the permutations are of length *n*, but of the *n*! possibilities, we only have the *n* cyclic permutations, say for *n* = 4:

{{1, 2, 3, 4}, {2, 3, 4, 1}, {3, 4, 1, 2}, {4, 1, 2, 3}}
As $n \to \infty$ we can consider this to limit to the Lie group U(1), corresponding to rotations by any angle $\theta$ on a circle.

It is not so clear [89][90] how to deal in more generality with collections of permutations, although one could imagine an analog of a manifold reconstruction procedure. To get an idea of how this might work, consider the inverse problem of approximating a Lie group by permutations. (Note that things would be much more straightforward if we could build up matrix representations, but this is not the setup we have.)

In some cases, there are definite known finite subgroups of Lie groups—such as the icosahedral group $A_5$ as a subset of the 3D rotation group SO(3). In such cases one can then explicitly consider the permutation representation of the finite group. It is also possible to imagine just taking a lattice (or perhaps some more general structure of the kind that might be used in symbolic dynamics [91][92]) and applying random elements of a particular Lie group to it, then in each case recording the transformation of lattice points that this yields. Typically these transformations will not be permutations, but it may be possible to approximate them as such. By inverting this kind of procedure, one can imagine potentially being able to go from a collection of permutations to an approximating Lie group.

## 6.13 Branchial Graphs and Multiway Causal Graphs

Consider the rule:

{{x, y}, {x, z}} → {{x, y}, {x, w}, {y, w}, {z, w}}

If we pick a foliation for the first few steps in the multiway graph for this rule

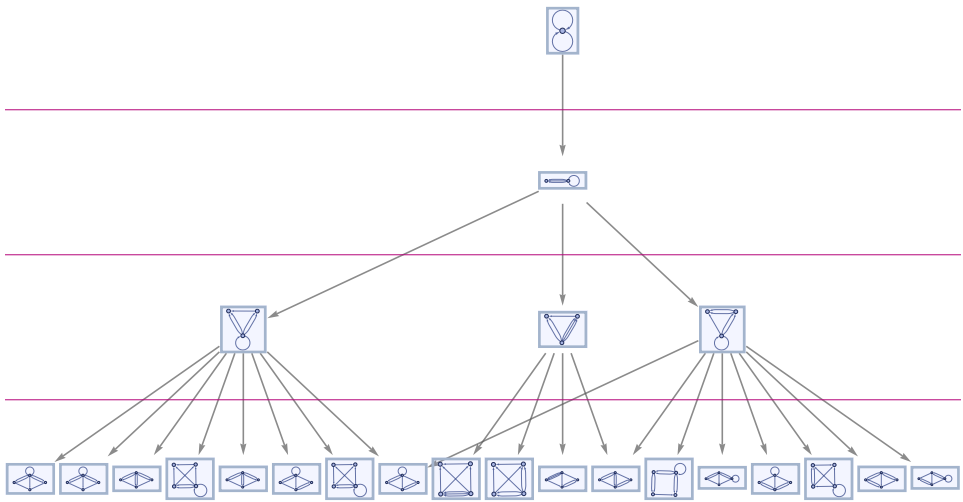

then just as in 5.15 for string substitution systems, we can generate branchial graphs that represent the connections defined by branch pairs between the states at each slice in the foliation:

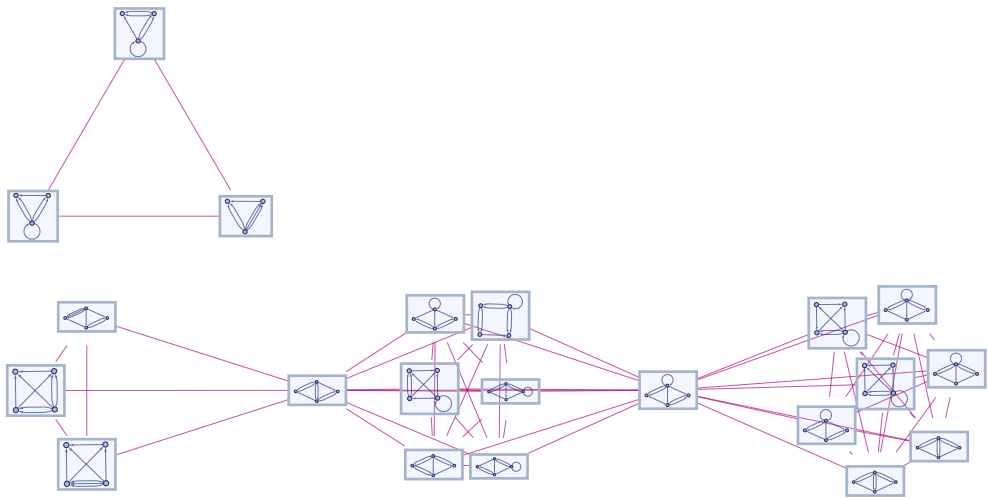

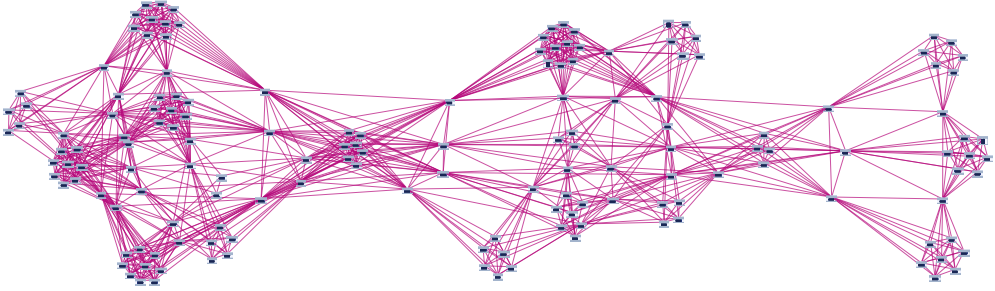

Branchial graphs provide one form of summary of the multiway evolution. Another summary is provided by the multiway causal graph, which includes causal connections between parts of hypergraphs both within a branch of the multiway system, and across different branches:

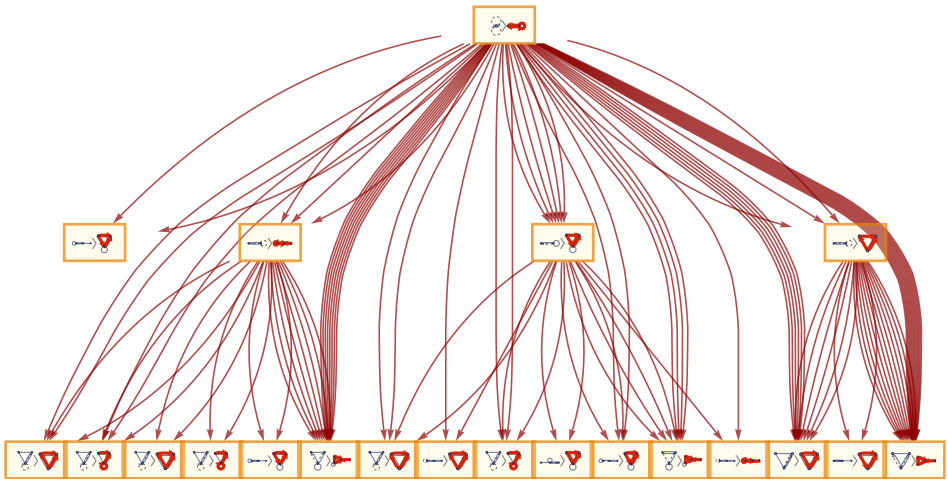

The multiway causal graph is in many respects the richest summary of the behavior of our models, and it will be important in our discussion of possible connections to physics.

In a case like the rule shown, the structure of branchial and multiway causal graphs is quite complex. As a simpler example, consider the causal invariant rule:

$\{\{x, y\}\} \to \{\{x, y\}, \{y, z\}\}$

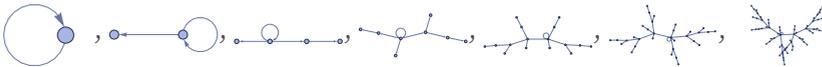

With this rule, the multiway graph has the form:

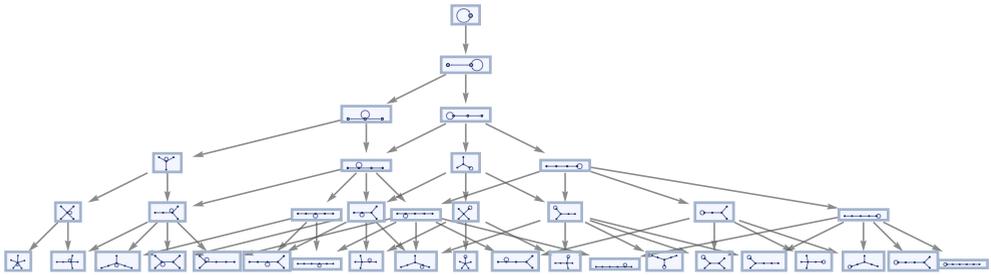

After more steps, and with a different rendering, the multiway graph is:

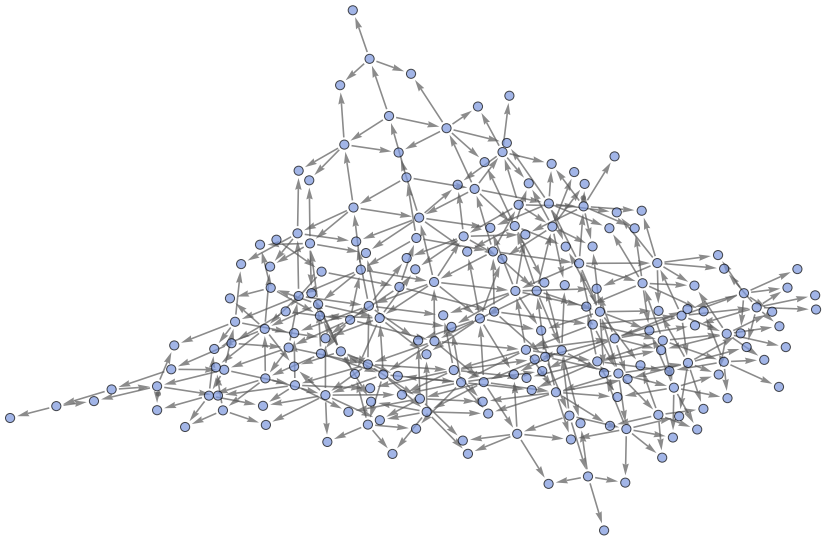

(In this case, the size of the multiway graph as measured by $\Sigma_t$ increases slightly faster than $2^t$.)

The branchial graphs with the standard layered foliation in this case are:

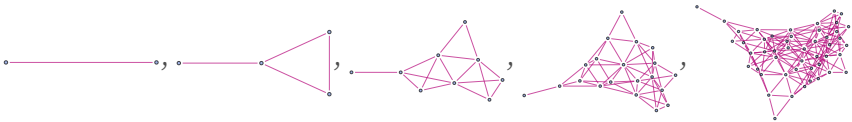

The volumes $B_t$ in the branchial graph grow on successive steps like:

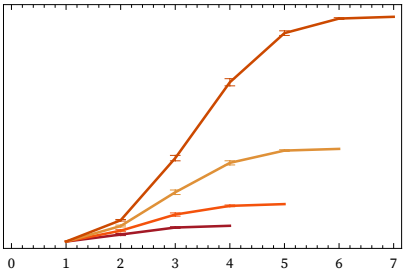

The multiway causal graph in this case is

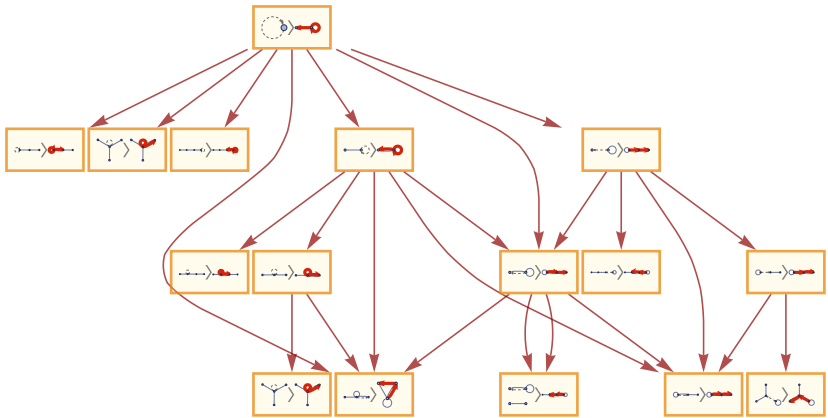

or with more steps and a different layout:

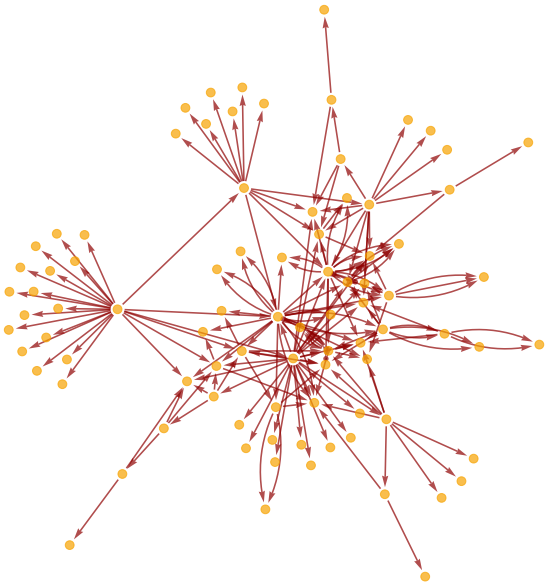

Note that in this case the ordinary multiway graph is simply:

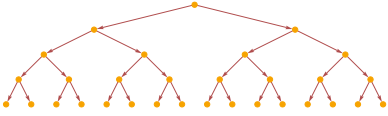

As a slightly more complicated example, consider the causal invariant $2_2 \to 3_2$ rule:

$\{\{x, y\}, \{x, z\}\} \to \{\{y, w\}, \{y, z\}, \{w, x\}\}$

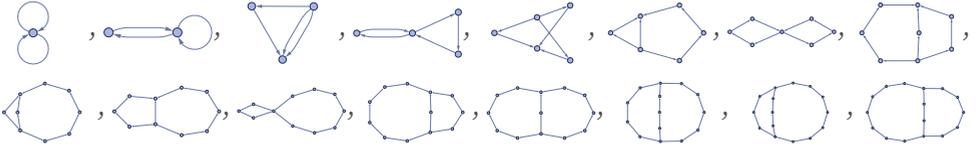

The multiway system in this case has the form:

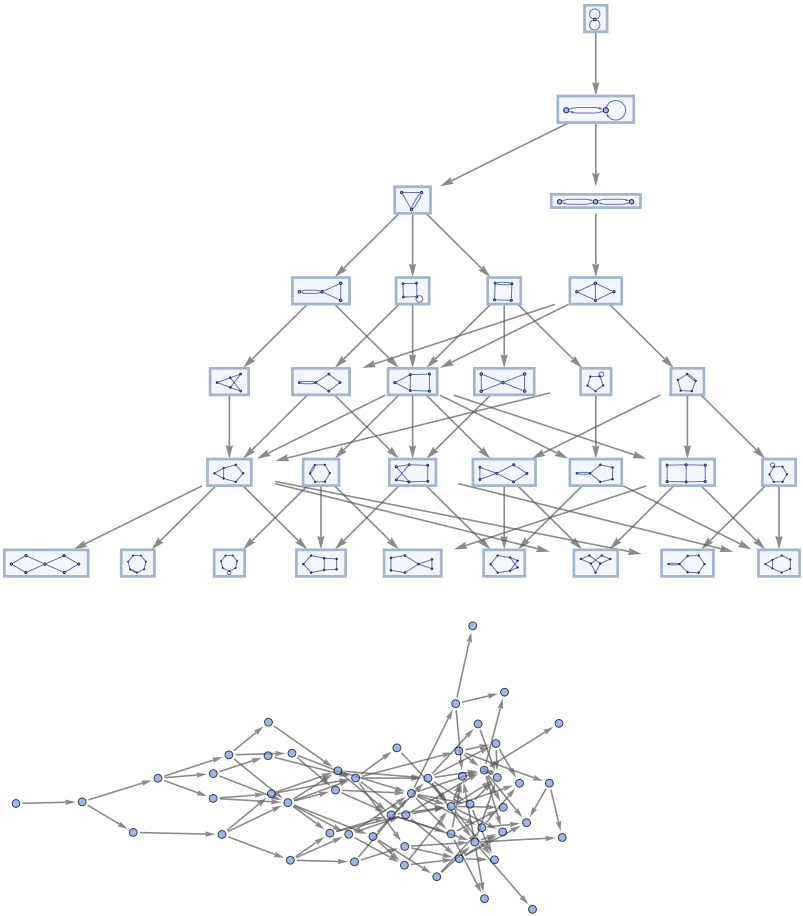

The sequence of branchial graphs in this case are:

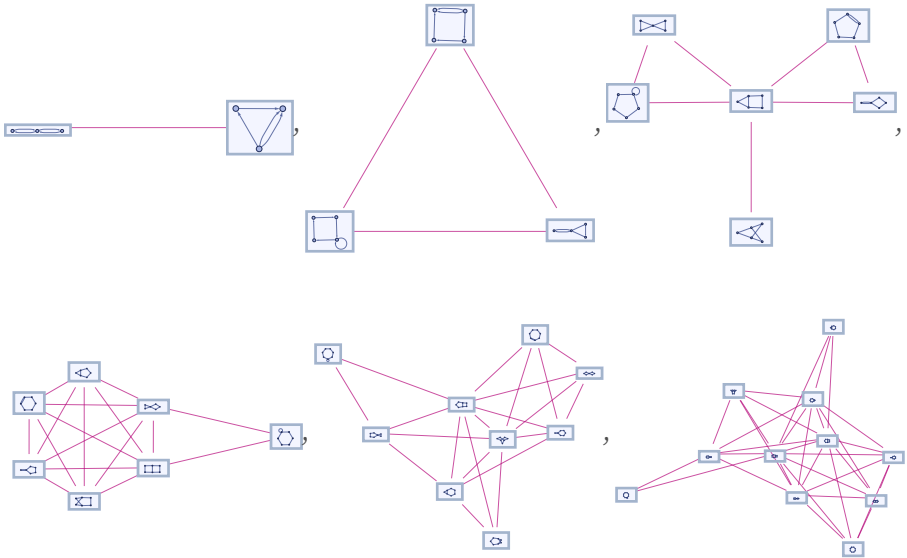

The causal graph for this rule is:

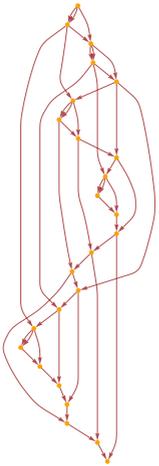

The multiway causal graph has many repeated edges:

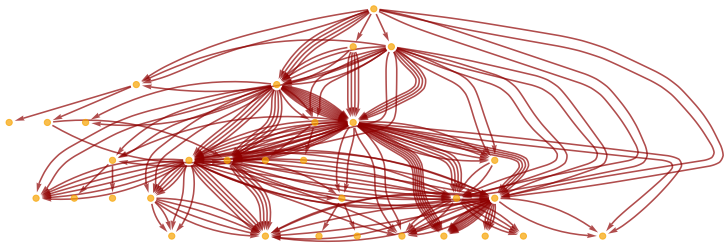

Here it is in a different rendering:

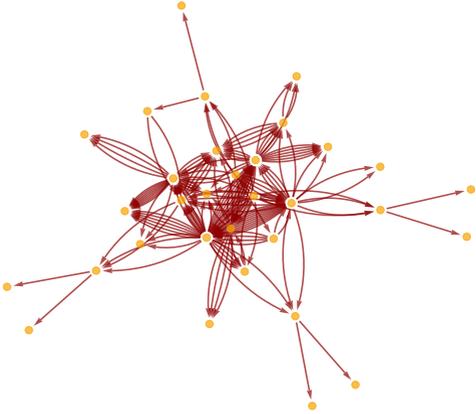

Note that in our models, even when the hypergraphs are disconnected, the branchial graphs can still be connected, as in the case of the rule:

{{x, y}} → {{y, z}, {z, w}}

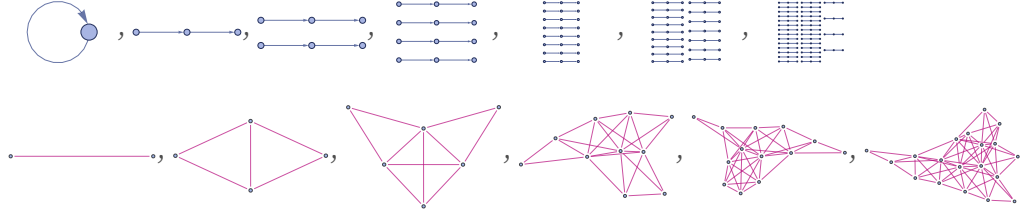

# 7 | Equivalence and Computation in Our Models

## 7.1 Correspondence with Other Systems

Our goal with the models introduced here is to have systems that are intrinsically as structureless as possible, and are therefore in a sense as flexible and general as possible. And one way to see how successful we have been is to look at what is involved in reproducing other systems using our models.

As a first example, consider the case of string substitution systems (e.g [1:3.5]). An obvious way to represent a string is to use a sequence of relations to set up what amounts to a linked list. The "payload" at each node of the linked list is an element of the string. If one has two kinds of string elements A and B, these can for example be represented respectively by 3-ary and 4-ary relations. Thus, for example, the string ABBAB could be (where the Bs can be identified from the "inner self-loops" in their 4-ary relations):

{{0, A1, 1}, {1, B2, B2, 2}, {2, B3, B3, 3}, {3, A4, 4}, {4, B5, B5, 5}}

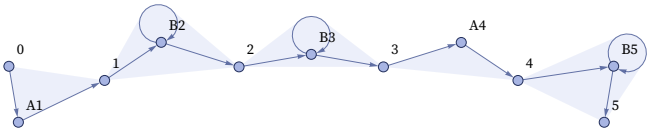

Note that the labels of the elements and the order of relations are not significant, so equivalent forms are

{{1, 7, 2}, {2, 9, 9, 3}, {3, 10, 10, 4}, {4, 8, 5}, {5, 11, 11, 6}}

or, with our standard canonicalization:

{{1, 2, 2, 3}, {3, 4, 4, 5}, {6, 7, 7, 8}, {5, 9, 6}, {10, 11, 1}}

A rule like

{A → BA, B → A}

can then be translated to

{{{0, A1, 1}} → {{0, B1, B1, −2}, {−2, A2, 1}}, {{0, B1, B1, 1}} → {{0, A1, 1}}}

or:

{{{x, y, z}} → {{x, u, u, v}, {v, w, z}}, {{x, y, y, z}} → {{x, u, z}}}

Starting with A, the original rule gives:

{A, BA, ABA, BAABA, ABABAABA, BAABAABABAABA, ABABAABABAABAABABAABA}

In terms of our translation, this is now:

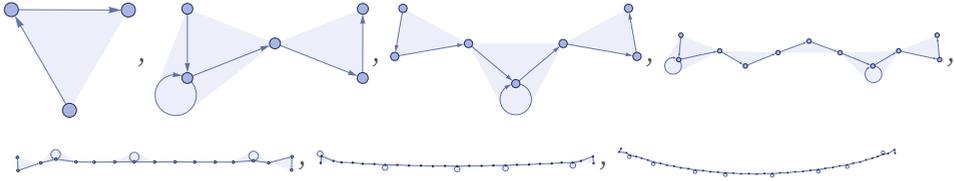

The causal graph for the rule in effect shows the dependence of the string elements

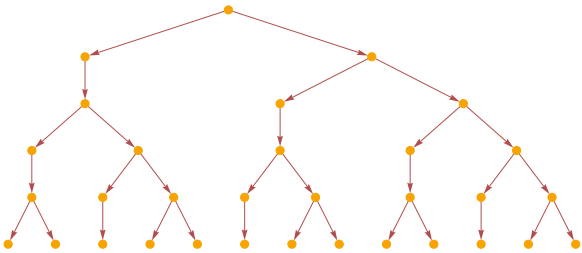

corresponding to the evolution graph (see 5.1):

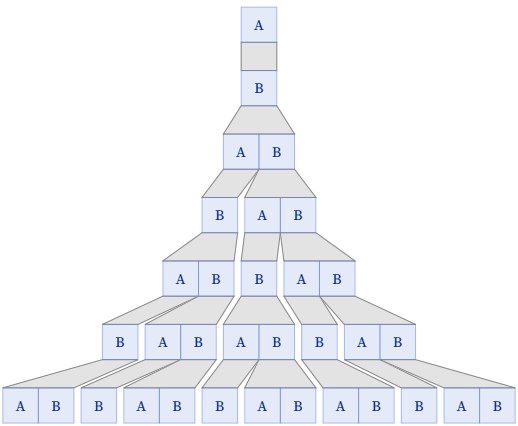

We can also take our translation of the string substitution system, and use it in a multiway system:

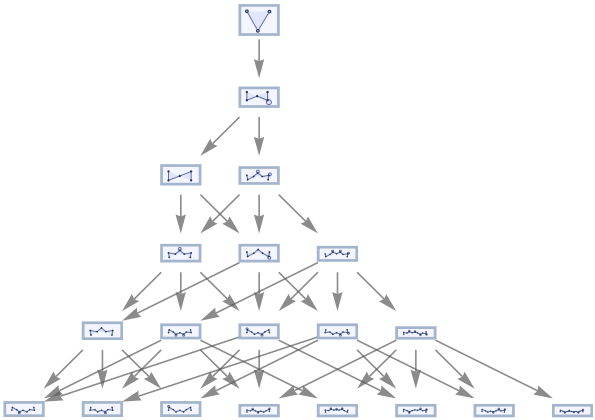

The result is a direct translation of what we could get with the underlying string system:

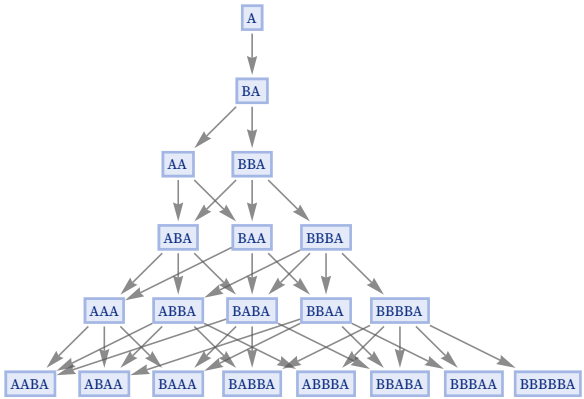

Having seen how our models can reproduce string substitution systems, we consider next the slightly more complex case of reproducing Turing machines [93].

As an example, consider the simplest universal Turing machine [1:p709][94][95] [96], which has the 2-state 3-color rule:

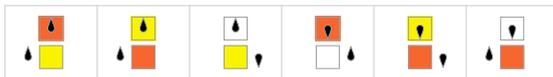

Given a tape with the Turing machine head at a certain position

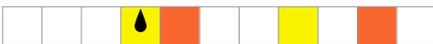

a possible encoding uses different-arity hyperedges to represent different values on the tape, and different states for the head, then attaches the head to a certain position on the tape, and uses special (in this case 6-ary) hyperedges to provide "extensible end caps" to the tape:

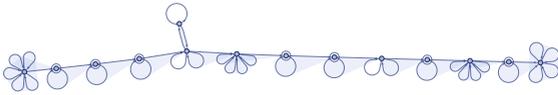

With this setup, the rule can be encoded as:

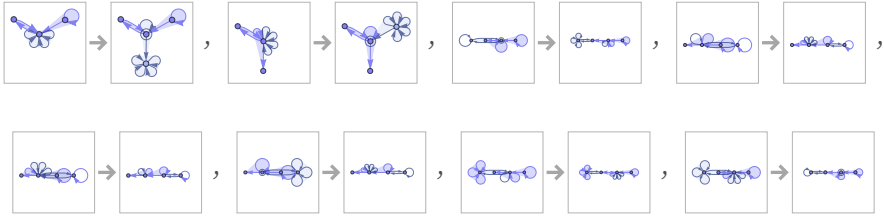

Starting from a representation of a blank tape, the first few steps of evolution are (note that the tape is extended as needed)

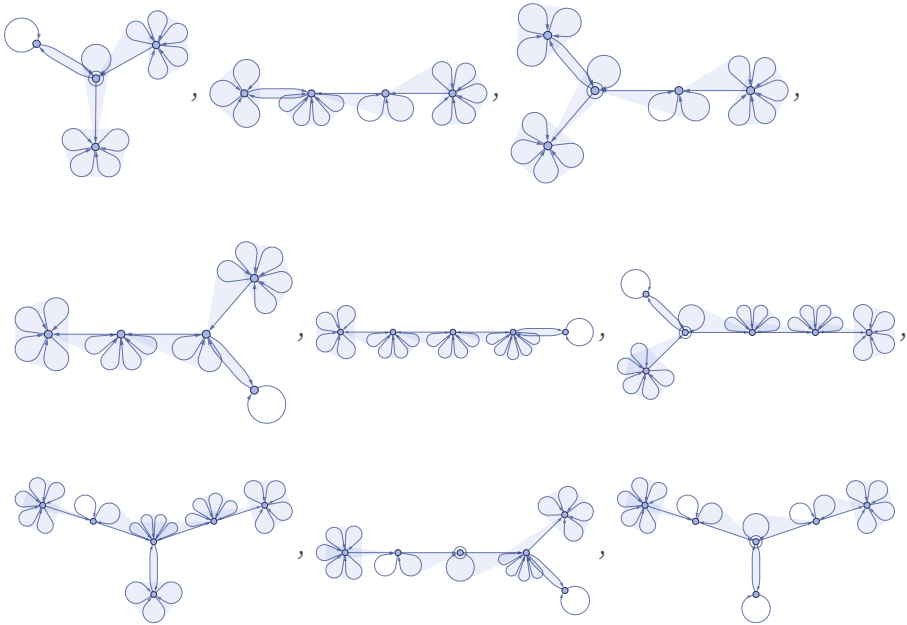

which corresponds to the first few steps of the Turing machine evolution:

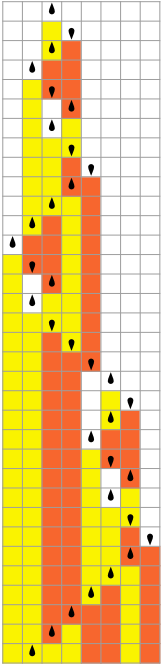

The causal graph for our model directly reflects the motion of the Turing machine head, as well as the causal connections generated by symbols "remembered" on the tape between head traversals:

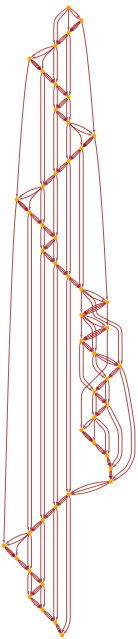

Re-rendering this causal graph, we see that it begins to form a grid:

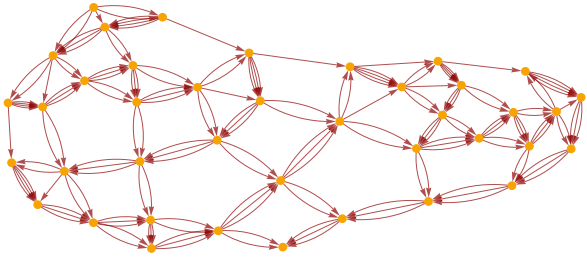

It is notable that even though in the underlying Turing machine only one action happens at each step, the causal graph still connects many events in parallel (cf. [1:p489]). After 1000 steps the graph has become a closer approximation to a flat 2D manifold, with the specific Turing machine evolution reflected in the detailed "knitting" of connections on its surface:

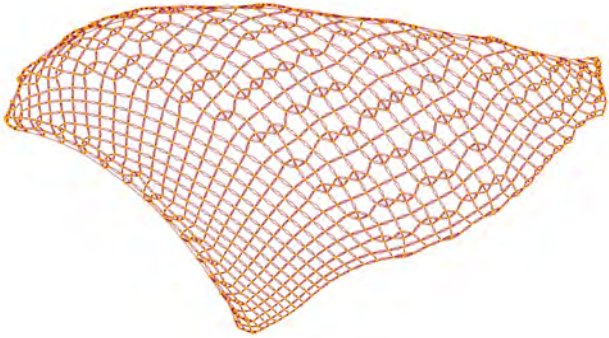

The rule we have set up allows only one thread of history, so the multiway system is trivial:

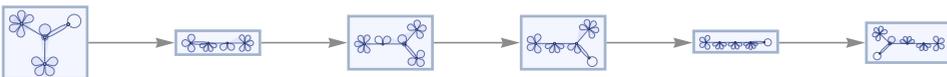

But with our model the underlying setup is general enough that it can handle not only ordinary deterministic Turing machines in which each possible case leads to a specific outcome, but also non-deterministic ones (as used in formulating NP problems) (e.g. [97]), in which there are multiple outcomes for some cases:

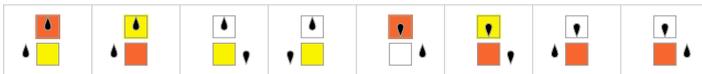

For a non-deterministic Turing machine, there can be multiple paths in the multiway system:

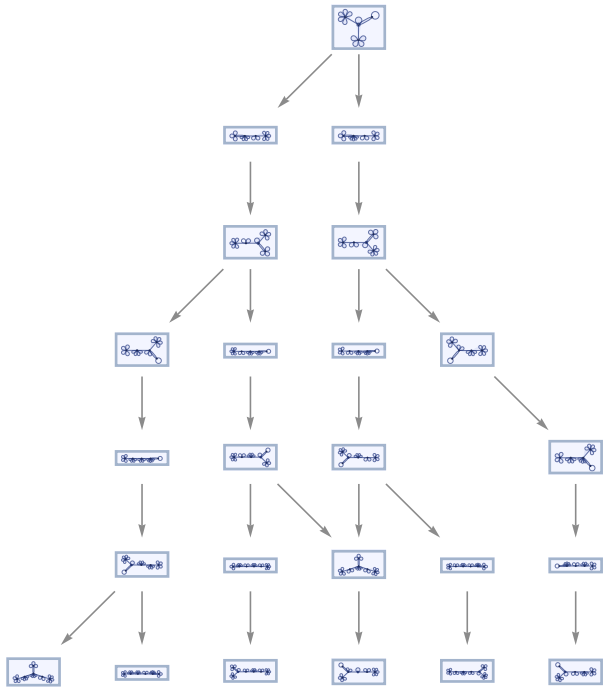

Continuing this, we see that the non-deterministic Turing machine shows a fairly complex pattern of branching and merging in the multiway system (this particular example is not causal invariant):

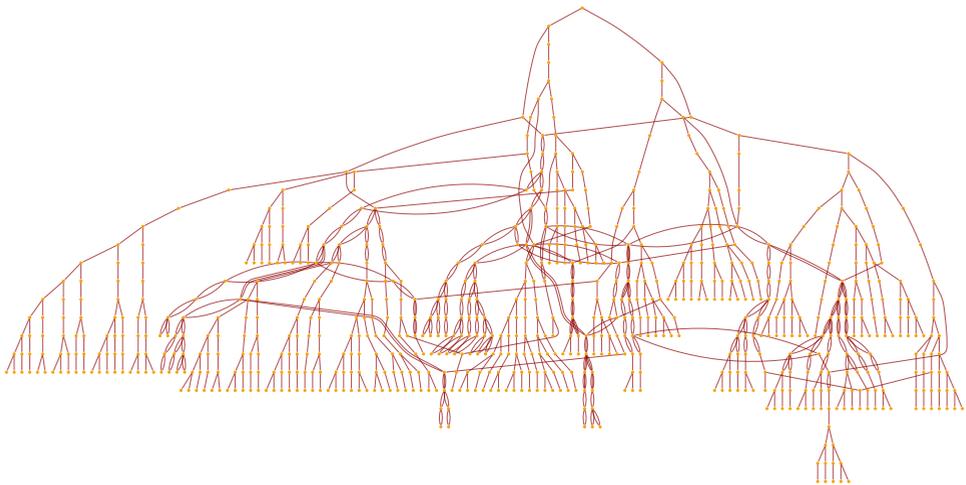

After a few more steps, and using a different rendering, the multiway system has the form:

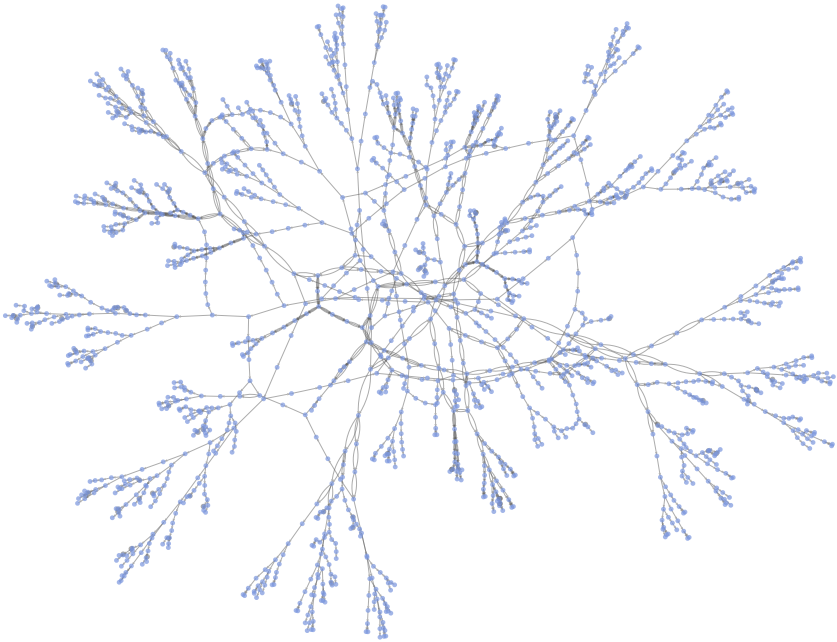

(Note that in actually using a non-deterministic Turing machine, say to solve an NP-complete problem, one needs to check the results on each branch of the multiway system—with different search strategies corresponding to using different foliations in exploring the multiway system.)

As a final example, consider using our models to reproduce cellular automata. Our models are in a sense intended to be as flexible as possible, while cellular automata have a simple but rigid structure. In particular, a cellular automaton consists of a rigid array of cells, with specific, discrete values that are updated in parallel at each step. In our models, on the other hand, there is no intrinsic geometry, no built-in notion of "values", and different updating events are treated as independent and "asynchronous", subject only to the partial ordering imposed by causal relations.

In reproducing a Turing machine using our models, we already needed a definite tape that encodes values, but we only had to deal with one action happening at a time, so there was no issue of synchronization. For a cellular automaton, however, we have to arrange for synchronization of updates across all cells. But as we will see, even though our models ultimately work quite differently, there is no fundamental problem in doing this with the models.

For example, given the rule 30 cellular automaton

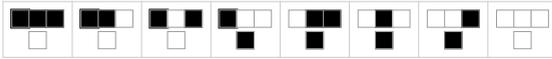

we encode a state like

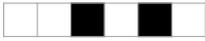

in the form

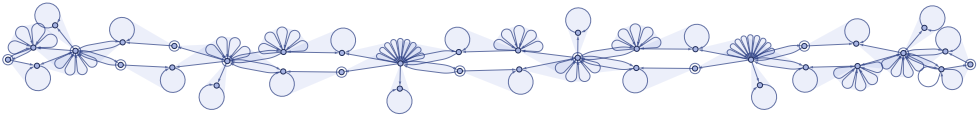

where the 6-ary self-loops represent black cells. Note that there is a quite complex structure that in effect maintains the cellular automaton array, complete with "extensible end caps" that allow it to grow.

Given this structure, the rule corresponding to the rule 30 cellular automaton becomes

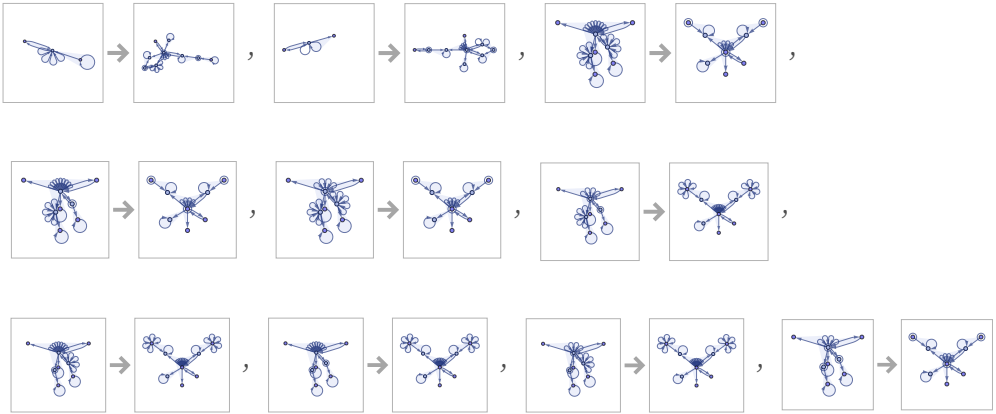

where the first two transformations relate to the end caps, and the remaining 8 actually implement the various cases of the cellular automata rule. Applying the rule for our model for a few steps to an initial condition consisting of a single black cell, we get:

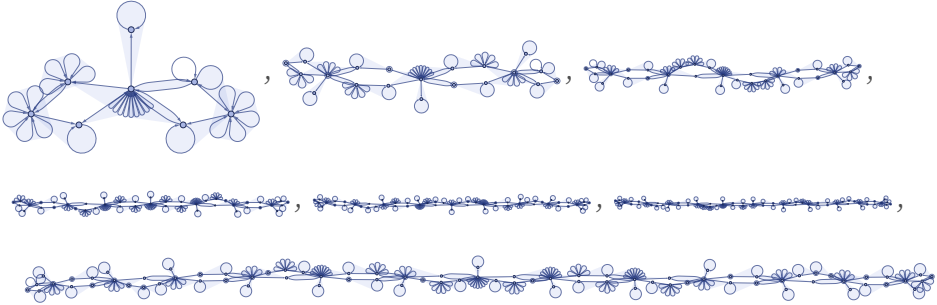

Each of these steps has information on certain cells in the cellular automaton at a certain step in the cellular automaton evolution. "Decoding" each of the steps in our model shown above, we get the following, in which the "front" of cellular automaton cells whose values are present at that step in our model are highlighted:

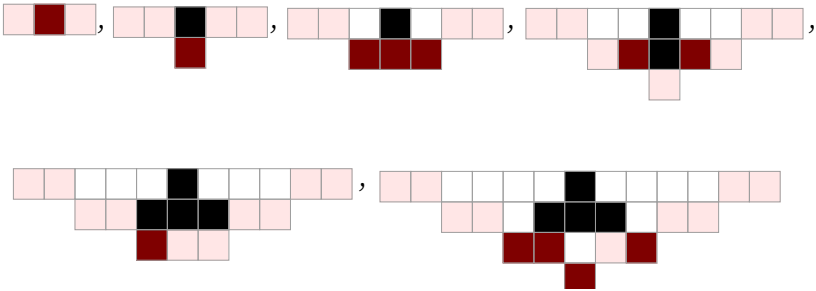

The particular foliation we have used to determine the steps in the evolution of our model corresponds to a particular foliation of the evolution of the cellular automaton:

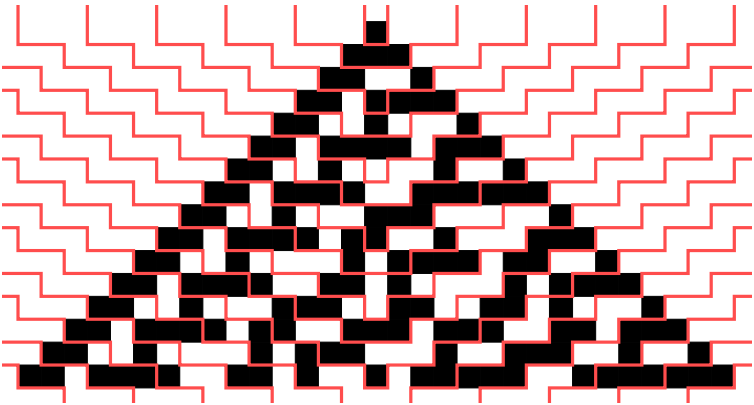

The final "spacetime" cellular automaton pattern is the same, but the foliation defines a specific order for building it up. We can visualize the way the data flows in the computation by looking at the causal graph (with events forming cells with different colors indicated):

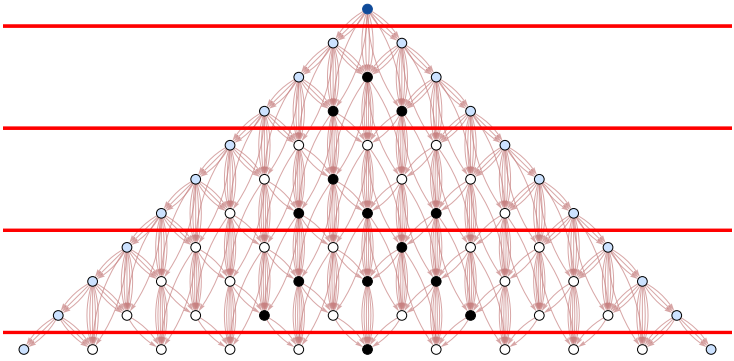

Here is the foliation of the causal graph that corresponds to each step in a traditional synchronized parallel cellular automaton updating:

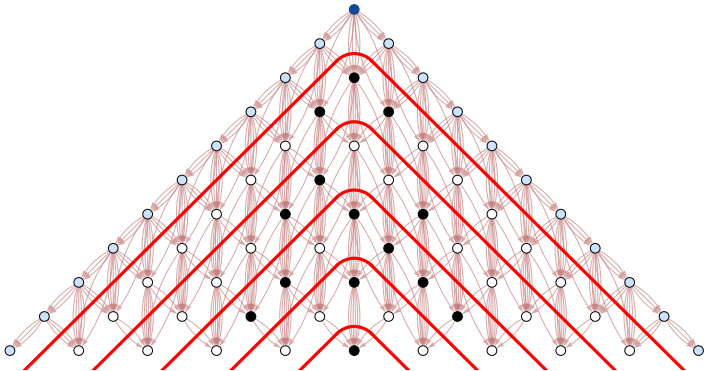

## 7.2 Alternative Formulations

We have formulated our models in terms of the rewriting of collections of relations between elements. And in this formulation, we might represent a state in one of our models as a list of (here 3-ary) relations

{{1, 2, 2}, {3, 1, 4}, {3, 5, 1}, {6, 5, 4}, {2, 7, 6}, {8, 7, 4}}

and the rule for the model as:

{{x, y, z}, {z, u, v}} → {{w, z, v}, {z, x, w}, {w, y, u}}

where $x, y, \ldots$ are taken to be pattern or quantified variables, suggesting notations like [98]

{{x_, y_, z_}, {z_, u_, v_}} → {{w, z, v}, {z, x, w}, {w, y, u}}

**331**

or [99]

$$\forall_{\{x,y,z,u,v\}} (\{\{x, y, z\}, \{z, u, v\}\} \to \{\{w, z, v\}, \{z, x, w\}, \{w, y, u\}\})$$

An alternative to these kinds of symbolic representations is to think—as we have often done here—in terms of transformations of directed hypergraphs. The state of one of our models might then be represented by a directed hypergraph such as

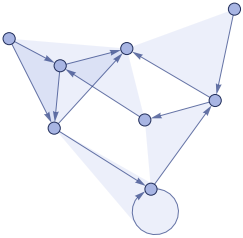

while the rule would be:

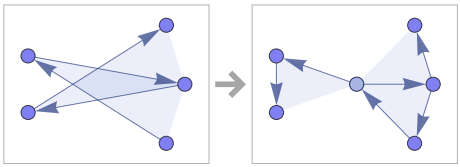

But in an effort to understand the generality of our models—as well as to see how best to enumerate instances of them—it is worthwhile to consider alternative formulations.

One possibility to consider is ordinary graphs. If we are dealing only with binary relations, then our models are immediately equivalent to transformations of directed graphs.

But if we have general *k*-ary relations in our models, there is no immediate equivalence to ordinary graphs. In principle we can represent a *k*-ary hyperedge (at least for *k* > 0) by a sequence of ordinary graph edges:

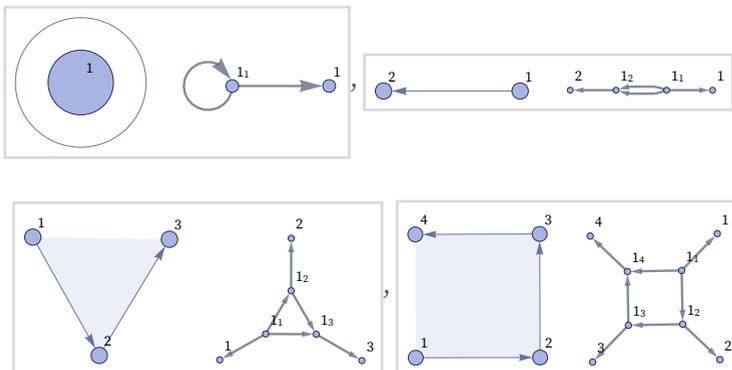

For the hypergraph above, this then yields:

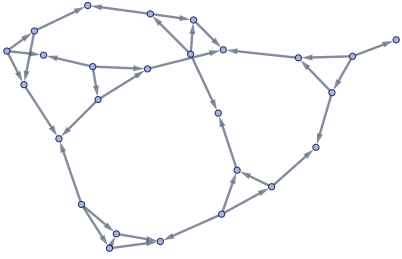

The rule above can be stated in terms of ordinary directed graphs as:

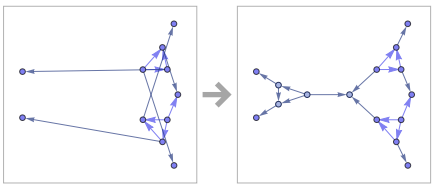

In terms of hypergraphs, the result of 5 and 10 steps of evolution according to this rule is

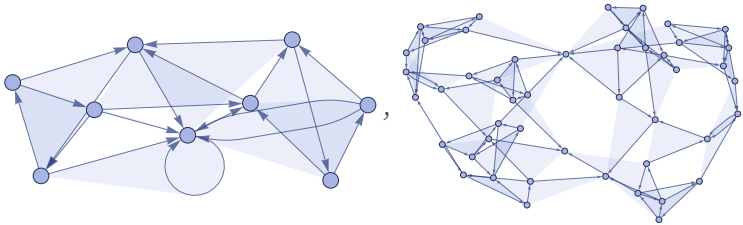

and the corresponding result in terms of ordinary directed graphs is:

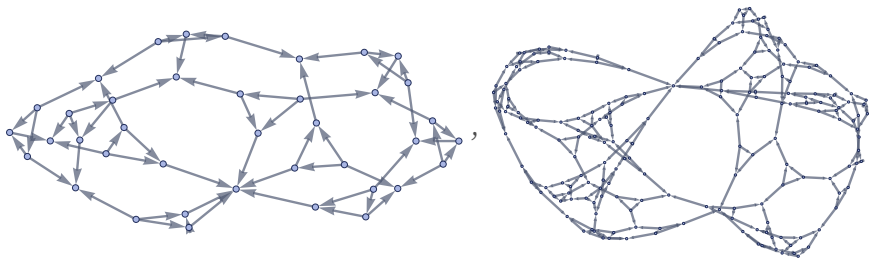

In thinking about ordinary graphs, it is natural also to consider the undirected case. And indeed—as was done extensively in [1:c9]—it is possible to study many of the same things we do here with our models also in the context of undirected graphs. However, transformations of undirected graphs lack some of the flexibility and generality that exist in our models based on directed hypergraphs.

It is straightforward to convert from a system described in terms of undirected graphs to one described using our models: just represent each edge in the undirected graph as a pair of directed binary hyperedges, as in:

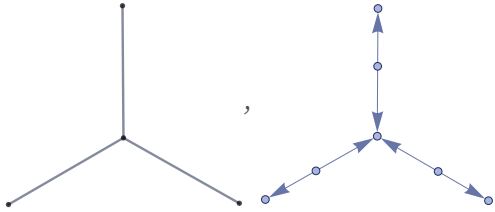

Transformations of undirected graphs work the same—though with paired edges. So, for example, the rule

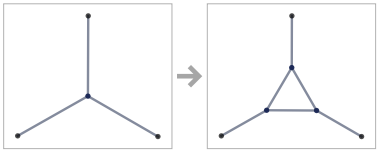

which yields

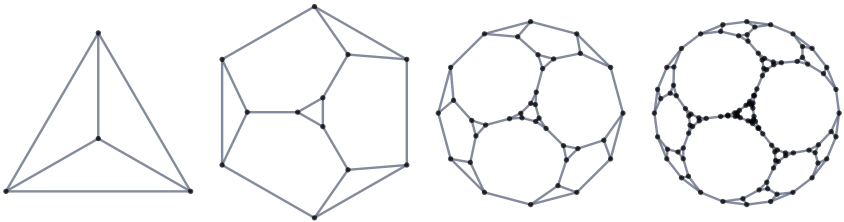

becomes

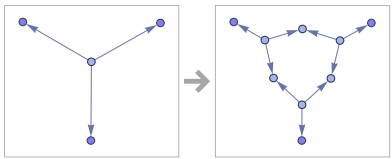

which yields:

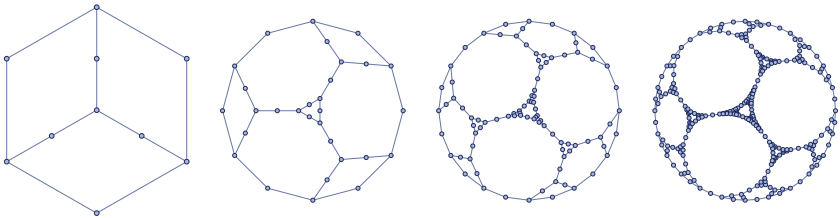

In dealing with undirected graphs—as in [1:c9]—it is natural to make the further simplification that all graphs are trivalent (or "cubic"). In the context of ordinary graphs, nothing is lost by this assumption: any higher-valence node can always be represented directly as a combination of trivalent nodes. But the point about restricting to trivalent graphs is that it makes the set of possible rules better defined—because without this restriction, one can easily end up having to specify an infinite family of rules to cover graphs of arbitrary valence that are generated. (In our models based on transformations for arbitrary relations, no analogous issue comes up.)

It is particularly easy to get intricate nested structures from rules based on undirected trivalent graphs; it is considerably more difficult to get more complex behavior:

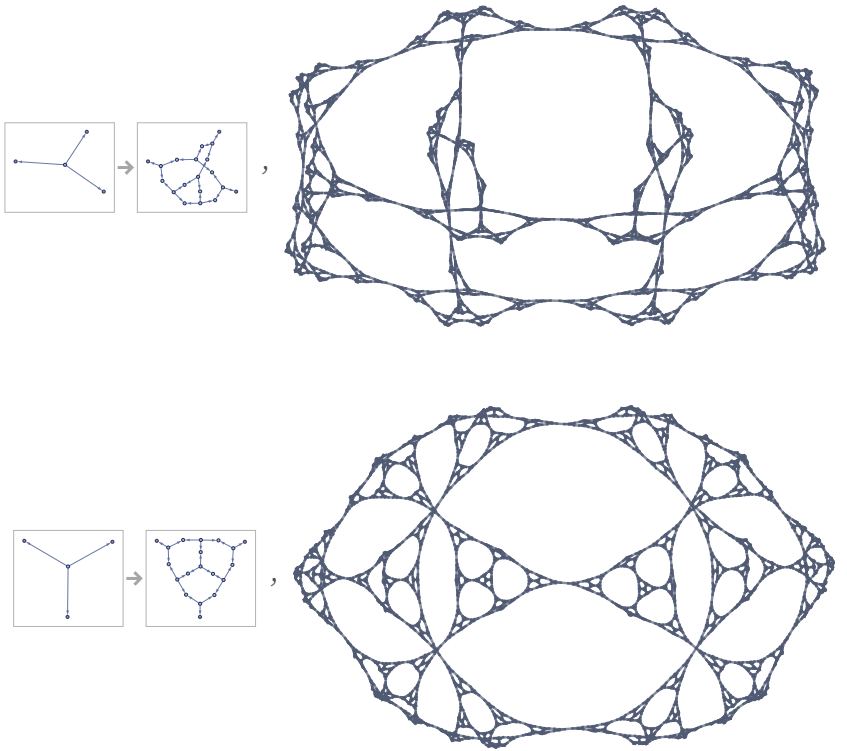

Another issue in models based on undirected graphs has to do with the fact that the objects that appear in their transformation rules do not have exactly the same character as the objects on which they act. In our hypergraph-based models, both sides of a transformation are collections of relations (that can be represented by hypergraphs)—just like what appears in the states on which these transformations act. But in models based on undirected graphs, what appears in a transformation is not an ordinary graph: instead it is a subgraph with "dangling connections" (or "half-edges") that must be matched up with part of the graph on which the transformation acts.

Given this setup, it is then unclear, for example, whether or not the rule above—stated in terms of undirected graphs—should be considered to match the graph:

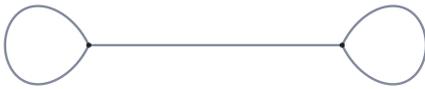

(In a sense, the issue is that while our models are based on applying rules to collections of complete hyperedges, models based on undirected graphs effectively apply rules to collections of nodes, requiring "dangling connections" to be treated separately.)

Another apparent problem with undirected trivalent graphs is that if the right-hand side of a transformation has lower symmetry than the left-hand side, as in then it can seem "undefined" how the right-hand side should be inserted into the final graph. Having seen our models here, however, it is now clear that this is just one of many examples where multiple different updates can be applied, as represented by multiway systems.

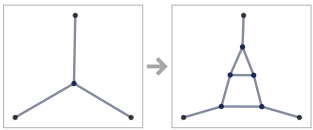

A further issue with systems based on undirected trivalent graphs has to do with the enumeration of possible states and possible rules. If a graph is represented by pairs of vertices corresponding to edges, as in

{{1, 2}, {1, 3}, {1, 4}, {2, 3}, {2, 4}, {3, 4}}

the fact that the graph is trivalent in a sense corresponds to a global constraint that each vertex must appear exactly three times. The alternate "vertex-based" representation

{1 → {2, 3, 4}, 2 → {1, 3, 4}, 3 → {1, 2, 4}, 4 → {1, 2, 3}}

does not overcome this issue. In our models based on collections of relations, however, there are no such global constraints, and enumeration of possible states—and rules—is straightforward. (In our models, as in trivalent undirected graphs, there is, however, still the issue of canonicalization.)

In the end, though, it is still perfectly possible to enumerate distinct trivalent undirected graphs (here dropping cases with self-loops and multiple edges)

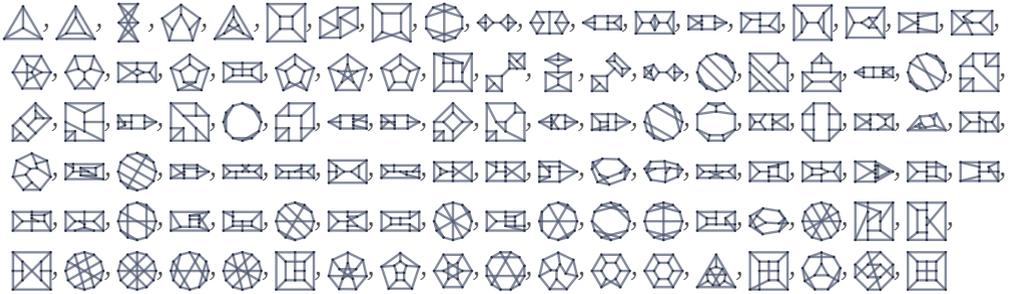

as well as rules for transforming them, and indeed to build up a rich analysis of their behavior [1:9.12]. Notions such as causal invariance are also immediately applicable, and for example one finds that the simplest subgraphs that do not overlap themselves, and so guarantee causal invariance, are [1:p515][87]:

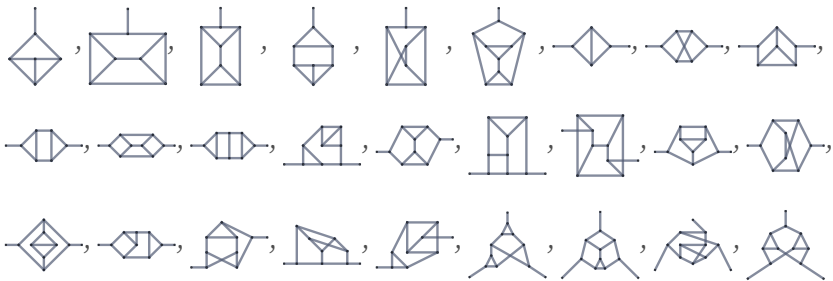

Directed graphs define an ordering for every edge. But it is also possible to have ordered graphs in which the individual edges are undirected, but an order is defined for the edges at any given vertex [87]. Trivalent such ordered graphs can be represented by collections of ordered triples, where each triple corresponds to a vertex, and each number in each triple specifies the destination in the whole list of a particular edge:

{{2, 1, 6}, {5, 4, 3}}

For visualization purposes one can "name" each element of each triple by a color

{<|■ → 2, ■ → 1, ■ → 6|>, <|■ → 5, ■ → 4, ■ → 3|>}

and then the ordered graph can be rendered as:

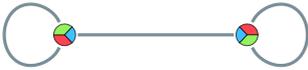

In the context of our models, an ordered trivalent graph can immediately be represented as a hypergraph with ternary hyperedges corresponding to the trivalent nodes, and binary hyperedges corresponding to the edges that connect these nodes:

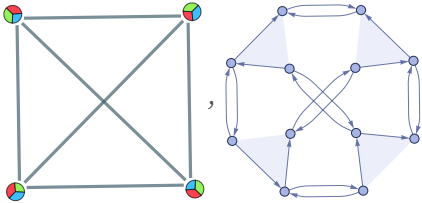

To give rules for ordered trivalent graphs, one must specify how to transform subgraphs with "dangling connections". Given the rule (where letters represent dangling connections)

{{4, *a*, *b*}, {1, *c*, *d*}} → {{4, 8, *a*}, {1, 11, *b*}, {10, 2, *c*}, {7, 5, *d*}}

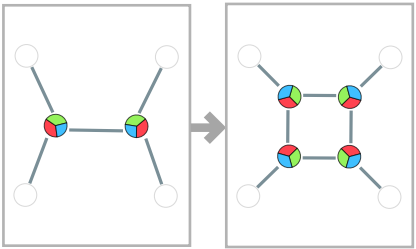

the evolution of the system is:

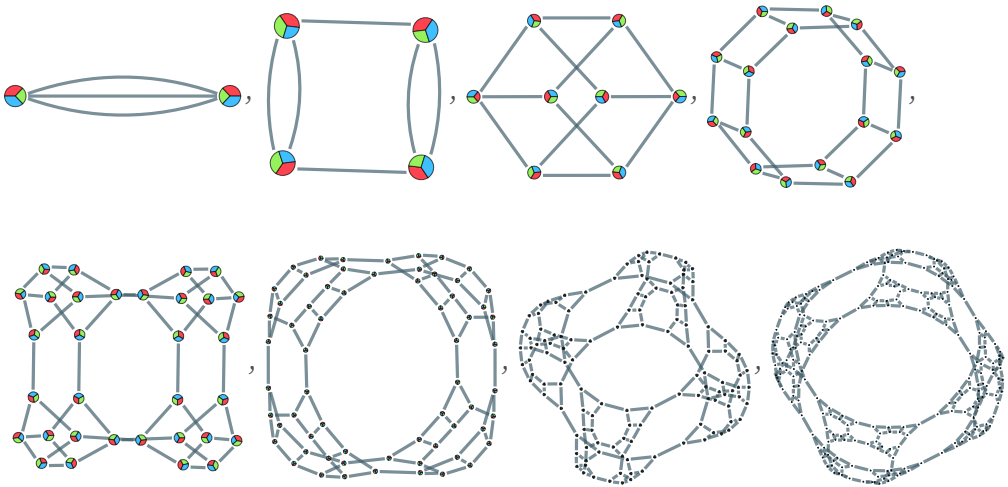

The corresponding rule for hypergraphs would be

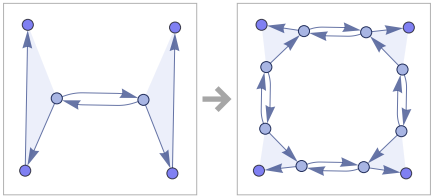

and the corresponding evolution is:

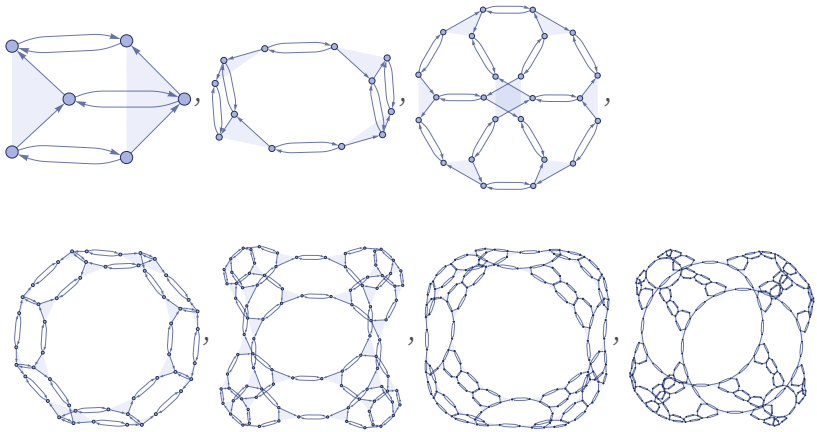

The rule just shown is example of a rule with 2 → 4 internal nodes and 4 dangling connections—which is the smallest class that supports growth from minimal initial conditions. There are altogether 264 rules of this type, with rules of the following forms (up to vertex orderings) [87]:

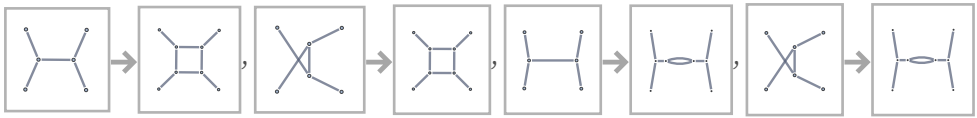

These rules produce the following distinct outcomes:

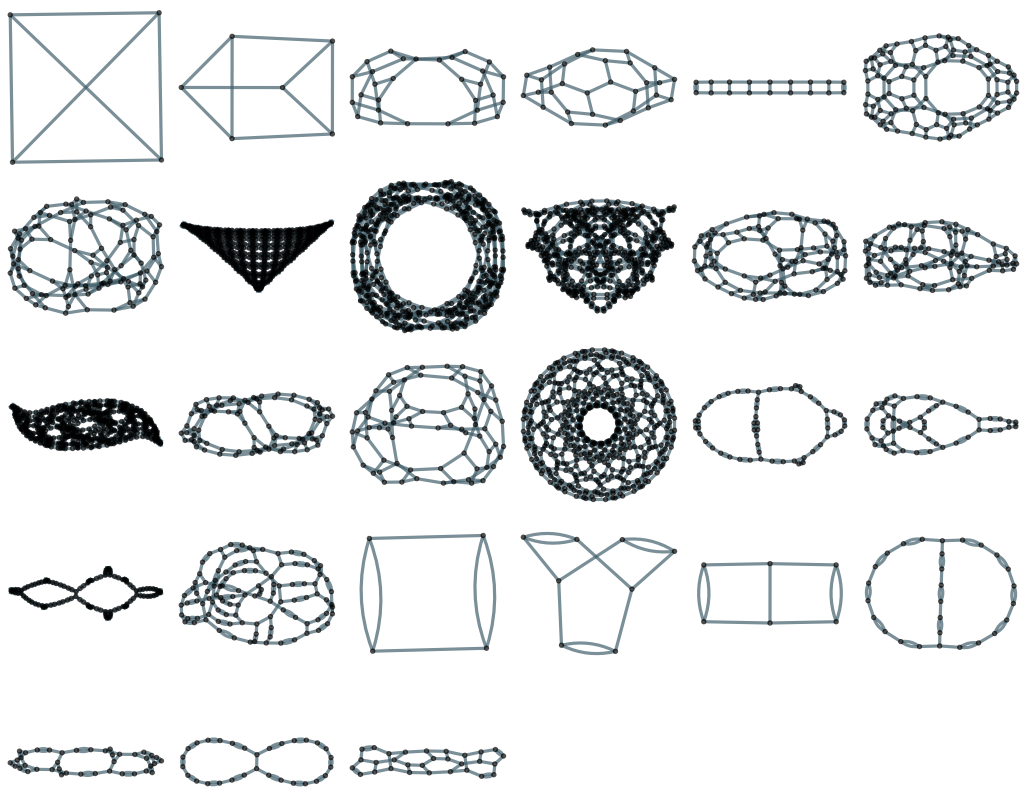

Even though there is a direct translation between ordered trivalent graphs and our models, what is considered a simple rule (for example for purposes of enumeration) is different in the two cases. And while it is more difficult to find valid rules with ordered trivalent graphs, it is notable that even some of the very simplest such rules generate structures with limiting manifold features that we see only after exploring thousands of rules in our models:

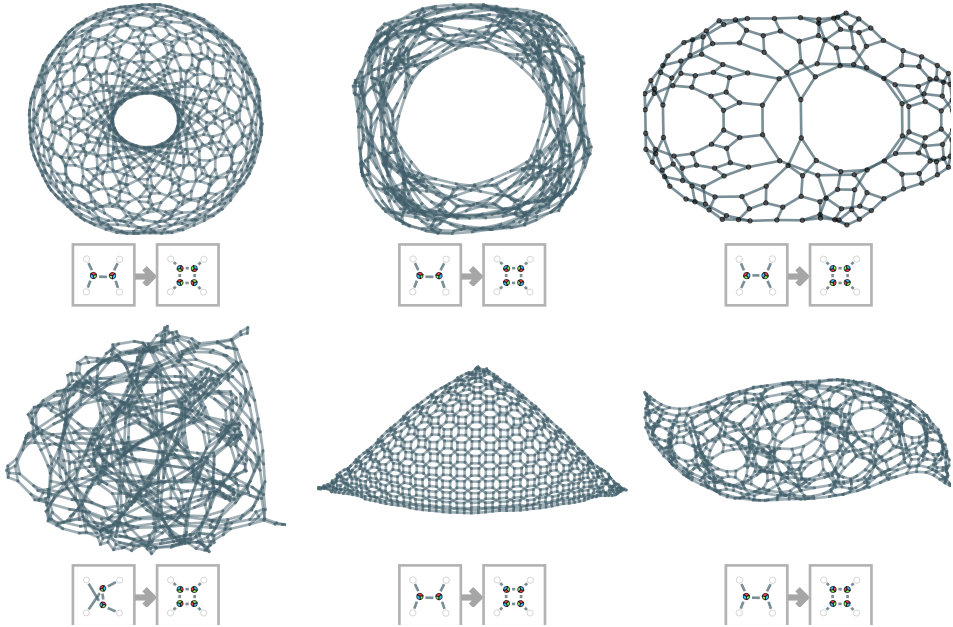

Our models are based on directed (or ordered) hypergraphs. And although the notion is not as natural as for ordinary graphs, one can also consider undirected (or unordered) hypergraphs, in which all elements in a hyperedge are in effect unordered and equivalent. (In general one can also imagine considering any specific set of permutations of elements to be equivalent.)

For unordered hypergraphs one can still use a representation like

{{1, 2, 3}, {1, 2, 4}, {3, 4, 5}}

but now there are no arrows needed within each hyperedge:

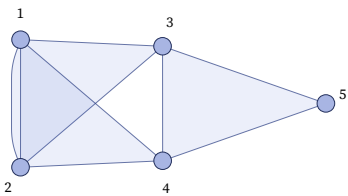

There are considerably fewer unordered hypergraphs with a given signature than ordered ones:

| | ordered | unordered |
|---|---|---|
| $1_2$ | 2 | 2 |
| $2_2$ | 8 | 4 |
| $3_2$ | 32 | 11 |
| $4_2$ | 167 | 30 |
| $5_2$ | 928 | 95 |
| $6_2$ | 5924 | 328 |
| $7_2$ | 40211 | 1211 |

| | ordered | unordered |
|---|---|---|
| $8_2$ | 293370 | 4779 |
| $9_2$ | 2255406 | 19902 |
| $10_2$ | 18201706 | 86682 |
| $1_3$ | 5 | 3 |
| $2_3$ | 102 | 15 |
| $3_3$ | 3268 | 107 |
| $4_3$ | 164391 | 1098 |

| | ordered | unordered |
|---|---|---|
| $1_4$ | 15 | 5 |
| $2_4$ | 2032 | 51 |
| $3_4$ | 678358 | 1048 |
| $1_5$ | 52 | 7 |
| $2_5$ | 57109 | 164 |
| $1_6$ | 203 | 11 |
| $2_6$ | 2089513 | 499 |

There is a translation between unordered hypergraphs and ordered ones, or specifically between unordered hypergraphs and directed graphs. Essentially one creates an incidence graph in which each node and each hyperedge in the unordered hypergraph becomes a node in the directed graph—so that the unordered hypergraph above becomes:

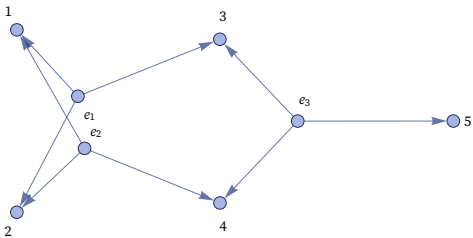

But despite this equivalence, just as in the case of ordered graphs, the sequence of rules will be different in an enumeration based on unordered hypergraphs from one based on ordered hypergraphs.

There are many fewer rules with a given signature for unordered hypergraphs than for ordered ones:

| | unordered | ordered |
|---|---|---|
| $1_2 \to 1_2$ | 5 | 11 |
| $1_2 \to 2_2$ | 19 | 73 |
| $1_2 \to 3_2$ | 71 | 506 |
| $1_2 \to 4_2$ | 296 | 3740 |
| $1_2 \to 5_2$ | 1266 | 28959 |
| $2_2 \to 1_2$ | 16 | 64 |
| $2_2 \to 2_2$ | 76 | 562 |
| $2_2 \to 3_2$ | 348 | 4702 |
| $2_2 \to 4_2$ | 1657 | 40405 |
| $2_2 \to 5_2$ | 7992 | 353462 |

| | unordered | ordered |
|---|---|---|
| $3_2 \to 1_2$ | 59 | 416 |
| $3_2 \to 2_2$ | 347 | 4688 |
| $3_2 \to 3_2$ | 1900 | 48554 |
| $4_2 \to 1_2$ | 235 | 3011 |
| $4_2 \to 2_2$ | 1697 | 42955 |
| $5_2 \to 1_2$ | 998 | 23211 |
| $1_3 \to 1_3$ | 22 | 178 |
| $1_3 \to 2_3$ | 257 | 9373 |
| $2_3 \to 1_3$ | 223 | 8413 |
| $1_4 \to 1_4$ | 84 | 3915 |

Here is an example of a $2_3 \to 3_3$ rule for unordered hypergraphs:

$\{\{x, y, z\}, \{u, v, z\}\} \to \{\{x, x, w\}, \{u, v, x\}, \{y, z, y\}\}$

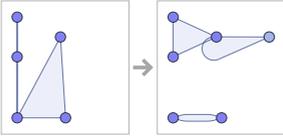

Starting from an unordered double ternary self-loop, this evolves as:

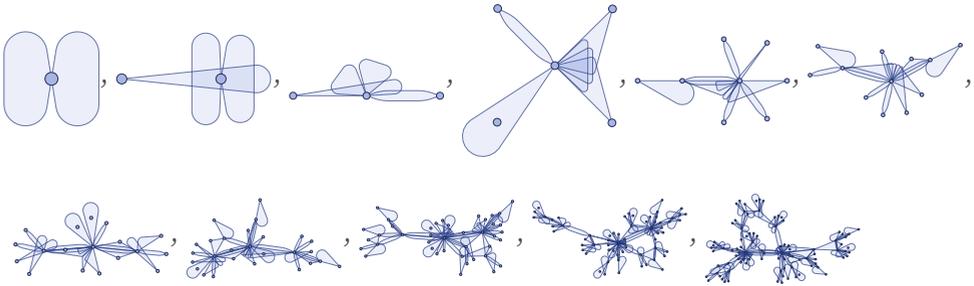

In general the behavior seen for unordered rules with a given signature is considerably simpler than for ordered rules with the same signature. For example, here is typical behavior seen with a random set of unordered $2_3 \to 3_3$ rules:

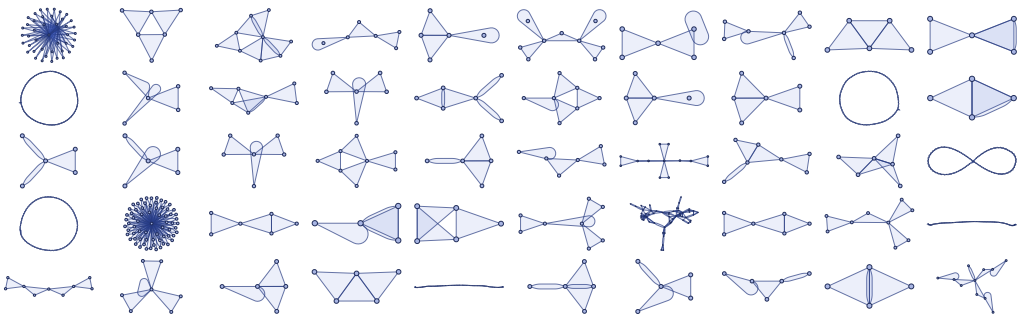

In ordered $2_3 \to 3_3$ rules, globular structures are quite common; in the unordered case they are not. Once one reaches $2_3 \to 4_3$ rules, however, globular structures become common even for unordered hypergraph rules:

$\{\{x, y, z\}, \{u, y, v\}\} \to \{\{x, w, w\}, \{x, s, z\}, \{z, s, u\}, \{y, v, w\}\}$

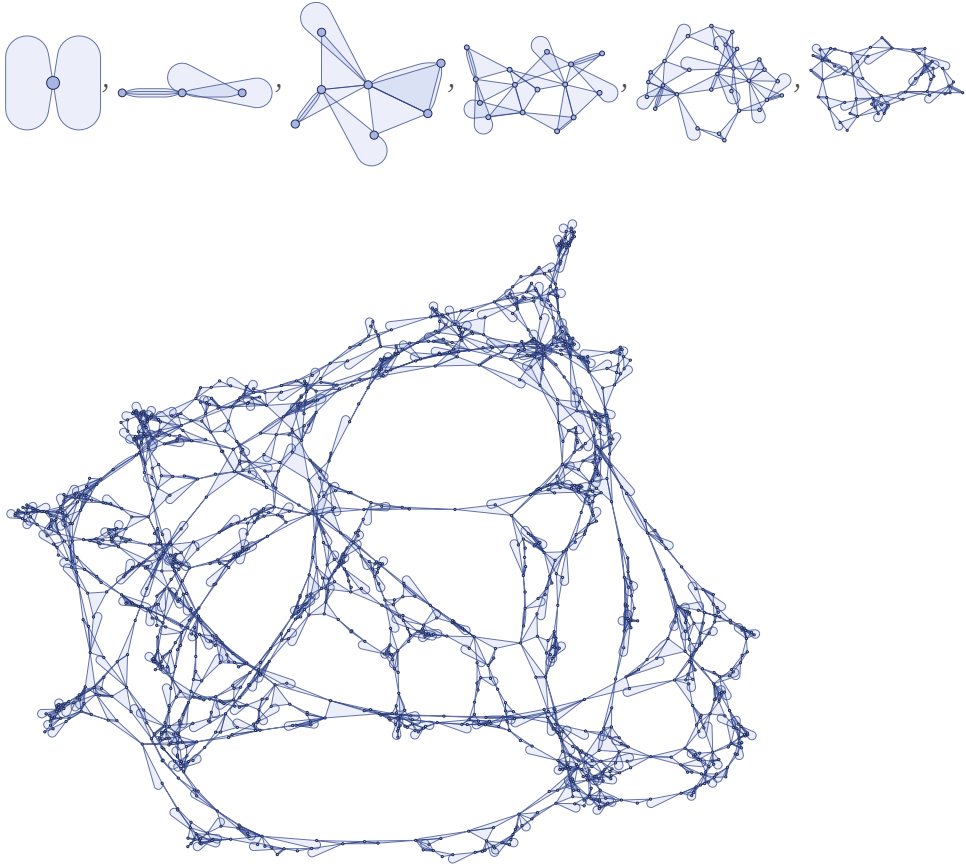

It is worth noting that the concept of unordered hypergraphs can also be applied for binary hyperedges, in which case it corresponds to undirected ordinary graphs. We discussed above the specific case of trivalent undirected graphs, but one can also consider enumerating rules that allow any valence.

An example is

$\{\{x, y\}, \{x, z\}\} \to \{\{x, w\}, \{y, z\}, \{y, w\}, \{z, w\}\}$

which evolves from an undirected double self-loop according to:

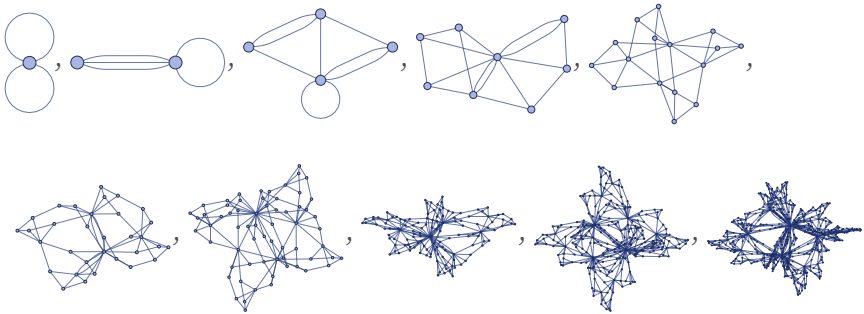

This rule is similar, but not identical, to a rule we have often used as an example:

{{*x*, *y*}, {*x*, *z*}} → {{*x*, *y*}, {*x*, *w*}, {*y*, *w*}, {*z*, *w*}}

Interpreting this rule as referring to undirected graphs, it evolves according to:

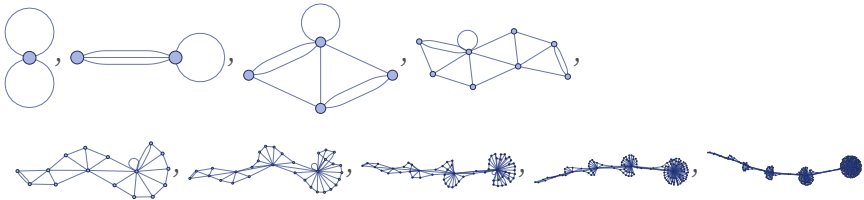

In general, rules for undirected graphs of a given signature yield significantly simpler behavior than rules of the same signature for directed graphs. And, for example, even among all the 7992 distinct $2_2 \to 5_2$ rules for undirected graphs, no globular structures are seen.

Hypergraphs provide a convenient approach to representing our models. But there are other approaches that focus more on the symbolic structure of the models. For example, we can think of a rule such as

{{*x*, *y*, *z*}, {*z*, *u*, *v*}} → {{*w*, *z*, *v*}, {*z*, *x*, *w*}, {*w*, *y*, *u*}}

as defining a transformation for expressions involving a ternary operator f together with a commutative and associative (*n*-ary) operator ∘:

f[x, y, z] ∘ f[z, u, v] → f[w, z, v] ∘ f[z, x, w] ∘ f[w, y, u]

In this formulation, the ∘ operator can effectively be arbitrarily nested. But in the usual setup of our models, f cannot be nested. One could certainly imagine a generalization in which one considers (much as in [98]) transformations on symbolic expressions with arbitrary structures, represented by pattern rules like

f[g[x_, y_], z_] ∘ f[h[g[z_, x_], x_]] → …

or even:

f[g[x_, y_], z_] ∘ f[h_[g[z_, x_], h_[x_]]] → …

And much as in the previous subsection, it is always possible to represent such transformations in our models, for example by having fixed subhypergraphs that act as "markers" to distinguish different functional heads or different "types". (Similar methods can be used to have literals in addition to pattern variables in the transformations, as well as "named slots" [100].)

Our models can be thought of as abstract rewriting (or reduction) systems that operate on hypergraphs, or general collections of relations. Frameworks such as lambda calculus [101][102] and combinatory logic [103][104] have some similarities, but focus on defining reductions for tree structures, rather than general graphs or hypergraphs.

One can ask how our models relate to traditional mathematical systems, for example from universal algebra [105][106]. One major difference is that our models focus on transformations, whereas traditional axiomatic systems tend to focus on equalities. However, it is always possible to define two-way rules or pairs of rules $X \to Y$, $Y \to X$ which in effect represent equalities, and on which a variety of methods from logic and mathematics can be used.

The general case of our models seems to be somewhat out of the scope of traditional mathematical systems. However, particularly if one considers the simpler case of string substitution systems, it is possible to see a variety of connections [1:p938]. For example, two-way string rewrites can be thought of as defining the relations for a semigroup (or, more specifically, a monoid). If one adds inverse elements, then one has a group.

One thinks of the strings as corresponding to words in the group. Then the multiway evolution of the system corresponds to starting with particular words and repeatedly applying relations to them—to produce other words which for the purposes of the group are considered equivalent.

This is in a sense a dual operation to what happens in constructing the Cayley graph of a group, where one repeatedly adds generators to words, always reducing by using the relations in the group (see 4.17).

For example, consider the multiway system defined by the rule:

{AB → BA, BA → AB}

The first part of the multiway (states) graph associated with this rule is:

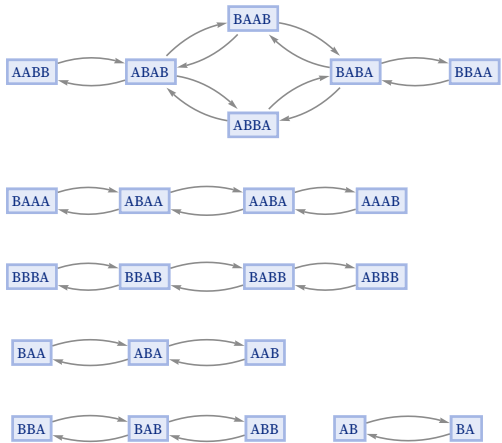

Ignoring inverse elements (which in this case just make double edges) the first part of the infinite Cayley graph for the group with relations AB↔BA has the form:

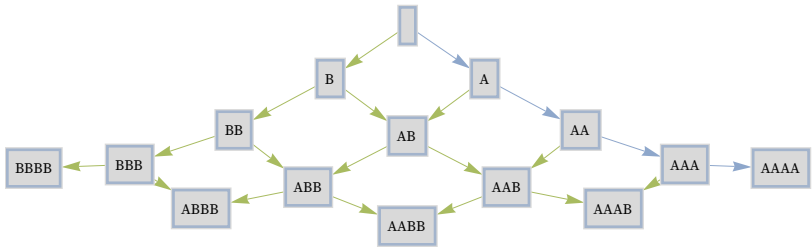

One can think of the Cayley graph as being created by starting with a tree, corresponding to the Cayley graph for a free group, then identifying nodes that are related by relations. The edges in the multiway graph (which correspond to updating events) thus have a correspondence to cycles in the Cayley graph.

As one further example, consider the (finite) group $S_3$ which can be thought of as being specified by the relations:

{ ↔ AA, AA ↔ BB, BB ↔ ABABAB}

The Cayley graph in this case is simply:

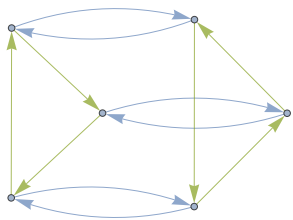

The multiway graph in this case begins:

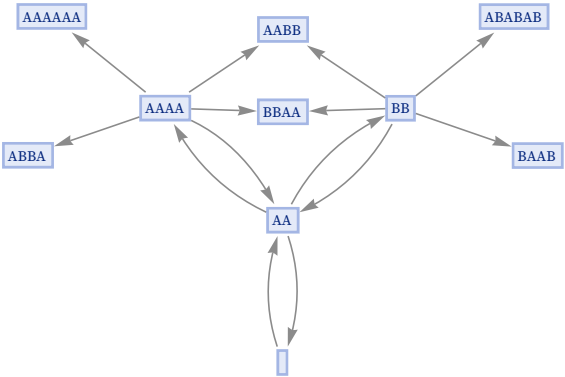

Continuing for a few more steps gives:

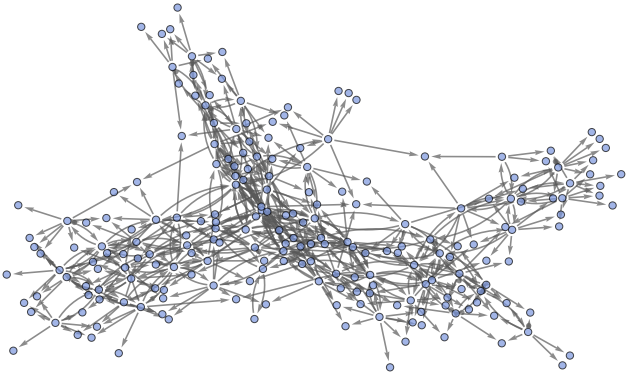

On successive steps, the volumes $\Sigma_t$ in these multiway graphs grow like:

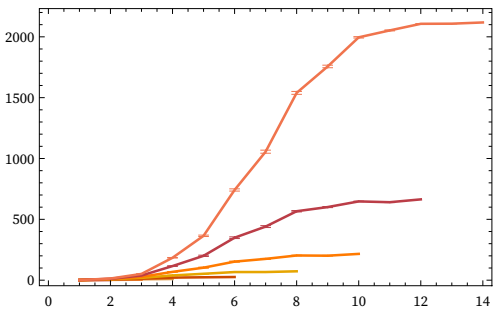

There does not appear to be any direct correspondence to quantities such as growth rates of Cayley graphs (cf. [22]).

## 7.3 Computational Capabilities of Our Models

An important way to characterize our models is in terms of their computational capabilities. We can always think of the evolution of one of our models as corresponding to a computation: the system starts from an initial state, then follows its rules, in effect carrying out a computation to generate a sequence of results.

The Principle of Computational Equivalence [1:c12] suggests that when the behavior of our models is not obviously simple it will typically correspond to a computation of effectively maximal sophistication. And an important piece of evidence for this is that our models are capable of universal computation.

We saw above that our models can emulate a variety of other kinds of systems. Among these are Turing machines and cellular automata. And in fact we already saw above how our models can emulate what is known to be the simplest universal Turing machine [1:p709][94][95] [96]. We also showed how our models can emulate the rule 30 cellular automaton, and we can use the same construction to emulate the rule 110 cellular automaton, which is known to be computation universal [1:11.8].

So what this means is that we can set up one of our models and then "program" it, by giving appropriate initial conditions, to make it do any computation, or emulate any other computational system. We have seen that our models can produce all sorts of behavior; what this shows is that at least in principle our models can produce any behavior that any computational system can produce.

But showing that we can set up one of our models to emulate a universal Turing machine is one thing; it is something different to ask what computations a random one of our models typically performs. To establish this for certain is difficult, but experience with the Principle of Computational Equivalence [1:c12] in a wide range of other kinds of systems with simple underlying rules strongly suggests that not only is sophisticated computation possible to achieve in our models, it is also ubiquitous, and will occur basically whenever the behavior we see is not obviously simple.

This notion has many consequences, but a particularly important one is computational irreducibility [1:12.6]. Given the simplicity of the underlying rules for our models, we might imagine that it would always be possible—by using some appropriately sophisticated mathematical or computational technique—to predict what the model would do after any number of steps. But in fact what the Principle of Computational Equivalence implies is that more or less whenever it is not obviously straightforward to do, making this prediction will actually take an irreducible amount of computational work—and that in effect we will not be able to compute what the system will do much more efficiently than by just following the steps of the actual evolution of the system itself.

Much of what we have done in studying our models here has been based on just explicitly running the models and seeing what they do. Computational irreducibility implies that this is not just something that is convenient in practice; instead it is something that cannot theoretically be avoided, at least in general.

Having said this, however, it is an inevitable feature of computational irreducibility that there is always an endless sequence of "pockets" of computational reducibility: specific features or questions that are amenable to computation or prediction without irreducible amounts of computational work.

But another consequence of computational irreducibility is the appearance of undecidability [107][108]. If we want to know what will happen in one of our models after a certain number of steps, then in the worst case we can just run the model for that many steps and see what it does. But if we want to know if the model will ever do some particular thing—even after an arbitrarily long time—then there can be no way to determine that with any guaranteed finite amount of effort, and therefore we must consider the question formally undecidable.

Will a particular rule ever terminate when running from a particular initial state? Will the hypergraphs it generates ever become disconnected? Will some branch pair generated in a multiway system ever resolve?

These are all questions that are in general undecidable in our models. And what the Principle of Computational Equivalence implies is that not only is this the case in principle; it is something ubiquitous, that can be expected to be encountered in studying any of our models that do not show obviously simple behavior.

It is worth pointing out that undecidability and computational irreducibility apply both to specific paths of evolution in our models, and to multiway systems. Multiway systems correspond to what are traditionally called non-deterministic computations [109]. And just as a single path of evolution in one of our models can reproduce the behavior of any ordinary deterministic Turing machine, so also the multiway evolution of our models can reproduce any non-deterministic Turing machine.

The fact that our models show computation universality means that if some system—like our universe—can be represented using computation of the kind done, for example, by a Turing machine, then it is inevitable that in principle our models will be able to reproduce it. But the important issue is not whether some behavior can in principle be programmed, but whether we can find a model that faithfully and efficiently reflects what the system we are modeling does. Put another way: we do not want to have to set up some elaborate program in the initial conditions for the model we use; we want there to be a direct way to get the initial conditions for the model from the system we are modeling.

There is another important point, particularly relevant, for example, in the effort to use our models in a search for a fundamental theory of physics. The presence of computation universality implies that any given model can in principle encode any other. But in practice this encoding can be arbitrarily complicated, and if one is going to make an enumeration of possible models, different choices of encoding can in effect produce arbitrarily large changes in the enumeration.

One can think of different classes of models as corresponding to different languages for describing systems. It is always in principle possible to translate between them, but the translation may be arbitrarily difficult, and if one wants a description that is going to be useful in practice, one needs to have a suitable language for it.

# 8 | Potential Relation to Physics

## 8.1 Introduction

Having explored our models and some of their behavior, we are now in a position to discuss their potential for application to physics. We shall see that the models generically show remarkable correspondence with a surprisingly wide range of known features of physics, inspiring the hope that perhaps a specific model can be found that precisely reproduces all details of physics. It should be emphasized at the outset that there is much left to explore in the potential correspondence between our models and physics, and what will be said here is merely an indication—and sometimes a speculative one—of how this might turn out.

(See also Notes & Further References.)

## 8.2  Basic Concepts

The basic concept of applying our models to physics is to imagine that the complete structure and content of the universe is represented by an evolving hypergraph. There is no intrinsic notion of space; space and its apparent continuum character are merely an emergent large-scale feature of the hypergraph. There is also no intrinsic notion of matter: everything in the universe just corresponds to features of the hypergraph.

There is also no intrinsic notion of time. The rule specifies possible updates in the hypergraph, and the passage of time essentially corresponds to these update events occurring. There are, however, many choices for the sequences in which the events can occur, and the idea is that all possible branches in some sense do occur.

But the concept is then that there is a crucial simplifying feature: the phenomenon of causal invariance. Causal invariance is a property (or perhaps effective property) of certain underlying rules that implies that when it comes to causal relationships between events, all possible branches give the same ultimate results.

As we will discuss, this equivalence seems to yield several core known features of physics, notably Lorentz invariance in special relativity, general covariance in general relativity, as well as local gauge invariance, and the perception of objective reality in quantum mechanics.

Our models ultimately just consist of rules about elements and relations. But we have seen that even with very simple such rules, highly complex structures can be produced. In particular, it is possible for the models to generate hypergraphs that can be considered to approximate flat or curved $d$-dimensional space. The dimension is not intrinsic to the model; it must emerge from the behavior of the model, and can be variable.

The evolving hypergraphs in our models must represent not just space, but also everything in it. At a bulk level, energy and momentum potentially correspond to certain specific measures of the local density of evolution in the hypergraph. Particles potentially correspond to evolution-stable local features of the hypergraph.

The multiway branching of possible updating events is potentially closely related to quantum mechanics, and much as large-scale limits of our hypergraphs may correspond to physical space, so large-scale limits of relations between branches may correspond to Hilbert spaces of states in quantum mechanics.

In the case of physical space, one can view different choices of updating orders as corresponding to different reference frames—with causal invariance implying equivalence between them. In multiway space, one can view different updating orders as different sequences of applications of quantum operators—with causal invariance implying equivalence between them that lead different observers to experience the same reality.

In attempting to apply our models to fundamental physics, it is notable how many features that are effectively implicitly assumed in the traditional formalism of physics can now potentially be explicitly derived.

It is inevitable that our models will show computational irreducibility, in the sense that irreducible amounts of computational work will in general be needed to determine the outcome of their behavior. But a surprising discovery is that many important features of physics seem to emerge quite generically in our models, and can be analyzed without explicitly running particular models.

It is to be expected, however, that specific aspects of our universe—such as the dimensionality of space and the masses and charges of particles—will require tracing the detailed behavior of models with particular rules.

It is already clear that modern mathematical methods can provide significant insight into certain aspects of the behavior of our models. One complication in the application of these methods is that in attempting to make correspondence between our models and physics, many levels of limits effectively have to be taken, and the mathematical definitions of these limits are likely to be subtle and complex.

In traditional approaches to physics, it is common to study some aspect of the physical world, but ignore or idealize away other parts. In our models, there are inevitably close connections between essentially all aspects of physics, making this kind of factored approach—as well as idealized partial models—much more difficult.

Even if the general structure of our models provides an effective framework for representing our physical universe at the lowest level, there does not seem to be any way to know within a wide margin just how simple or complex the specific rule—or class of equivalent rules—for our particular universe might be. But assuming a certain degree of simplicity, it is likely that fitting even a modest number of details of our universe will completely determine the rule.

The result of this would almost certainly be a large number of specific predictions about the universe that could be made even without irreducibly large amounts of computation. But even absent the determination of a specific rule, it seems increasingly likely that experimentally accessible predictions will be possible just from general features of our models.

## 8.3  Potential Basic Translations

As a guide to the potential application of our models to physics, we list here some current expectations about possible translations between features of physics and features of our models. This should be considered a rough summary, with every item requiring significant explanation and qualification. In addition, it should be noted that in an effort to clarify presentation, many highly abstract concepts have been indicated here by more mechanistic analogies.

### *Basic Physics Concepts*

**space**: general limiting structure of basic hypergraph

**time**: index of causal foliations of hypergraph rewriting

**matter (in bulk)**: local fluctuations of features of basic hypergraph

**energy:** flux of edges in the multiway causal graph through spacelike (or branchlike) hypersurfaces

**momentum**: flux of edges in the multiway causal graph through timelike hypersurfaces

**(rest) mass**: numbers of nodes in the hypergraph being reused in updating events

**motion**: possible because of causal invariance; associated with change of causal foliations

**particles**: locally stable configurations in the hypergraph

**charge, spin, etc.**: associated with local configurations of hyperedges

**quantum indeterminacy**: different foliations (of branchlike hypersurfaces) in the multiway graph

**quantum effects**: associated with locally unresolved branching in the multiway graph

**quantum states**: (instantaneously) nodes in the branchial graph

**quantum entanglement**: shared ancestry in the multiway graph / distance in branchial graph

**quantum amplitudes**: path counting and branchial directions in the multiway graph

**quantum action density (Lagrangian):** total flux (divergence) of multiway causal graph edges

## *Physical Theories & Principles*

**special relativity**: global consequence of causal invariance in hypergraph rewriting

**general relativity / general covariance**: effect of causal invariance in the causal graph

**locality / causality**: consequence of locality of hypergraph rewriting and causal invariance

**rotational invariance**: limiting homogeneity of the hypergraph

**Lorentz invariance**: consequence of causal invariance in the causal graph

**time dilation**: effect of different foliations of the causal graph

**relativistic mass increase**: effect of different foliations of the causal graph

**local gauge invariance**: consequence of causal invariance in the multiway graph

**lack of quantum cosmological constant**: space is effectively created by quantum fluctuations

**cosmological homogeneity**: early universe can have higher effective spatial dimension

**expansion of universe**: growth of hypergraph

**conservation of energy**: equilibrium in the causal graph

**conservation of momentum**: balance of different hyperedges during rewritings

**principle of equivalence**: gravitational and inertial mass both arise from features of the hypergraph

**discrete conservation laws**: features of the ways local hypergraph structures can combine

**microscopic reversibility**: limiting equilibrium of hypergraph rewriting processes

**quantum mechanics**: consequence of branching in the multiway system

**observer in quantum mechanics**: branchlike hypersurface foliation

**quantum objective reality**: equivalence of quantum observation frames in the multiway graph

**quantum measurements**: updating events with choice of outcomes, that can be frozen by a foliation

**quantum eigenstates**: branches in multiway system

**quantum linear superposition**: additivity of path counts in the multiway graph

**uncertainty principle**: non-commutation of update events in the multiway graph

**wave-particle duality**: relation between spacelike and branchlike projections of the multiway causal graph

**356**

**operator-state correspondence**: states in the multiway graph are generated by events (operators)

**path integral**: turning of paths in the multiway graph is proportional to causal edge density

**violation of Bell's inequalities, etc.**: existence of causal connections in the multiway graph

**quantum numbers**: associated with discrete local properties of the hypergraph

**quantization of charge, etc.**: consequence of the discrete hypergraph structure

**black holes / singularities**: causal disconnection in the causal graph

**dark matter**: (possibly) relic oligons / dimension changes in of space

**virtual particles**: local structures continually generated in the spatial and multiway graphs

**black hole radiation / information**: causal disconnection of branch pairs

**holographic principle**: correspondence between spatial and branchial structure

## *Physical Quantities & Constructs*

**dimension of space**: growth rate exponent in hypergraph / causal cones

**curvature of space**: polynomial part of growth rate in hypergraph / causal cones

**local gauge group**: limiting automorphisms of local hypergraph configurations

**speed of light ($c$)**: measure of edges in spatial graph vs. causal graph

**light cones**: causal cones in the causal graph

**unit of energy**: count of edges in the causal graph

**momentum space**: limiting structure of causal graph in terms of edges

**gravitational constant**: proportionality between node counts and spatial volume

**quantum parameter ($\hbar$)**: measure of edges in the branchial graph (maximum speed of measurement)

**elementary unit of entanglement**: branching of single branch pair

**electric/gauge charges**: counts of local hyperedge configurations

**spectrum of particles**: spectrum of locally stable configurations in the hypergraph

## *Idealizations, etc. Used in Physics*

**inertial frame**: parallel foliation of causal graph

**rest frame of universe**: geodesically layered foliation of causal graph **flat space**: uniform hypergraph (typically not maintained by rules)

**Minkowski space**: effectively uniform causal graph

**cosmological constant**: uniform curvature in the hypergraph

**de Sitter space**: cyclically connected hypergraph

**closed timelike curves**: loops in the causal graph (only possible in some rules)

**point particle**: a persistent structure in the hypergraph involving comparatively few nodes

**purely empty space**: not possible in our models (space is maintained by rule evolution)

**vacuum**: statistically uniform regions of the spatial hypergraph

**vacuum energy**: causal connections attributed purely to establishing the structure of space

**isolated quantum system**: disconnected part of the branchial/multiway graph

**collapse of the wave function**: degenerate foliation that infinitely retards branchlike entanglement

**non-interacting observer in quantum mechanics**: "parallel" foliation of multiway graph

**free field theory**: e.g. pure branching in the multiway system

**quantum computation**: following multiple branches in multiway system (limited by causal invariance)

**string field theory**: (potentially) continuous analog of the multiway causal graph for string substitutions

## 8.4 The Structure of Space

In our models, the structure of spacetime is defined by the structure of the evolving hypergraph. Causal foliations of the evolution can be used to define spacelike hypersurfaces. The instantaneous structure of space (on a particular spacelike hypersurface) corresponds to a particular state of the hypergraph.

A position in space is defined by a node in the hypergraph. A geometrical distance between positions can be defined as the number of hyperedges on the shortest path in the hypergraph between them. Although the underlying rules for hypergraph rewriting in our models depend on the ordering of elements in hyperedges, this is ignored in computing geometrical distance. (The geometrical distance discussed here is basically just a proxy for a true physical distance measured from dynamic information transmission between positions.) A shortest path on the hypergraph between two positions defines a geodesic between them, and can be considered to define a straight line.

The only information available to define the structure of space is the connectivity of the hypergraph; there is no predefined embedding or topological information. The continuum character of space assumed in traditional physics must emerge as a large-scale limit of the hypergraph (somewhat analogously to the way the continuum character of fluids emerges as a large-scale limit of discrete molecular dynamics (e.g. [1:p378][110] ). Although our models follow definite rules, they can intrinsically generate effective randomness (much like the rule 30 cellular automaton, or the computation of the digits of $\pi$). This effective randomness makes large-scale behavior typically approximate statistical averages of small-scale dynamics.

In our models, space has no intrinsic dimension defined; its effective dimension must emerge from the large-scale structure of the hypergraph. Around every node at position $X$ consider a geodesic ball consisting of all nodes that are a hypergraph distance not more than $r$ away. Let $V_r(X)$ be the total number of nodes in this ball. Then the hypergraph can be considered to approximate $d$-dimensional space if

$V_r(X) \sim r^d$

for a suitable range of values of $r$. Here we encounter the first of many limits that must be taken. We want to consider the limit of a large hypergraph (say as generated by a large number of steps of evolution), and we want $r$ to be large compared to 1, but small compared to the overall diameter of the hypergraph.

As a simple example, consider the hypergraph created by the rule

$\{\{x, y\}, \{x, z\}\} \rightarrow \{\{x, y\}, \{x, w\}, \{y, w\}, \{z, w\}\}$

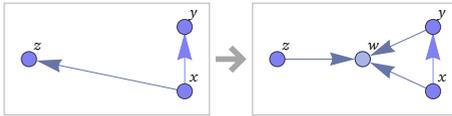

Starting from a minimal initial condition of two self-loops, the first few steps of evolution with our standard updating order are:

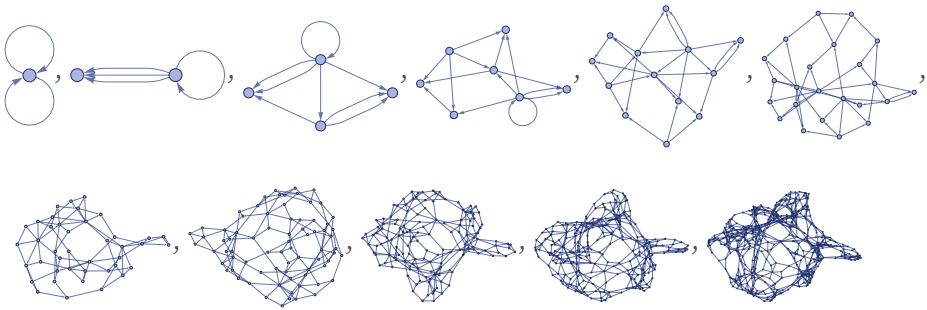

The hypergraph obtained after 12 steps has 1651 nodes and can be rendered as:

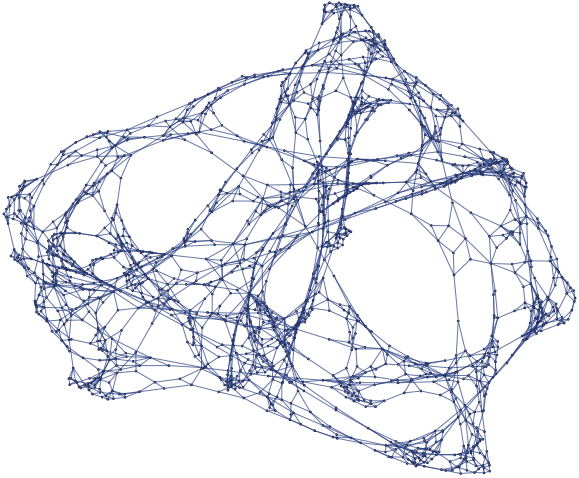

This plots the effective "dimension exponent" of $r$ in $V_r$ as a function of $r$, averaged over all nodes in the hypergraph, for a succession of steps in the evolution:

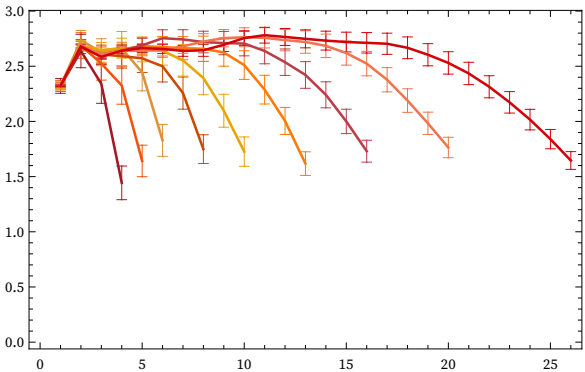

A constant limiting value $d$ indicates approximation to a "flat" $d$-dimensional space. For integer $d$, this corresponds to ordinary $d$-dimensional Euclidean space, but in our models $d$ often does not end up being integer valued, nor does it need to be constant at different positions, or through the course of evolution. It is also important to note that only some rules give $V_r \sim r^d$; exponential or more complex behavior is common.

Even when to leading order $V_r \sim r^d$, there are corrections. For small $r$ (measured, say, relative to the diameter of the hypergraph) one can consider a power series expansion in $r$. By comparison to ordinary manifolds one can then write (e.g. [24][1:p1050])

$$V_r \sim r^d \left(1 - \frac{r^2}{6(d+2)} R + O(r^4)\right)$$

where R can be identified as the (Ricci) scalar curvature [25][26] of the limiting space. The value of this curvature is again purely determined by the (limiting) structure of the hypergraph. (Note that particularly if one goes beyond a pure power series, there is the potential for subtle interplay between change in dimension and what one might attribute to curvature.)

It is also possible to identify other limiting features of the hypergraph. For example, consider a small stretch of geodesic (where by "small" we mean still large compared to individual connections in the hypergraph, but small compared to the scale on which statistical features of the hypergraph change). Now create a tube of radius $r$ by including every node with distance up to $r$ from any node on the geodesic. The growth rate of the number of nodes in this tube can then be approximated as [44]

$$\tilde{V}_r \sim r^d (1 + \frac{r^2}{6} R_{ij} \delta x^i \delta x^j + O(r^4))$$

where now $R_{ij} \delta x^i \delta x^j$ is the projection of the Ricci tensor along the direction of the geodesic. (The Ricci tensor measures the change in cross-sectional area for a bundle of geodesics, associated with their respective convergence and divergence for positive and negative curvature.)

In a suitable limit, the nodes in the hypergraph correspond to points in a space. A tangent bundle at each point can be defined in terms of the equivalence class of geodesics through that point, or in our case the equivalence class of sequences of hyperedges that pass through the corresponding node in the hypergraph.

One can set up what in the limit can be viewed as a rank-$p$ tensor field on the hypergraph by associating values with $p$ hyperedges at each node. When these values correspond to intrinsic features of the hypergraph (such as $V_r$), their limits give intrinsic properties of the space associated with the hypergraph. And for example the Riemann tensor can be seen as emerging from essentially measuring areas of "rectangles" defined by loops in the hypergraph, though in this case multiple limits need to be taken.

## 8.5  Time and Spacetime

In our models, the passage of time basically corresponds to the progressive updating of the hypergraph. Time is therefore fundamentally computational: its passage reflects the performance of a computation—and typically one that is computationally irreducible. It is notable that in a sense the progression of time is necessary even to maintain the structure of space. And this effectively forces the entropic arrow of time (reflected in the effective randomization associated with irreducible computation) to be aligned with the cosmological arrow of time (defined by the overall evolution of the structure of space).

At the outset, time in our models has a very different character from space. The phenomenon of causal invariance, however, implies a link which leads to relativistic invariance.

To see this, we can begin much as in the traditional development of special relativity [111] by considering what constitutes a physically realizable observer. In our model, everything is represented by the evolving hypergraph, including all of the internal state of any observer. One consequence of this is that the only way an observer can "sense" anything about the universe is by some updating event happening within the observer.

And indeed in the end all that any observer can ultimately be sensitive to is the causal relationships between different updating events that occur. From a particular evolution history of a hypergraph, we can construct a causal graph whose nodes correspond to updating events, and whose directed edges represent the causal relations between these events—in the sense that there is an edge between events A and B if the input to B involves output from A. For the evolution shown above, the beginning of the causal graph is:

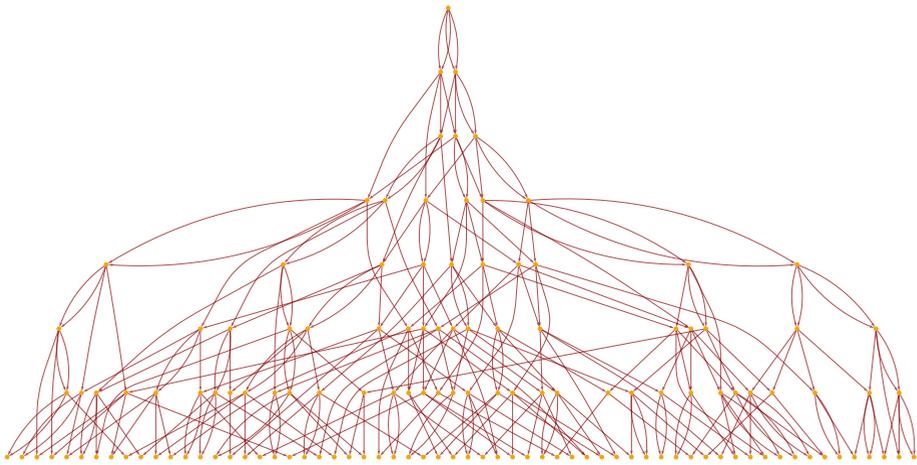

We can think of this causal graph as representing the evolution of our system in spacetime. The analog of a light cone is then the set of nodes that can be reached from a given node in the graph. Every edge in the graph represents a timelike relationship between events, and can be thought of as corresponding to a timelike direction in spacetime. Nodes that cannot be reached from each other by following edges of the graph can be thought of as spacelike separated. Just as for space with the "spatial hypergraphs" we discussed above, there is nothing in the abstract that defines the geometry of spacetime associated with the causal graph; everything must emerge from the pattern of connections in the graph, which in turn are generated by the operation of the underlying rules for our models.

In its original construction, a causal graph is in a sense a causal summary of a particular evolution history for a given rule, with a particular sequence of updating events. But when the underlying rule has the property of causal invariance, this has the important consequence that in the appropriate limit the causal graph obtained always has the same form, independent of the particular sequence of updating events. In other words, when there is causal invariance, the system in a sense always has a unique causal history.

The interpretation of this causal history in terms of a spacetime history, however, depends on what amount to definitions made by an observer. In particular, to define what can be interpreted as a time coordinate, one must set up a foliation of the causal graph, with successive slices corresponding to successive steps in time.

There are many such foliations that can be set up. The only fundamental constraint is that events in a given slice cannot be directly connected by an edge in the causal graph—or, in other words, they must be spacelike separated. The possible foliations thus correspond to possible sequences of spacelike hypersurfaces, analogous to those in standard discussions of spacetime.

(Note that the causal graph ultimately just defines a partial order on the set of events, and one could in principle imagine having arbitrarily complex foliations set up to imply any given total order of events. But such foliations are not realistic for macroscopic observers with bounded computational resources, and in our analysis of observable continuum limits we can ignore them.)

When one reaches a particular spacelike hypersurface, it represents a particular set of events having occurred, and thus a particular state of the underlying system having been reached, represented by a particular hypergraph. Different sequences of spacelike hypersurfaces thus correspond to different sequences of "instantaneous states" having been reached—corresponding to different evolution histories. But the crucial point is that causal invariance implies that even though the sequences of instantaneous states are different, the causal graphs representing the causal relationships between events that occur in them are always the same. And this is the essence of how the phenomena of relativistic invariance—and general covariance—are achieved.

## 8.6 Motion and Special Relativity

In the traditional formalism of physics, the principles of special relativity are in a sense introduced as axioms, and then their consequences are derived. In our models, what amount to these principles can in effect emerge directly from the models themselves, without having to be introduced from outside.

To see how this works, consider the phenomenon of motion. In standard physics, one thinks of different states of uniform motion as corresponding to different inertial reference frames (e.g. [111][112]). These different reference frames in turn correspond to different choices of sequences of spacelike hypersurfaces, or, in our setup, different foliations of the causal graph.

As a simple example, consider the string substitution system BA→AB, starting from …BABABA… The causal graph for the evolution of this system can be drawn as a grid:

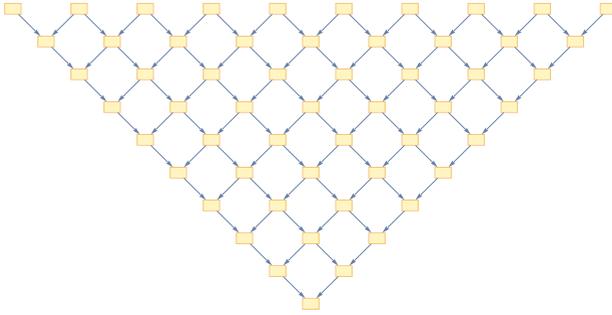

A simple foliation is just to form successive layers:

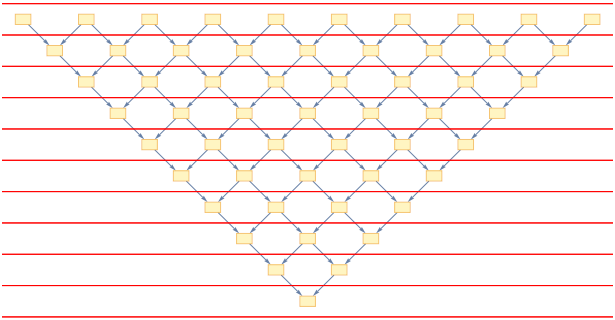

With this foliation, the sequence of states in the underlying string substitution system is:

| B | A | B | A | B | A | B | A | B | A | B | A | B | A | B | A | B | A |
|---|---|---|---|---|---|---|---|---|---|---|---|---|---|---|---|---|---|
| A | B | A | B | A | B | A | B | A | B | A | B | A | B | A | B | A | B |
| A | A | B | A | B | A | B | A | B | A | B | A | B | A | B | A | B | B |
| A | A | A | B | A | B | A | B | A | B | A | B | A | B | A | B | B | B |
| A | A | A | A | B | A | B | A | B | A | B | A | B | A | B | B | B | B |
| A | A | A | A | A | B | A | B | A | B | A | B | A | B | B | B | B | B |
| A | A | A | A | A | A | B | A | B | A | B | A | B | B | B | B | B | B |
| A | A | A | A | A | A | A | B | A | B | A | B | B | B | B | B | B | B |
| A | A | A | A | A | A | A | A | B | A | B | B | B | B | B | B | B | B |
| A | A | A | A | A | A | A | A | A | B | A | B | B | B | B | B | B | B |
| A | A | A | A | A | A | A | A | A | B | B | B | B | B | B | B | B | B |

In drawing our foliation of the causal graph, we can think of time as being vertical, and space horizontal. Now imagine we want to represent uniform motion. We can do this by making our foliation use slices with a slope proportional to velocity:

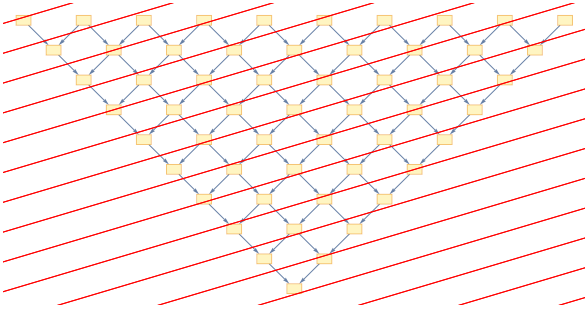

But imagine we want to show time vertically, while not destroying the partial order in our causal network. The unique way to do it (if we want to preserve straight lines) is to transform a point $\{t, x\}$ to $\{t - \beta x, x - \beta t\}/\sqrt{1 - \beta^2}$ :

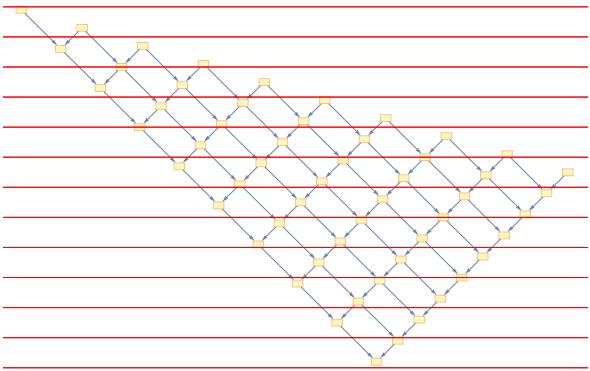

But this is precisely the usual Lorentz transformation of special relativity. And time dilation is then, for example, associated with the fact that to reach what corresponds to an event at slice *t* in the original foliation, one now has to go through a sequence of events that is longer by a factor of $\gamma = 1/\sqrt{1 - \beta^2}$ .

Normally one would argue for these results on the basis of principles supplied by special relativity. But the crucial point here is that in our models the results can be derived purely from the behavior of the models, without introducing additional principles.

Imagine simply using the transformed causal graph to determine the order of updating events in the underlying substitution system:

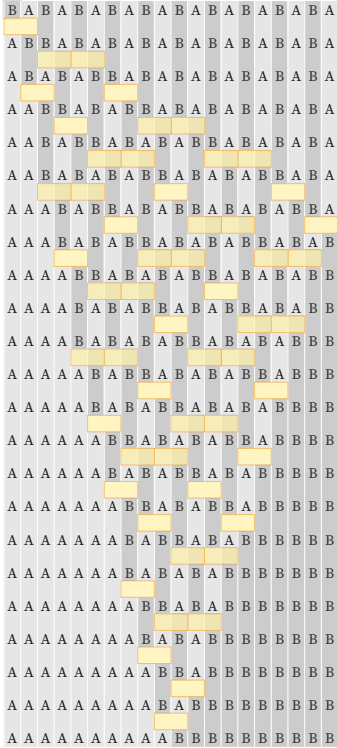

If we look vertically down the picture we see a different sequence of states of the system. But the crucial point is that the final outcome of the evolution is exactly the same as it was with the original foliation. In some sense the "physics" is the same, independent of the reference frame. And this is the essence of relativistic invariance (and here we immediately see some of its consequences, like time dilation).

But in the context of the string substitution system, we can now see its origin of the invariance. It is the fact that the underlying rule we have used is causal invariant, so that regardless of the specific order in which updating events occur, the same causal graph is obtained, with the same final output.

In our actual models based on infinitely evolving hypergraphs, the details are considerably more complicated. But the principles are exactly the same: if the underlying rule has causal invariance, its limiting behavior will show relativistic invariance, and (so long as it has limiting geometry corresponding to flat $d$-dimensional space) all the usual phenomena of special relativity.

(Note that the concept of a finite speed of light, leading effectively to locality in the causal graph, is related to the fact that the underlying rules involve rewriting hypergraphs only of bounded size.)

## 8.7 The Vacuum Einstein Equations

In discussing the structure of space, we considered how the volumes of geodesic balls grow with radius. In discussing spacetime, we want to consider the analogous question of how the volumes of light cones [1:p1052]) grow with time. But to do this, we have to say what we mean by time, since—as we saw in the previous subsection—different foliations can lead to different identifications.

Any particular foliation—with its sequence of spacelike hypersurfaces—provides at every point a timelike vector that defines a time direction in spacetime. So if we start at any point in the causal graph, we can look at the forward light cone from this point, and follow the connections in the causal graph until we have gone a proper time $t$ in the time direction we have defined. Then we can ask how how many nodes we have reached in the causal graph.

The result will depend on the underlying rule for the system. But if in the limit it is going to correspond to flat $(d + 1)$-dimensional spacetime, at any spacetime position $X$ it must grow like:

$C_t(X) \sim t^{d+1}$

If we include the possibility of curvature, we get to first order

$C_t(X) \sim t^{d+1}(1 - \frac{1}{6} \delta t^\mu \delta t^\nu R_{\mu\nu}(X) + \ldots)$

where $R_{\mu\nu}$ is the spacetime Ricci tensor, and $\delta t^\mu \delta t^\nu R_{\mu\nu}$ is effectively its projection along the infinitesimal timelike vector $\delta t^\mu$.

For any particular underlying rule, $C_t(X)$ will take on a definite form. But in making connections with traditional continuum spacetime, we are interested in its limiting behavior.

Assume, to begin, that we have scaled $t$ to be measured relative to the size of the whole causal graph. Then for small $t$ we can expand $C_t(X)$ to get the expression involving curvature above. But now imagine scaling up $t$. Eventually it is inevitable that the curvature term has the potential to affect the overall $t$ dependence, and potentially change the effective exponent of $t$. But if the overall continuum limit is going to correspond to a $(d + 1)$-dimensional spacetime, this cannot happen. And what this means is that at least a suitably averaged version of the curvature term must not in fact grow [1:9.15].

The details are slightly complicated [113], but suffice it to say here that the constraint on $R_{\mu\nu}$ is obtained by averaging over directions, then averaging over positions with a weighting determined by the volume element $\sqrt{g}$ associated with the metric $g_{\mu\nu}$ defined by our choice of hypersurfaces. The requirement that this average not grow when $t$ is scaled up can then be expressed as the vanishing of the variation of $\int R \sqrt{g}$, which is precisely the usual

Einstein–Hilbert action—thereby leading to the conclusion that $R_{\mu\nu}$ must satisfy exactly the usual vacuum Einstein equations [114][115][75][116]:

$$R_{\mu\nu} - \frac{1}{2} R \, g_{\mu\nu} = 0$$

A full derivation of this is given in [113]. Causal invariance plays a crucial role, ensuring for example that timelike directions $t^i$ associated with different foliations give invariant results. Much like in the derivation of continuum fluid behavior from microscopic molecular dynamics (e.g. [110]), one also needs to take a variety of fairly subtle limits, and one needs sufficient intrinsic generation of effective randomness [1:7.5] to justify the use of certain statistical averages.

But there is a fairly simple interpretation of the result above. Imagine all the geodesics that start at a particular point in the causal graph. The further we go, the more possible geodesic paths there will be in the graph. To achieve a power law corresponding to a definite dimension, the geodesics must in a sense just "stream outwards", evenly distributed in direction.

But the Ricci tensor specifically measures the rate at which bundles of geodesics change their cross-sectional area. And as soon as this change is nonzero, it will inevitably change the local density of geodesics and eventually grow to disrupt the power law. And so the only way a fixed limiting dimension can be achieved is for the Ricci curvature to vanish, just as it does according to the vacuum Einstein equations. (Note that higher-order terms, involving for example the Weyl tensor and other components of the Riemann tensor, yield changes in the shape of bundles of geodesics, but not in their cross-sectional area, and are therefore not constrained by the requirement of fixed limiting dimension.)

## 8.8 Matter, Energy and Gravitation

In our models, not only space, but also everything "in space", must be represented by features of our evolving hypergraphs. There is no notion of "empty space", with "matter" in it. Instead, space itself is a dynamic construct created and maintained by ongoing updating events in the hypergraph. And what we call "matter"—as well as things like energy—must just correspond to features of the evolving hypergraph that somehow deviate from the background activity that we call "space".

Anything we directly observe must ultimately have a signature in the causal graph. And a potential hypothesis about energy and momentum is that they may simply correspond to excess "fluxes" of causal edges in time and space. Consider a simple causal graph in which we have marked spacelike and timelike hypersurfaces:

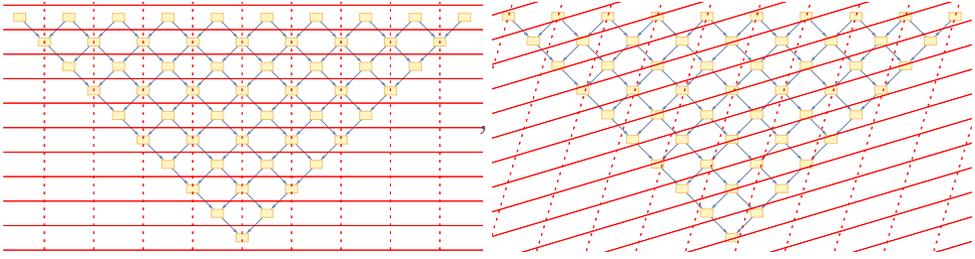

The basic idea is that the number of causal edges that cross spacelike hypersurfaces would correspond to energy, and the number that cross timelike hypersurfaces would correspond to momentum (in the spatial direction defined by a given hypersurface). Inevitably the results one gets would depend on the hypersurfaces one chooses, and so would differ from one observer to another.

And one important feature of this identification of energy and momentum is that it would explain why they follow the same relativistic transformations as time and space. In effect space and time are probing distances between nodes in the causal graph (as measured relative to a particular foliation), while momentum and energy are probing a directly dual property: the density of edges.

There is additional subtlety here, though, because causal edges are needed just to maintain the structure of spacetime—and whatever we measure as energy and momentum must just be some excess in the density of causal edges over the "background" corresponding to space. But even to know what we mean by density we have to have some notion of volume, but this is also itself defined in terms of edges in the causal graph.

But as a rough idealized picture, we might imagine that we have a causal graph that maintains the same overall structure, but adds some extra connections:

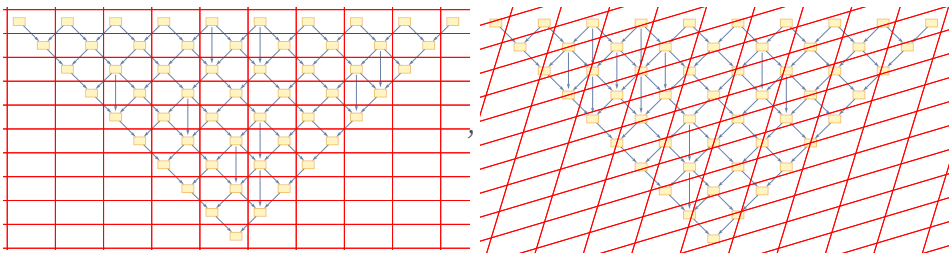

In our actual models, the causal graphs one gets are considerably more complicated. But one can still identify some features from the simple idealization. The basic concept is that energy and momentum add "extra causal connections" that are not "necessary" to define the basic structure of spacetime. In a sense the core thing that defines the structure of spacetime is the way that "elementary light cones" are knitted together.

Consider a causal graph like:

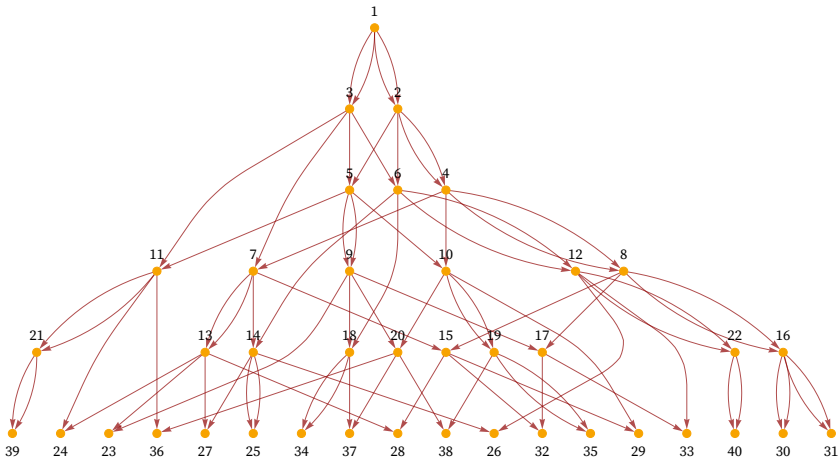

One can think of a set of edges like the ones indicated as in effect "outlining" the causal graph. But then there are other edges that add "extra connections". The edges that "outline the graph" in effect maximally connect spatially separated regions—or in a sense transmit causal information at a maximum speed. The other edges one can think of as having slower speeds—so they are typically drawn closer to vertical in a rendering like the one above.

But now let us return to our simple grid idealization of the causal graph—with additional vertical edges added. Now do foliations like the ones we used above to represent inertial frames, parametrized by a velocity ratio $\beta$ relative to the maximum speed (taken to be 1). Define $E(\beta)$ to be the density of causal edge crossing of the spacelike hypersurfaces, and $p(\beta)$ the corresponding quantity for timelike hypersurfaces. Then for speed 1 edges, we have (up to an overall multiplier) (cf. [111][112]):

$$E(\beta) = p(\beta) = \frac{1+\beta}{\sqrt{1-\beta^2}}$$

But in general for edges with speed $\alpha$ we have

$$E(\beta) = \frac{1-\alpha\beta}{\sqrt{1-\beta^2}}, \quad p(\beta) = \frac{\alpha-\beta}{\sqrt{1-\beta^2}}$$

which means that for any $\beta$

$$E(\beta)^2 - p(\beta)^2 = 1 - \alpha^2$$

thus showing that our crossing densities transform like energy and momentum for a particle with mass $\sqrt{1-\alpha^2}$. In other words, we can potentially identify edges that are not maximum speed in the causal graph as corresponding to "matter" with nonzero rest mass. Perhaps not surprisingly, this whole setup is quite analogous to thinking about world lines of massive particles in elementary treatments of relativity.

But in our context, all of this must emerge from underlying features of the evolving hypergraph. Causal connections that transfer information at maximum speed can be thought of as arising from updating events that involve maximally separate nodes, and that are somehow always entraining "fresh" nodes. But causal connections that transfer information more slowly are associated with sequences of updating events that in effect reuse nodes. So in other words, rest mass can be thought of as being associated with local collections of nodes in the hypergraph that allow repeated updating events to occur without the involvement of other nodes.

Given this setup, it is possible to derive other features of energy, momentum and mass by methods similar to those used in typical discussions of relativity. It is first helpful to include units in the quantities we have introduced. If an elementary light cone has timeline extent $T$ then we can consider its spacelike extent to be $c\,T$, where $c$ is the speed of light. Within the light cone let us say that there effectively $\mu$ causal edges oriented in the timelike direction. With the inertial frame foliations used above, the contribution of these causal edges to energy and momentum will be (the factor $c$ in the energy case comes from the spacelike extent of the light cone):

$$E(\beta) = c\,\frac{\mu}{\sqrt{1-\beta^2}} = c\,\mu\,(1 + \frac{\beta^2}{2} + O(\beta^4))$$

$$p(\beta) = \mu\,\frac{\beta}{\sqrt{1-\beta^2}} = \mu\,\beta\,(1 + O(\beta^2))$$

But if we define the mass $m$ as $\frac{\mu}{c}$ and substitute $\beta = \frac{v}{c}$, we get the standard formulas of special relativity [111][112], or to first order

$$E = m\,c^2 + \frac{1}{2}\,m\,v^2$$

$$p = m\,v$$

establishing in our model the relation $E = m\,c^2$ between energy and rest mass.

We should note that with our identification for energy and momentum, the conservation of energy becomes essentially the statement that the overall density of events in the causal network does not change as we progress through successive spacelike surfaces. And, as we will discuss later, if in effect the whole hypergraph is in some kind of dynamic equilibrium, then we can reasonably expect that this will be the case. Expansion (or, more specifically, non-uniform expansion) will lead to effective violations of energy conservation, much as it does for an expanding universe in the traditional formalism of general relativity [117][75].

In the previous subsection, we discussed the overall structure of spacetime, and we used the growth rate of the spacetime volume $C_t(X)$ as a way to assess this. But now let us ask about specific values of $C_t(X)$, complete with their "constant" multipliers. We can think of these multipliers as probing the local density of the causal graph. But deviations in this are what we have now identified as being associated with matter.

To compute $C_t(X)$ we ultimately need to be able to precisely count events in the causal graph. If the causal graph is somehow "uniform", then it cannot contain what can be considered to be "matter". In the setup we have defined, the presence of matter is effectively associated with "fluxes" of causal edges that reflect the non-uniform "arrangement" of nodes in the causal graph. To represent this, take $\rho(X)$ to be the "local density" of nodes in the causal graph. We can make a series expansion to probe deviations from uniformity in $\rho(X)$. And formally we can write

$$\rho(X) = \rho_0 \left( 1 + \sigma\, \delta t^\mu\, \delta t^\nu\, T_{\mu\nu} \ldots \right)$$

where the $t^\mu$ are timelike vectors used in the definition of $C_t$ and now $T_{\mu\nu}$ is effectively a tensor that represents "fluxes of edges" in the causal graph. But these fluxes are what we have identified as energy and momentum, and when we think about how causal edges traverse spacelike and timelike hypersurfaces, $T_{\mu\nu}$ turns out to correspond exactly to the standard energy-momentum tensor of general relativity.

So now we can combine our formula for the effect of local density with our formula for the effect of curvature from the previous section to get:

$$C_t(X) = \rho_0 (1 + \sigma\, \delta t^i\, \delta t^j\, T_{ij} + \ldots)\, t^{d+1} \left(1 - \frac{1}{6} \delta t^i\, \delta t^j\, R_{ij} + \ldots \right)$$

But if we apply the same argument as in the previous subsection, then to maintain limiting fixed dimension we get the condition

$$R_{\mu\nu} - \frac{1}{2} R\, g_{\mu\nu} = \sigma'\, T_{\mu\nu}$$

which has exactly the form of Einstein's equations in the presence of matter [114][115][75][116].

Just as we interpreted the curvature part of these equations in the previous subsection in terms of the change in area of geodesic bundles, we can interpret the "matter" part in terms of the change of geodesics associated with additional local connections. As an example, consider starting with a 2D hexagonal grid. Now imagine adding edges at each node. Doing this creates additional connections and additional geodesics, eventually producing something like the hyperbolic space examples in 4.2. So what the equation says is that any such effect, which would lead to negative curvature, must be compensated by positive curvature in the "background" spacetime—just as general relativity suggests.

## 8.9 Elementary Particles

Elementary particles are entities that—at least for some period—preserve their identity through space and time. In the context of our models, one can imagine that particles would correspond to structures in the hypergraph that are locally stable under the application of rules.

As an idealized example, consider rules that operate on an ordinary graph, and have the property of preserving planarity. Such rules can never remove non-planarity from a graph. But it is a basic result of graph theory [37][118] that any non-planarity can always be attributed to one of the two specific subgraphs:

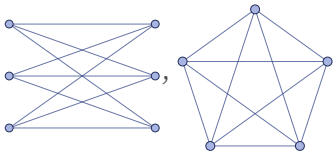

If one inserts such subgraphs into an otherwise planar graph, they behave very much as "particle-like" structures. They can move around, but unless they meet and annihilate, they are preserved:

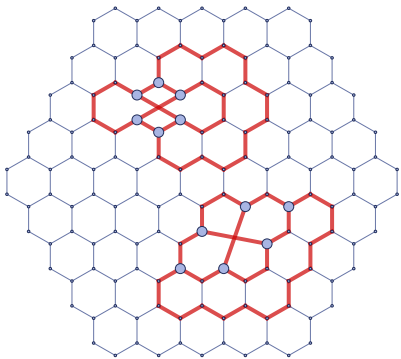

There are presumably analogs of this in hypergraph-rewriting rules of the kind that appear in our models. Given a particular set of rules, the expectation would be that a certain set of local sub-hypergraphs would be preserved by the rules. Existing results in graph theory do not go very far in elucidating the details.

However, there are analogs in other systems that provide some insight. Cellular automata provide a particularly good example. Consider the rule 110 cellular automaton [1:p32]. Starting from a random initial condition, the picture below shows how the system evolves to a collection of localized structures:

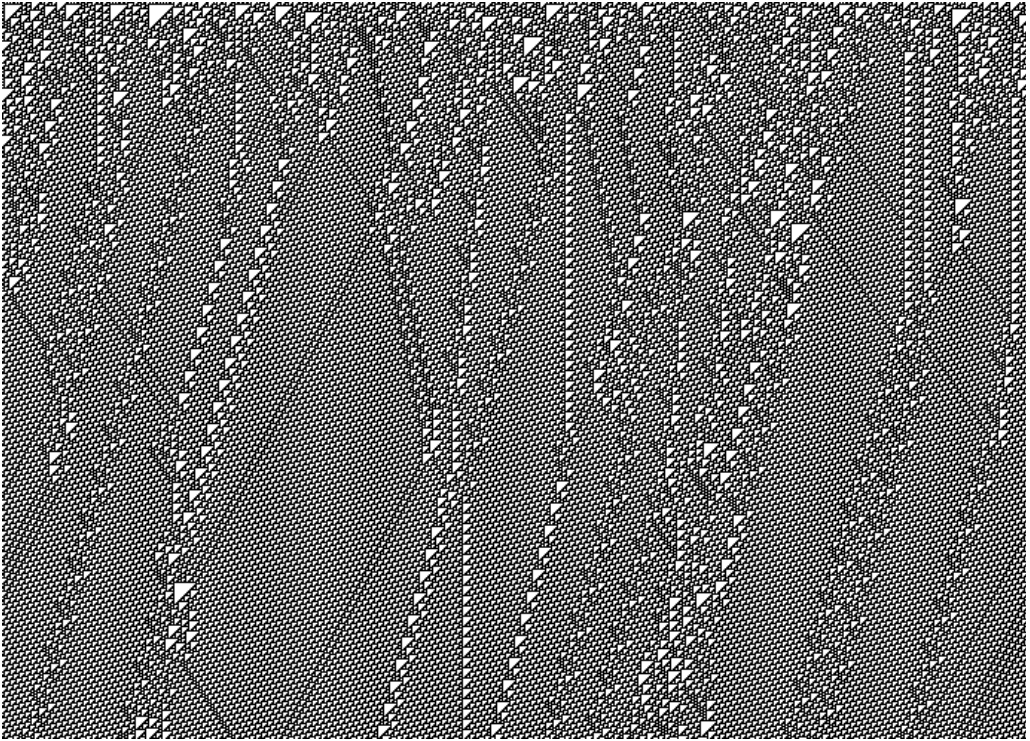

The form of these structures is hard to determine directly from the rule. (They are a little like hard-to-predict solutions to a Diophantine equation.) But by explicit computation one can determine for example that rule 110 supports the following types of localized structures [1:p292][119]

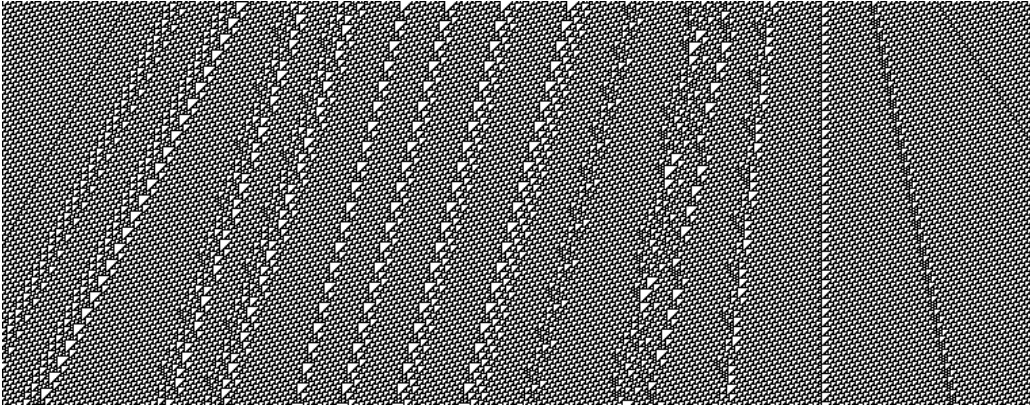

as well as the growing structure:

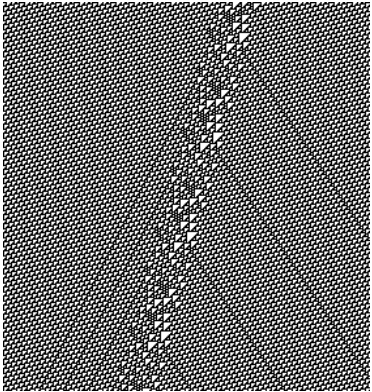

There is a complex web of possible interactions between localized structures, that can at least in some cases be interpreted in terms of conservation laws:

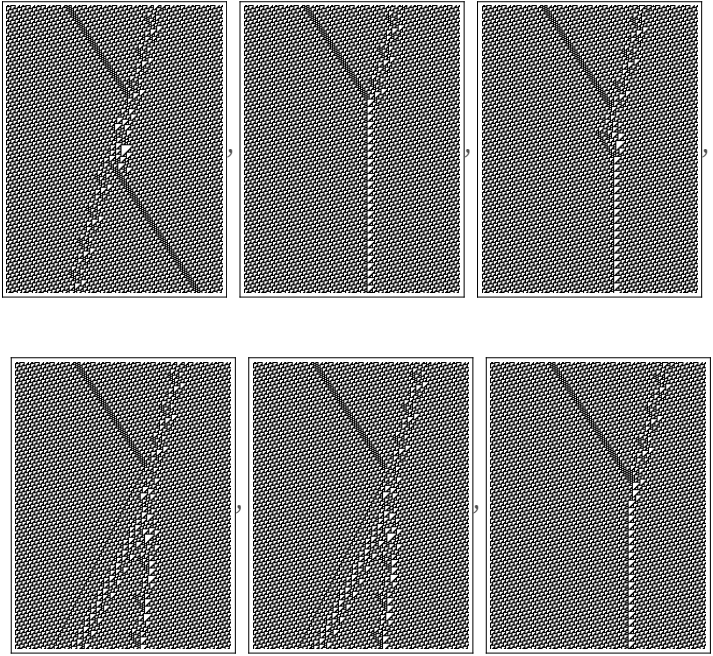

As in cellular automata, it is likely that not every one of our models will yield localized structures, although there is reason to think that some form of conserved structure will be more common in hypergraph rewriting than in cellular automata. But as in cellular automata, one can expect that with a given underlying rule, there will be a discrete set of possible localized structures, with hard-to-predict sizes and properties.

The particular set of localized structures will probably be quite specific to particular rules. But as we will discuss in the next subsection, there will often be symmetries that cause collections of similar structures to exist—or in fact force certain structures to exist.

In the previous subsection, we discussed the interpretation of energy and momentum in terms of additional edges in a causal graph. For particles, the expectation would be that there is a certain "core" structure that defines the core properties of a particle (like spin, charge, etc.), but that this structure is spread across a region of the hypergraph that maintains the "activity" associated with energy and momentum.

It is worth noting that even in an example like non-planarity, it is perfectly possible for topological-like features to effectively be spread out across many nodes, while still maintaining their discrete character.

In the previous subsection, we discussed the potential origin of rest mass in terms of "reuse" of nodes in the hypergraph. Once again, this seems to fit in well with our notion of the nature of particles—and to make it perfectly possible to imagine both "massive" and "massless" particles, associated with different kinds of structures in the evolving hypergraph.

In a system like the rule 110 cellular automaton, there is a clear "background structure" on which it is possible to identify localized structures. In some very simple cases, similar things happen in our models. For example, consider the rule:

$\{\{x, y\}, \{y, z, u, v\}\} \rightarrow \{\{x, y, z, u\}, \{u, v\}\}$

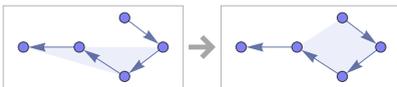

The evolution of this rule yields behavior like

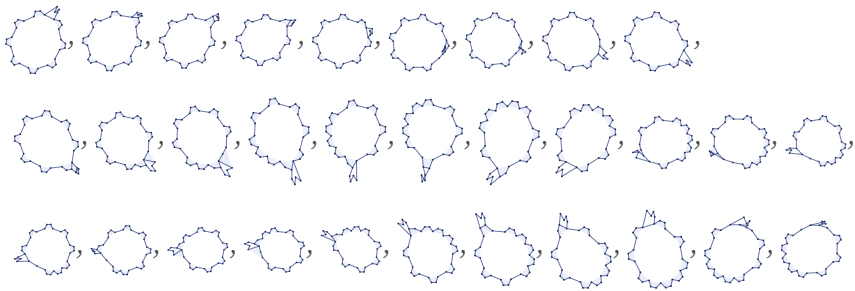

in which there is a circular "background", with a localized "particle-like" deformation. The causal graph (here generated for a larger case) also shows evidence of a particle-like structure on a simple grid-like background:

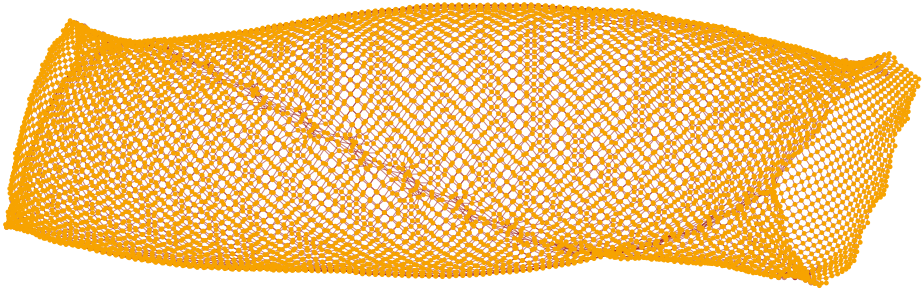

But in most of our models the "background" tends to be much more complicated, and so such direct methods for identifying particles cannot be used. But as an alternative, one can consider exploring the effect of perturbations, as in 4.14. In effect, one starts the system with a perturbation, then sees whether the perturbation somehow decomposes into quantized elements that one can identify as "particles". (The process is quite analogous to producing particles in a high-energy collision.)

Such quantized effects are at best rare in class 3 cellular automata, but they are a defining feature of class 4 cellular automata, and there is reason to believe that they will be at least fairly common in our models.

The defining feature of a localized "particle-like" structure is that it is capable of long-range propagation in the system. But the presence of even short-lived instances of particle-like structures will also potentially be identifiable—though with a certain margin of error—from detailed properties of the hypergraph in small regions. And in the "background" evolution of our models, one can expect that short-lived instances of particle-like structures will continually be being created and destroyed.

The process that in a sense "creates the structure of space" in our models can thus also be thought of as producing a "vacuum" full of particle-like activity. And particularly when this is combined with the phenomenon (to be discussed in a later subsection) that pairs of particle-like structures can be produced and subsequently merged in the multiway system, there is some definite similarity with the ubiquitous virtual particles that appear in traditional treatments of quantum field theory.

## 8.10 Reversibility and Irreversibility

One feature of the traditional formalism for fundamental physics is that it is reversible, in the sense that it implies that individual states of closed systems can be uniquely evolved both forward and backward in time. (Time reversal violation in things like $K^o$ particle decays show that the rule for going forward and backward in time can be slightly different. In addition, the cosmological expansion of the universe defines an overall arrow of time.)

One can certainly set up manifestly reversible rewriting rules (like A→B, B→A) in models like ours. And indeed the example of cellular automata [1:9.2] tends to suggest that most kinds of behavior seen in irreversible rules can also be seen—though perhaps more rarely—in reversible rules.

But it is important to realize that even when the underlying rules for a system are not reversible, the system can still evolve to a situation where there is effective reversibility. One way for this to happen is for the evolution of the system to lead to a particular set of "attractor" states, on which the evolution is reversible. Another possibility is that there is no such well-defined attractor, but that the system nevertheless evolves to some kind of "equilibrium" in which measurable effects show effective reversibility.

In our models, there is an additional complication: the fact that different possible updating orders lead to following different branches of the multiway system. In most kinds of systems, irreversible rules tend to be associated with the phenomenon of multiple initial states merging to produce a single final state in which the information about the initial state is lost. But when there is a branch in a multiway system, this is reversed: information is effectively created by the branch, and lost if one goes backwards.

When there is causal invariance, however, yet something different happens. Because now in a sense every branching will eventually merge. And what this means is that in the multiway system there is a kind of reversibility: any information created by a branching will always be destroyed again when the branches merge—even though temporarily the "information content" may change.

It is important to note that this kind of microscopic reversibility is quite unrelated to the more macroscopic irreversibility implied by the Second Law of thermodynamics. As discussed in [1:9.3] the Second Law seems to first and foremost be a consequence of computational irreducibility. Even when the underlying rules for a system are reversible, the actual evolution of the system can so "encrypt" the initial conditions that no computationally feasible measurement process will succeed in reconstructing them. (The idea of considering computational feasibility clarifies past uncertainty about what might count as a reasonable "coarse graining procedure".)

In any nontrivial example of one of our models, computational irreducibility is essentially inevitable. And this means that the model will tend to intrinsically generate effective randomness, or in other words, the computation it does will obscure whatever simplicity might have existed in its initial conditions.

There can still be large-scale features—or particle-like structures—that persist. But the presence of computational irreducibility implies that even at a level as low as the basic structure of space we can expect our models to show the kind of irreversibility associated with the Second Law. And in a sense we can view this as the reason that things like a robust structure for space can exist: because of computational irreducibility, our models show a kind of equilibrium in which the details are effectively random, and the only features that are computationally feasible to measure are the statistical regularities.

## 8.11 Cosmology, Expansion & Singularities

In our models the evolving hypergraph represents the whole universe, and the expansion of the universe is potentially a consequence of the growth of the hypergraph. In the minimal case of a model involving a single transformation rule, the growth of the hypergraph must be monotonic, although the rate can vary depending on the local structure of the hypergraph. If there are multiple transformation rules, there can be both increase and decrease in hypergraph size. (Even with a single rule, there is also still the possibility—discussed below—of effective size decrease as a result of pieces of the hypergraph becoming disconnected.)

In the case of uniform growth, measurable quantities such as length and energy would essentially all continually scale as the universe evolves. The core structure of particles—embodied for example in topological-like features of the hypergraph—could potentially persist even as the number of nodes "within them" increases. Since the rate of increase in size in the hypergraph would undoubtedly greatly exceed the measurable growth rate of the universe, uniform growth implies a kind of progressive refinement in which the length scale of the discrete structure of the hypergraph becomes ever more distant from any given measured length scale—so that in effect the universe is becoming an ever closer approximation to continuous.

In traditional cosmology, one thinks of the universe as effectively having exactly three dimensions of space (cf. [120]). In our models, dimension is in effect a dynamical variable. Possibly some of what is normally attributed to curvature in space can instead be reformulated as dimension change. But even beyond this, there is the potential for new phenomena associated, for example, with local change of dimension. In general, a change of dimension—like curvature—affects the density of geodesics. Changes of dimension generated by an underlying rule may potentially lead to effects that for example mimic the presence of mass, or positive or negative energy density. (There could also be dimension-change "waves", perhaps with some rather unusual features.)

In our models, the universe starts from some initial configuration. It could be something like a single self-loop hypergraph. Or in the multiway system it could be multiple initial hypergraphs. (Note that we can always "put the initial conditions into the rule" by adding a rule that says "from nothing, create the initial conditions".)

An obvious question is whether any traces of the initial conditions might persist, perhaps even through the whole evolution of the system. The effective randomness associated with computational irreducibility in the evolution will inevitably tend to "encrypt" most features of the initial conditions [1:9.3] to the point where they are unrecognizable. But it is still conceivable that, for example, some global symmetry breaking associated with the first few hypergraph updating events could survive—and the remote possibility exists that this could be visible today in the large-scale structure of the universe, say as a pattern of density fluctuations in the cosmic microwave background.

Our models have potentially important implications for the early universe. If, for example, the effective dimension of the universe was initially much higher than 3 (as is basically inevitable if the initial conditions are small), there will have been a much higher level of causal contact between different parts of the universe than we have deduced by extrapolating the 3D expansion of the universe today [1:p1055]. (In effect this happens because the volume of the past light cone will grow like $t^d$—or perhaps exponentially with $t$—and not just like $t^3$.)

As we discussed in 2.9, it is perfectly possible in our models for parts of the hypergraph to become disconnected as a result of the operation of the rule. But assuming that the rule is local (in the sense that its left-hand side is a connected hypergraph), pieces of the hypergraph that become disconnected can never interact again. Even independent of outright disconnection of the spatial graph, it is also possible for the causal graph to "tear" into disconnected parts that can never interact again (see 6.10):

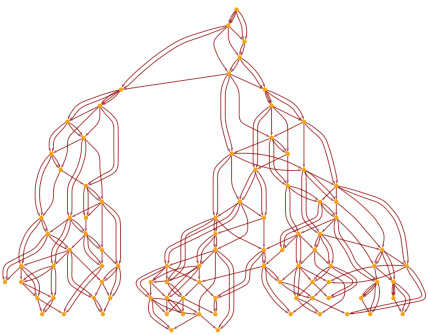

A disconnection in the causal graph corresponds to an event horizon in our system—that cannot be crossed by any timelike curve. (And indeed our causal graphs—consisting as they do of "elementary light cones knitted together"—are like microscopic analogs of the causal diagrams often used in studying general relativity.)

We can also ask about other extreme phenomena in spacetime. Closed timelike curves correspond to loops in the causal graph, and with some rules they can occur. But they do not represent any real form of "time travel"; they just correspond to the presence of states that are precisely repeated as a result of the evolution of the system. (Note that in our models, time effectively corresponds to the progression of computation, and has a very different underlying character from something like space.)

Wormholes and effective faster-than-light travel are not specifically excluded by the structure of our models, especially insofar as there can potentially be deviations in the effective local dimensionality of space. But insofar as the conditions to get general relativity as a limiting effective theory are satisfied, these will occur only in the circumstances where they do in that theory.

## 8.12 Basic Concepts of Quantum Mechanics

Quantum mechanics is a key known feature of physics, and also, it seems, a natural and inevitable feature of our models. In classical physics—or in a system like a cellular automaton—one basically has rules that specify a unique path of history for the evolution of a system. But our models are not set up to define any such unique path of history. Instead, the models just give possible rewrites that can be performed on hypergraphs—but they do not say when or where these rewrites should be applied. So this means that—like the formalism of quantum mechanics—our models in a sense allow many different paths of history.

There is, however, ultimately nothing non-deterministic about our models. Although they allow many different sequences of updating events—each of which can be viewed as a different path of history—the models still completely determine the overall set of possible sequences of updating events. And indeed at a global level, everything about the model can be captured in a multiway graph [1:5.6]—like the one below—with nodes in the graph corresponding to states of the system (here, for simplicity, a string substitution system), and every possible path through the graph corresponding to a possible history.

In the standard formalism of quantum mechanics, one usually just imagines that all one can determine are probabilities for different histories or different outcomes. But this has made it something of a mystery why we have the impression that a definite objective reality seems to exist. One possible explanation would be that at some level a branch of reality exists for every possible behavior, and that we just experience the branch that our thread of consciousness has happened to follow.

But our models immediately suggest another, more complete, and arguably much more scientifically satisfying, possibility. In essence, they suggest that there is ultimately a global objective reality, defined by the multiway system, and it is merely the locality of our experience that causes us to describe things in terms of probabilities, and all the various detailed features of the standard formalism of quantum mechanics.

We will proceed in two stages. First, we will discuss the notion of an observer in the context of multiway systems, and the relation of this to questions about objective reality. And having done this, we will be in a position to discuss ideas like quantum measurement, and the role that causal invariance turns out to play in allowing observers to experience definite, seemingly classical results.

So how might we represent a quantum observer in our models? The first key point is that the observer—being part of the universe—must themselves be a multiway system. And in addition, everything the observer does—and experiences—must correspond to events that occur in the model.

This latter point also came up when we discussed spacetime—and we concluded there that it meant we only needed to consider the graph of causal relationships between events. To characterize any given observer, we then just had to say how the observer would sample this causal graph. A typical example in studying spacetime is to consider an observer in an inertial reference frame—which corresponds to a particular foliation of the causal graph. But in general to characterize what any observer will experience in the course of time, we need some sequence of spacelike hypersurfaces that form a foliation which respects the causal relationships—and thus the ordering relations between events—defined by the causal graph.

But now we can see an analog of this in the quantum mechanical case. However, instead of considering foliations of the causal graph, what we need to consider now are foliations of the multiway graph:

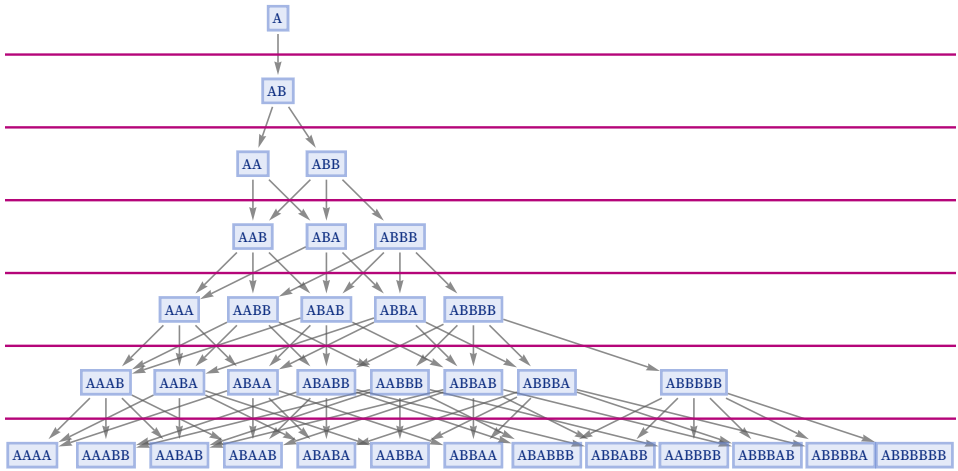

In the course of time, the observer progresses through such a foliation, in effect at each step observing some collection of states, with certain relationships between them. A different observer, however, might want to sample the states differently, and might effectively define a different foliation.

One can potentially think of a different foliation as being a different "quantum observation frame" or "quantum frame", analogous to the different reference frames one considers in studying spacetime. In the case of something like an inertial frame, one is effectively defining how an observer will sample different parts of space over the course of time. In a quantum observation frame one might have a more elaborate specification, involving sampling particular states of relevance to some measurement or another. But the key point is that a quantum observer can in principle use any quantum observation frame that corresponds to a foliation that respects the relationships between states defined by the multiway graph (and thus has a meaningful notion of time).

In both the spacetime case and the quantum case, the slices in the foliation are indexed by time. But while in the spacetime case, where each slice corresponds to a spacelike hypersurface that spans ordinary space, in the quantum case, each slice corresponds to what we can call a branchlike hypersurface that spans not ordinary space, but instead the space of states, or the space of branches in the multiway system. But even without knowing the details of this space, we can already come to some conclusions.

In particular, we can ask what observers with different quantum observation frames—and thus different choices of branchlike hypersurfaces—will conclude about relationships between states. And the point is that so long as the foliations that are used respect the orderings defined by the multiway graph, all observers must inevitably come to the same conclusions about the structure of the multiway graph—and therefore, for example, the relationships between states. Different observers may sample the multiway graph differently, and experience different histories, but they are always ultimately sampling the same graph.

One feature of traditional quantum formalism is its concept of making measurements that effectively reduce collections of states—as exist in a multiway system—to what is basically a single state analogous to what would be seen in a classical single path of evolution. From the point of view of quantum observation frames, one can think of such a measurement as being achieved by sculpting the quantum observation frame to effectively pick out a single state in the multiway system:

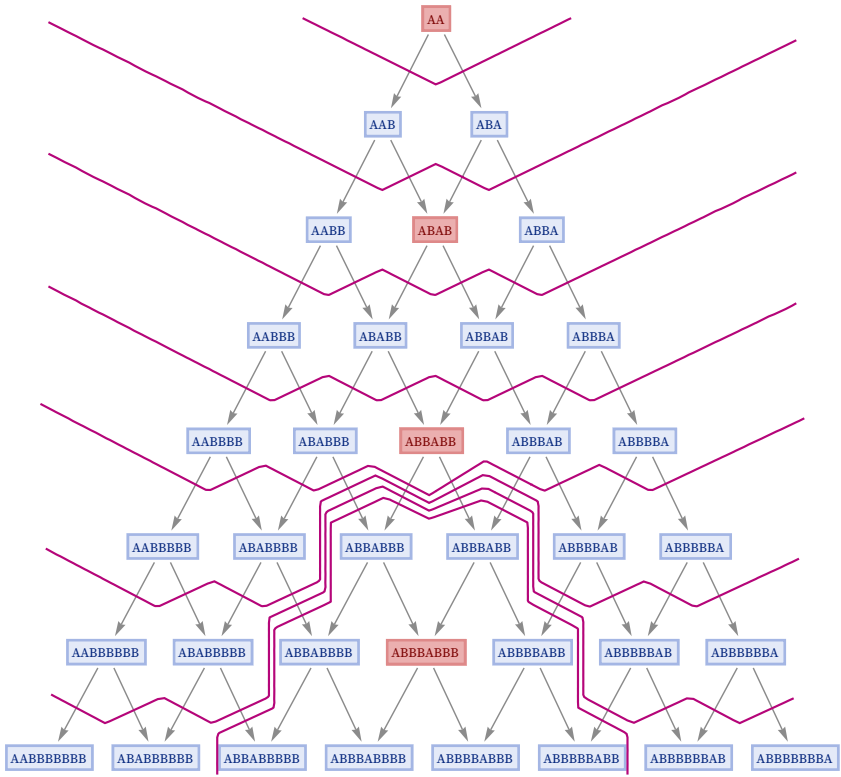

We will discuss this in more detail below. But the basic idea is as follows. Imagine that our universe is based on a simple string substitution system such as {A→AB}. If we start from a state AA, as in the picture above, the multiway evolution from this state immediately leads to multiple outcomes, associated with different updating events. But let us say that we just wanted some kind of "classical" summary of the evolution, ignoring all these different branches.

One thing we might do is not trace individual updates, but instead just look at "generational states" (5.21) in which all updates that can consistently be applied together have been applied. And with the particular rule shown here, we then get the unique sequence of states highlighted above. And as we will discuss below, we can indeed consider these generational states as corresponding to definite ("classical-like") states of the system, that can consistently be thought of as potential results of measurements.

But now let us imagine how this might work in something closer to a complete experiment. We are running the multiway system shown above. Multiple states are being generated. But at some moment we as observers notice that actually several states that have been produced (say ABBA and AABB) can be combined together to form a consistent generational state (ABBABB). But even though these states ultimately had a common ancestor, they now seem to be on different "branches of history".

But now causal invariance makes a crucial contribution. Because it implies that all such different branches must eventually converge. And indeed after a couple of steps, the fully assembled generational state ABBABB appears in the multiway system. To us as observers this is in a sense the state we were looking for (it is the "result of our measurement"), and as far as possible, we want to use it as our description of the system.

And by setting up an appropriate quantum observation frame, that is exactly what we can do. For example, as illustrated in the picture above, we can make the foliation we choose effectively freeze the generational state, so that in the description we use of the system, the state stays the same in successive slices.

The structure of the multiway system puts constraints on what foliations we can consistently set up. In the case shown here, it does allow us to freeze this particular state forever, but to do this consistently, it effectively forces us to freeze more and more states over time. And as we will see later, this kind of spreading of effects in the multiway graph is closely related to decoherence in the standard formalism of quantum mechanics.

In what we just discussed, causal invariance is what guarantees that states the observer notices can consistently be assembled to form a generational ("classical-like") state that will always actually converge in the multiway system to form that state. But it is worth pointing out that (as discussed in [121]) strict causal invariance is not ultimately needed for a picture like this to work.

Recall that the observer themselves is also a multiway system. So "within their consciousness" there will usually be many "simultaneous" states. Looked at formally from the outside, the observer can be seen to involve many distinct states. But one could imagine that the internal experience of the observer would be in effect to conflate these states.

Causal invariance ensures that branches in the multiway system will actually merge—just as a result of the evolution of the multiway system. But if the observer "experientially" conflates states, this in effect represents an additional way in which different branches in the multiway system will at least appear to merge [121]. Formally, one can think of this—in analogy to the operation of automated theorem-proving systems—as like the observer "adding lemmas" that assert the equivalence of branches, thereby allowing the system to be "completed" to the point where relevant branches converge. (For a given system, there is still the question of whether only a sufficiently bounded number of lemmas is needed to achieve the convergence one wants.)

Independent of whether there is strict causal invariance or not, there is also the question of what kinds of quantum observation frames are possible. In the end—just like in the space-time case—such frames reflect the description one is choosing to make of the world. And setting up different "coordinates", one is effectively changing one's description, and picking out different aspects of a system. And ultimately the restrictions on frames are computational ones. Something like an inertial frame in spacetime is simple to describe, and its coordinates are simple to compute. But a frame that tries to pick out some very particular aspect of a quantum system may run into issues of computational irreducibility. And as a result, much as happens in connection with the Second Law of thermodynamics [1:9.3], there can still for example be elaborate correlations that exist between different parts of a quantum system, but no realistic measurement—defined by a computationally feasible quantum observation frame—will succeed in picking them out.

## 8.13 Quantum Formalism

To continue understanding how our models might relate to quantum mechanics, it is useful to describe a little more of the potential correspondence with standard quantum formalism. We consider—quite directly—each state in the multiway system as some quantum basis state $|S>$.

An important feature of quantum states is the phenomenon of entanglement—which is effectively a phenomenon of connection or correlation between states. In our setup (as we will see more formally soon), entanglement is basically a reflection of common ancestry of states in the multiway graph. ("Interference" can then be seen as a reflection of merging—and therefore common successors—in the multiway graph.)

Consider the following multiway graph for a string substitution system:

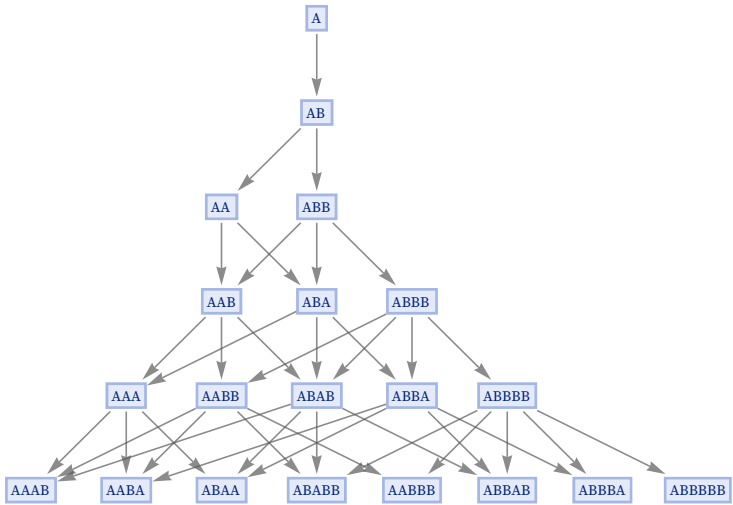

Each pair of states generated by a branching in this graph are considered to be entangled. And when the graph is viewed as defining a rewrite system, these pairs of states can also be said to form a branch pair.

Given a particular foliation of the multiway graph, we can now capture the entanglement of states in each slice of the foliation by forming a branchial graph in which we connect the states in each branch pair. For the string substitution system above, the sequence of branchial graphs is then:

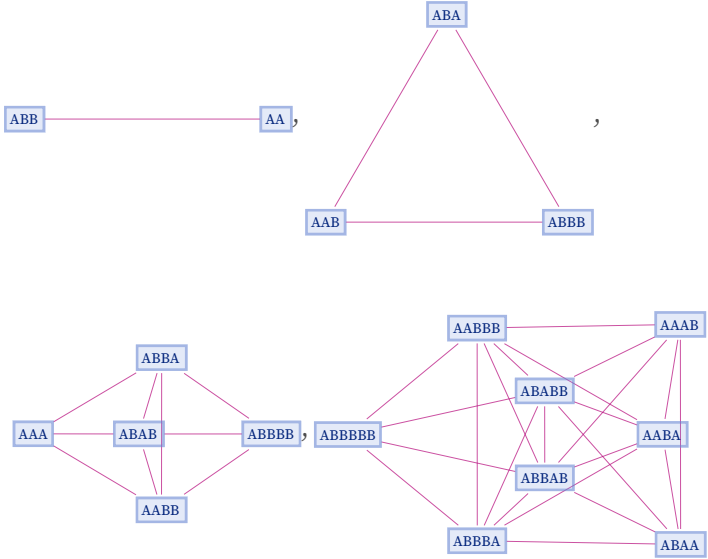

In physical terms, the nodes of the branchial graph are quantum states, and the graph itself forms a kind of map of entanglements between states. In general terms, we expect states that are closer on the branchial graph to be more correlated, and have more entanglement, than ones further away.

As we discussed in 5.17, the geometry of branchial space is not expected to be like the geometry of ordinary space. For example, it will not typically correspond to a finite-dimensional manifold. We can still think of it as a space of some kind that is reached in the limit of a sufficiently large multiway system, with a sufficiently large number of states. And in particular we can imagine—for any given foliation—defining coordinates of some kind on it, that we will denote $\overleftarrow{b}$. So this means that within a foliation, any state that appears in the multiway system can be assigned a position $(t, \overleftarrow{b})$ in "multiway space".

In the standard formalism of quantum mechanics, states are thought of as vectors in a Hilbert space, and now these vectors can be made explicit as corresponding to positions in multiway space.

But now there is an additional issue. The multiway system should represent not just all possible states, but also all possible paths leading to states. And this means that we must assign to states a weight that reflects the number of possible paths that can lead to them:

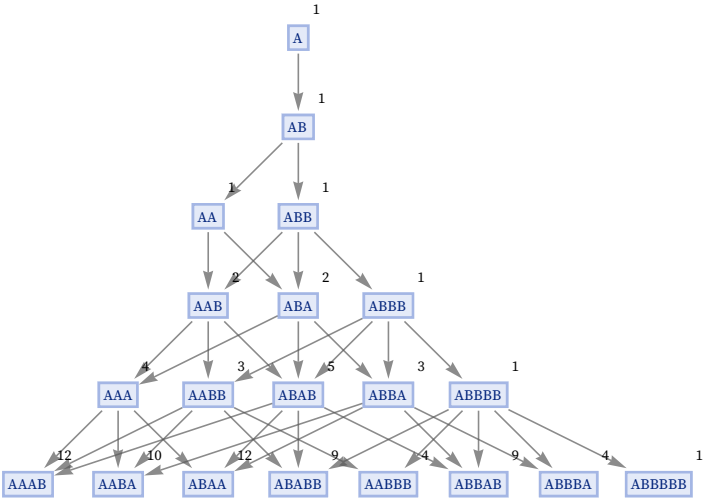

In effect, therefore, each branchlike hypersurface can be thought of as exposing some linear combination of basic states, each one with a certain weight:

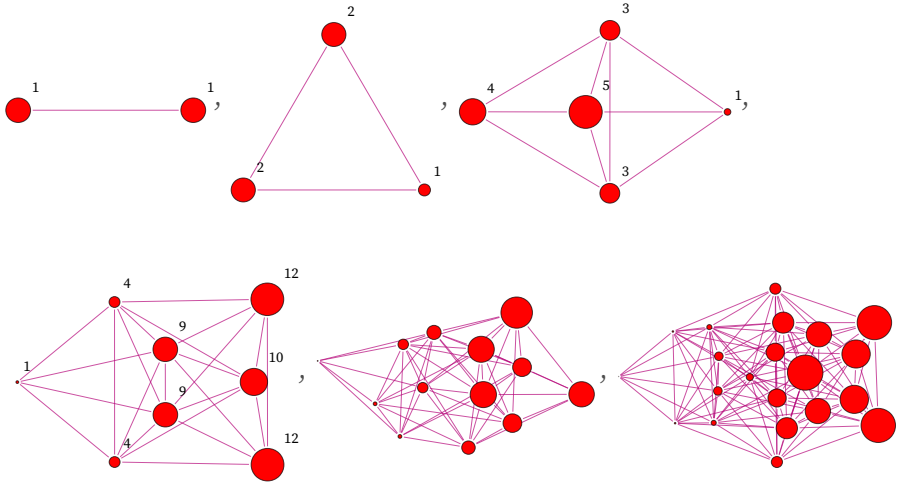

Let us say that we want to track what happens to some part of this branchlike hypersurface. Each state undergoes updating events that are represented by edges in the multiway graph. And in general the paths followed in the multiway graph can be thought of as geodesics in multiway space. And to determine what happens to some part of the branchlike hypersurface, we must then follow a bundle of geodesics.

A notable feature of the multiway graph is the presence of branching and merging, and this will cause our bundle of geodesics to diverge and converge. Often in standard quantum formalism we are interested in the projection of one quantum state on another < | >. In our

setup, the only truly meaningful computation is of the propagation of a geodesic bundle. But as an approximation to this that should be satisfactory in an appropriate limit, we can use distance between states in multiway space, and computing this in terms of the vectors $\xi_i = (t_i, b_i^?)$ the expected Hilbert space norm [122][123] appears: $(\xi_1 - \xi_2)^2 = \xi_1^2 + \xi_2^2 - 2\,\xi_1.\xi_2$.

Time evolution in our system is effectively the propagation of geodesics through the multiway graph. And to work out a transition amplitude $<i\,|\,S\,|\,f>$ between initial and final states we need to see what happens to a bundle of geodesics that correspond to the initial state as they propagate through the multiway graph. And in particular we want to know the measure (or essentially cross-sectional area) of the geodesic bundle when it intersects the branchlike hypersurface defined by a certain quantum observation frame to detect the final state.

To analyze this, consider a single path in the multiway system, corresponding to a single geodesic. The critical observation is that this path is effectively "turned" in multiway space every time a branching event occurs, essentially just like in the simple example below:

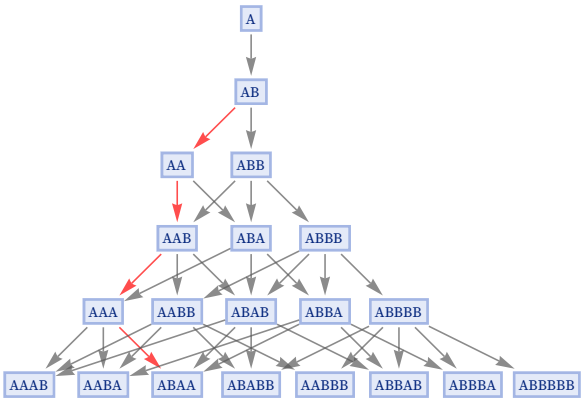

If we think of the turns as being through an angle $\theta$, the way the trajectory projects onto the final branchlike hypersurface can then be represented by $e^{i\theta}$. But to work out the angle $\theta$ for a given path, we need to know how much branching there will be in the region of the multiway graph through which it passes.

But now recall that in discussing spacetime we identified the flux of edges through spacelike hypersurfaces in the causal graph as potentially corresponding to energy. The spacetime causal graph, however, is just a projection of the full multiway causal graph, in which branchlike directions have been reduced out. (In a causal invariant system, it does not matter what "direction" this projection is done in; the reduced causal graph is always the same.) But now suppose that in the full multiway causal graph, the flux of edges across spacelike hypersurfaces can still be considered to correspond to energy.

Now note that every node in the multiway causal graph represents some event in the multiway graph. But events are what produce branching—and "turns"—of paths in the multiway graph. So what this suggests is that the amount of turning of a path in the multi-way graph should be proportional to energy, multiplied by the number of steps, or effectively the time. In standard quantum formalism, energy is identified with the Hamiltonian *H*, so what this says is that in our models, we can expect transition amplitudes to have the basic form $e^{iHt}$—in agreement with the result from quantum mechanics.

To think about this in more detail, we need not just a single energy quantity—corresponding to an overall rate of events—but rather we want a local measure of event rate as a function of location in multiway space. In addition, if we want to compute in a relativistically invariant way, we do not just want the flux of causal edges through spacelike hypersurfaces in some specific foliation. But now we can make a potential identification with standard quantum formalism: we suppose that the Lagrangian density $\mathcal{L}$ corresponds to the total flux in all directions (or, in other words, the divergence) of causal edges at each point in multiway space.

But now consider a path in the multiway system going through multiway space. To know how much "turning" to expect in the path, we need in effect to integrate the Lagrangian density along the path (together with the appropriate volume element). And this will give us something of the form $e^{iS}$, where *S* is the action. But this is exactly what we see in the standard path integral formulation of quantum mechanics [124].

There are many additional details (see [121]). But the correspondence between our models and the results of standard quantum formalism is notable.

It is worth pointing out that in our models, something like the Lagrangian is ultimately not something that is just inserted from the outside; instead it must emerge from actual rules operating on hypergraphs. In the standard formalism of quantum field theory, the Lagrangian is stated in terms of quantum field operators. And the implication is therefore that the structure of the Lagrangian must somehow emerge as a kind of limit of the underlying discrete system, perhaps a bit like how fluid mechanics can emerge from discrete underlying molecular dynamics (or cellular automata) [110].

One notable feature of standard quantum formalism is the appearance of complex numbers for amplitudes. Here the core concept is the turning of a path in multiway space; the complex numbers arise only as a convenient way to represent the path and understand its projections. But there is an additional way complex numbers can arise. Imagine that we want to put a metric on the full $(t, \vec{x}, \overleftarrow{b})$ space of the multiway causal graph. The normal convention for $(t, \vec{x})$ space is to have real-number coordinates and a norm based on $t^2 - x^2$—but an alternative is use $it$ for time. In extending to $(t, \vec{x}, \overleftarrow{b})$ space, one might imagine that a natural norm which allows the contributions of *t*, *x* and *b* components to be appropriately distinguished would be $t^2 - x^2 + ib^2$.

## 8.14 Quantum Measurement

Above we gave a brief summary of how quantum measurement can work in the context of our models. Here we give some more detail.

In a sense the key to quantum measurement is reconciling our notion that "definite things happen in the universe" with the formalism of quantum mechanics—or the branching structure of a multiway system.

But if definite things are going to happen, what might they be?

Here we will again consider the example of a string substitution system, although the core of what we say also applies to the full hypergraph case. Consider the rule

{A → AB, B → A}

We could imagine a simple "classical" procedure for evolving according to this rule, in which we just do all updates we can (say, based on a left-to-right scan) at each step:

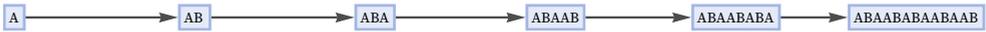

But in fact we know that there are many other possibilities, that can be represented by the multiway system:

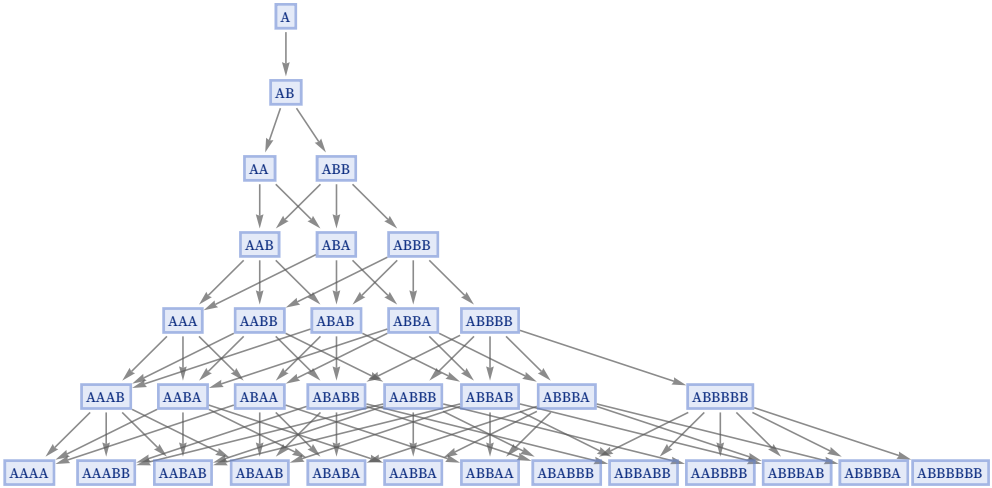

Most of the states that appear in the multiway system are, however, "unfinished", in the sense that there are additional "independent" updates that can consistently be done on them. For example, with the rule {A→BA} there are 4 separate updates that can be applied to AAAA:

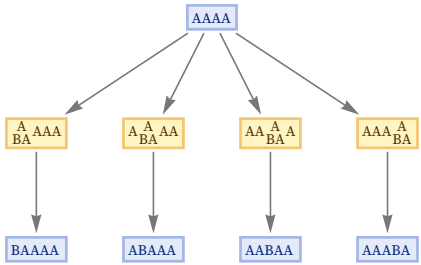

But none of these depend on the others, so they can in effect all be done together, giving the result BABABABA.

Put another way, all of these updates involve "spacelike separated" parts of the string, so they are all causally independent, and can all consistently be carried out at the same time. As discussed in 5.21, doing all updates across a state together can be thought of as evolving a system in "generational steps" to produce "generational states".

In some multiway cases, there may be a single sequence of generational states:

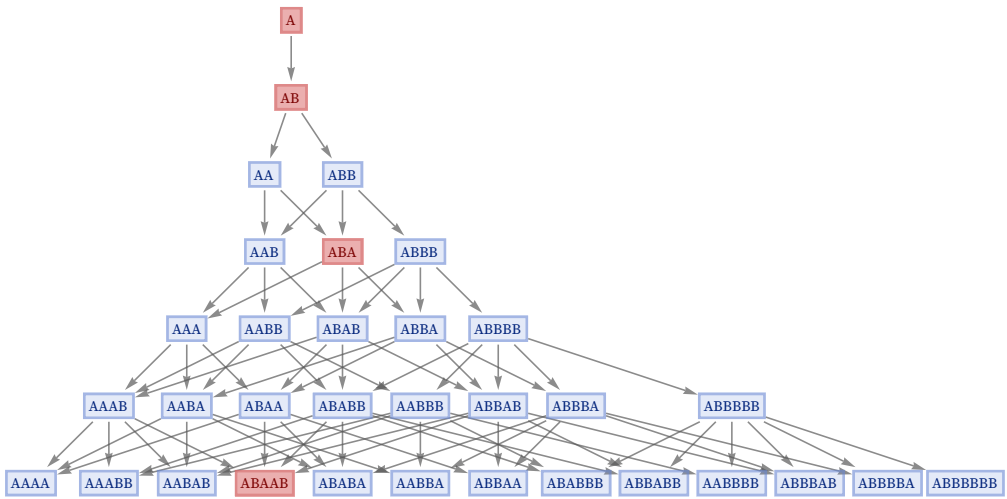

In other cases, there can be several branches of generational states:

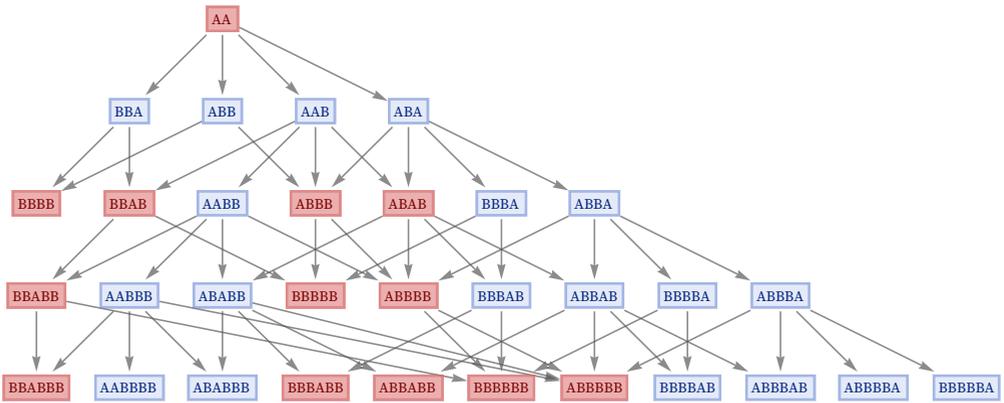

The presence of multiple branches is a consequence of having a mixture of spacelike and branchlike separated events that can be applied to a single state. For example, with the rule {A→AB,A→BB} the first and second updates here are spacelike separated, but the first and third are branchlike separated:

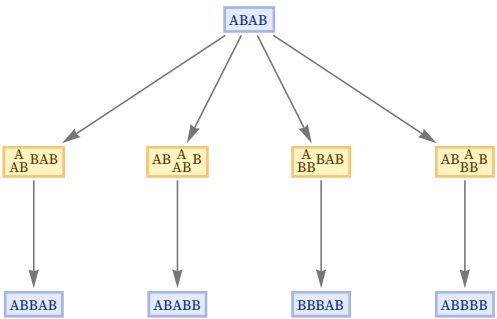

A view of quantum measurement is that it is an attempt to describe multiway systems using generational states. Sometimes there may be a unique "classical path"; sometimes there may be several outcomes for measurements, corresponding to several generational states.

But now let us consider the actual process of doing an experiment on a multiway system—or a quantum system. Our basic goal is—as much as possible—to describe the multiway system in terms of a limited number of generational states, without having to track all the different branches in the multiway system.

At some point in the evolution of a string substitution system we might see a large number of different strings. But we can view them all as part of a single generational state if they in effect yield only spacelike separated events. In other words, if the strings can be assembled without "branchlike ambiguity" they can be thought of as forming a consistent generational state.

In the standard formalism of quantum mechanics, we can think of the states in the multiway system as being quantum states. The construct we form by "assembling" these states can be thought of as a superposition of the states. Causal invariance then implies that through the evolution of the multiway system any such superposition will then actually become a single quantum state. In some sense the observer "did nothing": they just notionally identified a collection of states. It was the actual evolution of the system that produced the specific combined state.

In describing a quantum system—or a multiway system—one must in effect define coordinates, and in particular one must specify what foliation one is going to use to represent the progress of time. And this freedom to pick a "quantum observation frame" is critical in being able to maintain a view in which one imagines "definite things to happen" in the system.

With a foliation like the following, at any given time there is a mixture of different states, and no attempt has been made to find a way to "summarize" what the system is doing:

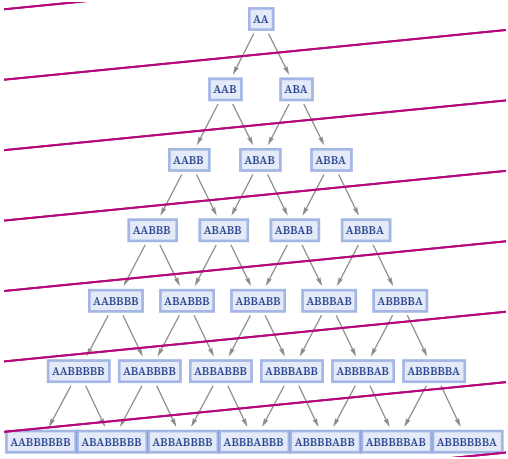

Consider, however, a foliation like the following:

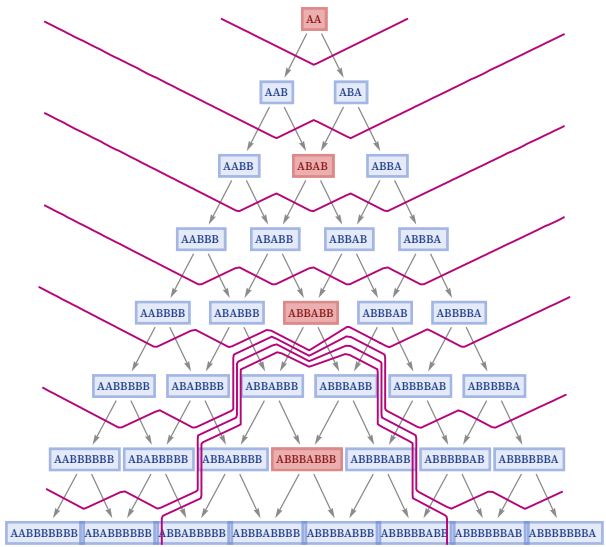

In this picture, generational states have been highlighted, and a foliation has been selected that essentially "freezes time" around a particular generational state. In effect, the observer is choosing a quantum observation frame in which there is a definite classical outcome for the behavior of the system.

"Freezing time" around a particular state is something an observer can choose to do in their description of the system. But the crucial point is that the actual dynamics of the evolution of the multiway system cause this choice to have implications.

In particular, in the case shown, the region of the multiway system in which "time is frozen" progressively expands. The choice the observer has made to freeze a particular state is causing more and more states to have to be considered as similarly frozen. In the physics of quantum measurement, one is used to the idea that for a quantum measurement to be considered to have a definite result, it must involve more and more quantum degrees of freedom. What we see here is effectively a manifestation of this phenomenon.

In freezing time in something like the foliation in the picture above what we are effectively doing is creating a coordinate singularity in defining our quantum observation frame. And there is an analogy to this in general relativity and the physics of spacetime. Just as we freeze time in our quantum frame, so also we can freeze time in a relativistic reference frame. For example, as an object approaches the event horizon of a black hole, its time as described by a typical coordinate system set up by an observer far from the black hole will become frozen—and just like in our quantum case, we will consider the state to stay fixed.

But there is a complicated issue here. To what extent is the singularity—and the freezing of time—a feature of our description, and to what extent is it something that "really happens"? This depends in a sense on the relationship one has to the system. In traditional thinking about quantum measurement, one is most interested in the "impressions" of observers who are in effect embedded in the system. And for them, the coordinate system they chose in effect defines their reality.

But one can also imagine being somehow "outside the system". For example, one might try to set up a quantum experiment (or a quantum computer) in which the construction of the system somehow makes it natural to maintain a "frozen time" foliation. The picture below shows a toy example in which the multiway system by its very construction has a terminal state for which time does not advance:

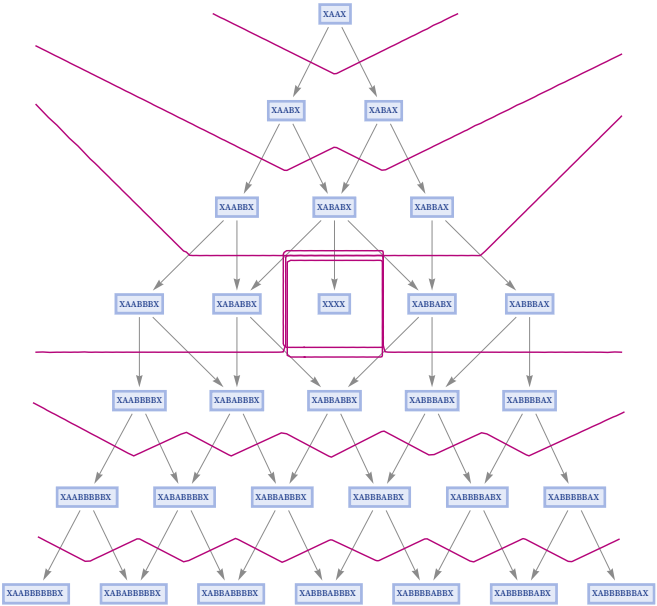

But now the question arises of what can be achieved in the multiway system corresponding to the actual physical universe. And here we can expect that one will not be able to set up truly isolated states, and that instead there will be continual inevitable entanglement. What one might have imagined could be maintained as a separate state will always become entangled with other states.

The picture below shows a slightly more realistic multiway system, with an attempt to construct a foliation that freezes time:

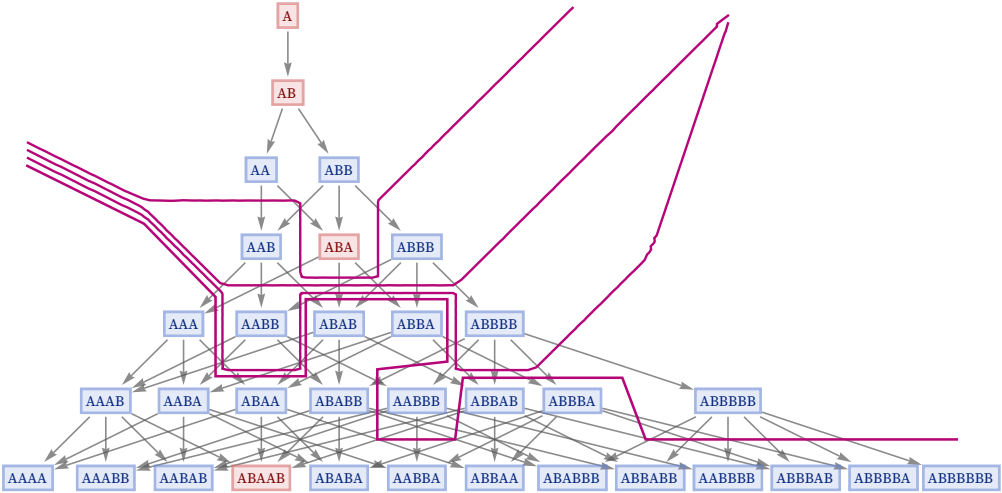

And what we see here is that in a sense the structure of the multiway graph limits the extent to which we can freeze time. In effect, the multiway system forces decoherence—or entanglement—-just by its very structure.

We should note that it is not necessarily the case that there is just a single possible sequence of generational states, corresponding in a sense to a single possible "classical path". Here is an example where there are four generational states that occur at a particular generational step. And now we can for example construct a foliation that—at least for a while—"freezes time" for all of these generational states:

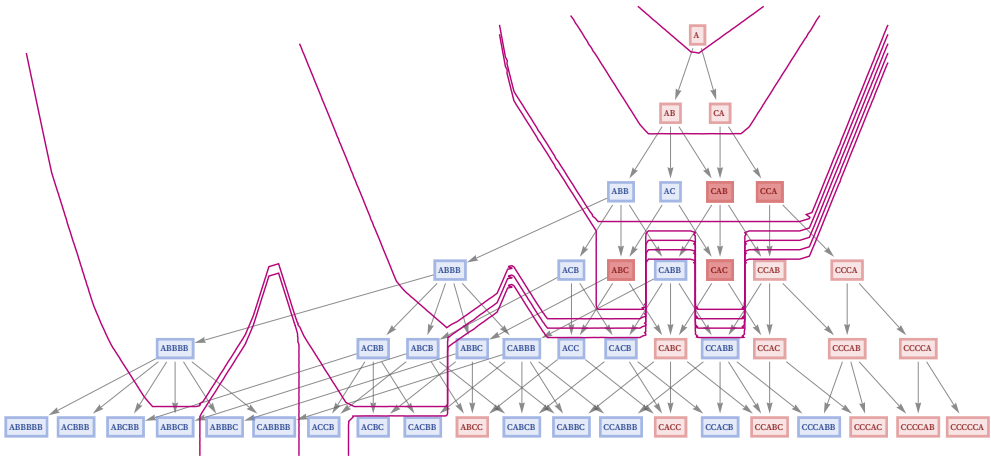

It is worth pointing out that if we try to freeze time for something that is not a proper generational state, there will be an immediate issue. A proper generational state contains the results of all spacelike separated events at a particular point in the evolution of a system. So when we freeze time for it, we are basically allowing other branchlike separated events to occur, but not other spacelike separated ones. However, if we tried to freeze time for a state that did not include all spacelike separated events, there would quickly be a mismatch with the progress of time for the excluded events—or in effect the singularity of quantum observation frame would "spill over" into a singularity in the causal graph, leading to a singularity in spacetime.

In other words, the fact that the states that appear in quantum measurement are generational states is not just a convenience but a necessity. Or, put another way, in doing a quantum measurement we are effectively setting up a singularity in branchial space, and only if the states we measure are in effect "complete in spacetime" will this singularity be kept only in branchial space; otherwise it will also become a singularity in physical spacetime.

In general, when we talk about quantum measurement, we are talking about how an observer manages to construct a description of a system that in effect allows the observer to "make a conclusion" about what has happened in the system. And what we have seen is that appropriate "time-freezing foliations" allow us to do this. And while there may be some restrictions, it is usually in principle possible to construct such foliations in a multiway system, and to have them last as long as we want.

But in practice, as the pictures above begin to suggest, after a while the foliations we have to construct can get increasingly complicated. In effect, what we are having to do in constructing the foliation is to "reverse engineer" the actual evolution of the multiway system, so that with our elaborate description we are still managing to maintain time as frozen for a particular state, carefully avoiding complicated entanglements that have built up with other states.

But there is a problem here. Because in effect we are asking the observer to "outcompute" the system itself. Yet we can expect that the evolution of the multiway system, say for one of our models, will usually correspond to an irreducible computation. And so we will be asking the observer to do a more and more elaborate computation to maintain the description they are using. And as soon as the computation required exceeds the capability of the observer, the observer will no longer be able to maintain the description, and so decoherence will be inevitable.

It is worthwhile to compare this situation with what happens in thermodynamic processes, and in particular with apparent entropy increase. In a reversible system, it is always in principle possible to recognize, say, that the initial conditions for the systems were simple (and "low entropy"). But in practice the actual configurations of the system usually become complicated enough that this is increasingly difficult to do. In traditional statistical mechanics one talks of "coarse-grained" measurements as a way to characterize what an observer can actually analyze about a system.

In computational terms we talk about the computational capabilities of the observer, and how computational irreducibility in the evolution of the system will eventually overwhelm the computational capabilities of the observer, making apparent entropy increase inevitable [1:9.3].

In the quantum case, we now see how something directly analogous happens. The analog of coarse graining is the effort to create a foliation with a particular apparent outcome. But eventually this becomes infeasible, and—just like in the thermodynamic case—we in effect see "thermalization", which we can now attribute to the effects of computational irreducibility.

## 8.15  Operators in Quantum Mechanics

In standard quantum formalism, there are states, and there are operators (e.g. [125]). In our models, updating events are what correspond to operators. In the standard evolution of the multiway system, all applicable operators are in effect "automatically applied" to every state to generate the actual evolution of the system. But to understand the correspondence with standard quantum formalism, we can imagine just applying particular operators by doing only particular updating events.

Consider the string substitution system:

{AB → ABA, BA → BAB}

In this system we effectively have two operators $O_1$ and $O_2$, corresponding to these two possible updating rules. We can think about building up an operator algebra by considering the relations between different sequences of applications of these operators.

In particular, we can study the commutator:

$[O_1, O_2] = O_1 O_2 - O_2 O_1$

In terms of the underlying rules, this commutator corresponds to:

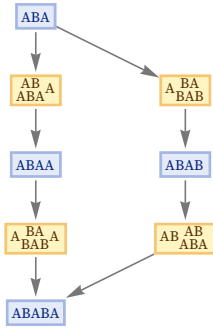

At the first step, the results of applying $O_1$ and $O_2$ to the initial state are different, and we can say that the states generated form a branch pair. But then at the second step, the branch pair resolves, and the branches merge to the same state. In effect, we can represent this by saying that $O_1$ and $O_2$ commute, or that:

$[O_1, O_2] = O_1 O_2 - O_2 O_1 = 0$

In general, there is a close relationship between causal invariance—and its implication for the resolution of all branch pairs—and the commuting of operators. And given our discussion above this should not be considered surprising: as we discussed, when there is causal invariance, it means that all branches can resolve to a single ("classical") state, just like in standard quantum formalism the commuting of operators is associated with seemingly classical behavior.

But there is a key point here: even if causal invariance implies that branch pairs (and similarly commutators) will eventually resolve, they may take time to do so. And it is this delay in resolution that is the core of what leads to what we normally think of as quantum effects.

Once a branch pair has resolved, there are no longer multiple branches, and a single state has emerged. But before the branch pair has resolved, there are multiple states, and therefore what one might think of as "quantum indeterminacy".

In the case where a branch pair has not yet resolved, the corresponding commutator will be nonzero—and in a sense the value of the commutator measures the branchlike distance between the states reached by applying the two different updates (corresponding to the two different operators).

In our model for spacetime, if a single event in the causal graph is connected in the causal graph to two different events we can ask what the spacelike separation of these events might be, and we might suppose that this spatial distance is determined by the speed of light $c$ (say multiplied by the elementary time corresponding to traversal of the causal edge).

In thinking now about the multiway system, we can ask what the branchlike separation of states in a branch pair might be. This will now be a distance on a branchial graph—or effectively a distance in state space—and we can suppose that this distance is determined by $\hbar$. And depending on our conventions for measuring branchial distance, we might introduce an $i$, yielding a setup very much aligned with traditional quantum formalism.

Another interpretation of the non-commuting of operators is connected to the entanglement of quantum states. And here we now have a very direct picture of entanglement: two states are entangled if they are part of the same unresolved branch pair, and thus have a common ancestor.

The multiway graph gives a full map of all entanglements. But at any particular time (corresponding to a particular slice of a foliation defined by a quantum observation frame), the branchial graph gives a snapshot that captures the "instantaneous" configuration of entanglements. States closer on the branchial graph are more entangled; those further apart are less entangled.

It is important to note that distance on the branchial graph is not necessarily correlated with distance on the spatial graph. If we look at events, we can use the multiway causal graph to give a complete map of all connections, involving both branchlike and spacelike (as well as timelike) separations. Ultimately, the underlying rule determines what connections will exist in the multiway causal graph. But just as in the standard formalism of quantum mechanics, it is perfectly possible for there to be entanglement of spacelike-separated events.

## 8.16 Wave-Particle Duality, Uncertainty Relations, Etc.

Wave-particle duality was an early but important concept in standard quantum mechanics, and turns out to be a core feature of our models, independent even of the details of particles. The key idea is to look at the correspondence between spacelike and branchlike projections of the multiway causal graph.

Let us consider some piece of "matter", ultimately represented as features of our hypergraphs. A complete description of what the matter does must include what happens on every branch of the multiway graph. But we can get a picture of this by looking at the multiway causal graph—which in effect has the most complete representation of all meaningful spatial and branchial features of our models.

Fundamentally what we will see is a bundle of geodesics that represent the matter, propagating through the multiway causal graph. Looked at in terms of spacelike coordinates, the bundle will seem to be following a definite path—characteristic of particle-like behavior. But inevitably the bundle will also be extended in the branchlike direction—and this is what leads to wave-like behavior.

Recall that we identified energy in spacetime as corresponding to the flux of causal edges through spacelike hypersurfaces. But as mentioned above, whenever causal edges are present, they correspond to events, which are associated with branching in the multiway graph and the multiway causal graph. And so when we look at geodesics in the bundle, the rate at which they turn in multiway space will be proportional to the rate at which events happen, or in other words, to energy—yielding the standard $E \propto \omega$ proportionality between particle energy and wave frequency.

Another fundamental phenomenon in quantum mechanics is the uncertainty principle. To understand this principle in our framework, we must think operationally about the process of, for example, first measuring position, then measuring momentum. It is best to think in terms of the multiway causal graph. If we want to measure position to a certain precision $\Delta x$ we effectively need to set up our detector (or arrange our quantum observation frame) so that there are $O(1/\Delta x)$ elements laid out in a spacelike array. But once we have made our position measurement, we must reconfigure our detector (or rearrange our quantum observation frame) to measure momentum instead.

But now recall that we identified momentum as corresponding to the flux of causal edges across timelike hypersurfaces. So to do our momentum measurement we effectively need to have the elements of our detector (or the pieces of our quantum observation frame) laid out on a timelike hypersurface. But inevitably it will take at least $O(1/\Delta x)$ updating events to rearrange the elements we need. But each of these updating events will typically generate a branch in the multiway system (and thus the multiway causal graph). And the result of this will be to produce an $O(1/\Delta x)$ spread in the multiway causal graph, which then leads to an $O(1/\Delta x)$ uncertainty in the measurement of momentum.

(Another ultimately equivalent approach is to consider different foliations, and to note for example that with a finer foliation in time, one is less able to determine the "true direction" of causal edges in the multiway graph, and thus to determine how many of them will cross a spacelike hypersurface.)

To make our discussion of the uncertainty principle more precise, we should consider operators—represented by sequences of updating events. In the ($t$, $x$, $b$) space of the multiway causal graph, the operators corresponding to position and momentum must generate events that correspond to moving at different angles; as a result the operators do not commute.

And with this setup we can see why position and momentum, as well as energy and time, form canonically conjugate pairs for which uncertainty relations hold: it is because these quantities are associated with features of the multiway causal graph that probe distinct (and effectively orthogonal) directions in multiway causal space.

## 8.17 Correspondence between Relativity and Quantum Mechanics

One of the surprising consequences of the potential application of our models to physics is their implications around deep relationships between relativity and quantum mechanics. These are particularly evident in thinking about the multiway causal graph. As a toy model, consider the graph:

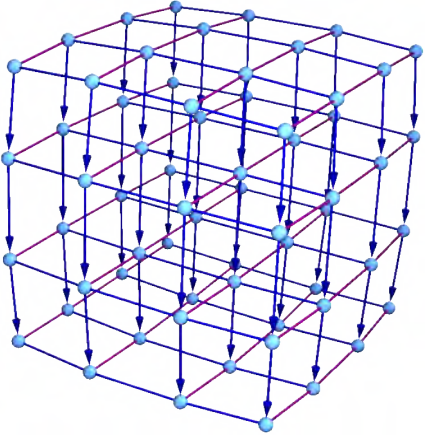

Timelike edges go down, but then in each slice there are spacelike and branchlike edges. A more realistic example of the very beginning of such a graph is:

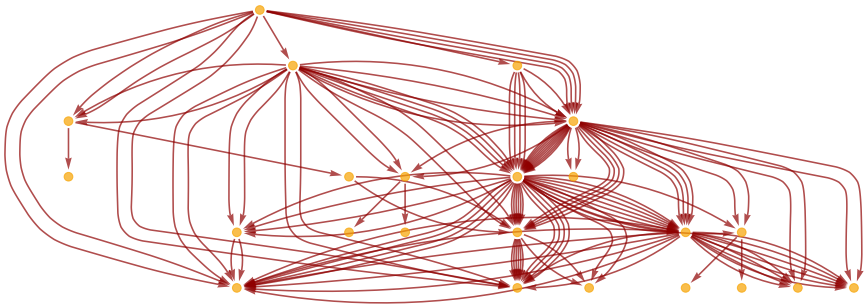

The multiway causal graph in a sense captures in one graph both relativity and quantum mechanics. Time is involved in both of them, and in our models it is an essentially computational concept, involving progressive application of the underlying rules of the system. But then relativity is associated with the structure formed by spacelike and timelike edges, while quantum mechanics is primarily associated with the structure formed by branchlike and timelike edges.

The spacelike direction corresponds to ordinary physical space; the branchlike direction is effectively the space of quantum states. Distance in the spacelike direction is ordinary spacetime distance. Distance in the branchlike direction reflects the level of quantum entanglement between states. When we form foliations in time, spacelike hypersurfaces represent in a sense the instantaneous configuration of space, while branchlike hypersurfaces represent the instantaneous entanglements between quantum states.

It should be emphasized that (unlike in the idealization of our first picture above) the detailed structure of the spacelike+timelike component of the multiway causal graph will in practice be very different from that of the branchlike+timelike one. The spacelike+timelike component is expected to limit to something like a finite-dimensional manifold, reflecting the characteristics of physical spacetime. The branchlike+timelike one potentially limits to an infinite dimensional space (that is perhaps a projective Hilbert space), reflecting the characteristics of the space of quantum states. But despite these substantial geometrical differences, one can expect many structural aspects and consequences to be basically the same.

We are used to the idea of motion in space. In the context of our models—and of the multiway causal graph—motion in space in effect corresponds to progressively sampling more spacelike edges in the graph. But now we can see a quantum analog: we can also have motion in the branchlike direction, in which, in effect, we progressively sample more branchlike edges, reaching more quantum states. Velocity in space is thus the analog of the rate at which additional states are sampled (and thus entangled).

In relativity there is a fairly well-developed notion of an idealized observer. The observer is typically represented by some some causal foliation of spacetime—like an inertial reference frame that moves without forces acting on it. One can also define an observer in quantum mechanics, and in the context of our models it makes sense—as we have done above—to parametrize the observer in terms of a quantum observation frame that consists not of a sequence of spacelike hypersurfaces, but instead of a series of branchlike ones.

A quantum observation frame in a sense defines a plan for how an observer will sample possible quantum states—and the analog of an inertial frame in spacetime is presumably a quantum observation frame that corresponds to a fixed plan that cannot be affected by anything outside. And in general, the analog in quantum mechanics of a world line in relativity is presumably a measurement plan.

In special relativity a key idea is to think about comparing the perceptions of observers in different inertial frames. But in the context of our models we can now do the exact same thing for quantum observers. And the analog of relativistic invariance then becomes a statement of perception or measurement invariance: that in the end different quantum observers (despite the branching of states) in a sense perceive the same things to happen, or, in other words, that there is at some level an objective reality even in quantum mechanics.

Our analogy between relativity and quantum mechanics suggests asking about quantum analogs of standard relativistic phenomena. One example is relativistic time dilation, in which, in effect, sampling spacelike edges faster reduces the rate of traversing timelike edges. The analog in quantum mechanics is presumably the quantum Zeno effect [126][127], in which more rapid measurement—corresponding to faster sampling of branchlike edges—slows the time evolution of a quantum system.

A key concept in relativity is the light cone, which characterizes the maximum rate at which causal effects spread in spacelike directions. In our models, spacetime causal edges in effect define elementary light cones, which are then knitted together by the structure of the (spacetime) causal graph. But now in our models there is a direct analog for quantum mechanics, visible in the full multiway causal graph.

In the multiway causal graph, every event effectively has a cone of causal influence. Some of that influence may be in spacelike directions (corresponding to ordinary relativistic light cone effects), but some of it may be in branchlike directions. And indeed, whenever there are branches in the multiway graph, these correspond to branchlike edges in the multiway causal graph.

So what this means is that in addition to a light cone of effects in spacetime, there is also what we may call an entanglement cone, which defines the region affected in branchial space by some event. In the light cone case, the spacelike extent of the light cone is set by the speed of light ($c$). In the entanglement cone case (as we will discuss below) the branchlike extent of the entanglement cone is essentially set by $\hbar$.

As we have mentioned, the definition of time is shared between spacelike and branchlike components of the multiway causal graph. Another shared concept appears to be energy (or in general, energy-momentum, or action). Time is effectively defined by displacement in the timelike direction; energy appears to be defined by the flux of causal edges in the timelike direction. In the relativistic setting, energy can be thought of as flux of causal edges through spacelike hypersurfaces; in the quantum mechanical setting, it can be thought of as a flux of causal edges through branchlike hypersurfaces.

An important feature of the spacetime causal graph is that it can potentially describe curved space, and reproduce general relativity. And here again we can now see that in our models there are analogs in quantum mechanics. One issue, though, is that whereas ordinary space is—at least on a large scale—finite-dimensional, comparatively flat, and well modeled by a simple Lorentzian manifold, branchial space is much more complicated, probably in the limit infinite–dimensional, and not at all flat.

At a mathematical level, we are in quantum mechanics used to forming commutators of operators, and in many cases finding that they do not commute, with their "deviation" being measured by $\hbar$. In general relativity, one can also form commutators, and indeed the Riemann tensor for measuring curvature is precisely the result of computing the commutator of two covariant derivatives. And perhaps even more analogously the Ricci scalar curvature gives the angle deficit for transport around a loop in spacetime.

In our context, therefore, the non-flatness of space is directly analogous to a core phenomenon of quantum mechanics: the non-commuting of operators.

In the general relativity case, we are used to thinking about the propagation of bundles of geodesics in spacetime, and the fact that the Ricci scalar curvature determines the local cross-section of the bundle. Now we can also consider the more general propagation of bundles of geodesics in the multiway causal graph. But when we look along branchlike directions, the limiting space we see tends to be highly connected, and effectively of high negative curvature. And what this means is that a bundle of geodesics can be expected to spread out rapidly in branchlike directions.

But this has an immediate interpretation in quantum mechanics: it is the phenomenon of decoherence, whereby quantum effects get spread (and entangled) across large numbers of quantum degrees of freedom.

In relativity, the speed of light $c$ sets a maximum speed for the propagation of effects in space. In quantum mechanics, our entanglement cones in essence also set a maximum speed for the propagation of effects in branchial space. In special relativity, there is then a maximum speed defined for any observer—or, in other words, a maximum speed for motion. In quantum mechanics, we can now expect that there will also be a maximum speed for entanglement, or for measurement: it is not possible to set up a quantum observation frame that achieves a higher speed while still respecting the causal relations in the multiway causal graph. We will call this maximum speed $\zeta$, and in 8.20 we will discuss its possible magnitude.

One may ask to what extent the correspondences between relativity and quantum mechanics that we have been discussing rely on our models. In principle, for example, one could imagine a kind of "multicausal continuum" that is a mathematical structure (conceivably related to twistor spaces [128]) corresponding to a continuum limit of our multiway causal graph. But while there are challenges in understanding the limits associated with our models, this seems likely to be even more difficult to construct and handle—and has the great disadvantage that it cannot be connected to explicit models that are readily amenable, for example, to enumeration.

## 8.18 Event Horizons and Singularities in Spacetime and Quantum Mechanics

Having discussed the general correspondence between relativity and quantum mechanics suggested by our models, we can now consider the extreme situation of event horizons and singularities.

As we discussed above, an event horizon in spacetime corresponds in our models to disconnection in the causal graph: after some slice in our foliation in time, there is no longer causal connection between different parts of the system. As a result, even if the system is locally causal invariant, branch pairs whose products go on different sides of the disconnection can never resolve. The only way to make a foliation in which this does not happen is then effectively to freeze time before the disconnection occurs.

When there is a true disconnection in the causal graph, there is no choice about this. But it is also perfectly possible just to imagine setting up a coordinate system that freezes time in a particular region of space—although it will typically take more and more effort (and energy) to consistently maintain such a coordinate singularity as other parts of the system evolve.

But now there is an interesting correspondence with quantum measurement. As we discussed in 8.14, in the context of our models, one can view a quantum measurement (or a "collapse of the wave function") as being associated with a foliation that freezes time for the state that is the outcome of the measurement. In essence, therefore, quantum measurement corresponds to having a coordinate singularity in a particular region of branchial space.

What about an event horizon? As we saw above, one way in which an event horizon can occur is if some branch of the multiway system simply terminates, so that in a sense time stops for it. Another possibility is that—at least temporarily—there can be a disconnected piece in the branchial graph. Consider for example the (causal invariant) string substitution system:

{A → BB, BBB → AA}

The multiway system for this rule is

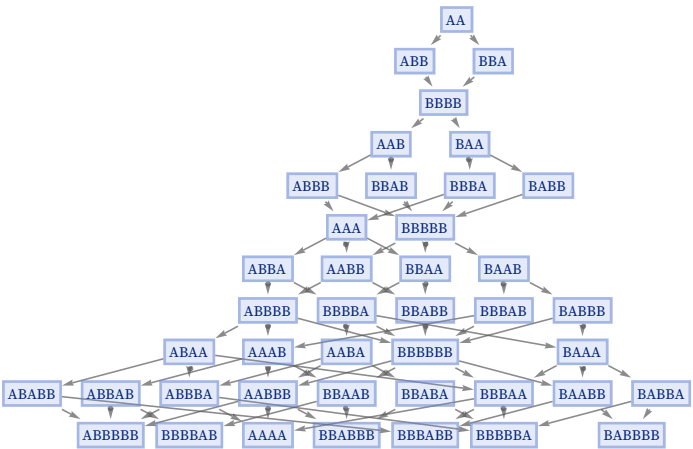

and the branchial graph shows temporary disconnections

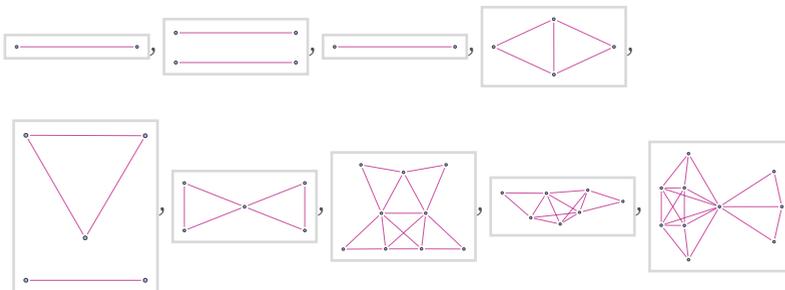

although the "spacetime" causal graph stays connected:

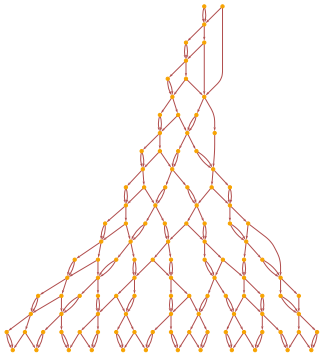

One can think of these temporary disconnections in the branchial graphs as corresponding to isolated regions of branchial space where entanglement at least temporarily cannot occur—and where some pure quantum state (such as qubits) can be maintained, at least for some period of time.

In some sense, one can potentially view such disconnections as being like black holes in branchial space. But the continued generation of branch pairs (in a potential analog to Hawking radiation [129]) causes the "black hole" to dissipate.

A different situation can occur when there is also disconnection in the causal graph—leading in our models to disconnection in the spatial hypergraph—and thus a spacetime event horizon. As a simple example, consider the string substitution system (starting from AA):

{AA → AAAB}

The causal graph in this case is

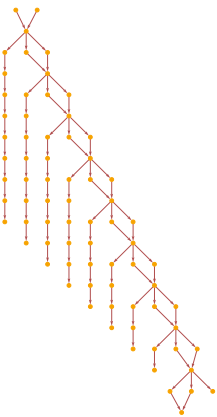

and the sequence of branchial graphs (with the standard foliation) is:

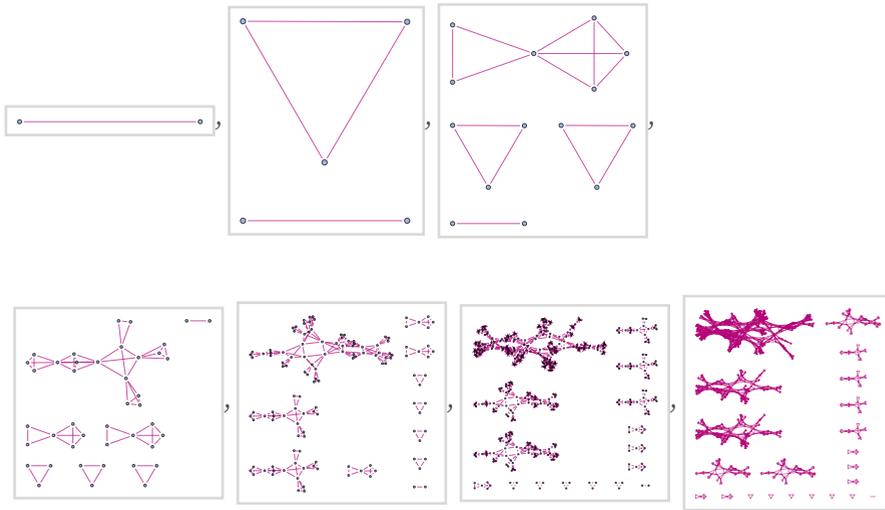

What has happened here is that there are event horizons both in physical space and in branchial space.

We can expect similar phenomena in our full models, and extrapolating this to a physical black hole what this represents is the presence of both a causal event horizon (associated with motion in space, propagation of light, etc.) and an entanglement event horizon (associated with quantum entanglement). The causal event horizon will be localized in physical space (say at the Schwarzschild radius [130]); the entanglement event horizon can be considered instead to be localized in branchial space.

It should be noted that these horizons are in a sense linked through the multiway causal graph, which in the example above initially has the form

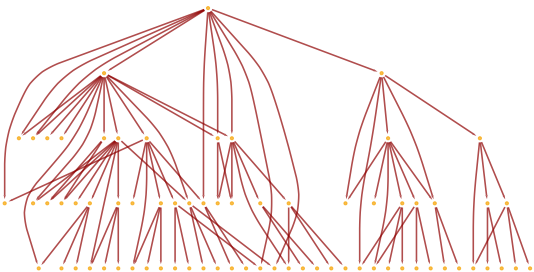

and after more steps builds up the structure:

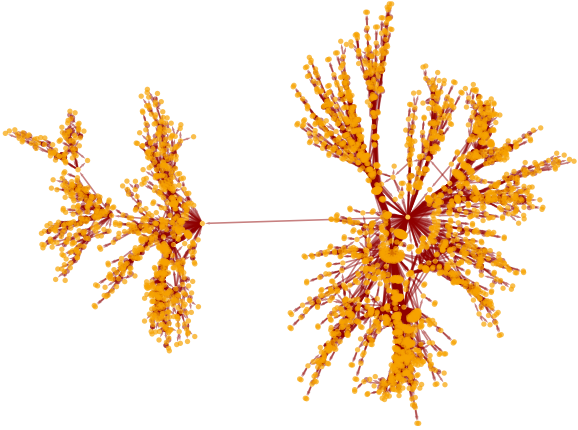

In this graph, there are both spacelike and branchlike connections, and here both of them exhibit disconnection, and therefore event horizons. And even though the geometrical structure of branchial space is very different from physical space, there are potentially further correspondences to be made between them. For example, while the speed of light $c$ governs the maximum spacelike speed, the maximum entanglement rate $v$ that we introduced above governs the maximum "branchlike speed", or entanglement rate.

When a disconnection occurs in the spacetime causal graph (and thus the spatial hypergraph), we can think of this as implying that geodesics in spacetime would have to exceed $c$ in order not to be trapped. When a disconnection occurs in the branchial graph, we can think of geodesics having to "exceed speed $v$" in order not to be trapped.

It is worth pointing out that the analog of a true singularity—and not just an event horizon—can occur in our models if there are paths in the multiway system that simply terminate, as for B, BB, etc. in:

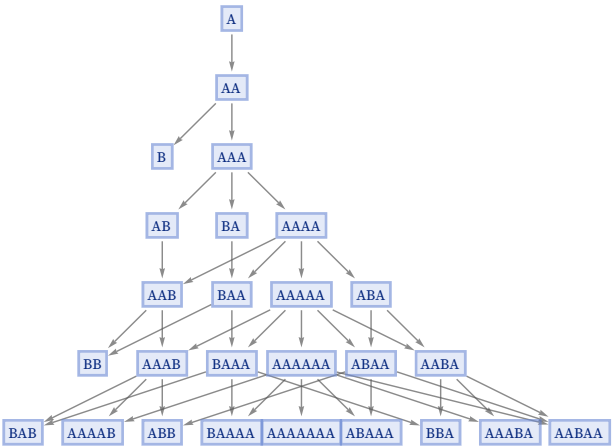

When this happens, there are many geodesics that in effect converge to a single point, like in spacetime singularities in general relativity. Here, however, we see that this can happen not only in physical space, but also in the multiway system, or, in other words, in branchial space. (In our systems, it is probably the case that singularities must be enclosed in event horizons, in the analog of the cosmic censorship hypothesis.)

Many results from general relativity can presumably be translated to our models, and can apply both to physical space and branchial space (see [121]). In the case of a black hole, our models suggest that not only may a causal event horizon form in physical space; also an entanglement horizon may form in branchial space. One may then imagine that quantum information is trapped inside the entanglement horizon, even without crossing the causal event horizon—with implications perhaps similar to recent discussions of resolutions to the black hole quantum information problem [131][132][133].

There is a simple physical picture that emerges from this setup. As we have discussed, quantum measurement can be thought of as a choice of coordinates that "freeze time" for some region in branchial space. For an observer close to the entanglement horizon, it will not be possible to do this. Much like an observer at a causal event horizon will be stretched in physical space, so also an observer at an entanglement horizon will be stretched in branchial space. And the result is that in a sense the observer will not be able to "form a classical thought": they will not successfully be able to do a measurement that definitively picks the branch of the multiway system in which something fell into the black hole, or the one in which it did not.

## 8.19 Local Gauge Invariance

An important phenomenon discussed especially in the context of quantum field theories is local gauge invariance (e.g. [134]). In our models this phenomenon can potentially arise as a result of local symmetries associated with underlying rules (see 6.12). The basic idea is that these symmetries allow different local configurations of rule applications—that can be thought of as different local "gauge" coordinate systems.

But the collection of all such possible configurations appears in the multiway graph (and the multiway causal graph)—so that a local choice of gauge can then be represented by a particular foliation in the multiway graph. But causal invariance then implies the equivalence of foliations—and establishes local gauge invariance.

As a very simple example, consider the rule:

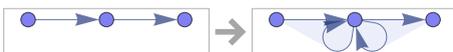

Starting from a square, this rule can be applied in two different ways:

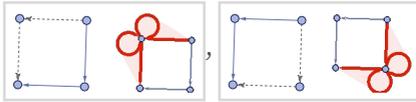

There is similar freedom if one applies the rule twice to a larger region:

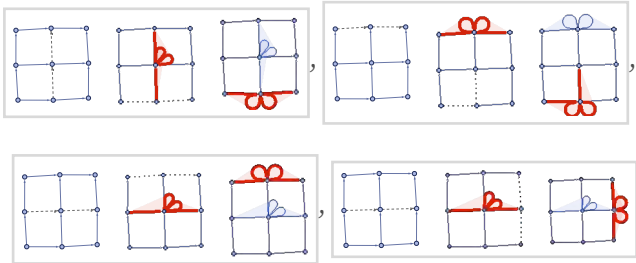

In both cases one can think of the freedom to apply the rule in different ways as being like a symmetry, for example characterized by the list of possible permutations of input elements.

But now imagine taking the limit of a large number of steps. Then one can expect to apply the resulting aggregate rule in a large number of ways. And much as we expect the limit of our spatial hypergraphs to be able to be represented—at least in certain cases—as a continuous manifold, we can expect something similar here. In particular, we can think of ourselves as winding up with a very large number of permutations corresponding to equivalent rule applications, which in the limit can potentially correspond to a Lie group.

Each different possible choice of how to apply the rule corresponds to a different event that is represented in the multiway graph, and the multiway causal graph:

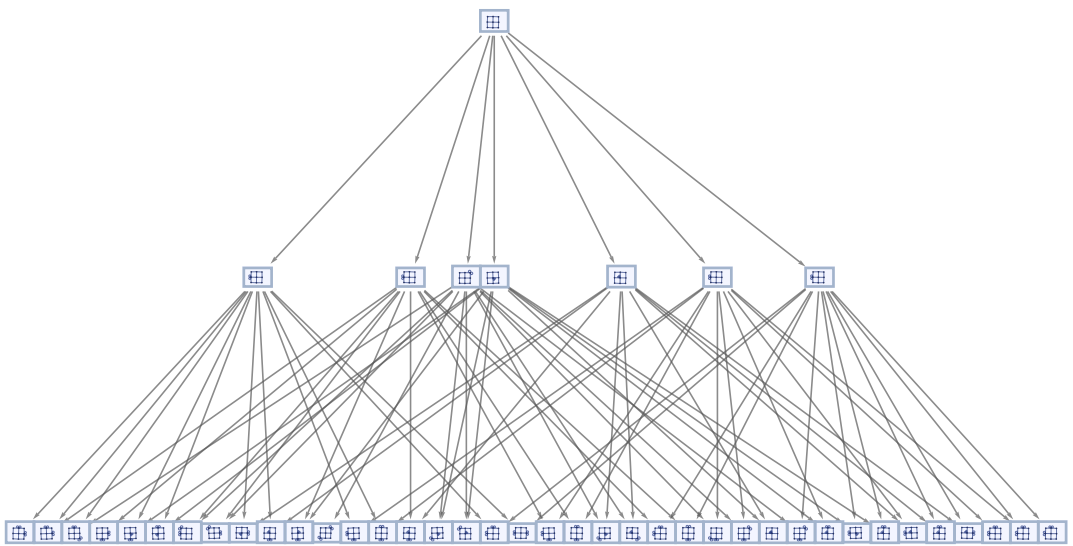

But the important point is that local choices of how the rule is repeatedly applied must always correspond to purely branchlike connections in the multiway causal graph.

The picture is analogous to the one in traditional mathematical physics. The spatial hypergraph can be thought of as a base space for a fiber bundle, then the different choices of which branchlike paths to follow correspond to different choices of coordinate systems (or gauges) in the fibers of the fiber bundle (cf. [135][136]). The connection between fibers is defined by the foliation that is chosen.

There is an analog when one sets up foliations in the spacetime causal graph—in which case, as we have argued, causal invariance leads to general covariance and general relativity. But here we are dealing with branchlike paths, and instead of getting general relativity, we potentially get gauge theories.

In traditional physics, local gauge invariance already occurs in classical theories (such as electromagnetism), and it is notable that for us it appears to arise from considering multiway systems. Yet although multiway systems appear to be deeply connected to quantum mechanics, the aggregate symmetry phenomenon that leads to gauge theories in effect makes slightly different use of the structure of the multiway causal graph.

But much as in other cases, we can think about geodesics—now in the multiway causal graph—and can study the properties of the effective space that emerges, with local phenomena (including things like commutators) potentially reflecting features of the Lie algebra.

In traditional physics an important consequence of local gauge invariance is its implication of the existence of fields, and gauge bosons such as the photon and gluon. In our models the mathematical derivations that lead to this implication should be similar. But by looking at the evolution of our models, it is possible to get a more explicit sense of how this works.

Consider a particular sequence of updates with the rule shown above:

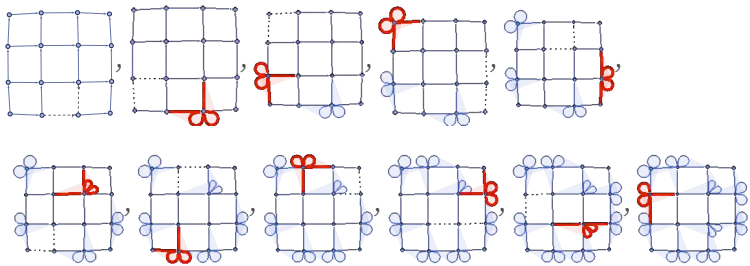

At the beginning, symmetry effectively allows many equivalent updates to be made. But once a particular update has been made, this has consequences for which of the possible updates—each independently equivalent on their own—can be made subsequently. These "consequences" are captured in the causal relationships encoded in the multiway causal graph—which have not only branchlike but also spacelike extent, corresponding in essence to the propagation of effects in what can be described as a gauge field.

## 8.20 Units and Scales

Most of our discussion so far has focused on how the structure of our models might correspond to the structure of our physical universe. But to make direct contact between our models and known physics, we need to fill in actual units and scales for the constructs in our models. In this section we give some indication of how this might work.

In our models, there is a fundamental unit of time (that we will call $\bar{T}$) that represents the interval of time corresponding to a single updating event. This interval of time in a sense defines the scale for everything in our models.

Given $\bar{T}$, there is an elementary length $\bar{L}$, determined by the speed of light $c$ according to:

$\bar{L} = c\,\bar{T}$

The elementary length defines the spatial separation of neighboring elements in the spatial hypergraph.

Another fundamental scale is the elementary energy $\bar{E}$: the contribution of a single causal edge to the energy of a system. The energy scale ultimately has both relativistic and quantum consequences. In general relativity, it relates to how much curvature a single causal edge can produce, and in quantum mechanics, it relates to how much change in angle in an edge in the multiway graph a single causal edge can produce.

The speed of light $c$ determines the elementary length in ordinary space, specifying in effect how far one can go in a single event, or in a single elementary time. To fill in scales for our models, we also need to know the elementary length in branchial space—or in effect how far in state space one can go in a single event, or a single elementary time (or, in effect, how far apart in branchial space two members of a branch pair are). And it is an obvious supposition that somehow the scale for this must be related to $\hbar$.

An important point about scales is that there is no reason to think that elementary quantities measured with respect to our current system of units need be constant in the history of the universe. For example, if the universe effectively just splits every spatial graph edge in two, the number of elementary lengths in what we call 1 meter will double, and so the elementary length measured in meters will halve.

Given the structure of our models, there are two key relationships that determine scales. The first—corresponding to the Einstein equations—relates energy density to spacetime curvature, or, more specifically, gives the contribution of a single causal edge (with one elementary unit of energy) to the change of $V_r$ and the corresponding Ricci curvature:

$$\frac{G}{c^4}\,\frac{\bar{E}}{\bar{L}^d} \approx \frac{1}{\bar{L}^2}$$

(Here we have dropped numerical factors, and G is the gravitational constant, which, we may note, is defined with its standard units only when the dimension of space $d$ = 3.)

The second key relationship that determines scales comes from quantum mechanics. The most obvious assumption might be that quantum mechanics would imply that the elementary energy should be related to the elementary time by $\bar{E} \approx \hbar/\bar{T}$. And if this were the case, then our various elementary quantities would be equal to their corresponding Planck units [137], as obtained with $G = c = \hbar = 1$ (yielding elementary length $\approx 10^{-35}$ m, elementary time $\approx 10^{-43}$ s, etc.)

But the setup of our models suggests something different—and instead suggests a relationship that in effect also depends on the size of the multiway graph. In our models, when we make a measurement in a quantum system, we are at a complete quantum observation frame—or in effect aggregating across all the states in the multiway graph that exist in the current slice of the foliation that we have defined with our quantum frame.

There are many individual causal edges in the multiway causal graph, each associated with a certain elementary energy $\bar{E}$. But when we measure an energy, it will be the aggregate of contributions from all the individual causal edges that we have combined in our quantum frame.

A single causal edge, associated with a single event which takes a single elementary time, has the effect of displacing a geodesic in the multiway graph by a certain unit distance in branchial space. (The result is a change of angle of the geodesic—with the formation of a single branch pair perhaps being considered to involve angle $\frac{\pi}{2}$.)

Standard quantum mechanics in effect defines $\hbar$ through $E = \hbar \omega$. But in this relationship $E$ is a measured energy, not the energy associated with a single causal edge. And to convert between these we need to know in effect the number of states in the branchial graph associated with our quantum frame, or the number of nodes in our current slice through the multiway system. We will call this number $\Xi$.

And finally now we can give a relation between elementary energy and elementary time:

$$\bar{E}\, \Xi \approx \frac{\hbar}{\bar{T}}$$

In effect, $\hbar$ sets a scale for measured energies, but $\hbar/\Xi$ sets a scale for energies of individual causal edges in the multiway causal graph.

This is now sufficient to determine our elementary units. The elementary length is given in dimension $d$ = 3 by

$$\bar{L} \approx \left(\frac{G\,\hbar}{c^3\, \Xi}\right)^{1/(d-1)} \approx \frac{l_P}{\sqrt{\Xi}} \approx \frac{10^{-35}\ \text{m}}{\sqrt{\Xi}}$$

$$\bar{T} \approx \frac{t_P}{\sqrt{\Xi}} \approx \frac{10^{-43} \text{ s}}{\sqrt{\Xi}}$$

$$\bar{E} \approx \frac{E_P}{\sqrt{\Xi}} \approx \frac{10^9 \text{ J}}{\sqrt{\Xi}} \approx \frac{10^{19} \text{ GeV}}{\sqrt{\Xi}}$$

where $l_P$, $t_P$, $E_P$ are the Planck length, time and energy.

To go further, however, we must estimate $\Xi$. Ultimately, $\Xi$ is determined by the actual evolution of the multiway system for a particular rule, together with whatever foliation and other features define the way we describe our experience of the universe. As a simple model, we might then characterize what we observe as being "generational states" in the evolution of a multiway system, as we discussed in 5.21.

But now we can use what we have seen in studying actual multiway systems, and assume that in one generational step of at least a causal invariant rule each generational state generates on average some number $\kappa$ of new states, where $\kappa$ is related to the number of new elements produced by a single updating event. In a generation of evolution, therefore, the total number of states in the multiway system will be multiplied by a factor $\kappa$.

But to relate this to observed quantities, we must ask what time an observer would perceive has elapsed in one generational step of evolution. From our discussion above, we expect that the typical time an observer will be able to coherently maintain the impression of a definite "classical-like" state will be roughly the elementary time $\bar{T}$ multiplied by the number of nodes in the branchlike hypersurface. The number of nodes will change as the multiway graph grows. But in the current universe we have defined it to be $\Xi$.

Thus we have the relation

$$\Xi \approx \kappa^{\frac{t_H}{\Xi \bar{T}}}$$

where $t_H$ is the current age of the universe, and for this estimate we have ignored the change of generation time at different points in the evolution of the multiway system.

Substituting our previous result for $\bar{T}$ we then get:

$$\Xi \approx \kappa^{\frac{t_H}{t_P} \frac{1}{\sqrt{\Xi}}} \approx \kappa^{\frac{10^{61}}{\sqrt{\Xi}}}$$

There is a rough upper limit on $\kappa$ from the signature for the underlying rule, or effectively the ratio in the size of the hypergraphs between the right and left-hand sides of a rule. (For most of the rules we have discussed here, for example, $\kappa \lesssim 2$.) The lower limit on $\kappa$ is related to the "efficiency" of causal invariance in the underlying rule, or, in effect, how long it takes branch pairs to resolve relative to how fast new ones are created. But inevitably $\kappa > 1$.

Given the transcendental equation

$$\Xi = \kappa^{\frac{\sigma}{\sqrt{\Xi}}}$$

we can solve for $\Xi$ to get

$$\Xi = e^{2\,W(\frac{1}{2}\sigma \log(\kappa))}$$

where $W$ is the product log function [138] that solves $w\,e^w = z$. But for large $\sigma \log(\kappa)$ (and we imagine that $\sigma \approx 10^{61}$), we have the asymptotic result [30]:

$$\Xi \approx \frac{(\sigma \log(\kappa))^2}{4 \log^2(\frac{1}{2}\sigma \log(\kappa))}$$

Plotting the actual estimate for $\Xi$ as a function of $\kappa$ we get the almost identical result:

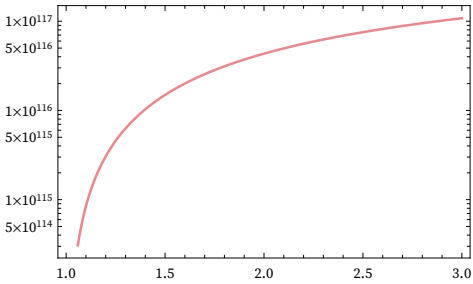

If $\kappa = 1$, then we would have $\Xi = 1$, and for $\kappa$ extremely close to 1, $\Xi \approx 1 + \sigma\,(\kappa - 1) + \ldots$ But even for $\kappa = 1.01$ we already have $\Xi \approx 10^{112}$, while for $\kappa = 1.1$ we have $\Xi \approx 10^{115}$, for $\kappa = 2$ we have $\Xi \approx 4 \times 10^{116}$ and for $\kappa = 10$ we have $\Xi \approx 5 \times 10^{117}$.

To get an accurate value for $\kappa$ we would have to know the underlying rule and the statistics of the multiway system it generates. But particularly at the level of the estimates we are giving, our results are quite insensitive to the value of $\kappa$, and we will assume simply:

$$\Xi \approx 10^{116}$$

In other words, for the universe today, we are assuming that the number of distinct instantaneous complete quantum states of the universe being represented by the multiway system (and thus appearing in the branchial graph) is about $10^{116}$.

But now we can estimate other quantities:

| | | |
|---|---|---|
| elementary length $\bar{L}$ | $\stackrel{?}{\approx} 10^{-93}$ m | $10^{-35}$ m $\Xi^{-\frac{1}{2}}$ |
| elementary time $\bar{T}$ | $\stackrel{?}{\approx} 10^{-101}$ s | $10^{-43}$ s $\Xi^{-\frac{1}{2}}$ |
| elementary energy $\bar{E}$ | $\stackrel{?}{\approx} 10^{-30}$ eV | $10^{28}$ eV $\Xi^{-\frac{1}{2}}$ |
| elementary lengths across current universe | $\stackrel{?}{\approx} 10^{120}$ | $10^{62}$ $\Xi^{\frac{1}{2}}$ |
| elements in spatial hypergraph | $\stackrel{?}{\approx} 10^{358}$ | $10^{184}$ $\Xi^{\frac{3}{2}}$ |
| elements in branchial graph $\Xi$ | $\stackrel{?}{\approx} 10^{116}$ | $\Xi$ |
| overall updates of universe so far | $\stackrel{?}{\approx} 10^{119}$ | $10^{61}$ $\Xi^{\frac{1}{2}}$ |
| individual updating events in universe so far | $\stackrel{?}{\approx} 10^{477}$ | $10^{245}$ $\Xi^2$ |

The fact that our estimate for the elementary length $\bar{L}$ is considerably smaller than the Planck length indicates that our models suggest that space may be more closely approximated by a continuum than one might expect.

The fact that the elementary energy $\bar{E}$ is much smaller than the surprisingly macroscopic Planck energy ($\approx 10^{19}$ GeV $\approx$ 2 GJ, or roughly the energy of a lightning bolt) is a reflection of the fact the Planck energy is related to measurable energy, not the individual energy associated with an updating event in the multiway causal graph.

Given the estimates above, we can use the rest mass of the electron to make some additional very rough estimates—subject to many assumptions—about the possible structure of the electron:

| | | |
|---|---|---|
| number of elements in an electron | $\stackrel{?}{\approx} 10^{35}$ | $10^{-23}$ $\Xi^{\frac{1}{2}}$ |
| radius of an electron | $\stackrel{?}{\approx} 10^{-81}$ m | $10^{-42}$ m $\Xi^{-\frac{1}{3}}$ |
| number of elementary lengths across an electron | $\stackrel{?}{\approx} 10^{12}$ | $10^{-8}$ $\Xi^{\frac{1}{6}}$ |

In quantum electrodynamics and other current physics, electrons are assumed to have zero intrinsic size. Experiments suggest that any intrinsic size must be less than about $10^{-22}$ m [139][140]—nearly $10^{60}$ times our estimate.

Even despite the comparatively large number of elements suggested to be within an electron, it is notable that the total number of elements in the spatial hypergraph is estimated to be more than $10^{200}$ times the number of elements in all known particles of matter in the universe—suggesting that in a sense most of the "computational effort" in the universe is expended on the creation of space rather than on the dynamics of matter as we know it.

The structure of our models implies that not only length and time but also energy and mass must ultimately be quantized. Our estimates indicate that the mass of the electron is > $10^{36}$ times the quantized unit of mass—far too large to expect to see "numerological relations" between particle masses.

But with our model of particles as localized structures in the spatial hypergraph, there seems no reason to think that structures much smaller than the electron might not exist—corresponding to particles with masses much smaller than the electron.

Such "oligon" particles involving comparatively few hypergraph elements could have masses that are fairly small multiples of $10^{-30}$ eV. One can expect that their cross-sections for interaction will be extremely small, causing them to drop out of thermal equilibrium extremely early in the history of the universe (e.g. [141][142]), and potentially leading to large numbers of cold, relic oligons in the current universe—making it possible that oligons could play a role in dark matter. (Relic oligons would behave as a more-or-less-perfect ideal gas; current data indicates only that particles constituting dark matter probably have masses $\gtrsim 10^{-22}$ eV [143].)

As we discussed in the previous subsection, the structure of our models—and specifically the multiway causal graph—indicates that just as the speed of light $c$ determines the maximum spacelike speed (or the maximum rate at which an observer can sample new parts of the spatial hypergraph), there should also be a maximum branchlike speed that we call $\zeta$ that determines the maximum rate at which an observer can sample new parts of the branchial graph, or, in effect, the maximum speed at which an observer can become entangled with new "quantum degrees of freedom" or new "quantum information".

Based on our estimates above, we can now give an estimate for the maximum entanglement speed. We could quote it in terms of the rate of sampling quantum states (or branches in the multiway system)

$$\frac{1}{\bar{T}} \approx \Xi \frac{\bar{E}}{\hbar} \approx 10^{102} / \text{second}$$

but in connecting to observable features of the universe, it seems better to quote it in terms of the energy associated with edges in the causal graph, in which case the result based on our estimates is:

$$\zeta \approx \frac{\bar{E}}{\bar{T}} \approx 10^{62} \text{ GeV/ second} \approx 10^{52} \ W \approx 10^5 \text{ solar masses / second}$$

This seems large compared to typical astrophysical processes, but one could imagine it being relevant for example in mergers of galactic black holes.

## 8.21 Specific Models of the Universe

If we pick a particular one of our models, with a particular set of underlying rules and initial conditions, we might think we could just run it to find out everything about the universe it generates. But any model that is plausibly similar to our universe will inevitably show computational irreducibility. And this means that we cannot in general expect to shortcut the computational work necessary to find out what it does.

In other words, if the actual universe follows our model and takes a certain number of computational steps to get to a certain point, we will not be in a position to reproduce in much less than this number of steps. And in practice, particularly with the numbers in the previous subsection, it will therefore be monumentally infeasible for us to find out much about our universe by pure, explicit simulation.

So how, then, can we expect to compare one of our models with the actual universe? A major surprise of this section is how many known features of fundamental physics seem in a sense to be generic to many of our models. It seems, for example, that both general relativity and quantum mechanics arise with great generality in models of our type—and do not depend on the specifics of underlying rules.

One may suspect, however, that there are still plenty of aspects of our universe that are specific to particular underlying rules. A few examples are the effective dimension of space, the local gauge group, and the specific masses and couplings of particles. The extent to which finding these for a particular rule will run into computational irreducibility is not clear.

It is, however, to be expected that parameters like the ones just mentioned will put strong constraints on the underlying rule, and that if the rule is simple, they will likely determine it uniquely.

Of all the detailed things one can predict from a rule, it is inevitable that most will involve computational irreducibility. But it could well be that those features that we have identified and measured as part of the development of physics are ones that correspond to computationally reducible aspects of our universe. Yet if the ultimate rule is in fact simple, it is likely that just these aspects will be sufficient to determine it.

In section 7 we discussed some of the many different representations that can be used for our models. And in different representations, there will inevitably be a different ranking of simplicity among models. In setting up a particular representation for a model, we are in effect defining a language—presumably suitable for interpretation by both humans and our current computer systems. Then the question of whether the rule for the universe is simple in this language is in effect just the question of how suitable the language is for describing physics.

Of course, there is no guarantee that there exists a language in which, with our current concepts, there is a simple way to describe the rule for our physical universe. The results of this section are encouraging, but not definitive. For they at least suggest that in the representation we are using, known features of our universe generically emerge: we do not have to define some thin and complicated subset to achieve this.

## 8.22  Multiway Systems in the Space of All Possible Rules

We have discussed the possibility that our physical universe might be described as following a model of the type we have introduced here, with a particular rule. And to find such a rule would be a great achievement, and might perhaps be considered a final answer to the core question of fundamental physics.

But if such a rule is found, one might then go on and ask why—out of the infinite number of possibilities—it is this particular rule, or, for example, a simple rule at all. And here the paradigm we have developed makes a potential additional suggestion: perhaps there is not just one rule being used after all, but instead in a sense all possible rules are simultaneously used.

In the multiway systems we have discussed so far, there is a single underlying rule, but separate branches for all possible sequences of updating events. But one can imagine a rule-space multiway system, that includes branches not only for every sequence of updating events, but also for every possible rule used to do the updating. Somewhat like with updating events, there will be many states reached to which many of the possible rules cannot apply. (For example, a rule that involves only ternary edges cannot apply to a state with only binary edges.) And like with updating events, branches with different sequences of rules applied may reach equivalent states, and thus merge.

Operationally, it is not so difficult to see how to set up a rule-space multiway system. All it really involves is listing not just one or a few possible rules that can be used for each updating event, but in a sense listing all possible rules. In principle there are an infinite number of such rules, but any rule that involves rewriting a hypergraph that is larger than the hypergraph that represents the whole universe can never apply, so at least at any given point in the evolution of the system, the number of rules to consider is finite. But like with the many other kinds of limits we have discussed, we can still imagine taking the limit of all infinitely many possible rules.

As a toy example of a rule-space multiway system, consider all inequivalent $2 \to 2$ rules on strings of As and Bs:

{AA → AA, AA → AB, AA → BB, AB → AA, AB → AB, AB → BA}

We can immediately construct the rule-space multiway graph for these rules (here starting from all possible length-4 sequences of As and Bs):

Different branches of the rule-space multiway system use different rules:

One can include causal connections:

But even removing multiedges, the full rule-space multiway causal graph is complicated:

The "branchial graph" of the rule-space multiway system, though, is fairly simple, at least after one step (though it is now really more of a "rule-space graph"):

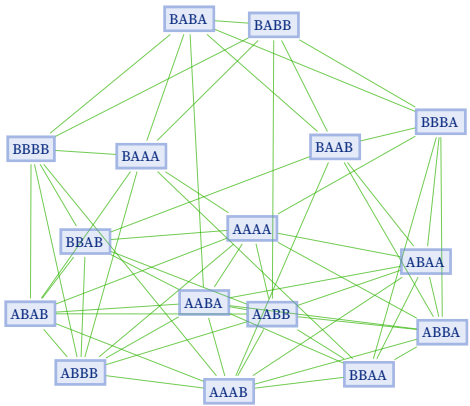

At least in this toy example, we already see something important: the rule-space multiway system is causal invariant: given branching even associated with using different rules, there is always corresponding merging—so the graph of causal relationships between updating events, even with different rules, is always the same.

Scaling up to unbounded evolutions and unbounded collections of rules involves many issues. But it seems likely that causal invariance will survive. And ultimately one may anticipate that across all possible rules it will emerge as a consequence of the Principle of Computational Equivalence [1:c12]. Because this principle implies that in the space of all possible rules, all but those with simple behavior are equivalent in their computational capabilities. And that means that across all the different possible sequences of rules that can be applied in the rule-space multiway system there is fundamental equivalence—with the result that one can expect causal invariance.

But now consider the role of the observer, who is inevitably embedded in the system, as part of the same rule-space multiway graph as everything else. Just as we did for ordinary multiway graphs above, we can imagine foliating the rule-space multiway graph, with the role of space or branchial space now being taken by rule space. And one can think of exploring rule space as effectively corresponding to sampling different possible descriptions of how the universe works, based on different underlying rules.

But if each event in the rule-space multiway graph is just a single update, based on a particular (finite) rule, there is immediately a consequence. Just like with light cones in ordinary space, or entanglement cones in branchial space, there will be a new kind of cone that defines a limit on how fast it is possible to "travel" in rule space.

For an observer, traveling in rule space involves ascribing different rules to the universe, or in effect changing one's "reference frame" for interpreting how the universe operates. (An "inertial frame" in rule space would probably correspond to continuing to use a particular rule.) But from the Principle of Computational Equivalence [1:c12] (and specifically from the idea of computation universality (e.g. [1:c11])) it is always possible to set up a computation that will translate between interpretations. But in a sense the further one goes in rule space, the more difficult the translation may become—and the more computation it will require.

But now remember that the observer is also embedded in the same system, so the fundamental rate at which it can do computation is defined by the structure of the system. And this is where what one might call the "translation cone" comes from: to go a certain "distance" in rule space, the observer must do a certain irreducible amount of computational work, which takes a certain amount of time.

The maximum rate of translation is effectively a ratio of "rule distance" to "translation effort" (measured in units of computational time). In a sense it probes something that has been difficult to quantify: just how "far apart" are different description languages, that involve different computational primitives? One can get some ideas by thinking about program size [144][145][146], or running time, but in the end new measures that take account of things, like the construction of sequences of abstractions, seem to be needed [147].

For our current discussion, however, the main point is the existence of a kind of "rule-space relativity". Depending on how an observer chooses to describe our universe, they may consider a different rule—or rather a different branch in the rule-space multiway system—to account for what they see. But if they change their "description frame", causal invariance (based on the Principle of Computational Equivalence) implies that they will still find a rule (or a branch in the rule-space multiway system) that accounts for what they see, but it will be a different one.

In the previous section, we discussed equivalences between our models and other formulations. The fact that we base our models on hypergraph rewriting (or any of its many equivalent descriptions) is in a sense like a choice of coordinate system in rule space—and there are presumably infinitely many others we could use.

But the fact that there are many different possible parametrizations does not mean that there are not definite things that can be said. It is just that there is potentially a higher level of abstraction that can be reached. And indeed, in our models, not only have we abstracted away notions of space, time, matter and measurement; now in the rule-space multiway system we are in a sense also abstracting away the very notion of abstraction itself (see also [2]).

# Notes & Further References

## Structure of Models & Methodology

The class of models studied here represent a simplification and generalization of the trivalent graph models introduced in [1:c9] and [87] (see also [148]).

The methodology of computational exploration used here has been developed particularly in [5][31][1]. Some exposition of the methodology has been given in [149].

The class of models studied here can be viewed as generalizing or being related to a great many kinds of abstract systems. One class is graph rewriting systems, also known as graph transformation systems or graph grammars (e.g. [150]). The models here are generalizations of both the double-pushout and single-pushout approaches. Note that the unlabeled graphs and hypergraphs studied here are different from the typical cases usually considered in graph rewriting systems and their applications.

Multiway systems as used here were explicitly introduced and studied in [1:p204] (see also [1:p938]). Versions of them have been invented many times, most often for strings, under names such as semi-Thue systems [151], string rewriting systems [152], term rewriting systems [65], production systems [153], associative calculi [154] and canonical systems [153][155].

## Connections to Physics Theories

An outline of applying models of a type very similar to those considered here was given in [1:c9]. Some additional exposition was given in [156][157][158]. The discussion here contains many new ideas and developments, explored in [159].

For a survey of ultimate models of physics, see [1:p1024]. The possibility of discreteness in space has been considered since antiquity [160][161][162][163]. Other approaches that have aspects potentially similar to what is discussed here include: causal dynamical triangulation [164][165][166], causal set theory [167][168][169], loop quantum gravity [170][171], pregeometry [172][173][174], quantum holography [175][176][177], quantum relativity [178], Regge calculus [179], spin networks [180][181][182][183][184], tensor networks [185], superrelativity [186], topochronology [187], topos theory [188], twistor theory [128]. Other discrete and computational approaches to fundamental physics include: [189][190][191][192][193][194][195][196].

The precise relationships among these approaches and references and the current work are not known. In some cases it is expected that conceptual motivations may be aligned; in others specific mathematical structures may have direct relevance. The latter may also be the case for such areas as conformal field theory [197], higher-order category theory [198], non-commutative geometry [199], string theory [200].

# Appendix: Implementation

## Tools Created for This Project

A variety of new functions have been added to the Wolfram Function Repository to directly implement, visualize and analyze the models defined here [201].

## Basic Direct Symbolic Transformation

The class of models defined here can be implemented very directly just using symbolic transformation rules of the kind on which the Wolfram Language [98] is based.

It is convenient to represent relations as Wolfram Language lists, such as {1,2}. One way to represent collections is to introduce a symbolic operator $\sigma$ that is defined to be flat (associative) and orderless (commutative):

*In[●]:=* SetAttributes[$\sigma$, {Flat, Orderless}]

Thus we have, for example:

*In[●]:=* $\sigma$[$\sigma$[a, b], $\sigma$[c]]

*Out[●]=* $\sigma$[a, b, c]

We can then write a rule such as

{{*x*, *y*}} → {{*x*, *y*}, {*y*, *z*}}

more explicitly as:

*In[●]:=* $\sigma$[{x_, y_}] :→ Module[{z}, $\sigma$[{x, y}, {y, z}]]

This rule can then be applied using standard Wolfram Language pattern matching:

*In[●]:=* $\sigma$[{a, b}] /. $\sigma$[{x_, y_}] :→ Module[{z}, $\sigma$[{x, y}, {y, z}]]

*Out[●]=* $\sigma$[{a, b}, {b, z$393804}]

The Module causes a globally unique new symbol to be created for the new node z every time it is used:

*In[●]:=* $\sigma$[{a, b}] /. $\sigma$[{x_, y_}] :→ Module[{z}, $\sigma$[{x, y}, {y, z}]]

*Out[●]=* $\sigma$[{a, b}, {b, z$393808}]

But in applying the rule to a collection with more than one relation, there is immediately an issue with the updating process. By default, the Wolfram Language performs only a single update in each collection:

*In[◦]:=* σ[{a, b}, {c, d}] /. σ[{x_, y_}] :→ Module[{z}, σ[{x, y}, {y, z}]]

*Out[◦]=* σ[{a, b}, {b, z$393812}, {c, d}]

As discussed in the main text, there are many possible updating orders one can use. A convenient way to get a whole "generation" of update events is to define an inert form of collection σ1 then repeatedly replace collections σ until a fixed point is reached:

*In[◦]:=* σ[{a, b}, {c, d}] //. σ[{x_, y_}] :→ Module[{z}, σ1[{x, y}, {y, z}]]

*Out[◦]=* σ[σ1[{a, b}, {b, z$393816}], σ1[{c, d}, {d, z$393817}]]

By replacing σ1 with σ at the end, one gets the result for a complete generation update:

*In[◦]:=* σ[{a, b}, {c, d}] //. σ[{x_, y_}] :→ Module[{z}, σ1[{x, y}, {y, z}]] /. σ1 → σ

*Out[◦]=* σ[{a, b}, {b, z$393821}, {c, d}, {d, z$393822}]

[NestList](#) applies this whole process repeatedly, here for 4 steps:

*In[◦]:=* evol = NestList[# //. σ[{x_, y_}] :→ Module[{z}, σ1[{x, y}, {y, z}]] /. σ1 → σ &, σ[{1, 1}], 4]

*Out[◦]=* {σ[{1, 1}], σ[{1, 1}, {1, z$393826}], σ[{1, 1}, {1, z$393826}, {1, z$393827}, {z$393826, z$393828}],
 σ[{1, 1}, {1, z$393826}, {1, z$393827}, {1, z$393829}, {z$393826, z$393828}, {z$393826, z$393830},
  {z$393827, z$393831}, {z$393828, z$393832}], σ[{1, 1}, {1, z$393826}, {1, z$393827},
  {1, z$393829}, {1, z$393833}, {z$393826, z$393828}, {z$393826, z$393830}, {z$393826, z$393834},
  {z$393827, z$393831}, {z$393827, z$393835}, {z$393828, z$393832}, {z$393828, z$393837},
  {z$393829, z$393836}, {z$393830, z$393838}, {z$393831, z$393839}, {z$393832, z$393840}]}

Replacing σ by a [Graph](#) operator, one can render the results as graphs:

*In[◦]:=* evol /. σ → (Graph[DirectedEdge @@@ {##}] &)

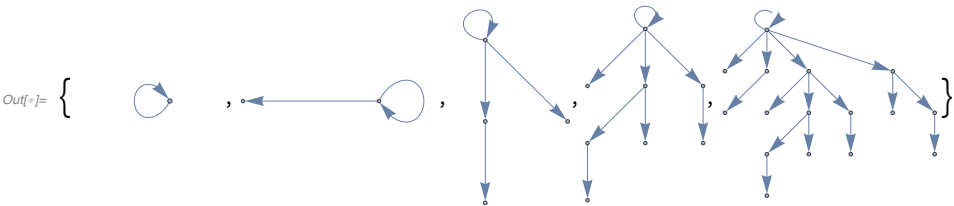

IndexGraph creates a graph in which nodes are renamed sequentially:

*In[ ]:=* evol /. σ → (IndexGraph[DirectedEdge @@@ {##}, VertexLabels → Automatic] &)

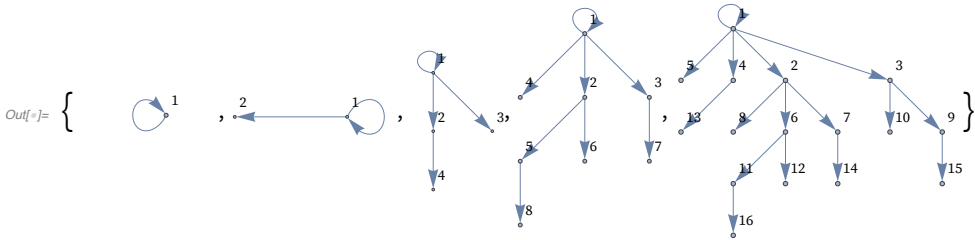

Here is the result with a different graph layout:

*In[ ]:=* evol /.
    σ → (IndexGraph[DirectedEdge @@@ {##}, GraphLayout → "SpringElectricalEmbedding"] &)

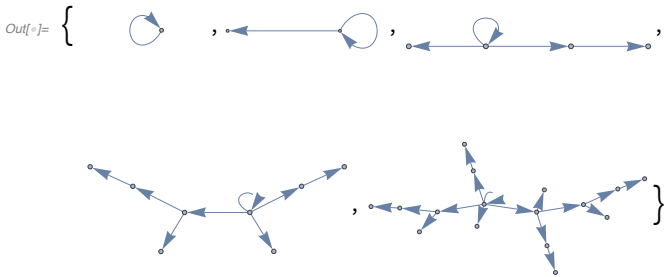

Exactly the same approach works for rules that involve multiple relations. For example, consider the rule:

{{x, y}, {x, z}} → {{x, z}, {x, w}, {y, w}, {z, w}}

This can be run for 2 steps using:

*In[ ]:=* NestList[# //.
        σ[{x_, y_}, {x_, z_}] :> Module[{w}, σ1[{x, z}, {x, w}, {y, w}, {z, w}]] /.
        σ1 → σ &, σ[{1, 1}, {1, 1}], 2]

*Out[ ]=* {σ[{1, 1}, {1, 1}], σ[{1, 1}, {1, w$393851}, {1, w$393851}, {1, w$393851}],
    σ[{1, w$393851}, {1, w$393851}, {1, w$393852}, {1, w$393852}, {1, w$393853},
      {w$393851, w$393852}, {w$393851, w$393853}, {w$393851, w$393853}]}

Here is the result after 10 steps, rendered as a graph:

*In[ ]:=* Nest[# //.
  σ[{x_, y_}, {x_, z_}] :⧴ Module[{w}, σ1[{x, z}, {x, w}, {y, w}, {z, w}]] /.
  σ1 → σ &, σ[{1, 1}, {1, 1}], 10] /. σ → (Graph[DirectedEdge @@@ {##}] &)

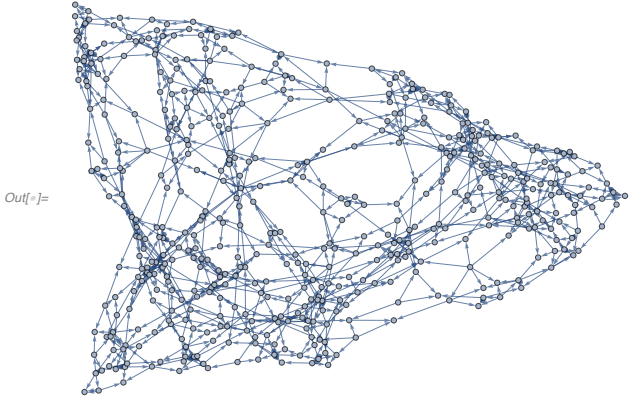

*Out[ ]=*

## Alternative Syntactic Representation

As an alternative to introducing an explicit head such as σ, one can use a system-defined matchfix operator such as AngleBracket (entered as :<:, :>:) that does not have a built-in meaning. With the definition

*In[ ]:=* SetAttributes[AngleBracket, {Flat, Orderless}]

one immediately has for example

*In[ ]:=* ⟨a, ⟨b, c⟩⟩

*Out[ ]=* ⟨a, b, c⟩

and one can set up rules such as

*In[ ]:=* ⟨{x_, y_}, {x_, z_}⟩ :⧴ Module[{w}, ⟨{x, z}, {x, w}, {y, w}, {z, w}⟩]

## Pattern Sequences

Instead of having an explicit "collection operator" that is defined to be flat and orderless, one can just use lists to represent collections, but then apply rules that are defined using OrderlessPatternSequence:

*In[ ]:=* {{0, 0}, {0, 0}, {0, 0}} /. {OrderlessPatternSequence[{x_, y_}, {x_, z_}, rest___]} :⧴
    Module[{w}, {{x, z}, {x, w}, {y, w}, {z, w}, rest}]

*Out[ ]=* {{0, 0}, {0, w$37227}, {0, w$37227}, {0, w$37227}, {0, 0}}

Note that even though the pattern appears twice, /. applies the rule only once:

*In[ ]:=* {{0, 0}, {0, 0}, {0, 0}, {0, 0}} /. {OrderlessPatternSequence[{x_, y_}, {x_, z_}, rest___]} :>
    Module[{w}, {{x, z}, {x, w}, {y, w}, {z, w}, rest}]

*Out[ ]=* {{0, 0}, {0, w$48054}, {0, w$48054}, {0, w$48054}, {0, 0}, {0, 0}}

## Subset Replacement

Yet another alternative is to use the function SubsetReplace (built into the Wolfram Language as of Version 12.1). SubsetReplace replaces subsets of elements in a list, regardless of where they occur:

*In[ ]:=* SubsetReplace[{a, b, b, a, c, a, d, b}, {a, b} → x]

*Out[ ]=* {x, x, c, x, d}

Unlike ReplaceAll (/.) it keeps scanning for possible replacements even after it has done one:

*In[ ]:=* SubsetReplace[{a, a, a, a, a}, {a, a} → x]

*Out[ ]=* {x, x, a}

One can find out what replacements SubsetReplace would perform using SubsetCases:

*In[ ]:=* SubsetCases[{a, b, c, d, e}, {_, _}]

*Out[ ]=* {{a, b}, {c, d}}

This uses SubsetReplace to apply a rule for one of our models; note that the rule is applied twice to this state (Splice is used to make the sequence of lists be spliced into the collection):

*In[ ]:=* SubsetReplace[{{0, 0}, {0, 0}, {0, 0}, {0, 0}},
    {{x_, y_}, {x_, z_}} :> Splice[Module[{w}, {{x, z}, {x, w}, {y, w}, {z, w}}]]]

*Out[ ]=* {{0, 0}, {0, w$55383}, {0, w$55383}, {0, w$55383}, {0, 0}, {0, w$55384}, {0, w$55384}, {0, w$55384}}

This gives the result of 10 applications of SubsetReplace:

*In[ ]:=* Nest[SubsetReplace[{{x_, y_}, {x_, z_}} :> Splice[Module[{w}, {{x, z}, {x, w}, {y, w}, {z, w}}]]],
    {{1, 2}, {1, 3}}, 10] // Short

*Out[ ]=* {{1, w$55543}, {1, w$55637}, {w$55490, w$55637}, ≪705≫, {w$55401, w$55452}, {2, w$55394}}

This turns each list in the collection into a directed edge, and renders the result as a graph:

*In[◦]:=* Graph[DirectedEdge@@@ %]

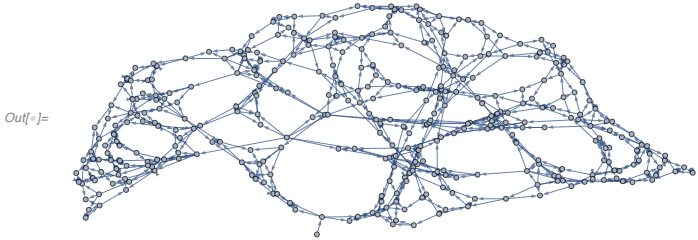

*Out[◦]=*

IndexGraph can then for example be used to relabel all elements in the graph to be sequential integers.

Note that SubsetReplace does not typically apply rules in exactly our "standard updating order".

## Parallelization

Our models do not intrinsically define updating order (see section 6), and thus allow for asynchronous implementation with immediate parallelization, subject only to the local partial ordering defined by the graph of causal relationships (or, equivalently, of data flows). However, as soon as a particular sequence of foliations—or a particular updating order—is defined, its implementation may require global coordination across the system.

# Appendix: Graph Types

A visual summary of the relationships between graph types is given in [202].

## Single-Evolution-History Graphs

*Graphs obtained from particular evolution histories, with particular sequences of updating events. For rules with causal invariance, the ultimate causal graph is independent of the sequence of updating events.*

## Spatial Graph

Hypergraph whose nodes and hyperedges represent the elements and relations in our models. Update events locally rewrite this hypergraph. In the large-scale limit, the hypergraph can show features of continuous space. The hypergraph potentially represents the "instantaneous" configuration of the universe on a spacelike hypersurface. Graph distances in the hypergraph potentially approximate distances in physical space.

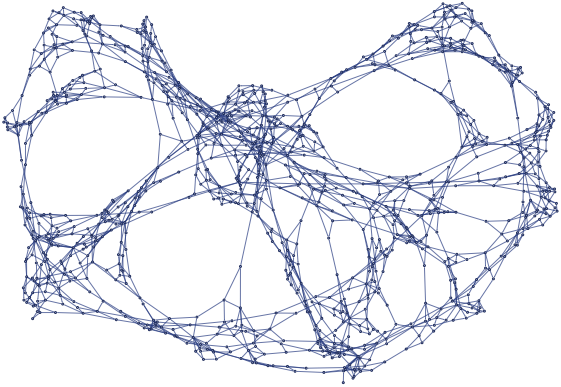

## Causal Graph ("Spacetime Causal Graph")

Graph with nodes representing updating events and edges representing their causal relationships. In causal invariant systems, the same ultimate causal graph is obtained regardless of the particular sequence of updating events. The causal graph potentially represents the causal history of the universe. Causal foliations correspond to sequences of spacelike hypersurfaces. The effect of an update event is represented by a causal cone, which potentially corresponds to a physical light cone. The translation from time units in the causal graph to lengths in the spatial graph is potentially given by the speed of light $c$.

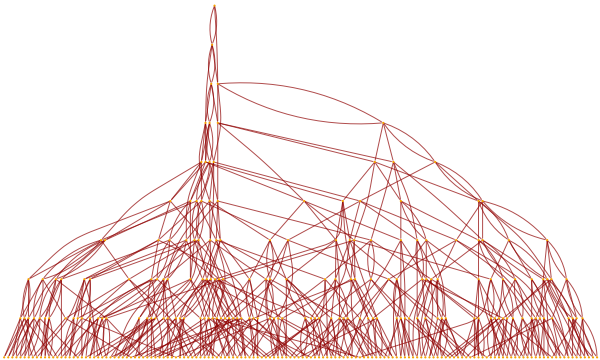

## Multiway-Evolution-Related Graphs

*Graphs obtained from all possible evolution histories, following every possible sequence of updating events. For rules with causal invariance, different paths in the multiway system lead to the same causal graph.*

# Multiway States Graph (Multiway Graph)

Graph representing all possible branches of evolution for the system. Each node represents a possible complete state of the system at a particular step. Each connection corresponds to the evolution of one state to another as a result of an updating event. The multiway graph potentially represents all possible paths of evolution in quantum mechanics. In a causal invariant system, every branching in the multiway system must ultimately reconverge.

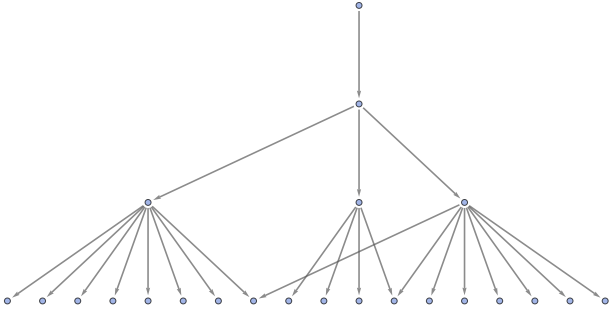

# Multiway States+Causal Graph

Graph representing both all possible branches of evolution for states, and all causal relationships between updating events. Each node representing a state connects to other states via nodes representing updating events. The updating events are connected to indicate their causal relationships. The multiway states+causal graph in effect gives complete, causally annotated information on the multiway evolution.

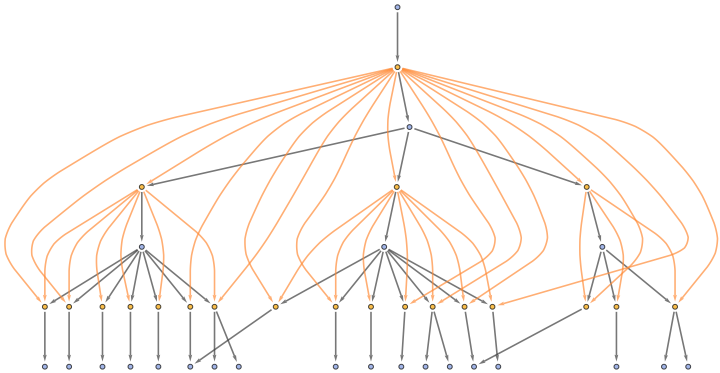

## Multiway Causal Graph

Graph representing causal connections among all possible updating events that can occur in all possible paths of evolution for the system. Each node represents a possible updating event in the system. Each edge represents the causal relationship between two possible updating events. In a causal invariant system, the part of the multiway causal graph corresponding to a particular path of evolution has the same structure for all possible paths of evolution. The multiway causal graph provides the ultimate description of potentially observable behavior of our models. Its edges represent both spacelike and branchlike relationships, and can potentially represent causal relations both in spacetime and through quantum entanglement.

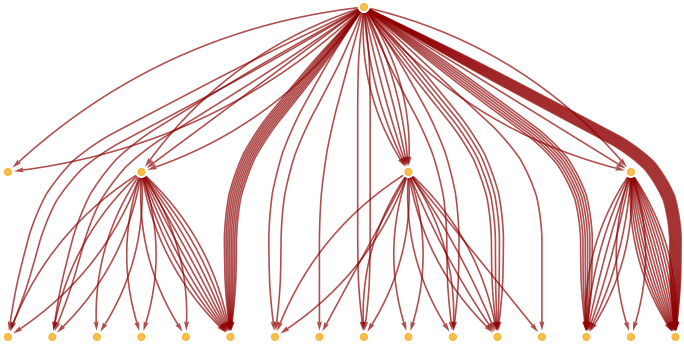

## Branchial Graph

Graph representing the common ancestry of states in the multiway system. Each node represents a state of the system, and two nodes are joined if they are obtained on different branches of evolution from the same state. To define a branchial graph requires specifying a foliation of the multiway graph. The branchial graph potentially represents entanglement in the "branchial space" of quantum states.

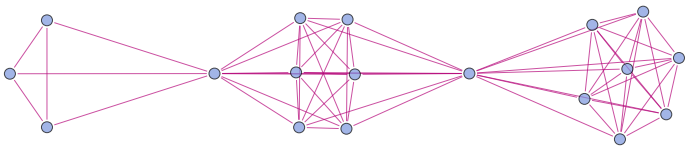

# Acknowledgements

I have been developing the ideas here for many years [203]. I worked particularly actively on them in 1995–1998, 2001 and 2004–2005 [148][1]. But they might have languished forever had it not been for Jonathan Gorard and Max Piskunov, who encouraged me to actively work on them again, and who over the past several months have explored them with me, providing extensive help, input and new ideas. For important additional recent help I thank Jeremy Davis, Sushma Kini and Ed Pegg, as well as Roger Dooley, Jesse Friedman, Andrea Gerlach, Charles Pooh, Chris Perardi, Toni Schindler and Jessica Wong. For recent input I thank Elise Cawley, Roger Germundsson, Chip Hurst, Rob Knapp, José Martin-Garcia, Nigel Goldenfeld, Isabella Retter, Oliver Ruebenkoenig, Matthew Szudzik, Michael Trott, Catherine Wolfram and Christopher Wolfram. For important help and input in earlier years, I thank David Hillman, Todd Rowland, Matthew Szudzik and Oyvind Tafjord. I have discussed the background to these ideas for a long time, with a great many people, including: Jan Ambjørn, John Baez, Tommaso Bolognesi, Greg Chaitin, David Deutsch, Richard Feynman, David Finkelstein, Ed Fredkin, Gerard 't Hooft, John Milnor, John Moussouris, Roger Penrose, David Reiss, Rudy Rucker, Dana Scott, Bill Thurston, Hector Zenil, as well as many others, notably including students at our Wolfram Summer Schools over the past 17 years. My explorations would never have been possible without the Wolfram Language, and I thank everyone at Wolfram Research for their consistent dedication to its development over the past 33 years, as well as our users for their support.

# Tools, Data & Source Materials

Extensive tools, data and source material related to this document and the project it describes are available at wolframphysicsproject.org.

This document is available in complete computable form as a Wolfram Notebook, including Wolfram Language input for all results shown. The notebook can be run directly in the Wolfram Cloud or downloaded for local use.

Specialized Wolfram Language functions developed for this project are available in the Wolfram Function Repository [204] for immediate use in the Wolfram Language. A tutorial of their use is given in [205].

The Registry of Notable Universes [8] contains results on specific examples of our models, including all those explicitly used in this document.

An archive of approximately 1000 working notebooks associated with this project from 1994 to the present is available at wolframphysicsproject.org. In addition, there is an archive of approximately 500 hours of recorded streams of working sessions (starting in fall 2019) associated with this project.